\documentclass[a4paper,12pt,titlepage]{article}
\usepackage{amssymb}

\title{                      HERA Collider Physics
}

\author{                   Halina Abramowicz
\thanks 
          {Raymond and Beverly Sackler Faculty of Exact Sciences,
           School of Physics, Tel Aviv University, Tel Aviv, Israel}
\and
                           Allen Caldwell
\thanks
              {Columbia University, Department of Physics, 
                        New York, New York, USA}
}
\date{\today\\ 
\vspace{5cm}
{\it submitted to Reviews of Modern Physics}
}

\usepackage{graphicx}

\newcommand\epsfigure[4][width=\hsize]{%
\begin{figure}%
  \begin{center}%
     \IfFileExists{#2.eps.bb}%
       {\includegraphics[draft,#1]{#2}}%
       {\includegraphics[#1]{#2}}%
  \end{center}%
\caption{\it #3}\label{fig:#4}%
\end{figure}%
}

\newcommand\tablecaption[1]{%
  \belowcaptionskip=\abovecaptionskip
  \abovecaptionskip=0pt
  \caption{\it #1}%
}

\def \q2{\ensuremath{Q^2}}
\def \r0{\ensuremath{\rho^0}}

\def\ETJ{\ensuremath{E_T^{\mathrm{jet}}}}

\def \as{\ensuremath{\alpha_S}}
\newcommand{\ycut}{\ensuremath{y_{\mathrm{cut{`}}}}}

\newcommand{\rpvio}{{\ensuremath{\textstyle\not\hspace{-.55ex}{R}_P}}}

\newcommand{\tauq}{\ensuremath{\tau q}}

\newcommand{\sTop}{{\tilde t}}
\newcommand{\suup}{{\tilde u}}
\newcommand{\sdown}{{\tilde d}}
\newcommand{\squark}{\ensuremath{{\tilde q}}}
\newcommand{\slepton}{\ensuremath{{\tilde l}}}

\newcommand{\MLQsq}{\ensuremath{M_{\scriptscriptstyle LQ}^2}}

\newcommand{\lepton}{\ell}
\newcommand{\ubar}{\ensuremath{\bar{u}}}
\newcommand{\dbar}{\ensuremath{\bar{d}}}
\newcommand{\ebar}{\ensuremath{\bar{e}}}

\newcommand{\nubar}{\ensuremath{\bar\nu}}

\newcommand{\Kpi}{\ensuremath{K\rightarrow \pi \nubar\nu }}

\newcommand{\BlnuX}{\ensuremath{B\rightarrow \lepton\nu X}}

\newcommand{\taupie}{\ensuremath{\tau \rightarrow \pi e}}
\newcommand{\tauKe}{\ensuremath{\tau \rightarrow Ke}}
\newcommand{\BtaueX}{\ensuremath{B\rightarrow \tau \ebar X}}
\newcommand{\tauegam}{\ensuremath{\tau \rightarrow e\gamma }}
\newcommand{\Kpinunu}{\ensuremath{K\rightarrow \pi \nubar\nu }}

\def\prop{\sim}

\def\as{\alpha_s}
\newcommand {\pom} {I\!\!P}
\newcommand {\pomsub} {{\scriptscriptstyle \pom}}
\newcommand {\reg} {I\!\!R}
\newcommand {\regsub} {{\scriptscriptstyle \reg}}

\newcommand {\xpom} {x_{\pomsub}}
\newcommand {\apom} {\alpha_{\pomsub}}
\newcommand {\areg} {\alpha_{\regsub}}
\newcommand {\aprime} {\alpha^\prime_\pomsub}
\newcommand {\deta} {\Delta\eta}
\newcommand {\aveapom} {\bar{\alpha}_\pomsub}

\newcommand{\yjb}{y_{\scriptscriptstyle JB}}

\newcommand{\gp}{\gamma p}
\newcommand{\gvp}{\gamma^* p}
\newcommand{\etamax}{\eta_{\mathrm{max}}}
\newcommand\units{\,\mathrm}
\newcommand{\gev}{\units{GeV}}
\newcommand{\gevtwo}{\units{GeV^2}}
\newcommand{\gevmtwo}{\units{GeV^{-2}}}
\newcommand{\mev}{\units{MeV}}

\newcommand{\nbi}{\units{nb^{-1}}}
\newcommand{\pbi}{\units{pb^{-1}}}
\newcommand{\ftwod}{F_2^D}
\newcommand{\ftwodthree}{F_2^{D(3)}}
\newcommand{\ftwodfour}{F_2^{D(4)}}
\newcommand{\ftwodthreearg}{F_2^{D(3)}\,(\beta,\,Q^2,\,\xpom)}
\newcommand{\ftwodfourarg}{F_2^{D(4)}\,(\beta,\,Q^2,\,\xpom,\,t)}
\newcommand{\ftwopom}{F_2^{\pomsub}}
\newcommand{\ftwopomarg}{\ftwopom\,(\beta,\,Q^2)}
\newcommand{\pomflux}{f_{\pomsub/p}}
\newcommand{\pomfluxarg}{\pomflux(\xpom)}
\def\cts{\cos\theta^{\ast}}
\def\xgobs{x_\gamma^{\scriptscriptstyle \mathrm{OBS}}}
\def\xpobs{x_p^{\scriptscriptstyle \mathrm{OBS}}}
\def\xg{x_\gamma}
\def\ETJ{E_T^{\mathrm{jet}}}
\def\ETAJ{\eta^{\mathrm{jet}}}

\usepackage[dcucite]{harvard}

\let\fragilecite\cite
\def\cite{\protect\fragilecite}
\let\fragileciteasnoun\citeasnoun
\def\citeasnoun{\protect\fragileciteasnoun}

\begin{document}

\maketitle

\begin{abstract}
  HERA, the first electron-proton collider, has been delivering
  luminosity since 1992.  It is the natural extension of an impressive
  series of fixed-target lepton-nucleon scattering experiments.  The
  increase of a factor ten in center-of-mass energy over that
  available for fixed-target experiments has allowed the discovery of
  several important results, such as the large number of slow partons
  in the proton, and the sizeable diffractive cross section at large
  $Q^2$.  Recent data point to a possible deviation from Standard
  Model expectations at very high $Q^2$, highlighting the physics
  potential of HERA for new effects.  The HERA program is currently in
  a transition period. The first six years of data taking have
  primarily elucidated the structure of the proton, allowed detailed
  QCD studies and had a strong impact on the understanding of QCD
  dynamics.  The coming years will bring the era of electroweak
  studies and high $Q^2$ measurements.  This is therefore an
  appropriate juncture at which to review HERA results.
\end{abstract}


\tableofcontents

\cleardoublepage


\section{Introduction}
\label{sec:intro}

The ultimate goal of research in high energy physics is to
understand and describe the structure of matter and its interactions.
The fundamental constituents of matter as we know them today, leptons
and quarks, are fermions arranged into generations characterized by
lepton numbers and quark flavors, respectively.  Leptons are free
particles that can be detected. Quarks, on the other hand, only exist in
bound states - hadrons. The existence of quarks can be inferred from
experimental measurements of the properties of particle interactions
and hadron production.

There are four known forces governing our world: gravitational, weak,
electromagnetic and strong.  Only the last three play a major role in
the microscopic world.  In the modern language of physics,
interactions are due to exchange of field quanta which determine the
properties of these interactions.  These field quanta correspond to
particles whose properties can be measured. All the known carriers of
forces are bosons: three vector bosons mediating the weak interactions
($W^{\pm}$, $Z^0$), the photon $\gamma$ mediating the electromagnetic
interactions and eight gluons $g$ mediating the strong interactions.
Each of them carries specific quantum numbers, as do the fundamental
constituents of matter.

There is no theory limiting the number of generations. The only
theoretical condition is that the number of lepton generations be
equal to that of the quarks. At present three generations of leptons
and quarks are known.  The leptons are the electron -- $e$, the muon
-- $\mu$ and the tau -- $\tau$, each one accompanied by a
corresponding neutrino, $\nu_e$, $\nu_{\mu}$ and $\nu_{\tau}$.  There
are six known quarks, the up -- $u$, down -- $d$, strange -- $s$,
charm -- $c$, bottom -- $b$ and top -- $t$ quarks.  Neutrinos, which
carry no electric charge, interact only weakly.  Charged leptons take
part in weak and electromagnetic interactions. Only quarks take part
in all the known interactions of the micro-world.

The theoretical framework which allows us to describe formally this
simple picture is based on gauge theories. The weak and
electromagnetic interactions are unified within the so-called
electroweak theory formulated by Glashow, Salam and
Weinberg~\cite{Glashow,Salam,Weinberg}. The strong interactions are
embedded in the framework of Quantum Chromodynamics~\cite{QCD}.  The
combination of the two constitutes what is generally known as the
Standard Model of particles and interactions. The experimental
evidence which led to this simple and elegant picture has been
provided by a multitude of experiments involving high energy
interactions of hadrons with hadrons, leptons with hadrons, and
leptons with leptons.

The description of electroweak interactions is based on the SU(2)
group of weak isospin and U(1) group of weak hyper-charge. This
symmetry is spontaneously broken at $\sim 100$~GeV by introducing in
the theory scalar mesons called Higgs particles. In the resulting
theory we find two charged and one neutral massive vector bosons --
the $W^{\pm}$ and the $Z^0$, which mediate weak interactions, and one
massless neutral vector boson -- the photon. While the existence of
the weak charged currents
was known (since the explanation of the $\beta$ decay of atomic nuclei
by Fermi in 1932), this theory predicted the existence of weak neutral
currents, which were subsequently discovered~\cite{gargamelle}.

The experimental and theoretical progress achieved in the electroweak
sector is tremendous. The $W^{\pm}$ and the $Z^0$ were discovered at the
CERN proton-antiproton collider~\cite{UA1,UA2} and with the advent of
two high energy electron positron colliders, LEP and SLC, the
electroweak parameters of the Standard Model have been determined to a
high precision~\cite{passarino,ee-review}. Suffice to say that the
experiment and the theory agree with each other at the level of
${\mathcal{O}}(10^{-3})$~\cite{pokorski}. The Higgs boson remains
the only missing link.  Global fits constrain
the Higgs mass at $95\%$ confidence level to be $m_H <
500$~GeV~\cite{ee-review}, with the minimum $\chi^2$ corresponding to
a value $m_H \simeq 115$~GeV.

The interactions of quarks and gluons are described by Quantum
Chromodynamics (QCD), a non-abelian gauge theory based on the SU(3)
color symmetry group.  Color constitutes the equivalent of the
electric charge in electromagnetic interactions. The quarks, each in
three colors, interact by exchange of electrically neutral vector
bosons - the gluons, which form a color charge octet. The gluons are
not color neutral and thus they themselves interact strongly.  A
consequence of this property is asymptotic freedom which states that
the interaction strength of two colored objects decreases the shorter
the distance between them.  The effective strong coupling constant
$\as$ depends on the scale at which the QCD process occurs.  The
solution of the renormalization group equation in leading order leads
to
\begin{equation}
\as(Q^2)=\frac{4\pi}{\beta_0 \ln (Q^2/\Lambda^2)} \ ,
\label{eq-int:alphas}
\end{equation}
where $Q^2$ denotes the scale at which $\as$ is probed and
$\Lambda$ is a QCD cut-off parameter. The parameter $\beta_0$ depends
on the number of quark flavors in the theory, $N_f$,
\begin{equation}
\beta_0 = 11 - \frac{2}{3}N_f \ .
\end{equation}
Since the known number of flavors is six, $\beta_0 > 0$, and the
coupling constant becomes smaller the larger the scale $Q^2$.  The
property of asymptotic freedom has been proven rigorously and allows
to make predictions for the properties of strong interactions in the
perturbative QCD regime, in which $\as$ is small.

Another property of QCD, which has not been proven rigorously, is
confinement, which keeps quarks bound into colorless hadrons and
prevents the observations of free quarks.  In QCD, the color degree of
freedom and confinement explain why the observed hadrons are made
either of $q \bar{q}$ or of $qqq$ ($\bar{q} \bar{q} \bar{q}$) states.
These combinations ensure that hadrons are colorless and have integer
electrical charge. It also explains why baryons made out of three
quark states are fermions while mesons made out of $q \bar{q}$ states
are bosons~\cite{gell-mann}. The model in which hadrons are viewed as
composed of free quarks $q$ or antiquarks $\bar{q}$ is called the
Quark Parton Model (QPM). In the presence of QCD, the naive QPM
picture of hadrons has to be altered to take into account the
radiation and absorption of gluons by quarks as well as the creation
of $q \bar{q}$ pairs by gluons. Thus in effect hadrons consist of
various partons, quarks and gluons. We know today that about 50\% of
the proton momentum is carried by gluons.

QCD has two properties which make it much more difficult to work with
theoretically than electroweak theory.  The first poperty is that the
coupling constant is large, making the use of perturbation theory
difficult.  The strong coupling constant depends on the scale, as
described above, and cross sections can only be calculated for
scatterings with a hard scale, for which $\as$ is small enough.  The
second property is the non-Abelian nature of the interaction.  Gluons
can interact with other gluons, leading to confinement of color.

The distribution of partons bound in hadrons cannot be calculated from
first principles. The calculations would have to be performed in a
regime of QCD where the perturbative approach breaks down.  However
the QCD factorization theorem~\cite{collins-soper} states that for
hard scattering reactions the cross section can be decomposed into the
flux of universal incoming partons and the cross section for the hard
subprocess between partons. The measurement of parton distributions in
hadrons becomes an essential element in testing the validity of QCD.

QCD has been tested in depth in the perturbative regime and describes
the measurements very well. However, because the observables are based
on hadron states rather than the partonic states to which perturbative
calculations apply the precision level which is achieved in testing
QCD is lower than in case of the electroweak interactions. In
addition, up to now there is no understanding 
within QCD of scattering processes in 
the non-perturbative regime, the so called soft regime, although this
is the regime which dominates the cross section for strong
interactions.

The soft hadron-hadron interactions are well described by the Regge
phenomenology~\cite{regge} in which the interaction is viewed as due
to exchanges of collective states called Regge poles. The Regge poles
can be classified into different families according to their quantum
numbers.  Among all possible families of Regge poles there is a
special one, with the quantum numbers of the vacuum, called the
pomeron ($\pom$) trajectory. The pomeron trajectory is believed to
determine the high energy properties of hadron-hadron interaction
cross sections. The link between Regge theory and QCD has not yet been
established.

QCD remains a largely unsolved theory and the justification for the
use of perturbative QCD rests to a large extent directly on
experiment. Every experiment in strong interactions involves a large
range of scales, over which the value of the strong coupling constant
changes radically. This, together with certain arbitrariness in
truncating the perturbative expansion, leads to uncertainties which
can only be successfully resolved if the gap between the perturbative
and non-perturbative approach is bridged.

The advantage of lepton-nucleon collisions in studying the structure
of matter lies in that leptons are point-like objects and their
interactions are well understood. The point-like, partonic
substructure of the nucleon was first firmly established in the
pioneering SLAC experiment~\cite{taylor1,taylor2} in which the
spectrum of electrons scattered off a nucleon target was measured.
This experiment was very similar in its essence to the famous
Rutherford experiment which established the structure of
atoms. In a scattering in which an electron of initial four momentum
$k$ emerges with four momentum $k'$, the exchanged virtual photon has a
mass $q^2=(k-k')^2=-Q^2$ and correspondingly a Compton wavelength of
$\hbar/\sqrt{Q^2}$. Thus for different values of $Q^2$ the interaction
is sensitive to structures at different scales.

The picture that has emerged from measurements of lepton-nucleon
scattering, in particular from deep inelastic scattering (DIS),
utilizing electron, muon and neutrino beams, confirmed the universality
of parton distributions as well as the validity of perturbative QCD
which predicts a change in the observed parton distributions as a
function of the scale at which they are probed (for a review
see~\citeasnoun{disreview}).

Charged lepton-nucleon interactions also are an ideal laboratory to
study photon-nucleon interactions.  When the lepton scattering angle
is very small the exchanged photon is almost real and the leptons can
be thought of as a source of photons that subsequently interact with
the nucleon. At high photon energies we can then study the properties
of photon interactions with hadronic matter. A simple guess would be
that photons, as gauge particles mediating electromagnetic
interactions, would interact only electromagnetically. However, given
the Heisenberg uncertainty principle, the photon can fluctuate into a
quark-antiquark pair, which can then develop further structure.  In
the presence of a hadronic target, the interaction can then be viewed
as hadron-hadron scattering.

In many respects the HERA accelerator, in which 27.5 GeV electrons or
positrons collide with 820 GeV protons, offers a unique possibility to
test both the static and dynamic properties of the Standard Model.
The center-of-mass energy of the electron-proton collisions is 300~GeV
and is more than a factor 10 larger than any previous fixed target
experiment. The available $Q^2$ range extends from $Q^2\simeq 0$ to
$Q^2 \simeq 9 \cdot 10^4$~GeV$^2$ and allows to probe structures down
to $10^{-16}$~cm, while partons can be probed down to very small
fractions of the proton momentum, $x = Q^2/(Q^2+W^2) \sim 10^{-5}$.
Two general purpose experiments, H1 and ZEUS, are dedicated to the
study of the HERA physics.

The first 0.02 pb$^{-1}$ of luminosity delivered by the HERA
accelerator has brought striking results which have opened a whole new
interest in QCD. In the deep inelastic regime, $Q^2 \ge 10$~GeV$^2$,
in which the partons can be easily resolved, it was found that the
number of slow partons increases steadily with decreasing
$x$~\cite{ref:H1F2_92,ref:ZEUSF2_92}. This observation is in line with
asymptotic expectations of perturbative QCD.  Further studies have
shown that an increase is observed even at $Q^2$ as low as 1
GeV~\cite{ref:ZEUSF2_BPC}, where it is not even clear that the parton
language is applicable. It was also found that a large fraction of DIS
events had a final state typical of diffractive scattering always
believed to be a soft phenomenon~\cite{ZEUSdiff1,H1diff1}.  This
opened the interesting possibility to explore the partonic nature of
the pomeron and provide some link between the Regge theory and QCD.

Before the advent of HERA the complicated nature of the photon was
inferred from low energy photon hadron interactions where the photon
behaved essentially as a vector meson, a bound state of a $q
\bar{q}$~\cite{ref:Bauer}. On the other hand in $\gamma \gamma$
interactions derived from $e^+e^-$ interactions, the photon behaved as
if it consisted mainly of $q \bar{q}$ states which could be calculated
in perturbative QCD~\cite{witten}.  The missing link was 
established at TRISTAN~\cite{AMY92} and at
HERA~\cite{H1ph1,ZEUSph1}. The cross section for photon
induced jet production could not be explained without a substantial
contribution from a photon consisting of partons. The presence of the
photon remnant after the hard collision was observed for the first
time with the HERA detectors~\cite{H1ph1,ZEUSph1}.

The investigation of hadronic final states in DIS and hard
photoproduction scattering demonstrated that we still do not have a
complete understanding of the transition from partonic states to the
hadronic ones
in the region affected by proton fragments~\cite{H1-hfsdis,ZEUS-hfsdis}.
This is unlike the description of the hadronic final states produced
in $e^+e^-$ interactions where models tuned to describe low energy
interactions hold very well in the increased phase
space~\cite{SLD-energy-flow}.  Here again the message is that simple
quark parton model approaches corrected for perturbative QCD effects
are not adequate when one of the initial particles has by itself a
complicated nature.

At very large $Q^2 >1000 \gevtwo$, where the contribution of $W$ and
$Z^0$ exchanges become important, the measured cross sections are
generally well described by the Standard Model. However, an excess of
events was reported in the region of large $x>0.4$ and very high
$Q^2>15000 \gevtwo$ by both the H1 and the ZEUS
experiment~\cite{ref:H1_highQ2,ref:ZEUS_highQ2}. The origin of these is
not yet understood, but the presence of these events underscores the
discovery potential at HERA for new physics beyond the Standard Model.

In the following chapters, we review HERA physics in some depth.
More detailed discussion of some aspects of HERA physics can be found
in recent reviews~\cite{devenish_rev,kuhlen,erdmann,ref:Crittenden} as well as
in selected lecture notes~\cite{wolf_rev,levy_rev}.

\section{Lepton-nucleon scattering} 
\label{sec:lepton-nucleon} 

The HERA physics program to date has primarily focused on testing our
understanding of the strong force.  Measurements have been performed
in kinematic regions where perturbative QCD calculations should be
accurate, as well as in regions where no hard scale is present and
non-perturbative processes dominate.  In the following sections, we
introduce the language of structure functions and the effects expected
from pQCD evolution.  This background serves as the base for
interpreting many of the physics results described in later sections.

\epsfigure[width=0.8\hsize]{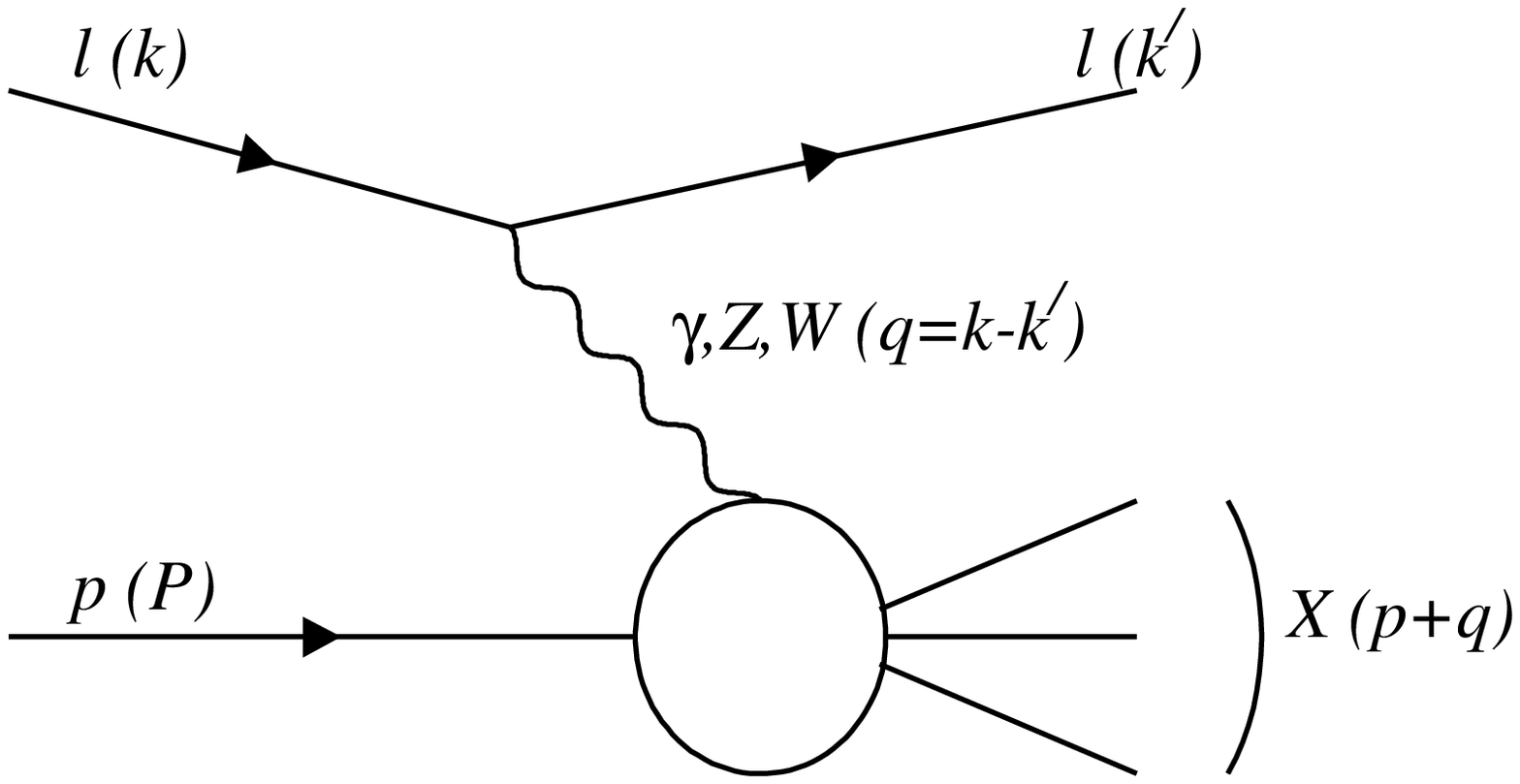}{
  Schematic diagram describing deep inelastic lepton nucleon
  scattering. The four vectors of the particles, or particles systems, 
  are given in parantheses.}{disdiagram}

In the most general case the lepton-nucleon interaction proceeds via
the exchange of a virtual vector boson as depicted in
Fig.~\ref{fig:disdiagram}.  Since the lepton number has to be
conserved, we expect the presence of a scattered lepton in the final
state, while the nucleon fragments into a hadronic final state $X$,
\begin{equation}
l N \to l' X \ .
\end{equation}
Assuming that $k$, $k'$, $P$, $P^\prime$ are the four vectors of the
initial and final lepton, of the incoming nucleon and of the outgoing
hadronic system, respectively (see Fig.~\ref{fig:disdiagram}), the
usual variables describing the lepton nucleon scattering are
\begin{eqnarray} 
\label{eq:disvariables}
Q^2 &=& -q^2 = -(k -k')^2 \ , \nonumber \\ 
s &=& (k + P)^2 \ , \nonumber \\ 
W^2 &=& (q + P)^2 =p_\prime^2 \ , \nonumber \\ 
x &=& \frac{Q^2}{2 P \cdot q} \ , \\ 
y &=&\frac{q \cdot P}{k \cdot P} \ , \nonumber \\
\nu &=& \frac{q \cdot P}{m_N} \ . \nonumber
\end{eqnarray}
The variables $s$ and $W^2$ are the center-of-mass energy squared of
the lepton-nucleon and intermediate boson-nucleon systems,
respectively.  The square of the four momentum transfer (the mass
squared of the virtual boson), $q^2 < 0$, determines the hardness of
the interaction, or in other words, the resolving power of the
interaction. The exchanged boson plays the role of a ``partonometer''
with a resolution $\Delta b$,
\begin{equation}
\Delta b \sim \frac{\hbar c}{\sqrt{Q^2}} = \frac{0.197}{\sqrt{Q^2}} 
\,\mbox{\rm GeV fm} \ ,
\end{equation}
where for convenience we introduce the positive variable $Q^2=-q^2$.
The meaning of $\nu$ and $y$ is best understood in the rest frame of
the target, in which $\nu$ is just the energy of the intermediate
boson and $y$ measures the inelasticity of the interaction and its
distribution reflects the spin structure of the interaction. The
variable $x$ is the dimensionless variable introduced by Bjorken.

\epsfigure[width=0.8\hsize]{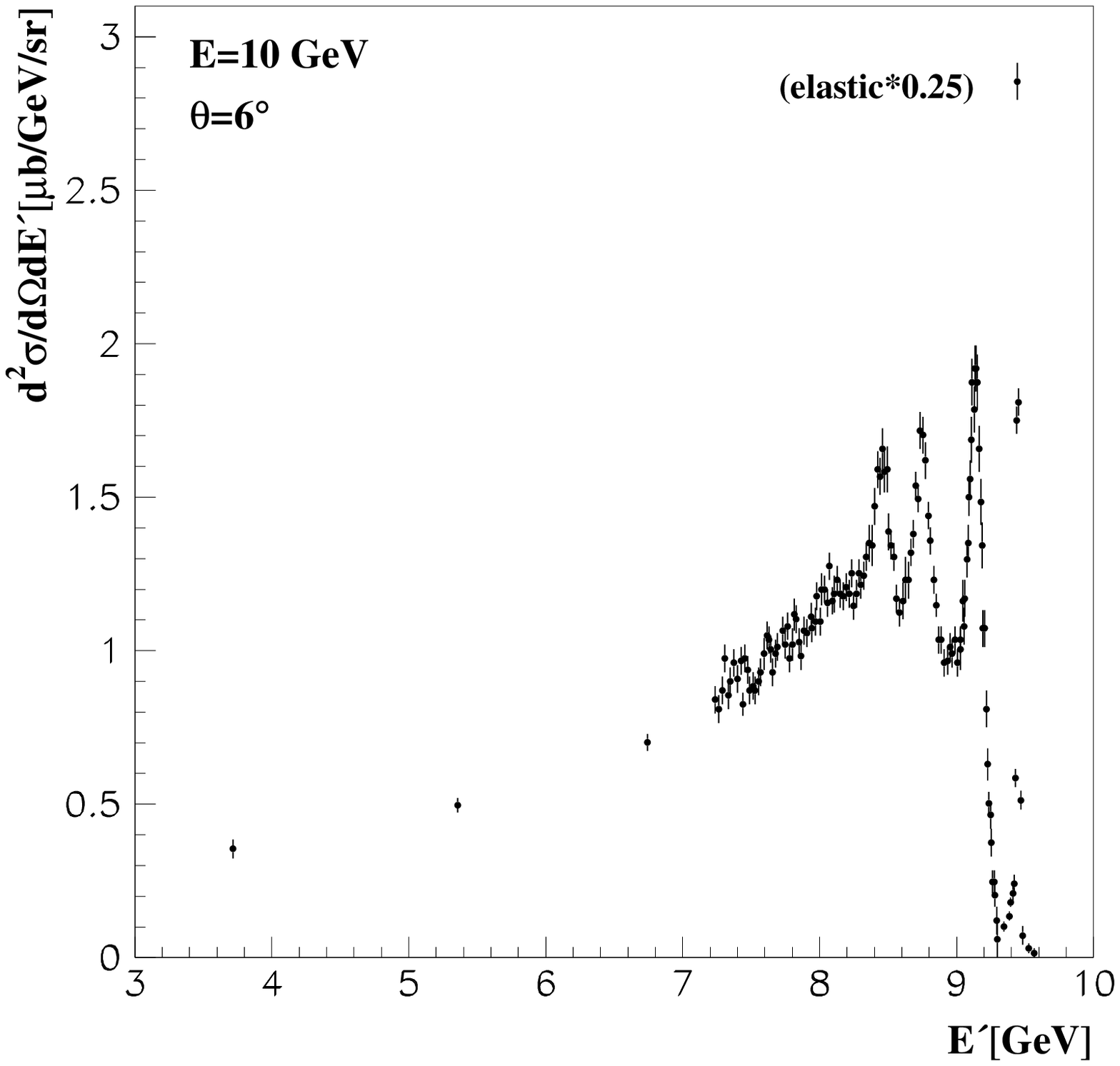}{ An example of the electron
  spectrum measured in electron proton scattering. From right to left,
  the direction of increasing inelasticity, $y$, we first observe a
  peak which corresponds to an elastic $ep \rightarrow ep$
  interaction~\protect\cite{elasticslac}, then a series of maxima
  which correspond to the proton excited states and then a continuum
  \protect\cite{taylor1}.}{eprime}

\epsfigure[width=0.8\hsize]{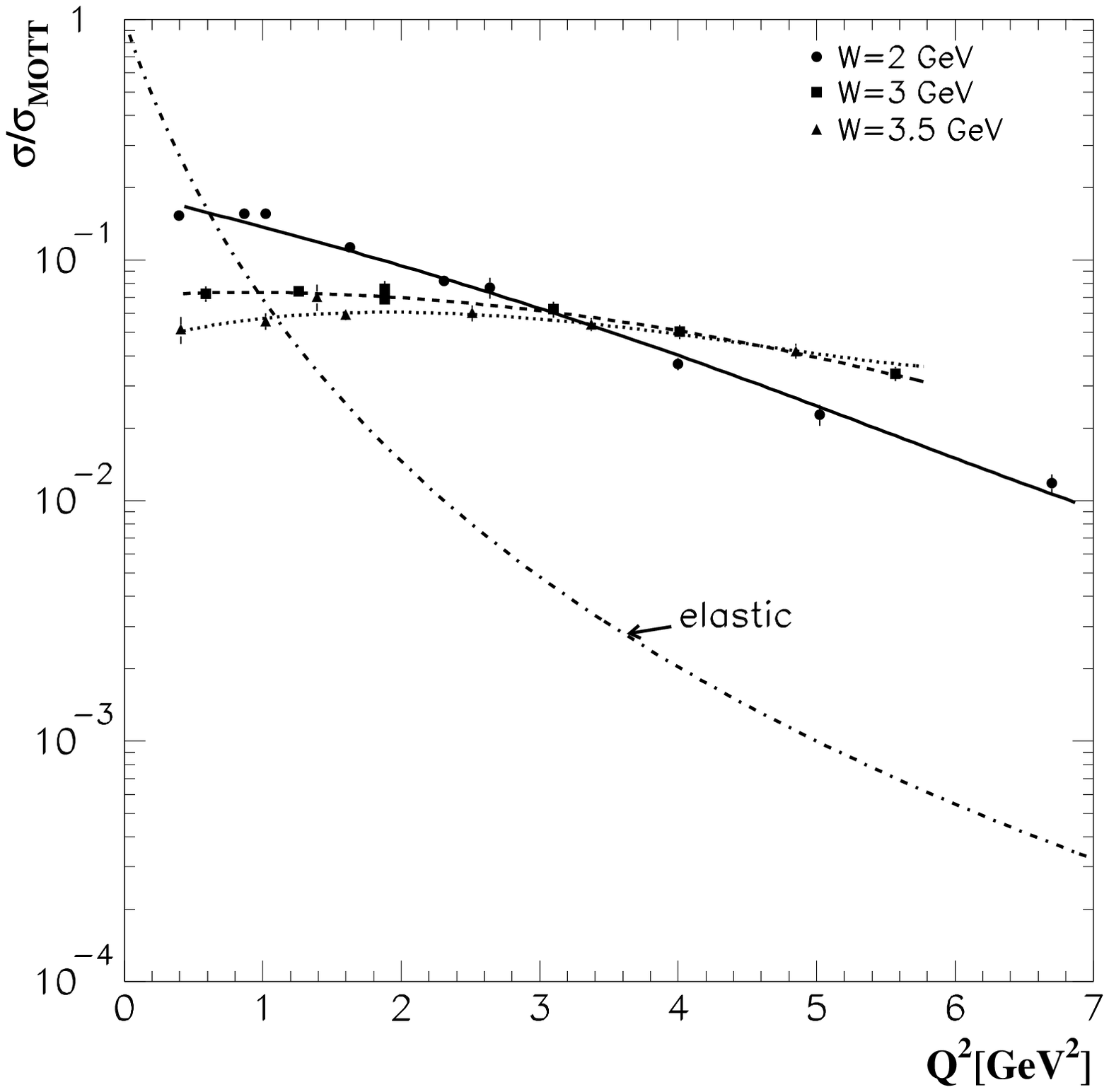}{
  The inelastic $ep\rightarrow eX$ cross section scaled by the Mott
  cross section for fixed $W$ values as a function of $Q^2$ as
  measured in the SLAC experiment~\protect\cite{taylor1}.  For
  comparison the elastic $ep\rightarrow ep$ cross section with dipole
  formfactors scaled by the Mott cross section is shown for electron
  scattering angle $\theta=10^o$ (dotted dashed line). The lines
  connecting the data points are to guide the eye (adapted from Fig.~1
  of \protect\citeasnoun{taylor2}).}{mott}

By selecting the outgoing lepton energy and angle we can vary the
values of $Q^2$ and $y$, thus probing the charge distribution
(electromagnetic or weak) within the nucleon. An example of the
scattered electron energy, $E'$, spectrum as measured in electron
proton collisions~\cite{taylor1} is shown in Fig.~\ref{fig:eprime}.
From right to left, the direction of increasing inelasticity $y$, we
first observe a peak which corresponds to an elastic $ep \rightarrow
ep$ interaction in which the proton remains intact~\cite{elasticslac}.
We then observe a series of maxima which correspond to the proton
excited states and then a continuum. Should the charge distribution in
the proton be continuous we would expect, in the region of continuum,
a decrease of the cross section with increasing $Q^2$ following the
Rutherford formula. The experimentally observed spectrum turned out to
be much flatter (see Fig.~\ref{fig:mott}). This was the first
indication that there is a substructure inside the nucleon.

The derivation of the formula for the inclusive scattering cross
section (which is beyond the scope of this review) is very similar to
that of $e \mu$ scattering.  The unknown couplings of the lepton
current to the nucleon are absorbed in the definition of the structure
functions, $F_i$, which can be thought of as Fourier transforms of the
spatial ``charge'' distribution.

The inclusive differential cross section, integrated over all possible
hadronic final states, is a function of two variables which uniquely
determine the kinematics of the events. These variables are most
easily recognizable as the energy and production angle of the
scattered lepton. However, in anticipation of the partonic structure
of hadrons, the differential cross section is usually expressed in
terms of two variables, $x$ and $Q^2$, defined in
Eq.~(\ref{eq:disvariables}),
\begin{equation}
\frac{d^2\sigma^{l(\bar{l})N}}{dx dQ^2} = A \left\{ \frac{y^2}{2} 
2 x F_1(x,Q^2) +  
(1-y)F_2(x,Q^2) \pm (y - \frac{y^2}{2})xF_3(x,Q^2) \right\} \ ,
\label{eq:disxsection}
\end{equation}
where, for $Q^2\ll M^2_{W,Z}$ (the mass squared of the intermediate
vector bosons), $A = G_F^2/2\pi x$ for neutrinos and anti-neutrinos
with $G_F$ the Fermi constant, and $A = 4\pi\alpha^2/xQ^4$ for
charged leptons with $\alpha$ the electromagnetic coupling
constant.  

The structure functions, $F_i$, may depend on the kinematics of the
scattering and the chosen variables are $x$ and $Q^2$. The reason for
this choice will become clearer in the next sections.  The structure
functions, $F_1$, $F_2,$ and $F_3$ are process dependent.  The $F_3$
structure function is non-zero only for weak interactions and is
generated by the parity violating interactions. In the following,
after discussing the kinematics of lepton-nucleon scattering, we will
concentrate on the interpretation and properties of the structure
functions.

\subsection{Kinematics of lepton-nucleon scattering}
\label{sec:kinematics}

The variables used in describing the properties of lepton-nucleon
scattering are defined by Eq.(~\ref{eq:disvariables}). Here we
would like to discuss in more detail their meaning. We will do so
assuming that the mass of the incoming and scattered leptons are
negligible and, in preparation for the HERA conditions, that the
nucleon is a proton with mass $m_p$.

The variable $s = (p+k)^2 \simeq m_p^2 + 2 p \cdot k $ is the square
of the center-of-mass energy. However, an energy variable which is
more appropriate at HERA is $W$, which is the invariant mass
of the system recoiling against the scattered lepton and can be
interpreted as the center-of-mass energy of the virtual boson-proton
system.
\begin{equation}
W^2 = (P+q)^2 = m_p^2 - Q^2 + 2 P \cdot q = ys - Q^2 + m_p^2 (1-y) \ ,
\label{eq:w2(y)}
\end{equation}  
where in the last step we have used the definition of $y$ (see
Eq.~(\ref{eq:disvariables})). The variable $y$ is an invariant and in
the proton rest frame the expression for $y$ reduces to
\begin{equation}
y=1- \frac{E'_l}{E_l} \, ,
\end{equation}
where $E_l$, $E'_l$ are the energies of the incoming and scattered
lepton in this frame, respectively. It is easy now to infer the most
general limits on $y$,
\begin{equation}
0\leq y \leq 1 \, .
\end{equation}
The variable $y$ is a measure of the fraction of the energy from the
electron transferred to the interaction. The limits on $x$ can be
readily deduced from the following:
\begin{equation}
x=\frac{Q^2}{2P\cdot q}=\frac{Q2}{W^2+Q^2-m_p^2} \ ,
\end{equation}
where we have used the relation~(\ref{eq:w2(y)}) in the last step.
Since $Q^2 \ge 0$ and $W^2$ cannot be smaller than $m_p^2$ the upper
limit on $x$ is $x \leq 1$. In practice the lower limit is determined
by the maximum $W^2$ available in the interaction, but formally $x$
can be infinitely small, although positive. Thus,
\begin{equation}
0 \leq x \leq 1 \ .
\end{equation}
The interpretation of $x$ is easiest in the QPM language. Define $z$
as the fraction of the proton momentum carried by the struck quark and
$p^\prime$ is the four momentum of the outgoing quark. If we assume
that the quark masses are zero as dictated by QPM (i.e.
$(zP)^2=p^{\prime 2}=0$) then
\begin{equation} 
p^{\prime 2}= (zP+q)^2= 2 zP\cdot q -Q^2=0 \, .
\end{equation}
It can be readily seen that $z=x$. Thus in the QPM $x$ is the fraction
of the proton momentum carried by the struck massless quark. Note also
that for $Q^2 \ll W^2$,
\begin{equation}
x \simeq \frac{Q^2}{W^2} \, ,
\end{equation}
and for fixed values of $Q^2$, the higher the $W$ the lower the $x$.
 
The value of $Q^2$ depends only on the lepton vertex and is given by
\begin{equation}
Q^2 = 2 E_l E'_l (1- \cos \theta) \,
\end{equation}
where $\theta$ is the angle between the initial and scattered
lepton. This expression is valid in all frames of reference. The
larger the scattering angle and the larger the energy of the scattered
lepton, the larger the $Q^2$. The maximum $Q^2$ is limited by $s$,
\begin{equation}
Q^2= xy (s - m_p^2) \, ,
\end{equation}
and occurs when both $x$ and $y$ tend to one. For a given $Q^2$
the lowest $x$ is achieved when $y = 1$ and the lowest $y$ when
$x=1$. Thus kinematically the small values of $x$ are associated with
large values of $y$ and vice versa. 

\epsfigure[width=\hsize]{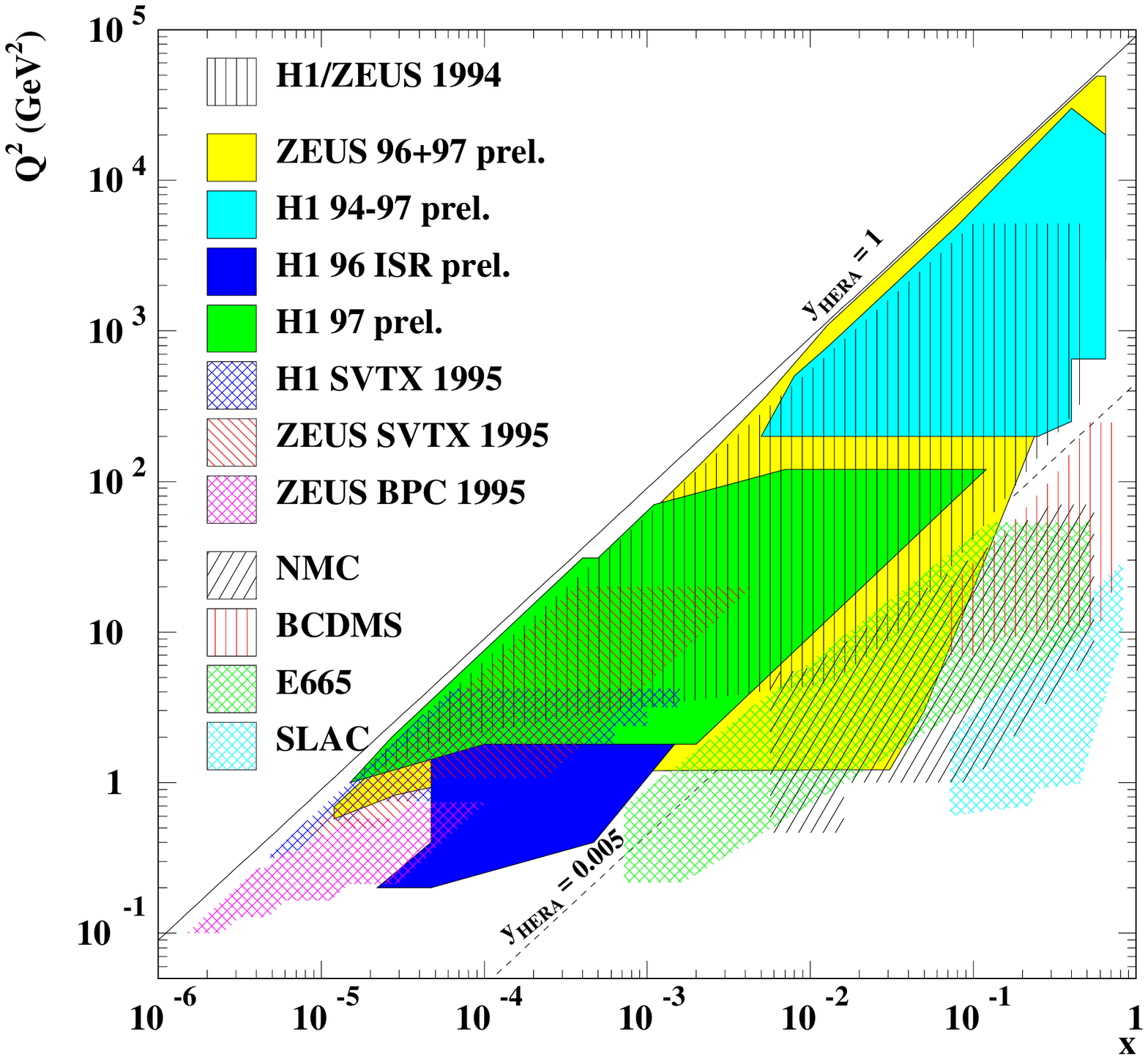}{ 
  The region in the $(x,Q^2)$ plane where measurements of $F_2$ have
  been performed by the fixed target and HERA experiments.  }{F2plane}

The kinematic plane available in $x$ and $Q^2$ for electron-proton
scattering at HERA is shown in Fig.~\ref{fig:F2plane}. 

\subsection{Structure functions in the Quark Parton Model}

In deep inelastic scattering (i.e. $Q^2 \gg 1 \gevtwo$) the nucleon is
viewed as composed of point like free constituents - quarks and
gluons. In the QPM the lepton nucleon interaction is described as an
incoherent sum of contributions from individual free quarks. To
justify this approach~\cite{Ioffe}, let's consider a virtual photon in
the frame in which its four momentum is purely space-like (in the so
called Breit frame). In this frame the four momenta of the photon,
$q_{\gamma}$ and of the proton, $p_p$, have the following form:
\begin{eqnarray}
q_{\gamma}&=&\big(0,0,0, -\sqrt{Q^2} \big ) \, , \nonumber \\
P_p&=& \left(\sqrt{\left(\frac{\nu^2}{Q^2}+1\right)m_p^2},0,0,
\frac{m_p\nu}{\sqrt{Q^2}} \right) \, .
\label{eq:breit}
\end{eqnarray}
In the Bjorken limit, when $Q^2 \sim m_p\nu \rightarrow \infty$ and
$x=Q^2/(2m_p\nu) \sim 1$, the proton momentum $m_p\nu/\sqrt{Q^2}
\rightarrow \infty$. The static photon field occupies a longitudinal
size $\sim 1/\sqrt{Q^2}$. Because of energy conservation, the photon
may only be absorbed by a quark with momentum equal to $ 0.5
|\sqrt{Q^2}|$. After absorption, the quark will change direction and
move with the same momentum value. The interaction time may be defined
as the overlap time between the quark and the field of the photon,
$t_i \sim 2m_q/Q^2$, where $m_q$ stands for some effective mass of the
quark. The lifetime of the quark is then estimated to be $t_q \sim
\sqrt{Q^2}/m_q^2 \gg t_i$. Thus at large $Q^2$, it is indeed justified to
consider the quark as free and to neglect possible interactions of the
photon with other partons.

The electroweak gauge bosons couple to quarks through a mixture of
vector ($v$) and axial-vector ($a$) couplings. The structure functions
can then be expressed in terms of quark distributions $q_i(x)$, where
$i$ stands for individual quark types. For non-interacting partons, as
is the case in QPM, Bjorken scaling (no $Q^2$ dependence) is expected.
\begin{eqnarray}
F_1(x)&=&\frac{1}{2}\sum_{i}q_i(x) (v_i^2 + a_i^2) \, , \nonumber \\
F_2(x)&=&\sum_{i}xq_i(x) (v_i^2 + a_i^2) \, ,\label{eq:stf} \\
F_3(x)&=&2\sum_{i}q_i(x) (v_i a_i) \, .\nonumber 
\end{eqnarray} 
The index $i$ runs over all flavors of quarks and antiquarks which are
allowed, by conservation laws, to participate in the interaction. For
the simplest case of electromagnetic interactions, $v_i=e_i$,
where $e_i$ is the charge of quark $i$ in units of the electron
charge, and $a_i=0$. For charged currents $v_i=a_i=1$ for
quarks and $v_i=-a_i=1$ for antiquarks. For neutral current
interactions mediated by the $Z^0$, $v_i=T_{i3}-2e_i \sin^2\Theta_W$
and $a_i=T_{i3}$, where $T_{i3}$ is the third component of the weak
isospin of quark $i$ and $\Theta_W$ denotes the Weinberg mixing
angle, one of the fundamental parameters of the Standard Model. The
couplings have a more complicated structure for neutral current
interactions in which the interference between the $Z^0$ and the
$\gamma$ play an important role. A direct consequence of
formulae~(\ref{eq:stf}),  derived for spin $1/2$ partons, is the
Callan-Gross relation~\cite{Callan-Gross}, i.e.
\begin{equation}
2xF_1(x)=F_2(x) \, .
\label{eq:callan-gross}
\end{equation}
For universal parton distributions in the proton, expected in the QPM and
QCD, formulae~(\ref{eq:stf}) can be used to relate DIS cross sections
obtained with different probes. In fact, many more relations and sum
rules can be derived assuming SU(3) or SU(4) flavor symmetry for
hadrons. Inversely, the validity of these assumptions can be tested
experimentally. A detailed discussion of these issues is beyond the
scope of this review.  The naive QPM approach, which allows the
construction of structure functions from quark distributions, has to
be altered to take into account some dynamical features predicted by
QCD, such as violation of scaling and of the Callan-Gross relation, as
well as higher twist effects.

Quarks are bound within the nucleon by means of gluons. We may thus
expect fluctuations such as emission and reabsorption of gluons as
well as creation and annihilation of $q \bar{q}$ pairs. Depending on
the resolving power of the probe and the time of the interaction, some
of these fluctuations can be seen and the partonic structure of the
hadron will change accordingly. The structure functions acquire a
$Q^2$ dependence.  This $Q^2$ dependence is encoded in QCD and the
measurement of the $Q^2$ dependence constitutes a test of perturbative
QCD at a fundamental level. The violation of the Callan-Gross relation
is also a consequence of QCD radiation.

Non-perturbative effects of QCD can contribute to the scale breaking
of structure functions. They are due, e.g., to scattering on coherent parton
states. These contributions vanish as inverse powers of $Q^2$.  The
theoretical understanding of higher twist effects is quite
limited~\cite{HTtheo1,HTtheo2}. Experimentally they are observed at
large $x$ and small $W^2$~\cite{Virchaux} and are also expected to affect
the very small-$x$ region~\cite{HTbartels}. The assumption that quarks
are massless certainly does not hold for the heavy quarks $c$, $b$ and
$t$, for which $m_c \simeq 1.5 \gev$, $m_b\simeq 4.5 \gev$ and $m_t
\simeq 175 \gev$, respectively. The radiation of heavy quarks will be
affected by threshold effects which may be substantial up to large
values of $Q^2$.

\subsection{QCD evolution equations} 
\label{sec:QCD}

\epsfigure[width=0.8\hsize]{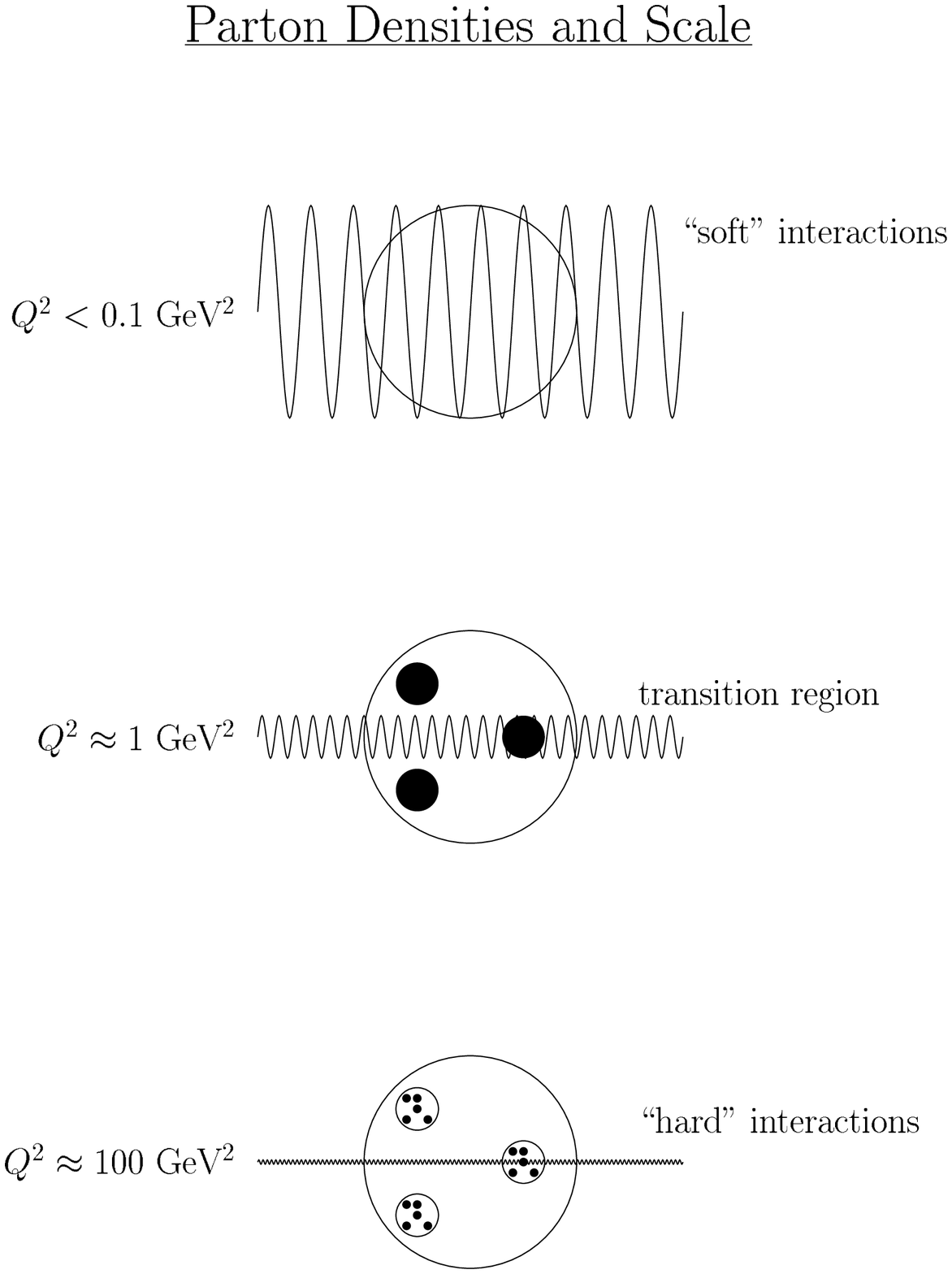}{ Schematic representation of
  photon-proton scattering for increasing photon virtuality $Q^2$ at
  fixed $W$.  As $Q^2$ increases, the photon probes smaller transverse
  distance scales and is able to resolve the structure of the proton.
  With further increases in $Q^2$, quarks are resolved into more
  quarks and gluons.}{proton_structure} 
\epsfigure[width=0.8\hsize]{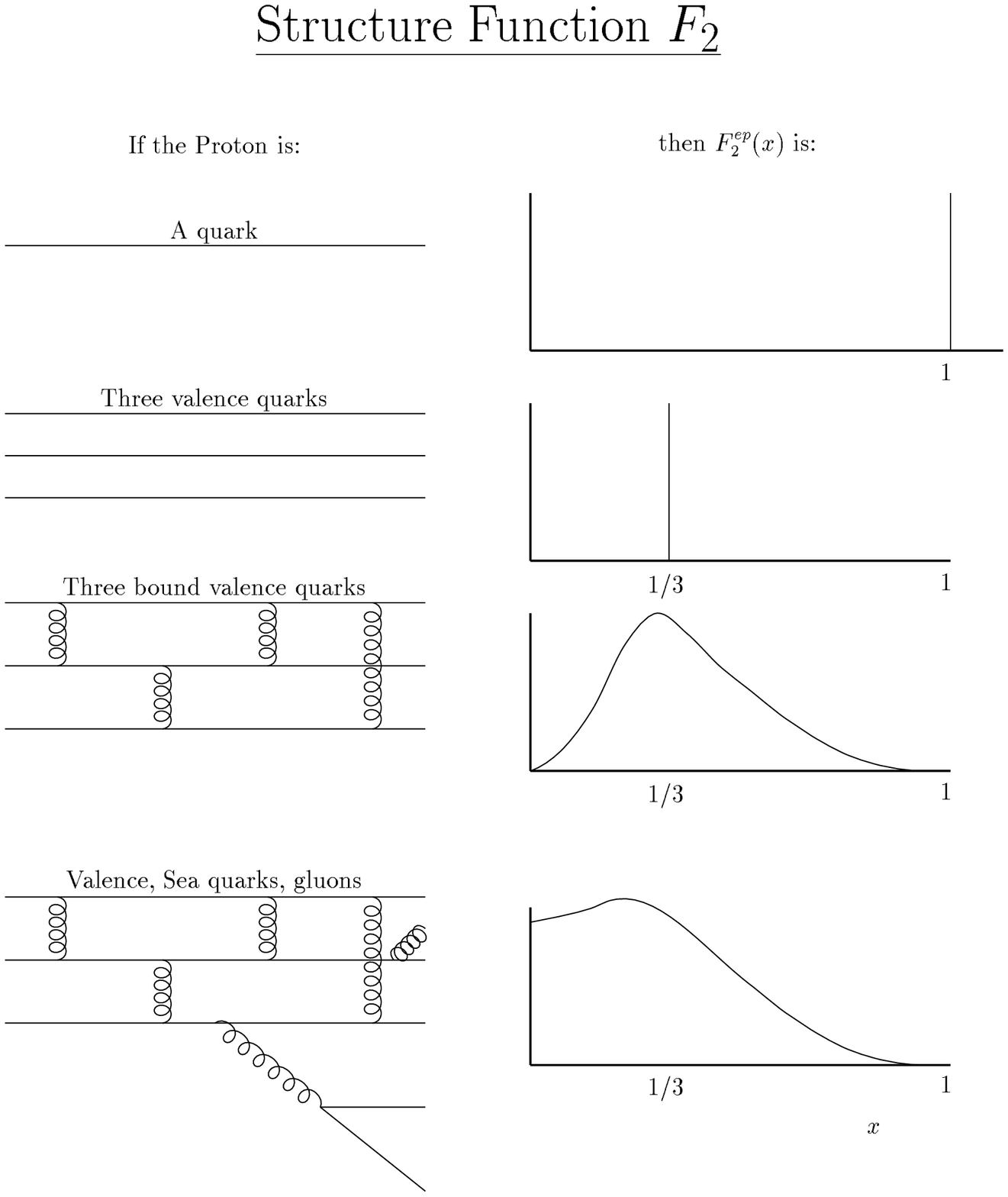}{ The expected behavior of the
  proton structure function $F_2$ in the QPM (top), and including the
  effects of QCD (adapted from figure~9.7
  of~\protect\citeasnoun{ref:HM}).}{lN_f2}

The parton distributions in the hadron cannot be calculated from first
principles. However, thanks to the QCD factorization
theorem~\cite{collins-soper} which allows to separate the long range
effects (such as the parton distribution at a small-$Q^2$ scale) from
the short range interactions, the $Q^2$ dependence of partons, called
parton evolution, can be calculated within perturbative QCD. The main
origin of this $Q^2$ dependence is that a quark seen at a scale
$Q^2_0$ as carrying a fraction $x_0$ of the proton momentum can be
resolved into more quarks and gluons when the scale $Q^2$ is
increased, as shown in Fig.~\ref{fig:proton_structure}.  These quarks
and gluons have $x<x_0$. Thus we can easily infer from this picture
that the number of slow quarks will increase and the number of fast
quarks will decrease when we increase the resolving power of the
probe. The implications for $F_2$ is depicted in Fig.~\ref{fig:lN_f2}.

In QCD, as in many gauge theories with massless particles, the loop
corrections to the quark gluon coupling diverge and the
renormalization procedure introduces a scale into the definition of
the effective coupling. The effective strong coupling constant $\as$
decreases with increasing scale relevant to the QCD subprocess (see
Eq.~(\ref{eq-int:alphas})) and when it becomes sufficiently small,
perturbative calculations can be performed.

The parton evolution equations derived on the basis of the
factorization theorem are known as the
Dokshitzer-Gribov-Lipatov-Altarelli-Parisi (DGLAP) evolution
equations~\cite{dglap_gl,dglap_d,dglap_ap}. The DGLAP equations
describe the way the quark $q$ and gluon $g$ momentum distributions in
a hadron evolve with the scale of the interactions $Q^2$.
\begin{equation} 
\label{dglap} 
{\partial \over \partial \ln
  Q^2} \left(\begin{array}{c} q \\ g \end{array}\right) = {\as (Q^2)
  \over 2 \pi} \left[\begin{array}{cc}
    P_{qq} & P_{qg}  \\
    P_{gq} & P_{gg}
\end{array}\right] \otimes
\left(\begin{array}{c}q \\ g\end{array}\right) \, ,
\end{equation}
where both $q$ and $g$ are functions of $x$ and $Q^2$.  The
splitting functions $P_{ij}$ provide the probability of finding parton
$i$ in parton $j$ with a given fraction of parton $j$ momentum.  This
probability will depend on the number of splittings allowed in the
approximation. Given a specific factorization and renormalization
scheme, the splitting functions $P_{ij}$ are obtained in QCD by
perturbative expansion in $\as$,
\begin{equation}
 \frac{\as}{2\pi}P_{ij}(x,Q^2)= \frac{\as}{2\pi}
P_{ij}^{(1)}(x)+ \left(\frac{\as}{2\pi}\right)^2 P_{ij}^{(2)}(x)+
\ldots \, .
\label{eq:pij}
\end{equation}
The truncation after the first two terms in the expansion defines the
next to leading order (NLO) DGLAP evolution. This approach assumes
that the dominant contribution to the evolution comes from subsequent
parton emissions which are strongly ordered in transverse momenta
$k_T$, the largest corresponding to the parton interacting with the
probe. 

It should also be noted that beyond leading order (LO) the splitting
functions depend on the factorization scale and thus the definition of
parton distributions is not unique. This affects the simple
relation~(\ref{eq:stf}) between quarks and structure functions. The
relation~(\ref{eq:stf}) is preserved in LO, but the parton
distribution functions acquire a $Q^2$ dependence. In NLO the
Callan-Gross relation is violated. The difference between $F_2-2xF_1$
is called the longitudinal structure function $F_L$ (its meaning will
be explained in section~\ref{sec:gammap}) and for virtual photon
exchange takes the following form in QCD:
\begin{equation}
\label{eq:FL}
F_L = \frac{\as(Q^2)}{\pi}\left[\frac{4}{3}\int_x^1\!\frac{dz}{z}
\left(\frac{x}{z}\right)\!^2 \!F_2(z,Q^2) + 
2\!\sum_{i=1,4} e_i^2 \!\int_x^1\! \frac{dz}{z}
\frac{x}{z}\left(1-\frac{x}{z}\right)zg(z,Q^2)\right] \; .
\end{equation}
Formula~(\ref{eq:disxsection}) for the deep inelastic cross section
remains valid to all orders.

\subsection{Structure functions in the $\gamma^* p$ system}
\label{sec:gammap}

The lepton-nucleon interaction cross section can also be described as
a convolution of a flux of virtual bosons with the absorption cross
section of a virtual boson by the nucleon. This is depicted in
Fig.~\ref{fig:ep_gammap} where for the sake of simplicity only $\gamma^*$
is considered.
\epsfigure[width=0.5\hsize]{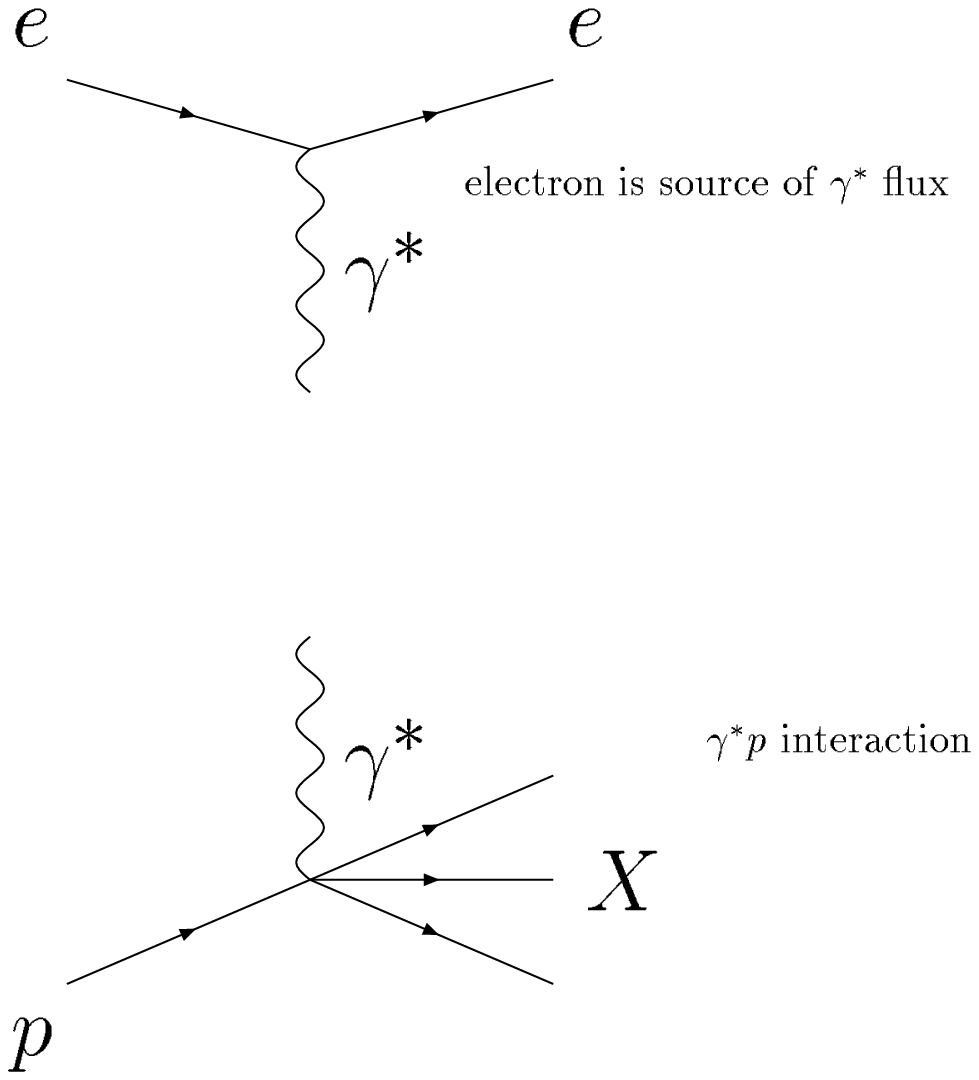}{
  At small $Q^2$, $ep$ scattering can be thought of as 1)~radiation of
  a virtual photon, $\gamma^*$, from the electron, 
  followed by 2)~$\gamma^* p$ scattering.}{ep_gammap}

The virtual photon is treated as a massive spin 1 particle and
acquires three polarization vectors corresponding to helicities
$\lambda =\pm 1,~0$. The absorptive cross section may depend on
helicity. In case of the virtual photon, parity invariance implies
that the cross sections corresponding to $\lambda=\pm 1$ have to be
equal. We will thus have two independent cross sections, one for
absorbing a transversely polarized photon $\sigma_T$ ($\lambda=\pm 1$)
and one for a longitudinally polarized photon $\sigma_L$
($\lambda=0$).  The relation between $\sigma_{T,L}$ and the structure
functions $F_{1,2}$ are as follows:
\begin{eqnarray}
2xF_1&=&\frac{K}{4\pi^2\alpha} \frac{Q^2}{\nu} \sigma_T \, , \\
F_2 &=& \frac{K}{4\pi^2\alpha} \frac{Q^2\nu}{Q^2+\nu^2}
(\sigma_T+\sigma_L)\, ,
\label{eq:defsig}
\end{eqnarray}
where $K$ stands for the virtual photon flux.  For virtual particles
there is no unambiguous definition of the flux factor. It is a matter
of convention. Here we will use the Hand convention~\cite{Hand} which,
in analogy to the real photon case where $K=\nu$, defines $K$ for
virtual photons as the energy that a real photon would need in order
to create the same final state.
\begin{equation}
K = \nu - \frac{Q^2}{2m_p} \, .
\label{eq:Hand}
\end{equation}

The ratio $R=\sigma_L/\sigma_T$ depends on the spin of the interacting
particles. A spin 1/2 massless particle cannot absorb a longitudinally
polarized photon in a head-on collision without breaking helicity
conservation. The early measurements in which $R$ was found to be
small~\cite{RSLAC74} gave support to the idea that partons in the
nucleon where in fact quarks. For scalar partons $R \rightarrow
\infty$. However, in a theory in which quarks are massive with mass
$m_q$ and/or have an intrinsic transverse momentum $k_T$, $R$ is
expected to be
\begin{equation}
R=\frac{\langle k_T^2 \rangle + \langle m_q^2 \rangle}{Q^2} \, .
\end{equation}
The parameter $R$ can be related to the longitudinal structure
function of the proton,
\begin{equation}
R = \frac{\left( \frac{Q^2}{\nu^2}+1 \right)F_2-2xF_1}{2xF_1} \simeq
\frac{F_L}{2xF_1} \, .
\end{equation}
As an outcome of QCD radiation partons in the proton acquire
transverse momentum, the more so the slower they are, and therefore
$F_L$ is non-zero.

The representation of structure functions in terms of absorption cross
sections turns out to be very useful in understanding some dynamical
properties of DIS as it creates a natural link between the
perturbative regime of QCD and the non-perturbative soft hadron-hadron
interactions.  The latter are best described in the framework of Regge
theory.

\subsection{Regge phenomenology}
\label{sec:reggeology}

The soft hadron-hadron interactions are well described by Regge
phenomenology~\cite{regge} in which the interaction is viewed as due
to exchanges of collective states called Regge poles. The Regge poles
can be classified into different families according to their quantum
numbers. The Regge poles with quantum numbers of mesons form linear
trajectories in the $m^2, l$ plane, where $m$ is the mass of the meson
and $l$ its spin. The continuation of a trajectory to negative values
of $m^2$ leads to a parameterization in terms of $t$, the square of
the four momentum transfer, as follows:
\begin{equation}
\alpha(t) = \alpha_0 + \alpha' \cdot  t \, ,
\label{eq:alpha}
\end{equation}
where $\alpha_0$ is the intercept and $\alpha'$ is the slope of the
trajectory.  
\epsfigure[width=0.8\hsize]{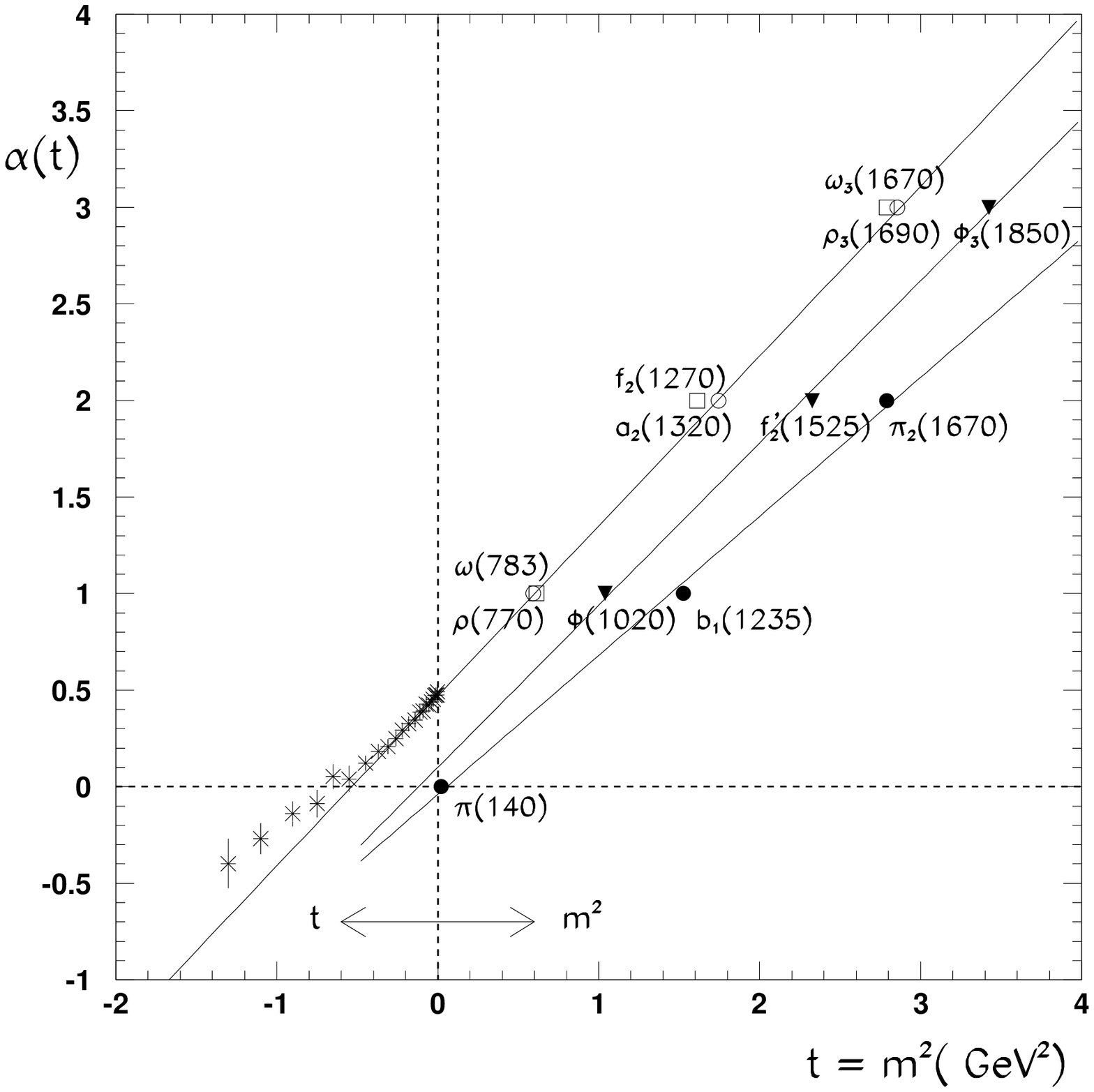}{
  Example of the $\rho$ (circles), $\omega$ (empty squares), $\phi$
  (triangles) and $\pi$ (dots) trajectories. Also shown is the
  continuation of the $\rho$ trajectory as measured in $\pi^-p
  \rightarrow \eta n$~\protect\cite{rhotrajectory}.}{trajectory}
An example of such trajectories, called reggeon
trajectories, is shown in Fig.~\ref{fig:trajectory}. Among all
possible families of Regge poles there is a special one, with the
quantum numbers of the vacuum, called the pomeron ($\pom$) trajectory.
There are no known hadronic bound states lying on this trajectory
(glue-balls would be expected to form this trajectory). Its parameters
have been determined
experimentally~\cite{DL-aprime,DL-flux,ref:DoLa_sigfit} to be
\begin{equation}
\alpha_{\pom} = 1.08 + 0.25 t \, .
\label{eq:pomeron_trajectory}
\end{equation}
In Regge theory the energy dependence of total and elastic cross
sections is derived from the analytic structure of the hadronic
amplitudes. In the limit $s \gg -t$, where $s$ is the square of the
center-of-mass energy of the scattering, the amplitude for elastic
scattering has the form $A(s,t) \propto s^{\alpha_{\pom}(t)}$. The
pomeron trajectory also provides the leading contribution to the high
energy behavior of the total cross section,
\begin{equation}
\sigma_{\mathrm{tot}} = s^{-1} {\it Im} A(s,t=0) \propto s^{\apom(0)-1}
\label{eq:Regge_xsection}
\end{equation}
The $s$ dependence of hadronic interactions fulfills this behavior
independently of the interacting particles~\cite{ref:DoLa_sigfit}
as expected from the universality of the exchanged trajectories.

Two types of soft interactions arise naturally in Regge theory:
elastic and diffractive scattering.  These are mediated by the
exchange of the pomeron. In elastic scattering, the square of the
momentum transfer, $t$, between the interacting hadrons is very small
and the only products of this interaction are the two hadrons which
emerge with little change in their initial directions. The
properties of the elastic scattering cross section determine the slope
of the pomeron trajectory. In diffractive scattering, the momentum
transfer between initial hadrons still remains very limited, but one
or both of the interacting hadrons may be excited into a state of
finite mass which then subsequently decays.  Single dissociation
occurs if only one hadron dissociates, while if both dissociate into
higher masses the scattering is called double dissociation. Typical of
diffractive scattering is the production of relatively low excited
masses and the mass spectrum is directly related to the properties of
the pomeron trajectory.

Regge phenomenology proved very successful in describing the energy
dependence of the total hadron-hadron interaction cross section as
well as in describing the properties of elastic and diffractive
scattering (for a review see~\citeasnoun{goulianos1}). 

\subsection{QCD dynamics at small $x$}
\label{sec:BFKL}

In deep inelastic scattering, the kinematic region which corresponds
to the Regge limit is that of small $x$ at fixed $Q^2$. In perturbative
QCD at small $x$, higher-loop contributions to the splitting functions
are enhanced,
\begin{equation}
P_{ij}^{(n)} \sim \frac{1}{x} \ln^{(n-1)} x \, .  
\end{equation}
The presence of these terms may spoil the convergence
of~(\ref{eq:pij}). The evolution equation which allows the resummation
in the expansion~(\ref{eq:pij}) of leading $(\as \ln x)^n$ terms is
known under the name of Balitsky-Fadin-Kuraev-Lipatov
(BFKL) equation~\cite{bfkl_1,ref:BFKL2,ref:BFKL3}. In the parton cascade
picture this evolution corresponds roughly to cascades with subsequent
emissions following a strong ordering in $x$ with no restriction on
the $k_T$. Here the evolution takes place from high longitudinal
momentum partons to low longitudinal momenta over a fixed transverse
area proportional to $1/Q^2$. The BFKL equation in its
original form does not address the $Q^2$ evolution of the parton
distributions.  The difference between DGLAP and BFKL evolution is
shown schematically in Fig.~\ref{fig:DGLAP_BFKL}.
\epsfigure[width=0.8\hsize]{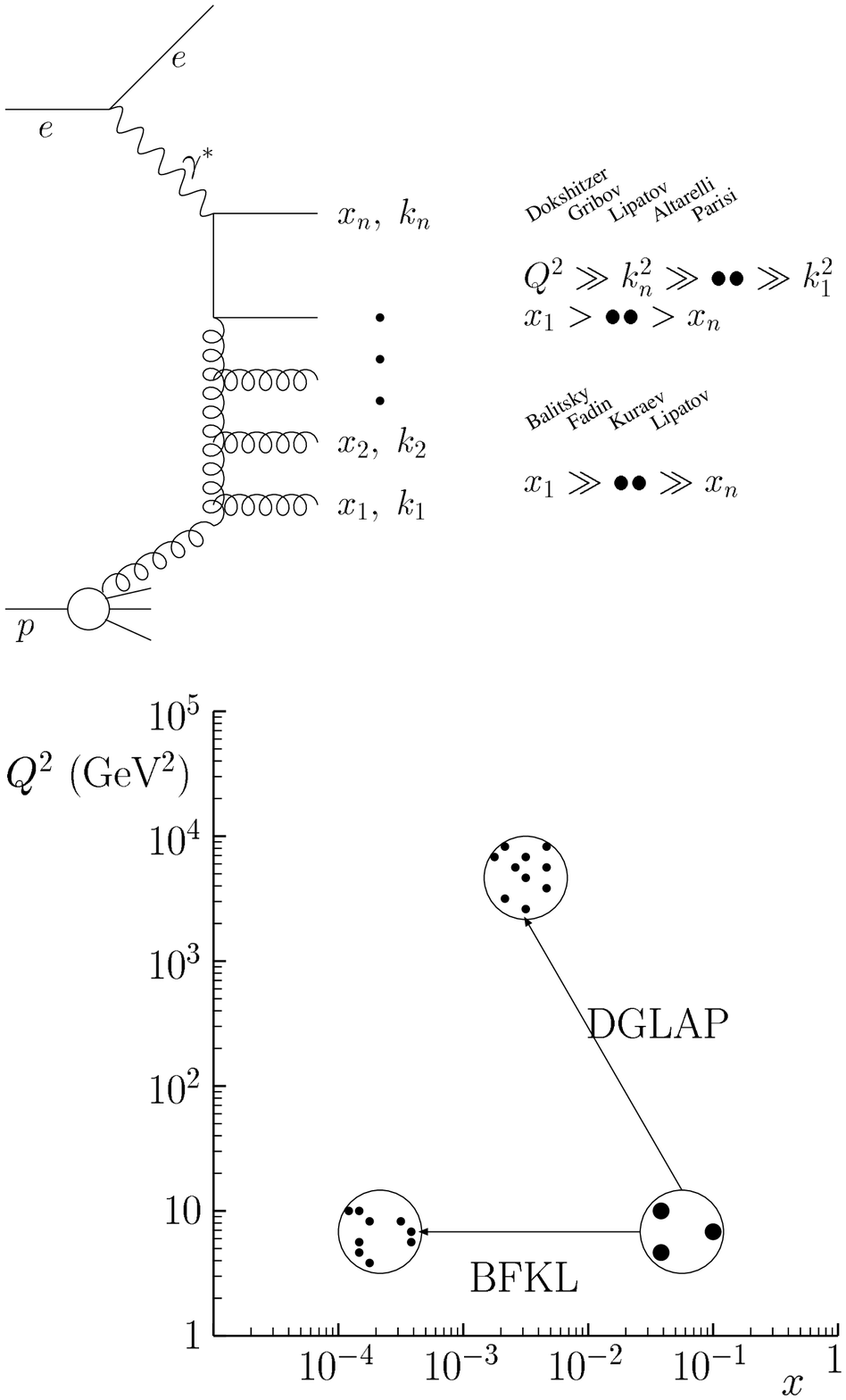}{
  Schematic description of the differences in the DGLAP and BFKL
  evolution equations.  The DGLAP equation has strong $Q^2$ ordering
  in the radiation sequence, while BFKL evolution has strong $x$
  ordering.}{DGLAP_BFKL}

The two approaches to parton evolution, DGLAP and BFKL, are embedded
in a single equation known as the CCFM
equation~\cite{ccfm_1,ccfm_2,ccfm_3} based on $k_T$ factorization and
angular ordering.

The solutions of the DGLAP equations and of the BFKL equation (in LO),
in the limit of very small $x$, where the dominant contribution to the
cross section is driven by gluon radiation, predict a rise of $F_2$
with decreasing $x$, 
\begin{equation} 
F_2^{DLL}(x,Q^2) \sim \exp\left(2
  \sqrt{\frac{C_A\as}{\pi} \ln \frac{1}{x} \ln
    \frac{Q^2}{Q^2_0}}\right)  \, , 
\label{eq:dlla}
\end{equation} 
\begin{equation}
F_2^{BFKL}(x,Q^2) \sim \sqrt{\frac{Q^2}{Q^2_0}}
x^{-\frac{4C_A\as}{\pi}\ln 2} \, ,
\label{eq:bfkl}
\end{equation}
where the superscript $DLL$ stands for the double leading logarithmic
approximation used in solving the DGLAP equation and $Q_0^2$ denotes
the starting scale of the evolution. In general the BFKL equation
predicts a faster increase of $F_2$ with decreasing $x$ and stronger
scaling violations in $Q^2$ as compared to the DGLAP evolution.
However the solution~(\ref{eq:bfkl}) is derived assuming a constant
$\as$ and higher order corrections are expected to tame the rise with
$1/x$ (for a discussion see~\citeasnoun{adm_rome}). In the BFKL
approach the concept of a QCD pomeron arises naturally.

The two solutions presented above~(\ref{eq:dlla}, \ref{eq:bfkl}) are
expected to violate unitarity at very small $x$~\cite{AFS}.  The fast
increase of parton densities at small $x$ expected in perturbative QCD
and confirmed experimentally (see section~\ref{sec:inclusive}) raise a
natural question, whether such high densities will not lead to
overcrowding of the proton. The annihilation and recombination of
partons could lead to saturation effects and would require corrections
to the known evolution equations~\cite{ref:GLR}.

\subsection{Perturbative QCD in the final states}

It is generally believed that the pattern of perturbative QCD
radiation should be observed in the hadronic final states. Although
colored partons cannot be observed directly, their fragmentation
produces jets of hadrons, collimated around the original direction of
the partons.  This is the principle of parton-hadron
duality~\cite{khoze}. In fact, while the need for gluons was inferred
from DIS measurements of structure functions, the actual proof of
their existence was first made in the $e^+e^-$ interactions (for a
review see~\citeasnoun{gluons}).  Hadron production in high energy
$e^+e^-$ interactions proceeds through the annihilation of leptons
into a photon (or a $Z^0$) with a subsequent production of a
$q\bar{q}$ pair. The fragmentation of the pair leads to a two jet
structure in the final state. However, each of the quarks (or both)
may emit a gluon with a large transverse momentum relative to the
parent quark.  Such a hard gluon will be a source of a third jet. The
probability of such a configuration can be calculated in perturbative
QCD.

A similar situation may arise in DIS. In a typical DIS interaction we
expect the final state to consist of a jet of hadrons originating from
the struck quark, called the current jet, which balances in transverse
momentum the scattered lepton. The remnant of the target also
fragments into hadrons which remain collimated around the direction of
the latter. The space between the fragmentation of the current jet and
the remnant is filled by radiation due to the color flow between the
struck quark and the remnant state of the target. Various approaches
exist to model this effect~\cite{lepto,ref:ARIADNE_2}.

\epsfigure[width=0.90\hsize]{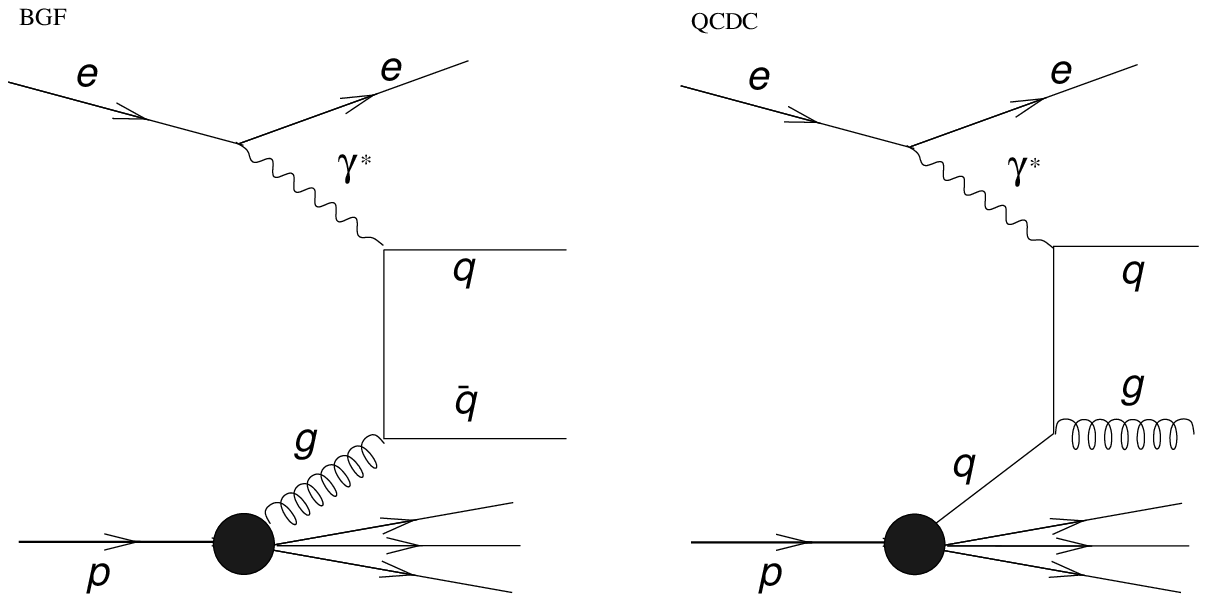}{
  Diagrams for boson-gluon fusion (BGF) and for QCD Compton
  (QCDC) processes in DIS.}{jets-disjets} 

A gluon may split into a pair of quarks with large relative transverse
momenta, before one of the quarks absorbs the virtual boson.  Two jets
will be observed in the final state. This process, called boson-gluon
fusion (BGF),is diagrammatically presented in
Fig.~\ref{fig:jets-disjets}.  Another possibility is that the quark
will emit a hard gluon before absorbing the virtual boson as shown in
Fig.~~\ref{fig:jets-disjets}b.  This process is called QCD Compton
scattering (QCDC).  The contribution of both diagrams to the DIS cross
section can be calculated in perturbative QCD~\cite{schuler-jets}.

In BGF the large transverse momenta partons may be replaced by heavy
quark production.  The latter in particular gives rise to the charm
content of the $F_2$ structure function. Both the QCDC and BGF
processes are included at some level in the evolution equation in the
NLO approximation~\cite{wukitung}. The BGF process is sensitive to the
gluon content of the nucleon.  The extraction of the gluon
distribution from a direct measurement of the BGF process is quite
challenging because of higher order QCD corrections, especially when
the square of transverse momenta of the jets is of the same order as
$Q^2$. However heavy quark production through the BGF mechanism is
easier to control theoretically~\cite{bgfcharm_1,bgfcharm_2}.

The perturbative calculation of cross sections for QCDC or BGF type of
processes does not require $Q^2$ to be large. In fact the presence of
one large scale, be it transverse momentum or heavy quark mass, is
sufficient to perform perturbative calculations, even for the case of
$Q^2 \simeq 0 \gevtwo$.

As has been mentioned earlier charged leptons are a natural source of a
flux of photons and the propagator effect favors photons with $Q^2
\simeq 0 \gevtwo$. At HERA, electroproduction events with $Q^2 \sim
0 \gevtwo$ are called photoproduction events. In QCD the production of
large transverse momentum jets in photoproduction is very similar to
jet production in hadron-hadron interactions, which are sensitive to
the parton distributions in the hadrons.  Thus, in lepton-hadron
interactions, the study of the structure of matter can be extended to
include the structure of the photon.

\subsection{Space time picture of $ep$ scattering at HERA}
\label{sec:spacetime}

In discussing the QPM picture of DIS we have presented arguments why a
virtual photon was able to resolve substructures inside the nucleon
and why the interaction could be viewed as an incoherent sum of
elastic $eq \rightarrow eq$ scattering. For the latter we argued in
the Breit frame where it is easy to depict DIS scattering. We could go
one step further and ask what would the hadronic final state look like
in such an approach. In the Breit frame the quark which absorbed the
virtual boson moves in the opposite direction to the remnant of the
target nucleon.  The two states are each colored and we expect
radiation, which eventually turns into hadrons, to fill up the
rapidity space between them.  Thus in the HERA frame of $ep$
interactions we expect the hadronic final state to consist of a jet of
hadrons originating from the fragmentation of the scattered quark
which balances the transverse momentum of the electron, a jet of
particles around the original direction of the proton and some
hadronic activity between the two.  This is schematically depicted in
Fig.~\ref{fig:colorflow}.

\epsfigure[width=0.95\hsize]{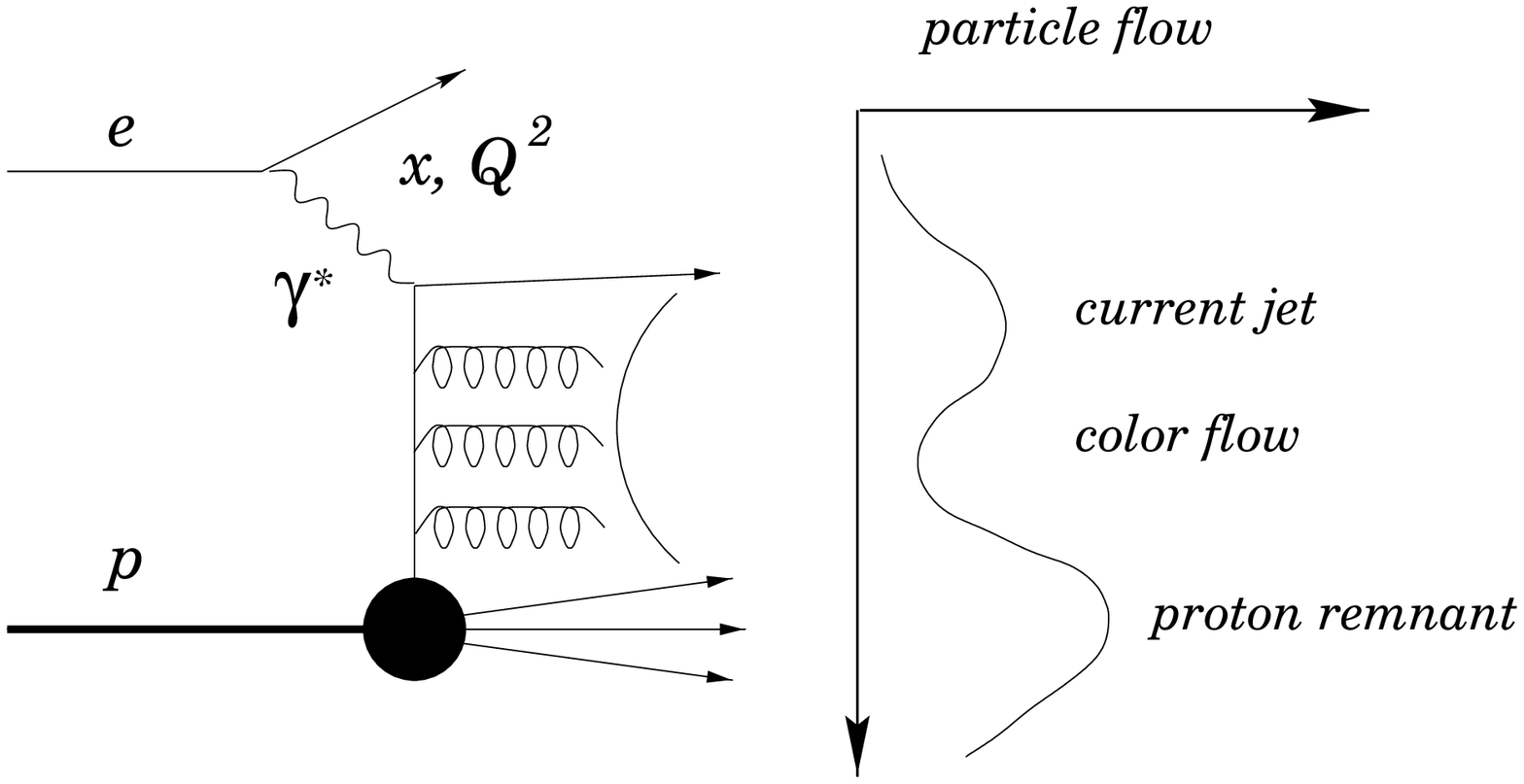}{
  Schematic representation of the expected energy flow in deep
  inelastic scattering.  The need for color flow between the struck quark
  and proton remnant generates particles distributed in rapidity (or
  pseudorapidity, $\eta$) between these.}{colorflow}

It is also of interest to consider $ep$ scattering in the rest frame
of the target. We will be mainly interested in the fate of the photon
in this frame and the coordinate system will be rotated such that the
virtual photon moves along the $z$ axis. The four momentum vector of
the photon in this frame is
\begin{equation}
q = (\nu,0,0,\sqrt{\nu^2+Q^2}) \, .
\end{equation}
According to quantum mechanics, we can think of the photon as
fluctuating with some probability into states of $q \bar{q}$,
$e^+e^-$, $\mu^+ \mu^-$, etc... The life time of such a quantum
mechanical fluctuation is given by the Heisenberg uncertainty
principle,
\begin{equation}
\tau \simeq \frac{1}{E_f-\nu} \, ,
\end{equation}
where $E_f$ is the energy of the state of mass $M_f$ in which the
photon happened to fluctuate. Here we will concentrate on the hadronic
fluctuation of the photon. For large $\nu$ the expression can be
approximated by
\begin{equation}
\tau \simeq \frac{2\nu}{M_f^2+Q^2} \, ,
\label{photonlifetime}
\end{equation}
where $M_f$ depends on the initial configuration of the $q\bar{q}$
system (for a derivation see for example~\citeasnoun{wusthoff-mf}), 
\begin{equation}
M_f^2=\frac{m_q^2+k_T^2}{z(1-z)} \, ,
\end{equation}
where $m_q$ is the mass of the quarks, $k_T$ their transverse momentum
relative to the photon and $z$ is the fraction of the photon momentum
carried by one of the quarks. 

The first observation we can make based on
expression~(\ref{photonlifetime}) is that the life time of a hadronic
fluctuation of the photon increases as its energy increases and
decreases as $Q^2$ increases. If the hadronic fluctuation lives long
enough to overlap with the size of the target the interaction of the
photon will proceed through its hadronic component. In the early days
it was natural to assume that the photon would turn into a vector
meson, preferably the $\rho^0$ meson as the one with the lowest mass.
This was the essence of the Vector Dominance Model
(VDM)~\cite{Sakurai} which explained why real photons behaved as
hadrons. The fact that with increasing $Q^2$ the life time of a
hadronic fluctuation would decrease made it natural to view the
virtual photon as a point like probe. However, it was also
realized~\cite{bjorken-cornell} that at sufficiently large $\nu$ a
virtual photon could acquire hadronic properties.

When QCD is turned on, the whole picture acquires even more substance.
As a consequence of the renormalizability of QCD, 
fluctuations with $M_f^2 \gg Q^2$ can be neglected and $M_f^2$ may be
approximated by $Q^2$~\cite{Ioffe,AFS}, in which case
expression~(\ref{photonlifetime}) can be reduced to the form
\begin{equation}
\tau \simeq \frac{1}{2m_p x} \, .
\end{equation}
To quantify this relation, at $x=0.1$ the longitudinal dimension of
the hadronic fluctuation is of the order of 1 fm, the typical size of
a hadronic target.  
Thus it becomes clear that at small $x$,
independently of $Q^2$, the hadronic fluctuations of the photon have
to be resurrected with important consequences for the physics
of small-$x$ DIS interactions. 


Until now, we have tacitly assumed that any fluctuation into a $q
\bar{q}$ pair will lead the photon to look like a hadron. This turns
out not to be the case in QCD.  For the sake of simplicity we consider
two extreme cases of the $q \bar{q}$ initial configuration, one where
the initial $k_T$ is small and one where it is large. If the $k_T$ is
small the $q\bar{q}$ form a large size object (for fixed $M_f$, small
$k_T$ implies very different $z$ and $(1-z)$ values such that the $q$
and $\bar{q}$ are moving at very different speeds.)  The color dipole
moment is large and given enough time the space will be filled by
gluon radiation. This fluctuation is likely to acquire hadron like
properties and interact with the target as in hadron-hadron
interactions. The slower quark is expected to interact with the target
while the faster one continues in the original direction of the photon
and fragments into hadrons.

If $k_T$ is large, then $z$ and $(1-z)$ have similar values and 
the $q$ and the $\bar{q}$ are spatially close to each other.  The
effective charge of such a dipole is very small and thus their
interaction cross section is expected to be small.  Such a small size
wave packet can resolve the partonic structure of the target hadron.
This is the essence of what is called the color transparency
phenomenon. It has been shown that the interaction cross section of
a small size colorless configuration is proportional to the gluon
distribution in the target~\cite{BBFS93},
\begin{equation} 
\sigma_{T}^{q \bar{q}} = \frac{\pi^2}{3} \as b^2 xG_{T}(x, 9/b^2),  
\label{eq:twoglu} 
\end{equation}
where $b$ is the transverse separation between the $q \bar{q}$ system
and $G_T$ stands for the gluon distribution in the target. Thus
effectively the contribution of small size configurations to the cross
section may be large when the density of gluons is large.

In summary, for $Q^2 \sim 0 \gevtwo$ the photon acquires a hadronic
structure. We thus expect features very similar to the ones observed
in hadron-hadron interactions, including hard scattering between the
parton content of the hadronic fluctuation of the photon and the
hadron target. These hard processes are called resolved photon
processes. An additional component is due to interactions of a photon
which fluctuated into a small size configuration, which gives rise to
the anomalous component of the photon structure.

The picture emerging in DIS is very similar to that derived from the
QPM in the Breit frame. Because of the dominance of small $k_T$
configurations to the cross section, corresponding to asymmetric
parton configurations in the photon fluctuation, the final state
consists of a current jet and proton remnant.  However, the hadronic
nature of the interaction implies that we should expect the same type
of contributions as in hadron-hadron interactions. In particular, the
presence of diffractive states, with large rapidity gaps separating
the photon and the proton fragmentation regions are very natural,
while they are hard to predict in the QCD improved parton model, where
the presence of large rapidity gaps is strongly suppressed.

QCD corrections to the simple QPM picture arise naturally when small
size $q\bar{q}$ configurations are allowed. In fact they give rise to
a special new class of perturbative interactions such as diffractive
production of jets or the exclusive production of vector mesons by
longitudinally polarized photons, mediated by two-gluon exchange.

In this approach, it also becomes clear that DIS scattering, in
particular at small $x$, is a result of an interplay of soft and hard
interactions.


\section{Experimental aspects} 
\label{sec:experimental} 

\subsection{The HERA accelerator} 
\label{sec:hera} 
The HERA (Hadron-Electron Ring Anlage) machine is the world's first
lepton-nucleon
collider~\cite{ref:HERA1,ref:HERA2,ref:HERA3,ref:HERA4}.  It is
located in Hamburg, Germany, and has been providing luminosity to the
colliding beam experiments H1 and ZEUS since the summer of 1992.  It
is schematically shown in Fig.~\ref{fig:HERA_layout}, along with the
pre-HERA accelerator elements. The basic HERA operational parameters
are given in Table~\ref{tab:HERA}.  HERA was approved in 1984, and was
built on schedule.  The electron machine was first commissioned in
1989, while the proton ring was first operated in March 1991.  First
electron-proton collisions were achieved in October 1991.  Following
this commissioning of HERA, the two colliding beam detectors ZEUS and
H1 moved into position to record data.

\epsfigure[width=0.8\hsize]{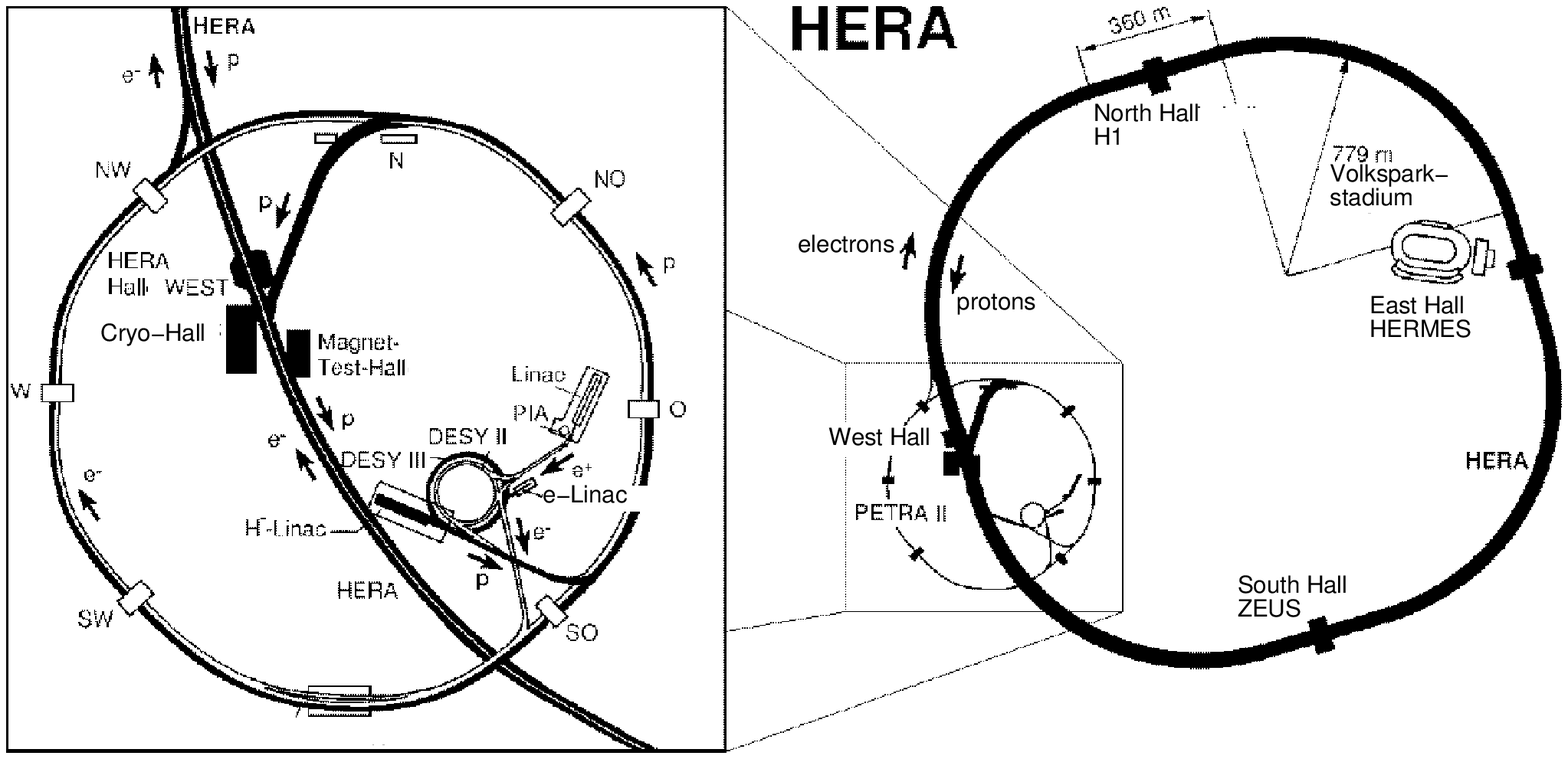} {The layout of HERA
  is shown on the right, along with the location of the different
  detector halls.  The pre-accelerators are shown in the blowup on the
  left.}  {HERA_layout}

\begin{table} 
\tablecaption{The main HERA parameters as
of the end of the 1997 running period.}
\label{tab:HERA}
\begin{center}
\begin{tabular}{ccc}
Parameter & Achieved \\
\hline
$E_p$ (GeV) & $820$ \\
$E_e$ (GeV) & $27.5$ \\
$I_e$ (mA) &  $40$    \\
$I_p$ (mA) &  $100$   \\
\# bunches &  $189$ \\
Time between crossings &  $96$~ns \\
$\sigma_x$ at IP ($\mathrm{\mu m}$) & $179$ \\
$\sigma_y$ at IP ($\mathrm{\mu m}$) & $48$ \\
$\sigma_z^{\mathrm{proton}}$ at IP (mm) & $200$ \\  
${\mathcal{L}}_{\mathrm{inst}}$ ($\mathrm{cm^{-2}\,s^{-1}}) $ 
                           & $1.4\cdot 10^{31}$ 
\end{tabular}
\end{center}
\end{table}

The HERA electron ring operates at ambient temperatures, while the
proton ring is super-conducting.  The two beam pipes merge into one at
two areas along the circumference. The beams are made to collide at
zero crossing angle to provide $ep$ interactions for the experiments
H1 and ZEUS.  These detectors will be described in more detail in
further sections.  The electrons (positrons) and protons are bunched,
with bunches within one bunch train separated by $96$~ns.  Some number
of bunches are left unpaired (i.e., the corresponding bunch in the
other beam is empty) for background studies.  The electron (positron)
beam is polarized up to $70$\,\% in the transverse direction via the
Sokholov-Ternov effect~\cite{ref:HERApolarization}.  A third
experiment, HERMES~\cite{ref:HERMES}, makes use of this polarized beam
by colliding it with a polarized proton gas jet to study the spin
structure of the proton.  A fourth experiment,
HERA-B~\cite{ref:HERAB}, is currently being assembled.  It uses wire
targets in the proton beam to study B hadron production and decay with
large statistics in an effort to find CP violation in the B hadron
sector.

\epsfigure[width=0.8\hsize]{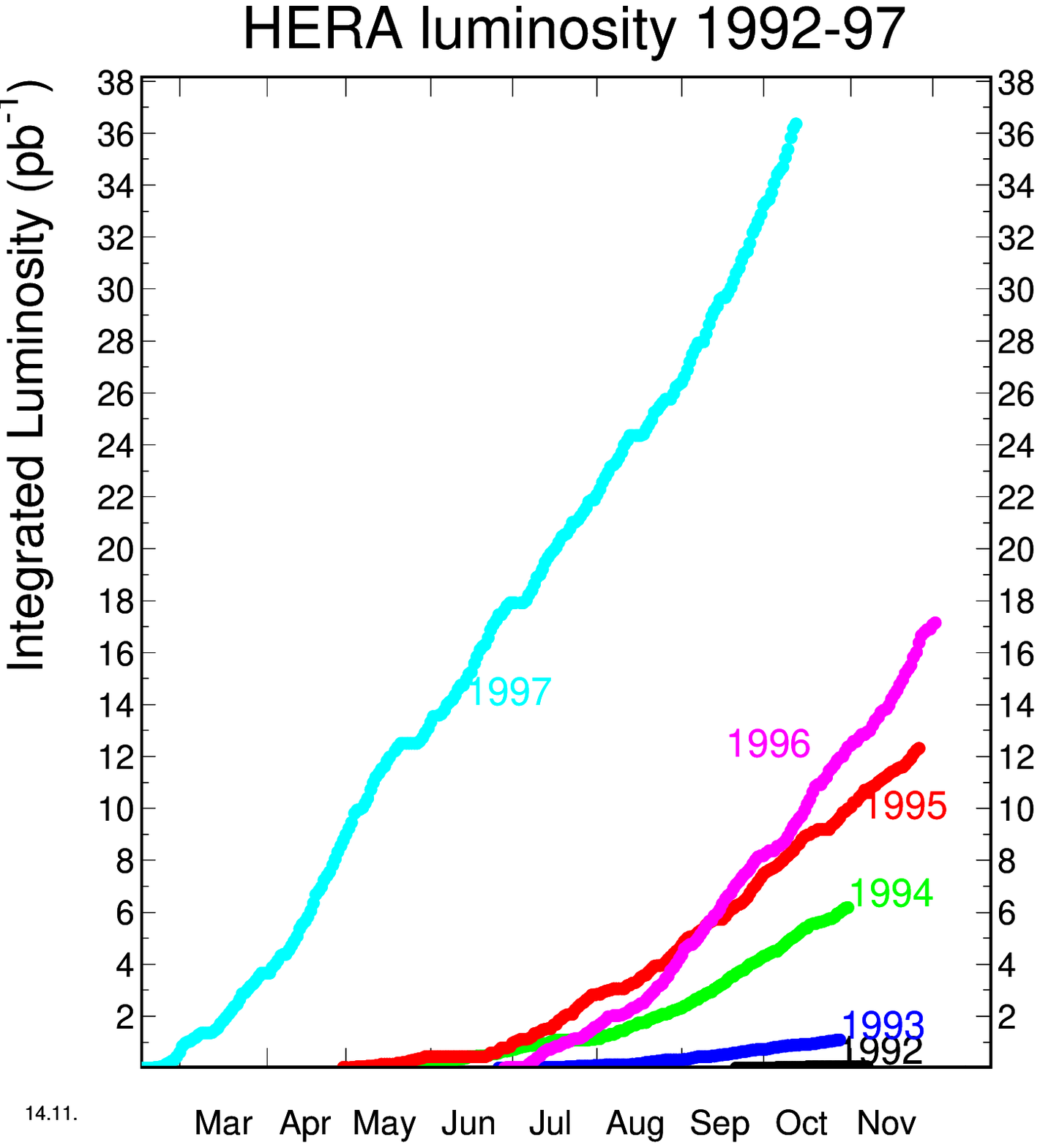} {The integrated
  luminosity delivered by HERA versus the date for the years since the
  start of HERA operation.}  {HERA_lumi}

The integrated luminosities per year are shown in
Fig.~\ref{fig:HERA_lumi} as a function of the day of the run.  These
luminosity profiles are comparable to those achieved at other
successful accelerator facilities such as LEP (CERN, Geneva) or the
Tevatron (FNAL, USA).  Note that HERA initially began as an
electron-proton collider, but switched to positron-proton collisions
in 1994 once it was determined that the electron lifetime was severely
limited at high currents.  It is thought that this is the result of
electrons interacting with positively charged ``macro''-particles in
the beam pipe.  New pumps have been installed in the electron ring in
the 1997/1998 shutdown to cure this problem.  In its 1997
configuration, HERA accelerated protons to $820$~GeV and positrons to
$27.5$~GeV.  The proton beam energy has been increased to $920$~GeV as
of August 1998.  The maximum instantaneous luminosity achieved so far
is ${\mathcal L}=1.4 \cdot 10^{31}$~cm$^{-2}$~s$^{-1}$ with 174
colliding bunches of electrons and protons, to be compared with a
design goal of ${\mathcal L} =1.5 \cdot 10^{31}$~cm$^{-2}$~s$^{-1}$
with 210 colliding bunches.  There is a significant luminosity upgrade
program planned for the HERA accelerator which should result in
luminosities of ${\mathcal L} = 7.5 \cdot 10^{31}$~cm$^{-2}$~s$^{-1}$.
This upgrade is currently planned for the 1999/2000 break.  Another
planned upgrade is the introduction of spin rotators to provide
longitudinal polarization to the colliding beam experiments.  These
upgrades are described in more detail in section~\ref{sec:future}.

\subsection{The detectors H1 and ZEUS} 
\label{sec:detectors} 
 In this section, we briefly review the properties of the two
colliding beam detectors H1 and ZEUS, run by the eponymous
collaborations.  Both are general purpose magnetic detectors with
nearly hermetic calorimetric coverage. They are differentiated
principally by the choices made for the calorimetry.  The H1
collaboration has stressed electron identification and energy
resolution, while the ZEUS collaboration has put its emphasis on
optimizing the calorimetry for hadronic measurements.  The detector
designs reflect these different emphases. The H1 detector has a large
diameter magnet encompassing the main liquid argon calorimeter, while
the ZEUS detector has chosen a uranium-scintillator sampling
calorimeter with equal response to electrons and hadrons.  The
detectors are undergoing continuous changes, with upgrades being
implemented and some detector components being removed or simply not
used.

We review here the capabilities of the two different detectors for
tracking of charged particles, energy measurements, and particle
identification.  For detailed information on the detectors, the reader
should refer to the technical proposals and status 
reports: H1~\cite{ref:H1_technical_proposal,ref:H1_technical_report} and
ZEUS~\cite{ref:ZEUS_technical_proposal,ref:ZEUS_status_report}.

\epsfigure[width=0.95\hsize]{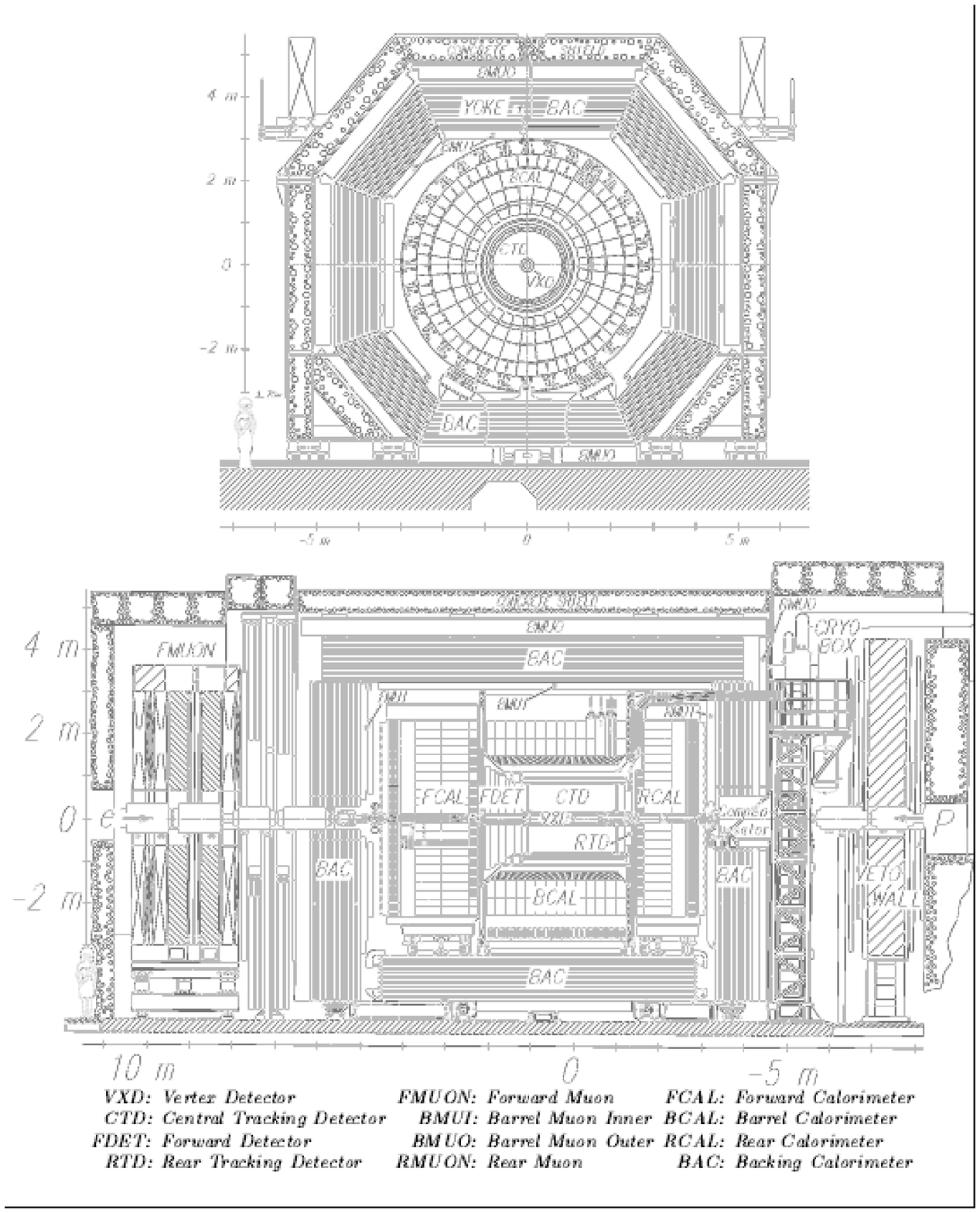} {Cross sectional
  views of the ZEUS detector.  } {ZEUS_detector}

\epsfigure[width=0.90\hsize]{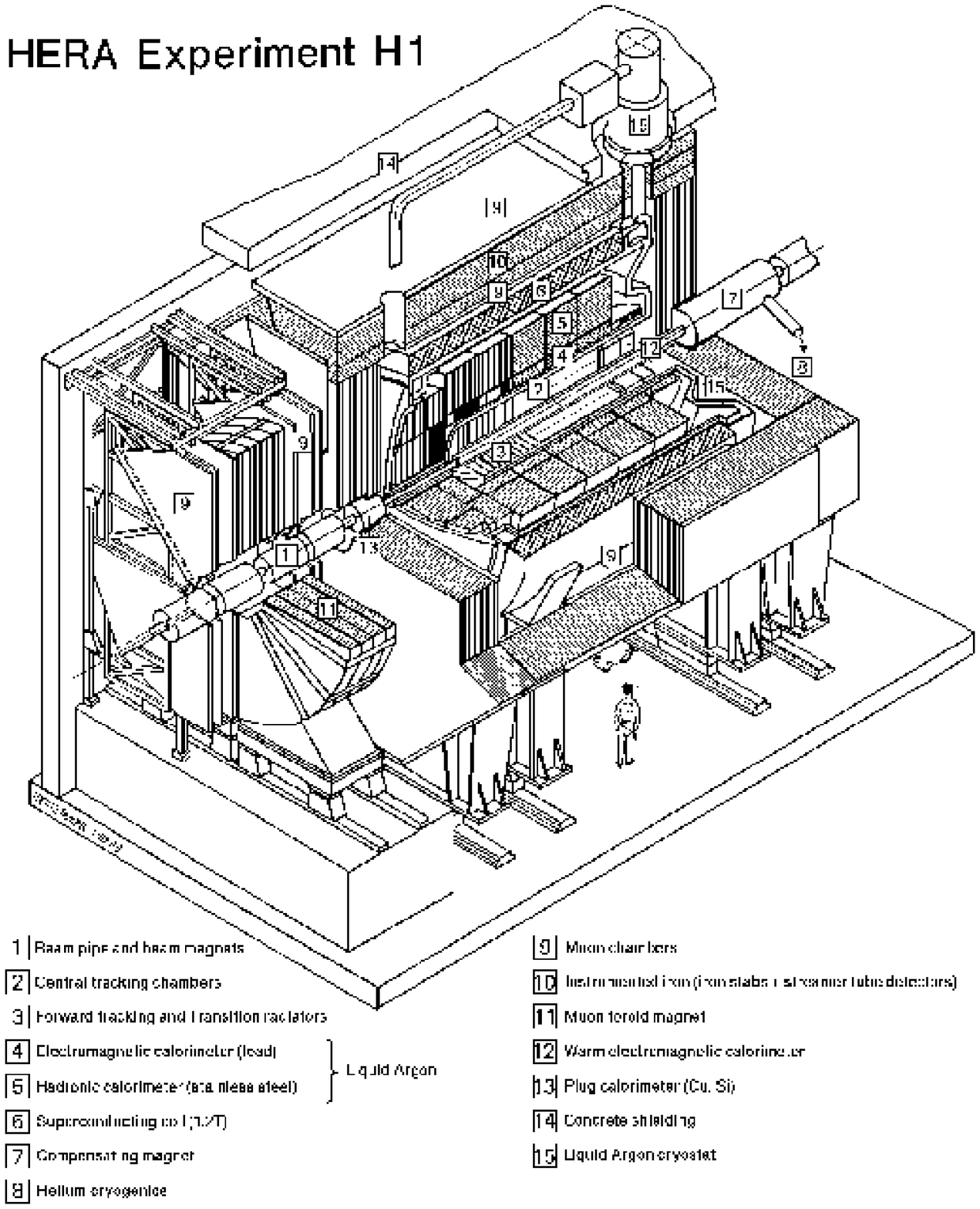} {Schematic drawing of
  the H1 detector.}  {H1_detector}
 
\begin{table} 
\tablecaption{The principal H1 central detector parameters (1997 status).}
\label{tab:H1_detector}
\begin{center}
\rotatebox{-90}{%
\begin{tabular}{clcp{5.5cm}}
Component & Parameter & Value & Comment \\
\hline
LAr  & Angular coverage & $4^{\circ} \leq \theta \leq 154^{\circ}$
& ref~\cite{ref:H1_LArcal}\\
Calorimeter& $\sigma/E$(EM showers) 
& $0.12/\sqrt{E(\mathrm{GeV})}\oplus 0.01$ & 
test beam~\cite{ref:H1_cal1} \\
&&& \cite{ref:H1_cal2} \\
& EM E scale uncertainty& $1-3$\,\% & ref~\cite{ref:H1_highQ2} \\
& $\sigma/E$(Hadronic showers) &$0.50/\sqrt{E(\mathrm{GeV})}\oplus 0.02$ & 
 test beam~\cite{ref:H1_cal1} \\
&&& \cite{ref:H1_cal2} \\
& had E scale uncertainty& $4$\,\% & ref~\cite{ref:H1_highQ2}  \\
& angular resolution & $2\units{mrad}$ & $\theta < 30^{\circ}$ \\
& angular resolution & $< 5\units{mrad}$ & 
$\theta > 30^{\circ}$ \\
\hline
SPACAL  & Angular coverage & $153^{\circ} \leq \theta \leq 177.8^{\circ}$
& ref~\cite{ref:H1_LArcal}\\
& $\sigma/E$(EM showers) & $0.075/\sqrt{E(\mathrm{GeV})}\oplus 0.025$ & 
in situ~\cite{ref:H1F2_95_SVTX} \\
& EM E scale uncertainty& $1\rightarrow 3$\,\% & 
 ref~\cite{ref:H1F2_95_SVTX} \\
& had E scale uncertainty& $7$\,\% & ref~\cite{ref:H1F2_95_SVTX}  \\
& spatial resolution & $4\units{mm}$ & ref~\cite{ref:H1F2_95_SVTX}  \\
& time resolution & $<1$~ns &  ref~\cite{ref:H1F2_95_SVTX} \\
\hline
Central & B-field (Tesla) & $1.15$ & \\
Tracking    & angular coverage & $15^{\circ} < \theta < 165^{\circ}$ & \\
                 & $\sigma/p_T$ & $0.01\cdot p_T(\mathrm{GeV})$ &
Full length tracks~\cite{ref:H1F2_95_SVTX} \\
\hline
Luminosity & normalization uncertainty & $1.5$\,\% & 
ref ~\cite{ref:H1F2_94} 
\end{tabular}
}
\end{center}
\end{table}

\begin{table}
\tablecaption{
  The principal ZEUS central detector parameters (1997 status).
}
\begin{center}
\rotatebox{-90}{%
\label{tab:ZEUS_detector}
\begin{tabular}{clcp{5.5cm}}
Component & Parameter & Value & Comment \\
\hline
Calorimeter & Angular coverage & $2.6<\theta<176.2^{\circ}$ &
Extended to $178.4^{\circ}$ in 1995 \\
& $\sigma/E$(EM showers) & 
$0.18/\sqrt{E(\mathrm{GeV})}\oplus 0.02$ & test beam~\cite{ref:ZEUS_cal2} \\
&&& \cite{ref:ZEUS_cal3} \\
& EM E scale uncertainty& $1-3$\,\% & ref~\cite{ref:ZEUS_highQ2} \\
& $\sigma/E$(Hadronic showers) &$0.35/\sqrt{E(\mathrm{GeV})}\oplus 0.03$ & 
 test beam~\cite{ref:ZEUS_cal2} \\
&&& \cite{ref:ZEUS_cal3} \\
& had E scale uncertainty& $3$\,\% & ref~\cite{ref:ZEUS_highQ2}  \\
& position resolution & $\sim 1\units{cm}$ & 
EM showers, ref~\cite{ref:ZEUSF2_94} \\
& time resolution & $<1\units{ns}$ & for $E>4.5\units{GeV}$ \\
\hline
Central  & B-field (Tesla) & $1.43$ & \\
Tracking    & angular coverage & $11^{\circ} < \theta < 168^{\circ}$ & \\
                 & $\sigma/p_T$ & $0.005\cdot p_T(\mathrm{GeV}) \oplus 0.016$ &
Full length tracks~\cite{ref:ZEUS_highQ2} \\
                 & $z$ vertex resolution & $0.4\units{cm}$ & 
Full length tracks, $p_T>5\units{GeV}$ \\
                 & $R-\phi$ vertex resolution & $0.1\units{cm}$ & 
Full length tracks, $p_T>5\units{GeV}$ \\
\hline
Luminosity & normalization uncertainty & $1.5$\,\% & 
ref ~\cite{ref:ZEUS_highQ2} 
\end{tabular}
}
\end{center}
\end{table}

A cross sectional view of the ZEUS detector is presented in
Fig.~\ref{fig:ZEUS_detector}. The H1 detector is shown in
Fig.~\ref{fig:H1_detector}.  The ZEUS detector consists of tracking
chambers inside a superconducting solenoidal magnet, surrounded by
calorimeters and muon chambers.   The H1 detector has several
different tracking detectors inside the calorimeter.  The
superconducting solenoid is placed outside the calorimeter to minimize
the amount of inactive material in the path of electrons.  Not shown
in the figures are luminosity detectors and electron detectors
downstream in the direction of the electron beam, and a proton
spectrometer and neutron calorimeter in the direction of the proton
beam.  The principal central detector parameters are given in
Tables~\ref{tab:H1_detector} and ~\ref{tab:ZEUS_detector}.

\subsubsection{The tracking detectors}

The ZEUS tracking detectors consist of a vertex detector 
(VXD), a central tracking detector 
(CTD), and forward and rear tracking detectors (FTD,RTD).

The VXD is a jet chamber with small layer spacing ($3$~mm),
which allows the measurement of 12 coordinates 
within the range $10.0 < R < 15.6$~cm,
and the slow drift gas (dimethyl ether) gives a resolution
$30$~$\mu$m at the center of the cell.  The VXD, in conjunction with
the CTD, has an impact parameter resolution of $40$~$\mu$m for high
momentum tracks.
The VXD was in operation since the beginning of data taking, but has 
been removed at the end of the 1995 running period.

The CTD is a drift chamber composed of 9 ``super'' layers, each
consisting of 8 wire layers. Of these, 5 are axial (along the $z$-axis
\footnote{The detector coordinate systems are chosen such that the
  proton beam points along the $z$-axis, the $y$-axis points vertical
  upward, and the $x$-axis points towards the center of the ring.  The
  nominal $ep$ interaction point is at $z=0$.})  superlayers and 4
stereo, allowing both an $R-\phi$ and a $z$ coordinate measurement.
The CTD has cells oriented at a $45^{\circ}$ angle to the radial
direction to produce drift lines approximately tangential to the
chamber azimuth in the strong axial magnetic field ($1.43$~T) provided
by the superconducting solenoid. This cell orientation also guarantees
that at least one layer per superlayer will have a drift time shorter
than the bunch crossing interval of 96~ns.  The CTD has a design
momentum resolution in a $1.7$~Tesla magnetic field of

\begin{equation}
\sigma/p = 0.002p(\mathrm{GeV}) \oplus 0.003
\end{equation}
at $\theta=90^{\circ}$, and a z coordinate resolution of $1$~mm from
the stereo wires.  In addition to its primary function of measuring
the momentum of charged particles, the CTD also provides particle
identification information via $dE/dx$.  A resolution of
\begin{equation}
\frac{\sigma(dE/dx)}{(dE/dx)} = 0.06
\end{equation}
is expected.

The FTD consists of 3 planar drift chambers and extends the tracking
region in the forward region to $7.5^{\circ} < \theta < 28^{\circ}$,
where high particle densities are expected due to the Lorentz boost in
the proton beam direction.  The transition radiation detector (TRD), a
tool for identifying electrons in the forward direction, is placed
between the FTD chambers.  The RTD consists of one plane of drift
chambers covering the angular range $160^{\circ}<\theta<170^{\circ}$.
Each of the FTD and RTD drift chambers consists internally of three
layers of drift cells, with the second and third wire layers rotated
by $+60^{\circ},-60^{\circ}$ with respect to the first layer.  The
design resolution is $120-130$~$\mu$m, and the two-track resolution is
$2.4$~mm.

The H1 tracking detectors consist of central jet chambers 
(CJC1,CJC2), central trackers for measuring the z 
coordinate (CIZ,COZ), forward tracking detectors 
(FTD), rear tracking detectors (BDC),
 and central and rear silicon microvertex detectors  (CST, BST).

The central jet chambers (CJC1,CJC2) are two large, concentric drift 
chambers.
The inner chamber, CJC1, has 24 layers of sense wires arranged in 30 phi 
cells, while CJC2 has 32 layers of sense wires in 60 phi cells.  The cells 
are at a 30$^{\circ}$ angle to the radial direction.  The point resolution 
is $170 \; \mu$m in the $R-\phi$ direction.  The $z$ coordinate is measured by charge division and has an accuracy of $22$~mm. Test beam results indicate a momentum resolution for the CJC of  
\begin{equation}
\sigma/p = 0.003p(\mathrm{GeV}) \, .
\end{equation}
The $dE/dx$ resolution is expected to be $6$~\%, as for the ZEUS CTD.

The CIZ and COZ are thin drift chambers with sense wires perpendicular to the 
beam axis, and therefore complement the accurate $R-\phi$ measurement provided
 by the CJC by providing accurate $z$ coordinates.  The CIZ is located inside
 the CJC1, while COZ is located between CJC1 and CJC2.  These two chambers 
deliver track elements with typically $300$~$\mu$m resolution in $z$.

The forward tracking detectors (FTD) are integrated assemblies of three 
supermodules, each including, in order of increasing $z$: three different 
orientations of planar wire drift chambers (each rotated by $60^{\circ}$ to 
each other in azimuth), a multiwire proportional chamber (FWPC) for fast 
triggering, a transition radiation detector and a radial wire drift chamber.  
The FTD is designed to give a momentum resolution of
\begin{equation}
\sigma/p = 0.003p(\mathrm{GeV})
\end{equation}
and track angular separation $\sigma_{\theta,\phi}<1$~mrad.

A backward proportional chamber (BPC) located just in front on the rear 
calorimeter provided an angular measurement of the electron, together with 
the vertex given by the main tracking detectors.  This detector has
been replaced in the 1994/95 shutdown by an eight layer drift chamber
(BDC) with a polar angle acceptance between $155.1 < \theta < 177.5^{\circ}$
~\cite{ref:BDC}.

\subsubsection{Calorimetry}  
The ZEUS tracking detectors are surrounded by a $^{238}$U-scintillator 
sampling calorimeter, covering the angular range  
$2.2^{\circ}<\theta<176.5^{\circ}$\footnote{The range was extended
in 1995 by placing the rear calorimeter closer to the beam, resulting
in the coverage $2.2^{\circ}<\theta<178.4^{\circ}$}.  
This calorimeter design was chosen to 
give the best possible energy resolution for hadrons.  The calorimeter 
consists of a forward part (FCAL), a barrel part (BCAL), and a rear part  
(RCAL), with maximum depths of $7.1 \lambda,\; 5.3 \lambda,\; {\rm and} \; 
4.0 \lambda$, respectively. The FCAL and BCAL are segmented longitudinally 
into an electromagnetic section (EMC), and two hadronic sections  (HAC1,2).  
The RCAL has one EMC and one HAC1 section.  The cell structure is formed by  
scintillator tiles;  cell sizes range from $5 \times 20$~cm$^2$  (FEMC) to 
$24.4 \times 35.2$~cm$^2$ at the front face of a BCAL HAC2 cell. The light 
generated in the scintillator is collected on both sides of the module by 
wavelength shifter (WLS) bars, allowing a coordinate measurement based on 
knowledge of the attenuation length in the scintillator.  The light is 
converted into an electronic signal by photomultiplier tubes (PMTs).

The performance of the calorimeter has been measured in detail in test beams, 
and some results are summarized in Table~\ref{tab:ZEUS_detector}. The 
signal from the $^{238}$U radioactivity has proven to be an extremely valuable 
calibration and monitoring tool.  The uranium activity signal is reproducible 
to better than $0.2$~\%. Test beam studies have shown that the inter calibration
 between cells of a module, and from module to module, is known at the $1$~\%
 level by setting the PMT gains in such a way as to equalize the uranium 
signal.  Despite the presence of the uranium activity, the calorimeter has 
very low noise (typically $10$~MeV for an EMC PMT and $20$~MeV for a HAC PMT).

The angular coverage in the electron beam direction was extended in the
1994/1995 shutdown with the addition of a small Tungsten-scintillator 
calorimeter (BPC) located behind the RCAL at $z=294$~cm, 
and within $4.4$~cm of the beam.
This calorimeter~\cite{ref:ZEUSF2_BPC} measures electrons in the angular range
$15$ to $34$~mrad.

Particle identification in the uranium calorimeter is enhanced by the addition
 of a silicon pad array (HES) near shower maximum in the 
RCAL and FCAL.  
All RCAL modules have been instrumented with these $3\times3.3$~cm$^2$ 
pads, while the FCAL is to be completed in the 1997/1998 shutdown.   
This detector is expected to improve electron recognition in jets by a 
factor of 10 to 20.

The high resolution calorimeter is surrounded by the backing calorimeter 
(BAC).  The BAC is formed by instrumenting the yoke used 
to guide the solenoidal field return flux, and consists of $40,000$ 
proportional tubes and $1,700$ pad towers, allowing an energy resolution of 
 $\sigma /E = 1.1/\sqrt{E}$.  The backing calorimeter allows for the 
correction or rejection of showers leaking from the uranium calorimeter. It 
is also useful for identifying muons.

The H1 detector places emphasis on electron recognition and energy 
measurement.  This led to placing the calorimeter inside the coil providing 
the axial field for the tracking detectors.  Liquid argon (LAr) was chosen 
because of its good stability, ease of calibration, fine granularity and 
homogeneity of response.  The LAr calorimeter covers the 
polar angle range between $4^{\circ}<\theta<154^{\circ}$.  The segmentation 
along the beam axis into ``wheels'' is eightfold, with each wheel segmented 
into octants in $\phi$. The hadronic stacks are made of stainless steel 
absorber plates with independent readout cells inserted between the plates.  
The orientation of the plates varies with $z$ such that particles always 
impact with angles greater than $45^{\circ}$.  The structure of the 
electromagnetic stack consists of a pile of G10-Pb-G10 sandwiches separated 
by spacers defining the LAr gaps.  The granularity ranges from 
$10\rightarrow 100$~cm$^2$ in the EMC section, to 
$50\rightarrow 2000$~cm$^2$ in the HAC section.
  Longitudinal segmentation is 3--4 fold in the EMC over 20--30 radiation 
lengths and 4--6 fold in the HAC over 5--8 interaction lengths.  The LAr
 calorimeter has a total of $45,000$ readout cells.  The noise per cell
 ranges from $15\rightarrow 30$~MeV. The resolution measured in the
test beam is given in Table~\ref{tab:H1_detector}.
The calorimeter is non-compensating, with the response to hadrons about 
$30$~\% lower than the response to electrons of the same energy.  An offline
 weighting technique is used to equalize the response and provide the optimal
 energy resolution.

The polar angle region $151^{\circ}<\theta<176^{\circ}$ of the H1 detector 
contains the backward electromagnetic calorimeter (BEMC). 
 This is a conventional lead-scintillator sandwich calorimeter used to 
measure the scattered electron for $Q^2 \leq 100$~GeV$^2$.  The calorimeter 
has a depth of $21.7 \; X_0$, or approximately $1$~hadronic interaction 
length, which on average contains 45~\% of the energy for a hadronic shower. 
 The detector is composed of $8 \times 8$~cm$^2$ stacks read out by 
wavelength shifter and photodiodes.  It has a two-fold segmentation in depth.
  The energy resolution is found to be

\begin{equation}
\sigma/E = 10~\%/\sqrt{E({\rm GeV})} \oplus 1.7~\% \; \; \; .
\end{equation}

The position resolution is $7$~mm for high energy electrons.  The
average noise per stack was measured to be $130$~MeV.  The
stack-to-stack calibration was performed using so-called kinematic
peak events, for which the electron has a well-defined energy, and is
better than $1$~\%.

The BEMC was replaced in the 1994/1995 shutdown with a lead/scinti\-llat\-ing
fiber calorimeter (SPACAL)~\cite{ref:BDC}.  The new calorimeter has both
electromagnetic and hadronic sections.  The angular region covered is
extended compared to the BEMC, and the calorimeter has very
high granularity (1192 cells) yielding a spatial resolution of about
$4$~mm.  Other parameters as measured with data are given in 
Table~\ref{tab:H1_detector}.

The LAr and BEMC calorimeters are  surrounded by a tail-catcher 
(TC) to measure hadronic energy leakage.  The TC is formed 
by instrumenting the iron yoke used to guide the solenoidal field with 
limited streamer tubes readout by pads.  The TC allows for
 the correction or rejection of showers leaking from the inner calorimeters. 
It is also useful for identifying muons. 

\subsubsection{Muon detectors}  
Recognition of muons is very important in the study of heavy quarks,
heavy vector mesons, $W$-production and
in the search for exotic physics. The ZEUS detector is surrounded by 
chambers to identify and measure the momentum of these 
muons.  The iron yoke making up the BAC is
 magnetized with a toroidal field of about $1.5$~T, and a momentum measurement
 is performed by measuring the angular deflection of the particle traversing 
the yoke.   In the barrel region and rear regions, LSTs are placed interior 
to, and  exterior to, the iron yoke.  A resolution of $20$~\% is expected for 
$10$~GeV muons. In the forward direction, where high muon momenta are 
expected, drift chambers (DC) and limited streamer tubes (LST) are used for 
 tracking. The momentum measurement, with a design goal of $20$~\% accuracy 
up to $100$~GeV, is enhanced with the aid of toroidal magnets residing outside
 the yoke.

The H1 detector measures muons in the central region by searching for 
particles penetrating the calorimeter and coil and leaving signals in the TC.
The TC is instrumented with 16 layers of LST.  Three are
 located before the first iron plate, and three after the last iron plate. 
 There is a double layer after four iron plates, and eight single layers in 
the remaining gaps between the iron sheets.  A minimum muon energy of 1.2~GeV
 is needed to reach the first LST, while 2~GeV muons just penetrate the iron.

In the very forward direction, a spectrometer composed of drift chambers 
surrounding a toroidal magnet with a field of 1.6~T is used to measure 
muons.  This spectrometer measures muons in the momentum range
 between $5 \rightarrow 200$~GeV.

\subsubsection{Forward detectors}

Both ZEUS and H1 have spectrometers downstream of the main detectors
in the proton beam direction to measure high energy protons, as well as
calorimeters at zero degrees to measure high energy neutrons.  These are
used in the study of diffractive scattering as well as in the study
of leading particle production.

The ZEUS leading proton spectrometer (LPS) is composed of 6 Roman pots 
containing silicon microstrip detectors placed within a few mm of the beam.  
The pots are located at distances from 24~m to 90~m from the 
interaction point (IP). The
track deflections induced by the proton beam magnets allow a momentum
analysis of the scattered proton.  The fractional momentum resolution
is $0.4$~\% for protons near the beam energy and $5$~MeV in the
transverse direction.  The effective transverse momentum resolution is
dominated by the beam divergence at the interaction point, 
which is about $40$~MeV in the horizontal
plane, and $90$~MeV in the vertical plane.
H1 installed a forward proton spectrometer (FPS) 
consisting of scintillating fibers in 1995 to detect
leading protons in the momentum range $500 < p < 760$~GeV and scattering
angles below $1$~mrad.

The ZEUS forward neutron calorimeter (FNC) is an iron-scintillator
sandwich calorimeter located $106$~m downstream of the interaction
point~\cite{ref:FNC}.  The calorimeter has a total depth of $10$
interaction lengths, and has a cross sectional area of 
$40\times30$~cm$^2$.  The FNC was calibrated using beam-gas data
with proton beams of different energies.
The H1 forward neutron calorimeter is located $107$~m 
downstream of the interaction point~\cite{ref:H1_FNC}.  The calorimeter
consists of interleaved layers of lead and scintillating fibers.  The
calorimeter has a total depth of $9.5$ 
interaction lengths.

\subsubsection{Luminosity detectors and taggers}
\label{sec:luminosity} 
The luminosity is measured at HERA via the  bremsstrahlung reaction

\begin{equation}
e \; p \rightarrow e \; p \; \gamma \; \; .
\end{equation}

Both ZEUS and H1 have constructed two detectors to measure this process: one to measure the electron, and the second to measure the photon.  The ZEUS collaboration has opted to use only the photon detector to measure luminosity, while the H1 collaboration uses a coincidence of photons and electrons.  The luminosity is determined from the  corrected rate of $\gamma$ or electron-$\gamma$ coincidence events, where the correction is found by measuring the rate induced from the unpaired electron bunches as follows:
\begin{equation}
R_{ep} = R_{tot} - R_{unp}\cdot \frac{I_{tot}}{I_{unp}} \; ,
\end{equation}

where $R_{tot}$ is the total rate within an energy window, $R_{unp}$ is
the rate measured in unpaired bunches, $I_{tot}$ is the total current and
$I_{unp}$ is the current in the unpaired electron bunches.

The ZEUS detectors consist of a $\gamma$ detector positioned $107$~m downstream from the IP, and an electron calorimeter  $35$~m from the IP.  The $\gamma$ detector consists of a carbon filter to absorb synchrotron radiation, an air filled Cerenkov counter to veto charged particles, and a lead scintillator sampling calorimeter.  The $\gamma$ detector has a geometrical acceptance of 98~\% independent of the photon energy for bremsstrahlung events. 
It also serves to measure the position and angular dispersion of the electron beam.  The electron detector is a lead-scintillator sampling calorimeter placed close to the beam line and measures electrons scattered at angles $\theta_e' < 6$~mrad, with an efficiency of greater than $70$~\%  for $0.35E_e < E_e' < 0.65E_e$~GeV.  This detector is also used to tag photoproduction events. 

The H1 detectors are a $\gamma$ detector positioned $103$~m downstream from the IP, and an electron calorimeter  $33$~m from the IP.  The $\gamma$ detector consists of a lead filter to absorb synchrotron radiation ($2 X_0$), a water filled Cerenkov counter ($1 X_0$), and a hodoscope of KRS-15 crystal Cerenkov counters with photomultiplier readout.
The $\gamma$ detector has a geometrical acceptance of 98~\% independent of the photon energy.  The electron detector has a similar construction, but without the lead and water absorbers and measures electrons  in the energy range $0.2E_e < E_e' < 0.8E_e$~GeV with an average efficiency of $48$~\%.  This detector is also used to tag photoproduction events.

In addition to the electron calorimeters of the luminosity systems, both ZEUS 
and H1 have small calorimeters located close to the beampipe at roughly $8$ and
$44$~m.  These measure scattered electrons in different energy ranges, and
thereby extend the kinematic range of tagged photoproduction events.

\subsubsection{Readout and triggering}  
The high bunch crossing frequency and large  background rates pose severe difficulties for the readout and  triggering.

The ZEUS trigger has three levels. The first level trigger must reach a decision time in $3$~$\mu$s and reduce the rate to less than $1$~kHz.  
Including various cable delays, the decision to keep an event must reach the front-end electronics within $5$~$\mu$s of the occurrence of the signal.  The first level trigger uses information from many detectors, and requires a global decision based on trigger information derived in parallel from the separate detectors.  Given the high background rates, some detectors have a significant chance for pile-up to occur during this interval, particularly in the regions near the beampipe. This requires a pipelining of the data for these detectors during the trigger decision time. Some pipelines are analog (e.g. in the calorimeter), where a dynamic range of 17~bits is required, while others are digital, where a lower dynamic range is required.  The pipelines are run synchronously with the HERA 10.4 MHz clock.  The minimum pipeline length is therefore about $55$ cells.  

Once a first level trigger is generated, data are read out from the
component pipelines into the second level of processing, where a
second level trigger decision based on more global event information
reduces the rate by a factor of $10$. As with the first level trigger,
the second level consists of parallel processing in component systems,
followed by a global decision in centralized processors.  The second
level trigger must be able to accept data at $1$~kHz, and come to a
decision within $10$~ms.  This requires substantial data buffering
(typically 16 events) in the component readout systems.  The remaining
events are then assembled and presented to an array of SGI processors.
At this point, a full event reconstruction is performed and events are
selected based on similar selection algorithms as used for the offline
analysis. The third level trigger typically reduces the rate by an
additional factor of $10$, such that data is written to offline
storage at typically $10$~Hz.

\epsfigure[width=0.8\hsize]{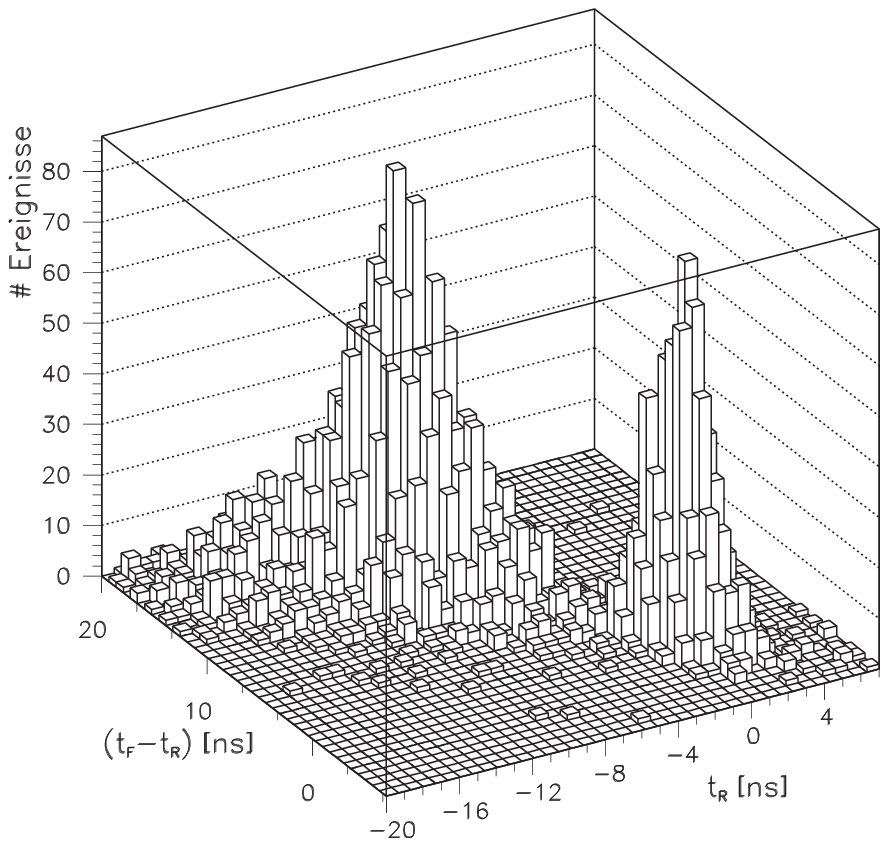} {The time measured in
  the rear part of the ZEUS calorimeter (RCAL) versus the difference
  in the times measured in the forward calorimeter (FCAL) and RCAL
  after the first level trigger.  A clear separation can be seen
  between $ep$ interactions, which are centered on $(0,0)$ and
  beam-gas interactions.}  {ZEUS_timing}
 
The precise timing available from the $^{238}U$-scintillator
calorimeter has been a particularly important tool for rejecting beam
induced backgrounds occurring upstream of the IP, as well as other
events which are asynchronous with the bunch crossing time (such as
cosmic rays and noise).  Figure~\ref{fig:ZEUS_timing} shows the
$t_{RCAL}$ versus $t_{FCAL} - t_{RCAL}$ distribution, where events
occurring upstream of the RCAL (giving $t_{FCAL} \approx 0$~ns,
$t_{RCAL} \approx -10.5$~ns) are clearly displaced from the events
occurring at the IP (with $t_{FCAL}, t_{RCAL} \approx 0$~ns).

The H1 trigger has four levels. The first level trigger has a decision delay of $2.5$~$\mu$s, which determines the minimum pipeline length needed to store the full detector information. As with the ZEUS trigger, correlations between the information from different detectors are used to make the trigger decision. H1 uses four different types of pipelines:

\begin{itemize}
\item Fast RAM (e.g., for drift chambers);
\item Digital shift registers (e.g., for systems readout with threshold discriminators);
\item Analog delay lines (in the BEMC);
\item Signal pulse shaping (in the LAr and TC, the pulse-shaping is adjusted in such a way that the maximum occurs at the time of the level one decision).
\end{itemize}

The second and third level triggers operate during the primary dead time of the readout.  They work on the same data as the first level trigger, and must reach a decision within $20$~$\mu$s and $800$~$\mu$s, respectively. The first three levels of triggering should not exceed 1~kHz, 200~Hz and 50~Hz, respectively. (The second and third level trigger systems were not in use in the first years of data taking, such that the first level trigger had to be limited to a rate of about 50~Hz.)

The fourth level of triggering is based on full event reconstruction in MIPS R3000 based processor boards.  Algorithms similar to the ones used for offline analysis are used to select valid events.  The fourth level filter rejects about 70~\% of the events, leading to a tape writing rate of about 15~Hz.

\subsection{Kinematics specific to HERA} 
\label{sec:HERAkin} 
HERA collides $27.5$~GeV electrons or positrons\footnote{In what follows, 
we will use the term electron to represent electrons or positrons, unless 
explicitly stated otherwise.} on $820$~GeV protons.  This leads to a 
center-of-mass energy squared
\begin{eqnarray}
s & = & (k+P)^2 \\
  & \approx & 4E_{e}E_{p} \\
  & \approx & (300\;\; {\rm GeV})^2
\end{eqnarray}
where $k$ and $P$ are the four-vectors of the incoming electron and proton, 
respectively.  To get an equivalent center-of-mass energy in a fixed target 
experiment would require a lepton beam of energy $E_{lepton}=s/2m_p$, or 
$E_{lepton}=47$~TeV, which is about two orders of magnitude beyond what can 
be achieved today.  It is therefore clear that HERA probes a very 
different kinematic regime to that seen by the fixed target experiments.  
For deeply inelastic scattering, the $Q^2$ range is extended to higher 
values by two orders of magnitude, while the range in the Bjorken-$x$ 
variable is increased by two orders of magnitude to smaller values for a 
fixed $Q^2$.  This allows measurements of the proton structure at much 
smaller transverse and longitudinal distance scales.

The HERA experiments H1 and ZEUS have almost hermetic detectors.  In the 
case of neutral current scattering, the kinematic variables can therefore 
be reconstructed from the electron, from the hadronic final state, or from 
a combination of the information from the electron and the hadrons.  The 
different techniques used to date are described in 
section~\ref{sec:methods}.

\epsfigure[width=0.8\hsize]{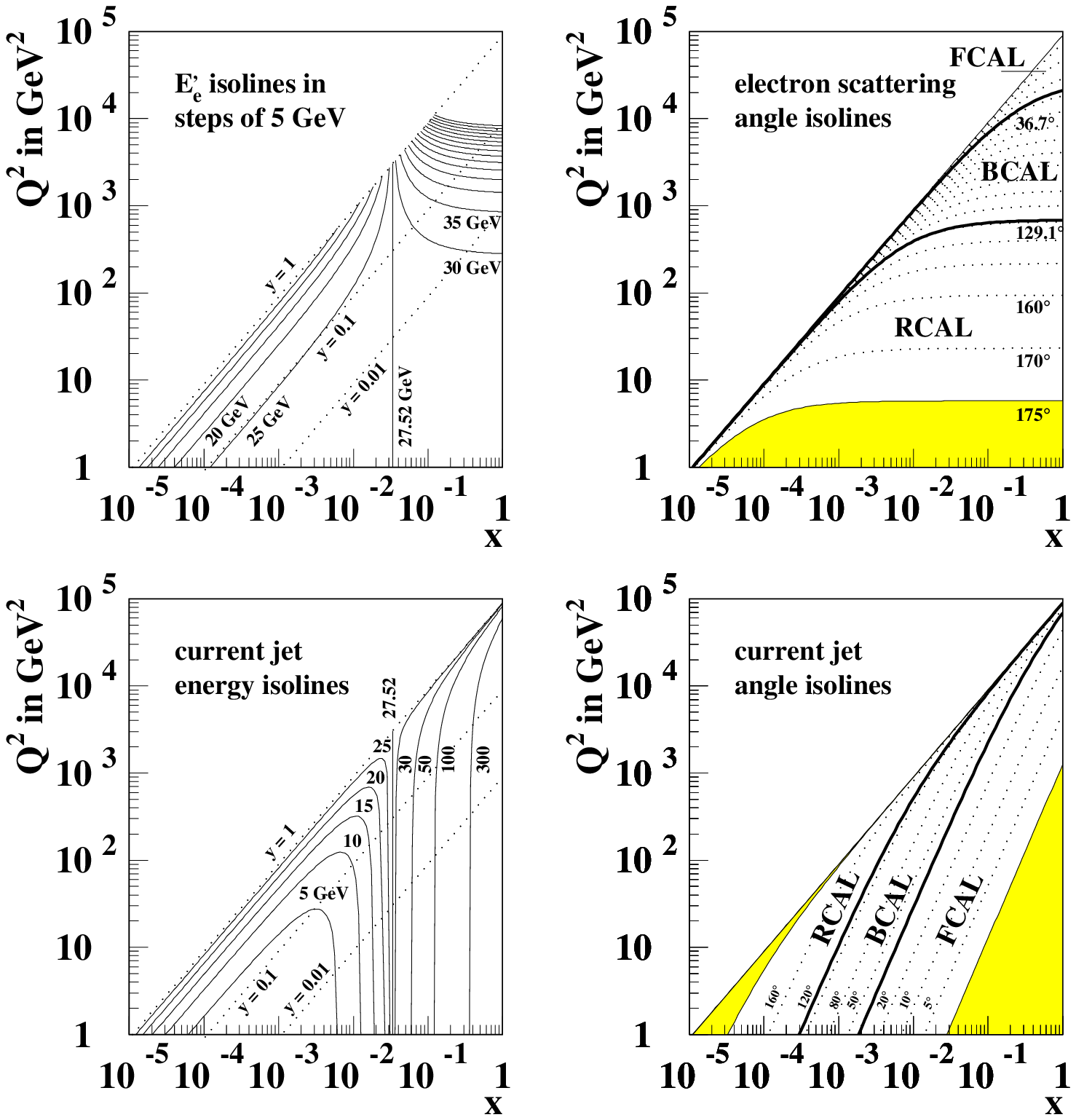} {Contour lines of
  fixed electron energy and angle, and fixed jet energy and angle in
  the ($x,Q^2$) plane.  The labels FCAL, BCAL and RCAL refer to the
  different sections of the ZEUS calorimeter.}  {HERA_kinematics}

In neutral current DIS, the interaction can be thought of as 
electron-quark elastic 
scattering.  The kinematics for DIS events are summarized in 
Fig.~\ref{fig:HERA_kinematics}, where the contours of constant electron 
energy and angle, and scattered quark energy and angle are drawn on the 
kinematic plane.  Most interactions involve small momentum transfers, 
and the electron is scattered at small angles.  In small-$x$ events the 
hadronic final state is generally boosted in the electron direction, while 
for large-$x$ events the hadronic final state is in the proton direction.  
There is an extended region around $x=E_e/E_p (=0.03)$ where the scattered 
electron has an energy close to the electron beam energy.  This region 
results in a ``kinematic peak'' in the electron energy spectrum, which is 
very useful for energy calibrations of the detectors.

Photoproduction is defined as the class of events where the square 
of the four-momentum transferred from the electron to the proton is 
very small (typically less than $10^{-2}$~GeV$^2$).  In this case, the 
electron is scattered at very small angles, and is not seen in the main 
detectors.  It is in some cases tagged by special purpose electron taggers 
(see section~\ref{sec:detectors}).  The exchanged gauge boson is then a 
quasi-real photon.  The kinematical variables most
 relevant for photoproduction are the center-of-mass energy of the
photon-proton system, $W$, and the transverse energy of the final
state, $E_T$.

\subsection{Kinematic variable reconstruction} 
\label{sec:methods} 
\subsubsection{Reconstruction of DIS variables}
The relevant kinematics of DIS events are specified with two variables,
as described in section~\ref{sec:kinematics}. 
There are several possible choices for these two variables.  Common choices 
are any two of ($x,y,Q^2,W$).  For the structure function measurements, the 
results are quoted in terms of $x$ and $Q^2$, while the natural variables to 
use for the total cross section measurement are $Q^2$ and $W$.  The 
experiments measure the energy, $E_e'$, 
and polar angle, $\theta_e$,  of the scattered electron, and the
longitudinal, $p_{z \; had}$ and transverse momentum of the hadronic final 
state, $p_{t \; had}$. There are many possible ways to combine these
measurements and reconstruct the kinematical variables.  
We review some of these here.

\begin{itemize}
\item[]Electron Method: 
\begin{eqnarray}
Q^2 & = & 2E_eE_e'(1+\cos\theta_e) \\
y & = & 1-\frac{E_e'}{2E_e}(1-\cos\theta_e) \\
x & = & \frac{Q^2}{sy}
\end{eqnarray}
This is the method which has historically been used in fixed target 
experiments.  It is in many ways the easiest method, since it only requires 
the measurement of one particle.  Its shortcomings are a seriously
degraded $x$ resolution at small $y$
and large radiative corrections.  The resolution is however very good
 at large $y$.

\item[]Hadron Method:
\begin{eqnarray}
\delta_{had} & = & \sum_{i=1}^{\# hadrons} E_i(1-\cos\theta_i) \\
             & = & E_{had} - p_{z \; had} \\ 
y & = & \frac{\delta_{had}}{2E_e} \\
Q^2 & = & \frac{p_{t \; had}^2}{1-y} \\
x & = & \frac{Q^2}{sy}
\end{eqnarray}
This method relies entirely on the hadronic system, and also goes under the 
name Jacquet-Blondel method~\cite{ref:Jacquet-Blondel}.  The sum indicated 
for the measurement of $\delta_{had}$ 
runs over all final state hadrons.  In practice, 
the final state hadrons are represented by tracks or calorimeter energy 
deposits. This method is stable against energy losses down the forward beam 
pipe since they contribute very little to $y$ or $p_T^2$.  However, it is 
sensitive to calorimeter noise at very small $y$, and is sensitive to energy 
losses in the rear direction at higher values of $y$.  It also requires a 
good understanding of energy scales and energy losses in inactive material.  
On the other hand, it is rather insensitive to radiative corrections.

\item[]Double Angle Method:
\begin{eqnarray}
\cos\gamma & = & \frac{p_{t\; had}^2 - \delta_{had}^2}{p_{t\; had}^2 + \delta_{had}^2}\\
Q^2 & = & 4E_e^2\frac{\sin\gamma(1+\cos\theta_e)}{\sin\gamma+\sin\theta_e-\sin(\theta_e+\gamma)} \\
x & = & \frac{E_e}{E_p}\frac{\sin\gamma+\sin\theta_e+\sin(\theta_e+\gamma)}{\sin\gamma+\sin\theta_e-\sin(\theta_e+\gamma)}
\end{eqnarray}
This method relies on the electron polar angle and the angle $\gamma$ which 
characterizes the hadronic final state~\cite{ref:Double-Angle}.  In a simple 
$eq \rightarrow eq$ picture with massless quarks, the angle $\gamma$ is the 
polar angle of the scattered quark. This reconstruction method has the 
advantage that it does not require precise 
knowledge of energy scales (variations 
in scale with polar angle will however distort the $\gamma$ measurement.)  
It also results in small radiative corrections.  However, the resolution is 
poor if $y$ is very small.

\item[]$\Sigma$--Method:  
\begin{eqnarray}
y & = & \frac{\delta_{had}}{\delta} \\
Q^2 & = & \frac{E_e'^2\sin^2\theta_e}{1-y} \\
x & = & \frac{Q^2}{sy}
\end{eqnarray}
This method~\cite{ref:H1F2_94} makes use of the longitudinal momentum 
conservation relation $E-P_z = 2E_e$.  The estimates of $E-P_z$ carried by 
the hadrons and electron respectively are $\delta_{had}$ and $\delta_e$, and 
$\delta=\delta_e+\delta_{had}$.  At small $y$, $\delta \approx \delta_e$, 
and is rather well measured, whereas 
at large $y$ both $\delta_{had}$ and $\delta_e$ are important.  $\delta_{had}$
suffers from significant energy 
losses (at the $15-20$~\% level), and a wide distribution. The 
$\Sigma$--method corrects in part for this loss.  The H1 collaboration has 
employed the $\Sigma$--method for $y<0.15$ in the measurement of $F_2$ from 
the 1994 data.  For higher $y$, they have used the electron method.  The 
$\Sigma$--method is rather insensitive to radiative corrections as it depends
 primarily on the hadronic variables.
  
\item[]$P_T$ Method:
\begin{eqnarray}
Q^2 & = & 4E_e^2\frac{\sin\gamma_{PT}(1+\cos\theta_e)}{\sin\gamma_{PT}+\sin\theta_e-\sin(\theta_e+\gamma_{PT})} \\
x & = & \frac{E_e}{E_p}\frac{\sin\gamma_{PT}+\sin\theta_e+\sin(\theta_e+\gamma_{PT})}{\sin\gamma_{PT}+\sin\theta_e-\sin(\theta_e+\gamma_{PT})} 
\end{eqnarray}
The $P_T$ method~\cite{ref:ZEUSF2_94} was developed by ZEUS in the analysis 
of the 1994 data.  The double-angle equations are used to calculate the 
kinematic variables, but with an improved estimate of the hadronic angle 
$\gamma$.  This estimate makes use of the $p_T$ balance in NC DIS events, 
whence its name.  The $p_T$ measured by the electron is used in the 
calculation of $\gamma$. It is also used to improve the measurement of 
$\delta_{had}$, as is longitudinal momentum balance in the form of the 
$\Sigma$--method.  The ZEUS collaboration has employed the $P_T$ method 
over the full kinematic range for the extraction of $F_2$ in the 1994 data.
\end{itemize}

\subsubsection{Reconstruction of photoproduction variables}
The primary variable used in photoproduction is the hadronic center-of-mass 
energy $W$.  In tagged photoproduction, this is calculated from the energy of 
the scattered electron:
\begin{equation}
W^2 \approx sy \approx s(1-E_e'/E_e) \; \; \; .
\end{equation}
The resolution is then given by the resolution of the electron calorimeters, 
which is typically $\sigma_E(GeV) \approx 0.25\cdot \sqrt{E(GeV)}$.  The 
energy scale is very well known as it is calibrated using Bremsstrahlung 
events.

In untagged events, the hadronic measurement of $y$ described above must be 
used.  This has much poorer resolution and has systematic shifts.  It requires 
good knowledge of the hadronic energy scale as well as a detailed 
understanding of the noise.

Jet transverse energies are needed for measurements in hard photoproduction. 
These require a precise knowledge of the hadronic energy scale as well as 
energy losses resulting from inactive material.  Energy smearing from the jet 
finding algorithm are also important.  Typical jet energy resolutions are 
$\sigma_E(GeV) \approx 1.0\cdot \sqrt{E(GeV)}$.  Uncorrected energies are 
about $20$~\% too low, and are corrected using Monte Carlo methods.

\subsubsection{Reconstruction of variables for exclusive final states}

The standard kinematical variables are often much better reconstructed in 
exclusive processes, where in some cases the final state is known and 
energy-momentum conservation can be used effectively.  An example is 
vector meson production ($ep \rightarrow epV$), described in
section~\ref{sec:VM}, where the vector meson $V$ is 
reconstructed with the tracking detectors (e.g., $\rho^0 \rightarrow 
\pi^+\pi^-$).  In this case, $y$ can be reconstructed using the hadronic 
method described above, using the measured energies and momenta, to high 
accuracy.  In DIS, the electron energy is then constrained if the scattering 
angle is known, via:
\begin{equation}
E_e' = \frac{2E_e-\delta_{had}}{1-\cos\theta_e} \; \; \; .
\end{equation}
In this case, the kinematic variables are reconstructed to an accuracy of a 
few \%.  Many other kinematic variables are needed to fully describe the 
exclusive processes.  These include the mass of the final hadronic state 
(excluding the proton), the square of the four-momentum transfer at the 
proton vertex, $t$, and various decay angles.  These will be described in 
more detail as they become relevant.

\subsection{Coverage of the phase space} 
\label{sec:coverage} 
\subsubsection{Phase space coverage in DIS}

The HERA measurements in DIS cover a vast kinematic range, as shown in
Fig.~\ref{fig:F2plane}.  For example, the measurements of $F_2$ with
the 1995 data sets extend down to $Q^2$ values of $0.1$~GeV$^2$ and up
to $Q^2=5000$~GeV$^2$.  For each $Q^2$, typically two orders of
magnitude are covered in $x$: from $10^{-5} \rightarrow 10^{-3}$ at
the lowest $Q^2$ to $10^{-2} \rightarrow 0.3$ at $Q^2=500$~GeV$^2$.
The region at large $y$ requires very efficient and pure electron
finding algorithms down to the lowest possible energies, while the
small-$y$, or large-$x$, region requires a precise understanding of
the hadronic final state.  Both collaborations have invested
tremendous efforts in developing analysis tools which could work in
these extreme regions and thus allow measurements over such a large
kinematic range.  The measurement of $F_2$ requires that the cross
section be measured in bins of $x$ and $Q^2$.  At the highest values
of $Q^2$, the measurements are limited by lack of event statistics.
It is possible to measure a single differential cross section
$d\sigma/dQ^2$ up to $Q^2$ values of $40,000$~GeV$^2$, for both
neutral and charged current events.  This allows a measurement of the
$W$ mass, as well as searches for exotic phenomena.

The main detectors of ZEUS and H1 were optimized primarily for the large 
$Q^2$ region of phase space.  At the lower values of $Q^2$, various detector 
improvement programs have enhanced the kinematical coverage.  The original 
calorimeter designs allowed the measurement of DIS NC events down to 
$Q^2\approx 4$~GeV$^2$ (with the vertex at $\overline{Z}\approx 0$~cm).  
In order to probe the transition region between photoproduction and DIS, 
the calorimeters had to be modified or enhanced by new devices.  These 
include a new rear calorimeter for the H1 experiment~\cite{ref:H1_spaghetti} 
which not only has higher precision than the BEMC, but also begins at a 
distance of $4$~cm from the beam, as opposed to $\approx 10$~cm.  This 
detector was installed in the 1994-1995 winter shutdown.  The ZEUS 
collaboration added a small angle electron calorimeter, the Beam Pipe 
Calorimeter (BPC), in the shutdown period 1994-1995 to access the $Q^2$ 
range $0.1 \rightarrow 1$~GeV$^2$.  In addition, the central modules of the 
Uranium-scintillator calorimeter were moved to within $4$~cm of the beam to 
reach down to $Q^2$ values of $1$~GeV$^2$.

In addition to the modifications to the detectors, the kinematic range 
covered was extended by several techniques.
\begin{enumerate}
\item Using data with a shifted vertex.  The angular coverage of the 
detectors extends to smaller values of the electron scattering angles as 
the vertex is moved away from the rear calorimeters.  The shifted vertex
 events were produced via dedicated HERA runs in which the primary vertex 
was moved to $\overline{Z}\approx+80$~cm.  These runs were used by ZEUS and 
H1 to measure down to $Q^2 = 0.4$~GeV$^2$ with their main detectors.  In 
addition, H1 used events from the standard running 
($\overline{Z}\approx 0$~cm) in the $+Z$ tail of the vertex distribution.  
A significant number of events are present at $Z\approx 80$~cm which result 
from protons in neighboring RF bunches.  These events access the same $Q^2$ 
range as the shifted vertex runs, but have larger normalization uncertainties.

\item Using data with hard initial state radiation.  A large fraction 
($\approx 30$~\%) of the photons from events with hard initial state 
radiation are measured in the photon calorimeters of the luminosity system.  
The resultant $ep$ collisions therefore occur at lower center-of-mass 
energies, and lower $Q^2$ values can be attained.  This method was employed 
both by the ZEUS and H1 experiments.
\end{enumerate}

\subsubsection{Phase space coverage in photoproduction}
\label{sec:e_taggers}

Photoproduction events at HERA are very small $Q^2$ events where the scattered
 electron is either measured by an electron calorimeter near the electron 
beamline, or are antitagged as DIS via the absence of an electron in the main 
detectors.  These two techniques allow very different coverage of the phase 
space.  We consider each in turn.

\paragraph{Tagged photoproduction}

In addition to the main calorimeter, the ZEUS detector currently has 
electron calorimeters positioned at $Z=-8,-35$, and, $-44$~m.  These 
calorimeters are placed very close to the beamline, and are positioned 
after magnets of the HERA accelerator. They require that the electron 
scattering angle be very small, and analyze different regions of the 
electron energy spectrum. E.g., the $35$~m tagger limits 
$Q^2<10^{-2}$~GeV$^2$, and measures electrons in the energy range 
$5<E<18$~GeV, corresponding to a $W$ range $180<W<270$~GeV.  The recently 
added $8$ and $44$~m taggers cover higher and lower $W$ ranges, 
respectively, as shown in Fig.~\ref{fig:ZEUS_taggers}.  The H1 experiment 
currently has taggers at $Z=33,44$~m and is planning a tagger at $Z=8$~m. 
 These cover essentially the same kinematic range as the ZEUS detectors.

\epsfigure[width=0.8\hsize]{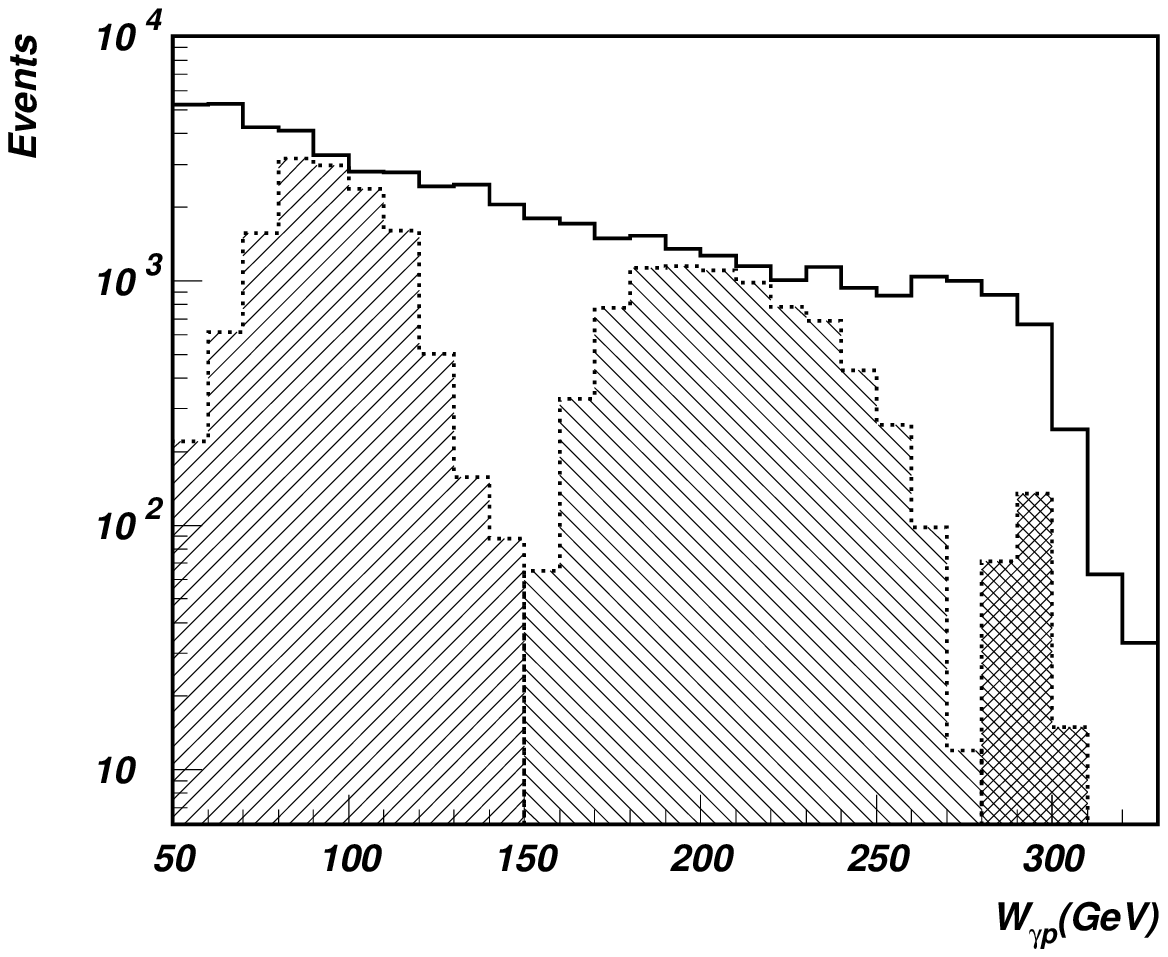} {The $W$ range tagged
  by the different electron calorimeters located along the electron
  beamline of the ZEUS detector.  The $8$-m tagger tags photons at the
  large $W$ range, the $35$-m tagger tags the intermediate $W$ range,
  and the $44$-m tagger tags the small $W$ range. The open histogram
  shows the $W$ distribution of all photoproduction events.}
{ZEUS_taggers}

\paragraph{Untagged photoproduction}

It is also possible to tag photoproduction by requiring the absence of 
the scattered electron in the main detector.  This limits $Q^2<4$~GeV$^2$
($Q^2<1$~GeV$^2$ as of the 1995 data sets).  
The median $Q^2$ depends on the process studied, but is typically 
$10^{-5}$~GeV$^2$~\cite{ref:ZEUS_photo_rho1}.  This method allows measurements
over a much wider range of $W$.  However, it requires that $W$ be measured 
from the hadronic system in the main detector which has much lower 
precision than determining it directly from the scattered electron.

\subsubsection{Phase space coverage in the forward direction}

Many of the physics results coming from HERA in the first years of operation 
deal with diffraction ($ep \rightarrow epX$, viewed as photon diffraction on
 the proton).  Ideally, this physics would be done solely with events where 
the scattered proton is tagged.  Both ZEUS and H1 have leading proton 
spectrometers, with high acceptance (for the ZEUS LPS) 
in the proton elastic peak, $x_l=1$, for $0.2<p_T<0.6$~GeV.  The overall
acceptance for elastic photoproduction of $\rho^0$ mesons  is about $6$~\%.

The acceptance of the leading proton spectrometers is quite small, and it 
is therefore profitable to also study diffraction with the other detector 
components.  There are different techniques for doing this 
(see section \ref{sec:diffraction}). The most important detector parameter 
is the rapidity coverage of the forward calorimeters and tagging devices. 
Forward rapidity coverage is needed both for the reconstruction of 
large mass diffractive final states as well as for rejecting backgrounds.

The ZEUS Uranium-scintillator forward calorimeter extends to within 
$10$~cm of the beam and begins at $Z=2.2$~m, which implies a rapidity 
\footnote{Pseudorapidity, $\eta=-\ln tan (\theta/2)$ is 
used as an estimate for rapidity $y=\ln(E-P_z/E+P_z)$.}
cutoff near $\eta=4$. Additional 
information is available from the proton remnant tagger (PRT), which has 
scintillator planes located at $Z=5$~m and $Z=22$~m.  The PRT allows the 
tagging of charged particles in the rapidity intervals $4< \eta< 7$.  

The H1 LAr calorimeter extends to $\eta\approx 3.7$.  This is complemented by
 the forward plug calorimeter, FPC, with coverage $3.5<\eta<5.5$.  Collimators
 are placed between the interaction point and the FPC.  This implies a minimum
 energy cutoff for particles to reach the FPC.  However, the scattering and 
showering produced by the collimator can be used to advantage.  The forward 
muon chamber is sensitive to charged particles from these processes.  The 
resulting coverage of the H1 detector is $5.0<\eta<6.5$.  The H1 PRT covers
the region $6.0<\eta<7.5$.

\subsection{Radiative corrections} 
\label{sec:radiation} 
To compare measured cross sections with theoretical calculations it is 
necessary to include the effects of QED radiative 
processes.  This is typically done by correcting 
the measured cross sections for these effects, either by using event 
simulation packages which include radiative effects, or from analytical 
calculations.  The difference between the radiative cross sections and the 
Born cross sections are used to apply correction factors. 

\epsfigure[totalheight=0.95\hsize,angle=270]{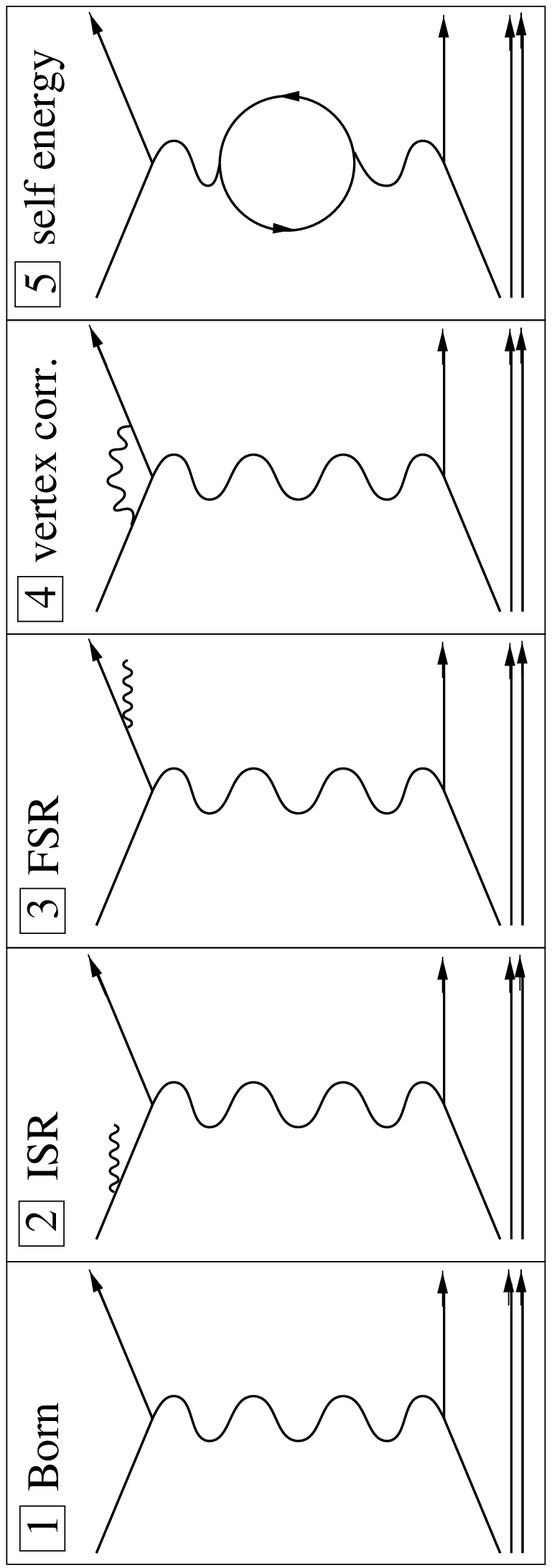}
{Diagrams showing the different LO QED corrections to the Born
  diagram.}  {QED_radiation}

First order QED radiative processes from the electron are shown in 
Fig.~\ref{fig:QED_radiation}.  They include initial state radiation 
(ISR), final state radiation (FSR) and virtual loop corrections.  There 
are also radiative effects for the proton, or quarks.  These are typically
 not corrected, and the measured structure functions therefore include QED 
radiation.

For ISR, the energy of the incoming electron is effectively lowered.  
The {\it apparent} $x$ and $Q^2$ calculated with the electron variables are 
no longer the same as the propagator $x$ and $Q^2$.  QED radiation therefore 
causes shifts in the mean reconstructed kinematic variables.  The shifts 
depend on the choice of reconstruction method.  Methods relying more on the 
hadronic system typically suffer smaller shifts.  Note that the bulk of the 
radiated photons are collinear with the incoming electron.  They can 
therefore be measured by the photon calorimeters of the luminosity systems 
from ZEUS and H1.  Measuring this spectrum is a very good check that ISR is 
simulated correctly.  FSR is also typically collinear, with the outgoing 
photon at small angles to the scattered electron.  These are usually not 
distinguished as separate clusters
by the calorimeters, and therefore do not results in shifts in 
the reconstructed kinematic variables.  However, there is a tail to larger 
opening angles, which will produce shifts in the kinematic variables if not
 correctly simulated.  These effects are best studied 
with Monte Carlo event simulations which take QED radiation into account.  
In the Monte Carlo program HERACLES~\cite{ref:HERACLES_1,ref:HERACLES_2}, 
the 4--momentum of 
the radiative photon is sampled according to the 5--fold differential cross 
section for radiative photon emission 
$d\sigma/dxdydE_{\gamma}d\theta_{\gamma}d\phi_{\gamma}$ from the electron and 
is saved together with the other stable particles.  These are then traced 
through the detector.

The shift of the apparent variables from the true depends strongly on the 
reconstruction method chosen for the kinematic variables.  The shift can be
 effectively limited by appropriate event selection cuts.  For example, 
requiring $\delta>0.7 (2E_e)$ limits the maximum ISR photon energy to
$E_{\gamma}<0.3E_e$, which in turn limits the amount of shift possible in 
the reconstruction of the kinematic variables. 
Note that the correction to the measured 
cross section is typically highest at large $y$, as this corresponds to small 
scattered electron energy. 

In addition to producing measured kinematic variables which are shifted from 
the propagator values, QED radiation also changes the cross section for a 
given $x$ and $Q^2$.  The cross section must therefore also be corrected to 
the Born level (no QED radiation).  
These corrections are typically small, since the QED coupling
 constant appears, and can usually be treated as a multiplicative correction.
  For example, the differential cross section for NC DIS can be approximated
 as
\begin{equation}
\frac{d^2\sigma^{NC}}{dxdQ^2} = \frac{d^2\sigma^{Born}}{dxdQ^2}(1+
\delta^r(x,Q^2)) \; \; \; .
\end{equation}

\subsection{Event modeling and unfolding of detector effects} 
\label{sec:modelling} 
Detailed Monte Carlo simulations are indispensable for the extraction of
 physical quantities from the measurements.  The use of Monte Carlo event 
simulators was already mentioned in the previous section for the study
of QED radiative effects.  Monte Carlo simulations are also used to
correct for limited geometrical acceptance of the detectors and to 
understand the effects of the measurement resolution on kinematical
variables.

Monte Carlo simulations can also be used to study the relationship
between measured quantities and the physically fundamental quantities.  
An example of this is the relation between the jets measured at the 
detector level and the partons which originally produced the jets. 

Monte Carlo event simulations are used extensively in developing the 
data selection cuts.  The Monte Carlo programs can help determine which 
variables are particularly useful for separating signal from background.  
They also indicate where certain variables are measured accurately, and 
where they are not well measured. In developing the cuts and determining 
the acceptance corrections, it is necessary that the Monte Carlo simulations 
accurately describe the data.  In practice, this means that many different 
Monte Carlo programs must be used, each tuned for a specific class of 
processes.  It also means that detailed simulations of the detector apparatus 
must be performed, including inactive as well as active materials.

 We describe briefly a few of the more commonly used event generators for 
DIS and photoproduction.  The many special purpose generators, e.g. for 
diffractive processes or exotic processes, are beyond the scope of this 
report.  We finish this section with a brief description of the detector 
simulation packages.

\subsubsection{Generators for DIS}
Several generators are used together to describe DIS events at HERA.  The 
program HERACLES~\cite{ref:HERACLES_1,ref:HERACLES_2} 
is used to simulate the effects of 
electroweak radiation from the electron and quark.  HERACLES is interfaced 
with either ARIADNE~\cite{ref:ARIADNE_1,ref:ARIADNE_2}, 
HERWIG~\cite{ref:HERWIG} or 
LEPTO~\cite{ref:LEPTO} for the simulation of the hard scattering and the 
simulation of the initial and final QCD radiation from the partons.  
ARIADNE is based on the color dipole model~\cite{ref:color_dipole}, where 
the struck quark-diquark system act as a radiating color antenna.  In the 
program HERWIG, coherent parton showers are produced which form color neutral
 clusters, while the LEPTO package produces parton showers according to the
 DGLAP evolution equation.  Perturbative QCD corrections are implemented
by taking into account the full ${\mathcal O}(\as)$ or 
${\mathcal O}(\as^2)$  matrix elements.
In the case of ARIADNE and LEPTO, the program
 package JETSET~\cite{ref:JETSET} is then used for the fragmentation of the 
resulting partons into hadrons, and for their decay.  JETSET is based
on the Lund string model of fragmentation~\cite{ref:LUND}. These 
programs are under constant revision as new data become available, and 
different versions are used for different analyses.

\subsubsection{Generators for photoproduction}
The two principal generators used to simulate photoproduction reactions
are the PYTHIA program~\cite{ref:PYTHIA} and the HERWIG 
program~\cite{ref:HERWIG}.  PYTHIA is a Lund type Monte Carlo 
program.  It allows for both resolved as well as direct photon
reactions.  The program allows for both initial as well as final state
QCD radiation, and uses the JETSET package for hadronization.  The
hard scattering cross sections are simulated in LO.  The H1 collaboration
has also made extensive use of the PHOJET generator
\cite{ref:PHOJET1,ref:PHOJET2}.

\subsubsection{Detector simulation} 
The output of the event generation step is fed into the detector simulation 
programs.  These are based on the GEANT~\cite{ref:GEANT} package which is in 
common use in high energy physics experiments.  Detailed simulation of the 
geometry and materials of the detector are performed, and the output of the 
programs are compared with measurements from test beams as well as from $ep$ 
interaction data.  The individual programs used by the ZEUS and H1 experiments
represent enormous efforts.  They are constantly being refined to give 
the best possible reproduction of the response of the detector.

 
\section{Inclusive cross section measurements} 
\label{sec:inclusive} 
Inclusive cross section measurements are of fundamental importance.
Considerable debate on the behavior of the $ep$ cross section at high
center-of-mass energies was fueled in the late 1980's and early 1990's
by the approaching turn-on of HERA.  In $e^+e^-$ scattering, the total
cross section as a function of the center-of-mass energy is calculable
in electroweak theory, and the measurements are in excellent agreement
with the predictions.  The behavior of the hadron-hadron scattering
cross sections as a function of the center-of-mass energy are not
calculable since they depend on the structure of the hadrons.
However, cross sections have been measured for many different
combinations of hadrons (proton-proton, proton-antiproton,
proton-pion, ...).  It was found experimentally that these cross
sections all have a similar energy dependence at high
energies~\cite{ref:DoLa_sigfit}, given by
\begin{equation}
\sigma \sim s^{0.08} \, ,
\end{equation}
where $s$ is the hadron-hadron center-of-mass energy.  What would the
data show in the case of $ep$ scattering~?

Two different regimes are generally differentiated: 
\begin{enumerate}
\item The small $Q^2$ regime, where electron-proton scattering
  proceeds via photon exchange.  When $Q^2 \rightarrow 0$, the photon
  is almost real, and we can consider the scattering as photon-proton
  scattering.  The transverse distance scale defined by $Q^2$ is much
  larger than the size of the proton, so that the details of the
  proton structure are not expected to be important for the total
  cross section. It has been known for some time that photoproduction
  can be viewed as hadron-hadron scattering, where the photon has
  fluctuated into a vector meson.  It was therefore expected that the
  cross section at $Q^2\approx 0$ would have a similar behavior to the
  hadron-hadron scattering cross sections.
  
\item The deep inelastic scattering regime (DIS), where
  electron-proton scattering proceeds via $\gamma, Z \; \rm{or} \; W$
  exchange (photon exchange dominates until $Q^2\approx M_{Z,W}^2$).
  In this case, the transverse distances probed in the interaction are
  a fraction of the proton radius, and the cross section will depend
  on the details of the proton structure.  No experimental data
  existed pre-HERA for center-of-mass energies of the exchanged
  boson-proton system larger than about $25$~GeV.  Many different
  predictions existed for the behavior of the cross section, ranging
  from a small energy dependence similar to that seen in hadron-hadron
  scattering~\cite{ref:DoLa_F2} to a very steep energy
  dependence~\cite{ref:BFKL1,ref:BFKL2,ref:BFKL3}.  The measurement of
  this cross section (or equivalently $F_2$) is one of the most
  exciting results to come from HERA.
\end{enumerate}
In the following sections, we first describe the measurements of the
$\gamma p$ and $\gamma^*p$ cross section at HERA.  We then describe
the structure function measurements and discuss the results.

\subsection{Total photoproduction cross section} 
\label{sec:photoprod} 

\epsfigure[width=0.8\hsize]{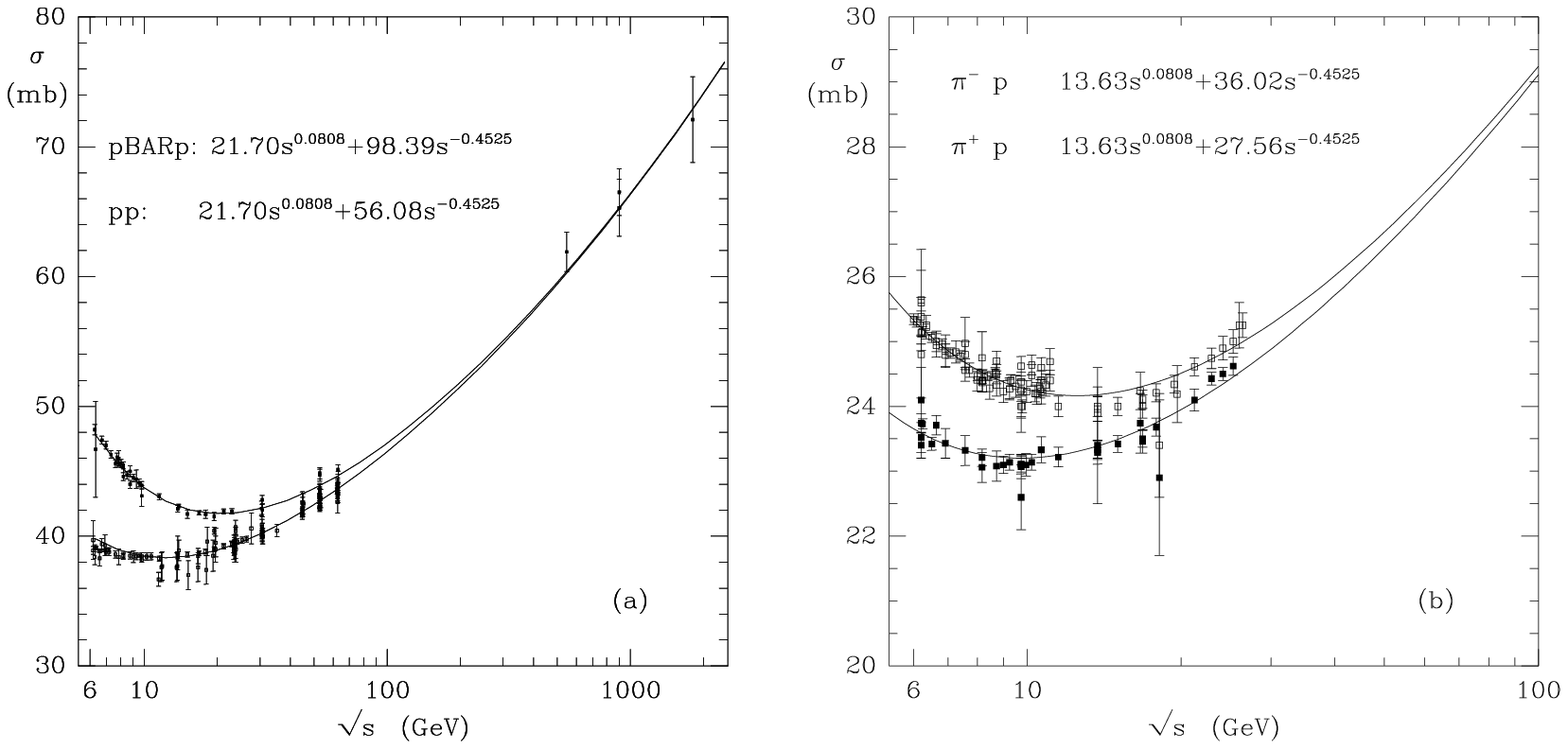} {The total cross
  section measured in hadronic scattering as a function of the
  center-of-mass energy for $pp,\; p\overline{p},\; \pi^+ p,\; \pi^-
  p$ scattering.  The cross sections show a universal rise at high
  energies of the form $\sigma \sim
  s^{0.08}$~\protect\cite{ref:DoLa_sigfit}.}  {hadronic_sigtot}

As mentioned above, the total hadronic cross sections rise with
center-of-mass energy in a universal way.  The energy dependence for
$\overline{p} p, pp, \pi^- p,\rm{and}\; \pi^+p$ scattering is shown in
Fig.~\ref{fig:hadronic_sigtot}.  The solid lines shown in the figure
are fits which include a component dying rapidly with $s$, the
center-of-mass energy, and a second component which persists at high
energies~\cite{ref:DoLa_sigfit}.  The second component was found to be
universal, with the cross sections rising as $s^{0.08}$.  The
corresponding measurements for $\gamma p$ scattering only extended to
$\gamma p$ center-of-mass energies of about $25$~GeV before the advent
of HERA, and the high energy dependence was therefore not well known.
Predictions based on perturbative QCD varied widely, from a soft
energy dependence to a very steep energy
dependence~\cite{ref:gamma_QCD1,ref:gamma_QCD2,ref:gamma_QCD3,ref:gamma_QCD4,ref:gamma_QCD5}.

\subsubsection{From $ep$ scattering to $\gamma p$ scattering}

In the Standard Model, the reaction $ep \rightarrow eX$ can proceed
via $\gamma$ or $Z^0$ exchange. At small enough $Q^2$, $Z^0$ exchange
can be neglected.  Furthermore, we can decompose the cross section
into a term representing the flux of photons from the electron, and a
term representing the photon-proton scattering process.  This is
depicted schematically in Fig.~\ref{fig:ep_gammap}.  The differential
cross section can then be written in the form
\begin{equation}
\frac{d^2\sigma(ep)}{dWdQ^2} = \Gamma(\sigma_T^{\gamma p} + \epsilon \sigma_L^{\gamma p}) \; ,
\end{equation}
where $\Gamma$ represents the flux factor of virtual photons,
$\sigma_T, \sigma_L$ are the absorption cross sections for transverse
and longitudinally polarized photons, and $\epsilon$ is the ratio of
the flux for longitudinal to transverse photons.

If we restrict ourselves to photoproduction, $Q^2 \approx 0$, then the
photon is transversely polarized and only the first term in the cross
section expression will contribute.  Given that the flux factor
$\Gamma$ is known (see discussion in section~\ref{sec:gammap}), a
measurement of the differential $ep$ cross section allows the
extraction of the total photoproduction cross section.

\subsubsection{Measurements at HERA}
The measurements at HERA rely on the electron calorimeters described
in section~\ref{sec:e_taggers}.  These detectors measure the scattered
electrons in the energy range $10 - 22$~GeV, which corresponds to a
$W$ range $150-250$~GeV. Observing the electron in these detectors
limits $Q^2$ to a maximum of $0.02$~GeV$^2$.  Loose requirements were
placed on the energy seen in the main detectors.  Even so, the
acceptance depends crucially on the detailed composition of the
hadronic final state, so that acceptance corrections relied heavily on
Monte Carlo programs.  The results from ZEUS~\cite{ref:ZEUS_sigtot2},
$\sigma_{tot}=143\pm 4 \pm 17$~$\mu$b, and H1~\cite{ref:H1_sigtot2},
$\sigma_{tot}=165\pm 2 \pm 11$~$\mu$b, are shown in
Fig.~\ref{fig:HERA_sigtot}.  The expectations based on the energy
dependence measured in hadronic interactions are in good agreement
with the HERA results, confirming the hadronic behavior of real
photons.

\epsfigure[width=0.8\hsize]{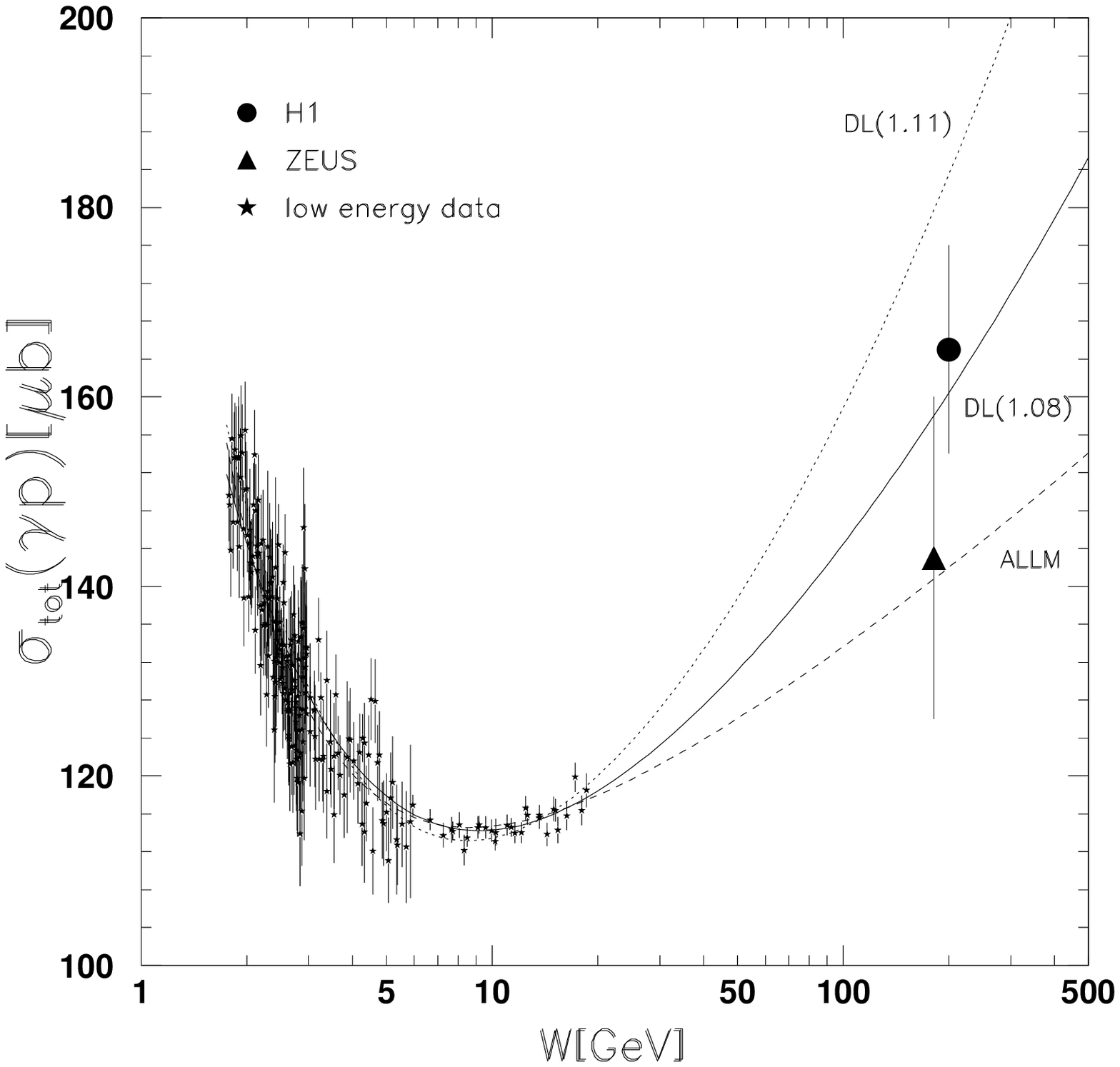} {The dependence of
  the total photoproduction cross section as a function of the $\gamma
  p$ center-of-mass energy, $W$. The HERA results are compatible with
  an extrapolation of the low energy measurements using the energy
  dependence found from hadronic scattering, indicating that the
  photon behaves similarly to a hadron.  The curves are from
  \protect\citeasnoun{ref:DoLa_sigfit} with $\alpha_0=1.08$ and
  $\alpha_0=1.11$ and from the ALLM
  parameterization~\protect\cite{ref:ALLM92}.}  {HERA_sigtot}

\subsection{Total DIS cross section} 
\label{sec:DISprod} 
The measurements of the total photoproduction cross section can be
extended to higher $Q^2$, where a virtual photon (or weak boson) is
involved.  There are several important differences to consider once
the virtuality increases:

\begin{itemize}
\item The photon can acquire a longitudinal polarization, so that both
  the transverse as well as the longitudinal cross sections are
  important.
\item The flux of virtual photons is not uniquely defined, and
  requires a convention.  The Hand convention is normally chosen (see
  Eq.~(\ref{eq:Hand})).
  
\item The photon probes smaller transverse distances.  The transverse
  distance scale probed is given by
\begin{equation}
d \approx \frac{0.2}{Q({\rm GeV})}\; {\rm fm} \; \; \; .
\end{equation}
For $Q^2 > 1$~GeV$^2$, transverse distances at the level of a fraction
of the proton radius are probed.

The longitudinal distances over which the photon fluctuations occur
depend on $1/x$, as described in section~\ref{sec:spacetime}.
\end{itemize}

\epsfigure[width=0.95\hsize]{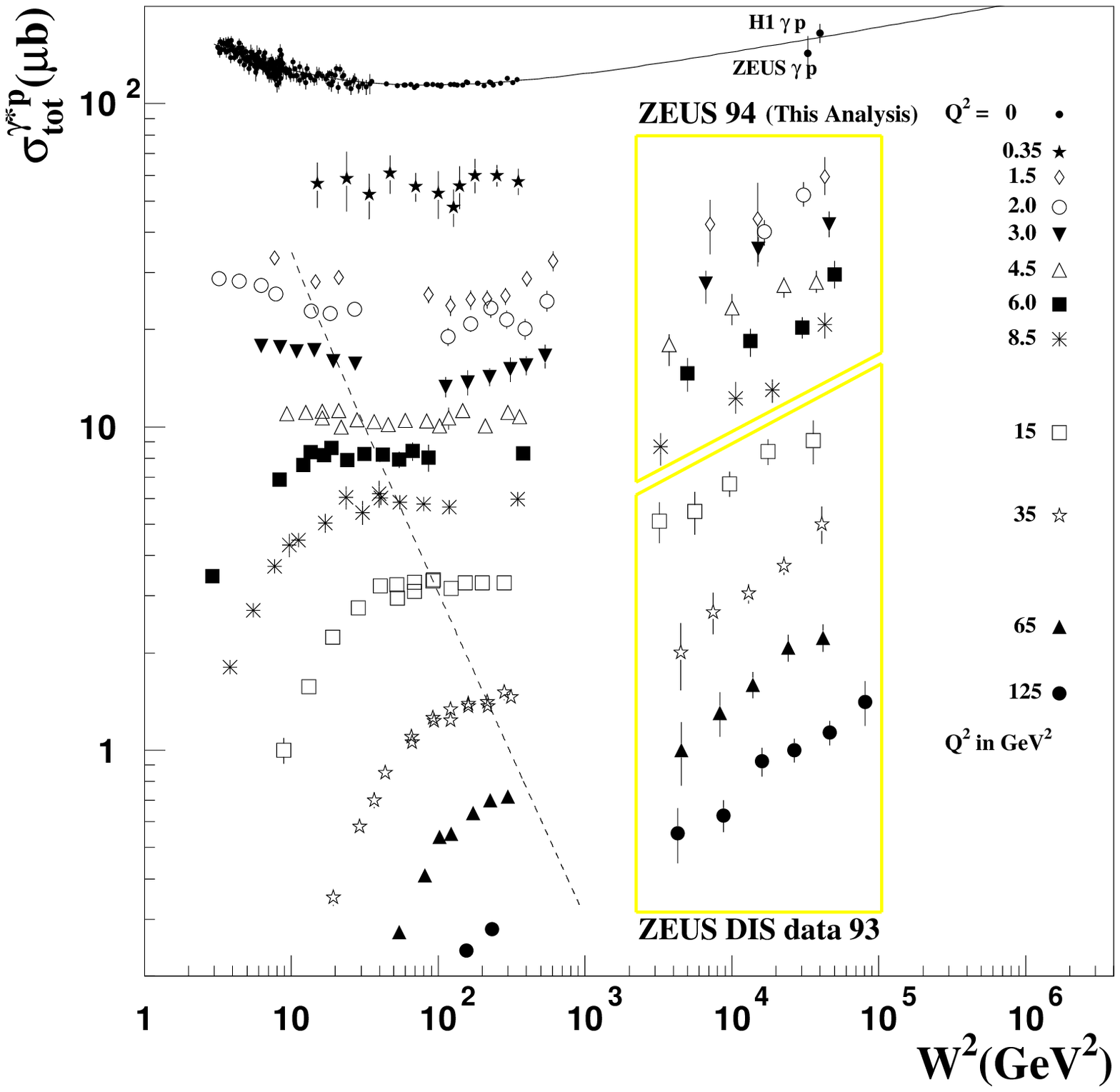} {The $\gamma^*p$
  cross section versus $W^2$ for small $Q^2$ data from the
  \protect\citeasnoun{ref:ZEUSF2_94_lowQ2}.  The data are compared to
  results from fixed target experiments as well as to the
  photoproduction results shown in the previous figure.}  {dis_sigtot}

The total $\gamma^* p$ cross sections are extracted from the measured
values of $F_2$ (discussed in the next sections) using
equation~\ref{eq:defsig}.  They are given as a function of $W^2$ in
Fig.~\ref{fig:dis_sigtot} for $Q^2$ values in the range
$1.5<Q^2<125$~GeV$^2$, and are compared with the photoproduction
results as well as with results from fixed target experiments. As can
be seen clearly in this figure, the DIS cross section has a $W^2$
dependence which is considerably steeper at large $W^2$ than that
found in photoproduction.  The $W^2$ dependence is also seen to vary
strongly with $Q^2$.  The region to the right of the dashed line
corresponds to $x<0.1$, where the photon lifetime, in the proton rest
frame, is large enough so that it typically survives for distances
larger than the proton radius (see section%
~\ref{sec:spacetime}).

The rise of the cross section with energy, for a fixed $Q^2$, is
expected to come to an end at some $W^2$, in analogy with the
Froissart bound for hadron-hadron scattering~\cite{ref:Froissart}:
\begin{equation}
\sigma_{tot} \leq C \ln^2 \frac{s}{s_0} \;\; .
\end{equation}
The form of this bound is not known precisely in DIS~\cite{ref:GLM97},
but it is expected that there is some analogous limit.  The present
HERA data show no signs of a flattening of the cross section at large
$W$, indicating that unitarity constraints do not play a strong role
in the kinematic region presently probed.

In a geometrical picture of the interaction, the constituents of the
proton are viewed as being resolved into sub-constituents as the
transverse and longitudinal distance scales are reduced.  A schematic
representation of this effect is shown in
Fig.~\ref{fig:proton_structure}.  In this picture, all the structure
is assigned to the proton.  The structure can also be thought of as
belonging to the photon, as discussed in section~\ref{sec:spacetime}.
In the proton rest frame, the photon has a long lifetime, and can
fluctuate into a quark-antiquark pair, and perhaps develop further
hadronic structure, before scattering from the proton.  In this case,
the scattering cross section is increased because of the increased
flux of partons from the photon scattering off the proton.

\subsection{HERA structure functions} 
\label{sec:structure_functions} 
We now discuss the measured cross sections from the perspective of
proton structure functions.  As has been discussed in
section~\ref{sec:kinematics}, the collider kinematics allow to reach
much larger values of $Q^2$ and smaller values of $x$.  The
measurements from fixed target experiments are limited to $x>0.01$ for
$Q^2>10$~GeV$^2$, and therefore do not probe the very small values of
$x$ in the regime of perturbative QCD.  However, the measurements from
the fixed target experiments fix the valence quark densities, and
constrain the form of the sea quark and gluon densities for $x \geq
0.01$.  The HERA measurements probe similar regions of $x$, but at
much larger $Q^2$, thereby testing the $Q^2$ evolution of the
structure functions.  They also probe much smaller values of $x$ for a
fixed $Q^2$, thereby allowing precise measurements on the behavior of
the parton densities for very small proton momentum fraction.  The
behavior of the parton densities for $x<0.01$ is of great theoretical
interest, and could lead the way to a better understanding of
perturbative QCD.  It is expected that parton densities become large
at small $x$, while the coupling strength, $\as$ remains small,
representing a new physical situation for QCD calculations. As was
discussed in section~\ref{sec:QCD}, there are different approaches to
the perturbative solution of QCD leading to different predictions on
the behavior of the structure function at small $x$ (NLO DGLAP versus
LO BFKL).
  
In the following sections, we first review the definition of the
structure functions and the expectations pre-HERA.  We then describe
how the structure function measurements are performed at HERA, and
discuss the results.  Finally, the methods used to extract the gluon
density in the proton are described.

A comprehensive review of structure functions has recently been
published by \citeasnoun{devenish_rev}.

\subsubsection{Structure functions in neutral and charged current scattering}
\label{sec:sf_formulae}

\paragraph{Neutral current cross sections}

The differential cross section for $e^{-}p$ neutral current (NC)
scattering is conventionally written in terms of the NC structure
functions $F_2,\; F_L$ and $F_3$ as shown in
Eq.~(\ref{eq:disxsection}).  We rewrite the cross section here and
discuss explicitly the dependence on polarization.

\begin{equation}
\frac{d^2\sigma^{e^{-}p}}{dxdQ^2} = \frac{2\pi\alpha^2}{xQ^4}\left[ Y_+ 
F_2^{L,R}(x,Q^2) - y^2 F_L^{L,R}(x,Q^2) + Y_- xF_3^{L,R}(x,Q^2)\right] \; ,
\label{eq:NCdiff}
\end{equation}
with $x$ and $Q^2$ defined at the vector boson-quark vertex. Note that
the structure functions are process dependent.  For example, one has
to make the replacement $F_3^{L,R} \rightarrow -F_3^{R,L}$ for $e^+p$
scattering, where $L,R$ represent the polarization of the lepton beam.
The parity-violating $xF_3$ term is only important at large $x$ and
$Q^2$ where it substantially reduces the $e^{+}p$ cross section, while
increasing the cross section for $e^{-}p$ scattering.  The
contribution from $F_L$ is important at large $y$ and small $x$.  It
is expected to be negligible at large $x$ and $Q^2$.  The structure
functions depend on the lepton beam polarization as described below.
We assume here that the proton beam is unpolarized.

The explicit $y$ dependence, which is due to the
helicity dependence of electroweak interactions, is contained in the
functions
\begin{equation}
Y_{\pm }(y)=1\pm (1-y)^2\  \label{ydep} \; ,
\end{equation}
with $y$ related to the electron scattering angle in the $eq$
rest frame, $\theta^*$, as
\begin{equation}
y = \frac{1-cos\theta^*}{2} \; .
\label{eq:eqangle}
\end{equation}

The dependence on the quark structure of the proton, and on the $Z^0$ 
propagator, is absorbed in the structure functions.  To LO,
these are
\begin{equation}
\left( 
\begin{array}{c}
F_2^{L,R}(x,Q^2) \\ 
xF_3^{L,R}(x,Q^2)
\end{array}
\right) =x\sum_{q={\rm quarks}}\left( 
\begin{array}{c}
C_2^{q\; L,R}(Q^2)[q(x,Q^2)+\overline{q}(x,Q^2)] \\ 
C_3^{q\; L,R}(Q^2)[q(x,Q^2)-\overline{q}(x,Q^2)]
\end{array}
\right)
\end{equation}
written in terms of the quark densities in the proton
($q=u,\;d,\;c,\;s,\;t,\;b$)
and the corresponding antiquark densities $\overline{q}$.
The structure function $F_L$ is discussed in section~\ref{sec:FL}.

The $Q^2$ dependent coefficient functions, 
$C_2^q$ and $C_3^q$, are given in terms of the vector and
axial vector couplings
\begin{eqnarray}
C_2^{q \; L,R}(Q^2) &=& |V_q^{L,R}|^2+|A_q^{L,R}|^2 \; ,\\
C_3^{q \; L,R}(Q^2) &=& \pm 2 V_q^{L,R} A_q^{L,R} \; ,
\end{eqnarray}
which depend on the polarization of the $e^{\pm}$ beam ($L=+, R=-$)
\begin{equation}
\begin{array}{rrll}
V_q^{L,R} =& e_q & + & e_e(v_e\pm a_e)v_q\chi_Z \; ,\\
A_q^{L,R} =&     & & e_e(v_e\pm a_e)a_q\chi_Z \; .
\end{array}
\label{eq:va}
\end{equation}
Expanding the expressions for $C_2,C_3$ yields
\begin{equation}
\begin{array}{rrll}
C_2^q(Q^2)=&e_q^2 & + & 2e_ee_qv_q(v_e+Pa_e)\chi_Z+
(v_q^2+a_q^2)(v_e^2+a_e^2+2v_ea_eP)\chi _Z^2 \; , \\ 
C_3^q(Q^2)=&      & & 2e_ee_qa_q(a_e+Pv_e)\chi_Z+
e_e^2(2v_qa_q)(2v_ea_e+Pv_e^2+Pa_e^2)\chi_Z^2 \; ,
\end{array}\label{eq:coefun}
\end{equation}
where 
$e_e$ is the charge of the electron ($-1$) and
$e_q$ is the quark charge in units of the positron charge.
Note that the charge does not change signs for antiparticles
(e.g., $e_e=-1$ for positrons, and $e_{\bar{u}}=2/3$).
$P$ is the degree of left-handed longitudinal polarization,
\begin{equation}
P = \frac{N_L - N_R}{N_L+N_R} \; ,
\end{equation}
where $N_L, \; N_R$ are the number of left-handed and right-handed
electrons (or positrons).
The vector and axial vector coupling of the fermion are given by
\begin{equation}
\begin{array}{rcl}
v_f & = & (T_{3}^f-2e_f\sin^2\theta _w) \; ,\\
a_f & = & T_{3}^f \; ,
\end{array}
\end{equation}
with $\theta _w$ the weak mixing angle,
$T_3^f$ the third component of the weak isospin and $e_f$ the electric
charge in units of the positron charge.
The factor $\chi_Z$ is given by
\begin{equation}
\chi _Z=\frac 1{4\sin ^2\theta _w\cos ^2\theta _w}\frac{Q^2}{Q^2+M_Z^2} \; .
\label{eq:chiZ}
\end{equation}
where $M_Z$ is the $Z^0$ mass.
All relevant electroweak parameters have been
measured to high precision~\cite{ref:EWpar}.

For $Q^2 \ll M_Z^2$, the structure functions in LO reduce to
\begin{equation}
\left( 
\begin{array}{c}
F_2(x,Q^2) \\ 
xF_3(x,Q^2)
\end{array}
\right) =x\sum_{q={\rm quarks}}\left( 
\begin{array}{c}
e_q^2[q(x,Q^2)+\overline{q}(x,Q^2)] \\ 
0
\end{array}
\right) \; .
\end{equation}

\paragraph{Charged current cross sections}

The differential cross section for $e^{\pm}p$ 
charged current (CC) scattering can be written
in terms of the of the CC structure functions $F_2,\; F_L$ and 
$F_3$ as 

\begin{equation}
\frac{d^2\sigma^{e^{\pm}p}}{dxdQ^2} = \frac{G_F^2}{4\pi x}
\left(\frac{M_W^2}{Q^2+M_W^2}\right)^2\left[ Y_+ 
F_2(x,Q^2) - y^2 F_L(x,Q^2) \mp Y_- xF_3(x,Q^2)\right] \; \; .
\label{eq:CCdiff}
\end{equation}
The Fermi constant, $G_F$, can be expressed as
\begin{equation}
G_F = \frac{\pi \alpha}{\sqrt{2} \sin^2\theta_W M_W^2} \; ,
\label{eq:fermi}
\end{equation}
with $M_W$ the mass of the $W^{\pm}$.

As in the NC case, the parity-violating $xF_3$ term substantially
reduces the $e^{+}p$ cross section at large $Q^2$, while increasing
the cross section for $e^{-}p$ scattering.  The contribution from
$F_L$ is expected to be negligible and is ignored.  In LO pQCD, we can
specify the quark flavors which enter into the scattering.  For $e^-p$
scattering, and left-handed electron polarization,
\begin{eqnarray}
F_2 & = & 2x\left[
 u(x,Q^2) + c(x,Q^2) + \bar{d}(x,Q^2) + \bar{s}(x,Q^2)\right] \; , \\
xF_3 & = & 2x\left[
 u(x,Q^2) + c(x,Q^2) - \bar{d}(x,Q^2) - \bar{s}(x,Q^2)\right] \; ,
\end{eqnarray}
while $F_2=xF_3=0$ for right-handed electron polarization.
For $e^+p$ scattering, and right-handed positron
polarization, 
\begin{eqnarray}
F_2 & = & 2x\left[
 d(x,Q^2) + s(x,Q^2) + \bar{u}(x,Q^2) + \bar{c}(x,Q^2)\right]\; , \\
xF_3 & = & 2x\left[
 d(x,Q^2) + s(x,Q^2) - \bar{u}(x,Q^2) - \bar{c}(x,Q^2)\right]\; ,
\end{eqnarray}
while $F_2=xF_3=0$ for left-handed positron polarization.

\subsubsection{Pre-HERA expectations}
It is interesting to review the pre-HERA expectations for the
structure functions.  These expectations are summarized in the
proceedings of the `Physics at HERA' workshop from
1991~\cite{ref:HERA_91}.  We focus on the expectations for the neutral
current structure function $F_2$.  Most of the initial speculation
concerned the behavior of the parton densities at small $x$, and the
consequences of the anticipated rise of $F_2$ with decreasing $x$.

\paragraph{$F_2$ at small $x$}

It was widely believed that $F_2$ would rise with decreasing $x$ in the
HERA regime, as this is a consequence of both the DGLAP and BFKL evolution
equations.  However, this view was not universally held.
 Some authors expected the structure function to remain rather
flat with decreasing $x$~\cite{ref:DoLa_F2}, in contrast to
BFKL inspired approaches which
predicted a steep rise of the form $x^{-0.5}$ at small $x$
(as discussed in section~\ref{sec:BFKL}).
Parton density parameterizations existed with a range of possible $F_2$ 
behaviors within these limits~\cite{ref:CTEQ3,ref:MRS92_1,ref:MRS92_2,%
  ref:Yndurain1,ref:Yndurain2}.  One group of authors, Gl\"uck, Reya
and Vogt (GRV), predicted a steep rise of $F_2$ due purely to QCD
radiative processes.  They parameterized the parton densities as
essentially valence like at a very small scale $Q_0^2=0.34$~GeV$^2$,
and predicted $F_2$ at larger values of $Q^2$ using NLO DGLAP
evolution.  I.e., they assumed $F_2$ was calculable from pQCD with
valence-like parton densities defined at a much smaller scale than
that used by other authors.

\paragraph{Saturation at small $x$}

The question of the validity of the DGLAP evolution at small $x$ was
widely discussed.  One issue was whether the BFKL evolution, which is
based on an expansion in $\ln(1/x)$, was more appropriate at small
$x$. A perhaps more fundamental issue was whether the standard
evolution equations based on parton splitting would be valid at all,
or whether parton recombination would play a role at high parton
densities. \citeasnoun{ref:GLR} suggested that such effects could be
represented by nonlinear terms in the evolution equations. The result
would be a `saturation' of the parton densities at small $x$.  It was
expected that these saturation effects would be seen at HERA (see,
e.g., the article by Bartels and Feltesse in the Proceedings of the
1991 HERA workshop).  In this case, neither the DGLAP nor the BFKL
evolution equations would be valid.

\paragraph{Hot Spots at HERA}

The issue of saturation was further discussed in terms of the possibility
of `hot spots', or localized regions with high parton densities.  These
were predicted as a possible outcome of the BFKL evolution 
equation.  The presence of `hot spots' would be signaled
by an early onset of the saturation of $F_2$.  Other signals for 
`hot spots' formation were discussed by Mueller~\cite{ref:Mueller,%
  ref:Mueller_Navelet}.  Here it was proposed to look at the rate of
forward jet production with $p_T^{jet} \approx Q$ in small-$x$ events.
A high rate for such events would indicate a strong evolution in $x$
at fixed $Q^2$, a feature of the BFKL evolution equations.

\subsubsection{Structure function measurements at HERA}
\label{sec:HERA_strufu}

We now discuss the measurement of the structure function $F_2$.  We
first review the experimental procedure used to extract $F_2$, before
turning to a discussion of the results.  A discussion of the
measurement of $F_L$ and $F_2^{c\bar{c}}$ follows.

The extraction of the structure function $F_2$ relies on an accurate
determination of the inclusive neutral current differential cross
section.  This implies the need for well understood event selections
and precise kinematic reconstruction. Detailed detector simulations
are required to transform the measured data into $ep$ cross sections.
The measured cross sections must then be corrected for electroweak
radiative effects before $F_2$ can be extracted.

\begin{table} 
\tablecaption{ Summary of the published data sets
  which have been used in the extraction
 of $F_2$ at HERA.  The symbols ($\mathrm{NOM, SVTX, ISR, SAT, BPC}$) refer to
 (nominal vertex data, shifted vertex data, initial state radiation data,
 satellite data and beam pipe calorimeter data) respectively.}
\label{tab:F2_sets}
\begin{center}
\rotatebox{-90}{%
\begin{tabular}{c|c|c|c|c|c|c}
Data  & Data & $\cal{L}$ & Kin.  & $Q^2$  & $x$  & Typical \\
year & set &             & recon.&  range &  range  & Syst.  \\
     &     &             &        &            &           &  Error \\
     &     & (pb$^{-1}$) &        & (GeV$^2$) &            & (\%) \\
\hline
1992 & & & & & & $15$ \\
\cite{ref:H1F2_92} & NOM  & $0.02$ & el, had & $8.5-60$ & 
$2.\; 10^{-4} - 0.01$&  \\
\cite{ref:ZEUSF2_92} & NOM  & $0.02$ & DA & $15-1000$ & 
$4.\; 10^{-4} - 0.03$ &  \\
 & & & & & & \\
1993 & & & & & & $5-15$\\
\cite{ref:H1F2_93} & NOM,SAT & $0.27$ & el, $\Sigma$ & $4.5-1600$ & 
$2.\;10^{-4} - 0.13$ \\
\cite{ref:ZEUSF2_93} & NOM & $0.54$ & DA & $8.5-5000$ & 
$4.\;10^{-4} - 0.16$ & \\
 & & & & & & \\
1994 & & & & & & \\
\cite{ref:H1F2_94} & 
NOM,ISR & $2.7$ & 
el, $\Sigma$ & $1.5-5000$ & $3.\;10^{-5} - 0.32$ & 
$3-5$ \\
 & 
SVTX,SAT & $0.07$ & 
el, $\Sigma$ &  &  & $^{(*)}$ \\
\cite{ref:ZEUSF2_94_lowQ2} &  ISR & $ 2.5$
 & el & $1.5-15$ & 
$3.5.\;10^{-5} - 0.0015$ & 
$10-15$\\
 & SVTX, ISR & $0.06$
 & el &  & & \\
\cite{ref:ZEUSF2_94} & NOM & $2.5$ & DA & $3.5-5000$ & 
$6.\;10^{-5} - 0.20$ & $3-5$ \\
 & & & & & & \\
1995 & & & & & & \\
\cite{ref:H1F2_95_SVTX} & SVTX & $0.12$ & el, 
$\Sigma$ & $0.35-3.5$ & 
$6.\;10^{-6} - 4.\;10^{-4}$ & 
$15$ \\
\cite{ref:ZEUSF2_BPC} & BPC & $1.65$
 & el & $0.16-0.57$ & $3.\;10^{-6} - 3.\;10^{-5}$ & 
$7-15$ \\
\cite{ref:ZEUSF2_97} & SVTX & $0.24$
 & el & $0.6-17$ & $1.2\;10^{-5} - 1.9\;10^{-3}$ & 
$5-10$ \\ \hline
\multicolumn{7}{l}{$^{(*)}$Note that the SVTX, ISR and SAT have larger
systematic errors, with typical values around 20~\%.}
\end{tabular}
}
\end{center}
\end{table}

As described in section~\ref{sec:coverage}, the kinematic plane is
covered using many different data sets, and different detectors.
These data sets are summarized in Table~\ref{tab:F2_sets}, where the
methods used to extract $F_2$, the integrated luminosity of the data
sets, the kinematic ranges covered, and the typical systematic errors
are listed.  There is a steady trend to increased coverage and smaller
errors as the years progress.

\paragraph{Event Selection}

The neutral current event selection relies primarily on the observation 
of the scattered electron.  Observing the electron 
in the main calorimeters typically limits the range to $Q^2>4$~GeV$^2$%
~\footnote{As of 1995, the kinematic range of the main detectors have
  been extended to $Q^2>1.5$~GeV$^2$, as described in
  section~\ref{sec:coverage}.}.  Electron selection algorithms achieve
high efficiency and purity for energies above about $E_e'=5$~GeV.  The
various analyses of the structure function used electron energy cuts
in the range $E_e'>5-12$~GeV.  A typical observed electron energy
spectrum is shown in Fig.~\ref{fig:electron_spectrum}.  The
expectations from the Monte Carlo simulation are compared with the
data, showing very good agreement.  This type of agreement between
simulation and data is necessary for a precise determination of the
structure functions.

\epsfigure[width=0.8\hsize]{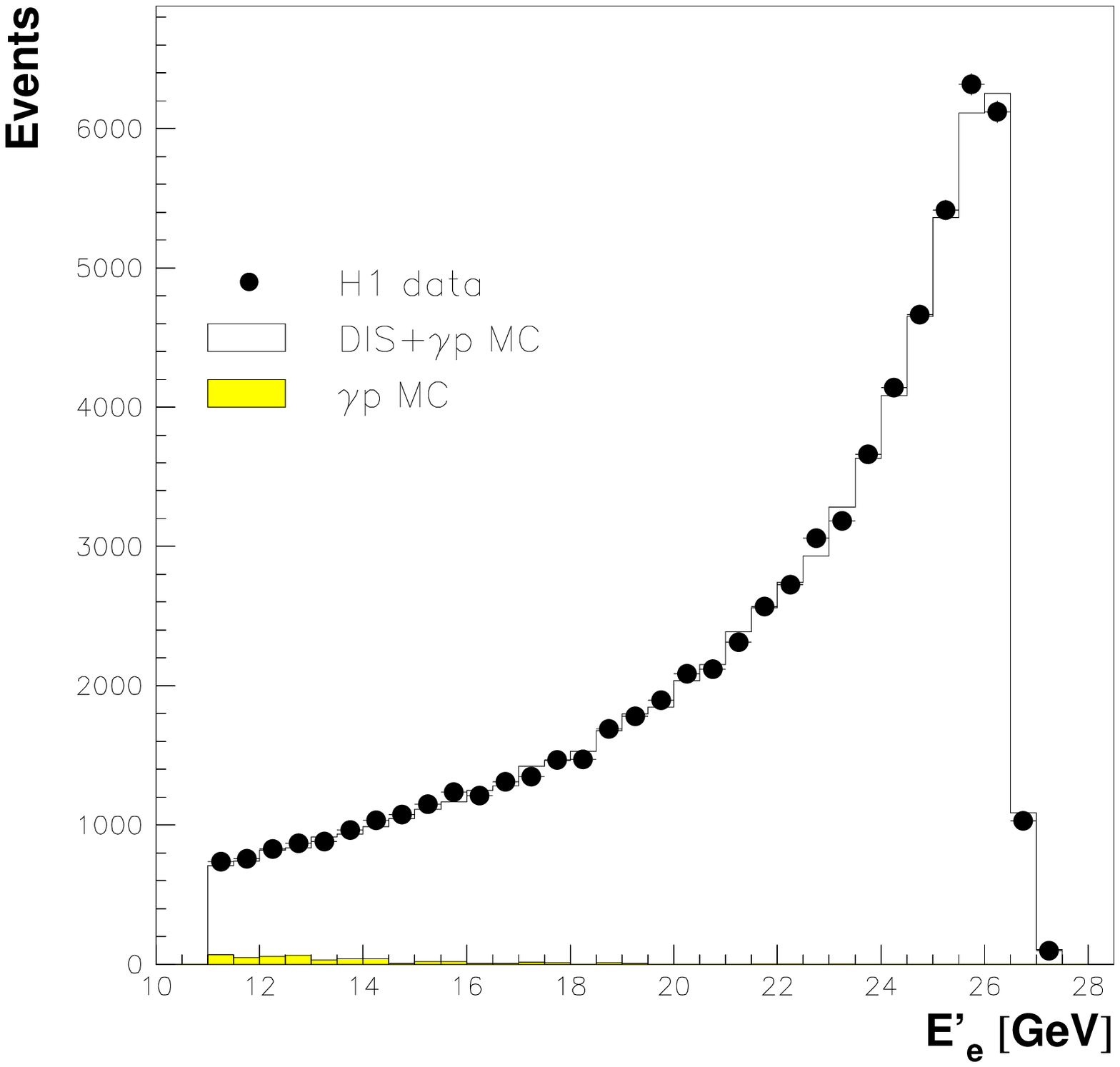} {The energy spectrum
  of the scattered electron from the 1993 H1 DIS data used in the
  extraction of $F_2$.  The expectations from a Monte Carlo simulation
  are also shown, normalized to luminosity.  The shaded histogram
  shows the estimation of the photoproduction background.}
{electron_spectrum}

Another important requirement is the observation of approximate
longitudinal energy-momentum conservation.  The variable 
$\delta = (E-p_Z)_{Observed}$ is calculated as

\begin{equation}
\delta = \sum_{i} E_i (1-cos\theta_i) \; ,
\end{equation}
where the sum runs over clusters measured in the calorimeter and/or
track momenta. The measured $\delta$ should equal $2E_e$ for perfect
detector resolution and in the absence of energy leakage in the
electron direction, which can arise from initial state QED radiation
or from final state particle losses.  A typical requirement is
$\delta>35$~GeV. A $\delta$ spectrum as measured by ZEUS is shown in
Fig.~\ref{fig:E-pz}.

\epsfigure[width=0.8\hsize]{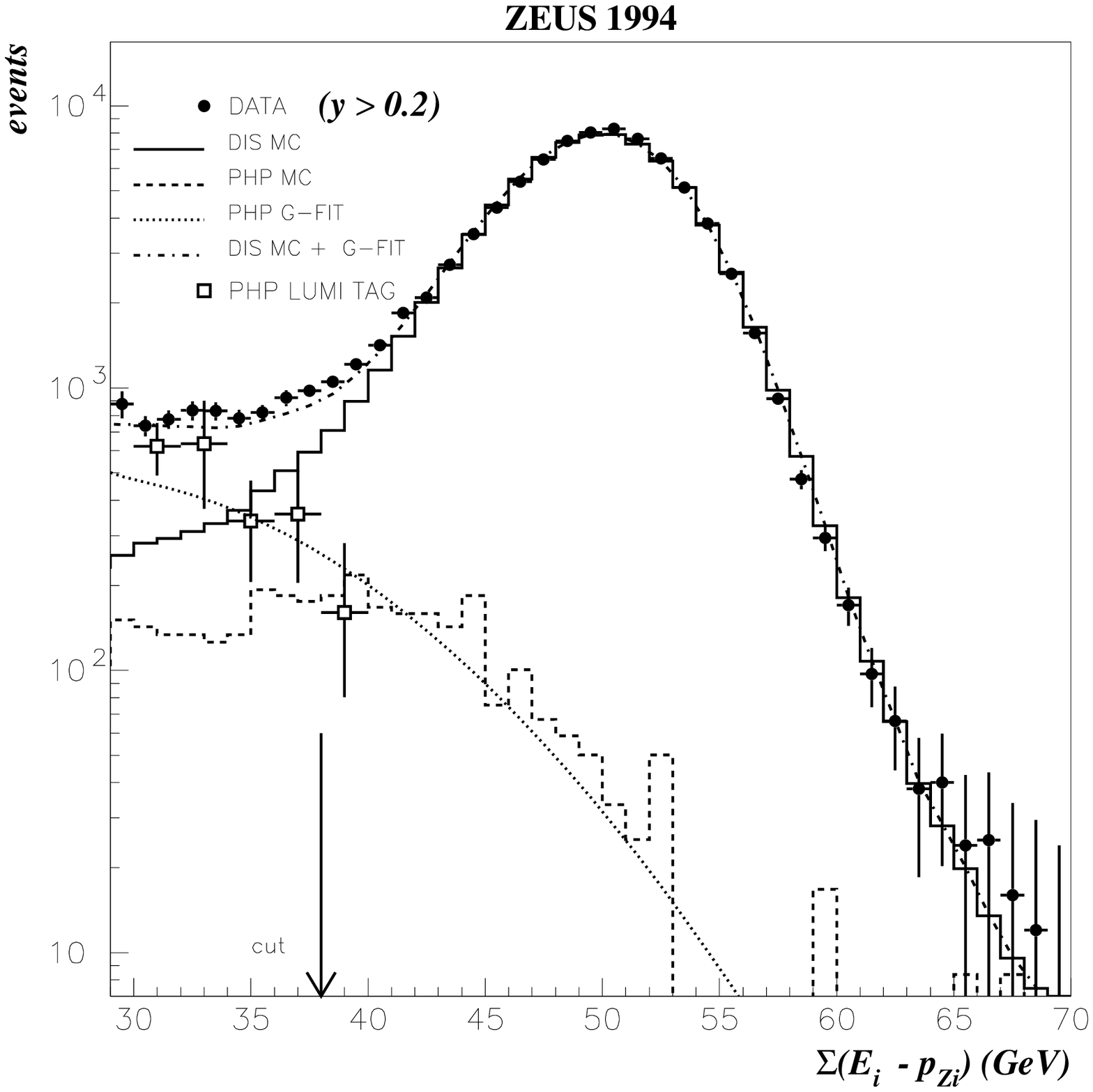} {The $\delta=E-P_Z$
  distribution measured by the ZEUS experiment in the analysis of the
  1994 DIS data.  The solid points show the data which pass the cuts
  applied for the measurement of $F_2$ (except the $\delta$ cut).  The
  open points show the $\delta$ distribution for photoproduction
  background events where the scattered electron has been measured in
  the luminosity electron detector.  The DIS Monte Carlo simulation
  result is shown as the solid histogram while the dashed histogram
  shows the expectations from the photoproduction Monte Carlo
  simulation.  The result of a fit to the data is shown as the
  dashed-dotted line.  The contribution in the fit from
  photoproduction is shown separately as the dotted curve.}  {E-pz}

A minimum hadronic activity is also typically required in the
analysis.  This requirement is usually phrased in terms of the
estimator of the inelasticity, $y$, calculated from the hadronic
system.  This estimator is calculated as
follows~\cite{ref:Jacquet-Blondel}:

\begin{equation}
y = \frac{\delta_{had}}{2E_e} \; ,
\end{equation}
where $\delta_{had}$ is the contribution to $\delta$ from the hadronic
part of the event (i.e., excluding the scattered electron).  The
requirement on $y$ depends strongly on the type of kinematic
reconstruction used.

Other selection criteria often employed are:

\begin{itemize}
\item A reconstructed event vertex within a defined range;
\item Specific cuts to reject cosmic ray induced events or events initiated
from muons in the proton beam halo;
\item Timing cuts to reject events coming from upstream interactions of the 
proton or electron.
\end{itemize}

\paragraph{Backgrounds}

There are several sources of background in the sample of DIS events chosen by the selection 
cuts listed above.  These are:

\begin{itemize}
\item Photoproduction events where a calorimeter cluster has been
  falsely identified as the scattered electron (the scattered electron
  escapes down the beampipe). Photoproduction events are usually
  easily removed by requiring the observation of a high energy
  electron and a large $\delta$. However, these events can still pose
  a problem because of the very large photoproduction cross section
  (roughly a factor $100$ larger than the cross section for DIS with
  $Q^2>10$~GeV$^2$).  While electron identification is usually
  reliable, there are certain kinematical regions where it is
  difficult to distinguish photoproduction events from DIS events.
  The problem is most serious at large $y$, where the scattered
  electron and the fragmentation products of the scattered quark
  overlap spatially.  Algorithms aiming for a high efficiency for
  selecting these events invariably find large background from
  photoproduction processes.  Algorithms aiming for small backgrounds
  have low efficiencies in this kinematic region.  Typical backgrounds
  range from 10~\% in the larger $y$ bins, decreasing rapidly to a
  negligible level at $y \approx 0.5$.
  
\item Scattering of beam particles with rest gas in the beam pipe.
  These types of interactions are both easily studied, using the
  unpaired electron and proton bunches, and also rather easily
  removed.  They produce background below 1~\%.
  
\item QED Compton events ($ep \rightarrow ep\gamma$), where the final
  state electron and photon both end up in the main detector.  They
  form part of the radiative corrections, but were not included in the
  event simulations used for the extraction of $F_2$.  They must
  therefore be removed by algorithms specifically designed to select
  this final state.

\end{itemize}

The only significant background is from photoproduction.  Great care
must be taken in evaluating this background since it has a strong $y$
dependence, and could easily distort the measured $x$ dependence of
the structure functions at small $x$.  There are however several ways
of estimating this background (see for example the $\delta$ spectrum
in Fig.~\ref{fig:E-pz}).

\paragraph{Cross section measurement}

The cross section is determined in bins of $x$ and $Q^2$ (sometimes
$y$ and $Q^2$) which are commensurate with the resolution.  At small
$Q^2$, bin sizes are chosen from a Monte Carlo study such that the
fraction of events reconstructed in the same bin as that determined
from the generated variables is larger than about $30$~\%.  This
limits the size of the corrections which arise from unfolding the
detector response.  At larger $Q^2$, the bin sizes are generally set
by requiring a minimum number of expected events in the bin.  The
cross section is determined from the number of events measured in a
bin after corrections for background, inefficiencies, acceptances and
resolution smearing, and the measured luminosity.  The uncertainties
in the luminosity measurements have steadily improved and are now at
the $1.5$~\% level.

\paragraph{Extraction of the structure function}

The differential cross section for $e^-p$ 
NC scattering is given in terms of the 
of the structure functions $F_2, F_L$ and $F_3$ as 

\begin{equation}
\label{eq:cross}
\frac{d^2\sigma_{NC}}{dxdQ^2} = \frac{2\pi\alpha^2}{xQ^4}\left[ Y_+ 
F_2(x,Q^2) - y^2 F_L(x,Q^2) + Y_- xF_3(x,Q^2)\right] \; .
\end{equation}

The variables $x$ and $Q^2$ are defined at the vertex of the Feynman
graph where the gauge boson couples with the quark from the proton. 
The data are corrected for acceptance, 
resolution effects and radiative corrections using Monte Carlo methods
before the differential cross sections are extracted.

The correction for $F_L$ is performed using the QCD prescription for
$F_L$ given in Eq.~(\ref{eq:FL}).  Target mass and higher twist
effects are expected to be small and are ignored.

Note that $F_2=F_2^{EM}+F_2^{Weak}+F_2^{Int}$, where $F_2^{EM}$
represents the contribution from $\gamma$-exchange, $F_2^{Weak}$ the
contribution from $Z^0$-exchange, and $F_2^{Int}$ the contribution
from $\gamma-Z^0$ interference (see section~\ref{sec:sf_formulae}).
The value of $F_2^{EM}$ is often quoted separately.  The contributions
from $F_2^{Weak}\; {\rm and}\; F_2^{Int}$ become larger than a few
percent above $Q^2=1000$~GeV$^2$, as does the contribution to the
cross section from $F_3$.  The effects of $W$ and $Z$ exchange are
subtracted using standard parton density parameterizations.  In the
largest $Q^2$ bin measured, $Q^2=5000$~GeV$^2$, the $F_3$ correction
reaches $22$~\% for $e^+p$ scattering in the smallest $x$ bin
measured, while the $Z^0$ contribution to $F_2$, including
interference term, is $8$~\%.  It is expected that each of these
contributions will eventually be individually measured at HERA.

\subsubsection{HERA structure function results}

\epsfigure[width=0.8\hsize]{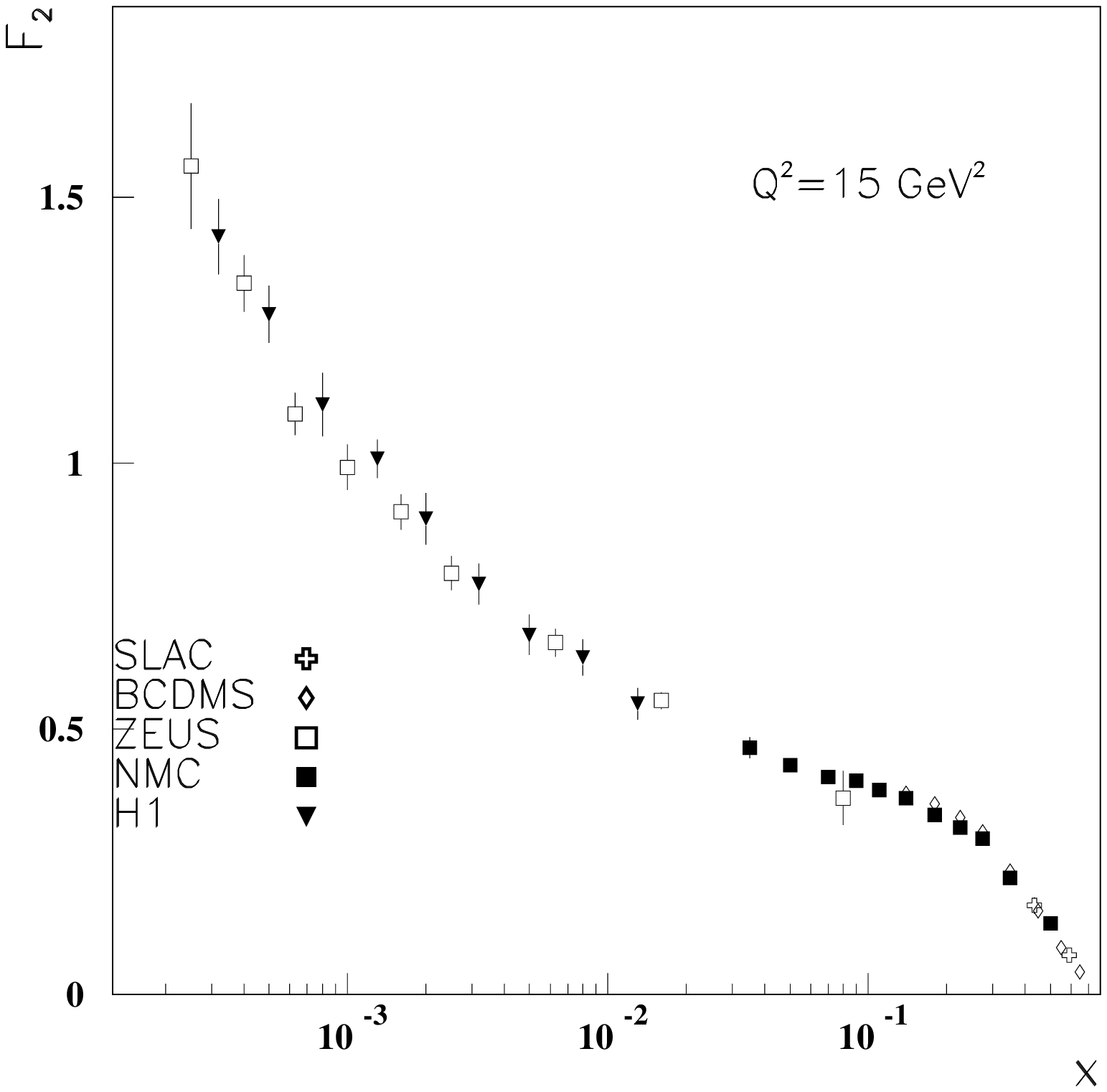} {Comparison of the
  fixed target and HERA data at $Q^2=15$~GeV$^2$.  The data span $3.5$
  orders of magnitude in $x$ and match in a smooth way.}  {F2_Q215}

\epsfigure[width=0.8\hsize]{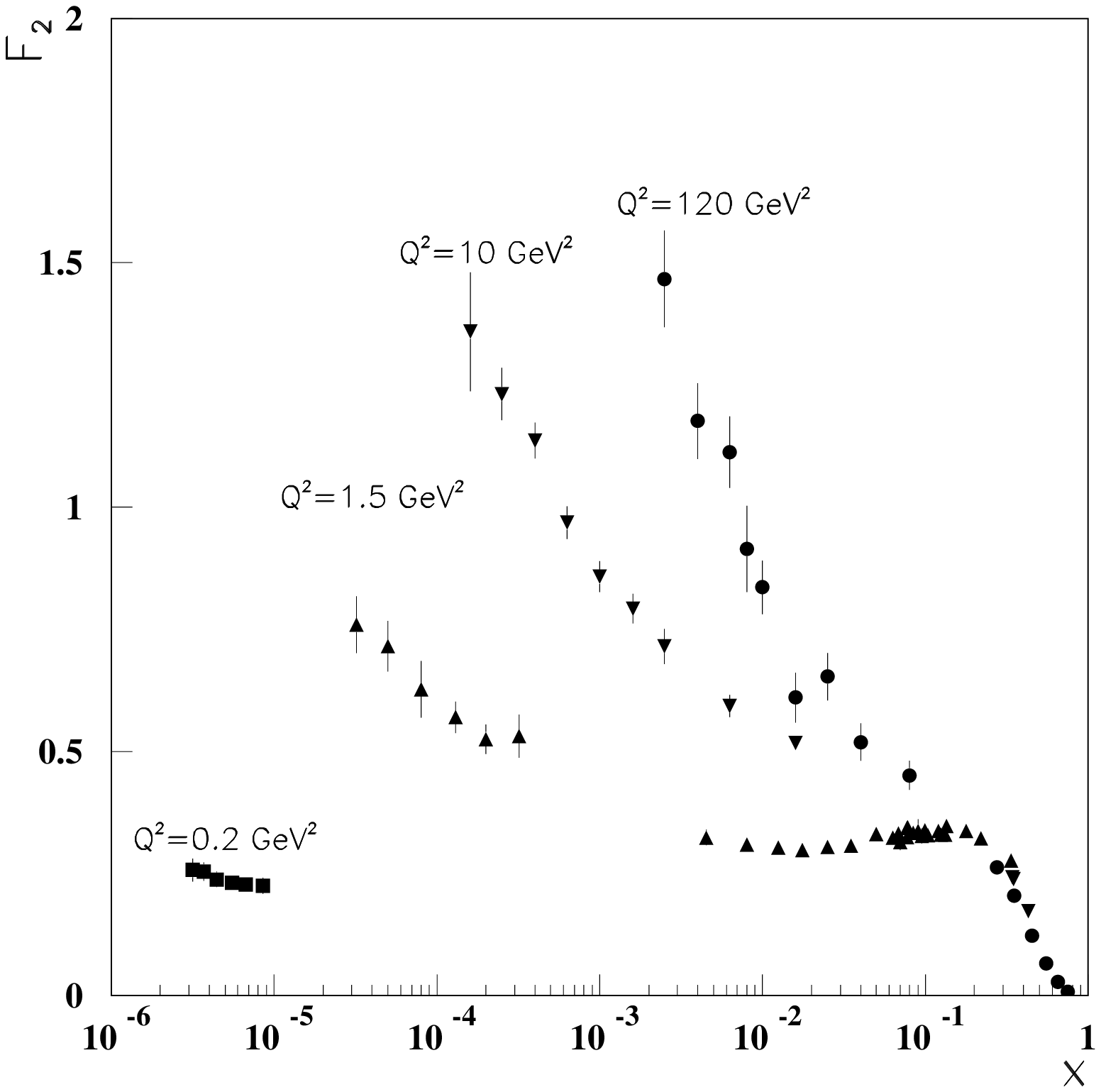} {The $F_2$ data from
  HERA and from fixed target experiments is plotted vs $x$ for
  different values of $Q^2$.  The rise of $F_2$ towards smaller $x$
  depends strongly on $Q^2$.}  {F2vsx}

The HERA accelerator started delivering luminosity in 1992, and the
experiments ZEUS and H1 have now each published several sets of proton
structure function measurements (see Table~\ref{tab:F2_sets}).  The
first measurements showed the now well known rise of $F_2$ with
decreasing $x$ at small $x$.  These results were confirmed with the
1993 and 1994 data.  The $(x,Q^2)$ plane covered by the fixed target
experiments, and by HERA, is shown in Fig.~\ref{fig:F2plane}.  As can
be seen in this figure, the HERA results extended the $Q^2$ coverage
at fixed $x$ by two orders of magnitude, and extended the $x$ range at
fixed $Q^2$ by two orders of magnitude.  The current data is of very
high quality, as can be seen in the example shown in
Fig.~\ref{fig:F2_Q215}.  Here, the HERA $F_2$ data is plotted vs $x$
for $Q^2=15$~GeV$^2$ and compared to the results from fixed target
experiments.  The HERA data matches on smoothly to the fixed target
data.  In this $Q^2$ range, the $F_2$ data now span nearly four orders
of magnitude in $x$.  The rise of $F_2$ with decreasing $x$ depends
strongly on $Q^2$, as shown in Fig.~\ref{fig:F2vsx}.  Increasing $Q^2$
implies that smaller transverse distance scales are probed.  The
number of quarks and antiquarks in the proton seen by the probe
increases with decreasing momentum fraction, $x$, more rapidly as the
distance scale is reduced.

\epsfigure[width=0.95\hsize]{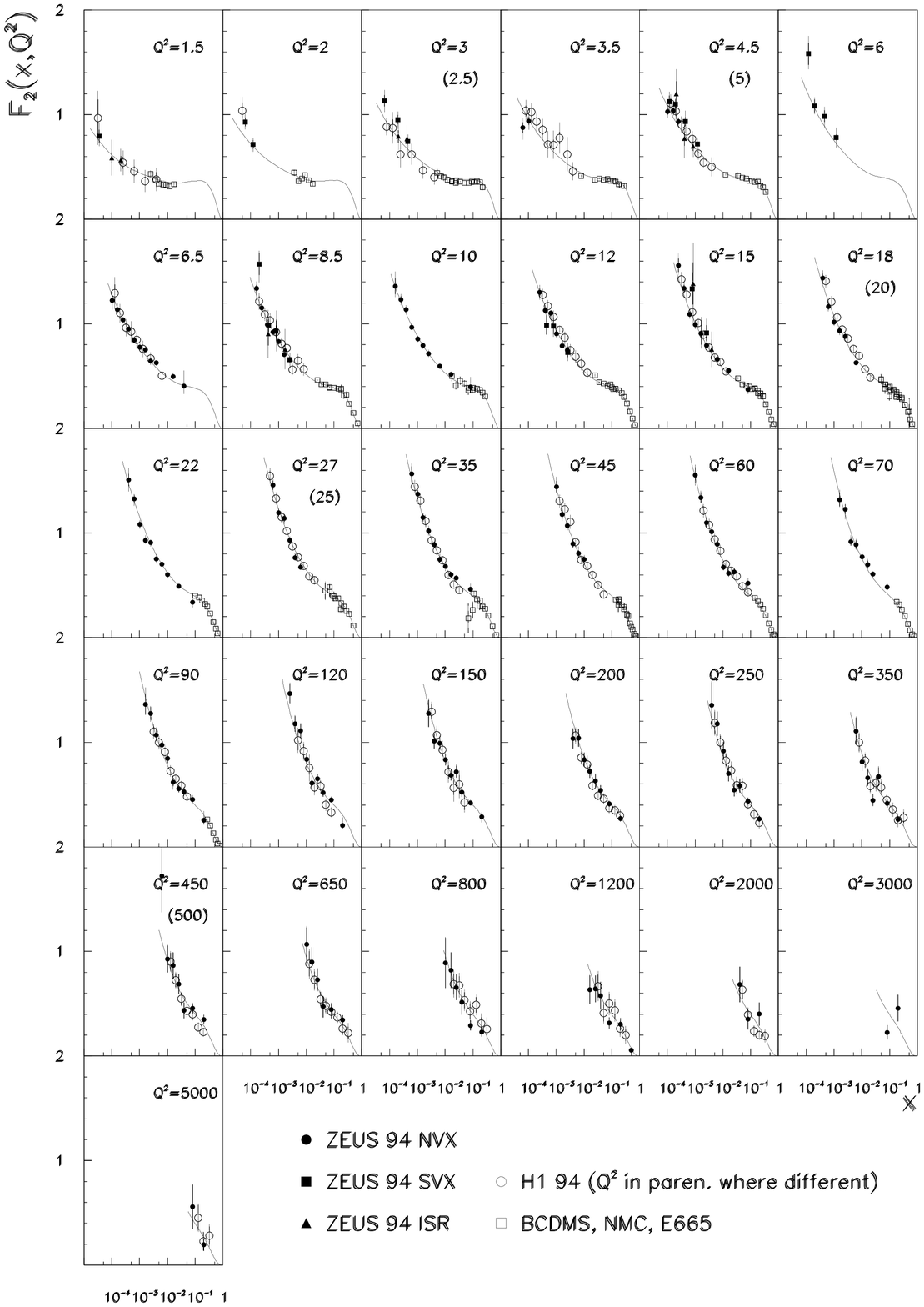} {$F_2$ is plotted vs
  $x$ for the 1994 HERA data, and compared to results from fixed
  target experiments.  $F_2$ shows a strong rise towards smaller $x$,
  with the strength of the rise increasing with $Q^2$.  The curve
  represents the result of an NLO QCD fit performed by the ZEUS
  collaboration.  It is able to reproduce the data down to
  $Q^2=1.5$~GeV$^2$.}  {F2_94}

The published HERA structure function measurements over the full
kinematic range covered by the 1994 data are shown in
Fig.~\ref{fig:F2_94}.  This figure includes the data from the
different methods used to extract $F_2$ (nominal, shifted vertex and
initial state radiation events).  In some cases, several measurements
from the same experiment are shown in overlapping kinematical ranges.
The HERA data are in good agreement with each other, and with the
fixed target data.  The curves in the plots represent the results of a
fit based on perturbative QCD (the NLO DGLAP evolution equation,
described in section~\ref{sec:QCD}).  It is found to reproduce the
data well down to the smallest values of $Q^2$ in this plot.  The
success of the pQCD fit down to small $Q^2$ was quite surprising, and
has generated a lot of interest in the small $Q^2$ region.

\epsfigure[width=0.95\hsize]{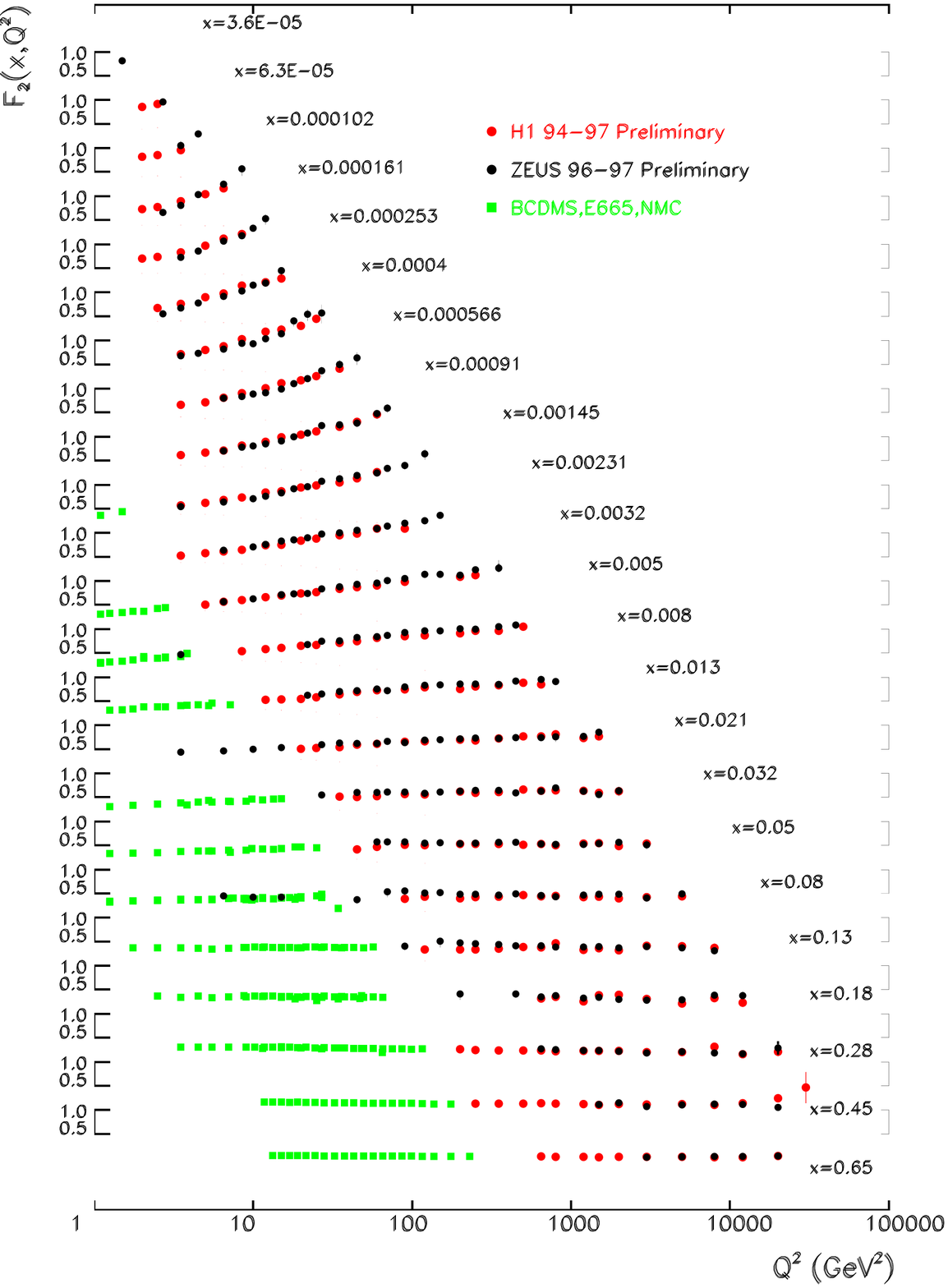} {$F_2$ is plotted vs
  $Q^2$ for fixed $x$ values for the preliminary 1994-97 HERA data,
  and compared to results from fixed target experiments. $F_2$ shows
  strong scaling violations, particularly at small $x$. } {F2_94_Q2}

The preliminary 1994-97 $F_2$ results from HERA are shown versus
$Q^2$, for fixed $x$, in Fig.~\ref{fig:F2_94_Q2}.  As can be seen in
the figure, the more recent HERA data allow an extension of the $F_2$
measurements up to $Q^2=30,000$~GeV$^2$.  Strong scaling violations
are seen: $F_2$ increases with $Q^2$ at small $x$, while it decreases
with increasing $Q^2$ at large $x$.  This pattern of scaling
violations is expected in QCD from gluon radiation from the quarks,
and gluon splitting into quark-antiquark pairs.  This radiation is
made visible when the distance scale is reduced, and large $x$ quarks
are resolved as a composite of quarks, antiquarks and gluons.  The
prediction of scaling violations is a success of QCD.  The strong
scaling violations at small $x$ point to the presence of a large gluon
density; in fact, the measurement of the scaling violations leads to a
determination of the gluon density in the proton, as explained in
section~\ref{sec:partons}.

\paragraph{Discussion of HERA Structure Function Results}

The early measurements indicated that $F_2$ indeed rises quickly at
small $x$, but not as quickly as predicted by the first BFKL
approaches.  The prediction of GRV was remarkably close to the HERA
data down to much smaller $Q^2$ than initially expected.  These
developments had several effects:
\begin{enumerate}
\item The need for BFKL evolution could not be demonstrated, and it
  now seems unlikely that $F_2$ measurements will be able to separate
  BFKL from DGLAP evolution.  Recent BFKL type
  calculations~\cite{ref:BFKL_recent1,ref:BFKL_recent2} indicate a
  less steep rise of $F_2$ at small $x$, making it more difficult to
  separate from DGLAP evolution.
\item NLO DGLAP evolution, with appropriate parton distributions, is
  able to fit all structure function data down to values of
  $Q^2\approx 1$~GeV$^2$.  There is no indication in $F_2$ of the
  onset of saturation in the kinematic range measured.  The data are
  now precise enough that clear deviations are seen from the GRV
  prediction, but the accuracy of this prediction gave the first
  indication that the standard NLO DGLAP evolution could reproduce the
  bulk of the HERA structure function data.
\item The success of perturbative QCD down to small values of $Q^2$
  has inspired new research into the predictions of the DGLAP
  equations. One such prediction goes by the name of double asymptotic
  scaling~\cite{ref:DAS1,ref:DAS2}.  The DGLAP evolution equations are
  rewritten in terms of the variables

\begin{eqnarray}
\sigma & = & \sqrt{\ln\left(\frac{x_0}{x}\right)
\ln\left(\frac{\alpha_S(Q_0)}{\alpha_S(Q)}\right)} \; ,\\
\rho & = & \sqrt{\frac{\ln\left(\frac{x_0}{x}\right)}
{\ln\left(\frac{\alpha_S(Q_0)}{\alpha_S(Q)}\right)}} \; ,
\end{eqnarray}

where $Q_0$ and $x_0$ are parameters which must be determined
experimentally.  The measured values of $F_2$ are rescaled by
functions $R_F(\sigma,\rho)$ and $R_F'(\sigma,\rho)$ to remove the
part of the leading sub-asymptotic behavior which can be calculated in
a model independent way.  The results for the H1~\cite{ref:H1F2_94}
$F_2$ measurements are shown in Fig.~\ref{fig:H1_DAS}.  Scaling is
found to set in for $\rho \geq 2$ and $Q^2>5$~GeV$^2$.  The
measurement requires a value of $Q_0^2\approx 2.5$~GeV$^2$, while a
value of $x_0=0.1$ was found to be a good choice.  Although the
physical meaning of these scaling variables is not clear, the results
can be interpreted to indicate that the DGLAP formalism is valid down
to small values of $Q^2$.
\end{enumerate}

\epsfigure[width=0.8\hsize]{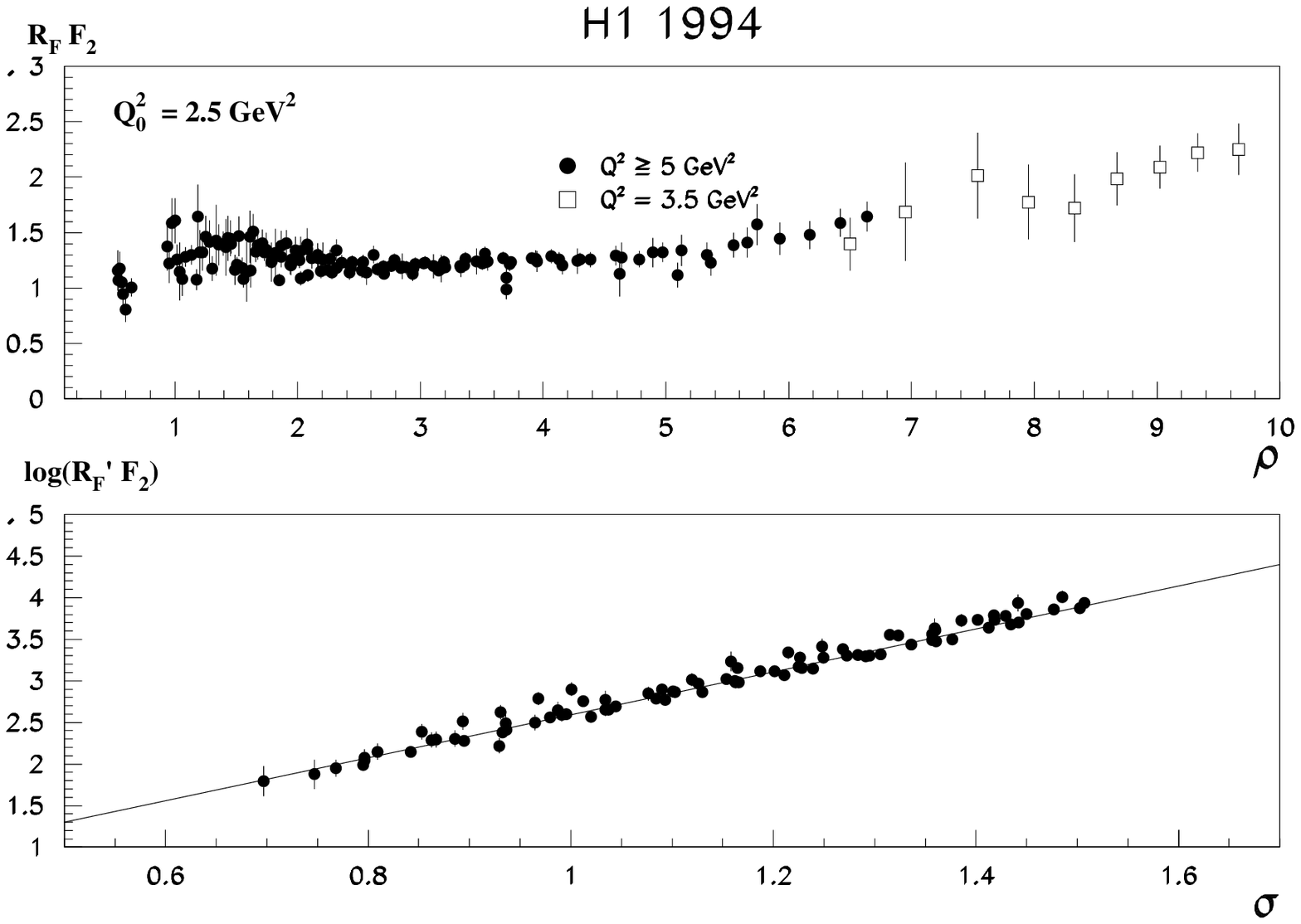} {The rescaled
  structure functions: top, $R_FF_2$ versus $\rho$; bottom,
  $log(R_F'F_2)$ versus $\sigma$ (see text).  Only data with
  $Q^2\geq5$~GeV$^2$ and $\rho>2$ are shown in the lower plot.}
{H1_DAS}

Deviation from NLO DGLAP evolution are now searched for at the edges
of the available phase space.  These include measurements down to
$Q^2=0.11$~GeV$^2$ which probe the transition from non-perturbative to
perturbative processes, measurements at large $y$, and measurements at
very large $Q^2$.  These are all discussed in the following sections.

Before we turn to these more detailed descriptions of the HERA data,
we first review one outstanding inconsistency in the fixed target
data, and see what HERA measurements can contribute to the debate.

\subsubsection{Discrepancy in small-$x$ fixed target data}

One outstanding discrepancy in the fixed target data involves the
comparison of the $F_2$ data from the NMC~\cite{ref:NMC_F2},
E665~\cite{ref:E665_F2} and CCFR~\cite{ref:CCFR_F2} experiments at
$x\approx 0.01$.  The NMC and E665 data is for a muon beam on a proton
target.  The CCFR experiment used a neutrino beam on an iron target,
where the data were corrected for nuclear effects as well as for the
different propagator (photon versus $W,Z$)~\cite{ref:seligman}.  At
the smallest values of $x$ probed by these experiments, $x\approx
0.01$, the CCFR data lie about 15~\% above the NMC data.  This
represents a much larger difference than can be accounted for by the
normalization uncertainties quoted.  At larger $x$, the discrepancy
vanishes.  Many theoretical ideas have been discussed for the
discrepancy, such as effects related to the strange
sea~\cite{ref:NN_strange,ref:BM_strange}, or charm threshold
effects~\cite{ref:Gluck}.  As can be seen in Fig.~\ref{fig:F2plane},
the HERA data has some overlap with the fixed target data in this
region, and can be compared to the NMC, E665 and CCFR data.  This is
done in Fig.~\ref{fig:NMC_CCFR_HERA} for two different $x$ values.
The HERA data tend to agree better with the CCFR data, despite
experimental conditions which are more similar to those of the NMC and
E665 experiments.  These results need confirmation with more precise
data from future measurements at HERA.  Any theoretical explanation of
the NMC, E665 and CCFR difference should also account for the HERA
data.

\epsfigure[width=0.95\hsize]{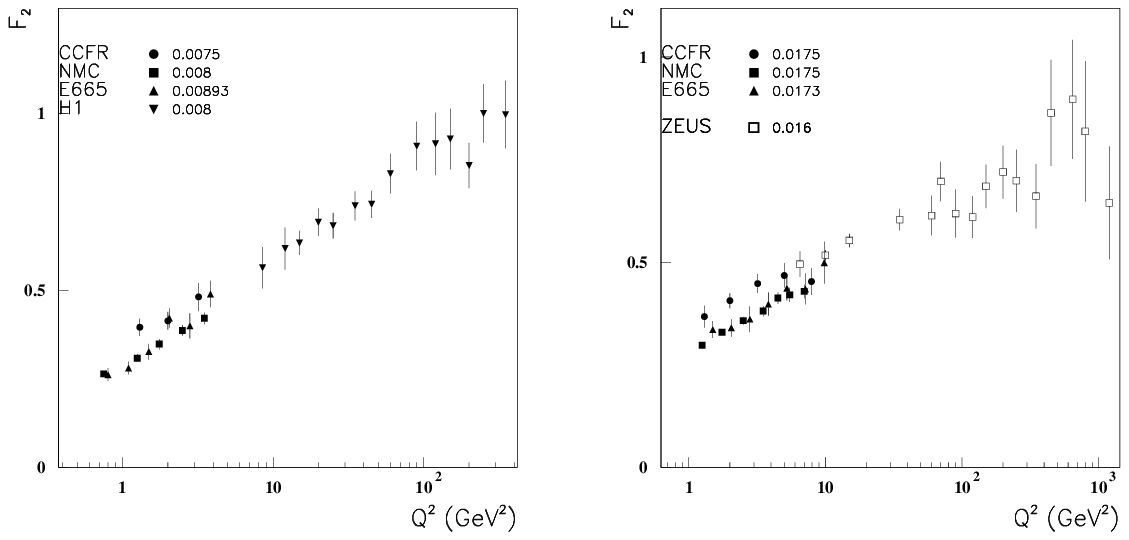} {The HERA data
  compared to that from NMC, CCFR and E665 for $x\approx 0.01$.  The
  $x$ value of the different data sets is given in the plot.  Note
  that the expected change in $F_2$ from the differences in $x$ are
  much smaller than the NMC/CCFR difference.  The NMC and E665 data
  are below the CCFR data by about 15~\% in this $x$ range.  The HERA
  data just begin to overlap the fixed target data, and are more
  compatible with the results from CCFR than those from NMC and E665.}
{NMC_CCFR_HERA}

\subsubsection{$F_2$ at small $Q^2$}

We now turn to a more detailed discussion of the small $Q^2$ structure
function data at HERA.  In photoproduction, it was found that the
total cross section has a weak energy dependence similar to that found
in hadron-hadron scattering.  In contrast, a steep energy (or $x$)
dependence was found in DIS.  It is therefore very interesting to
consider the transition region between these two different energy
behaviors.  As mentioned above, the ability of pQCD fits (in
particular the success of the GRV parameterization, which predated
these results) to reproduce the DIS data down to the lowest values of
$Q^2$ initially available, $1.5$~GeV$^2$, inspired the ZEUS and H1
collaborations to make measurements at still smaller $Q^2$.

Several approaches have been used to access smaller $Q^2$, as
described in section~\ref{sec:coverage}.  These include using data
sets with the interaction vertex shifted in the proton beam direction,
thereby increasing the angular acceptance of the detectors at smaller
electron scattering angles, and the addition of dedicated detectors at
small angles to the electron beam direction such as the ZEUS beam pipe
calorimeter (BPC).  Events where the effective electron energy is
reduced through tagged initial state radiation have also been
employed, but these analyses tend to suffer from higher statistical
and systematic uncertainties.

\epsfigure[width=0.95\hsize]{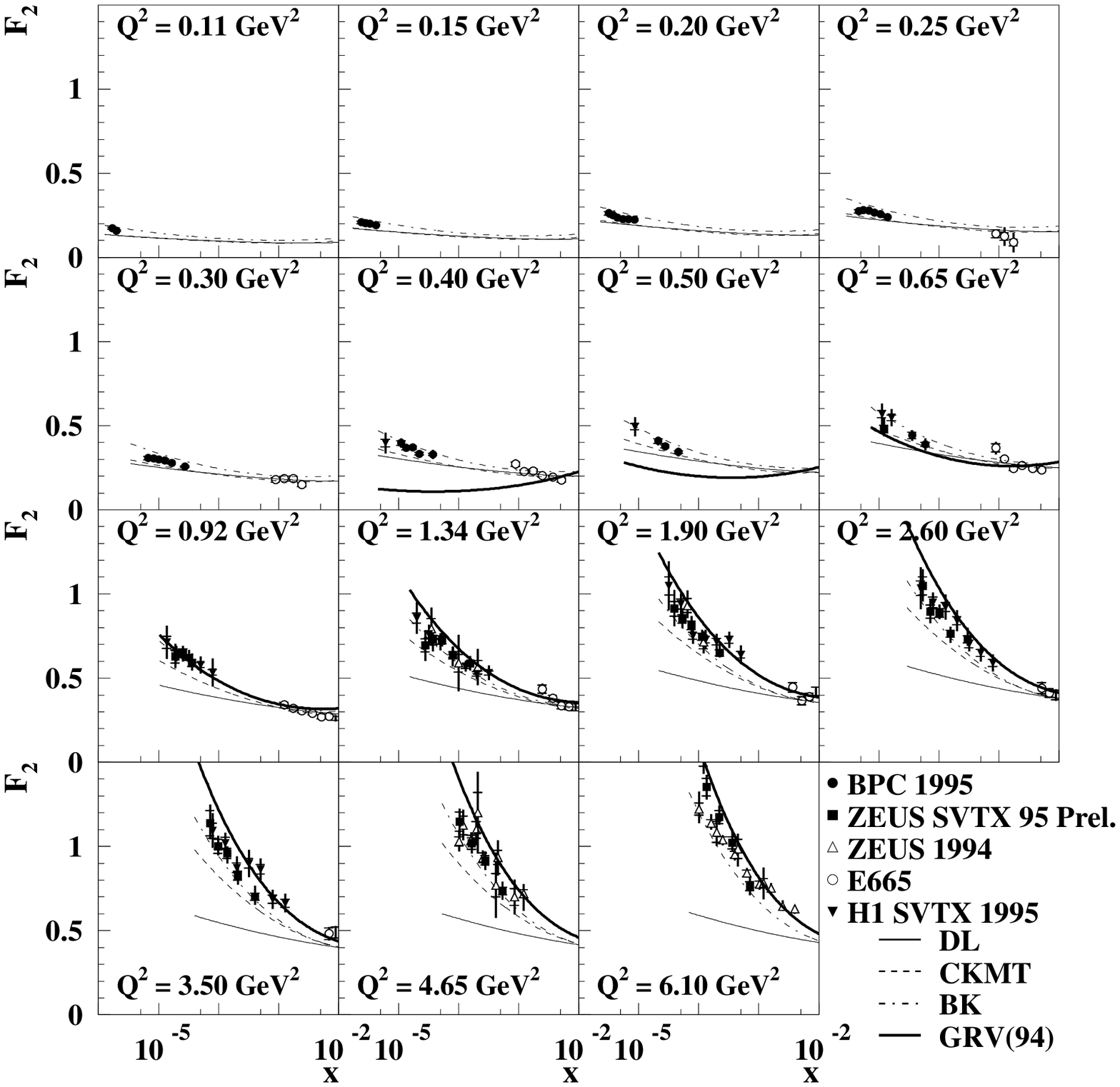} {The $F_2$ data
  at the lowest $Q^2$ values measured at HERA, compared to fixed
  target data from E665. $F_2$ is relatively flat at lower $Q^2$
  values, and starts to show a rise at small $x$ around $Q^2\approx
  1$~GeV$^2$.  The $F_2$ data are compared to different
  parameterizations (see text for details).}  {F2_lowQ2}

The $F_2$ data at the smallest $Q^2$ values measured at HERA are shown
in Fig.~\ref{fig:F2_lowQ2}. $F_2$ is relatively flat at the smallest
$Q^2$ values, and starts to show a rise at small $x$ for $Q^2\approx
1$~GeV$^2$.  The data are compared to different parameterizations
which we now describe.

\begin{enumerate}
  
\item The simplest of these parameterizations, from Donnachie and
  Landshoff~\cite{ref:DoLa_F2}, is labeled DL.  It is an extension of
  a Regge based parameterization which was very successful in
  explaining inclusive cross section measurements in hadron-hadron and
  photoproduction data, as well as earlier fixed target DIS data.  It
  clearly fails to reproduce the rise of $F_2$, and is systematically
  below the data at the smaller values of $Q^2$.
  
\item A more complicated Regge-type parameterization has been put
  forward by CKMT~\cite{ref:CKMT}, which has a pomeron intercept
  changing with $Q^2$.  The parameterization follows the rise of
  $F_2$, but lies below the data for $Q^2>1$~GeV$^2$.
  
\item The curve BK~\cite{ref:BK1,ref:BK2} represents the predictions
  of a generalized vector dominance model (GVDM), matched to the GRV
  prediction at larger $Q^2$.  It lies above the data at the smallest
  $Q^2$, and below the data for $Q^2>1$~GeV$^2$.
  
\item The curve labeled GRV is the prediction of Gl\"uck, Reya and
  Vogt.  In their approach, the parton densities are parameterized at
  a very small $Q^2_0$ as consisting of valence quarks and
  gluons~\cite{ref:GRV94}.  The rise of the parton densities at small
  $x$ results from QCD radiative processes, which are calculated with
  the NLO DGLAP equation.  The GRV parameterization lies below the
  data near its starting point of $Q_0^2=0.34$~GeV$^2$, and rises
  quickly with increasing $Q^2$ such that it lies above the data for
  $Q^2>1.3$~GeV$^2$.
\end{enumerate} 

\epsfigure[width=0.9\hsize]{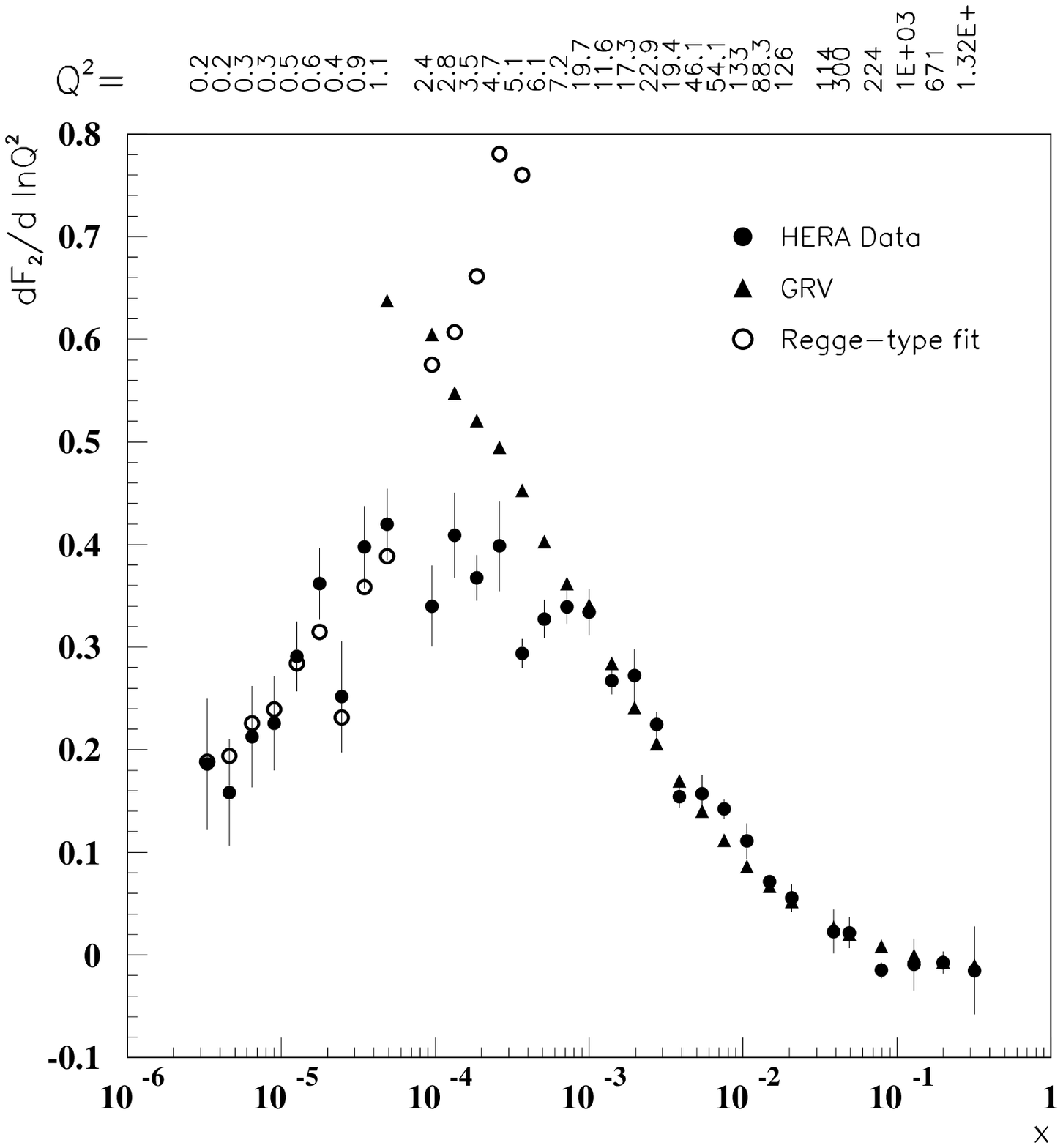} {The values of
  $dF_2/d\ln Q^2$ for the HERA published data, compared to the
  prediction from the GRV parameterization~\protect\cite{ref:GRV94}.
  The fits to the HERA data have been performed by binning the data in
  $x$, and fitting a line $F_2 = a + b \ln Q^2$.  The central $x$
  value of the bin is given on the horizontal axis.  The central $\ln
  Q^2$ value of the bin is given at the top of the plot.  The data
  slopes are also compared to the prediction from a Regge
  parameterization (see text) .}  {hera_scaling}

The behavior of the GRV curve indicates that the scaling violations
predicted by this pQCD approach are much too strong at small $Q^2$.
The scaling violations can be studied by plotting the data versus
$Q^2$ in bins of fixed $x$.  For this purpose, we have taken the
published HERA data (ZEUS F2(BPC), F2(1994) and H1 F2(SVTX), F2(1994))
as of the fall of 1997.  The statistical errors from the measurements
were added in quadrature with the systematic errors, and a fit of the
form
\begin{equation}
F_2 = a + b \cdot \ln Q^2
\end{equation}
was performed in bins of $x$.  The slope $b$ is plotted in
Fig.~\ref{fig:hera_scaling} and compared to the prediction from GRV.
The slope in the data increases with decreasing $x$ down to $x\approx
10^{-4}$, at which point it turns over.  The slope from GRV follows
the data at larger $Q^2$, but overshoots the data for $Q^2$ smaller
than about $4$~GeV$^2$, and shows no sign of decreasing at smaller
$x,Q^2$.  The underlying physics explanation for the turnover has yet
to be determined.  The slope $dF_2/d\ln Q^2$ is in leading order
directly related to the gluon momentum density in the proton.  The
turnover at $(x,Q^2) \approx (10^{-4},5~\rm{GeV}^2)$ indicates that
the gluons are starting to decrease in number as $(x,Q^2)$ decrease.
The reasons for this interesting observation are currently under
investigation -- one possibility is that this is a sign of shadowing
corrections in DIS~\cite{ref:Levinetal}.

The behavior of Regge type parameterizations was studied by fitting a 
function of the form
\begin{equation}
F_2 \propto \frac{M_V^2 Q^2}{M_V^2+Q^2}  W^{2\epsilon} \; ,
\end{equation}
where
\begin{equation}
\epsilon=\epsilon_0+\epsilon_1\cdot \ln (Q^2+Q^2_0) \; ,
\end{equation}
to the ZEUS BPC data.  The slope of $F_2$ from this type of
parameterization is compared to the data in
Fig.~\ref{fig:hera_scaling}.  The parameterization works well up to
about $Q^2=1$~GeV$^2$.  Beyond this value of $Q^2$, the predicted
slope is much greater than seen in the data.

\epsfigure[width=0.8\hsize]{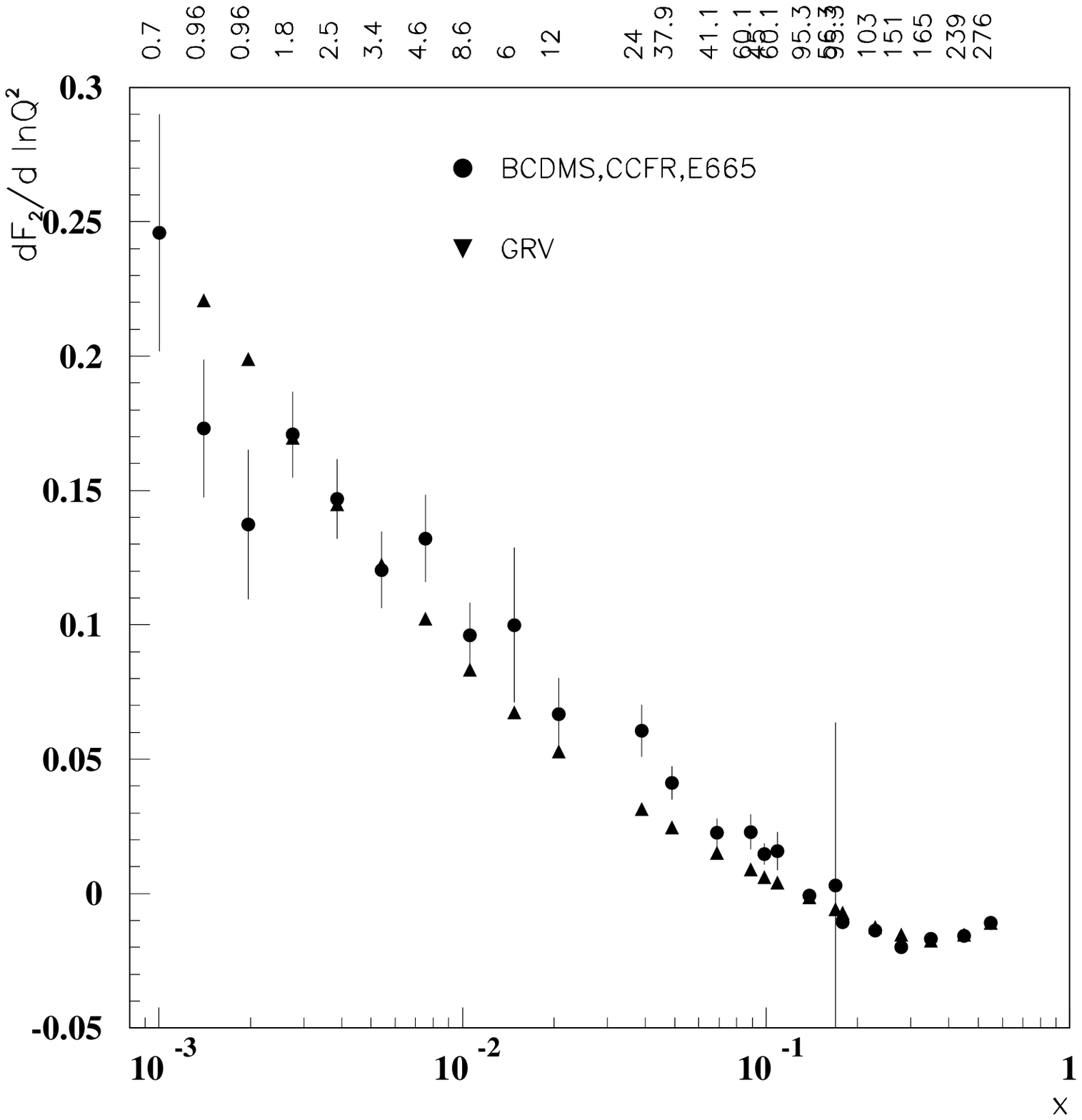} {The values of
  $dF_2/d\ln Q^2$ for fixed target data.}  {scaling_fixed}

Note that the turnover of $dF_2/d\ln Q^2$ at small $(x,Q^2)$ is not
purely due to a $Q^2$ effect. This can be seen in
Fig.~\ref{fig:scaling_fixed}, where $dF_2/d\ln Q^2$ for fixed target
data is plotted in a similar way to HERA data.  Similar values of
$Q^2$ are reached, but at larger $x$.  There is no indication for a
turnover of the slopes.

The $dF_2/d\ln Q^2$ measurements can be reproduced by a general fit using
the NLO DGLAP equations~\cite{ref:ZEUSF2_97}, albeit with rather curious
parton densities.  It is found from the fit that the gluon density
vanishes at small-$x$ near $Q^2=1$~GeV$^2$.  This is discussed in more
detail in section~\ref{sec:gluonfits}.

\epsfigure[width=0.8\hsize]{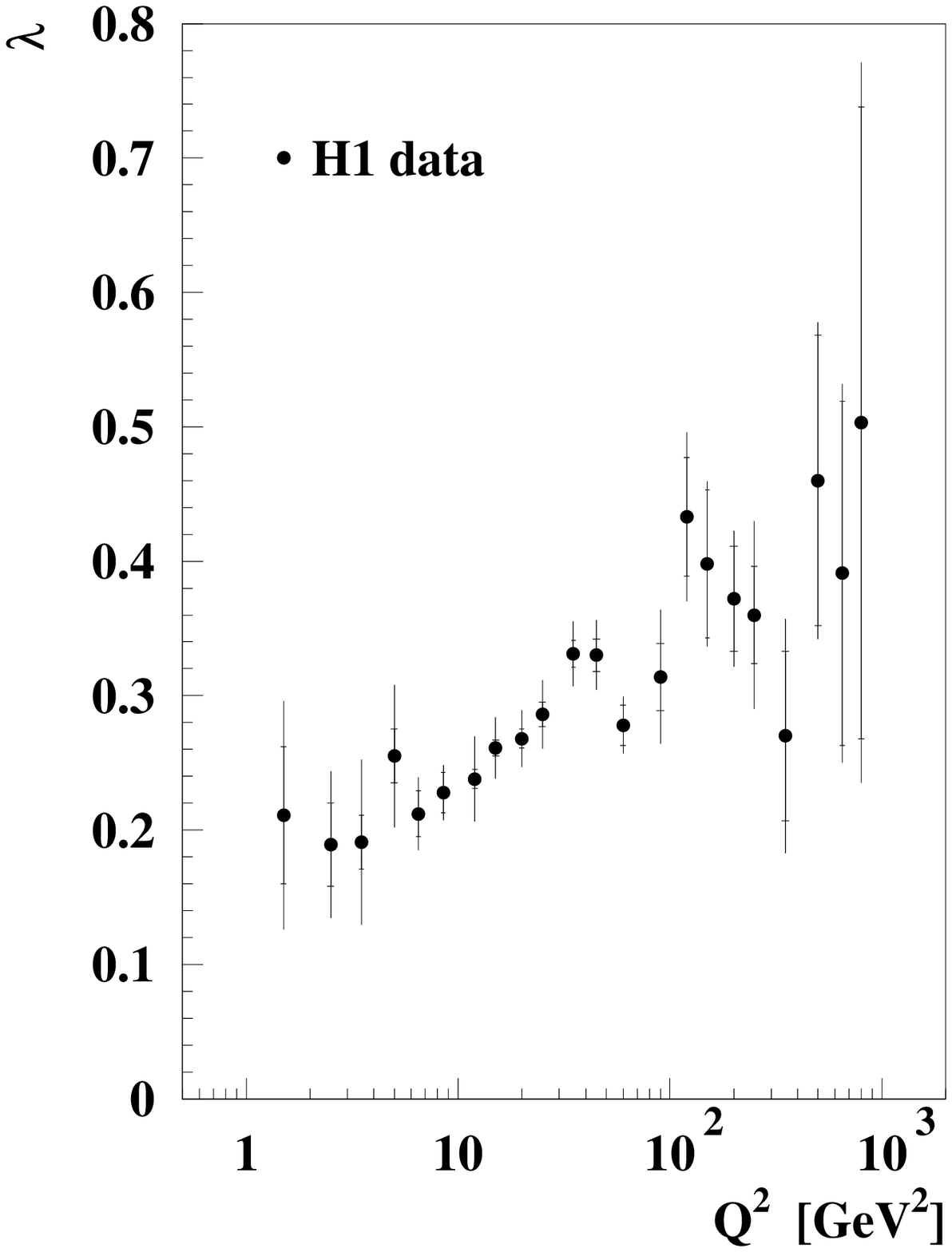} {Variation of the
  exponent $\lambda$ from fits to the H1
  data~\protect\cite{ref:H1F2_94} of the form $F_2 \sim x^{-\lambda}$
  at fixed $Q^2$ values and $x<0.1$.}  {H1_lambda_vs_Q2}

In Fig.~\ref{fig:F2_94} and \ref{fig:F2_lowQ2}, the slope of $F_2$
with decreasing $x$ is seen to increase with $Q^2$ in a smooth way.
This behavior is quantified in Fig.~\ref{fig:H1_lambda_vs_Q2}, where
the power $\lambda$, from a fit of the type $F_2 \sim x^{-\lambda}$,
is shown to decrease smoothly with $Q^2$.  It reaches values near the
photoproduction result ($0.16$) near $Q^2=1$~GeV$^2$. If we associate
the steep rise of the structure function with the presence of
perturbative QCD processes, then this result also indicates that pQCD
processes become important around $Q^2=1$~GeV$^2$.

These results lead to the conclusion that non-perturbative
effects dominate for $Q^2 < 1$~GeV$^2$, whereas pQCD processes
dominate the cross section above $Q^2\approx 5$~GeV$^2$.  The HERA
data span this transition, and may therefore yield insight into
non-perturbative QCD processes.

\subsubsection{$F_L$}
\label{sec:FL}

The measurement of $F_L$ is very important to test the reliability of
pQCD calculations, since its value is predicted from the gluon density
and $F_2$ (see Eq.~(\ref{eq:FL})).  In addition, the uncertainty in
$F_L$ limits the accuracy with which $F_2$ can be determined.  As can
be seen in Eq.~(\ref{eq:NCdiff}), the contribution to the differential
cross section from $F_L$ is suppressed by $y^2/Y_+$ relative to $F_2$.
Given the bound $F_L\leq F_2$, the $F_L$ contribution will be small at
small $y$.  However, the particularly interesting small-$x$ region
corresponds to large $y$, where the $F_L$ contribution cannot be
ignored.  The usual procedure adopted experimentally for the
extraction of $F_2$ is to use, in the region of moderate $x$, the
parameterization for the ratio $R\approx F_L/(F_2-F_L)$ based on the
dedicated SLAC measurements~\cite{ref:slac_r90} and, at small $x$, the
QCD expectations based on existing parton distributions.

The experimental determination of $F_L$ requires the measurement of
the differential cross section at different center-of-mass energies.
This allows a different value of $y$ for the same $(x,Q^2)$, and
therefore a separation of $F_2$ and $F_L$.  This is unfortunately a
very difficult measurement, as it requires the difference of cross
sections.  The measurement at HERA would require the reduction of the
proton and/or electron beam energy, and will likely be performed
before HERA running is completed.

\epsfigure[width=0.8\hsize]{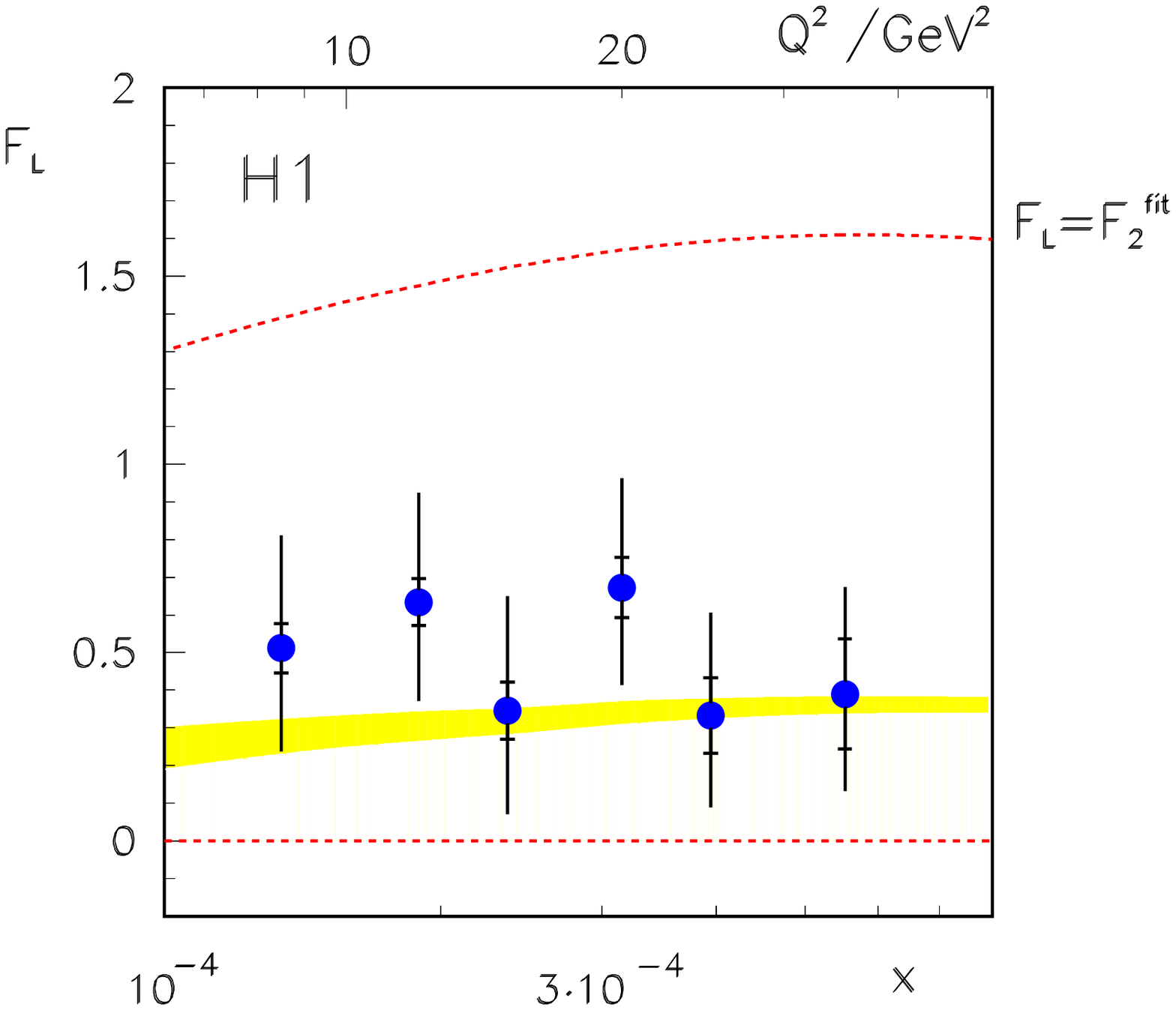} {$F_L$ as a function
  of $x$ (and $Q^2$) as determined by the
  \protect\citeasnoun{ref:H1_FL}.  The shaded band corresponds to NLO
  QCD expectations using the parton densities extracted by H1.  }
{H1_Fl}

The H1 Collaboration~\cite{ref:H1_FL} has estimated $F_L$ by assuming
that the DGLAP evolution equations are capable of describing the
evolution of $F_2$.  In this method, the cross section, and $F_2$, are
initially extracted at small $y$ where the contribution from $F_L$ is
small.  The NLO DGLAP equations are then used to extrapolate $F_2$ to
large $y$, where the cross section contribution from $F_2$ is compared
to the measured cross section.  The difference in cross sections is
attributed to $F_L$.  The results of this procedure are shown in
Fig.~\ref{fig:H1_Fl}.  Within errors, there is good agreement between
the extracted value of $F_L$ and the expected value from pQCD.  The
expectations for $F_L$ obtained from a QCD fit and assuming that
$F_L=F_2$ is also shown.  This value of $F_L$ is clearly ruled out.

\subsubsection{$F_{2}^{c\bar{c}}$}
\label{sec:F2_charm}

\epsfigure[width=0.5\hsize]{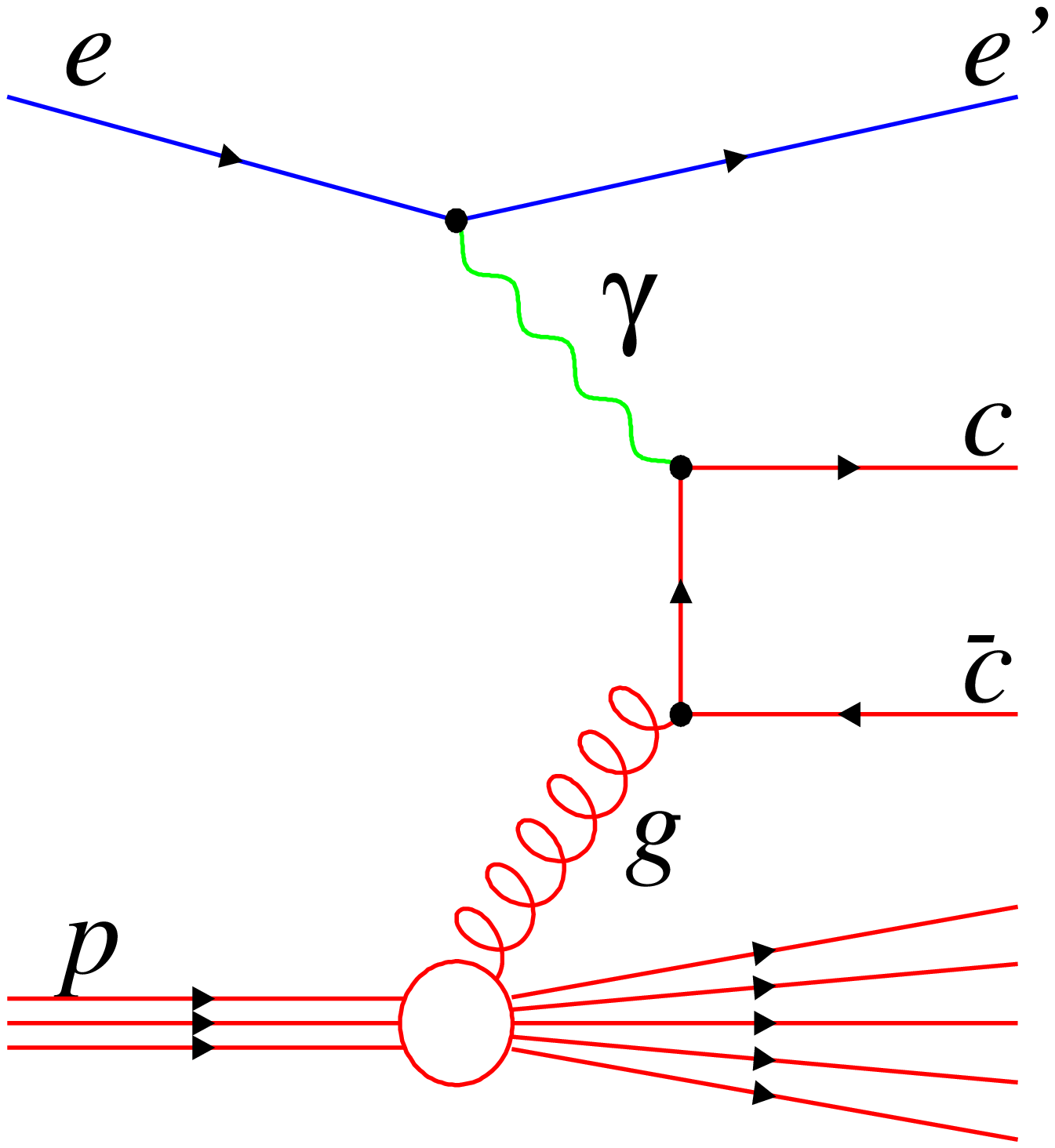} {The leading order
  diagram for charm production via photon-gluon fusion.}  {BGF_charm}

The measurement of $F_2$ is based on the inclusive cross section for
$ep$ scattering.  It is also possible to define the contributions to
$F_2$ which result in particular final states.  This is useful in
cases where the relevant cross section can be calculated
theoretically.  One such example is for charm production in DIS.  The
reaction $ep \rightarrow ecX$, where $c$ represents a charm quark or
anti-quark, is expected theoretically to proceed primarily via
photon-gluon fusion, $\gamma g \rightarrow c\bar{c}$.  This process is
shown in Fig.~\ref{fig:BGF_charm} for the leading order diagram.
Calculations for this process exist theoretically to
NLO~\cite{ref:NLOcharm1,ref:NLOcharm2,ref:NLOcharm3}.  The cross
section depends directly on the gluon density in the proton.  The
gluon density from the inclusive analysis can be used in the
calculation, and the results compared to the measured charm cross
sections.  This gives a powerful cross check on the pQCD calculations.

There are other processes beside photon-gluon fusion which can produce
charm in DIS: diffractive heavy flavor production~\cite{ref:BIS},
scattering off charmed sea quarks~\cite{ref:IJN}, charmed hadron production
from bottom quarks~\cite{ref:Ali}, charm production in fragmentation%
~\cite{ref:MNS1,ref:MNS2,ref:MNS3}, and intrinsic charm in the
proton~\cite{ref:Brodsky}. These processes have either been measured
to have small cross sections relative to the photon-gluon fusion
mechanism, or are not expected to contribute in the relevant kinematic
regions.

\epsfigure[width=0.8\hsize]{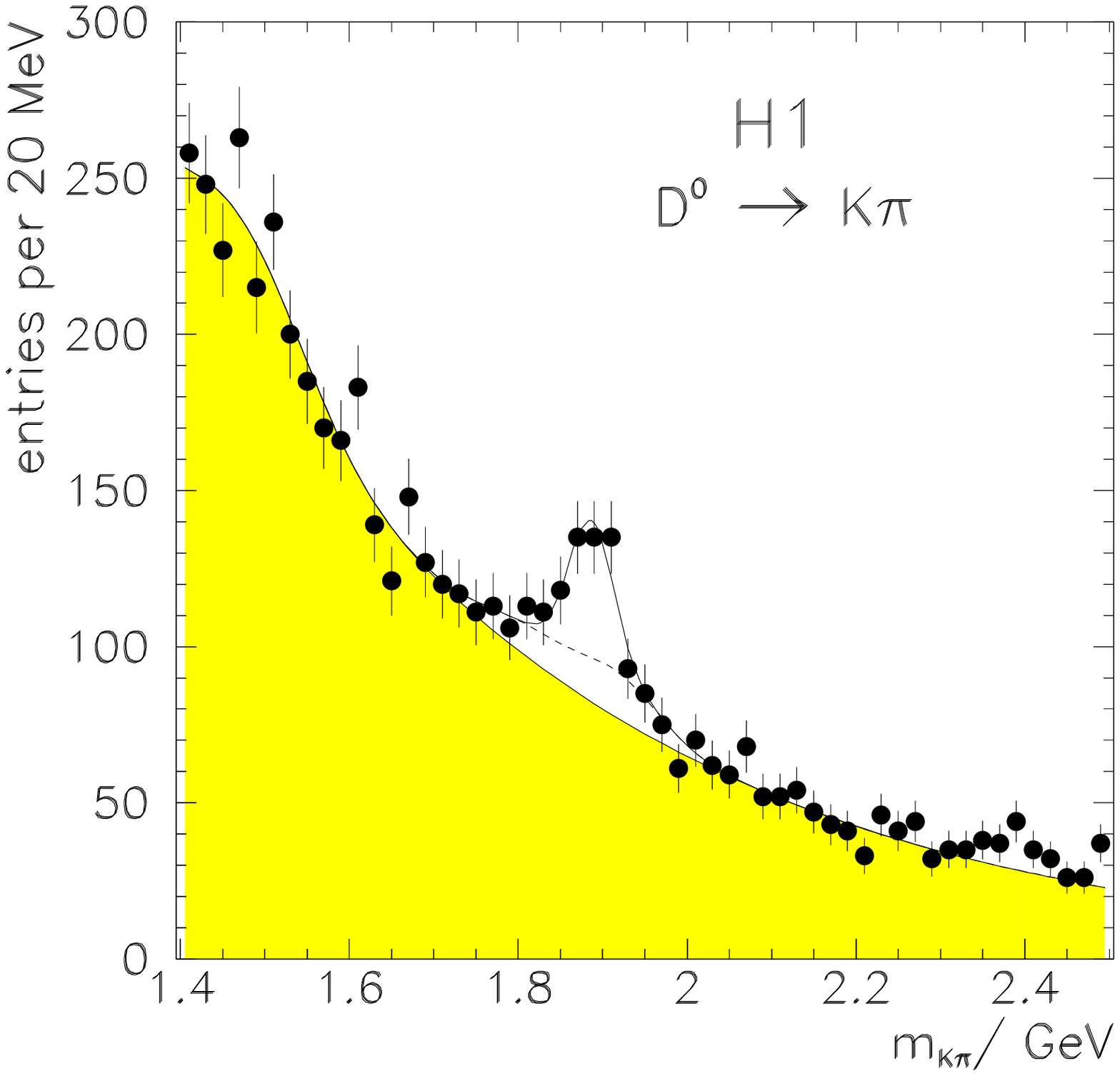} {The two particle
  invariant mass spectrum obtained when assigning the $K^{\pm}$ mass
  to one of the particles and the $\pi^{\pm}$ mass to the other, as
  measured by the \protect\citeasnoun{ref:H1_f2charm}.}  {H1_d0}

\epsfigure[width=0.8\hsize]{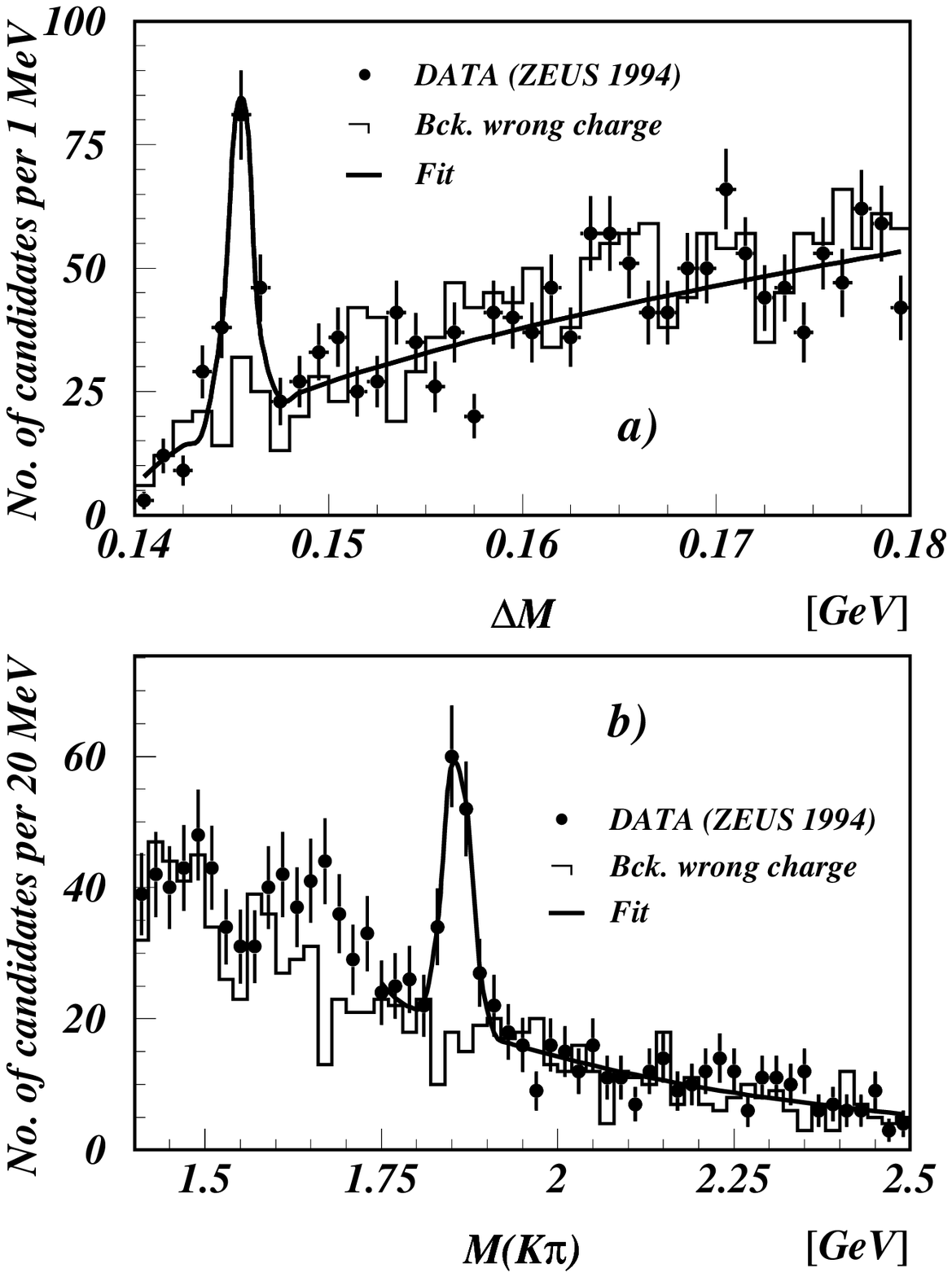} {a) The $\Delta M =M
  (K\pi\pi_s)-M(K\pi)$ distribution for $M(K\pi)$ in the $D^0$ signal
  region, where $\pi_s$ denotes the softest pion. b) The $M(K\pi)$
  distribution for events in the $\Delta M$ signal region.  The
  results are from the \protect\citeasnoun{ref:ZEUS_charm2}.  }
{ZEUS_dstar}

Results on charm production in DIS have been presented by
H1~\cite{ref:H1_f2charm} and
ZEUS~\cite{ref:ZEUS_charm1,ref:ZEUS_charm2}.  Charm production is
tagged by reconstructing the $D^0$ (H1) or the $D^*$ (H1 and ZEUS)
mesons and their charge conjugates.  The signal for $D^0 \rightarrow
K\pi$ as seen in the H1 detector is presented in Fig.~\ref{fig:H1_d0}.
The $D^*$ decays are identified by the decay chain $D^* \rightarrow
D^0\pi \rightarrow K\pi\pi$, taking advantage of the tight kinematic
constraint imposed by the small mass difference $\Delta M =
m_{D^*}-m_{D^0}=145.5\pm0.15$~MeV.  As an example the signal observed
in the $\Delta M$ distribution by ZEUS is shown in
Fig.~\ref{fig:ZEUS_dstar}.

As described above, the dominant mechanism for charm production in DIS
at HERA energies is expected to be the photon-gluon fusion process.
Indeed, the AROMA~\cite{ref:aroma} Monte Carlo simulation program
based on this process is found to reproduce properly the shapes of the
transverse momentum and pseudorapidity spectra for the $D$ mesons, as
well as the overall $W$ and $Q^2$ dependences.  An upper limit of
$5$~\% on a possible contribution of the charm sea has been estimated
by H1.

The charm production cross section has been measured by the H1
collaboration in the kinematic range $10 \leq Q^2 \leq 100$~GeV$^2$
and $0.01<y<0.7$.  The measurements are performed within
experimentally accessible regions (various cuts are applied on the
tracks from the charmed hadrons, leading to $p_T$ and $\eta$
restrictions) and are then extrapolated to the full phase space using
the AROMA Monte Carlo simulation.  The result is found to be somewhat
larger than predicted.  This is true for all gluon density
parameterizations used, including the gluon density measured by H1
from the scaling violations of $F_2$.  The latter gluon density gives
the best agreement with the measurement.  For $m_c=1.5$~GeV, the
prediction using the H1 gluon density is $13.6$~nb, while the measured
cross section is $17.1\pm2.3$~nb.

The ZEUS experiment finds, for the same kinematic region, a measured
cross section of $12.5\pm3.9$~nb, in reasonable agreement with the H1
measurement.  They find for the NLO prediction, with $m_c=1.5$~GeV and
the GRV NLO gluon density~\cite{ref:GRV94}, a value of $11.1$~nb, and
conclude that this is in good agreement with the data.  The NLO
predictions are also able to reproduce the shapes of the $p_T,\; W$
and $Q^2$ distributions, and is consistent with the $\eta$
distribution measured in the data.  ZEUS then extrapolates into the
unmeasured region using NLO
calculations~\cite{ref:HarrisSmith1,ref:HarrisSmith2}.

$F_2^{c\bar{c}}$ is calculated from the measured charm cross section as
follows:
\begin{enumerate}
\item The cross section for $c\bar{c}$ production is calculated from
  the $D^*$ cross section (extrapolated to the full phase space),
  using
\begin{equation}
\sigma(ep \rightarrow ec\bar{c}X) = \frac{1}{2}
\frac{\sigma(ep \rightarrow eD^*X)}{P(c\rightarrow D^*)} \; ,
\end{equation}
where $P(c\rightarrow D^*)$ is the probability that a charm quark will
produce a $D^*$ meson (about 25~\%).
\item
The measurements are made at small $Q^2$, such that $F_3$ can be neglected.
Also, the possible contribution from $F_L$ has been estimated to be small 
and is ignored. $F_2^{c\bar{c}}$ is then extracted using
\begin{equation}
\frac{d^2\sigma(ep \rightarrow ec\bar{c}X)}{dxdQ^2} =
\frac{2\pi\alpha^2}{xQ^4}\left[Y_+F_2^{c\bar{c}}(x,Q^2)\right] \; .
\end{equation}

\end{enumerate}

\epsfigure[width=0.8\hsize]{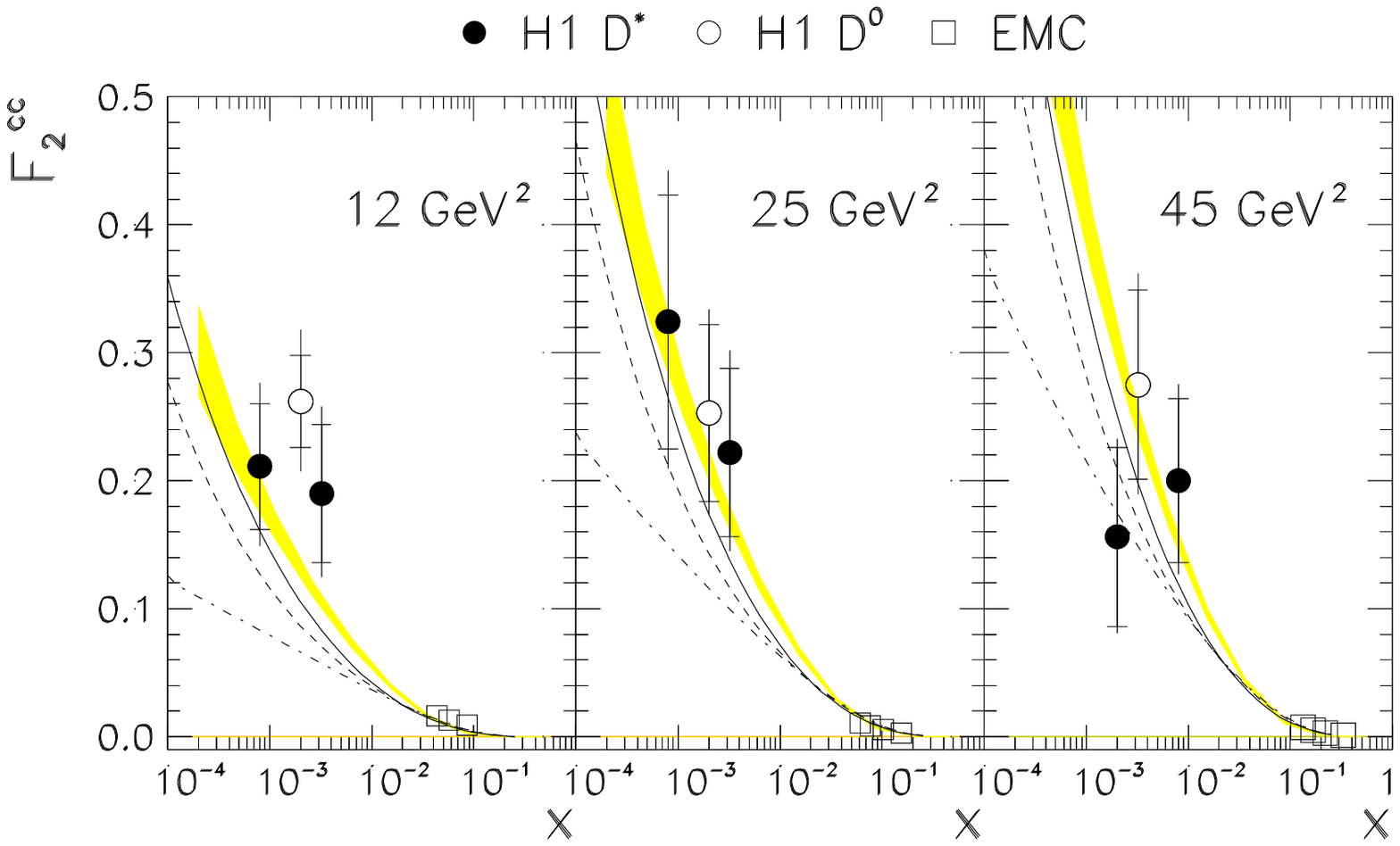} {$F_2^{c\bar{c}}$ as
  measured by the H1 experiment~\protect\cite{ref:H1_f2charm}.  The
  curves correspond to different parton parameterizations: GRV-H0
  (full line), MRSH (dashed line), and MRSD0' (dashed-dotted line).
  The shaded area is the expectation based on the H1 NLO QCD fit. The
  EMC data are also shown as open boxes~\protect\cite{ref:EMC_charm}.}
{h1_f2charm}

\epsfigure[width=0.95\hsize]{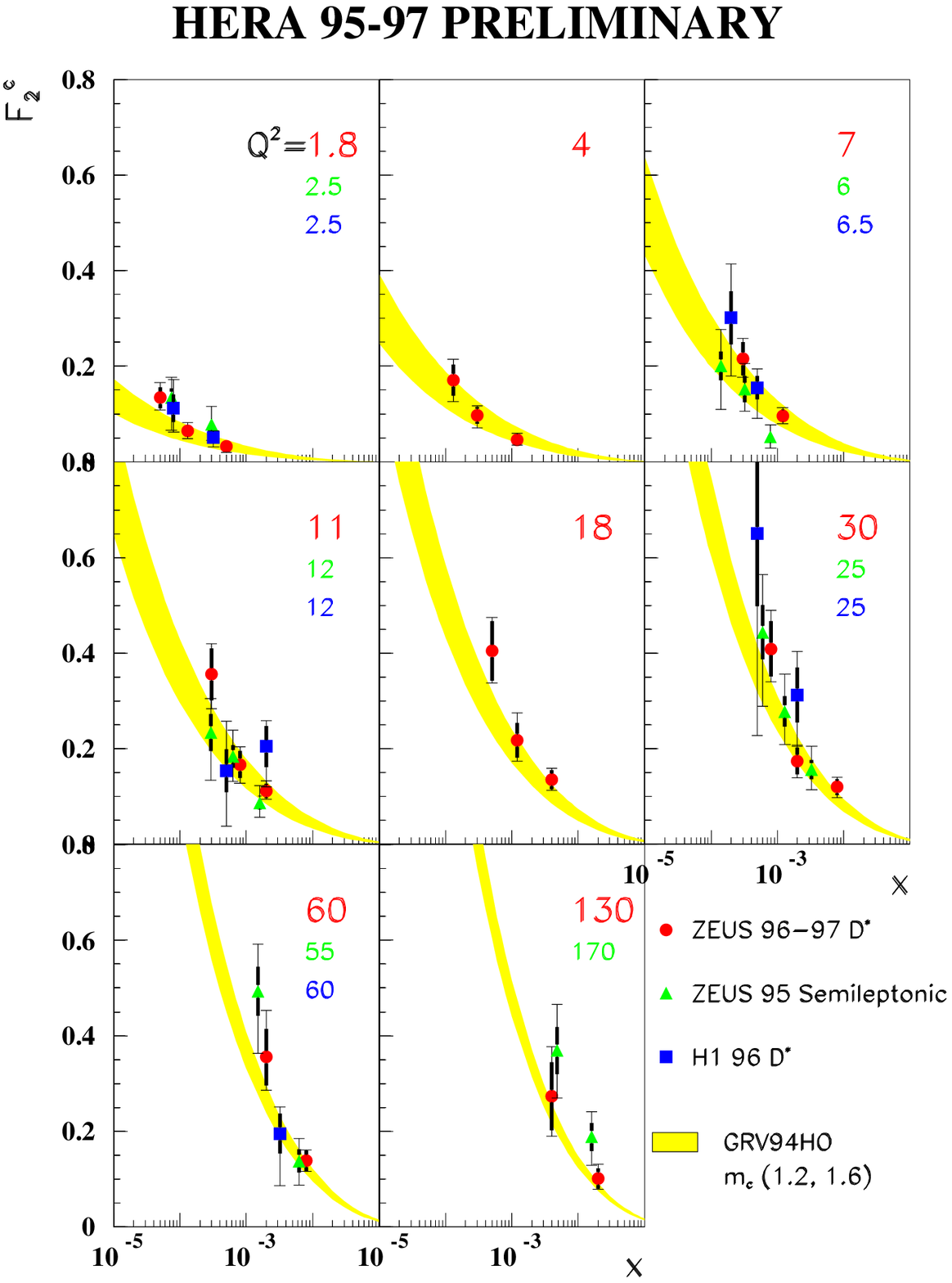} {$F_2^{c\bar{c}}$ as
  measured by the ZEUS experiment from the 1995-97 data
  sets~\protect\cite{ref:ZEUS_charm2}. The $Q^2$ values in the figure
  are in GeV$^2$. The shaded area is the expectation using the
  GRV94(HO) parton density set, allowing for a range of charm quark
  masses between $1.2<m_c<1.6$~GeV.}  {ZEUS_f2charm}

The first results on $F_2^{c\bar{c}}$ as measured by the H1
collaboration are shown in Fig.~\ref{fig:h1_f2charm}, and compared
with the EMC~\cite{ref:EMC_charm} measurements.  The data reveal a
steep rise of $F_2^{c\bar{c}}$ as $x$ decreases from $x= 0.1$ to
$x=10^{-3}$.  Averaged over the kinematical range of H1, a ratio
$<F_2^{c\bar{c}}/F_2>=0.237\pm0.021^{+0.043}_{-0.039}$ is obtained,
one order of magnitude larger than at large $x$.

The latest preliminary results from
HERA~\cite{ref:ZEUS_charm97,ref:H1_charm97} on $F_2^{c\bar{c}}$, based
on data up to 1997, is shown in Fig.~\ref{fig:ZEUS_f2charm}.  The
measurements are seen to be in good agreement. They are compared to
the predictions from NLO
calculations~\cite{ref:HarrisSmith1,ref:HarrisSmith2} using as input
the gluon density parameterization from GRV.  The agreement is good,
indicating that the pQCD description of the data is self-consistent.

The data available so far are in good agreement with expectations from
NLO pQCD calculations. However, the statistical errors are still quite
large, and the data span a rather limited kinematic range.  A large
increase in statistics is forthcoming, and it is expected that the
$F_2^{c\bar{c}}$ measurements will eventually reach a precision level
similar to that for the inclusive $F_2$ measurements presented in this
review.  The evolution of $F_2^{c\bar{c}}(x,Q^2)$ is a matter of some
theoretical debate. Some authors~\cite{ref:MRSP,ref:CTEQ_charm} start
with zero charm content in the proton below some scale $Q^2_0\sim
m_c^2$, and then evolve charm as a massless quark above this
threshold.  A second approach is to generate charm exclusively via
photon-gluon fusion taking into account the charm quark
mass~\cite{ref:GRV_charm}.  A third approach consists of having a
variable flavor number scheme~\cite{ref:ACOT}.  More precise data will
allow much more stringent tests of the pQCD calculations, and should
allow to discriminate between these, and possibly other, approaches to
calculating the charm content of the proton.

\subsection{The gluon density in the proton} 
\label{sec:partons} 
The behavior of many of the cross sections measurable at HERA is
driven by the gluon density in the proton.  This results from the fact
that, due to experimental cuts, most cross sections are measured in
the large-$W$, or small-$x$, domain.  Small-$x$ quarks are therefore
involved in the reaction.  The distribution of the small-$x$ quarks is
determined principally by the gluon density in the proton via $g
\rightarrow q\bar{q}$ splitting.  The cross sections for diverse
reactions such as open charm production, jet production, vector meson
production, as well as inclusive cross sections, can be used to
measure the gluon density in the proton.  The consistency of the gluon
densities extracted via these different methods provides a measure of
how well the pQCD expansions can currently be used for calculational
purposes.  The measurement of the gluon density therefore serves a
two-fold purpose: it is necessary to know the gluon density to
calculate cross sections for current as well as future experiments;
and, assuming the validity of QCD, the consistency of different
methods of measuring the gluon density is a measure of calculational
ability.  Discrepancies in the gluon densities beyond theoretical and
experimental uncertainties would be a sign that QCD does not provide a
complete description of the strong interactions.

There are very different levels of theoretical sophistication for the
various methods proposed for the extraction of the gluon density.  The
extraction based on the total DIS cross section (fits to $F_2$) has a
long experimental as well as theoretical history, and is the most
advanced.  Many NLO evolution programs exist which give consistent
results at the fraction of a percent level~\cite{ref:QCD_programs}.
Most other methods are based on LO QCD calculations, and therefore
suffer from large normalization uncertainties (factors of two are
possible).  Generally, the $x$- or $W$-dependence of the cross
sections are more firmly predicted, such that the shape of the gluon
density is then constrained.  In the following sections, we review in
some detail the extraction of the gluon density from the inclusive
cross section measurement, and then briefly describe other proposed
methods.

\subsubsection{Gluon density extraction from $F_2$}
\label{sec:gluonfits}

The DGLAP evolution equations relate the change with $Q^2$ of the quark 
densities in the proton, $q(x,Q^2)$,  or equivalently the structure function 
$F_2(x,Q^2)$, to the density of gluons, $g(x,Q^2)$. For 
example~\cite{ref:Buras}, 

\begin{eqnarray}
 \label{eq:df2dlnq2} 
\frac{dF_2(x,Q^2)}{d\ln Q^2} &=& \frac{\as (Q^2)}{2\pi}            
\left[\int_x^1\frac{dz}{z}\frac{x}{z}P_{qq}\left(\frac{x}{z}\right)
F_2(z,Q^2) \right. \nonumber \\         
& & \mbox{ }\left. +2\sum_q e_q^2\int_x^1\frac{dz}{z}\frac{x}{z}P_{qg}
\left(\frac{x}{z}\right)zg(z,Q^2)\right] \; ,
\end{eqnarray}
 where the sum runs over quark flavors, and $e_q$ is the quark charge. 
$P_{qq}$ and $P_{qg}$ are the splitting functions to quarks of the quark and 
gluon, respectively.  To leading order they are given 
by (see, e.g., \cite{ref:HM,ref:Roberts,ref:Renton})
\begin{eqnarray}
 P_{qq}(z)    &=& C_F\left[\frac{1+z^2}{1-z}\right] \; ,\\ 
P_{qg}(z)    &=& D_F[z^2+(1-z)^2] \; , 
\end{eqnarray} 
where the multiplicative constants are color factors. As a result, the
$Q^2$ dependence of the parton densities may be calculated provided
their $x$ dependence is known from data at some lower $Q^2$. A common
approach is to parameterize at fixed $Q^2=Q^2_0$ the $x$ dependence of
the parton densities by
\begin{equation}
  x\cdot{\rm density}(x,Q^2_0) = Ax^{\delta}(1-x)^{\eta}\cdot{\rm 
polynomial}(\sqrt x) \; , 
\end{equation}
with a distinct set of parameters for each type of parton.  The parton
densities are then determined by applying the evolution equations to
perform a global fit of all the parameters to deep inelastic
scattering data.  This approach has been used by the
ZEUS~\cite{ref:ZEUS_glue,ref:ZEUSF2_97} and
H1~\cite{ref:H1glue,ref:H1F2_94} collaborations using their 1993, 1994
and 1995 $F_2$ measurements.

\paragraph{H1 DGLAP fits}
The H1 collaboration~\cite{ref:H1F2_94} performed NLO DGLAP fits to
$F_2$ data by defining parton densities in the $\overline{MS}$
renormalization scheme.  Three light flavors of quarks were taken into
account.  The charm quark contribution was generated dynamically using
the photon-gluon fusion prescription given in~\cite{ref:GHR,ref:GRS},
with scale $\sqrt{Q^2+4m_c^2}$.  The charm quark mass was set to
$m_c=1.5$~GeV.  Beauty quark contributions are expected to be small
and were neglected.

The specific functional forms used by H1 were
\begin{equation} 
\label{eq:H1par}
\begin{array}{rcl}
 xg(x,Q^2_0) & = & A_gx^{\delta_g}(1-x)^{\eta_g} \; , \\
 xu_v(x,Q^2_0) & = & A_ux^{\delta_u}(1-x)^{\eta_u}(1+B_ux+C_u\sqrt{x})\; ,  \\
 xd_v(x,Q^2_0) & = & A_dx^{\delta_d}(1-x)^{\eta_d}(1+B_vx+C_d\sqrt{x})\; ,  \\
 xS(x,Q^2_0) & = & A_Sx^{\delta_S}(1-x)^{\eta_S}(1+B_Sx+C_S\sqrt{x}) \; , 
\end{array}
\end{equation} 
where the glue, valence $u$ and $d$ and sea, $S\equiv\bar{u}+\bar{d}$, 
quarks have distinct parameters.
The quark and antiquark components of the sea are assumed equal, and
$\bar{u}$ is set equal to $\bar{d}$.  The strange quark density is set
to be $S/4$~\cite{ref:Bazarko}.  The normalizations of the valence quark
densities were fixed by the counting rules
\begin{equation}
\begin{array}{rcl}
\int_{0}^{1}u_v(x)dx & = & 2 \; , \\
\int_{0}^{1}d_v(x)dx & = & 1 \; .
\end{array}
\end{equation}
The normalization of the gluon density, $A_g$, was obtained via the
momentum sum rule.  The further constraint $\delta_u = \delta_d$ was 
imposed.

The fits were performed with $Q_0^2=5$~GeV$^2$ and
$\Lambda_4^{\overline{MS}}= 263$~MeV.  In addition to the H1 data, the
proton and deuteron results from BCDMS \cite{ref:BCDMS} and
NMC~\cite{ref:NMC_F2} were used to constrain the parton densities at
large $x$.  Data in the range $Q^2>5$~GeV$^2$ were fit, except for
data with $x>0.5$ and $Q^2<15$~GeV$^2$.

\epsfigure[width=0.95\hsize]{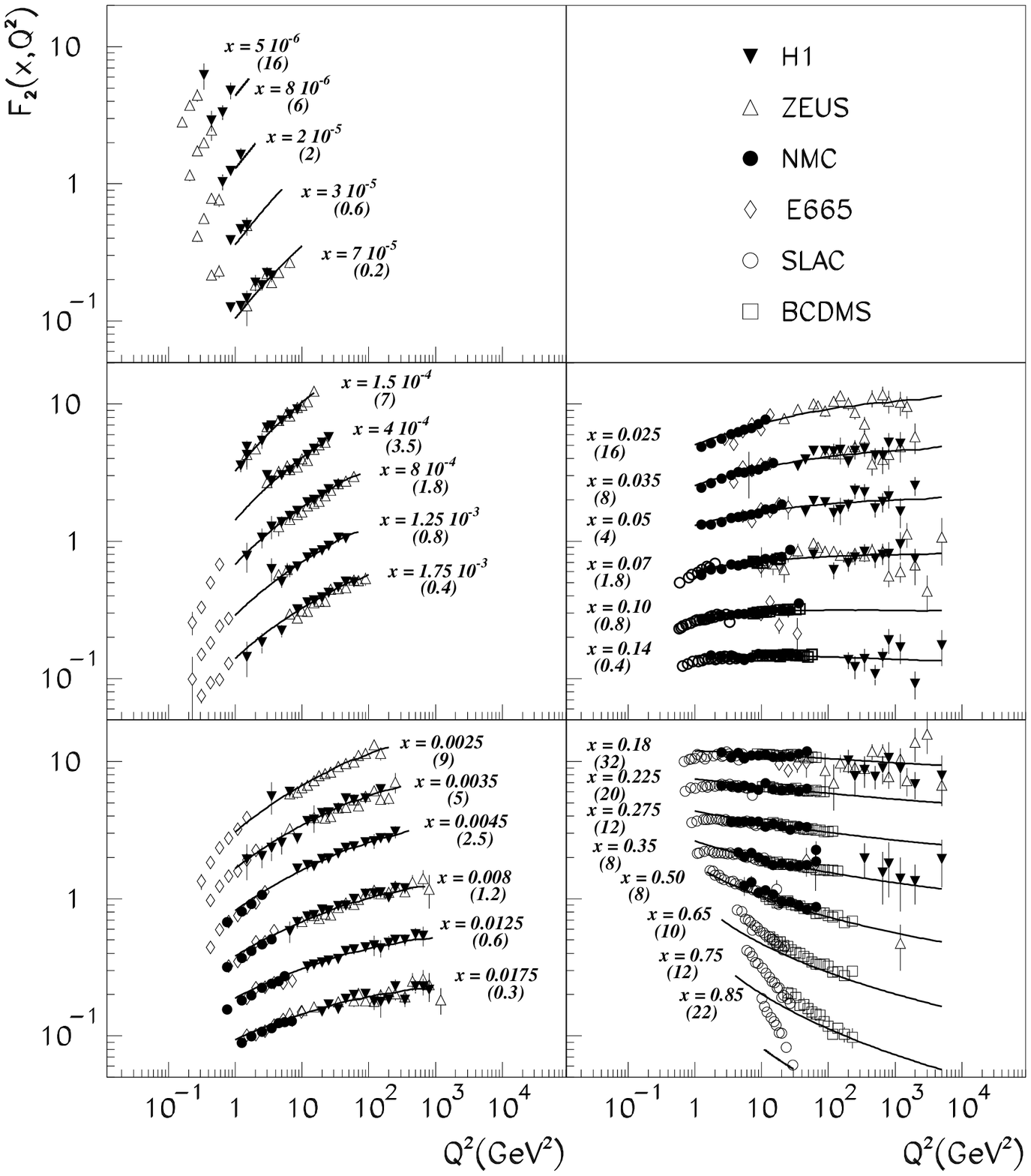} {Compilation of
  measurements of $F_2$ as a function of $Q^2$ for selected values of
  $x$ as denoted on the figure. The numbers in parentheses are the
  scaling factors by which the value of $F_2$ has been multiplied in
  the plot. The overlayed curves are the result of an NLO QCD fit
  performed by the H1 experiment to H1, NMC and BCDMS data.}
{F2_fits}

The fits to the $F_2$ data are shown in Fig.~\ref{fig:F2_fits}
compared to HERA and fixed target data.  Data not used in the fitting
procedure are also shown.  As can be seen in this figure, the fits
reproduce the data quite well.  The values of the parameters extracted
in the fits are given in Table~\ref{tab:F2_glue}.  The value of
$\delta$ is negative for both the gluon and sea quark density,
indicating that these densities rise with decreasing $x$.

\begin{table} 
\tablecaption{ The values of the parameters for the parton density 
parameterizations
as extracted by H1~\cite{ref:H1F2_94} and ZEUS~\cite{ref:ZEUSF2_97} from
NLO DGLAP fits to structure function data.  The parameters are given in the
H1 case for $Q_0^2=5$~GeV$^2$, and in the ZEUS case for $Q_0^2=7$~GeV$^2$. 
The form of the parameterizations are given in the text.}
\label{tab:F2_glue}
\renewcommand\arraystretch{1.3}
\begin{center}
\rotatebox{-90}{%
\begin{tabular}{c|ccccc|ccccc}
 & \multicolumn{5}{c|}{H1 fit result} & \multicolumn{5}{c}{ZEUS fit result} \\
\hline
      &  $A$   & $\delta$ & $\eta$ & $B$ & $C$ &  $A$   & $\delta$ 
& $\eta$ & $B$ & $C$ \\ \hline
$xg$ &  $2.24$  & $-0.20$ & $8.52$ &    &     &  $1.77$ & $-0.225$ 
& $9.07$ & $3.00$ & \\
$xu_v$ & $2.84$ & $0.55$ & $4.19$ & $4.42$ & $-1.40$ & & & & & \\
$xd_v$ & $1.05$ & $0.55$ & $6.44$ & $-1.16$ & $3.87$ & & & & & \\
$x(\bar{u}+\bar{d})$ & $0.27$ & $-0.19$ & $1.66$ & $0.16$ & $-1.00$ 
& & & & & \\
$x(u+\bar{u}-d-\bar{d})$ & & & & & & $6.07$ & $1.27$ & $3.68$ & &  \\
$2x(\bar{u}+\bar{d}+\bar{s})$ & & & & & & $0.52$ & $-0.24$ & $8.60$ 
& $3.27$ & $0.29$ 
\end{tabular}
}
\end{center}
\end{table}

\paragraph{ZEUS DGLAP fits}

The most recent results from the ZEUS collaboration are presented
in~\cite{ref:ZEUSF2_97}.  The fits were performed on the ZEUS 1994 F2
data as well as the 1995 shifted vertex data presented in
~\cite{ref:ZEUSF2_97}. NMC~\cite{ref:NMC_97} and
BCDMS \cite{ref:BCDMS,ref:BCDMS1} results are used to constrain the
large-$x$ parton densities.  ZEUS defined the parton densities at an
input scale $Q^2_0=7$~GeV$^2$, and fit data down to $Q^2=1$~GeV$^2$.
The gluon, sea quark $S = 2(\bar{u}+\bar{d}+\bar{s})$, and difference
of up and down quarks in the proton $\Delta$ were parameterized as
\begin{equation} 
\label{eq:ZEUSpar}
\begin{array}{rcl}
 xg(x,Q^2_0) & = & A_gx^{\delta_g}(1-x)^{\eta_g}(1+B_gx) \; , \\
 xS(x,Q^2_0) & = & A_{S}x^{\delta_{S}}(1-x)^{\eta_{S}}
(1+B_{S}x+C_{S}\sqrt{x})  \; ,\\
 x\Delta_{ud}(x,Q^2_0) & = & A_{\Delta}x^{\delta_{\Delta}}(1-x)^{\eta_{\Delta}}
\; .
\end{array}
\end{equation} 

The input valence distribution $xu_v=x(u-\bar{u})$ and
$xd_v=x(d-\bar{d})$ were taken from the parton distribution set
MRS(R2)~\cite{ref:MRSR1}.  The strange quark density was assumed to be
$20$~\% of the sea at $Q^2=1$~GeV$^2$.  The gluon density
normalization was fixed using the momentum sum rule.  The input value
of the strong coupling constant was set to $\as(M_Z^2)=0.118$.

The parton densities are defined in the $\overline{MS}$ scheme, and
were evolved using the NLO DGLAP equation with three light flavors.
The charm contribution was calculated in NLO using the calculations of
Riemersma et al.~\cite{ref:NLOcharm3} with the charm quark mass set to
$m_c=1.5$~GeV. The values of the fit parameters are given in
Table~\ref{tab:F2_glue}.

\epsfigure[width=0.95\hsize]{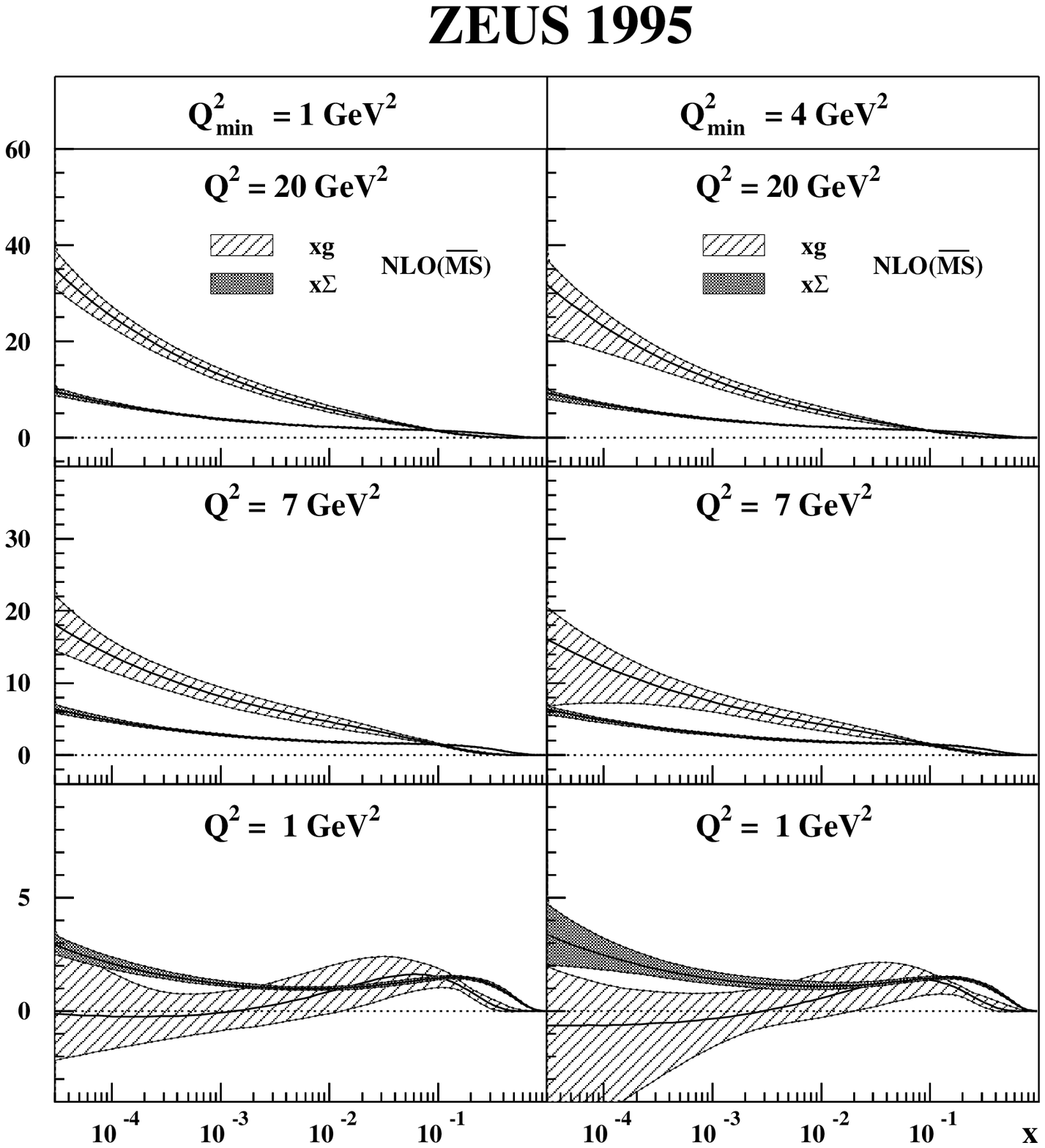} {The quark singlet momentum
  distribution, $x\Sigma$ (shaded), and the gluon distribution, $xg$
  (hatched), as functions of $x$ from the ZEUS NLO QCD fits. The three
  left-hand plots show the results including $F_2$ data with
  $Q^2>1$~GeV$^2$; the three right-hand plots show the corresponding
  results for data with $Q^2>4$~GeV$^2$.}  {97gluons_zeus}

As mentioned previously, the ZEUS DGLAP fit is able to reproduce the
turnover in the scaling violations plot.  However, this is only
possible with a rather non-intuitive result for the parton densities.
The gluon density is seen to dominate at small-$x$ for moderate $Q^2$,
but decreases sharply as $Q^2$ approaches $1$~GeV$^2$, at which point
it is smaller than the sea quark density, and essentially
valence-like.  The gluon density found by ZEUS is compared to the sum
of the quark distributions, $x\Sigma$, in
Fig.~\ref{fig:97gluons_zeus}.  It is at this point not clear whether
the parton densities at these small values of $Q^2$ are meaningful
(i.e., could be used to predict a cross section for a different
process), or result from non-perturbative contributions to the cross
section.  It is not possible to determine from the existing data where
the DGLAP fit is no longer valid.  The flexibility in the parton
density parameterizations is such that the fits find solutions.  Other
criteria must be found to determine whether the results are sensible.

\paragraph{Discussion}

\epsfigure[width=0.8\hsize]{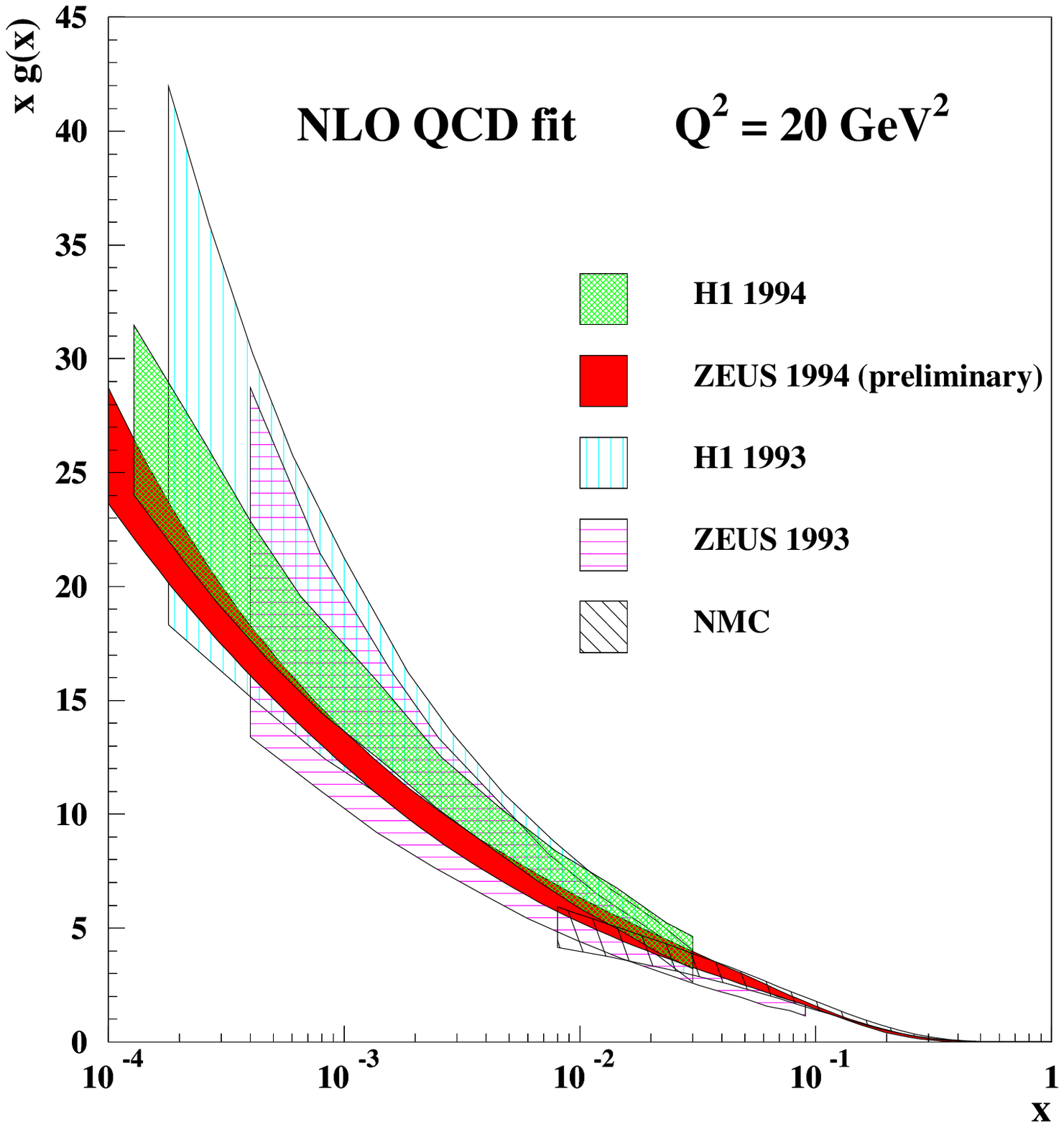} {The gluon
  distribution in the proton as determined from NLO QCD fits to the
  $F_2$ data.  The results from the 1993 data are compared to those
  from the 1994 data.}  {F2gluons}

The results for the extracted gluon density for ZEUS and H1 are
compared to those from NMC in Fig.~\ref{fig:F2gluons}.  The NLO gluon
density is now known at the $10$~\% level from these fits.  This is
currently the most precise determination of the gluon density in the
proton.

There are several points to be made regarding this method  of
extracting  the
gluon density:
\begin{itemize} 
\item[-] The DGLAP fits are performed on the extracted $F_2$, which
  were determined from the measured differential cross section.  In
  the process of extracting $F_2$, the contribution from $F_L$ is
  calculated assuming some gluon density and subtracted.  Different
  parameterizations for the gluon density are used to test the
  sensitivity of $F_2$ to this choice.  The resulting change in the
  extracted gluon density was found to be small.
  
\item[-] The HERA data alone are not sufficient to constrain the gluon
  density. The normalization, $A_g$, and large-$x$ behavior, $\eta_g$,
  are primarily determined by the fixed target data.  Relative
  normalization differences in different experimental data sets and
  correlated errors must be handled carefully.
 
\item[-] The extracted gluon density depends on $\as$, which also
evolves with $Q^2$.  The value of $\as$ is taken from
other experiments.

\item[-] The validity of the DGLAP evolution equation across a
  transition from a flat parton density (at some small value of $Q^2$)
  to a steep parton density has been called into
  question~\cite{ref:EKL}.  This could limit the range in $Q^2$ over
  which the fits can be performed to the $Q^2$ region where the parton
  densities have been shown to be steep.

\end{itemize}  

Despite these caveats, it should be stressed that the gluon density is
now known rather precisely at small-$x$.  Given the factorization
theorem described in section~\ref{sec:QCD}, this allows precise
predictions for cross sections of many processes at HERA, and for
other experiments utilizing protons.

\subsubsection{Other gluon density extraction methods}

\paragraph{Heavy quark production}

The production of heavy quarks, both in photoproduction and DIS, is
sensitive to the gluon density inside the proton.  At HERA, this
principally means charm quark production.  We can distinguish between
open charm production, where hadrons with net charm quantum numbers
are produced, and the production of $c\overline{c}$ bound states
($J/\psi$ and excited states.) The latter are described in
section~\ref{sec:VM}, while open charm production in DIS has been
described above in section~\ref{sec:F2_charm}. It is believed that
photoproduction reactions can also be used for this purpose.  In this
case, the sizeable charm mass as well as the transverse energy of the
charm particles set the scale at which the gluon density is probed.

Full NLO calculations for charm photoproduction cross sections are
available.  They are distinguished in the method in which the charm
mass is handled.  In the so-called `massive-charm' scheme, the mass of
the charm quark acts as a cutoff in the perturbative
calculation~\cite{ref:Frixione1,ref:Frixione2}.  In the
`massless-charm' scheme, the charm is treated as one of the active
flavors in the proton and
photon~\cite{ref:Kniehl,ref:Binnewies,ref:Cacciari1,ref:Cacciari2}.  A
significant difference between the two schemes is that in the massive
scheme the cross section is dominated by photon-gluon fusion, while in
the massless scheme the contribution from resolved photons is of
comparable size.

\epsfigure[width=0.8\hsize]{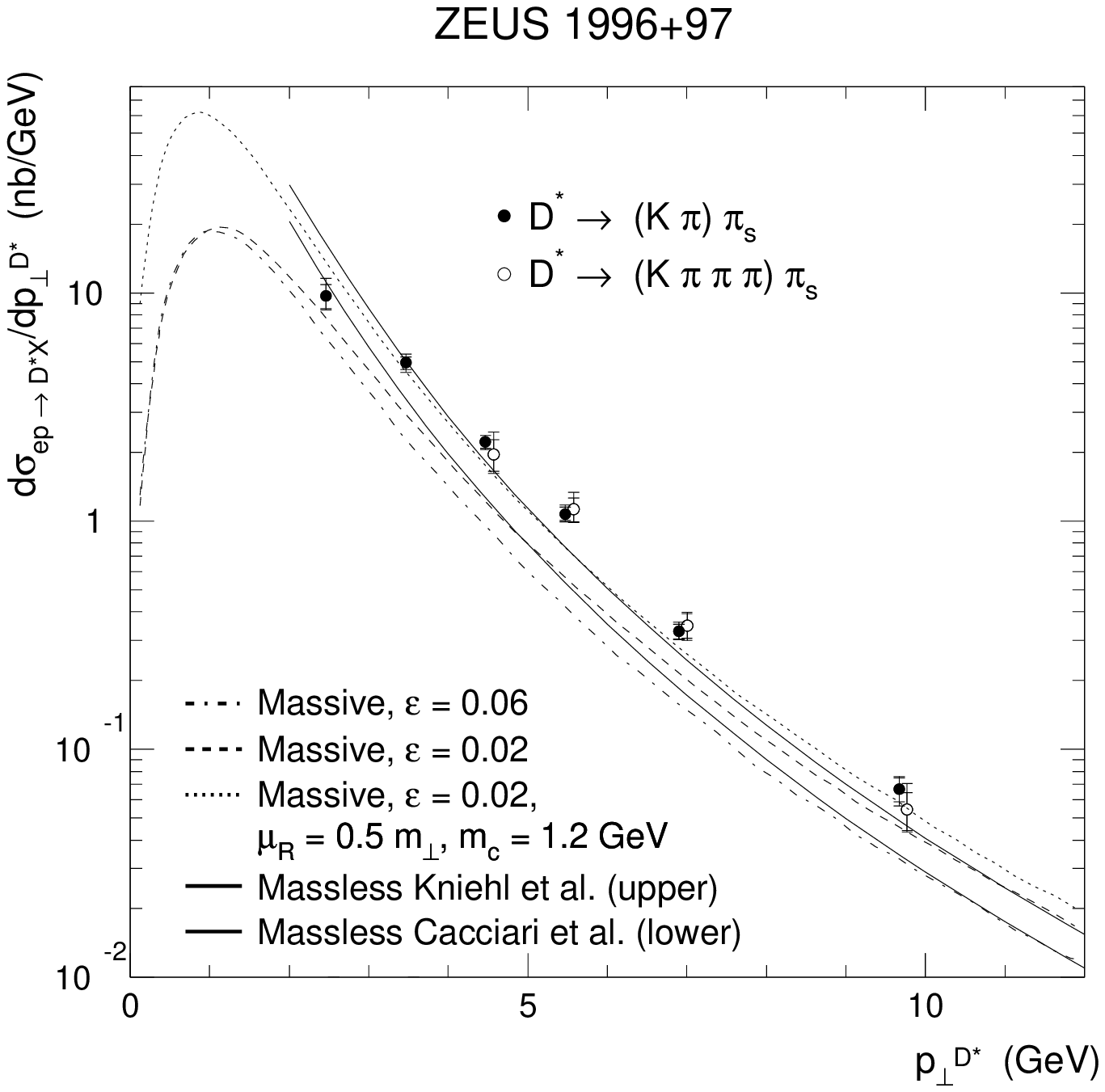} { The differential
  cross section $d\sigma/dp_{\perp}^{D^*}$ for $D^*$ photoproduction
  in the kinematic range $Q^2<1$~GeV$^2$, $130<W<280$~GeV and
  $-1.5<\eta^{D^*}<1.0$ from the
  \protect\citeasnoun{ref:ZEUS_charm97}.  The data are compared to
  various calculations as described in the text.}  {PT_photo_charm}

\epsfigure[width=0.95\hsize]{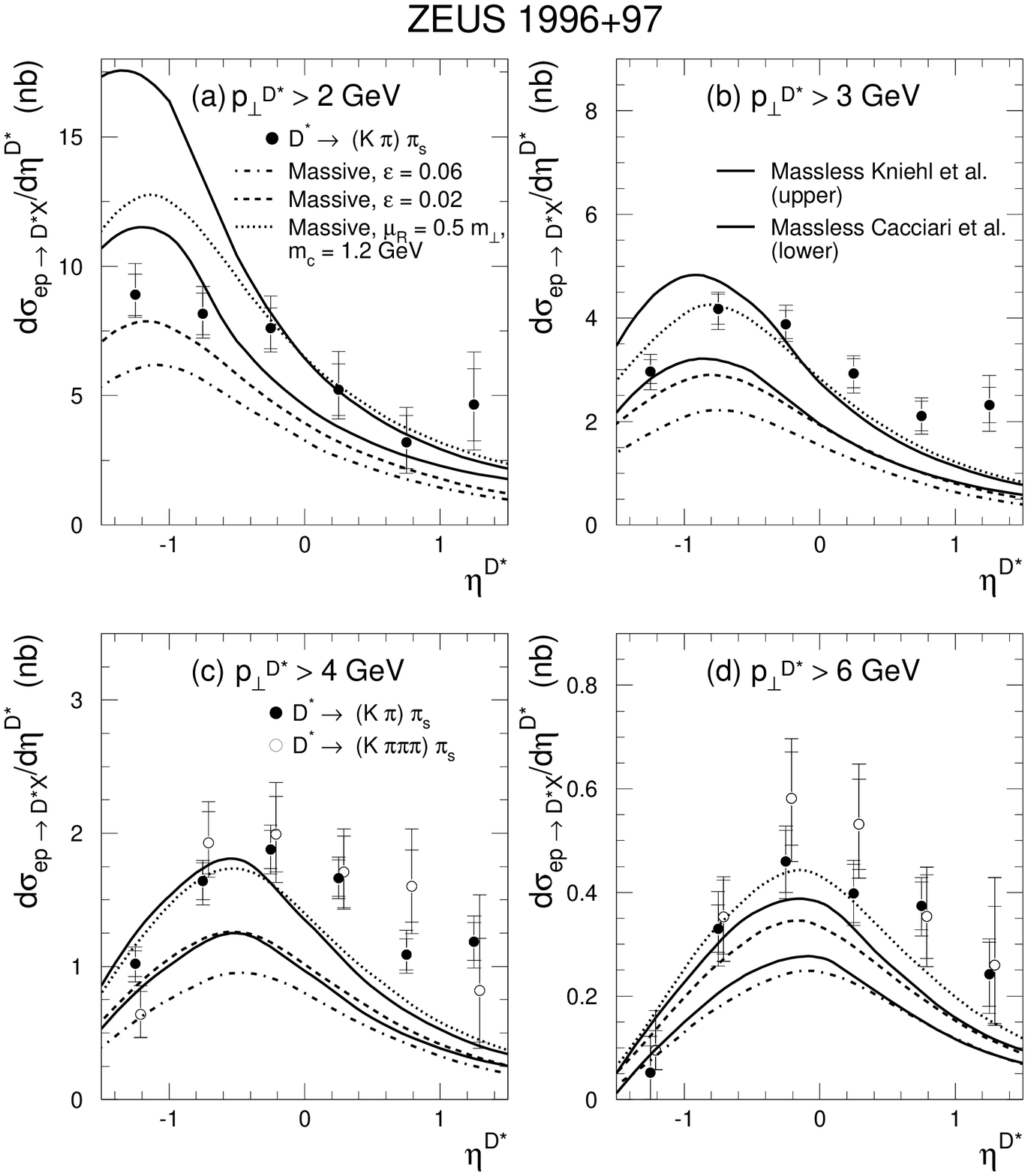} { The differential
  cross section $d\sigma/d\eta^{D^*}$ for $D^*$ photoproduction in the
  kinematic range $Q^2<1$~GeV$^2$, $130<W<280$~GeV and different
  minimum $p_{\perp}^{D^*}$.  The curves represent the NLO
  calculations under the same conditions as for
  figure~\ref{fig:PT_photo_charm}.}  {eta_photo_charm}

Recent data~\cite{ref:ZEUS_dstar97} for $d\sigma_{ep\rightarrow
  D^*X}/dp_T,\; d\sigma_{ep\rightarrow D^*X}/d\eta$.  and
$d\sigma_{ep\rightarrow D^*X}/dW$, measured by the ZEUS collaboration,
are shown in Figs.~\ref{fig:PT_photo_charm},\ref{fig:eta_photo_charm}
and compared to different theoretical calculations.  Both calculations
using standard settings of the charm mass and renormalization scales
are low relative to the data, although the absolute magnitude of the
cross section calculated in the massless approach agree better with
the data.  The deficit of the calculations appears to reside primarily
at large $\eta$.  There are indications that this may be due to an
intrinsic charm content in the photon, which has very weak
experimental constraints.

It is clear that a better control of the theoretical and experimental
issues must be obtained before this method can be used to extract the
gluon density in the proton. The corrections with respect to leading
order are substantial, as is the remaining scale uncertainty in the
NLO calculations. There is also sensitivity to the structure of the
photon, particularly in the massless scheme, and theoretical
calculations of the cross section suffer from a number of
uncertainties, such as the appropriate value of the charm mass and the
choice of scale for the process.

\paragraph{Vector meson production}

The sensitivity of exclusive vector meson production to the gluon
density in the proton is described in detail in section~\ref{sec:VM}.
The strength of this method is that the cross sections depend, in
pQCD, on the square of the gluon density, and are therefore very
sensitive to the $x$ dependence of the gluon density.  However, full
calculations are not available in NLO and therefore normalization
uncertainties are large.  The scale at which the gluon density should
be evaluated is not clear, at least at small $Q^2$, and questions have
been raised relating to the correlation function between the gluons
involved in the scattering.  If these theoretical issues can be
successfully addressed, then elastic vector meson production may
indeed become the method of choice for the gluon density measurement.

\epsfigure[width=0.6\hsize]{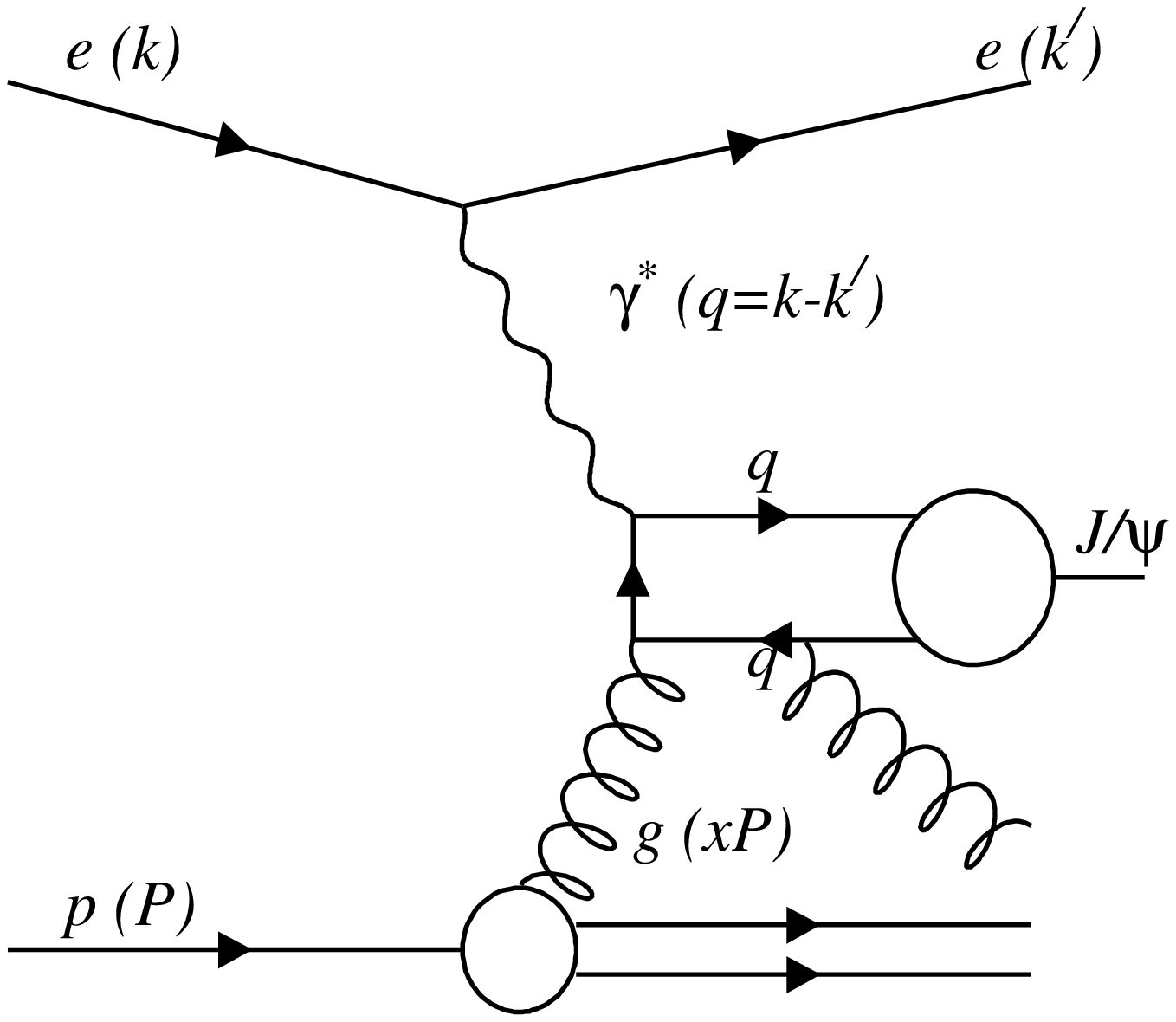} {Inelastic $J/\psi$
  production via photon-gluon fusion.  This is one of several possible
  diagrams where a $J/\psi$ can be produced in an inelastic reaction.}
{jpsiin}

Inelastic $J/\psi$ production has also been proposed as a means for
extracting the gluon density. In this process, as with open charm
production, the incoming photon couples to a single gluon via a
virtual $c\bar{c}$ pair (Fig.~\ref{fig:jpsiin}).  Full NLO pQCD
calculations exist for this
process~\cite{ref:NLO_JpsiX1,ref:NLO_JpsiX2} in the framework of the
color singlet model~\cite{ref:color_singlet}, in which $J/\psi$
production proceeds via photon-gluon fusion followed by the emission
of a hard gluon in the final state.  The color-octet
model~\cite{ref:color_octet} includes, additionally, the production of
the $J/\psi$ in a color-octet state, which then becomes a
color-singlet via the emission of a soft gluon.  This model, developed
to explain hadroproduction
data~\cite{ref:Jpsi_hadroproduction1,ref:Jpsi_hadroproduction2},
attempts to include non-perturbative effects and requires parameter
tuning.  Other mechanisms for inelastic $J/\psi$ production are
diffractive production with proton dissociation, and production via
resolved photons.  The latter two can be effectively isolated
experimentally by cutting on $z$, the fraction of the photon energy,
evaluated in the proton rest frame, carried by the $J/\psi$.  Resolved
photon production is expected to populate the low $z$ region, while
diffractive production peaks at $z=1$.  Color-octet contributions are
also expected to populate the large $z$ region.

\epsfigure[width=0.8\hsize]{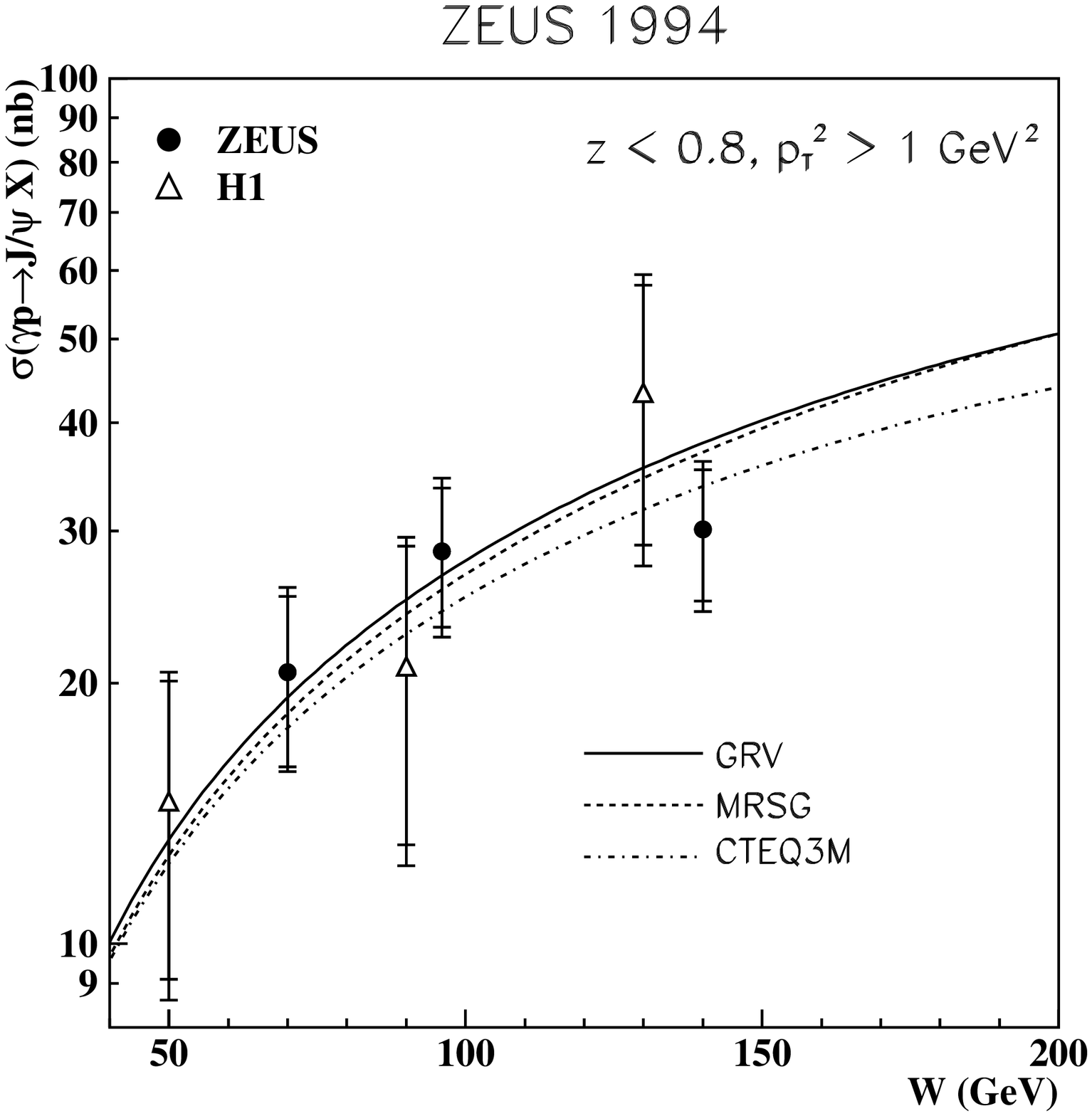} {Inelastic $J/\psi$
  production cross section as a function of $W$ from
  ZEUS~\protect\cite{ref:ZEUS_JpsiX} and
  H1~\protect\cite{ref:H1_JpsiX}.  The lines correspond to the NLO
  pQCD prediction from~\protect\cite{ref:NLO_JpsiX1,ref:NLO_JpsiX2}
  for the three different parton density sets shown in the plot.}
{inel_psi}

\epsfigure[width=0.8\hsize]{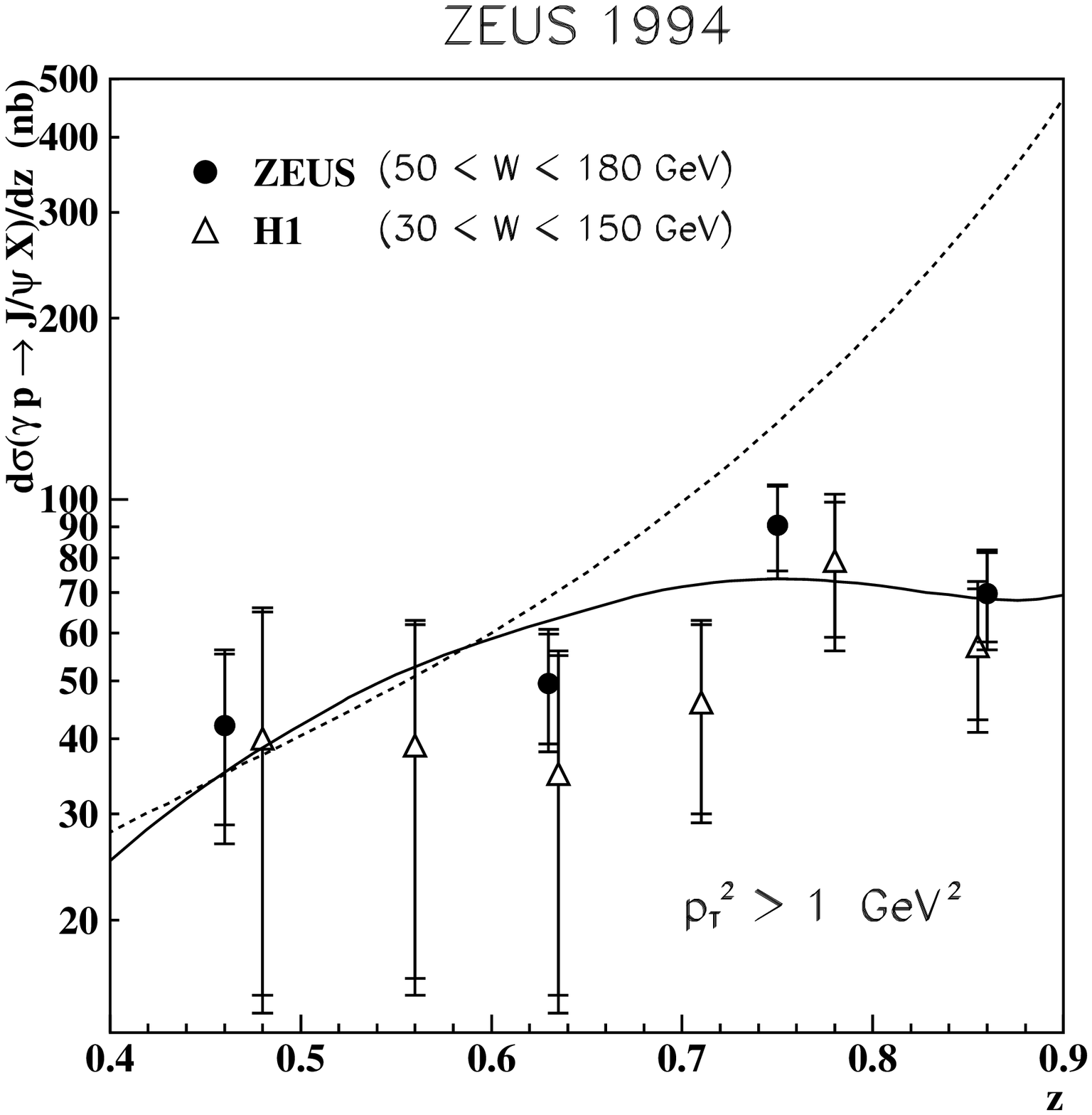} {The differential
  cross section versus $z$ for inelastic $J/\psi$ production at HERA
  in the kinematic range $50 < W < 180$~GeV and $p_T^2>1$~GeV$^2$. The
  solid line is the prediction from the color-singlet model using as
  input the GRV parton densities.  The dashed line represents the sum
  of the color-singlet and color-octet contributions.}  {inel_psi_Ptz}

The experimentally measured cross
sections~\cite{ref:H1_JpsiX,ref:ZEUS_JpsiX} are shown as a function of
$W$ in Fig.~\ref{fig:inel_psi}, where they are compared to NLO
calculations~\cite{ref:color_singlet} with different gluon density
parameterizations.  There is good agreement between the data and the
NLO calculations.  It is also clear from the figure that it will be
very difficult to use this process to distinguish different gluon
densities.  This results in large part from the experimental cuts
needed to select the sample - the requirement of a minimum $p_T$ of
the $J/\psi$ limits the $x$ range which can be probed at HERA.  The
HERA data are plotted as a function of $z$ in
Fig.~\ref{fig:inel_psi_Ptz}, and clearly show a preference for a pure
color singlet contribution.  However, it should be mentioned that the
color-octet contributions are non-perturbative, and that considerable
theoretical work is ongoing to refine the predictions.

\paragraph{Jet production}

The cross section for the production of jets in photoproduction and
DIS also depends on the gluon density via the photon-gluon fusion
mechanism.  It is important to separate the contribution to the cross
section from photon-gluon fusion from that due to QCD-Compton
scattering.  These processes are shown in LO in
Fig.~\ref{fig:jets-disjets}.  The fractional momentum carried by the
parton entering in the hard interaction is given by
\begin{equation}
\xi = x\frac{Q^2+\hat{s}}{Q^2} \; ,
\end{equation}
where $x$ is the usual Bjorken scaling variable and $\hat{s}$ is the
invariant mass of the two jets.  The density of partons can be
extracted from the measured rate of dijet events and the calculable
hard scattering reaction cross section.

\epsfigure[width=0.8\hsize]{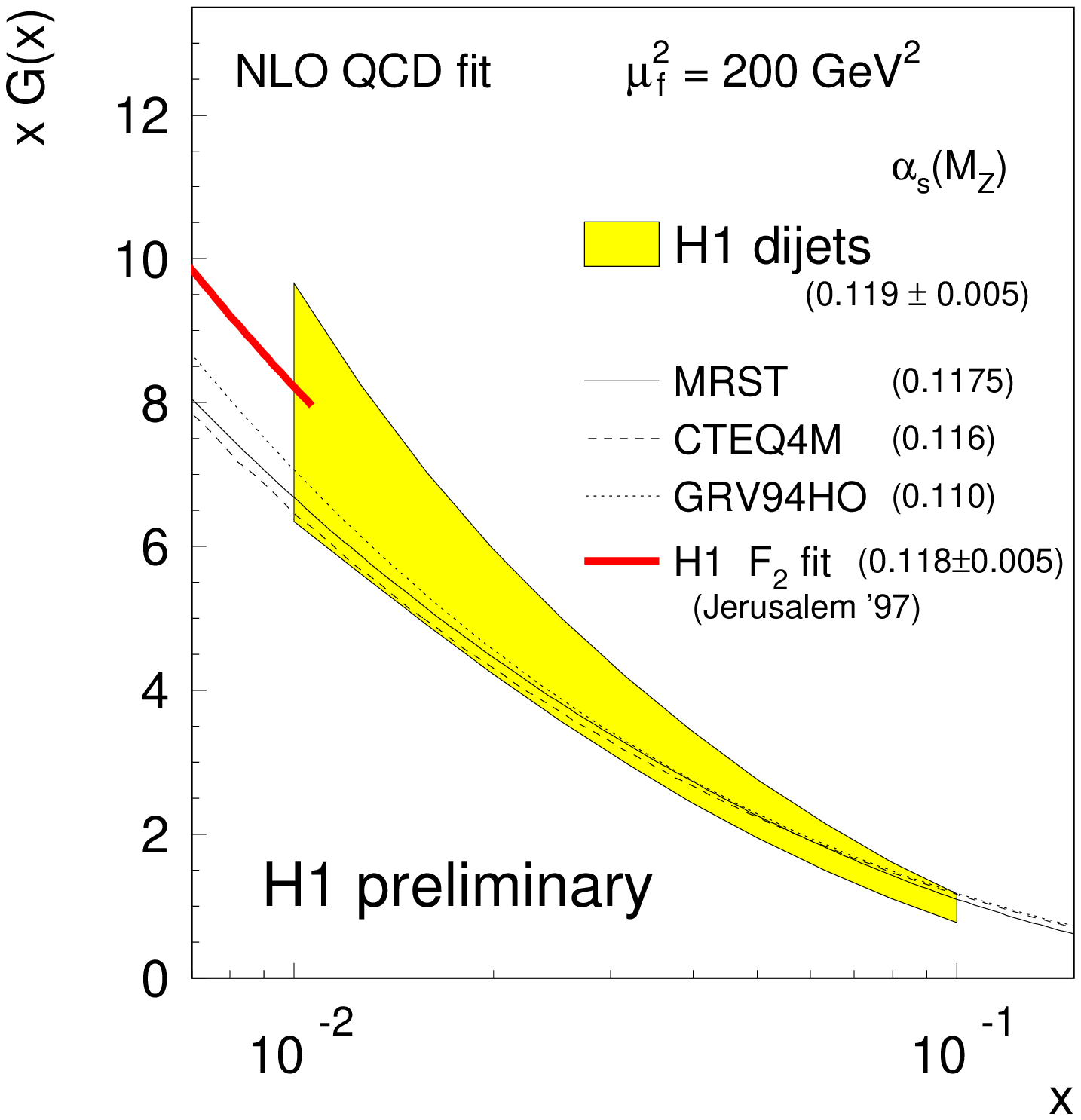} {The gluon momentum
  density in NLO measured from dijet rates in DIS by the
  \protect\citeasnoun{ref:H1_jet_98}.  The gluon density is probed at
  an average $\mu^2$ of $200$~GeV$^2$ and is for a value of
  $\alpha_S(M_Z^2)=0.119\pm0.005$.  The result is compared to the
  gluon density extracted by H1 from a fit to $F_2$, and to other
  parton density parameterizations.}  {H1_jet_glue}

The H1 collaboration has extracted a LO value for the gluon density by
studying dijet production in DIS~\cite{ref:H1_jet_glue}.  A NLO
extraction of the gluon density from this method is quite desirable.
The latter has been attempted by the H1
collaboration~\cite{ref:H1_jet_98}.  The results of this analysis are
shown in Fig.~\ref{fig:H1_jet_glue}.  The data are consistent with a
steep rise of the gluon density found from the analysis of $F_2$.
There NLO extraction of the gluon distribution is complicated by
several factors:
\begin{itemize}
\item The simple form for the momentum fraction carried by the parton
  initiating the hard reaction given above is not valid in NLO.  The
  extraction of the gluon density therefore requires a fit via Monte
  Carlo techniques or a sophisticated unfolding of the data.
\item The jets measured at the hadron level must be related to jets at
  the parton level to allow comparisons with theoretical calculations.
  This introduces a large uncertainty from the hadronization process.
\item The NLO calculations are not able to reproduce the measured
  dijet cross sections.  It is therefore questionable whether the
  extracted gluon densities will be accurate.
\end{itemize}
Large increases in data sets may allow for strong cuts on the data
such that more reliable measurements are possible.

\subsection{NC and CC cross sections at large $Q^2$} 
\label{sec:high_Q2} 
The large $Q^2$ region accessible at HERA is a completely new
kinematic regime for DIS scattering, and opens up the possibility for
novel effects beyond Standard Model expectations.

\epsfigure[width=0.95\hsize]{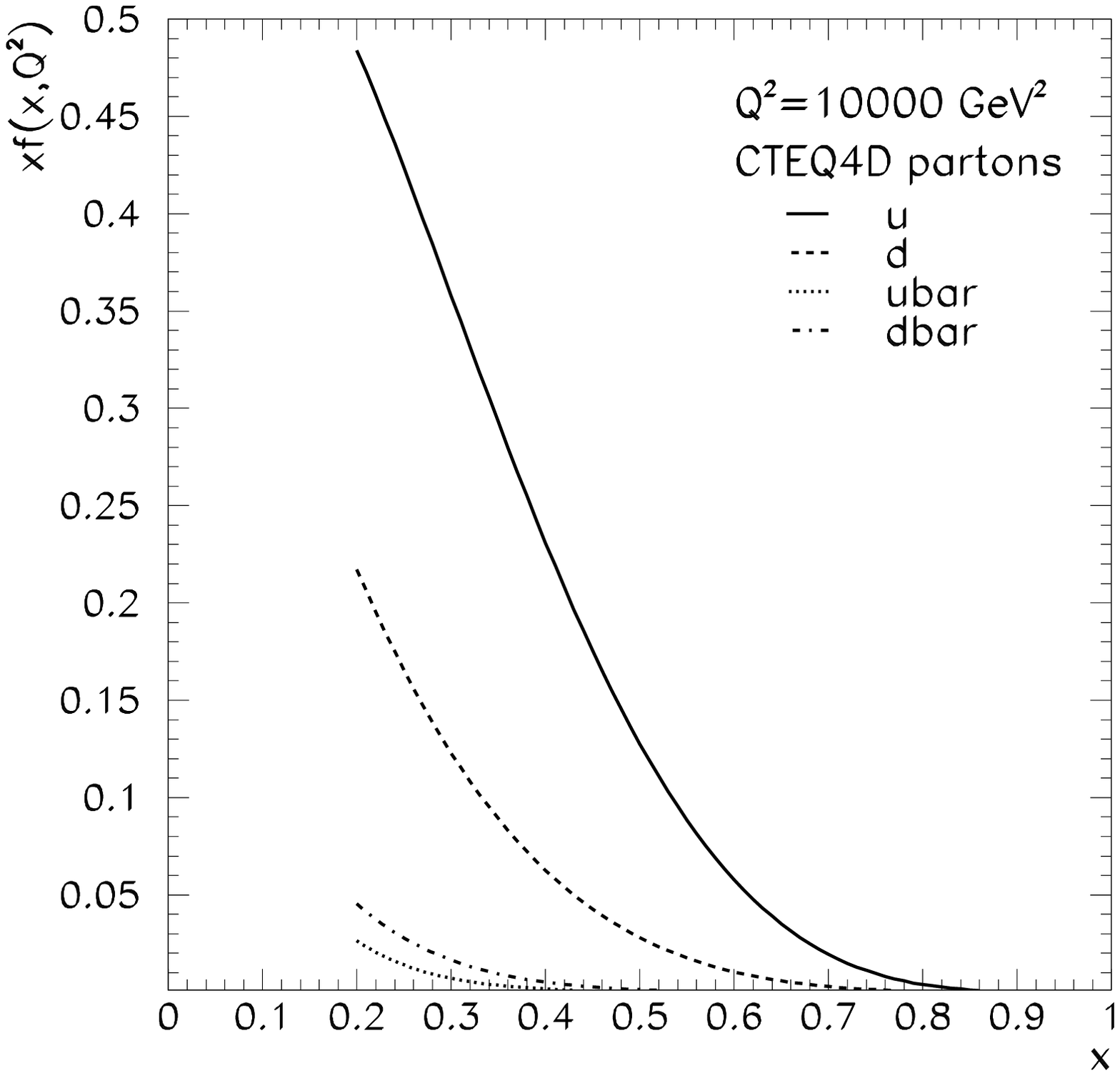} {The up and down
  quark and antiquark momentum densities for $x>0.2$ at
  $Q^2=10000$~GeV$^2$ from the CTEQ4D
  parameterizations~\protect\cite{ref:CTEQ4D}.}  {highQ2_quarks}

The NC double differential cross sections, and $F_2$, have been
published by the H1 and ZEUS experiments up to $Q^2=5000$~GeV$^2$.
The results are shown in Fig.~\ref{fig:F2_94}.  The statistical errors
at large $Q^2$ are large, and show no deviation from the Standard
Model expectations.  The larger data sets available in 1996 and 1997
have allowed both ZEUS~\cite{ref:ZEUSF2_98,ref:ZEUS_NChighQ2} and
H1~\cite{ref:H1_NCCChighQ2} to extend the measurements to higher $Q^2$
and higher $x$, and to reduce the statistical error on the
measurements. The cross sections at large-$x$ are primarily determined
by the valence quark densities (up quark for NC scattering, and down
quark for $e^+p$ CC scattering).  The expectations for the parton
densities at high $Q^2$ are shown in Fig.~\ref{fig:highQ2_quarks}.
The high statistics data from HERA start to constrain these large-$x$
parton densities.

\epsfigure[width=0.8\hsize]{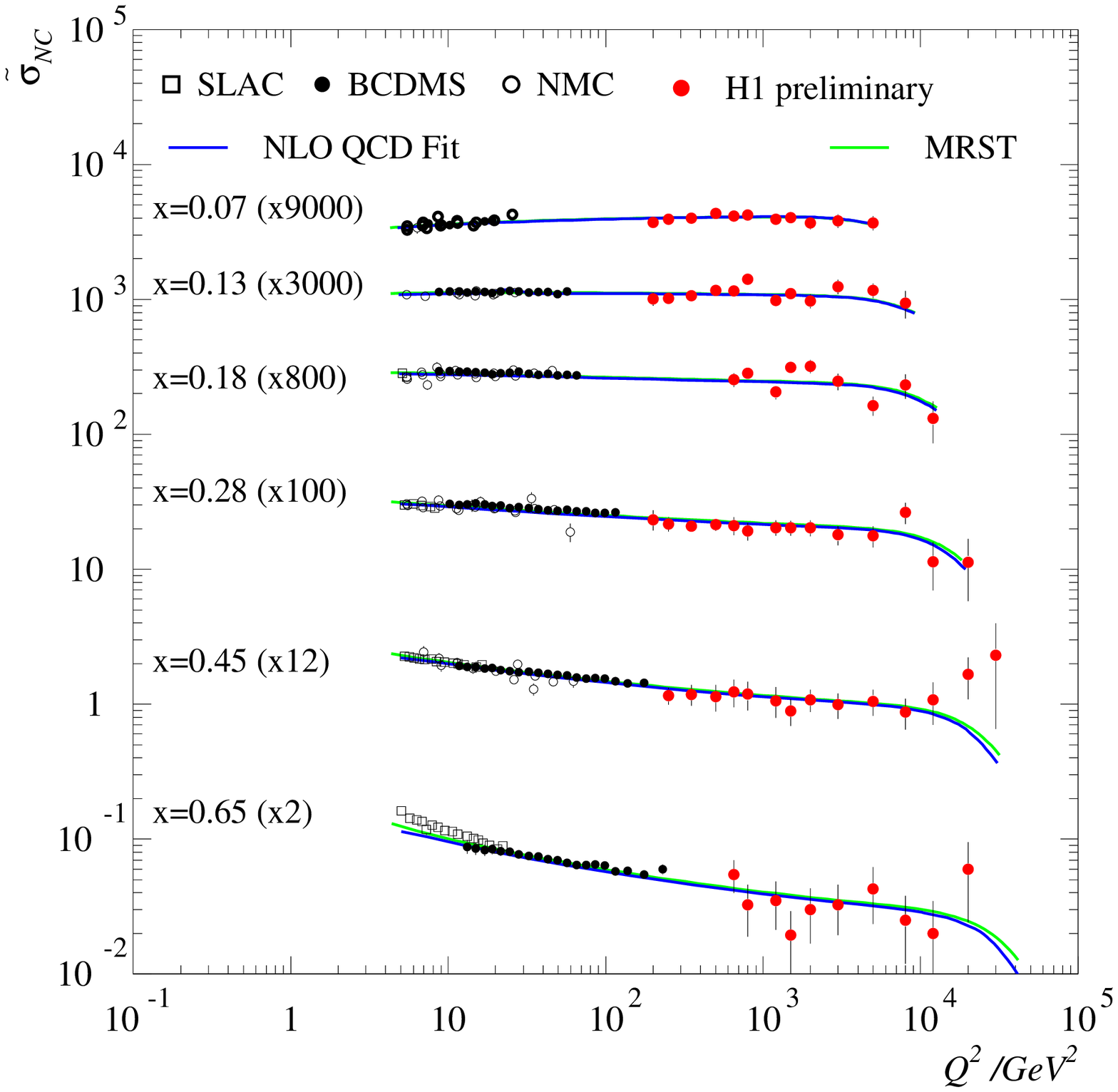} {The reduced neutral
  current cross section at high-$x$ compared with Standard Model
  predictions using the MRST parton density parameterization (lightly
  shaded line) and parton densities resulting from an H1 NLO QCD fit
  (solid line). } {H1_reduced97}

Recent data on the NC cross sections are shown in
Fig.~\ref{fig:H1_reduced97}, where the preliminary data from
H1 at large-$x$ for the reduced cross section
\begin{equation}
\tilde\sigma^{NC} \equiv \frac{xQ^4}{2\pi\alpha^2}\frac{1}{Y_+}
\frac{d^2\sigma}{dxdQ^2}
\end{equation}
are plotted as a function of $Q^2$.  The cross sections presented now
extend to $x=0.65$ and $Q^2=30,000$~GeV$^2$. 

\epsfigure[width=0.8\hsize]{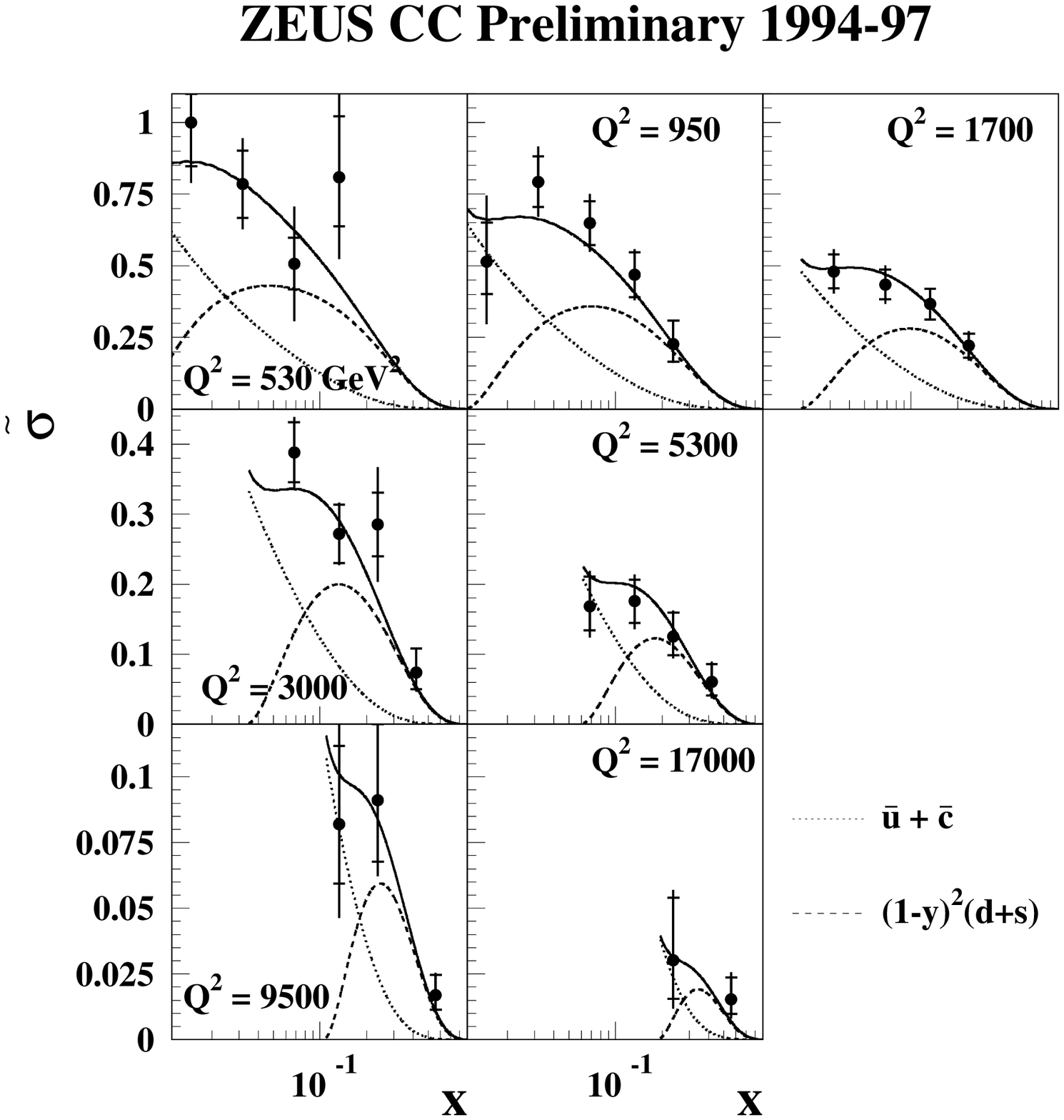} {The reduced charged
  current cross section compared with Standard Model predictions using
  the CTED4D parton density parameterization.  The contributions of
  the PDF combination $\bar{u}+\bar{c}$ and $(1-y)^2(d+s)$ evaluated
  at leading order are shown by the dotted and dashed lines,
  respectively. } {ZEUS_CC97}

Preliminary double differential cross sections from the 1994-97 data
are now also available for the CC
reactions~\cite{ref:H1_NCCChighQ2,ref:ZEUS_CChighQ2}.  The results
from the ZEUS collaboration are shown in Fig.~\ref{fig:ZEUS_CC97}.
The results are again shown for the reduced cross section
\begin{equation}
\tilde\sigma^{CC} \equiv \frac{2\pi x}{G_F^2}\left(\frac{M_W^2+Q^2}{M_W^2}\right)^2
\frac{d^2\sigma}{dxdQ^2} \; .
\end{equation}
In LO, the reduced cross section is simply related to the following quark
densities
\begin{equation}
\tilde\sigma^{CC} = (\bar{u}+\bar{c}) + (1-y)^2(d+s) \; .
\end{equation}
The separate contributions from the quarks and anti-quarks are shown
in the figure.  The quarks dominate at large-$x$, while the antiquark
contribution is dominant at smaller $x$.

\epsfigure[width=0.8\hsize]{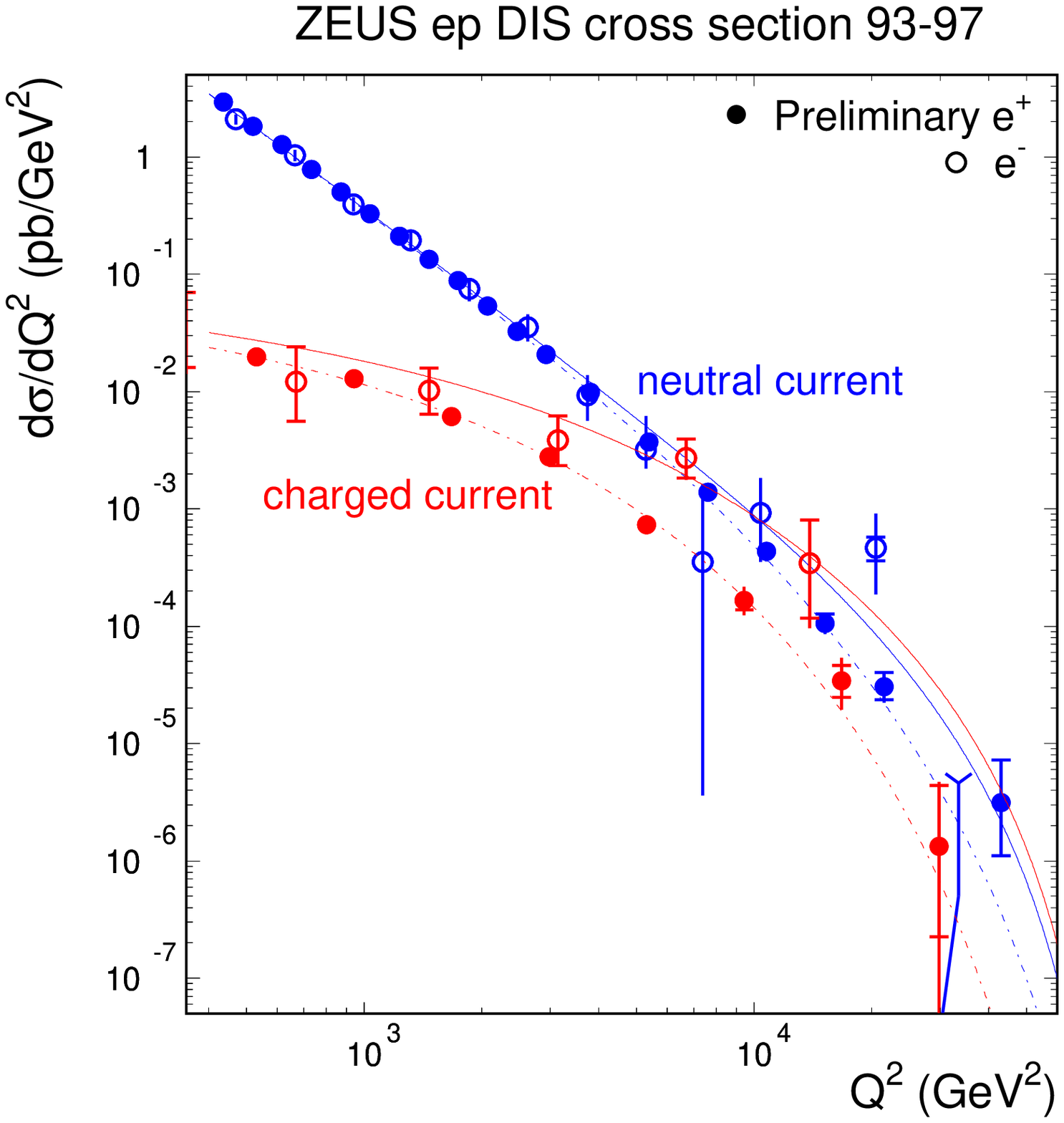} {The differential
  cross section versus $Q^2$ for neutral and charged current
  scattering as measured by ZEUS in the 1993-1997 data sets.  The
  results are compared to Standard Model expectations separately for
  $e^-p$ and $e^+p$ scattering.  The cross sections all become
  comparable near $Q^2=10000$~GeV$^2$, as expected from electroweak
  unification.}  {highQ2_95}

The single differential cross sections from ZEUS for both NC and CC
scattering for the data through 1993-97 data are summarized in
Fig.~\ref{fig:highQ2_95}.  As can be seen in this figure, the $e^-$
and $e^+$ cross sections at small $Q^2$ are very similar, as expected
since photon exchange dominates this region.  At higher $Q^2$, the
expectation for the $e^-p$ cross section is higher than that for the
$e^+p$ cross section as discussed above.  The data are in agreement
with the Standard Model within the experimental uncertainty.  The CC
cross sections for $e^-p$ are higher than for $e^+p$, as expected, and
below the NC cross sections up to $Q^2\approx 10000$~GeV$^2$, at which
point the cross sections become of comparable size.  This is direct
experimental evidence for electroweak unification.  At this scale, the
electromagnetic and weak interactions are of similar strength.

The two experiments H1 and ZEUS reported an excess of events above
Standard Model expectations in February, 1997.  This excess occurs at
very large $Q^2>15000$~GeV$^2$, and will be discussed in
section~\ref{sec:excess}.  This recent observation clearly emphasizes
the importance of measurements in this previously unexplored kinematic
regime.

\section{Jet production}
\label{sec:jets}

In QCD hadrons are composed of many quarks and gluons.  When
two hadrons collide at high energy, two partons can undergo a
large angle scattering generating two or more final state partons at
large momentum transverse to the initial beam direction.  Each of the
partons, released as a result of the hard scattering, will radiate
more partons and evolve into a "spray" of hadrons labeled as a
jet. The nature of QCD radiation is such that the radiated partons, and
subsequently the formed hadrons, will remain collimated around the
original parton direction. This property is used to reconstruct the
underlying partonic interaction.  

The properties of jet production such as cross sections, transverse
momentum distributions, angular distributions and jet shapes, are all
sensitive to the properties of QCD and parton distributions in the
interacting particles. This allows to study QCD at various levels and
in various interactions.

In $ep$ interactions at HERA, jets can be produced as a result of a
hard scattering between a quasi-real photon and the proton as well as
from higher order QCD processes in deep inelastic scattering.
While presumably there is a common denominator to all these processes,
they are usually treated differently within perturbative QCD (pQCD) in
order to improve the quality of the theoretical predictions. The
guideline is set by the factorization theorem which prescribes which
part of the cross section can be calculated within pQCD and which part
is to be absorbed into the non-perturbative universal component. The
general rule is that the perturbative process is the one which occurred
at the hardest scale in the interaction.

Following this guideline, we will first discuss jet production in
photoproduction and then in deep inelastic scattering.  The subject is
of great interest and very wide. Here we will only concentrate on some
highlights typical of $ep$ scattering. For a full account, the
interested reader is referred to two recent
reviews~\cite{erdmann,kuhlen}.

\subsection{Defining the jets}

\epsfigure[width=0.8\hsize]{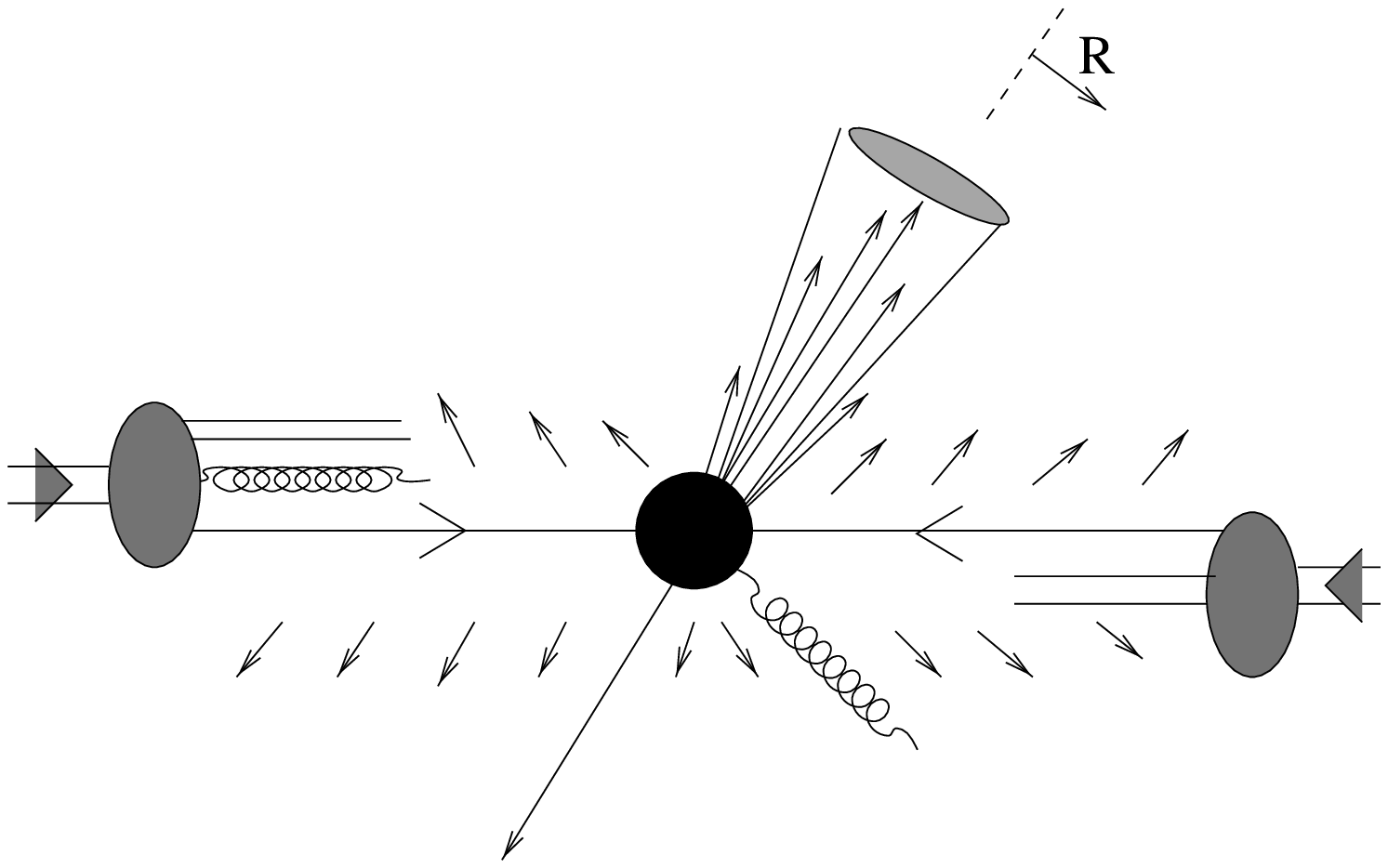}
{ A schematic representation of a hard scattering between two hadrons
  (from~\protect\citeasnoun{sdellis}).  The arrows represent the expected
  particle flow after the interaction. Hadrons produced as a result of
  fragmentation of one of the outgoing partons are contained within a
  cone of radius $R$.}{jets-underlyingevent}

Jets of particles, produced around partons emerging from the hard
scattering, are accompanied by extra hadronic activity resulting from
fragmentation (initial and final state parton radiation) and
hadronization (hadron formation) of the spectator partons (see
Fig.~\ref{fig:jets-underlyingevent}). This extra activity, both
correlated and uncorrelated with the hard scattering, is called the
underlying event. The underlying event consists primarily of low
transverse momentum particles which overlap with the jets. As a result
there is no unique way of assigning final state particles to the
original partons. This is true both from the theoretical and
experimental point of view. For quantitative measurements jets have to
be somehow defined. This is achieved by constructing jet algorithms
which prescribe how to combine hadrons close in phase space into jets.
The experimental procedure has to match the theoretical calculations.
There exist several jet algorithms and their use depends on particular
applications.

\subsubsection{Jet search algorithms}
\label{sec:jetalgorithm}

The jet search search algorithms are all based on the notion that
particles emitted as a result of parton fragmentation and
hadronization should be close to each other in phase space. The more
sophisticated ones also take into account the dynamical features of
parton radiation. In addition, since the underlying event is soft and
particles have limited transverse momentum, the presence of jets will
result in high transverse energy deposits.  The variety of jet
algorithms reflects the need to minimize hadronization effects, which
cannot be taken into account in the theoretical calculations.

\paragraph{Cone algorithm}

In the cone algorithm particles are combined in the pseudo-rapidity,
$\eta$, and azimuthal angle, $\phi$, phase space. The variable
$\eta=-\ln \tan \frac{\theta}{2}$ is defined through the angle
$\theta$ relative to the interaction axis and $\phi$ is the angle
around the interaction axis. All particles within a cone of radius R,
\begin{equation}
R=\sqrt{\Delta \eta^2+\Delta \phi^2} \leq R_0 \, ,
\end{equation}
are combined into a jet of transverse energy $E_T$
\begin{equation}
E_T=\sum_i E_{Ti} \, ,
\label{eq-jets:etjet}
\end{equation}
where $i$ runs over all particles in the cone and $E_{Ti}=E_i \sin
\theta_i$. The value of $R_0$ usually varies between $R_0=0.7$ to $R_0=1$. 
If the resulting $E_T$ is above a certain threshold the jet axis is
defined by
\begin{eqnarray}
\eta_J &=& \frac{1}{E_T} \sum_{i} E_{Ti} \eta_i \, ,\label{eq-jets:etajet}
\\
\phi_J &=& \frac{1}{E_T} \sum_{i} E_{Ti} \phi_i \, .
\label{eq-jets:phijet}
\end{eqnarray}
In practice this procedure is applied on pair of particles and implies
some number of iterations until the quantities defined in
Eq.~(\ref{eq-jets:etjet}),( \ref{eq-jets:etajet}), (\ref{eq-jets:phijet})
are stable with the jet cone remaining fixed.

The cone algorithm has an equivalent theoretical definition. Its
advantage is that it can easily be applied to calorimetric
measurements where energy deposits in calorimeter cells are treated as
single particles. The cone algorithm was adopted as standard jet
definition, called Snowmass Accord~\cite{snowmass}, because of its ease
of implementation and reliable results at all orders in perturbation
theory.

\paragraph{JADE algorithm}
\label{jets:jade}
In the JADE algorithm~\cite{JADEalgorithm} one defines for each pair
of particles or clusters $i$ and $j$ the quantity,
\begin{equation} 
y_{ij}=\frac{2E_i E_j (1-\cos \theta_{ij})}{W^2}\simeq \frac{m_{ij}}{W^2} \, ,
\label{eq-jets:jade}
\end{equation}
where $m_{ij}$ is the invariant mass of the objects $i$ and $j$ and
$W$ is the center-of-mass energy of the interaction. If $y_{ij}$ is
smaller than a resolution parameter $y_{\mathrm{cut}}$ the objects are
combined.  The final set of jets is obtained when no further merging
is possible. No requirement on $E_T$ of jets is applied. In the
modified version of the algorithm, particularly suitable for deep
inelastic scattering, the reference scale $W^2$ is replaced by $Q^2$.
The number of jets found in an event depends on the $y_{\mathrm{cut}}$
parameter. This dependence is in principle predicted by the theory. In
hadronic interactions the parent hadrons are included in the
clustering as pseudo-particles. The jet containing the pseudo-particle
is called the remnant jet.

\paragraph{$k_T$ algorithm}

In the $k_T$ algorithm~\cite{ktalgorithm} it is the relative
transverse momentum of two particles, rather than their invariant
mass, that is used for clustering particles. Two particles, or at a
later stage two clusters, $i$ and $j$ are merged if the transverse
momentum, $k_T$, of the least energetic of the two objects is smaller
than a predefined resolution scale, $k^2_{\mathrm{cut}}$,
\begin{equation}
k_T^2 = 2 {\mathrm{min}}(E_i^2,E_j^2)(1-\cos \theta_{ij}) < k^2_{\mathrm{cut}} 
\, .
\label{eq-jets:kt}
\end{equation}
In the original algorithm~\cite{ktalgorithm} which was designed for
deep inelastic scattering $k^2_{\mathrm{cut}} = Q^2$. The treatment of
the remnant jet is similar to the JADE algorithm. The $k_T$ algorithm
is expected to be least affected by hadronization
effects~\cite{WebberParis}.

\subsection{Jet production in photoproduction}

One of the interests in studying jet production in photoproduction is
to probe the hadronic nature of the photon.  The photon is one of the
gauge particles of the Standard Model and as such has no intrinsic
structure. However it acquires a structure in its interactions with
matter and in that sense it is a prototype for studying the formation
of a partonic state.  

In $ep$ scattering at HERA cross section considerations (see
section~\ref{sec:photoprod}) favor interactions with exchanged photon
virtuality $Q^2 \simeq 0$. These interactions can be treated as due to
a convolution of a flux of quasi-real photons from the electron and
the photon-proton interactions,
\begin{equation}
\frac{d\sigma_{ep}(y,Q^2)}{dy} = \sigma_{\gp}(y) f_{\gamma/e}(y,Q^2)\, ,
\end{equation}
where $f_{\gamma/e}$ denotes the photon flux. The photon flux is assumed
to be given by the Weizs\"acker-Williams
approximation~\cite{weizsacker,williams}.
For a range $Q^2_{min}<Q^2<Q^2_{max} \ll 1 \gevtwo$, the 
flux is given by
\begin{equation}
f_{\gamma/e}= \frac{\alpha}{2\pi} \left[ \frac{1+(1-y)^2}{y} 
\ln \frac{Q^2_{max}}{Q^2_{min}} - 2 m_e^2 \left( \frac{1}{Q^2_{min}}-
\frac{1}{Q^2_{max}} \right) \right] \, .
\label{eq-jets:wwa}
\end{equation}

\subsubsection{Theoretical framework}
\label{sec:photon_structure}

In photon-nucleon interactions at low energies it was established that
the photon behaves as if it fluctuated into a vector meson before
interacting (for a review see~\citeasnoun{ref:Bauer}).  When QCD is
turned on the variety of hadronic fluctuations of the photon
increases, small and large partonic configurations can be formed, and
their presence has implications for the interactions of the photon
with hadronic matter.

The momentum distribution of partonic fluctuations of the photon can
be measured directly in deep inelastic scattering of a lepton on a
photon target. These measurement are performed in $e^+e^-$ colliders
in which one of the electrons is the source of the probing photon and
the other of the target photon. The notion of the photon structure
function, and subsequently of parton distributions, can be
consistently introduced.

 \epsfigure[width=0.8\hsize]{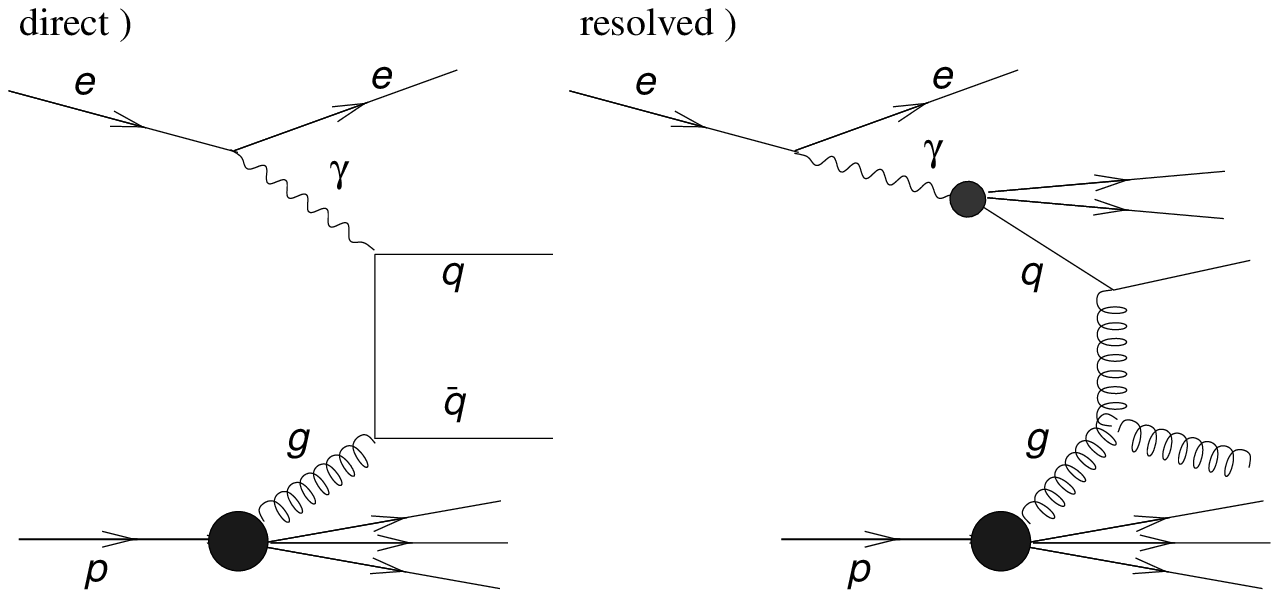}
{ Examples of diagrams which contribute to jet production in
  photoproduction: direct photon and resolved photon
  contributions as denoted in the figure.}{jets-photoprd-diag}

Once the photon is replaced by a flux of partons, the QCD
factorization theorem~\cite{collins-soper} can be applied to calculate
cross sections for jet production in photon-proton interactions. These
processes are called resolved photon processes. An example is shown in
Fig.~\ref{fig:jets-photoprd-diag}.

Partonic fluctuations of the photon in which the transverse momentum
of the $q\bar{q}$ pair is larger than the virtuality of the probing
photon are not included in the usual structure function formalism.
They do however contribute to the jet production cross section and
induce the direct photon processes. The name derives from the fact
that these processes look as if the photon was probing directly the
parton structure of the proton. This is depicted schematically in
Fig.~\ref{fig:jets-photoprd-diag}.

In summary, QCD predictions for photon proton interactions are based on
two essential ingredients, the partonic nature of the photon and the
underlying QCD hard subprocesses.  We discuss them in more
details below.

\paragraph{The photon structure function}

The structure functions of the photon are defined in DIS $e \gamma
\rightarrow e X$ through the interaction cross section,
\begin{equation}
\frac{d^2\sigma}{dxdQ^2}=\frac{2\pi\alpha^2}{xQ^4}\left[ \left( 1+(1-y)^2
\right) F_2^{\gamma} - y^2F_L^{\gamma} \right] \, .
\label{eq-jets:xsec_dis_ph}
\end{equation}

 \epsfigure[width=0.4\hsize]{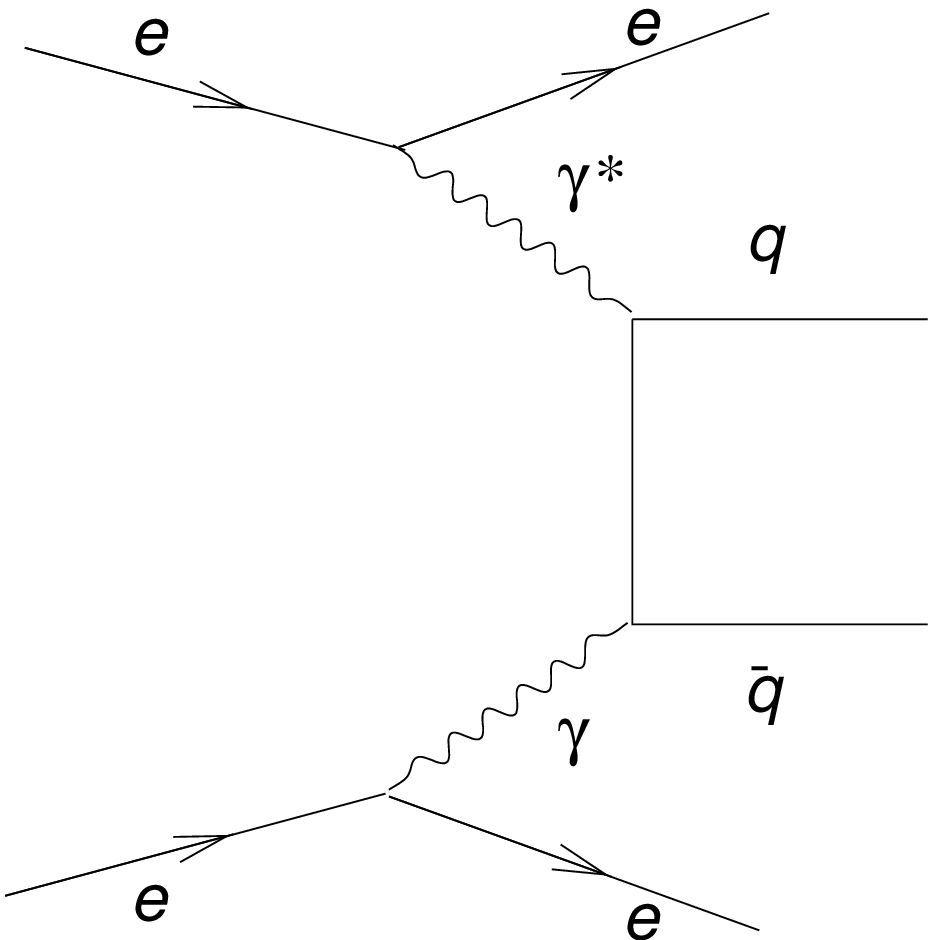}
{Basic box diagram which contributes to the structure function of the
  photon in $\gamma^* \gamma$ collisions.}{jets-box-diag}

In the quark-parton model (QPM) the contribution to this cross section comes
from the so called box diagram $\gamma^* \gamma \rightarrow q
\bar{q}$ (see Fig.~\ref{fig:jets-box-diag}),
\begin{equation}
F_2^{\gamma}(x,Q^2) = 3 \frac{\alpha}{\pi}\sum_{n_f} e^4_q
x \left( x^2+(1-x)^2 \right) 
\ln \frac{Q^2}{m^2_q} \, ,
\label{eq-jets:box}
\end{equation}
where $\alpha$ is the electromagnetic coupling constant, $e_q$ is the
charge of quark $q$ in units of electron charge and $m_q$ is the mass
of the quark. The sum runs over all active quark flavors, $n_f$. 

Should this be the only contribution, the structure function of the
photon would be fully calculable in QCD~\cite{witten}. In QCD, to
leading order, the result (\ref{eq-jets:box}) is preserved with the
$m_q^2$ term replaced by the QCD cut-off parameter $\Lambda^2_{QCD}$.
In the language of photon fluctuations the box diagram corresponds to
rather symmetric $q\bar{q}$ fluctuations of the photon with a
point-like coupling of quarks to the photon.  The resulting
contribution to the photon structure is called point-like or anomalous
(as it does not appear for real hadrons).  In reality the photon can
fluctuate into more complicated partonic states and in particular into
vector meson states. This generates a non-perturbative contribution to
$F_2^{\gamma}$.  Theoretically the two contributions cannot be
disentangled. To parameterize $F_2^\gamma (x,Q^2)$ it is therefore
convenient, by analogy to the proton case, to define parton
distributions in the photon and to use the DGLAP evolution equations,
appropriately extended to the photon
case~\cite{dglapph,dglappherratum}. The evolution equation for the
quark density becomes inhomogeneous to account for the splitting of
the photon into a $q\bar{q}$ pair.  The latter is responsible for the
fact that $F_2^{\gamma}(x,Q^2)$ is large at large $x$ and increases
with $Q^2$ at any value of $x$.

Experimentally $F_2^{\gamma}$ is measured in $e^+e^-$
interactions, by requiring that one of the electrons scatters under
small angles and remains undetected (the source of the target photon)
while the second one scatters under a large angle, providing the
probing virtual photon.

 \epsfigure[width=.95\hsize]{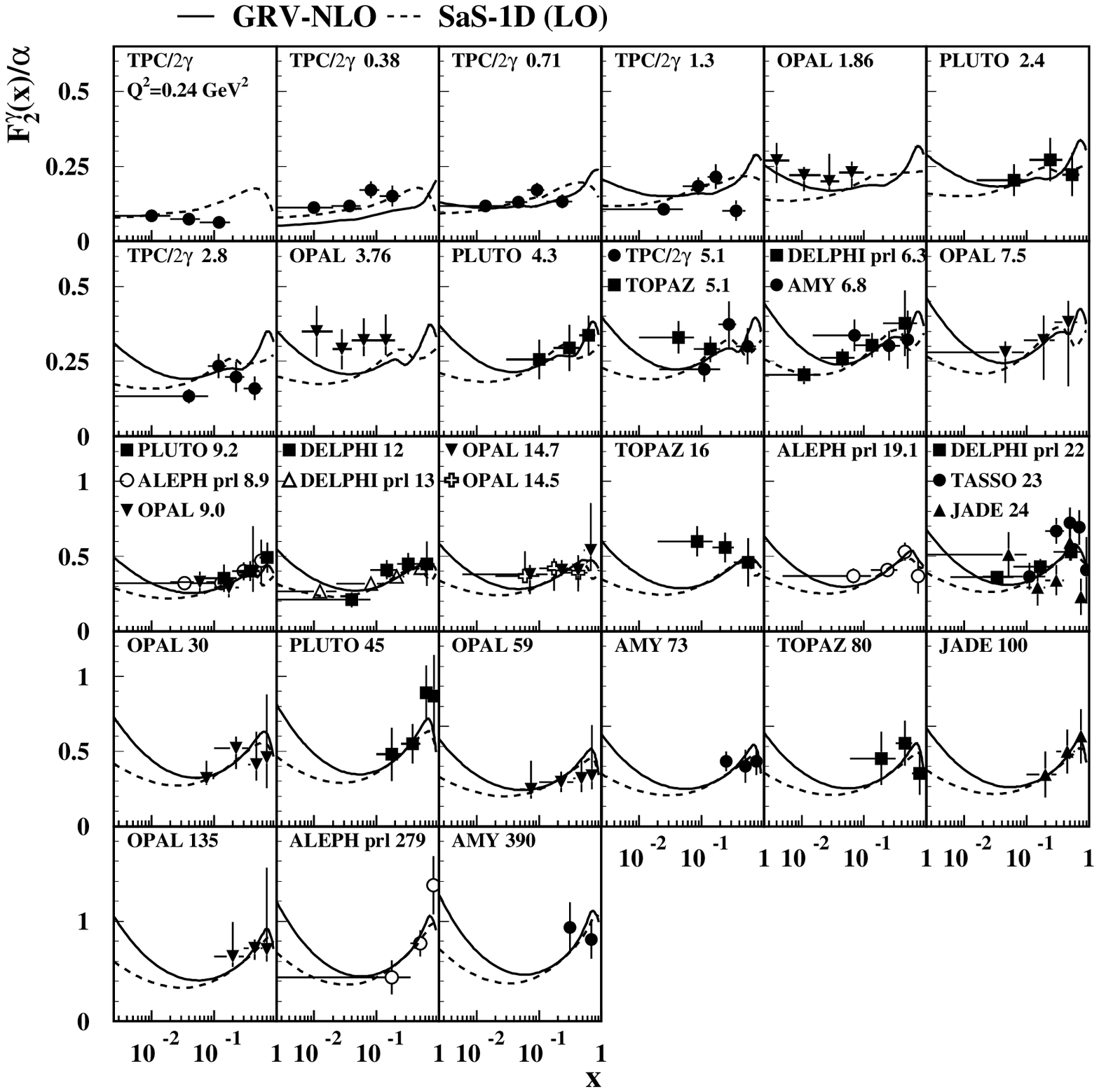}
{  The structure function of the photon $F_2^\gamma/\alpha$ as a
  function of $x$ for various $Q^2$ values as denoted in the figure 
  (taken from~\protect\citeasnoun{ref:rembold}.
  The symbols correspond to measurements in $e^+e^-$ interactions.
  Two selected parton parameterizations are also shown:
  GRV~\protect\cite{grv_f2ph} (full line) and
  SaS~\protect\cite{sas_f2ph95} (dashed line).}{jets-f2gamma}

The $x$ dependence of $F_2^{\gamma}$ in bins of $Q^2$ is shown in
Fig.~\ref{fig:jets-f2gamma}.  For the sake of comparison with the proton
structure function the scale of $x$ was chosen logarithmic. The
measurements are compared to two commonly used sets of parton
distributions in the photon~\cite{grv_f2ph,sas_f2ph95}. At
higher $x$, where data are available, all the parameterizations
describe $F_2^{\gamma}$ well. The largest differences are seen at low
$x$, where gluons play an important role.  The uncertainty in the
gluon content of the photon is thought to be worse than for the proton
since no momentum sum rule is used to constrain the gluon
distribution.  Recently such a sum rule has been
derived~\cite{gamma_sumrule}. The uncertainty in the gluon
distribution can be partly reduced by the hard photoproduction data
from the HERA experiments.

It is of interest to consider what happens to the photon structure
function when the virtuality increases. 
We will denote by $P^2$ the virtuality of the photon and keep $Q^2$ to
denote the scale at which the virtual photon is probed. It has been
shown~\cite{walsh1,walsh2} that in the limit $\Lambda^2_{QCD}\ll P^2
\ll Q^2$ the virtual photon structure function can be calculated in
QCD. The result is similar to the QPM result (\ref{eq-jets:box})
with the mass of the quark replaced by the virtuality of the photon,
\begin{equation}
F_2^{\gamma^*}(x,Q^2,P^2) = 3 \frac{\alpha}{\pi}\sum_{n_f} e^4_q
x \left( x^2+(1-x)^2 \right) 
\ln \frac{Q^2}{P^2} \, .
\label{eq-jets:virtualph}
\end{equation}
The details of how the non-perturbative part behaves with $P^2$ cannot
be calculated in QCD. One can only speculate that its contribution
disappears faster with increasing $P^2$ than the point-like part.
However the non-perturbative contributions to the photon structure may
remain sizeable up to $P^2=1 \gevtwo$~\cite{grs_f2ph}.

\paragraph{Predictions of perturbative QCD}
\label{sec:dijet_xsec}
In leading order of perturbative QCD, jet production in photon-proton
interactions is attributed to two hard processes, called the direct
and the resolved photon processes. The calculation of the cross
sections is based on the QCD factorization
theorem~\cite{collins-soper}. It allows to decompose (factorize) the
calculation into the cross section for the elementary hard subprocess,
calculable in QCD, and fluxes of partons entering the hard scattering.
The hard scale in the interaction is provided by the transverse
momenta of the outgoing partons.  For the production of two jets with
transverse momentum $p_T$, the $\gamma p$ cross section can be written
symbolically as
\begin{equation}
\sigma_{\gp}(p_T^2,W) =  
\sum_{i,j} \sum_{k,l} f_{i/\gamma}(x_\gamma,p_T^2) f_{j/p}(x_p,p_T^2) 
\hat{\sigma} (i+j \rightarrow k+l)(p_T^2,\hat{s}) \, ,
\label{eq-jets:factorization}
\end{equation}
where $x_\gamma$ and $x_p$ denote the fraction of the photon and
proton momentum carried by the interacting partons and
$\hat{s}=x_\gamma x_p W^2$ with $W$ the center-of-mass energy of the
$\gp$ interaction. An implicit integration over $\xg$ and $x_p$ is
assumed. 

For resolved photon processes the hard scattering is modeled as taking
place between a parton in the photon and a parton in the
proton (see 
Fig.~\ref{fig:jets-photoprd-diag}). Many elementary processes
contribute to this cross section. A quark or a gluon from the photon
may interact with a quark or a gluon in the proton. In the final state
we expect two back-to-back jets, this time accompanied by the remnants
of the proton and of the photon.

In direct photon processes it is convenient to view the hard
scattering as occurring between the photon and a parton from the
proton (see Fig.~\ref{fig:jets-photoprd-diag}). In
formula (\ref{eq-jets:factorization}) parton $i$ is replaced by a
photon and $f_{\gamma/\gamma}(x_\gamma)=\delta (1-x_\gamma)$.  The
process $\gamma q\rightarrow q g$ is called QCD Compton scattering
(QCDC), while $\gamma g \rightarrow q\bar{q}$ is called boson-gluon
fusion (BGF).  In leading order we expect the final state to consist
of two back-to-back jets and the remnant of the proton.

When higher order QCD corrections are included, extra hard parton
radiation is present and the division of photon induced processes into
direct and resolved is no longer possible. However, it is
always possible to enhance one of the contributions because the direct
photon contribution is expected to dominate at $x_\gamma \simeq 1$
while the resolved processes are expected at lower $x_\gamma$. 

The parton densities in the proton are well constrained by other
measurements (see section~\ref{sec:structure_functions}), so that
photoproduction of jets can be used to constrain the parton content in
the photon.

\subsubsection{pre-Hera results}

Little was known experimentally about hard scattering in
photoproduction before HERA data became available. The first
indication for the presence of hard scattering, in excess of what
would be expected if the photon were just a meson, was observed by the
NA14 experiment at CERN~\cite{NA1486}. A similar excess was observed
later by the WA69 experiment at CERN~\cite{OME89} at a center-of-mass
energy $\sqrt{s_{\gp}}=18 \gev$. In both cases the excess was assigned
to the direct photon contribution. The presence of the resolved photon
component could not be established.

The first experimental evidence for the presence of a hard resolved
photon component was reported by the AMY experiment~\cite{AMY92},
based on the study of the hadronic final states in $\gamma\gamma$
interactions at $4<\sqrt{s_{\gamma\gamma}}<20 \gev$.

\subsubsection{Selection of HERA results}

The production of events with large transverse energy in $\gp$
interactions has been extensively studied at HERA. The concept of 
resolved and direct photon contributions as well as the validity of
the QCD approach to photoproduction have been fully confirmed. First
attempts have been made to constrain the parton distributions in the
photon. After discussing the kinematic reconstruction of hard
photoproduction events, we describe these results below.

\paragraph{Event kinematics}

The experimental signature for photoproduction events is either the
presence of an electron scattered under a very small angle in the
luminosity detector (tagged photoproduction with $Q^2<10^{-2}
\gevtwo$) or the lack of the scattered electron in the main HERA
detectors (untagged photoproduction, $Q^2<4 \gevtwo$ and $\langle Q^2
\rangle \simeq 10^{-3} \gevtwo$). Hard photoproduction events are
selected by requiring that the total transverse energy $E_T$ in the
event, measured with respect to the beam axis, be large (typically
$E_T > 10 \gev$).

For tagged photoproduction, the center-of-mass energy of the $\gp$
interaction is obtained using the initial, $E_e$, and the scattered
electron, $E_e^\prime$, energies,
\begin{equation}
W^2 = y_{el} s = \left(1-\frac{E_e^\prime}{E_e} \right)s \, .
\label{eq-jets:yel}
\end{equation}
In the untagged photoproduction $W$ is determined using the hadronic
final state,
\begin{equation}
W^2 = \yjb s = \frac{\sum_i(E_i - p_{zi})}{2E_e} s \, ,
\label{eq-jets:yjb}
\end{equation}
where the sum runs over all the objects (particles and/or clusters)
reconstructed in the detector.

To describe the event kinematics we define $x_\gamma$ as the fraction
of the photon momentum, $q$, and $x_p$ as that of the proton, $P$,
carried by the partons initiating the hard scattering. We denote by
$p_1$ and $p_2$ the four-momenta of the scattered partons.  For a two
body process we expect,
\begin{equation}
x_\gamma q + x_p p = p_1+p_2 \, ,
\end{equation}
By multiplying (scalar product) both sides, once by $P$
and once by $q$ and manipulating the two equations, we obtain
\begin{eqnarray}
x_\gamma &=& \frac{(p_1+p_2) \cdot P}{q \cdot P} \, ,
\label{eq-jets:xgammadef} \\ 
x_p &=& \frac{(p_1+p_2) \cdot q}{q \cdot P} \, ,
\label{eq-jets:xprotondef}
\end{eqnarray}
where the approximation, $q^2=0$ and $P^2=0$ was made.  This
approximation holds well for high energy photoproduction.  Another
representation, more convenient for measurements, is through the
transverse momentum, $p_{Ti}$, and the pseudo-rapidity, $\eta_i$, of
the partons,
\begin{eqnarray}
x_\gamma &=& \frac{\sum_{i=1,2} p_{Ti} e^{-\eta_i}}{2yE_e} \, ,
\label{eq-jets:xgamma} \\ 
x_p &=& \frac{\sum_{i=1,2} p_{Ti} e^{\eta_i}}{2E_p} \, .
\label{eq-jets:xproton}
\end{eqnarray}

Since variables defined at the parton level are not observables,
experimental estimators are introduced, $\xgobs$, and $\xpobs$. In the
first step a jet algorithm is applied. For a cluster of particles to
be defined as a jet, its transverse energy relative to the interaction
axis, is required to be typically $E_T^{jet} > 4
\gev$. To determine $\xgobs$ or $\xpobs$ at least two jets
are required.  If more than two jets are found in an event, usually
the two with the highest transverse energy are used.  The variables
$\xgobs$ and $\xpobs$ are determined using
formulae~(\ref{eq-jets:xgamma}) and (\ref{eq-jets:xproton}), after
replacing the $p_T$ and $\eta$ of the partons by the corresponding jet
variables which are $\ETJ$ and $\ETAJ$ as determined from the jet
algorithm (see section~\ref{sec:jetalgorithm}).

\paragraph{The resolved and direct components}

\epsfigure[width=0.8\hsize]{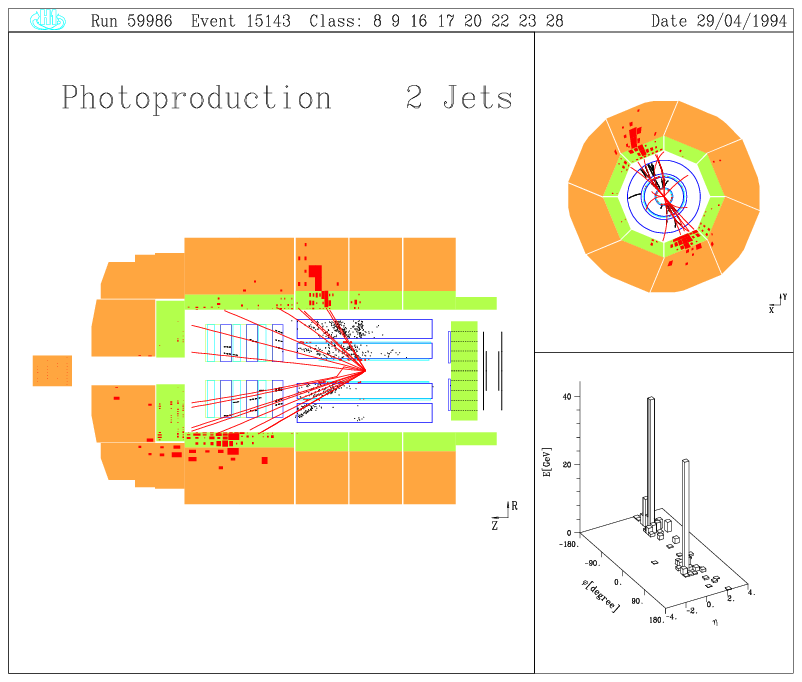}{ Example of a
  photoproduction event presumably mediated by a direct photon as seen
  in the H1 detector.  The electrons come from the left, the protons
  from the right.  Two jets are clearly seen. There is no activity to
  the right of the detector, the presumed direction of the emitted
  photon.}{jets-h1event-direct}

\epsfigure[height=0.8\hsize,angle=-90]{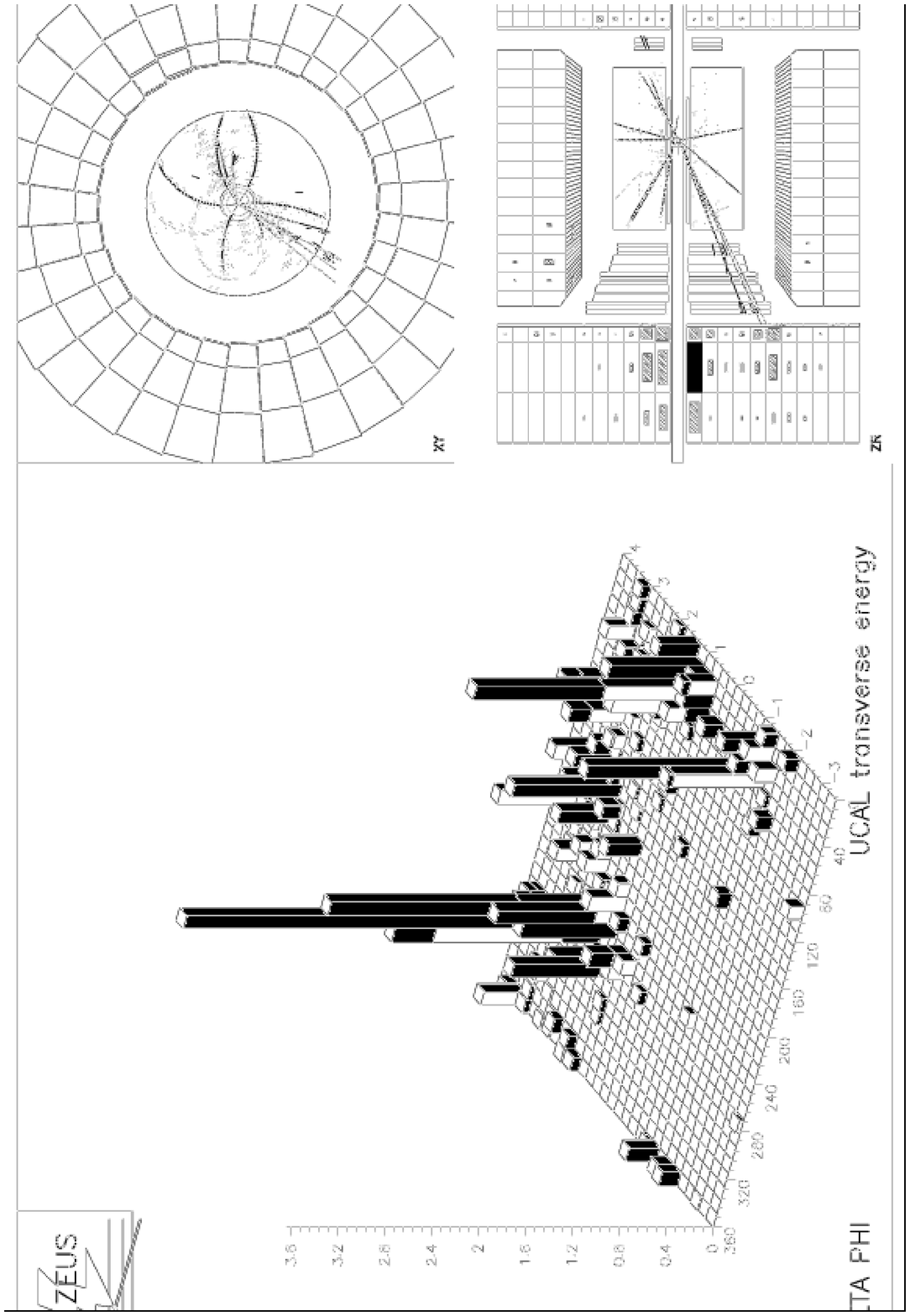}{ Example
  of a photoproduction event due to a resolved photon, as seen in the
  ZEUS detector. The electrons come from the left, the protons from
  the right. In addition to the proton remnant, there is clear 
  jet structure. Most likely, the jet at smallest $\eta$ is the photon
  remnant.}{jets-zeusevent-resolved}

Jet production in photoproduction was established for the first time 
by the HERA experiments~\cite{ZEUSph1,H1ph1}. The jet structure in the
events can be seen by eye. An example of a direct photon event is
shown in Fig.~\ref{fig:jets-h1event-direct}. Two jets in the detector are
clearly seen and there is no activity in the rear region which
corresponds to the photon fragmentation, as if the whole photon was
absorbed in the interaction. In Fig.~\ref{fig:jets-zeusevent-resolved} a
resolved photon event is shown. A third jet with much less transverse
energy is seen in the rear detector - this is the remnant of the photon.

\epsfigure[width=0.8\hsize]{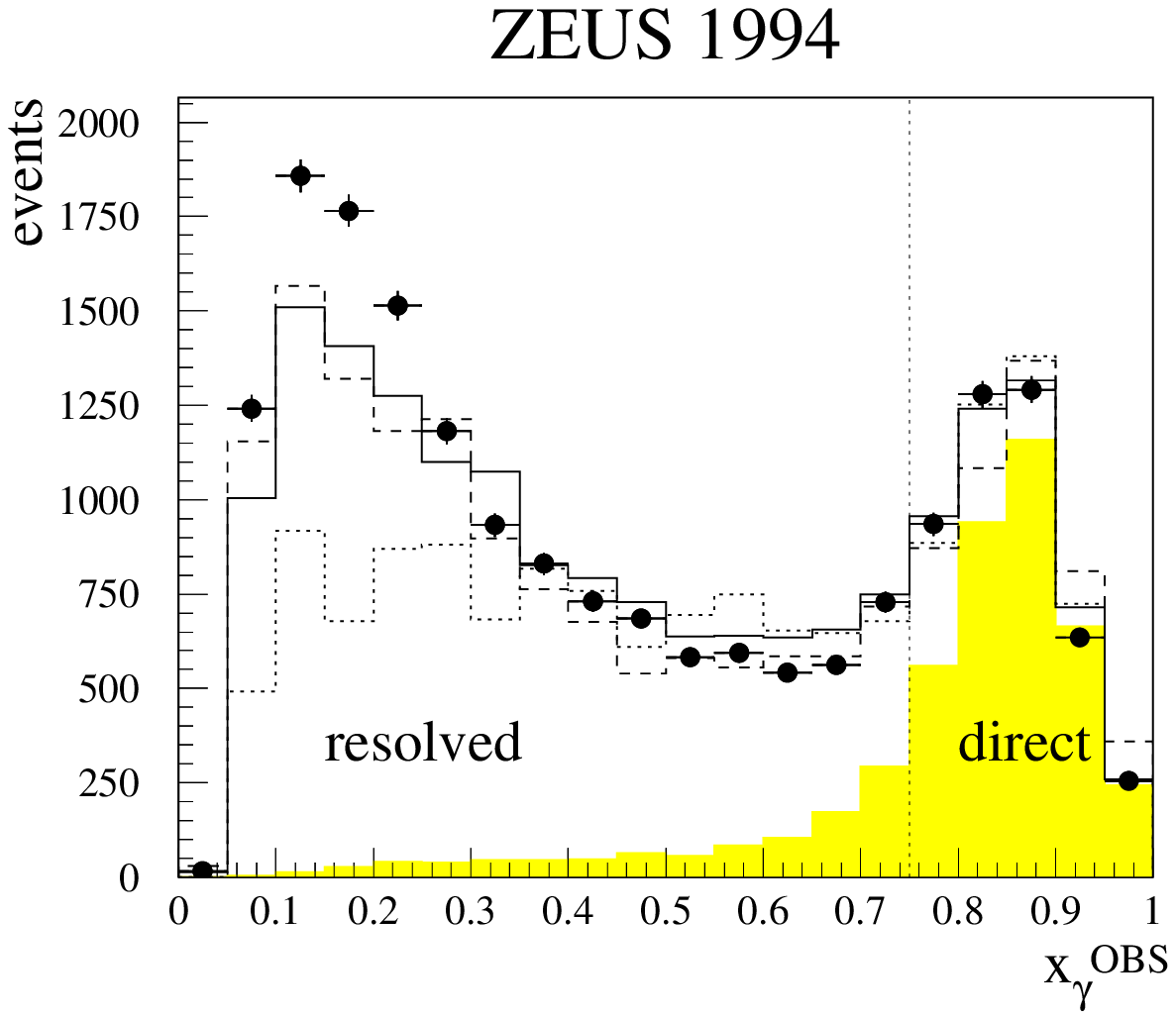}{ $\xgobs$
  distribution for jets with $\ETJ > 6$~GeV, $-1.375 < \ETAJ <1.875$,
  where $\xgobs$ is calculated using corrected variables.  The data
  (black dots) are compared to the results of the HERWIG MC model with
  multiple interactions (solid line) and without (dotted line) and
  PYTHIA MC model with multiple interactions (dashed line).  The
  shaded area represents the direct photon contribution expected in
  the HERWIG MC model.}{jets-xgamma}

The $\xgobs$ distribution measured for a sample of di-jet events with
$\ETJ>6 \gev$ and $-1.375<\ETAJ<1.875$ is shown in
Fig.~\ref{fig:jets-xgamma}~\cite{ZEUSdijet}.  The events dominated by
the direct photon processes cluster at large $\xgobs$. A clear peak of
events with $\xgobs>0.75 $ is observed and the distribution is well
reproduced by a MC model of direct photon processes
(HERWIG~\cite{herwig}). The majority of events have $\xgobs <0.75$ and
are attributed to the resolved photon processes.  MC models which
include the simulation of both processes at leading order of QCD and
include in addition multiple interactions reproduce the overall shape
of the data but fail to describe the small $\xgobs$ region.

\epsfigure[width=0.8\hsize]{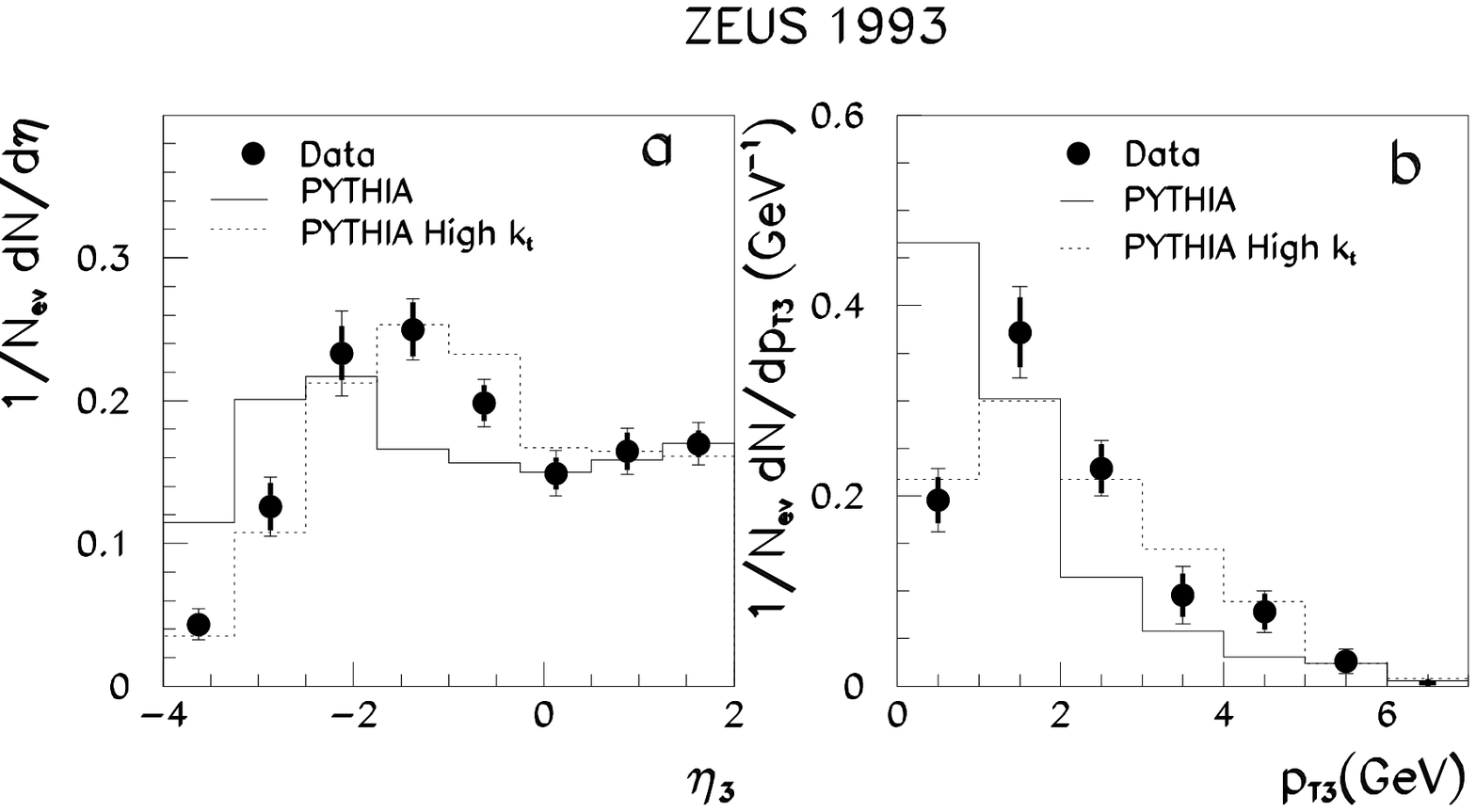}{ (a) Pseudorapidity,
  $\eta_3$, distribution of the third, lowest transverse momentum jet
  in a three jet event. (b) The transverse momentum distribution of
  that jet, $p_{T3}$, if $\eta_3<-1.0$. The data are compared to the
  expectations of two approaches in the PYTHIA MC model: in one, the
  intrinsic transverse momentum of partons in the photon is negligible
  (solid line), in the other, a power like dependence is assumed
  (dotted line).}{jets-remnant}

The properties of the photon remnant have been studied by the ZEUS
experiment~\cite{ZEUSremnant}. The $k_T$ cluster algorithm was used to
decompose the event into three jets. If two jets had transverse energy
$\ETJ > 6 \gev$ and rapidities well into the forward region, the third
cluster on average was found in the photon fragmentation region as
shown in Fig.~\ref{fig:jets-remnant}a. The two leading jets were
interpreted as due to the hard scattering, while the third jet was
interpreted as originating from the photon remnant. The third jet was
found to be well collimated, with a relatively large transverse
momentum, as can be seem in Fig.~\ref{fig:jets-remnant}b.  This is in
contrast with expectations for hadron-hadron interactions where the
remnants (with spectator partons) remain with low transverse momenta.
However, the anomalous component of the photon structure, which is due
to small size configurations, leads to initial partons with relatively
large transverse momenta. The requirement of the presence of three
jet-like structure favors events from the anomalous photon
fluctuation. Note that the average transverse energy of the third jet
remains low compared to $E_T^{jet}$ of the two leading jets (see
Fig.~\ref{fig:jets-remnant}b).

\paragraph{Angular distributions}

The angular distribution of the jets in their center-of-mass system
reflects the underlying parton dynamics.  To leading order in QCD, the
direct photon processes proceed via either boson-gluon fusion or QCD
Compton scattering.  These processes involve a quark propagator in the
$s$, $t$ or $u$ channel, with $t$ and $u$ channel processes
dominating.  In resolved processes the dominant subprocesses, e.g. $q
g \rightarrow q g$, $g g \rightarrow g g$ and $q q \rightarrow q q$,
have $t$-channel gluon exchange diagrams.  

The angular dependence of the cross section for resolved processes
with a spin-1 gluon propagator is approximately $\propto ( 1 - |\cts|
)^{-2}$ (as in Rutherford scattering), where $\theta^\ast$ is
the scattering angle of one of the jets in their center-of-mass.  This
cross section rises more steeply with increasing $|\cts|$ than that
for direct processes with a spin-$\frac{1}{2}$ quark propagator, where
the angular dependence is approximately $\propto ( 1 - |\cts| )^{-1}$.
After inclusion of all LO diagrams, QCD predicts that the angular
distribution of the outgoing partons in resolved processes will be
enhanced at large $|\cts|$ with respect to direct photon processes.
This property is expected to be preserved in next-to-leading order
(NLO) calculations~\cite{owens}.  

\epsfigure[width=0.8\hsize]{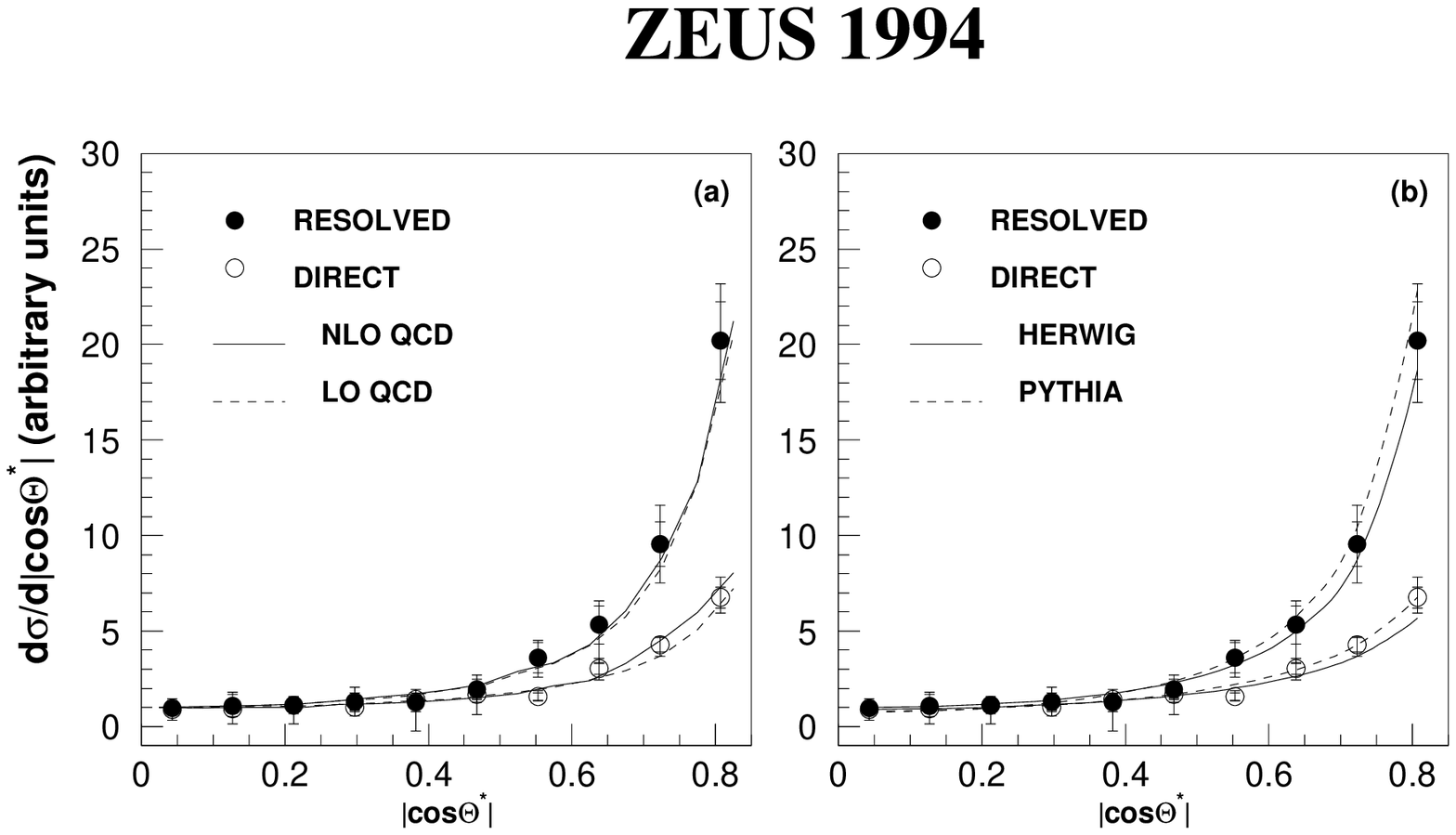}{The cross section
  $d\sigma/d|\cos \theta^*| $ normalized to one at center-of-mass
  scattering angle $\theta^*=90^o$, for resolved (black dots) and
  direct (open circles) photoproduction. The data are compared to (a)
  NLO and LO QCD calculations as well as (b) expectations of the
  HERWIG and PYTHIA MC models.}{jets-angle}

By dividing dijet events into samples with $\xgobs >0.75$ and $\xgobs
<0.75$ it is possible to obtain samples enriched in direct and
resolved processes respectively~\cite{ZEUSangle}. The $\cts$
distributions thus obtained are shown in Fig.~\ref{fig:jets-angle}.
The resolved photon processes have as expected a steeper angular
distribution than the direct processes.  The distribution is well
reproduced by LO and NLO QCD calculations as well as MC models.

\paragraph{Structure of the real photon}

The measured jet production cross section can be confronted with QCD
expectations. These depend on the momentum distribution of partons in
the photon and in the proton (see Eq.~(\ref{eq-jets:factorization})). The
procedure for unfolding the parton distributions is not as simple as
in the case of deep inelastic scattering. The measured cross section
is determined at the hadron level and the information on the hard
scattering process is obscured by higher order QCD radiation and
hadronization effects. It may further be obscured by the presence of
other hard scatters occurring in the same event~\cite{H1mpi}. From a
purely probabilistic approach, parton densities in the HERA kinematic
regime are high enough so that the chance for more than one scatter in
an event becomes significant~\cite{Sjostrand-mpi}. 

\epsfigure[width=0.8\hsize]{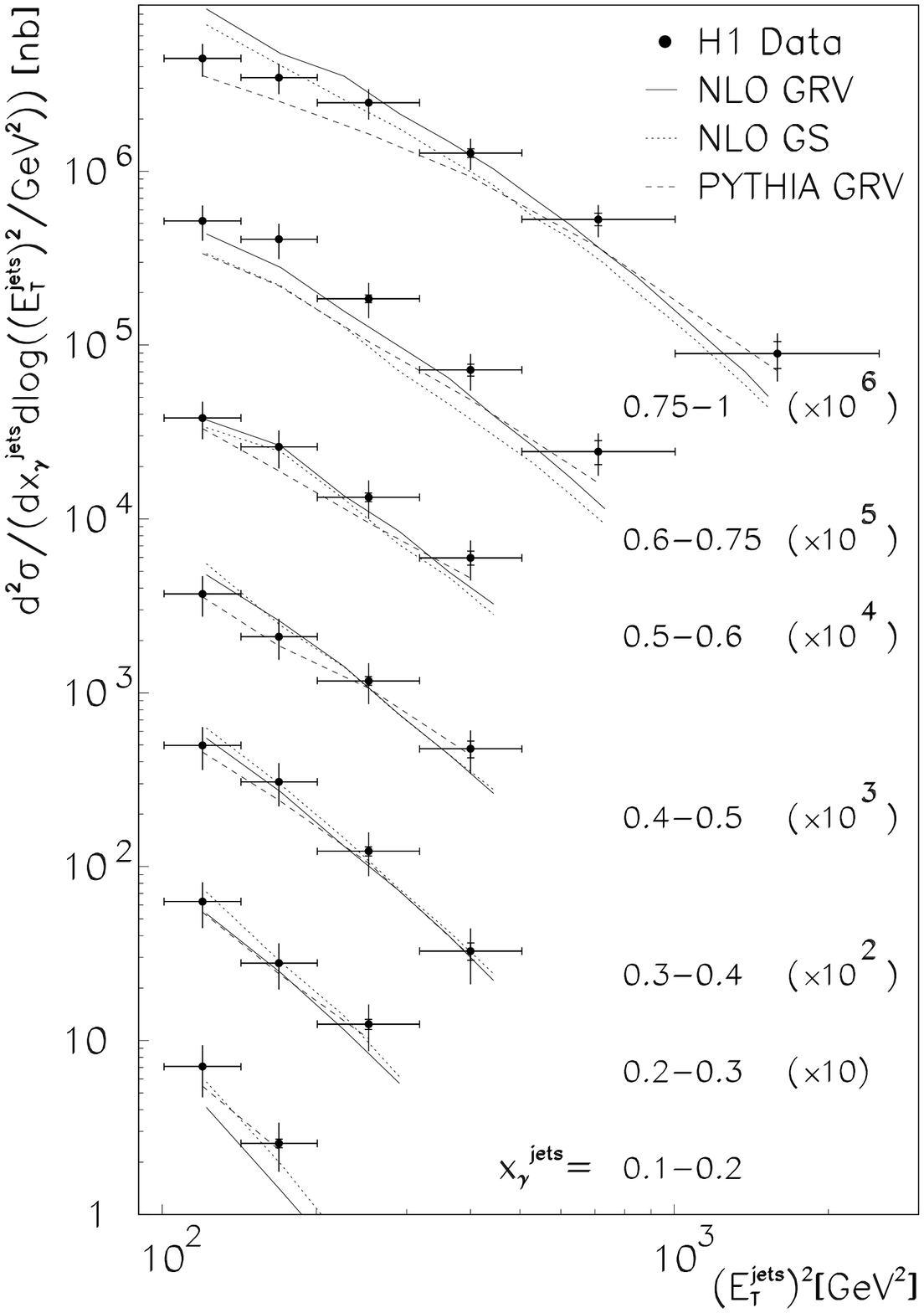}{ The inclusive dijet
  cross section as a function of the squared jet transverse energy
  $(E_T^{\mathrm jets})^2$ for ranges of the reconstructed parton
  fractional energy $x_\gamma^{\mathrm jets}$ as denoted in the
  figure. The data are compared to NLO QCD calculations for two
  parameterizations of parton densities in the photon,
  GRV~\protect\cite{grv_f2ph} and GS~\protect\cite{gs_f2ph96}. Also
  shown is the expectation of the PYTHIA MC model in which the LO
  order GRV parameterization was used.  } {jets-dijetsection}

The additional effects can be modeled within a particular MC generator
and if a good description of the data is achieved , then, within this
particular model, the correlation between the partons from the hard
scattering and the jets can be established. One can then deduce what
parton distributions in the photon are needed to successfully describe
the observed jet rates. This procedure was carried out by the H1
experiment~\cite{H1partoninph}.  The measured inclusive dijet cross
section as a function of $(E_T^{jet})^2$ and $\xg$ is very well
reproduced by the PYTHIA MC model~\cite{ref:PYTHIA} used to unfold the
parton densities, as shown in Fig.~\ref{fig:jets-dijetsection}. It is
also in good agreement with NLO QCD calculations~\cite{nlodijet}.

\epsfigure[width=0.8\hsize]{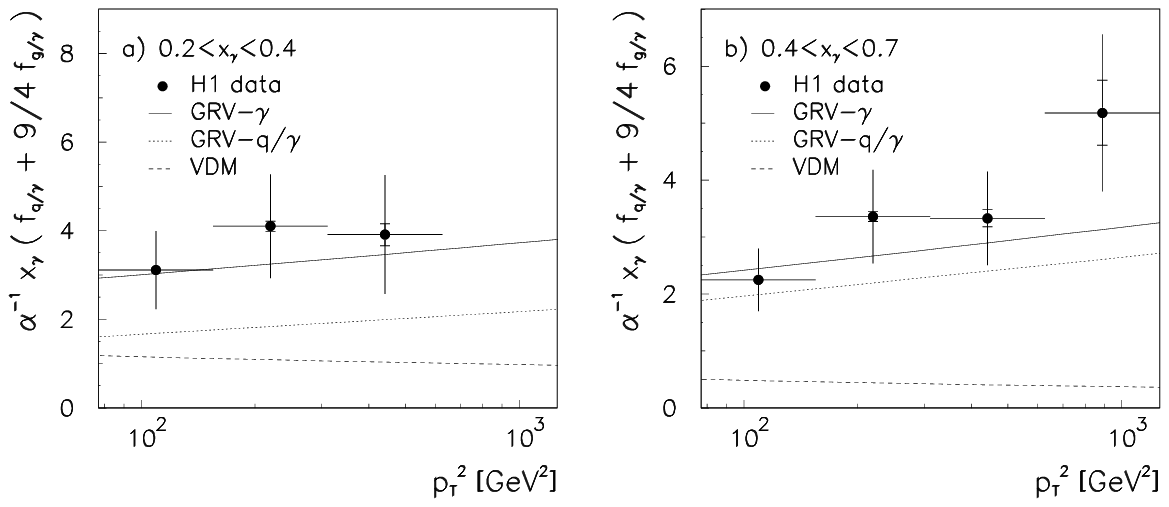}{Leading order
  effective parton distribution of the photon $x_\gamma
  \bar{f}_\gamma(x_\gamma,p_T^2)=x_\gamma
  (f_{q/\gamma}(x_\gamma,p_T^2)+9/4 f_{9/\gamma}(x_\gamma,p_T^2))$,
  divided by the fine structure constant $\alpha$, shown as a function
  of the squared parton transverse momentum $p_T^2$ for two ranges of
  the parton fractional energy $x_\gamma$. The full line represents
  the equivalent function derived from the LO GRV parameterization of
  the photon structure function. Also shown are the contribution of
  the quark component alone (dotted line) and the expectation if the
  photon was a meson (dashed line - VDM). } {jets-partons}

The effective parton distributions were unfolded from the cross
section in two bins of $\xg$ as a function of the hard scattering
scale, assumed to be $p_T^2$ of the jets. The result is shown in
Fig.~\ref{fig:jets-partons} and compared to the expectations of the
GRV~\cite{grv_f2ph} parameterization. As predicted for the photon
case, parton densities rise with increasing $p_T^2$ both for medium
and large $\xg$.

Since the calculation is carried out at leading order it is possible
to disentangle the various contributions. This is also shown in the
figure. The non-perturbative part of the photon structure, expected to
contribute at small $\xg$ constitutes only a small fraction of parton
densities for $\xg>0.2$. The quark component of the photon is not
sufficient to explain the observed effective parton densities, a
substantial contribution of gluons is necessary. 

\paragraph{Structure of the virtual photon}

The more the photon is off-shell the more likely it becomes that its
coupling to quarks becomes point-like. The structure function of the
photon can then be calculated in QPM and QCD \cite{walsh1,rossi}. The
expectation is that in hard photoproduction induced by a virtual
photon, the hadronic contribution will decrease and the resolved
photon processes will be suppressed compared to the direct photon
processes~\cite{drees,grs-ep,florian}.

At large virtuality of the photon, we move into the deep inelastic
scattering regime, $Q^2$ becomes the largest scale in the interaction,
and application of the factorization theorem leads to viewing the
scattering as due to direct photon coupling to the quarks in the
proton. However if, as a result of QCD Compton or BGF processes, jets
are produced with transverse energy squared $(\ETJ)^2 > Q^2$, the largest
scale will be in the hard scatter leading to jets and the photon, once
again, will develop a partonic
structure~\cite{frixione,grs-ep,klasen}. This apparent contradiction
is in fact simple to understand; the perturbative QCD evolution
formalism applied to the proton structure function does not include
the region $E_T^2 > Q^2$. Their contribution to the total DIS cross
section is assumed to be negligible.

\epsfigure[width=0.8\hsize]{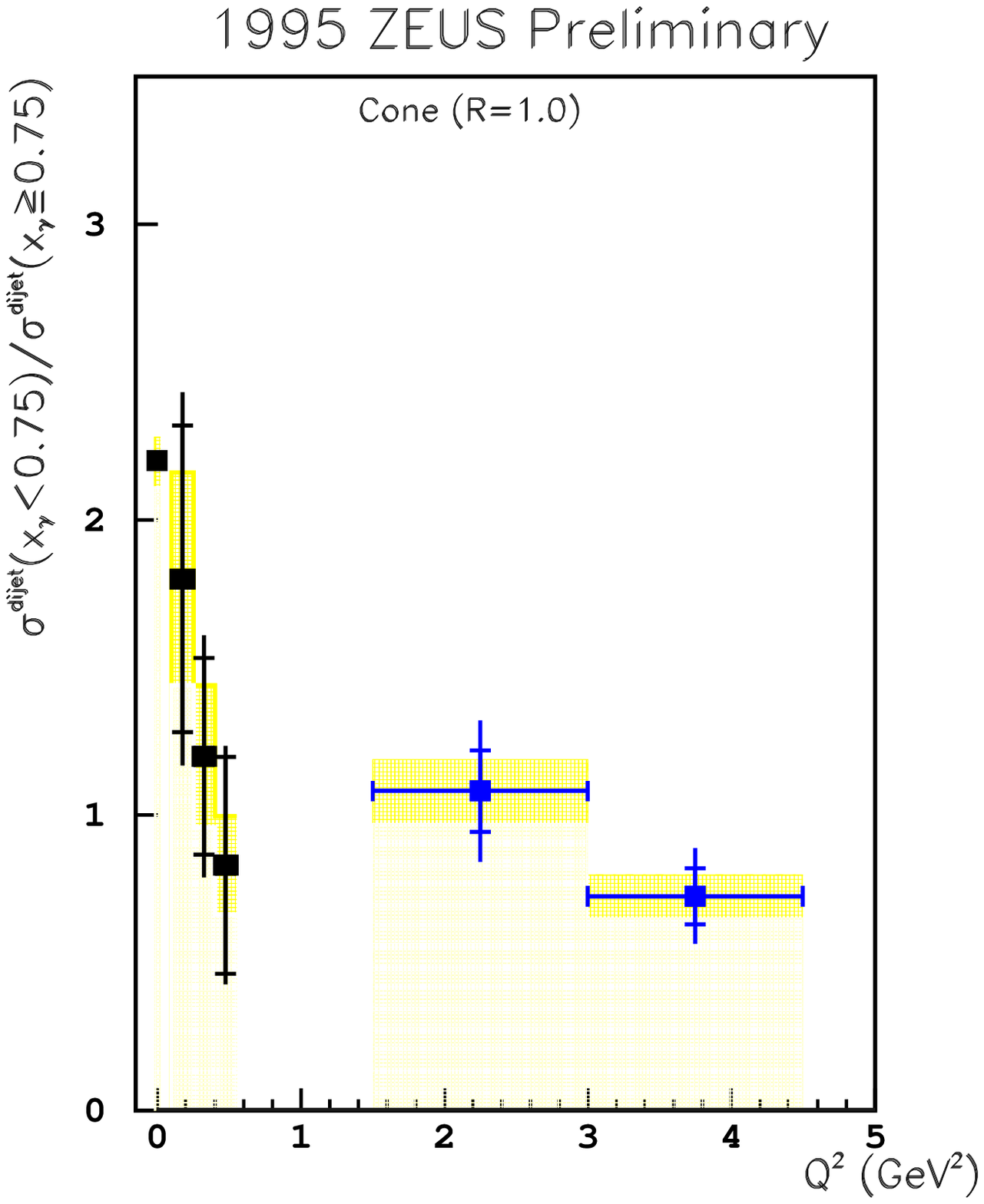}{ The ratio of
  resolved to direct photon cross sections, $\sigma_R/\sigma_D$, as a
  function of photon virtuality $Q^2$.  The shaded bands represent the
  systematic error due to jet energy scale uncertainties.}{jets-zvirtual}

The ZEUS experiment has studied dijet production in virtual
photon-proton scattering in three $Q^2$ ranges, $Q^2 \simeq 0
\gevtwo$ (untagged $\gp$), $0.1<Q^2<0.7 \gevtwo$ ($e$ tagged in the
BPC) and $1.5<Q^2<4.5 \gevtwo$ ($e$ tagged in the
UCAL)~\cite{ZEUSvirtual}.
The jets were required to have $\ETJ> 6.5 \gev$ and
$-1.125<\ETAJ<1.875$. In each $Q^2$ range the sample was divided into
events with $\xgobs > 0.75$ (direct) and $\xgobs < 0.75$ (resolved).
The ratio of resolved over direct dijet cross sections was measured as
a function of $Q^2$ and is shown in Fig.~\ref{fig:jets-zvirtual}. As
expected, the fraction of events dominated by resolved photon
processes decreases relative to the direct photon ones as a function
of $Q^2$. This trend is not reproduced by a MC model in which no $Q^2$
dependence of the parton distributions in the photon is assumed.

\epsfigure[width=0.8\hsize]{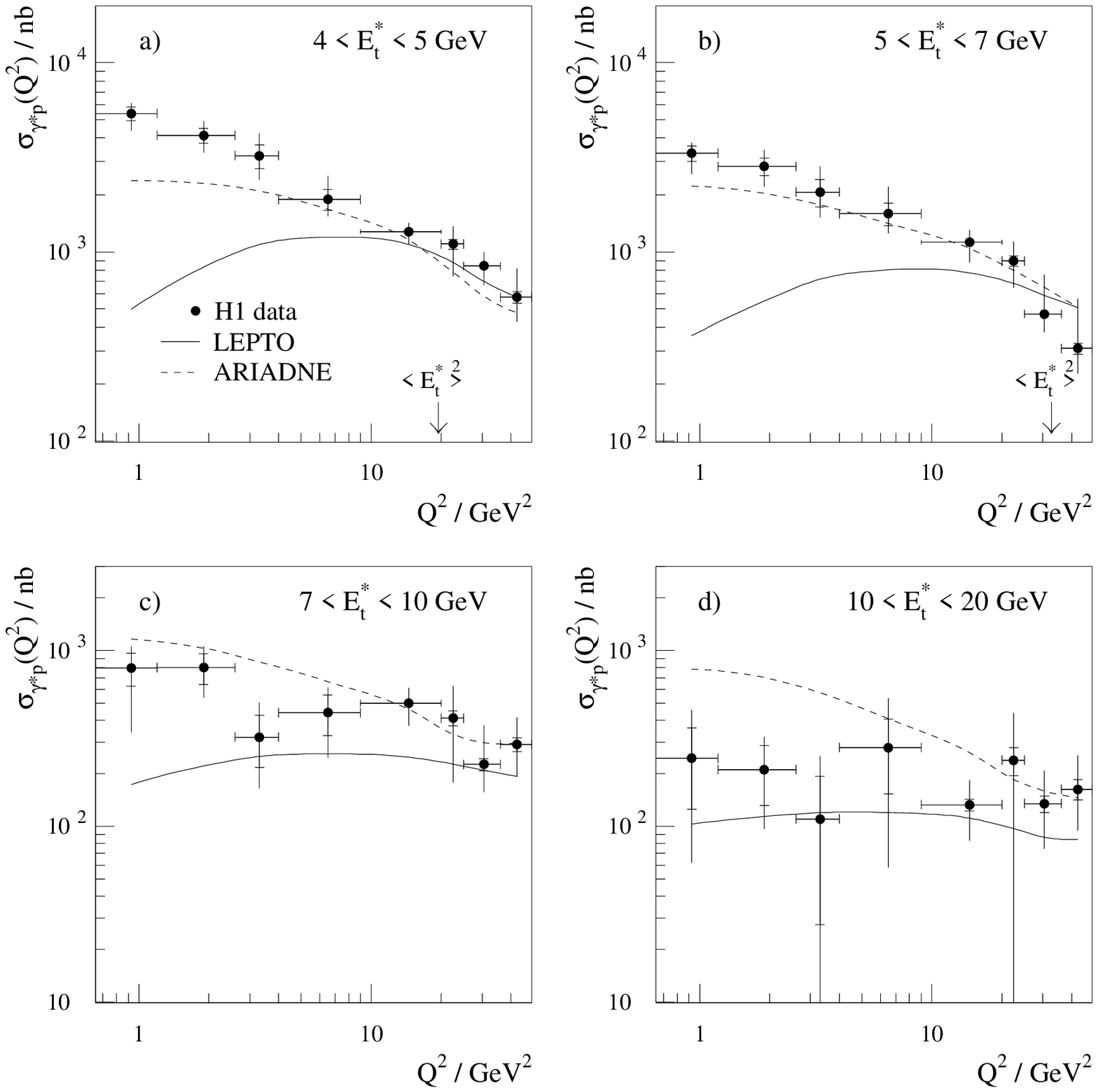} { The inclusive
  $\gamma^*p$ jet cross section $\sigma_{\gamma^*p}(Q^2)$ as a
  function of $Q^2$, for jets with transverse energy $E_T^*$, measured
  in the $\gamma^*p$ system, as denoted in the figure. The lines
  correspond to expectations from two MC models for final states in
  DIS, LEPTO (solid line) and ARIADNE (dashed
  line).}{jets-h1virtualdis}

\epsfigure[width=0.8\hsize]{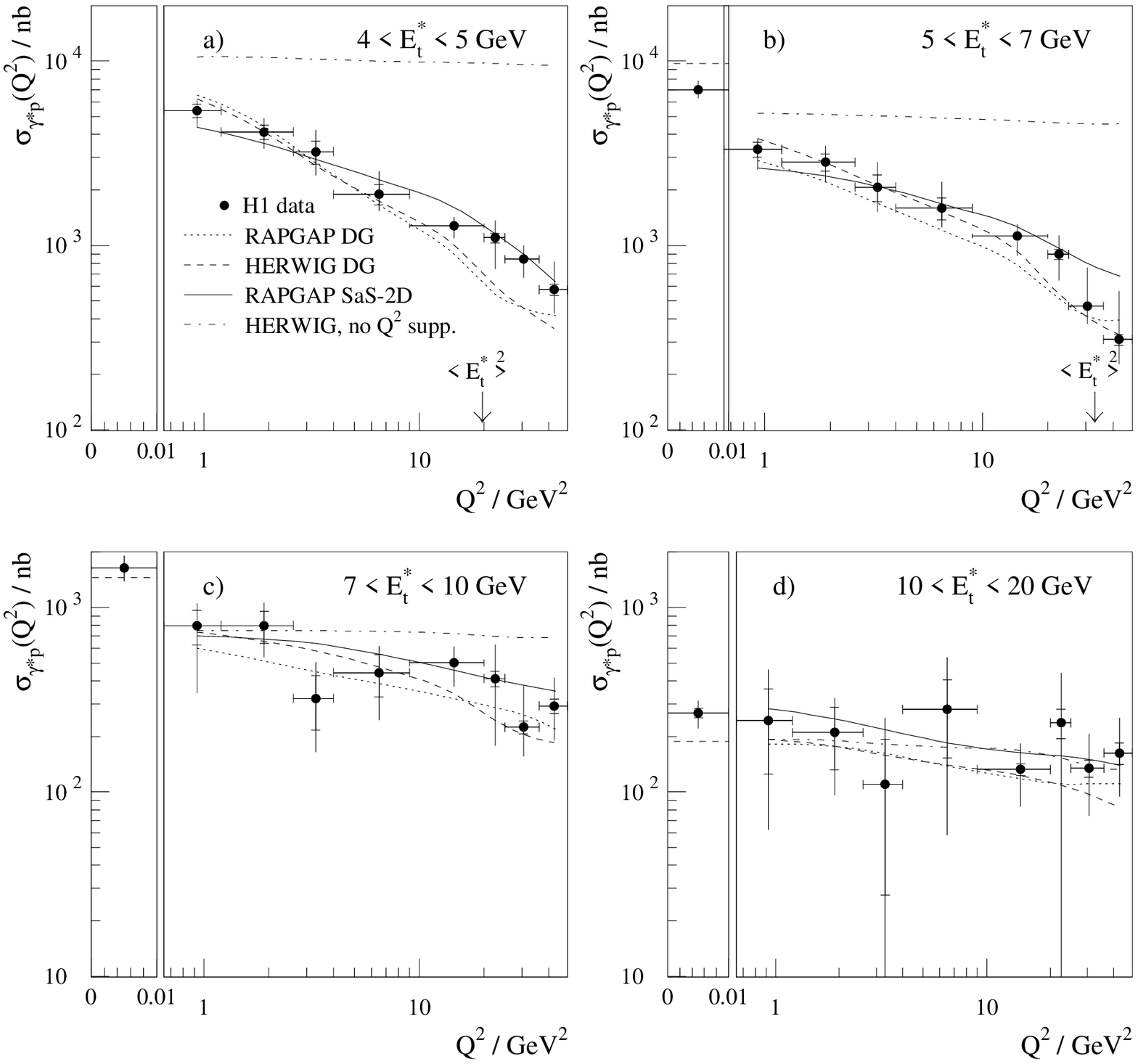}{The inclusive
  $\gamma^*p$ jet cross section $\sigma_{\gamma^*p}(Q^2)$ as a
  function of $Q^2$, for jets with transverse energy $E_T^*$, measured
  in the $\gamma^*p$ system, as denoted in the figure. The lines
  correspond to expectations from MC models (as shown in the figure)
  in which the virtual photon induces direct and resolved processes.
  Different parameterizations of the virtual photon structure function
  are shown. Also shown is the expectation if the structure of the
  virtual photon would not be suppressed with increasing
  $Q^2$.}{jets-h1virtualphp}

The H1 analysis~\cite{h1virtual} covers the range $0<Q^2<49 \gevtwo$
and $0.3<y<0.6$. The inclusive cross section for jet production with
$\ETJ>4 \gev$ was studied as a function of $Q^2$ and compared with MC
models with and without structure assigned to the virtual photon.  As
seen in Fig.~\ref{fig:jets-h1virtualdis}, models such as ARIADNE and
LEPTO, which describe well the inclusive features of deep inelastic
scattering, fail to reproduce the cross section when $(\ETJ)^2>Q^2$.
However, as shown in Fig.~\ref{fig:jets-h1virtualphp},
models~\cite{herwig,RAPGAP} in which a photon structure, dependent on
its virtuality, is assumed, reproduce the data very well.  This
observation establishes contributions to $ep$ scattering cross
section, which are not taken into account in the perturbative
expansion up to NLO.
In fact, for $Q^2>10 \gevtwo$, H1 has observed that the
production rate of dijets with $(\ETJ)^2 \geq Q^2$ is well reproduced by
QCD calculations in which the jet cross section is evaluated up to
NLO~\cite{H1r2virtual}.

In summary, HERA results have established that even a virtual photon
develops a hadronic structure when probed at a scale larger than its
virtuality.

\subsection{Jet production in deep inelastic scattering}

Jet production in deep inelastic scattering is the manifestation of
QCD beyond leading order approximation and therefore a sensitive probe
of QCD dynamics. At HERA, clean jet structures can develop thanks to
the large available phase space. There are many features of jet
production in deep inelastic scattering which are important for
understanding perturbative and non-perturbative aspects of QCD. Here
we will concentrate on those aspects which make HERA unique.

\subsubsection{Theoretical framework}

Jets in DIS result from the scattered quark and from QCD
radiation.  The rates of jet
production can be calculated in QCD and depend on the parton densities
in the proton as well as on the value of the strong coupling constant
$\as$. The pattern of radiation depends on the details of the
evolution dynamics in QCD and can be exploited in searching for
deviations from the DGLAP evolution. This is particularly important at
small $x$ when many radiation steps have to be integrated over.

\paragraph{Higher order QCD processes}

Multi-jet production rates are at the present the only features of DIS
final states that have been fully predicted in perturbative QCD. In
leading order of QCD, jet production in DIS is due to QCD Compton
scattering (QCDC) and boson-gluon fusion (BGF) (see
Fig.~\ref{fig:jets-disjets}). The corresponding diagrams are the same
as for direct photon processes (see
Fig.~\ref{fig:jets-photoprd-diag}). The cross section for the
production of jets have been calculated in QCD up to
NLO~\cite{ref:PROJET,koerner,mirkes,catanijets}.

The cross section for the production of $n$ jets plus the remnant,
$\sigma_{n+1}$, can be represented as
\begin{equation}
\sigma_{n+1}(x, Q^2; y_{\mathrm{cut}}) = \as^{n-1}(Q^2) A_n(x, Q^2;
y_{\mathrm{cut}}) +  \as^{n} (Q^2) B_n(x, Q^2;y_{\mathrm{cut}}) + \ldots \, ,
\label{eq-jets:njet}
\end{equation}
where $A_n$ and $B_n$ are calculable in terms of parton distribution
functions of the proton. Here $y_{\mathrm{cut}}$ denotes the resolution
parameter used in the jet algorithm. Measurements of jet rates provide
a method to determine the strong coupling constant.

The QCDC and BGF processes result in the presence of two jets in the
hadronic final state, in addition to the proton remnant.  At HERA, the
rate of two-jet production is used to measure $\as$ as a function of
$Q^2$.  QCDC scattering depends on the quark densities in the proton.
The BGF process is sensitive to the gluon content of the proton.
Schematically,
\begin{equation}
\sigma_{2+1}=\as(Q^2)\left[A_{_{\mathrm{QCDC}}} q(x_{q/p},Q^2) +
                           A_{_{\mathrm{BGF}}} g(x_{g/p},Q^2) \right] \, ,
\label{eq-jets:protonpdf}
\end{equation}
where $x_{q/p}$ and $x_{g/p}$ are the fractional momenta carried by
the incoming quark and gluon respectively. Since the quark
distributions are well constrained by the measurements of the $F_2$
structure function (see section~\ref{sec:structure_functions}) the
two-jet production cross section can be used to unfold the gluon
distribution in the proton.

The main uncertainty in confronting experimental data with theory stems
from hadronization effects which cannot be calculated perturbatively.

\paragraph{BFKL dynamics}

The main difference between the BFKL evolution and the DGLAP evolution
is in the pattern of radiation assumed to be dominant in parton
evolution.  The process of parton evolution can be schematically
represented by the diagram of Fig.~\ref{fig:DGLAP_BFKL}. The
distance in rapidity space of parton $i$ relative to the proton can be
expressed as
\begin{equation}
\Delta y_i = y_p - y_i = \ln \frac{2E_p}{m_p} - \ln \frac{2x_iE_p}{k_{Ti}} 
\sim \ln \frac{1}{x_i} \, ,
\end{equation}
where $E_p$ is the energy and $m_p$ the mass of the proton. Here $x_i$
denotes the fraction of the proton momentum carried by the $i$-th
parton and $k_{Ti}$ is its transverse momentum relative to the proton.

The evolution equations derived in perturbative QCD are meant to
describe the $Q^2$ dependence of global quantities such as the DIS
cross section.  It is however believed that also the final states will
remember the radiation pattern which leads to evolution. This is
introduced in MC models such as LEPTO~\cite{lepto} or
HERWIG~\cite{herwig}.

In the DGLAP evolution, the strong ordering in transverse momenta
(relative to the proton beam) $k_{T1}<k_{T2}< \ldots <k_{Tn} \simeq Q$
of the subsequent radiation requires that partons with the highest
$k_T$ be radiated close to the photon absorption point. The QCDC and
BGF processes are the results of such an evolution.  Close to the
proton, the partons are expected to have small $k_T$.  In the BFKL
evolution, no ordering in $k_T$ is implied and one expects an enhanced
parton activity in the central and forward region, between the current
system and the proton remnant.

Assuming parton-hadron duality~\cite{khoze}, deviations from the DGLAP
evolution would result in an increase of the transverse energy in the
central rapidity range~\cite{bfkl-et1,bfkl-et2}.  It should also be
possible to observe jets with $k_t \simeq Q$ for large
$x_i$~\cite{mueller-jets,tang-jets,bartels-jets}, which are not
allowed in the DGLAP evolution.  In fact the rate of such jets, if
they exist, should be enhanced for small Bjorken $x$ since in the BFKL
approach, at least to leading order, the cross section is expected to
rise very fast ($\sigma \sim (x/x_i)^{-0.5}$)~\cite{bartels-bfkl}.

To summarize, the footprint of BFKL dynamics should manifest itself in
partons in the final state with transverse momenta larger than
expected from the DGLAP evolution. Assuming duality between partons
and hadrons the same effects should be present in the hadronic final
state.  Furthermore the rate of jet production with large fraction of
the proton momentum should be enhanced for small-$x$ DIS.  

One of the problems encountered in looking for BFKL symptoms in the
final states is the lack of MC models for hadron production which
incorporate fully the BFKL dynamics. The predictions that exist are
calculated at the parton level and the effect of hadronization, which
may be strong, precludes direct comparisons with BFKL based
calculations.

The data are usually compared to three MC models of hadronic final
states in DIS. Leading log DGLAP parton showers are implemented in the
LEPTO~\cite{lepto} and HERWIG~\cite{herwig} generators. The main
difference in the two generators is in the hadronization stage.  The
ARIADNE generator~\cite{ref:ARIADNE_2} parton showers are generated as a
result of radiation from color dipoles formed by the color charges
after the interaction. The gluons emitted by the dipoles do not obey
strong ordering in $k_T$. All of these MC generators provide a
satisfactory overall description of current DIS final state
data~\cite{carli}.

\subsubsection{Selection of HERA results}

From the multitude of studies which are carried out for the hadronic
final states in DIS~\cite{kuhlen}, we report here on the study of
jets shapes, the determination of $\as$ and on the search for the BFKL
dynamics.

\paragraph{Jet shapes}

The internal structure of jets is expected to depend on the type of
primary parton, quark or gluon, from which the hadronic jet evolved.
The formation of a jet from a parton is driven mainly by gluon
emission. As a consequence of a larger gluon-gluon coupling strength
compared to that of quark-gluon coupling, the gluon jets are expected
to be broader than quark jets. This expectation can be tested by
comparing jet shapes in different reactions in which in the final
state jets are predominantly initiated by quarks or gluons.

At HERA, in deep inelastic neutral and charged current scattering,
the dominant jet production mechanism is that of one quark jet. The
same is true for $e^+e^-$ annihilation. In $p\bar{p}$ and $\gp$
scattering gluon initiated jets are expected to give a substantial
contribution. 

Most suitable for jet shape studies is the cone algorithm in which
relatively little is assumed about the energy flow within a cone of
radius $R=1$ in the pseudo-rapidity $\eta$ and azimuthal angle $\phi$
space. 

The integrated jet shape, $\psi(r)$ is defined by
\begin{equation}
\psi(r)=\frac{1}{N_{jets}}\sum_{jets}\frac{E_T(r)}{E_T(R)} \, ,
\label{eq-jets:jetshape}
\end{equation}
where $E_T(r)$ denotes the transverse energy contained in a cone of
radius $r$ around the axis of the jet and $N_{jets}$ is the total
number of jets in the sample. By definition $\psi(R)=1$.

\epsfigure[width=0.8\hsize]{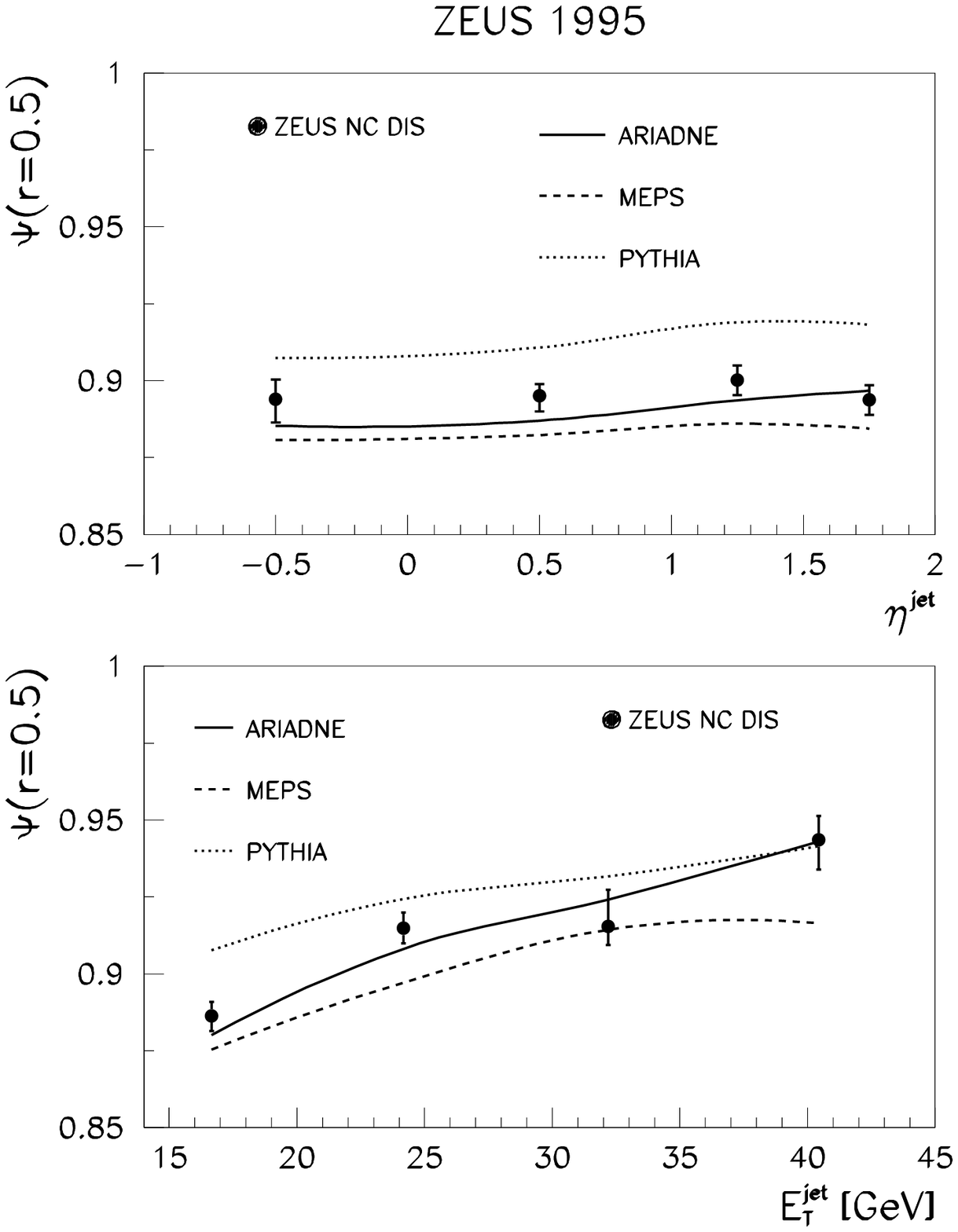} {Integrated jet shape
  $\psi$ at a fixed value of $r=0.5$, as a function of $\ETAJ$ (upper
  plot) and $\ETJ$ (lower plot) in neutral current DIS at $Q^2>100
  \gevtwo$. The expectations of different MC models are shown for
  comparison.  } {jets-jetshapenc}

\epsfigure[width=0.8\hsize]{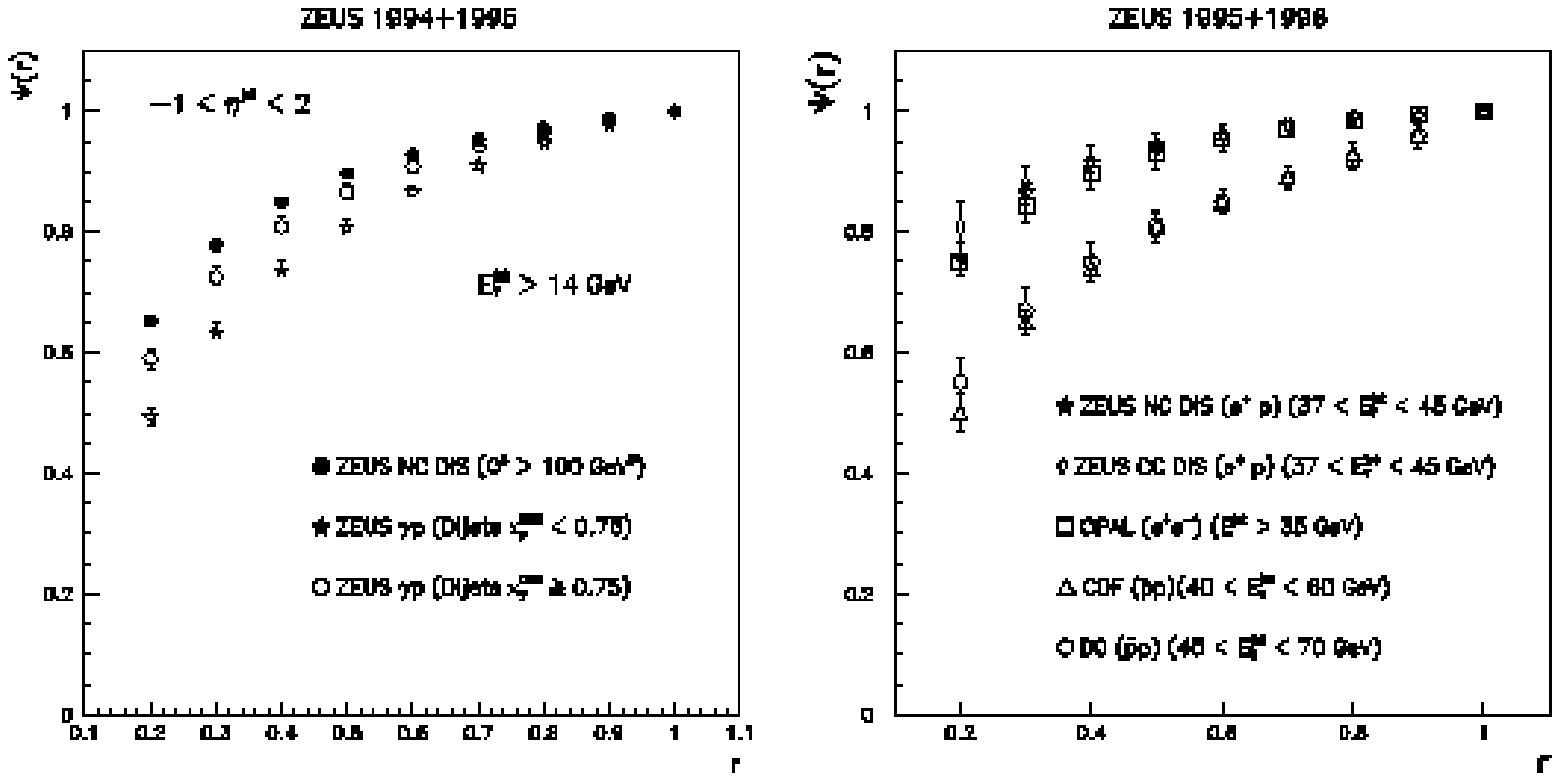} { Integrated jet
  shape $\psi$ as function of the radius of integration $r$, in
  neutral current DIS at $Q^2>100 \gevtwo$ compared to jet shapes in
  direct and resolved photoproduction (left plot).  Comparison between
  jet shapes in neutral current, charged current DIS and $e^+e^-$ and
  $p\bar{p}$ interactions (right plot).  } {jets-jetshapes}

The integrated jet shape $\psi(0.5)$ for jets found in neutral current
DIS at $Q^2>100 \gevtwo$, with $E_T^{jet}>14 \gev$ and
$-1<\eta^{jet}<2$~\cite{zeusjetshape}, is shown in
Fig.~\ref{fig:jets-jetshapenc}. The jet shape does not depend on
$\eta^{jet}$, but the jets become more collimated as $E_T^{jet}$
increases. For jets with $E_T^{jet} \simeq 40 \gev$, almost $95\%$ of
the transverse energy is contained within a cone of radius $r=0.5$.

When compared to jets seen in hard photoproduction, the NC jets are
found to be more collimated as shown in
Fig.~\ref{fig:jets-jetshapes}. The difference between jet shapes in
processes dominated by direct photon and resolved photon processes
is also apparent in the figure. The latter are dominated by production
of gluon jets. 

In Fig.~\ref{fig:jets-jetshapes} the jet shapes are compared for
various processes. The jets in NC, CC DIS and in
$e^+e^-$~\cite{opaljetshape} interactions, which are predominantly
initiated by quarks are less broad than jets with similar $E_T^{jet}$
measured in $p\bar{p}$ experiments~\cite{cdfjetshape,d0jetshape}.

These observations confirm the expectation that gluon initiated jets
are broader than quark initiated jets. The similarity in the jet
shapes also indicates that the pattern of QCD radiation close to the
primary parton is to a large extent independent of the hard scattering
process.

\paragraph{Measurements of $\as$}

The strong coupling constant $\as$ is a fundamental parameter of the
Standard Model and the measurement of its value in different
reactions, sensitive to different aspects of the theory,
is of utmost importance.  At HERA, $\as$ can be extracted in several
ways, as outlined below.

The decrease in the quark densities at large $x$ with increasing scale
$Q^2$ is understood in QCD as due to the radiation of gluons from
the quarks.  A measurement of the scaling violations therefore yields 
information on $\as$.  Fixed target DIS experiments
have extracted $\as$ via this method and have achieved very precise
results ($\delta_{exp}\as =0.003$).  At small $x$, where HERA
structure function measurements have been performed,
the scaling
violations depend not just on $\as$ but also on the gluon density in
the proton (the number of quarks increase due to the process
$g \rightarrow q\bar{q}$).  
A combined analysis is quite difficult, and will require
much larger data sets and considerably smaller systematic errors than
are currently available.  The extraction of $\as$ at large-$x$
values is statistically limited.

It should be noted that $\as$ can be extracted in the context of
different QCD based models.  For instance, in the double asymptotic scaling
approach~\cite{ref:DAS1,ref:DAS2}, $\as$ appears as a parameter but the gluon
density does not.  A fit to the H1 data~\cite{ref:H1_DAS_alphas} 
with this model yields
\begin{equation}
\as(M_Z) = 0.113 \pm 0.002 \pm 0.006 \; .
\end{equation}
This method yields very small errors, but the theoretical uncertainties
must still be worked out.

Dijet production in DIS proceeds in LO via QCDC scattering or
BGF, as depicted in Fig.~\ref{fig:jets-disjets}.  The matrix
elements for these processes have been calculated in NLO, and the rates
for dijet events can be compared to expectations to extract $\as$.
It is imperative to use calculations at least at NLO to reduce uncertainties
due to different scales (factorization, renormalization).

A measurement of the rate of dijet events to single jet events as a
function of $y_{\mathrm{cut}}$, the resolution parameter in the jet
definition, is sensitive to the value of $\as$ (see
Eq.~(\ref{eq-jets:protonpdf})). In NLO the dijet rate can be expressed
as
\begin{eqnarray}
R_{2+1}(Q^2,y_c) &=& \frac{\sigma_{2+1}(Q^2,y_c)}{\sigma_{tot}(Q^2)} \\
                 &=& a(Q^2,y_c)\as(Q^2) + b(Q^2,y_c)\as^2(Q^2) \, ,
\end{eqnarray}
where $y_c=y_{\mathrm{cut}}$ and the coefficients $a$ and $b$ are given by
perturbation theory.

The measurements at HERA have been performed with the JADE algorithm
for finding jets.  Jets are found in the detector, and must then be
corrected to the parton level to allow comparisons to NLO
calculations. For this to be possible, different Monte Carlo event
simulation programs are used to evaluate the effect of hadronization.
In principle, the measurement should be performed at small $Q^2$ as
$\as$ decreases with $Q^2$.  However, given the fixed $W$ range which
is accessible to the detectors, small $Q^2$ corresponds to small $x$.  It
is found that the small-$x$ region must be avoided for the measurement
to be reliable.  There are two principle reasons for this: small-$x$
events can have considerable QCD radiation in the initial state which
can lead to unwanted jets at the detector level, and the parton
densities at small $x$ are not well known, leading to a large
uncertainty in the measurement.  Given these constraints, the
measurements have been performed at relatively large $Q^2$ and $x$,
where these uncertainties are small.

\epsfigure[width=0.8\hsize]{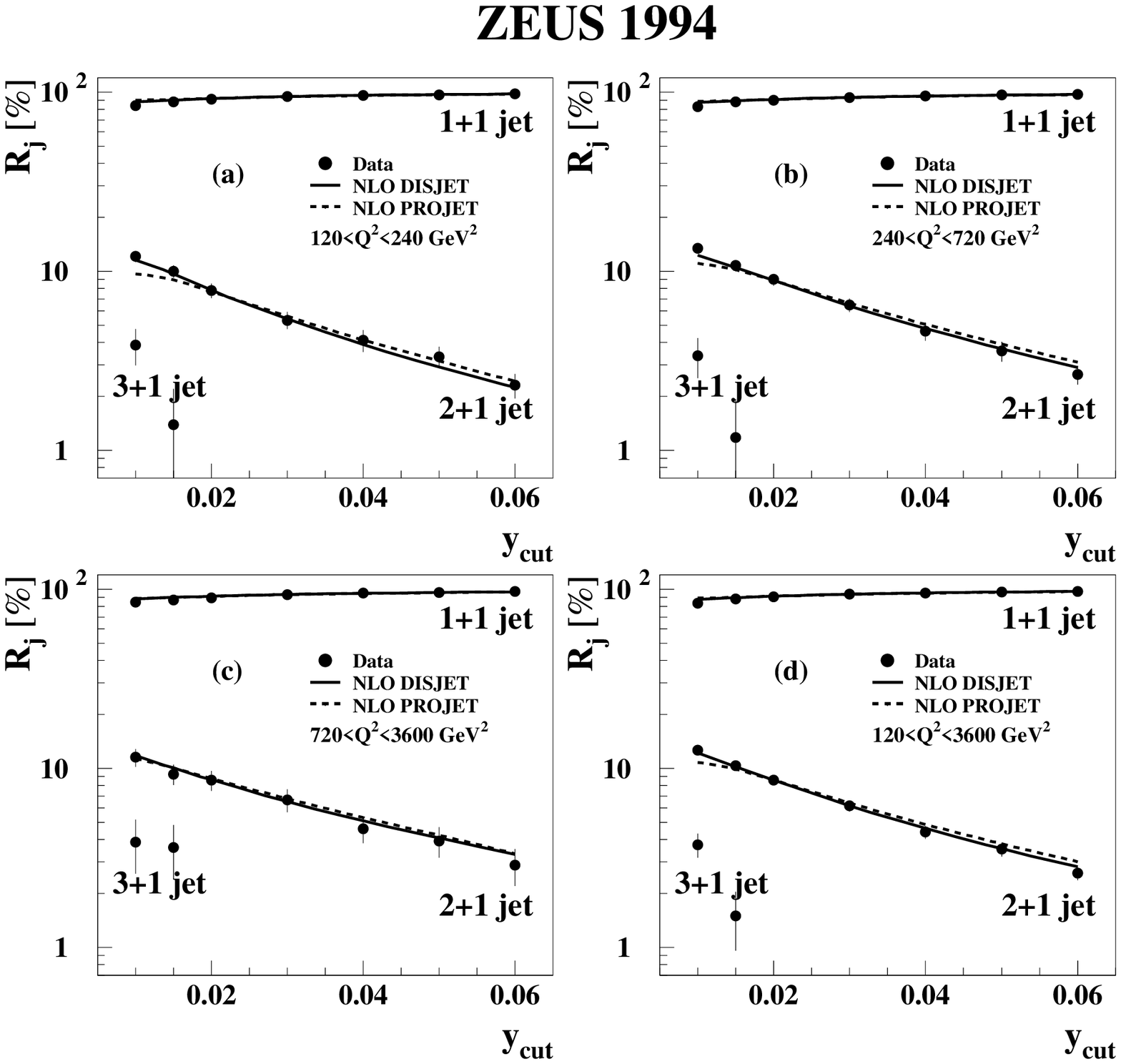} { Jet production
  rates $R_j$ as a function of the jet resolution parameter \ycut\ for
  $Q^2$ in the range (a) $120<Q^2<240$~GeV$^2$, (b)
  $240<Q^2<720$~GeV$^2$, (c) $720<Q^2<3600$~GeV$^2$, and (d)
  $120<Q^2<3600$~GeV$^2$.  Only statistical errors are shown.  Two NLO
  QCD calculations, DISJET and PROJET, each with the value of
  $\Lambda_{\overline{MS}}$ obtained from the fit at \ycut=0.02, are
  also shown.}{jetrates}

The ZEUS measured rates~\cite{ref:ZEUS_alphas} of $1+1$, $2+1$ and
$3+1$ jet events are displayed in Fig.~\ref{fig:jetrates} as a
function of the value of $y_c$.  The results are compared to NLO
calculations.  The NLO calculations must be performed in the same
kinematic range and with the same cuts as used by the experiments, and
are done via the Monte Carlo integration programs
DISJET~\cite{ref:DISJET} and PROJET~\cite{ref:PROJET}.  As can be seen
in the figure, there is good agreement between these calculations and
the measured jet rates.  The value of $\as$ is then determined by
varying $\Lambda_{\overline{MS}}^{(5)}$ (corresponding to five active
flavors) in the QCD calculations until the best fit is achieved for
the ratio $R_{2+1}$.

\epsfigure[width=0.8\hsize]{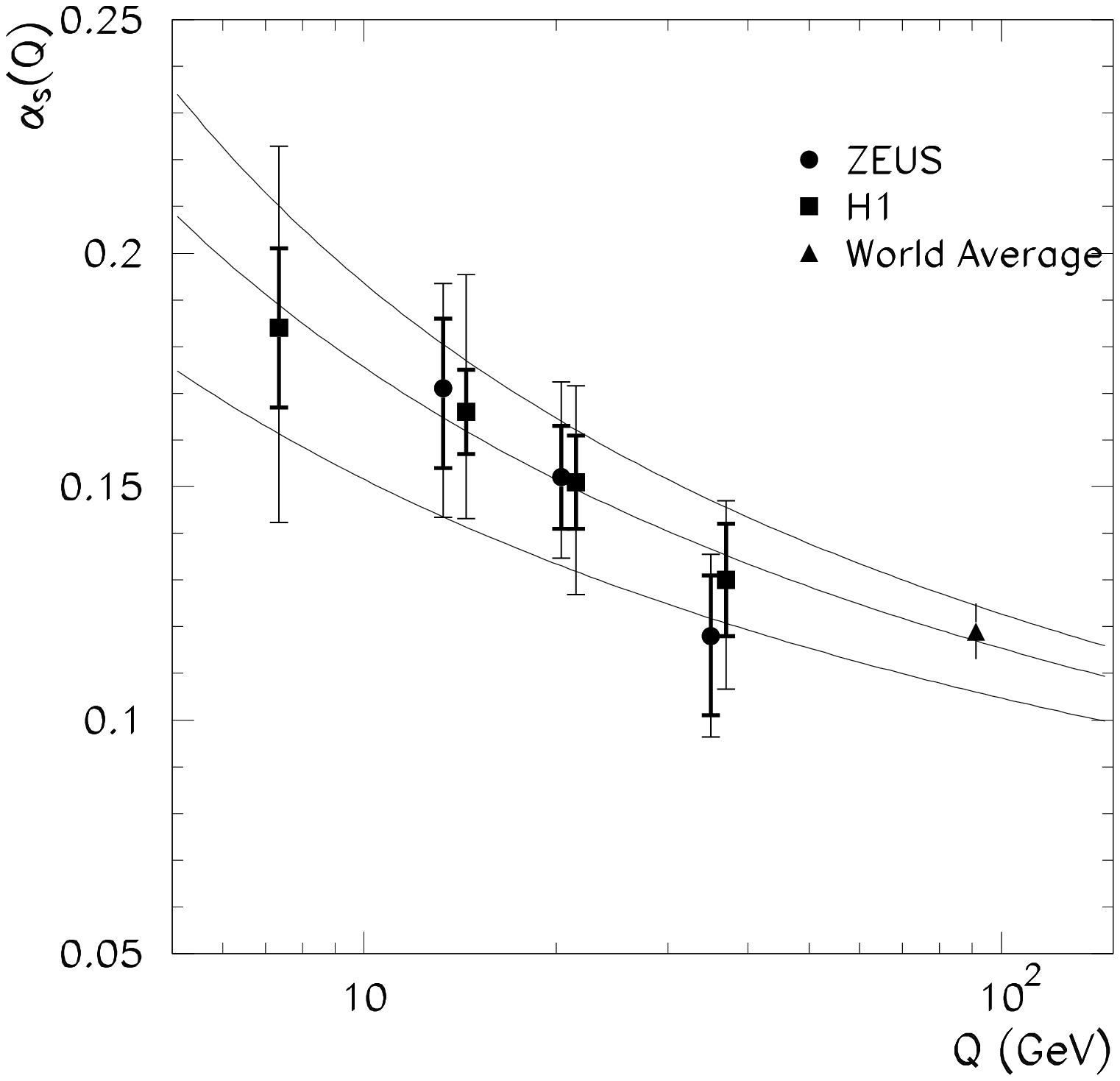} { Measured values of
  $\as(Q)$ at HERA.  The statistical error corresponds to the inner
  bar and the outer bar reflects the statistical and systematic
  (including theoretical) error added in quadrature.  Note that the
  systematic errors are strongly correlated.  The curves represent
  $\as$ expectations with $\Lambda^{(5)}_{\overline{MS}} = 100$ (lower
  curve), 200 and 300~MeV (upper curve).  The triangle at $Q=M_Z$
  indicates the world average from 1997~\protect\cite{Bethke} $\as
  (M_Z^2)=0.119 \pm 0.006$. }{alphas}

The extracted values of $\as$ from ZEUS and the most recent H1
measurements~\cite{ref:H1_alphas_97} are displayed in Fig.~\ref{fig:alphas}.
The values from the two experiments are in good agreement.  The value of
$\as$ extrapolated to $Q^2=M_Z^2$ are
\begin{eqnarray}
\as^{ZEUS}(M_Z) & = & 
0.117 \pm 0.005^{+0.004}_{-0.005} \pm 0.007  \, ,\\
\as^{H1}(M_Z) & = & 
0.117 \pm 0.003^{+0.009}_{-0.013} \pm 0.006 \, ,
\end{eqnarray}
where the third error in the ZEUS case corresponds to the theoretical error,
while the third error in the H1 case represents the error from the 
recombination scheme used in the jet finder.  The theoretical error has
been included by H1 in the second error term.

These measurements are compared to the world average 
value $\as=0.119\pm0.006$
~\cite{Bethke} in Fig.~\ref{fig:alphas}, and good agreement
is found.  The systematic errors are dominated on the experimental side by 
the hadronic energy scales of the calorimeters, and on the theoretical
side by the scale dependence and hadronization model used.  There is
certainly some room for improvements in these areas in the future.

\paragraph{Search for BFKL dynamics} 

Observation of an excess of transverse energy in the central rapidity
region of hadronic final states~\cite{h1diset}, compared to DGLAP
based MC models, was not conclusive. The perturbative phase could be
compensated by hadronization
effects~\cite{edin,ingelman-sci2}.  However studies of MC
generators showed that the production of high transverse momentum
particles from hadronization is suppressed, while it is sensitive to
parton radiation.

\epsfigure[width=0.8\hsize]{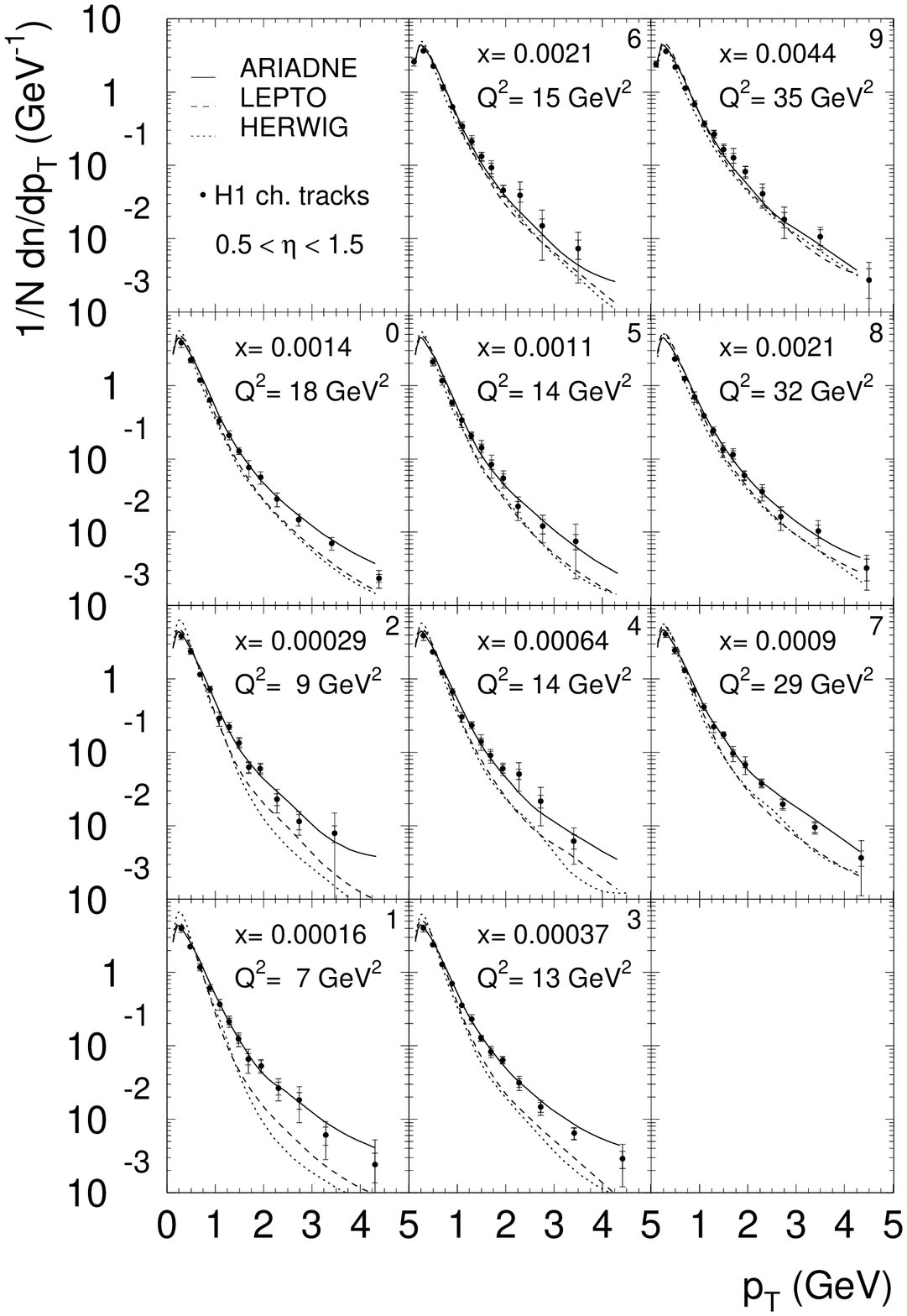} { The $x$ and $Q^2$
  dependence of the transverse momentum, $p_T$, for charged particles.
  The measurements are performed in the $\gamma^*p$ center-of-mass
  system in the pseudorapidity interval $0.5<\eta<1.5$. The combined
  sample is shown in bin 0. For comparison expectations of different
  MC models are overlaid. }{jets-ptspectrum}

The H1 experiment has measured the $p_T$ spectra of charged particles
as a function of $x$ and $Q^2$~\cite{h1ptcharged}. The spectra were
then compared to the expectations of MC generators. The results, as a
function of $x$ and $Q^2$, for particles with $0.5<\eta<1.5$ in the
center-of-mass system, are shown in Fig.~\ref{fig:jets-ptspectrum}.
For $x>0.0011$ a tail of large $p_T$ particles is observed which
cannot be accommodated by MC generators based on DGLAP parton showers
while it is well reproduced by the ARIADNE MC with no implicit $k_T$
ordered parton radiation.  This is an indication that in part of the
phase space, the pattern of parton radiation does not match that of
the DGLAP evolution. 

\epsfigure[width=0.8\hsize]{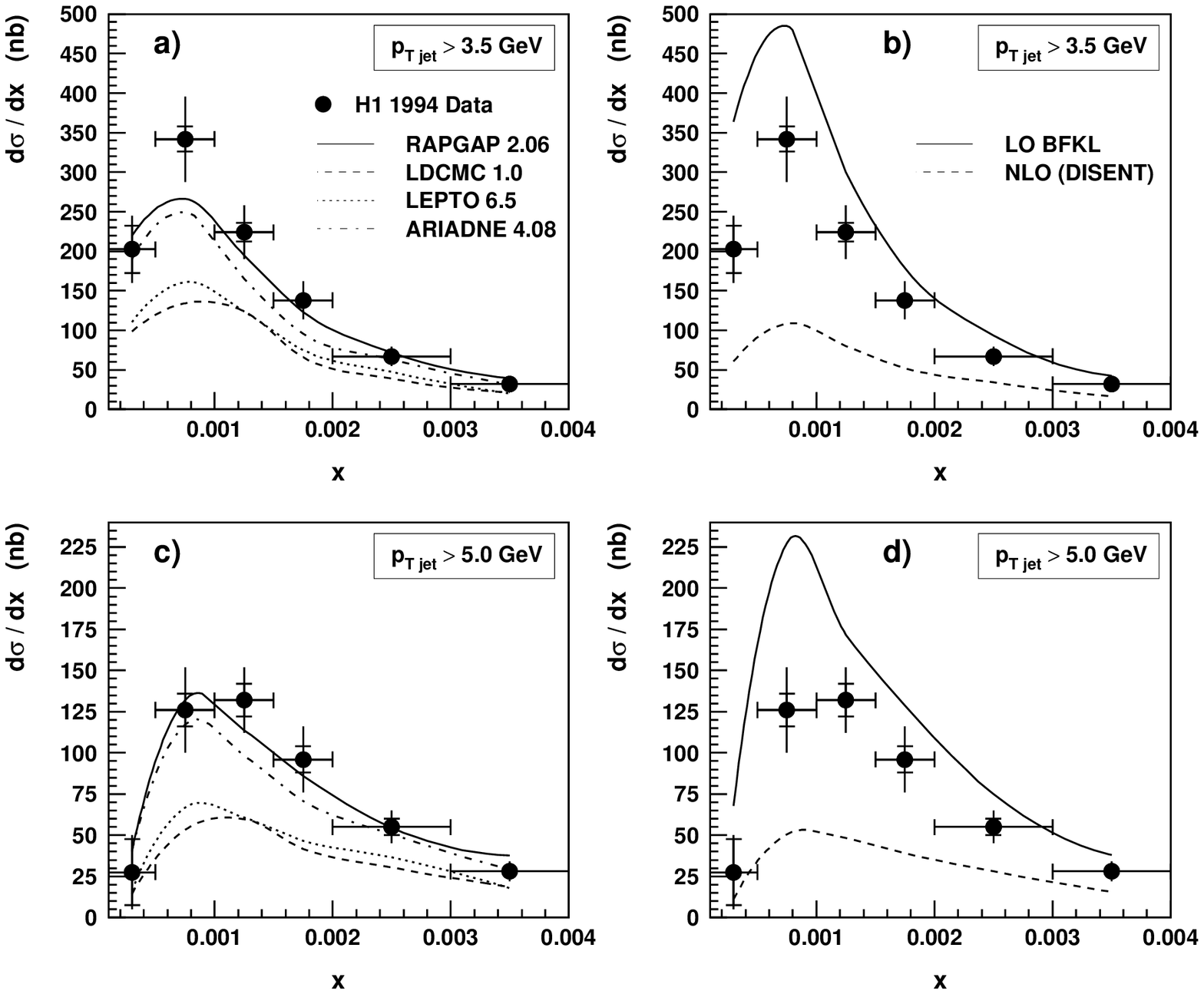} { Forward jet
  production cross section as a function of $x_{Bj}$, for jets with
  $x_{jet}>0.035$, $0.5<p_T^2/Q^2<2$ and $p_T^{jet}>3.5 (5) \gev$, as
  measured by the H1 experiment. In (a) and (c) the measured cross
  sections, corrected to the hadron level, are compared to
  expectations of ARIADNE (CDM), LEPTO (MEPS),
  LDCMC~\protect\cite{ldcmc} and RAPGAP (with resolved $\gvp$) MC
  models.  In (b) and (d) the same cross sections are compared to
  perturbative QCD calculations (see text for details).}
{jets-fwdjets}

A further evidence that leading order DGLAP parton showers are not
sufficient to describe the final states was obtained by studying the
cross section for forward jet production~\cite{h1fwdjets,zfwdjets}.
The jets with energy $E_{jet}$ and transverse momentum $p_T$ are
required to have $p_T>3.5 (5)\gev$ in H1 and $p_T>5 \gev$ in ZEUS,
$x_{jet}=E_{jet}/E_p > 0.035$ and $0.5<p_T^2/Q^2<2$. The cross section
as a function of Bjorken $x$ is then compared to MC generators and
numerical calculations (see for example Fig.~\ref{fig:jets-fwdjets}).
Among the standard DIS generators only ARIADNE describes the data
adequately.  An even better description of the data is obtained with
the RAPGAP MC which includes the resolved contribution of the virtual
photon.  The numerical calculations based on LO BFKL dynamics
overestimate the cross section at small $x$~\cite{bartels-bfkl}, while
NLO calculations underestimate it~\cite{catanijets,mirkes}.

The results of these studies cannot be taken as proof for the BFKL
dynamics. They show however that higher order QCD effects, which are
not important in the total DIS cross section measurements, may be
important for the description of the hadronic final state. One
possible way of modeling these higher order effects is by introducing
contributions from a virtual resolved photons~\cite{h1virtual}.
This would lead to enhanced parton activity and the production of
forward jets with $p_T > Q$.

\subsection{Summary}

Jet production in $ep$ scattering is an important laboratory for
studying the properties of perturbative QCD in a multi-partonic
environment. First indications for the importance of higher order QCD
effects have been established. A full understanding of HERA results
still requires more theoretical and experimental investigations.


\section{Diffractive hard scattering}
\label{sec:diffraction}

One of the surprises which came along with the first results from HERA
was the observation of DIS events with a large rapidity gap in the
hadronic final state, located between the photon and the proton
fragmentation regions~\cite{ZEUSdiff1,H1diff1}.  The observed fraction
of events with large rapidity gaps is of the order of $10 \%$, fairly
independent of $W$ and $Q^2$.  In QCD, the fragmentation process
driven by parton radiation leads to an exponential suppression of
large rapidity gaps between hadrons~\cite{basics_pqcd}. Large rapidity
gap formation, with little dependence on $W$, is typical of
diffractive scattering and invokes the notion of the pomeron. The fact
that the diffractive exchange can be probed with large $Q^2$ virtual
photons means that its structure can be studied much the same way as
the partonic structure of the proton.

In hadron-hadron scattering, one usually identifies three types of
interactions which differ in the characteristics of the final states.
In elastic scattering, both hadrons emerge relatively unscathed and no
other particles are produced. In diffractive scattering, the energy
transfer between the two interacting hadrons remains small, but one
(single dissociation) or both (double dissociation) hadrons dissociate
into multi-particle final states preserving the  quantum
numbers of the initial hadrons. The remaining configurations
correspond to so called inelastic interactions.

Characteristic of high energy elastic and diffractive scattering is the
exponential behavior of the cross section as a function of the square
of the momentum transfer, $t$, between the two hadrons. As this
property is reminiscent of the diffraction of light by a circular
aperture, diffractive scattering in strong interactions derives
its name from the optical analogy~\cite{amaldi}.  

Since the slope of the $t$ distribution is a measure of the radius of
interaction, diffractive scattering is inherently related to the
properties of the exchanged fields in strong interactions.
Understandably, the appearance of diffractive phenomena in the
presence of a large scale has steered a lot of renewed interest in
diffraction and its interpretation in the language of the QPM and QCD. We
will concentrate on single diffractive dissociation
phenomena and, in particular, on the dissociation of real and virtual
photons on a proton target in the presence of at least one large
scale.

\subsection{Diffractive scattering in soft interactions}

We will start with a short overview of what is known about diffractive
scattering in soft hadron-hadron interactions and with a review of the
Regge phenomenology which is used in the description of diffractive
processes. 

\subsubsection{Properties of diffractive scattering}

There is no precise definition of diffraction.  In what is called
single diffraction, where only one of the interacting hadrons
dissociates, one expects that the diffracted state preserves the
intrinsic quantum numbers of the initial hadron. However, this does
not have to be the case for the spin and parity as the momentum
transferred in the interaction may affect the internal motion of the
hadron while the total angular momentum is preserved. An additional
constraint is imposed by the fact that the 'absorbing' hadron should
preserve its identity - this leads to the so called coherence
condition which limits the invariant mass of the diffracted state. For
a particle of mass $m$ and momentum $p$ to diffract on a stationary
target into a state of mass $M_X$, the minimum momentum transfer square
required is
\begin{equation}
t_{\rm min} = \left( \frac{M_X^2 - m^2}{2p} \right)^2 \, .
\end{equation}
From the uncertainty principle, the coherence of the absorbing hadron
of radius $R$ will be preserved if
\begin{equation}
(M_X^2-m^2) \lesssim \frac{2p}{R} \, .
\end{equation}
Based on the above considerations, diffractive processes are expected
to have the following key features:
\begin{enumerate}
\item the differential cross section has a pronounced forward peak
  (with a distribution $\sim e^{bt}$) with a slope which is related to
  the typical size of hadrons (for $R=1$ fm the slope
  $b \simeq 8$~GeV$^{-2}$),
\item the diffracted final state will be well separated in phase space
  from the target, because of the large forward peak required by the
  first condition,  
\item the incoming momentum has to be large enough to allow coherence
  over the dimensions of the stationary target hadron,
\item large mass diffractive states will be suppressed. To preserve 
coherence $M_X^2/s \lesssim 0.15$, where $s=2mp$ is the centre of mass
energy squared.
\end{enumerate}
In addition, the energy dependence of the diffractive processes is
expected to be similar to that of the inelastic cross section
(diffraction is a shadow of inelastic interactions).

Experimentally, the double differential cross section $d^2\!\sigma/dt
dM_X^2$ is studied as a function of the centre of mass energy squared
$s$ and for different interacting hadrons. A detailed review of
experimental data is beyond the scope of this paper, but can be found
in~\cite{goulianos1,goulianos2}. The experimentally established
properties of single diffraction are in line with the expectations:\\
$\bullet$ For fixed values of $s$, the $t$ distribution is well
described by a functional form $e^{bt}$ for $t < 0.2 \gevtwo$. The
value of $b$ depends on the interacting hadrons and the nature of the
interaction.  Typical values of $b$ for single diffractive
dissociation off a proton target are $b\simeq 7 \gevmtwo$.  In the
mass region dominated by resonance production, $b$ is typically larger
and mass dependent.  The value of $b$ is found to increase
logarithmically with $s$ - the
forward diffractive peak shrinks with energy.\\
$\bullet$ The mass distribution, above the resonance region and
integrated over $t$, follows a $1/M_X^2$ distribution.\\
$\bullet$ The total single diffraction cross section is a slow
function of $s$.\\
$\bullet$ The ratio of the diffractive cross section to the total
cross section was found to be fairly independent of the dissociating
hadron for the same target hadron suggesting factorization properties.

\subsubsection{Pomeron exchange and triple Regge formalism}

As mentioned in the introduction~(see~\ref{sec:reggeology}), the
energy dependence of the cross sections for hadron-hadron interactions
is well described by assuming that the interaction is due to the
exchange of Regge trajectories, of which the pomeron, with quantum
numbers of vacuum, dominates at high energy~\cite{ref:DoLa_sigfit}.
Many properties of diffractive scattering, in particular the observed
factorization properties, find a natural explanation if one assumes
that diffraction is mediated by the exchange of a universal pomeron
trajectory with a coupling strength depending on the interacting
hadrons. The universality of the exchanged trajectory has been
proposed originally by Gribov and Pomeranchuk~\cite{gribovfac} and is
usually referred to as Regge factorization.

Regge theory supplemented by Regge factorization provides a framework
on the basis of which many features of high energy hadronic
interactions find a simple explanation. The total, elastic and single
diffractive cross sections are expressed in terms of Regge
trajectories, $\alpha_i(t)$, and their couplings to hadrons,
$\beta(t)$, called residue functions. In the so called Regge limit
(for $t\ll M_X^2 \ll s$), the following formulae hold for
the interaction of hadrons $a$ and $b$:
\begin{eqnarray}
\sigma^{ab}_{\mathrm{tot}}&=& \sum_k \beta_{ak}(0) \beta_{bk}(0)
s^{\alpha_k(0)-1}  \label{eq-diff:reggetot} \\
\frac{d\sigma^{ab}_{\mathrm{el}}} {dt}&=&\sum_k \frac{\beta_{ak}^2(t) 
\beta_{bk}^2(t)} {16\pi} s^{2(\alpha_k(t)-1)}  \label{eq-diff:reggeel} \\
\frac{d^2\sigma^{ab}_{\mathrm{diff} } } {dt dM_X^2}&=&\sum_{k,l} 
\!\frac{\beta_{ak}^2(t) \beta_{bl}(0) g_{kkl}(t)} {16\pi} \frac{1}{M_X^2} 
\left(\!\!\frac{s}{M_X^2}\!\!\right)^{2(\alpha_k(t)-1)}
\!\!\!(M_X^2)^{\alpha_l(0)-1}.
\label{eq-diff:3regge}
\end{eqnarray}
The last formula~(\ref{eq-diff:3regge}) is based on Mueller's
generalization of the optical theorem~\cite{mueller-3reg}, which
relates the total cross section of two-body scattering with the
imaginary part of the forward elastic amplitude ($ab \rightarrow ab$),
to the case of three body scattering.  It can be extended to any
inclusive process of the type $a+b\rightarrow c+X$.  The term
$g_{kkl}$ is called the triple-Regge coupling.  The diagrams
corresponding to the expressions above are presented in
Fig.~\ref{fig:diff-regge}.

\epsfigure[width=\hsize]{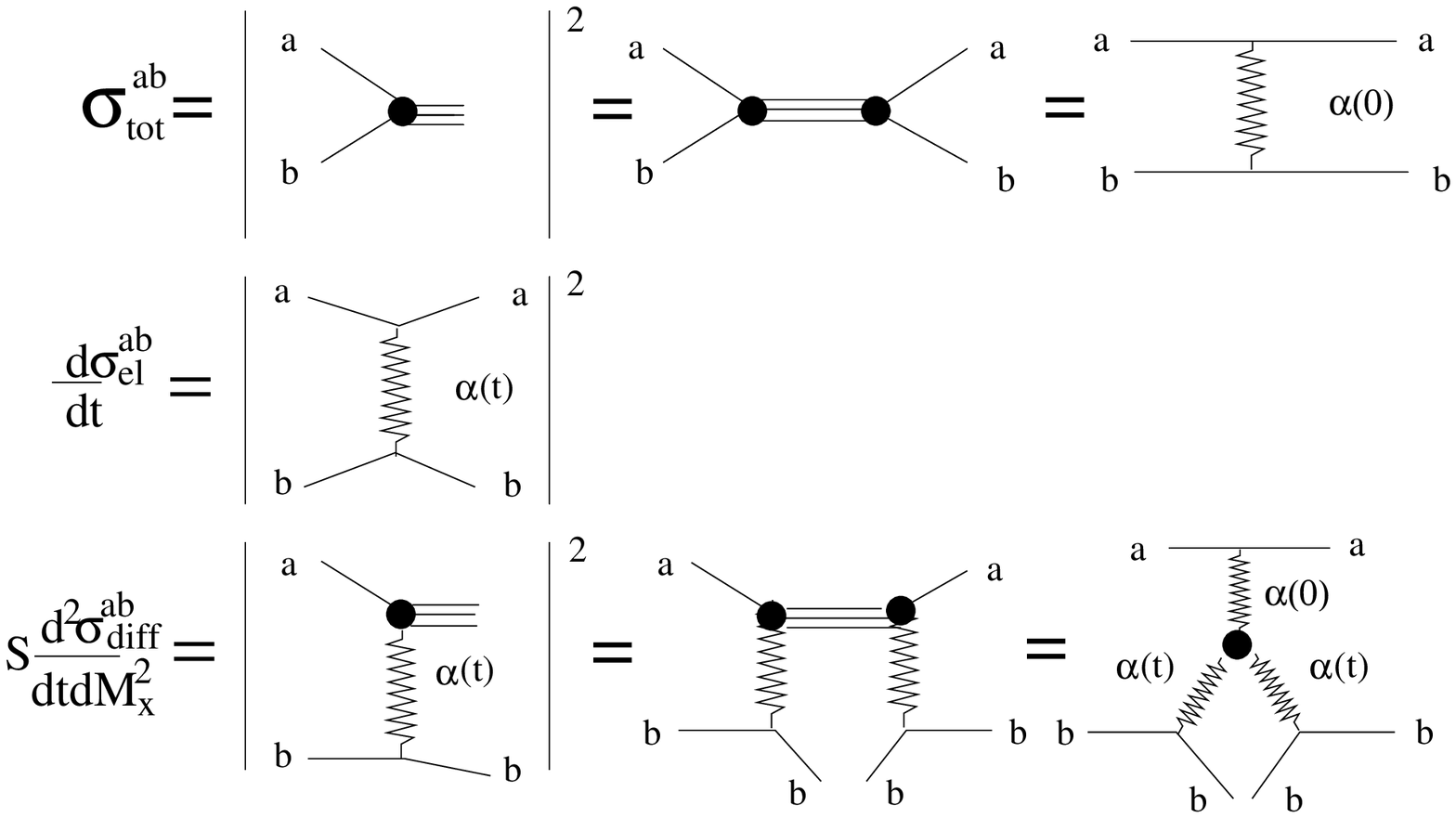}{ Regge diagrams
for (top to bottom) the total, elastic and single diffractive
scattering in hadron-hadron collisions.  } {diff-regge}

The expressions~(\ref{eq-diff:reggeel}) and~(\ref{eq-diff:3regge})
have been written in such a way that they allow for contributions
other than the pomeron exchange as long as the quantum numbers are
preserved. A source of big confusion is the fact that also
the $f$ trajectory can contribute to the elastic and diffractive
scattering.  This contribution has a very different $s$ and
$M_X^2$ dependence and is expected to be negligible at large $s$ and
small $M_X^2/s$. The $f$ trajectory has the form,
\begin{equation}
\alpha_f(t)\simeq 0.5 + t \, ,
\label{eq-diff:ftrajectory}
\end{equation}
typical of Reggeon exchange, while the pomeron trajectory as
determined by Donnachie and Landshoff~\cite{ref:DoLa_sigfit,DL-aprime}
is given by
\begin{equation}
\apom(t)= 1.085 + 0.25 t \, .
\label{eq-diff:pomeron}
\end{equation} 
However, in principle $k=\pom,~f$, while $l = \pom,~\reg$.

For the sake of simplicity we assume that at high energy, only the
pomeron exchange contributes and we will take
\begin{equation}
  \apom(t) = 1+\epsilon + \aprime t \, .
\end{equation}
Furthermore, since the $t$ distribution in elastic scattering is to a
good approximation exponential, we assume that
\begin{equation}
\beta_{a\pomsub}(t) = \beta_{a\pomsub}(0)\cdot e^{b_a t}\, ,
\end{equation}
where $b_a$ is an effective slope of the elastic form-factor of
particle $a$ and is related to the average radius squared of the
density distribution. In a region of pomeron dominance,
\begin{equation}
\frac{d^2\sigma^{ab}_{\mathrm{diff}}}{dt dM_X^2}=\frac{\beta_{a\pomsub}^2(0) 
\beta_{b\pomsub}(0) g_{\pomsub \pomsub \pomsub}(t)}{16\pi} \frac{1}{M_X^2} 
\left(\frac{s}{M_X^2}\right)^{2\epsilon}
\left(M_X^2\right)^{\epsilon}\, e^{b_{\mathrm{sd}}t} \, ,
\label{eq-diff:3pom}
\end{equation}
where
\begin{equation}
b_{\rm sd} = 2b_a + 2\aprime \ln \frac{s}{M_X^2} \, .
\label{eq-diff:shrinkage}
\end{equation}
It is usually assumed that the $t$ dependence of the triple-pomeron
coupling, $g_{\pomsub\pomsub\pomsub}$, is mild compared to the $t$
dependence of the elastic form-factors of hadrons.  If we take
$\epsilon=0$, we recuperate in formula~(\ref{eq-diff:3pom}) all the
properties assigned to diffraction, i.e. no energy dependence of the
cross section, $1/M_X^2$ dependence on the dissociated mass and an
exponential slope in the $t$ distribution, shrinking with $s$. Note
that for triple-pomeron exchange the $M_X^2$ dependence is related to
the $s$ dependence.

The factorization properties implemented in the triple-Regge formula
allow the diffractive differential cross section to be decomposed into
two terms,
\begin{equation}
\frac{d^2\sigma^{ab}_{\mathrm{diff}}}{dt dM_X^2}= f_{\pomsub/a}(\xpom,t)\cdot
\sigma_{b\pomsub}(M_X^2) \, ,
\label{eq-diff:pomeronfluxdef}
\end{equation}
where $\xpom = M_X^2/s$. The first term on the RHS of
Eq.(~\ref{eq-diff:pomeronfluxdef}) depends only on $\xpom$ and $t$
and is called the pomeron flux, while the second term can be
thought off as the total cross section for $b \pom$ interactions.  The
separation into these terms is arbitrary as far as constant
factors are concerned as the $\pom$ is not a real particle. The
pomeron flux factor is usually defined as
\begin{equation}
f_{\pomsub/a}(\xpom,t)= \frac{N} {16\pi}\beta_{a\pomsub}^2(t)
\xpom^{1-2 \apom (t)}\, ,
\label{eq-diff:pomeronflux}
\end{equation}
with $N=1$~\cite{ingelman-schlein} or $N=2/\pi$~\cite{DL-flux} (for a
discussion see~\citeasnoun{collins-huston}).

It should be noted that, while the triple-Regge approach is very
useful in parameterizing the data and sets-up a framework in which
diffraction is being described, its theoretical standing is not
strong. For a supercritical pomeron ($\epsilon>0$), the implied energy
dependence of single diffraction is such that already at the Tevatron
energies, the expected cross section is by factor five to ten larger
than the measured one~\cite{goulianos3} and it could become larger
than the total cross section at around 40 TeV~\cite{schuler}.

\subsection{Diffraction in the Quark Parton Model}

The requirement of a separation of the diffracted final state from the
target leads, at high energy, to the presence of a large rapidity gap
between the two systems. In the QPM, there is no mechanism for producing
large rapidity gaps other than by fluctuations in the hadronization
process. Therefore diffractive dissociation as such has to be
introduced by hand.

\subsubsection{The partonic pomeron}

The idea of Ingelman and Schlein~\cite{ingelman-schlein} was
to postulate that the pomeron has a partonic structure which may be
probed in hard interactions in much the same way as the partonic
structure of hadrons. They suggested that the partonic
structure of the pomeron would manifest itself in the production of
high transverse momentum jets associated with single diffractive
dissociation, for example in $pp$ scattering. The trigger for such a
reaction would consist of a quasi elastically scattered proton and the
presence of high $p_T$ jets in final state of the the dissociated
hadron. The jets would be accompanied by remnants of the pomeron
and of the diffracted hadron. 

For numerical estimates they assumed that in the presence of a hard
scale Regge factorization holds and the pomeron flux is the same as in
soft hadron interactions (with $N=1$). The pomeron was assumed to
consist of gluons and two extreme cases were considered.  One case was
inspired by the Low-Nussinov QCD pomeron~\cite{low,nussinov}
consisting of two gluons and the other by the gluon distribution in
the proton. In both cases it was assumed that the momentum sum rule
applies to the pomeron as to a normal hadron.

In such an approach it was also natural to expect diffractive
dissociation in deep inelastic scattering.

\subsubsection{The Aligned Jet Model}

The presence of diffractive dissociation in deep inelastic scattering
was predicted by Bjorken~\cite{bjorken-cornell} based on the Aligned
Jet Model (AJM) \cite{bjorken-AJM}. The AJM is a generalization of the
vector dominance model~\cite{ref:Bauer} used to describe $\gamma p$
interactions to the $\gamma^*p$ interactions in the quark-parton
model.

In the AJM, deep inelastic $ep$ scattering is considered in the rest
frame of the target proton. In this frame the virtual photon emitted
by the electron fluctuates into a quark-antiquark ($q\bar{q}$) pair
which then interacts with the proton. If all configurations of the
hadronic fluctuation were to contribute to the total cross section in
a universal manner, as observed for hadron-hadron scattering, the
cross section for $\gamma^*p$ interaction would be $Q^2$ independent
leading to violation of scaling properties of the structure functions
($F_2 \prop Q^2\sigma(\gamma^*p)$). Bjorken solved this puzzle by
assuming that only configurations with a very asymmetric distribution of
the longitudinal momentum of the $q\bar{q}$, and thus limited relative
transverse momenta, were allowed while the others were sterile. In the
allowed configurations, the quark which takes most of the momentum of
the initial virtual photon emerges in the final state as the current
jet, while the slow quark interacts with the target almost as a
hadron. The latter interaction, being of hadronic origin, can proceed
through diffractive dissociation. This will happen only if the
$q\bar{q}$ lives long enough to evolve into a hadronic state - that
is at high energies. Therefore Bjorken was able to predict that at
large $\gamma^*p$ energy, even in the QPM diffraction will
reappear.

In the AJM, diffractive scattering remains a soft phenomenon and thus
the Regge phenomenology applies as for hadron-hadron scattering,
corrected for the presence of a virtual photon.

\subsubsection{Formalism of diffractive DIS in QPM}

\epsfigure[width=0.5\hsize]{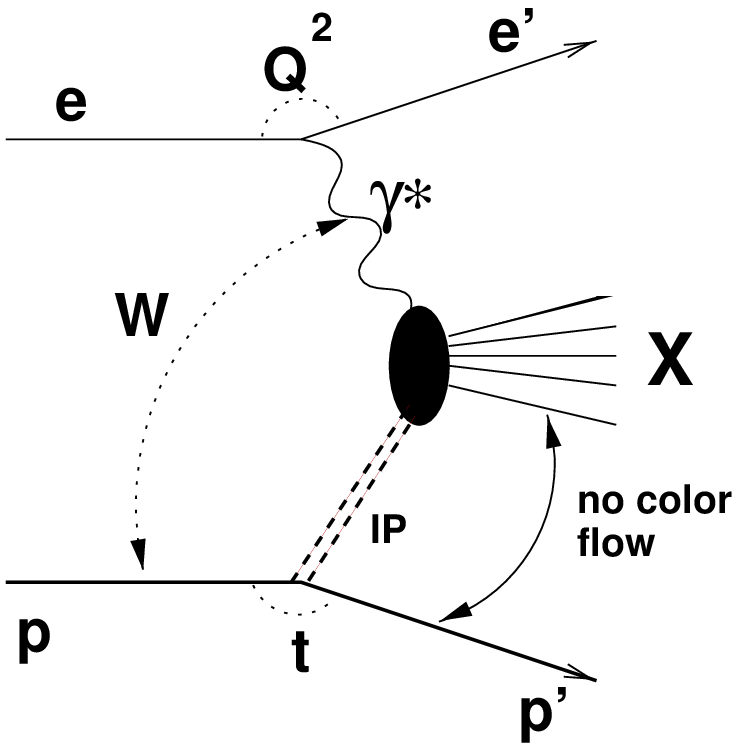}
{Diagram representing DIS scattering on a pomeron.  }
{diff-diffdisdiag}

In the Ingelman-Schlein approach, deep inelastic $ep$ diffractive
scattering proceeds in two steps. First, a pomeron is emitted from the
proton and then the virtual photon is absorbed by a constituent of the
pomeron. This is depicted schematically in
Fig.~\ref{fig:diff-diffdisdiag}. In order to describe this process
we have to introduce new variables, in addition to the ones used to
describe the inclusive DIS such as $Q^2$, $W$, $x$ and $y$.  The flux
of the pomeron depends on the fraction of the proton four-momentum
carried by the pomeron, $\xpom$, and on the square of the momentum
transfer, $t$, 
\begin{eqnarray}
\xpom &=& \frac{q\cdot (P-P^\prime)}{q \cdot P} \, , 
\label{eq-diff:defxpom} \\
t &=& (P-P^\prime)^2 \, ,
\label{eq-diff:deft}
\end{eqnarray}
where $P$ and $P^\prime$ denote the initial and final
proton four-momenta respectively. If we note that by definition $M_X^2
= (q+P-P^\prime)^2$ and $W^2 = (q+p)^2$ then it is easy to show that 
\begin{equation}
\xpom = \frac{M_X^2+Q^2-t}{W^2+Q^2-m_p^2}\simeq \frac{M_X^2+Q^2}{W^2+Q^2} 
\, ,
\label{eq-diff:xpom}
\end{equation}
where $m_p$ stands for the mass of the proton and the last relation on
the RHS holds for small $t$ and large $W$, a good approximation for
the kinematic region of high energy diffractive scattering.

Since the deep inelastic scattering now takes place on the pomeron, we
need to introduce a variable equivalent to the Bjorken $x$, but
relative to the pomeron momentum. This variable is usually called
$\beta$,
\begin{equation}
\beta = \frac{Q^2}{2q \cdot (P-P^\prime)} = \frac{x}{\xpom}
\simeq \frac{Q^2}{Q^2+M_X^2} \, ,
\label{eq-diff:beta}
\end{equation}
where again in the last step the contribution from $t$ was neglected.
Note that the definitions of the new variables are very general and do
not rely on the dynamics of diffraction as assumed in the
Ingelman-Schlein model.

In the same spirit one can extend the DIS cross section to include the
dependence on two additional variables. It is customary to retain the
variables $t$, $\xpom$, $x$ and $Q^2$. The four-fold differential
cross section for $ep$ scattering can be written as
\begin{equation}
\frac{d^4\sigma^D}{ d\,\xpom d\,t d\,Q^2 d\,x}
=\frac{2\pi \alpha^2}{x Q^4} \left[ 1+(1-y)^2\right] 
F_2^{D(4)}(x,Q^2,\xpom,t) \, ,\label{eq-diff:f2d4} 
\end{equation}
where for the sake of simplicity we have omitted the contribution from
the longitudinal structure function. The superscript $D$ denotes the
diffractive process and the number in parenthesis in the superscript of
the $F_2^D$ is a reminder that the units of the structure
function have changed. Keeping in mind the relation between $F_2$ and
the absorption cross section $\sigma(\gamma^* p)$ we can postulate that,
\begin{eqnarray}
F^{D(4)}_2 &\equiv& \frac{d^2 F_2^D(x,Q^2,\xpom,t)}{d\,\xpom d\,t} \\ 
\nonumber
&=&\frac{N}{16\pi}\beta_{a\pomsub}^2(t)\xpom^{1-2\apom(t)}
 \ftwopom(x,Q^2,\xpom,t)\, .
\label{eq-diff:f2d4fromf2}
\end{eqnarray}
If Regge factorization holds we expect that
\begin{equation}
\ftwopom(x,Q^2,\xpom,t) = \ftwopom(\frac{x}{\xpom},Q^2) \, ,
\label{eq-diff:f2d4factorization}
\end{equation}
and one can think of $\ftwopom$ as the structure function of the
pomeron.  Because the pomeron is self-charge-conjugate and an
isoscalar the density of any flavor of $\bar{q}$ is expected to be
equal to the density of the corresponding $q$,
\begin{equation}
f_{i/\pomsub}(x)=f_{\bar{i}/\pomsub}(x)\, ,
\label{eq-diff:SCCpom}
\end{equation}
and the densities of the $u$ and $d$ quarks are also equal,
\begin{equation}
f_{u/\pomsub}(x)=f_{d/\pomsub}(x)=f_{\bar{u}/\pomsub}(x)=f_{\bar{d}/\pomsub}(x)\,.
\label{eq-diff:isoscalarpom}
\end{equation}
To first approximation one can assume that there are only two parton
densities in the pomeron, the quark and gluon densities. It would be
tempting to assume that the strange quark density is the same as that
of the $u$ and $d$ quarks. However, the $Kp$ total cross section is
smaller than the $\pi p$ cross section, suggesting a possible
suppression of strange quarks. As for heavier quarks, it may be
assumed that their density can be dynamically generated by QCD
evolution. Obviously the issue of the strange and heavy quark
component of the gluon depends of the real nature of the pomeron, a
subject which will be discussed in the next section.

\subsection{Diffraction in QCD}

Some properties of diffractive dissociation make it a very interesting
study ground for QCD. Diffractive processes lead to the presence of
large rapidity gaps in the final states while in QCD, in which
fragmentation is driven by parton radiation, large rapidity gaps are
exponentially suppressed~\cite{basics_pqcd}. To understand diffraction
in QCD one has to invoke new coherent phenomena.

In QCD the concept of the pomeron still remains mysterious.  The ideas
about the nature of the pomeron range from a simple diagram with two
gluon exchange~\cite{low,nussinov} to a sophisticated gluon exchange
whose properties depend on the nature of the interaction (for a review
see~\citeasnoun{HERAWS}). The BKFL
pomeron~\cite{ref:BFKL1,ref:BFKL2,ref:BFKL3}, called sometimes hard or
perturbative pomeron, would consist a ladder of (reggeized) gluons
with a very special radiation pattern (strong $x$ ordering and no
$k_t$ ordering of the gluons inside the ladder). From the theoretical
point of view it becomes essential to analyze various processes in
terms of which type of pomeron could be probed. The QCD picture of
the pomeron points to its non-universal nature~\footnote{The
non-universal nature of the pomeron exchange has prompted many
theoreticians to abandon calling the mechanism of large rapidity gap
production, pomeron exchange. This is most probably the correct
approach. However it is still very much the habit to talk about the
pomeron in the context of diffractive dissociation - a habit which we
will reluctantly perpetrate in this article.}.


\subsubsection{Interplay of soft and hard physics}

Together with the high parton density physics of the small-$x$ regime
of $ep$ scattering at HERA came the realization that the hard physics
studied till now is the result of an interplay of hard and soft
phenomena. In case of deep inelastic scattering the unknown soft
physics is hidden in the initial parton distributions which are
parameterized at a relatively small scale $Q_0^2 \sim 1 \gevtwo$.
The lack of a dynamical picture of the proton structure
leads to a large uncertainty about the region of phase space which has
not been probed as yet.  This uncertainty propagates itself in QCD
predictions for high energy hard scattering at future colliders.

The ability to separate clearly the regimes dominated by soft or by
hard processes is essential for exploring QCD both at a quantitative
and qualitative level.  A typical example of a process dominated by
soft phenomena is the interaction of two large size partonic
configurations such as two hadrons. A process which would lend itself
to a fully perturbative calculation, and therefore hard, is the
scattering of two small size heavy onium-states each consisting of a
pair of heavy $q\bar{q}$ pair~\cite{muelleronium}.

In deep inelastic scattering the partonic fluctuations of the virtual
photon can lead to configurations of different sizes. The size of the
configuration will depend on the relative transverse momentum $k_T$ of
the $q\bar{q}$ pair. Small size configurations (large $k_T\sim Q/2$)
are favored by phase space considerations (the phase space volume
available is proportional to $k_T^2/Q^2$). In the QPM, in order to
preserve scaling, it was necessary to suppress their presence by
making them sterile. In QCD there is a simple explanation for this
suppression - the effective color charge of a small size $q\bar{q}$
pair is small due to the screening of one parton by the other and
therefore the interaction cross section will be
small~\cite{BBFS93,FMS}. This phenomenon is known under the name of
color transparency.

At small $x$ the smallness of the cross section is compensated by the
large gluon density (see Eq.~\ref{eq:twoglu}). 
The dominant mechanism for diffractive scattering of a small size
$q\bar{q}$ pair is two gluon exchange and the cross section can be
calculated in perturbative QCD.

For large size configurations, as noted previously, one expects to be
in the regime of soft interactions, modified by the typical QCD
evolution.  Here the Ingelman-Schlein type model would be applicable.

This qualitative picture based on QCD considerations leads to a
picture of diffraction very different from the one expected from the
QPM. The origin of a large rapidity gap may be either of perturbative
nature or due to soft processes.  The pomeron exchange is
non-universal and exhibits very different energy dependences for
different initial photon configurations.  Processes dominated by hard
scattering are expected to have a faster energy dependence than those
dominated by soft processes.  The establishment of the validity of
this approach has important consequences for the theoretical
understanding and possible observation of unitarization effects, which
is considered one of the main challenge of strong interactions.

\subsubsection{Perturbative hard diffractive scattering}

By perturbative hard diffractive scattering we mean a process in which
the large rapidity gap is of perturbative origin. The cross section
for these processes are calculable in perturbative QCD, with the
parton distributions in the proton as input. For this to happen, the
photon has to fluctuate into a large $k_T$, $q\bar{q}$ pair.
Furthermore, for this process to be dominated by a two-gluon exchange,
the invariant mass $M_X^2 \ll Q^2$. By minimizing the amount of energy
transferred from the proton to the virtual photon, extra radiation is
suppressed. The price for forcing $M_X^2 \ll Q^2$ is a suppression of
the cross section by extra powers of $Q^2$.

To differentiate between diffractive processes with a hard scale but a
soft component and those which can be calculated in perturbative QCD,
we will call the latter perturbative diffractive scattering.  Three
classes of processes have been identified as possible candidates for
perturbative diffractive scattering in $ep$ interactions. One class is
small mass diffractive dissociation of a longitudinally polarized
virtual photon, $\gamma^*_L$, which fluctuates preferentially into a
small size $q\bar{q}$ configuration~\cite{ref:BFGMS}.  A possible
trigger for a $\gamma^*_L$ induced diffractive interaction is the
production of a longitudinally polarized vector meson, such as $\rho$,
$\omega$, $\phi$, $J/\psi$ or $\Upsilon$, or any of their radial
excitations. Exclusive vector meson production is discussed in detail
in section~\ref{sec:VM}.  The second class consists of diffractive,
exclusive dijet production, where the invariant mass of the two jets
saturates the invariant mass of the photon dissociation
system~\cite{bartels2jet1,diehljet}, independently of the polarization
of the virtual photon.  In the same category is the diffractive
production of heavy flavors~\cite{martincharm,diehlcharm}.  The third
process is large $t$ diffractive production of vector
mesons~\cite{ref:Forshaw,ref:Bartels} where $t$ is of the order of few
GeV. The contribution from the soft pomeron exchange should be highly
suppressed due to its steeply falling $t$ distribution. At large $t$
the proton is expected to dissociate into a multi-hadronic final
state.

In perturbative diffractive processes, $d\sigma/dt|_{t=0} \prop
[xG(x,Q^2_{\rm eff})]^2$, with the effective hard scale, $Q^2_{\rm
eff}$, depending on the process.  For vector meson
production~\cite{ref:Koepf}
\begin{equation}
Q^2_{\mathrm{eff}}=Q^2\frac{\langle b^2_L
\rangle}{\langle b^2_V \rangle} \ll Q^2 \, ,
\label{eq-diff:q2effvm}
\end{equation}
where $\langle b^2_L \rangle$ and $\langle b^2_V \rangle$ are the
effective sizes of the $q\bar{q}$ pair for the longitudinal photon and
the vector meson respectively, while for jets with transverse momentum
$p_T$
\begin{equation}
Q^2_{\mathrm{eff}}= \frac{p_T^2}{(1-\beta)} \geq p_T^2 \, .
\label{eq-diff:q2effjets}
\end{equation}
Therefore the cross section for hard diffractive processes is expected
to rise with $W$ with a scale dependent power and considerably faster
than for soft processes.

In QCD the slope of the pomeron trajectory, $\aprime$, is related to
the average $k_T$ of the partons in the
exchange~\cite{gribovdiffusion}. For hard processes $k_T$ is expected
to be large and $\aprime \prop 1/Q^2$. Therefore another distinctive
feature of perturbative diffractive processes is that the slope of the
$t$ distribution should be universal and independent of
energy~\cite{ref:Bartels,AFS}.

The interest in studying perturbative diffractive processes lies in
that at very large energy (small $x$), when the density of gluons
becomes very large, unitarity corrections are expected to set-in and
soft phenomena will become dominant again.  However because the
coupling constant will remain small, it will be possible to study the
approach to the unitarity limit from the perturbative regime.

\subsubsection{QCD inspired models of LRG production}

Only a special class of hard diffractive scattering processes lends
itself to perturbative calculations. These processes are
usually suppressed by extra powers of $Q^2$ compared to the total
cross section (they are called higher twist contributions) and
populate the region of large $\beta$. The experimental evidence, on the
other hand, points to the presence of a large fraction of DIS events
with LRG which remains relatively constant as a function of $Q^2$ and
$W$~\cite{ZEUSdiff1,H1diff1} and populates evenly the $\beta$ phase
space. It is thus of interest to review the phenomenological
approaches which try to explain such a large fraction of
diffractive-like events within QCD.

Since the first results of HERA were presented, the scientific
community is burgeoning with ideas about the possible origin of hard
diffraction and probable consequences. It is therefore not possible to
give a complete and fair review. Instead we will present a broad
classification and concentrate on discussing in more details the most
popular approaches.

\paragraph{Regge factorization and QCD}

There is a class of models which follow the original idea of Ingelman
and Schlein~\cite{ingelman-schlein}. The cross section for diffractive
scattering is assumed to factorize into a pomeron flux and the pomeron
is assumed to consist of partons whose number densities have to be
determined directly from the data, usually by applying the standard
DGLAP QCD evolution.

Only recently the QCD factorization theorem was proven to hold for the
leading twist diffractive structure function, $\ftwod$, of the proton
measured in DIS~\cite{collins}, giving support to this approach. Prior
to that, it was conjectured that one could introduce the concept of
diffractive parton densities~\cite{trentadue-veneziano,berera-soper1}
(also called differential fracture functions) and that these new parton
densities would factorize similarly to the DIS parton densities and
satisfy the DGLAP evolution
equation~\cite{kunszt-stirling,berera-soper1}. The Regge factorization
of these diffractive parton densities is an additional assumption.

A large group of physicists have explored such an
approach~\cite{DL-flux,berger,KKKdiff,CKMTdiff,GBK1,GS,VBLY,alvero}.
They differ in the way one of the three major ingredients are treated:
the pomeron flux, the evolution equation and the initial parton
distributions.  In this Ingelman-Schlein type of approach it is also
natural to incorporate the exchange of sub-leading trajectories in the
same way as the pomeron exchange~\cite{GBK2,h1_f2d2}.

\paragraph{Color dipole interactions}

Since in small-$x$ interactions it is rather the $q\bar{q}$
fluctuation of the photon which interacts with the target proton, it
is natural to view the interaction as that of a color dipole with the
proton~\cite{NikZak1,NikZak2,wusthoff,bialas}.  \citeasnoun{bialas}
have extended the dipole approach assuming that the proton also
consists of color dipoles.

For diffractive scattering the dipole interacts with the proton
through two-gluon exchange. The models differ in the way the two-gluon
exchange is handled in QCD. Since there is no guarantee that the
approach can be fully perturbative, the uncertainties are absorbed
into effective parameterizations, whose parameters are derived from
the inclusive DIS scattering in particular.

In these models the pomeron is non-universal and cannot be represented
by a single flux. Common to this approach is the prediction of a
dominant contribution of the longitudinal photon to the large $\beta$
spectrum. The small-$\beta$ spectrum is populated by a $q\bar{q} g$
configurations of the photon, while the $q \bar{q}$ configuration of
the transversely polarized photon populates the mid $\beta$ region.

\paragraph{Perturbative QCD approach}

In this approach diffractive scattering proceeds through the coupling
of two-gluons to the
photon~\cite{ryskinf2d,bartelsf2d,gotsman,bartels-kowalski}.  For low
mass diffraction the final state consists of a $q \bar{q}$ pair, while
large mass diffraction includes the production of $q\bar{q} g$ final
states. The dynamical content of these models differ in the treatment
of QCD corrections and in the choice of the gluon density.  In many
respects the results are similar to the ones obtained with the dipole
approach.

This approach is used to calculate high $p_T$ jets~\cite{bartelsjets}
and charm production in diffractive
scattering \cite{ryskinf2d,martincharm,diehlcharm}. In the perturbative
approach the charm yield is expected to be large. The only exception
is the model~\cite{diehlcharm} based on non-perturbative two-gluon
exchange of~\citeasnoun{nachtmann}.

\paragraph{Semi-classical approach and soft color interactions}
In the semi-cla\-ss\-ical approach partonic fluctuations of the virtual
photon are scattered off the proton treated as a classical color field
localized within a sphere of radius $1/\Lambda$~\cite{BH1}. A final
state color singlet partonic configuration is assumed to lead to a
diffractive event while a color non-singlet configuration yields an
ordinary non-diffractive event. This simple physical picture, which is
a generalization of the AJM model, leads to a number of
predictions~\cite{BDH1,BDH2} which are independent of the details of
the proton color field. In this approach the notion of pomeron does
not really appear and large rapidity gaps are generated as a result of
color rearrangement in the final state.

The phenomenology of the semi-classical approach is qualitatively
similar to the Ingelman-Schlein approach, with a pomeron which is
predominantly gluonic. The dominant partonic process is boson-gluon
fusion~\cite{BHbgf} and non-perturbative soft color
interactions~\cite{edin} cause the formation of a color singlet final
state.

\subsubsection{Summary}

To summarize, a tremendous theoretical progress has been achieved
since the first appearance of large rapidity gaps in DIS. While the
number of models may seem overwhelming, in fact in many respects they
follow the same pattern and their validity is limited to specific
regions. Their variety reflects the problem of the interplay of soft
and hard QCD in diffraction, as well as the interplay of leading and
non leading twist effects. The latter have been pretty much ignored in
the region of small $x$, however they may turn out to play an
important role~\cite{HTbartels,NikZakHT} and improve our understanding
of the small-$x$ physics in general.

Many of the presented models have predictions which can be tested
experimentally, such as Regge factorization, the $Q^2$ and $W$
dependence as a function of $\beta$ as well as the $t$ dependence. The
characteristics of the final states is another probe for the validity
of the presented ideas. The experimental program at HERA barely
started and the present achievements will be summarized below.

\subsection{Pre-HERA experimental results}

The first evidence for the existence of a hard component in
diffractive scattering was reported by the UA8
experiment~\cite{bonino,brandt}. It was based on the presence of large
$p_T$ jets in $p\bar{p}$ interaction at $\sqrt{s}=630 \gev$ in which a
proton (or anti-proton respectively) was found carrying a large
fraction of the beam momentum, $0.90<x_L<.97$. The measurements were
performed at relatively large values of $t$, $t >0.9 \gevtwo$. Based
on a sample of two jet events, the pomeron internal structure was
found to be hard~\cite{brandt}, with about $30\%$ of the sample
consisting of jets carrying away all of the pomeron momentum.

\subsection{Diffractive dissociation in photoproduction}

Before embarking on a more detailed discussion of hard diffractive
scattering at HERA, it is of interest to establish whether diffractive
dissociation initiated by a real (or quasi-real) photon follows
the properties observed in hadron-hadron interactions. 

In fixed target experiments, diffractive dissociation of the photon was
studied in $\gamma p$ interactions at centre of mass energies
$\sqrt{s} \simeq 14 \gev$~\cite{chapin}. The double-differential cross
section $d^2\sigma/d\,t d\,M_X^2$ was measured for $M_X^2/s<0.1$ and
$0.02<t<0.2 \gevtwo$. The differential cross section was found to be
dominated by the production of the $\rho$ meson. The large mass
distribution was found to follow an $1/M_X^2$ dependence consistent
with a large triple-pomeron presence in the diffractive amplitude. The
$t$ distribution was found to be exponential with a slope $b=10.6 \pm
1.0$ for $\rho$ production, while for $M_X^2>4 \gevtwo$ the slope was
found to be mass independent and roughly half of the value measured
for the $\rho$. A comparison with $\pi p$ interactions with $\pi$
dissociation, performed under similar conditions showed a good
agreement with Regge factorization.

Both HERA collider experiments, H1~\cite{diff:h1_photon} and
ZEUS \cite{diff:zeus_photon} have studied photon diffractive
dissociation, using very small $Q^2$ (typically $Q^2 < 0.01 \div 0.02
\gevtwo$) electroproduction, in which the electron is scattered under
a small angle and the interaction can be thought off as proceeding
through a beam of quasi-real photons scattering of a proton target.
The energy range covered by the HERA experiments is an order of
magnitude larger than for the fixed target experiment.  The
experimental procedures used to extract the diffractive samples will
be discussed in more details in the context of hard diffractive
processes. Here we just report the main findings.

H1 has performed a measurement of single photon dissociation at two
energies, $W=187 \gev$ and $231 \gev$. Both from the study of the
energy dependence, which included the results of~\citeasnoun{chapin},
and of the $M_X$ dependence, it is concluded that the triple-Regge
formalism describes well the data. The extracted value of
\begin{equation}
\apom(0)=1.068 \pm 0.016 \pm 0.022 \pm 0.041\, ,
\label{eq-diff:H1alphaph}
\end{equation}
with the last error due to MC model uncertainties, agrees with the one
from hadron-hadron interactions.

The ZEUS
experiment studied the $M_X$ distribution at a fixed value of $W=200
\gev$ and obtained a value of
\begin{equation}
\apom(0)=1.12 \pm 0.04 \pm 0.08 \, ,
\label{eq-diff:ZEUSalphaph}
\end{equation}
again in good agreement with expectations. The percentage of inelastic
photon dissociation (i.e. excluding the mass region $M_X< m_{\phi}$)
in the total cross section was found to agree with expectations of
Regge factorization. In addition, the LPS detector of ZEUS was used to
measure the $t$ distribution associated with photon
dissociation~\cite{diff:zeus_tph}. In the energy interval $176<W<225
\gev$ and for masses $4<M_X<32 \gev$, the $t$ distribution has an
exponential behavior with
\begin{equation}
b=6.8 \pm 0.9 ^{+\textstyle 1.2}_{-\textstyle 1.1} \gevmtwo \, .
\label{eq-diff:ZEUSbph}
\end{equation}

It can be concluded that photon diffractive dissociation in
photoproduction follows a similar pattern as single diffractive
dissociation in hadron-hadron interactions.

\subsection{Diffractive deep inelastic scattering at HERA}

The diagram corresponding to diffractive DIS, $ep\rightarrow epX$, is
presented in Fig.~\ref{fig:diff-diffdisdiag}. The dissociating
particle is the virtual photon emitted by the electron. The final
state consists of the scattered electron and hadrons which populate
the photon fragmentation region. The proton is scattered in the
direction of the initial beam proton with little change in the
momentum and angle.

Two major difficulties arise when studying diffractive scattering at
HERA. The first difficulty is encountered in the selection of the
diffractive sample. At HERA, the highly sophisticated detectors cover
predominantly the photon fragmentation region, leaving out, for
precise measurements, most of the proton fragmentation region. The
second difficulty arises because not all events which have properties
typical of diffractive dissociation are due to pomeron exchange.

In the following we first discuss the kinematic configurations of
diffractive events and then the experimental procedures leading to the
selection of diffractive candidate events. We then discuss the
measurements of the $\ftwodthree$ structure function ($\ftwodfour$
integrated over $t$) and the properties of the associated final
states.

\subsubsection{Kinematics of diffractive final states}

We first consider a typical kinematic configuration for the $\gamma^*p
\rightarrow Xp$ reaction with mass $M_X$ at centre of mass energy $W$
and with $M_X \ll W$. For simplicity we will start by describing this
configuration in the $\gamma^*p$ centre of mass system (cms) and
assume $t=0 \gevtwo$. The kinematics are very much like that of a two
body scattering with transverse momentum $p_T=0$. The proton and the
system $X$ move in opposite directions with longitudinal momentum $p_L
\simeq W/2$. The respective
rapidities of the proton, $y_p$, and of the system $X$, $y_X$, are
\begin{eqnarray}
y_p&=&\frac{1}{2}\ln \frac{E_p+p_L}{E_p-p_L} 
\simeq \frac{1}{2}\ln \frac{W^2}{m_p^2} \, , \\
y_X&=&\frac{1}{2}\ln \frac{E_X-p_L}{E_X+p_L} 
\simeq -\frac{1}{2}\ln \frac{W^2}{M_X^2} \, ,
\label{eq-diff:rapidity}
\end{eqnarray}
where we have used the coordinate system of HERA with the proton
moving in the positive $z$ direction and have neglected the masses
compared to the energies. The rapidity separation between the proton
and system $X$ is
\begin{equation}
\bigtriangleup y = y_p - y_X = \ln \frac{W^2}{m_p M_X} \, .
\end{equation}
For $W=200 \gev$ and $M_X=20 \gev$ the rapidity separation
$\bigtriangleup y \simeq 7.7$. Obviously the system $X$ will fragment
into hadrons and will typically occupy a region in $\Delta y$ $\prop
\ln M_X \simeq 3$. The proton and the fragments of system $X$ will be
separated by a rapidity gap larger than 4 units. From soft
hadron-hadron interactions, the typical density correlation length in
rapidity is 2 units of rapidity. Therefore typical of diffractive
scattering at high energy is a large rapidity gap (LRG) between the
proton and the remaining hadronic system (see Fig.~\ref{fig:diff-LRGkin}).

\epsfigure[width=0.8\hsize]{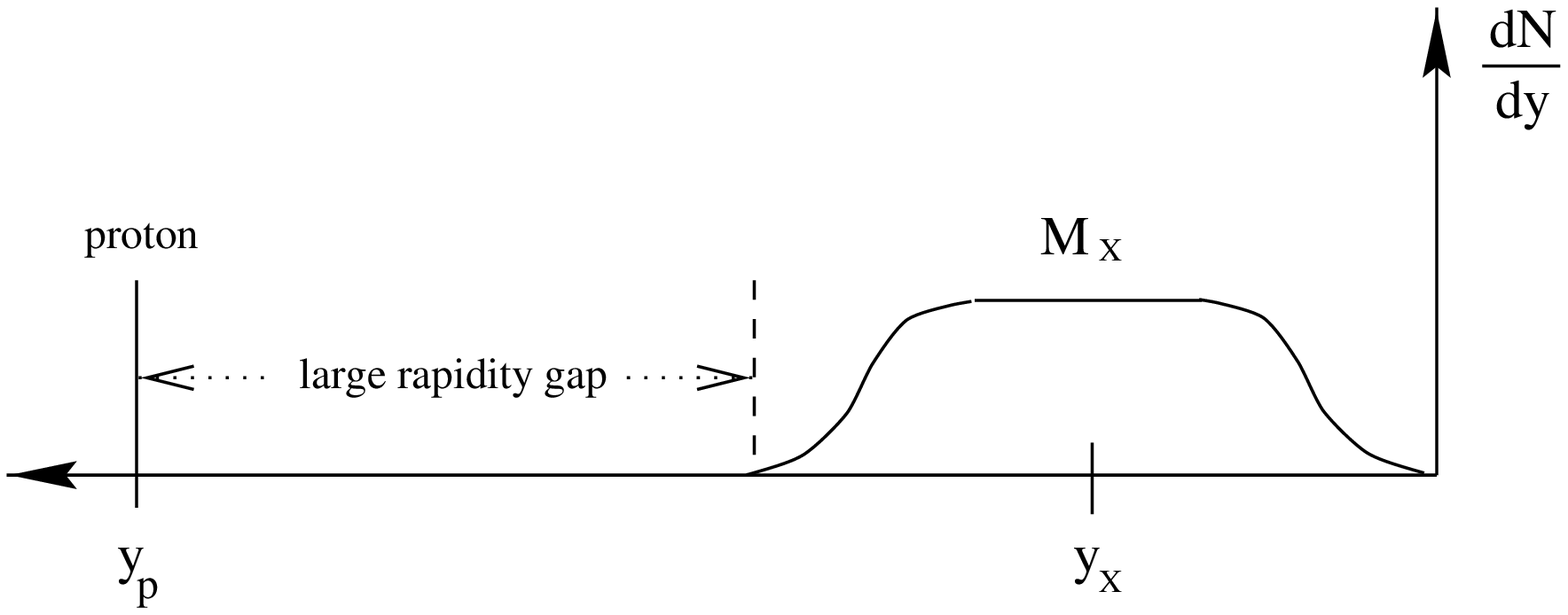} {Schematic
  representation of the particle distribution in the rapidity ($y$)
  space for a single diffractive dissociation event.}  {diff-LRGkin}

For moderate $Q^2$ interactions at HERA, the $\gamma^* p$ system is to
a good approximation boosted longitudinally with respect to the
laboratory system. A longitudinal boost causes all the rapidities to
be shifted by the same amount and thus the basic rapidity gap
structure is preserved. A proton with momentum $p_p = 820 \gev$ has
rapidity $y_p^{\rm lab} \simeq 7.5$. In our example the $M_X$ system
will have $y_X^{\rm lab} \simeq 0$. The coverage of the main
calorimeters in rapidity space is typically $-3.5<y<4$, therefore the
$M_X$ system will be measured in the main detector, while the proton
will escape detection.

Note that in events in which the proton dissociates as well, it may
happen that some of the hadrons from the dissociation will be visible
in the calorimeter. The presence of such events may destroy the
rapidity gap between the two systems. 

For calorimetric measurements of the hadronic energy flow, the
pseudo-rapidity variable $\eta$ is used instead of rapidity defined as
\begin{equation}
\eta= -\ln \tan\frac{\theta}{2}
\label{eq-diff:pseudorap}
\end{equation}
with $\theta$ the polar angle measured with respect to the proton
direction. The pseudo-rapidity is a very good approximation to
rapidity for particles with small mass ($E\simeq |p|$).

\subsubsection{Experimental selection of diffractive samples}

The selection of diffractive events in DIS proceeds in two steps.  The
events are first selected based on the presence of the scattered
electron in the detector. The procedure follows exactly the one
described for the $F_2$ measurements in section~\ref{sec:DISprod}.  For
the diffractive selection three different methods have been used at
HERA: 
\begin{enumerate} 
\item a reconstructed proton track was required in the leading proton 
spectrometer (LPS) with a fraction of the initial proton momentum
$x_L>0.97$~\cite{zeus_lps}; 
\item the hadronic system $X$ measured in the central detector was 
required to be separated by a large rapidity gap from the rest of the
hadronic final state~\cite{zeus_f2d1,h1_f2d1,h1_f2d2}; 
\item the diffractive events were identified as the excess 
of events populating the small mass $M_X$ region, in the inclusive $\ln
M_X^2$ distribution~\cite{zeus_mx1}.
\end{enumerate}

The remaining background is estimated using Monte Carlo simulation of
background processes, tested on control samples. The acceptance and
resolution corrections are calculated using samples of diffractive
Monte Carlo simulated events tuned to reproduced the data. For the
diffractive DIS analysis, the RAPGAP~\cite{RAPGAP} generator, based on
the Ingelman-Schlein model, is widely used.  It is very similar to
non-diffractive DIS generators, in which the proton is replaced by a
beam of pomerons with a partonic content. Other diffractive models
are used for systematic checks.

\paragraph{Selection based on the LPS}

As in fixed target hadron-hadron interactions the cleanest selection
of diffractive events with photon dissociation is based on the
presence of a leading proton in the final state. By leading proton we
mean a proton which carries a large fraction of the initial beam
proton momentum. The spectrum of protons in a sample of DIS events has
been measured in the LPS of the ZEUS experiment~\cite{zeus_lps} and is
shown in Fig.~\ref{fig:diff-LPSspectrum} as a function of
$x_L=|p_f|/|p_i|$, where the subscripts $f$ and $i$ denote the final
and initial protons respectively and $|p|$ stands for the absolute
value of the momentum.

\epsfigure[width=0.8\hsize]{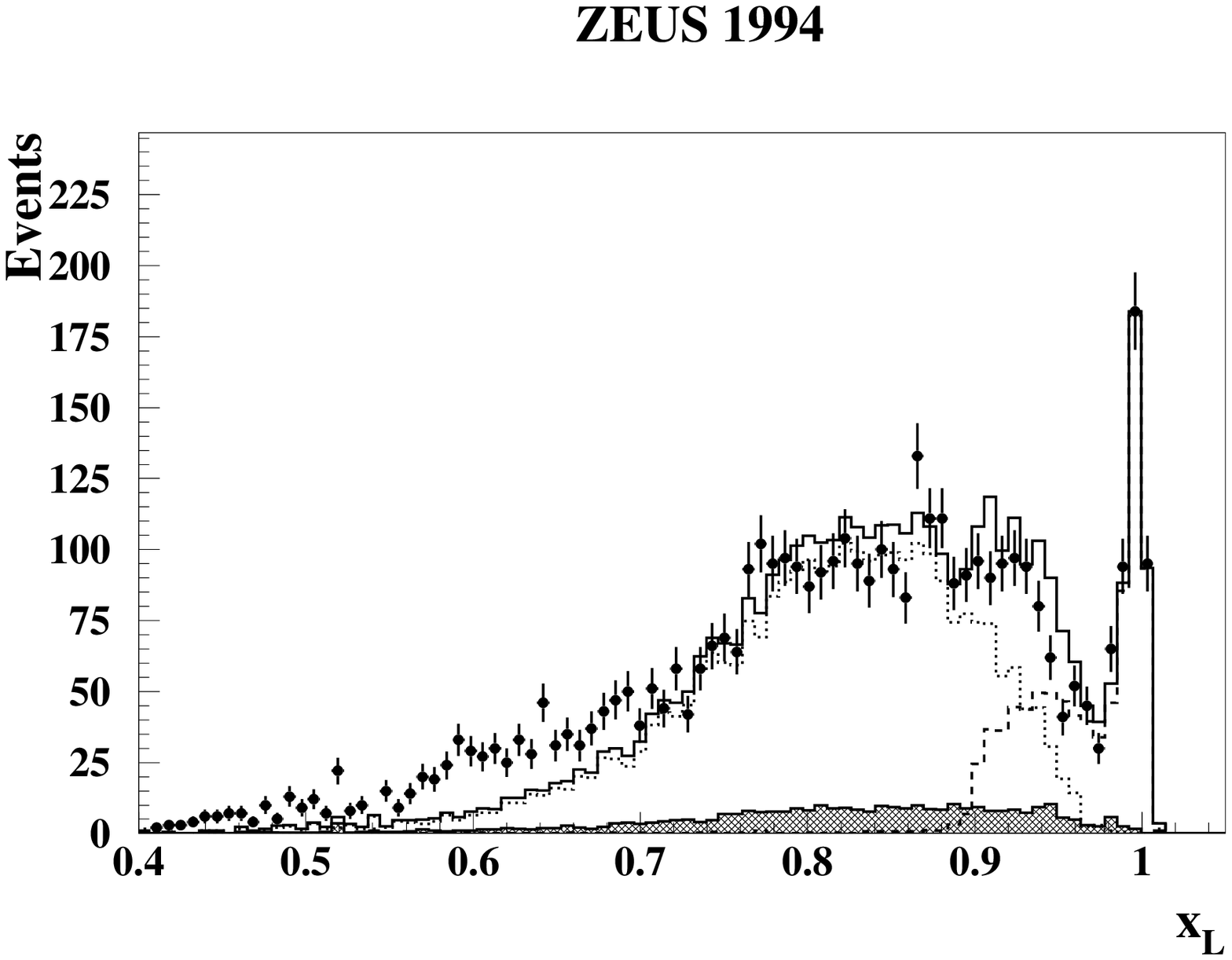} {Observed
  spectrum of the fraction of the proton beam momentum carried by the
  scattered proton, $x_L$.  Overlaid (full line) is the result of
  fitting this distribution with a sum of contributions from MC
  generators for double dissociation (shaded area), for pion exchange
  (dotted line) and for photon diffractive dissociation (dashed line).
  } {diff-LPSspectrum}

A characteristic peak is observed at $x_L
\simeq 1$ which corresponds to photon dissociation events. The
geometrical acceptance of the LPS is of the order of $10\%$. A clean
sample of diffractive events is obtained by requiring $x_L>0.97 $.
Studies based on MC simulation of background processes indicate that
the contamination by processes such a $\pi$ trajectory exchange or
proton dissociation remains below $3\%$. The measurement of the
momentum vector of the scattered proton allows to determine the $t$
value and to study the $t$ distribution in inclusive diffractive
dissociation.

\paragraph{Selection based on large rapidity gaps}

As discussed in the section on kinematic properties of diffractive photon
dissociation, a large rapidity gap is expected in the hadronic final
state. For inelastic DIS events, the rapidity phase space is
populated evenly by final state particles.

\epsfigure[width=0.8\hsize]{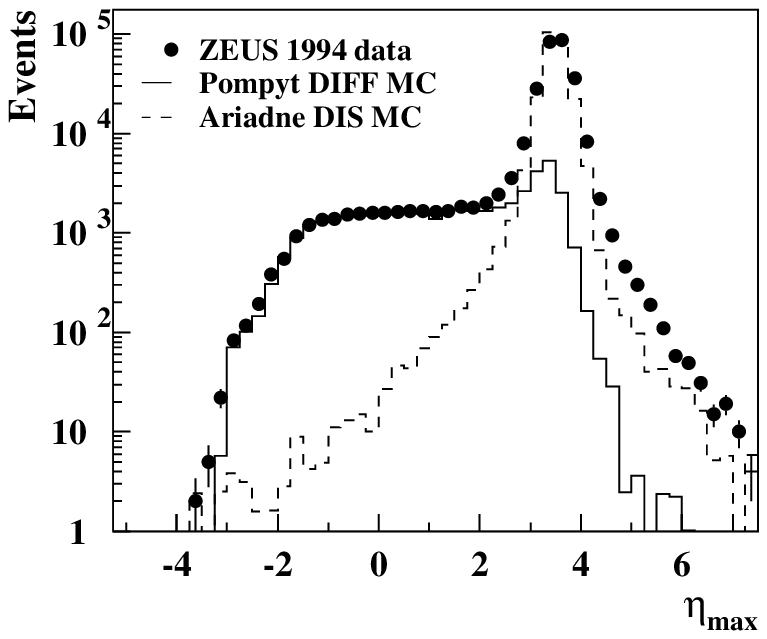}{ Distribution of
  $\etamax$ for DIS events. The solid circles are the data points, the
  dashed histogram is the result of the non-diffractive (ARIADNE) MC
  simulation and the full histogram is that of the DIS diffractive MC
  (POMPYT) } {diff-etamax}

One way of establishing the presence of events with a large rapidity gap
is shown in Fig.~\ref{fig:diff-etamax}. Each event is assigned a
variable $\etamax$ defined as the $\eta$ of the energy deposit in the
central detector above 400 MeV closest to the proton direction.  The
distribution of $\etamax$ for DIS NC events is compared to the
expectations of a DIS Monte Carlo based on the Color Dipole Model
(ARIADNE~\cite{ref:ARIADNE_2}) for the production of the inelastic
hadronic final state. A clear excess of events with large rapidity
gaps is observed in the region of $\etamax<2$, which corresponds to an
effective rapidity gap of more than 2 units of $\eta$ in the central
calorimeter and possibly more than 5 units of $\eta$ relative to the
initial proton. In the non-diffractive MC the LRG events are strongly
suppressed in this region.

In the selection based on a rapidity gap starting in the central
detector, events in which the proton has also dissociated may be
erroneously assigned to single photon dissociation. This background is
estimated to be $\lesssim 10\%$ for the H1 detector, for which good
tagging efficiency for the hadronic final states extends into the
forward region up to $\eta \simeq 7$. Only events in which the proton
dissociated into a mass $M_Y \lesssim 1.6 \gev$ escape the tagging
devices~\cite{h1_f2d2}. In the ZEUS detector, which relies mainly on
the coverage of the uranium calorimeter, masses $M_Y$ up to about $5.5
\gev$ remain undetected. The background is therefore larger and
estimated to be $\sim 30\%$.

The requirement that the system $X$ be contained in the central
detector limits the acceptance for large $\xpom$ values. For H1 $\xpom
< 0.05$, while for ZEUS it is $\xpom<0.03$. Note that for $\xpom >
0.01$ color singlet exchanges other than the pomeron can also
contribute.

\paragraph{Selection based on the $M_X$ distribution}

The $M_X$ method to extract the diffractive contribution is based on
the observation that the spectrum of the invariant mass $M_X$ measured
in the calorimeter consists of two components with very different
behavior~\cite{zeus_mx1}. The diffractive contribution is identified as
the excess of events at small $M_X$ above the exponential fall-off of
the non-diffractive contribution with decreasing $\ln M^2_X$ (see
Fig.~\ref{fig:diff-lnmx}). The exponential fall-off, expected in
QCD~\cite{basics_pqcd}, permits the subtraction of the non-diffractive
contribution and therefore the extraction of the diffractive
contribution without assuming the precise $M_X$ dependence of the
latter. 

\epsfigure[width=0.8\hsize]{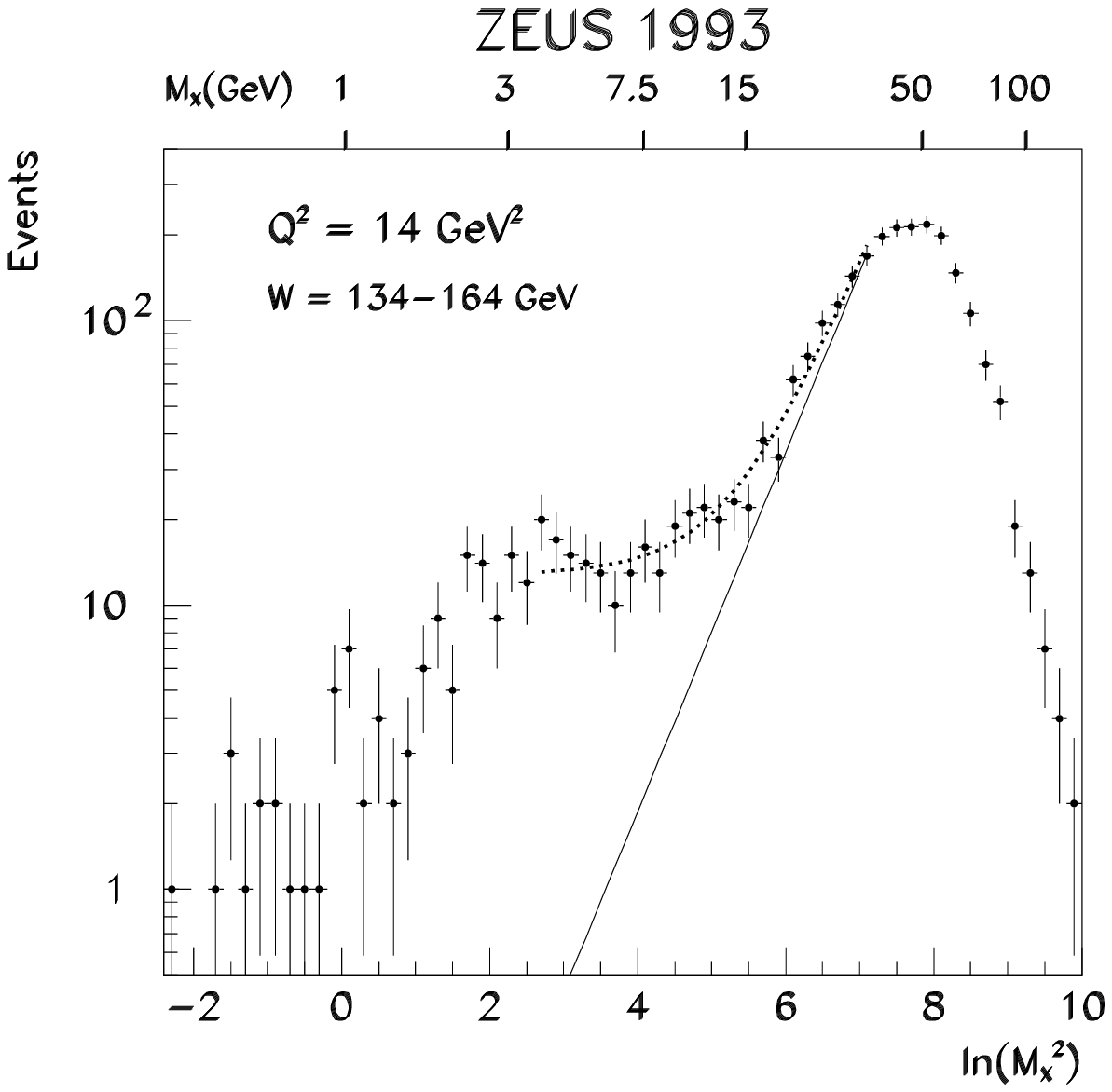}
{Distribution of $\ln M^2_X$ for $W$ and $Q^2$ as denoted in the
  figure.  The solid line shows the exponential fall-off of the
  non-diffractive contribution resulting from fitting the data with
  Eq.~(\protect\ref{eq-diff:lnmx}).  }{diff-lnmx}

The $M_X$ distribution is expected to be of the form
\begin{eqnarray} 
\frac{d N}{d\ln M_X^2} = D \;+ \;
 c\exp (b \ln {M_X^2}) \, .
\label{eq-diff:lnmx}
\end{eqnarray}
Here, $D$ denotes the diffractive contribution. Assuming an $M_X^{-2}$
dependence for photon dissociation, expected to hold for $M_X^2\gg
Q^2$~\cite{DL-flux,NikZak1}, $D$ is a constant.  The second term on the
RHS represents the non-diffractive contribution.
Expression~(\ref{eq-diff:lnmx}) is fitted to the measured $\ln M_X^2$
distribution in bins of $Q^2$ and $W$. The fit parameters are $D$, $b$
and $c$.  The fits are performed to the data in a limited range of
$M_X^2$, in which the expression~(\ref{eq-diff:lnmx}) is expected to
hold to a good approximation.

The parameters of the fit are very well constrained at large $W$,
where the separation of the diffractive and non-diffractive events is
clearly visible. Monte Carlo studies have shown that the value of $b$
does not depend on $W$ or $Q^2$. The same tendency is observed in the
data. As a result a single value of $b$, as determined at large $W$
and $Q^2$, is assumed for the fits performed in other regions.

From the results of the fitting procedure performed in each $W,~Q^2$
interval, only the parameterization of the non-diffractive background
is retained. This parameterization is then extrapolated to the small
$\ln M^2_X$ region and subtracted from the experimental distribution
giving the diffractive contribution in the data.  The results of the
fit are also presented in Fig.~\ref{fig:diff-lnmx}.

Given the size of the background this method limits the measurements
to small masses, $M_X<15 \gev$ and $\xpom \lesssim 0.01$, dominated by
pomeron exchange. The background from events in which the proton also
dissociated remains the same as for the LRG method.

\subsubsection{Measurements of $F_2^D$}

The measurements of $\ftwodthree$ have been performed by H1 and
ZEUS. The early results~\cite{h1_f2d1,zeus_f2d1} were compatible with
Regge factorization. However, the values of $\apom$ determined from the
$\xpom$ dependence for fixed values of $Q^2$ and $\beta$ were
inconclusive as to the nature of the pomeron probed in DIS. The lack
of $Q^2$ dependence established that in diffractive scattering the
virtual photon was scattering off a point-like particle.

\epsfigure[width=0.9\hsize]{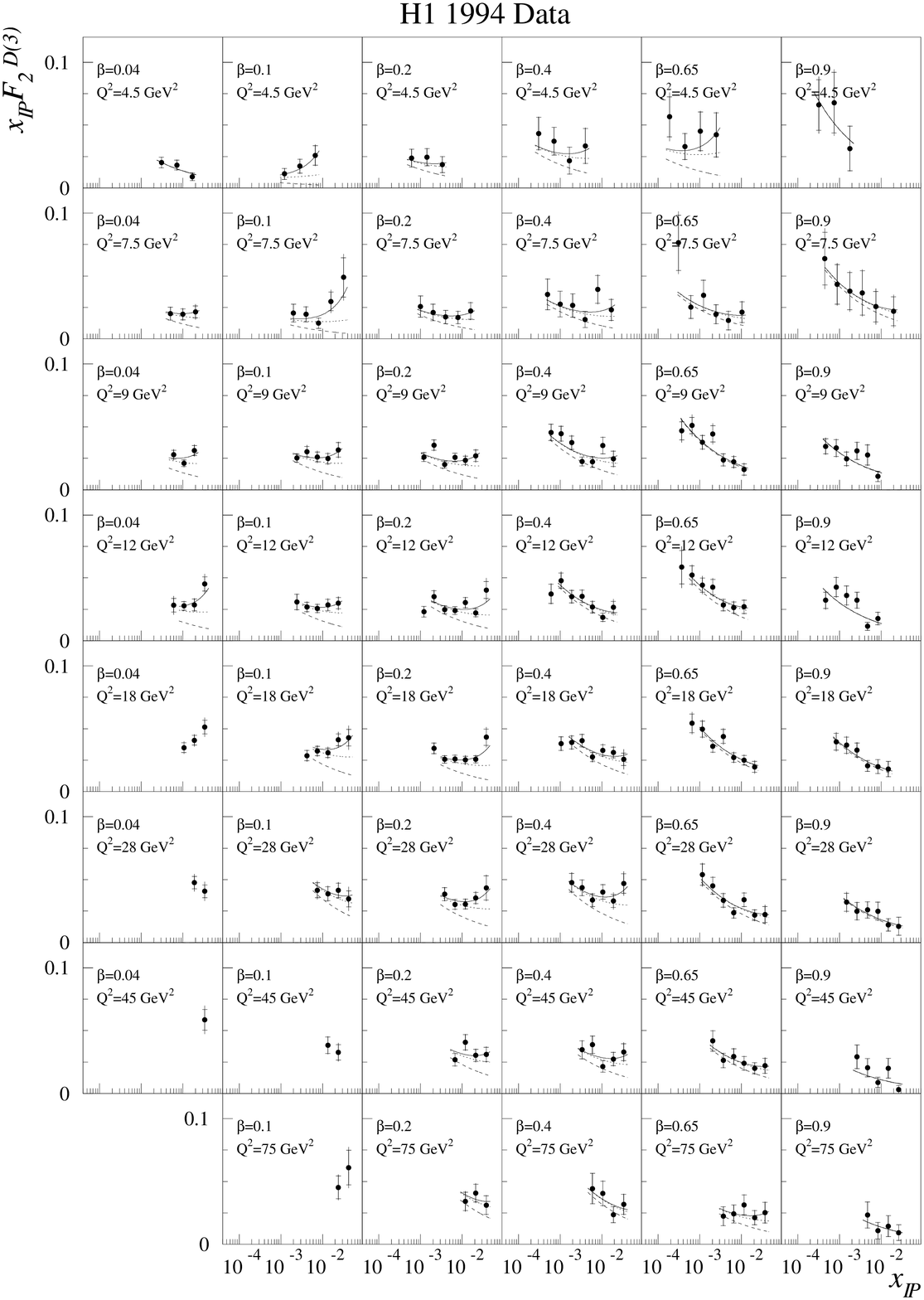}
{The diffractive structure function $\ftwodthreearg$ multiplied by
  $\xpom$ as a function of $\xpom$ (solid points) for fixed values of
  $\beta$ and $Q^2$ as denoted in the figure. The solid line is the
  result of a Regge fit in which the pomeron and the reggeon exchange
  contribute with maximum interference. Also shown is the pomeron
  contribution alone (dashed line) and including the interference
  (dotted line).  } {diff-h1f2d3}

The recent measurements of $\ftwodthree$ by the H1
experiment~\cite{h1_f2d2} are based on a ten-fold increase in
statistics (integrated luminosity of $1.96 \pbi$) and cover the
kinematic region, $4.5<Q^2<75 \gevtwo$, $2\cdot 10^{-4}<\xpom <0.04$
and $0.04<\beta <0.9$. The analysis requires the presence in the event
of a large rapidity gap, of at least 4 units in rapidity. The hadronic
final state is required to be contained in the main calorimeter which
provides, together with the tracking detectors, the measurement of the
invariant mass $M_X$. The results of these measurements are shown in
Fig.~\ref{fig:diff-h1f2d3}, where the values of $\xpom
\ftwodthreearg$ are displayed as a function of $\xpom$ for different
values of $\beta$ and $Q^2$. The $\xpom$ dependence of $\ftwodthree$
is seen to change with $\beta$ and remains fairly independent of
$Q^2$.

\epsfigure[width=0.8\hsize]{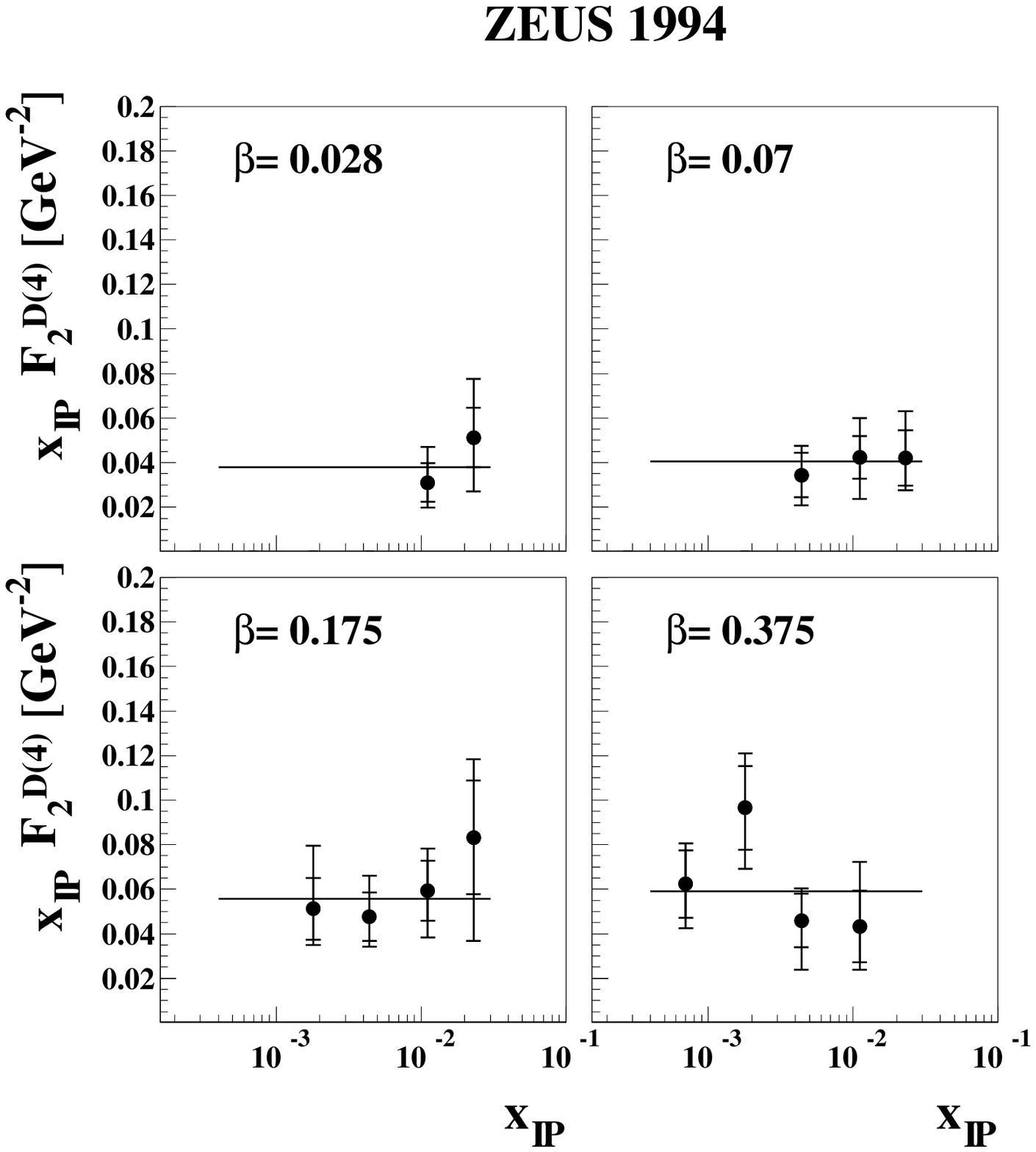}
{The diffractive structure function of $\ftwodfourarg$ multiplied by
  $\xpom$ as a function of $\xpom$ (solid points) for fixed values of
  $\beta$ as denoted in the figure, $Q^2=8 \gevtwo$ and $|t|=0.17
  \gevtwo$. Also shown is the result of the fit assuming an
  $(1/\xpom)^a$ dependence, independent of $\beta$.  } {diff-f2d4lps}

The ZEUS experiment has used a sample of LPS tagged DIS events to
measure $\ftwodfour$ in the range $5<Q^2<20 \gevtwo$, $x_L>0.97$, 
$0.015<\beta<0.5$ and $0.073<|t|<0.4 \gevmtwo$. The sample corresponds
to an integrated luminosity of $900 \nbi$. The results of the
measurement at $Q^2=8 \gevtwo$ and $|t|=0.17 \gevmtwo$ are displayed
in Fig.~\ref{fig:diff-f2d4lps}. The measurement are confined to
relatively large values of $\xpom$ especially at small $\beta$. The data
are compatible with Regge factorization, albeit the 
significance is limited by the dominant statistical error. The same
data sample was used to determine the $t$ distribution, which was
found to follow an exponential behavior as seen in
Fig.~\ref{fig:diff-tdislps}.

\epsfigure[width=0.8\hsize]{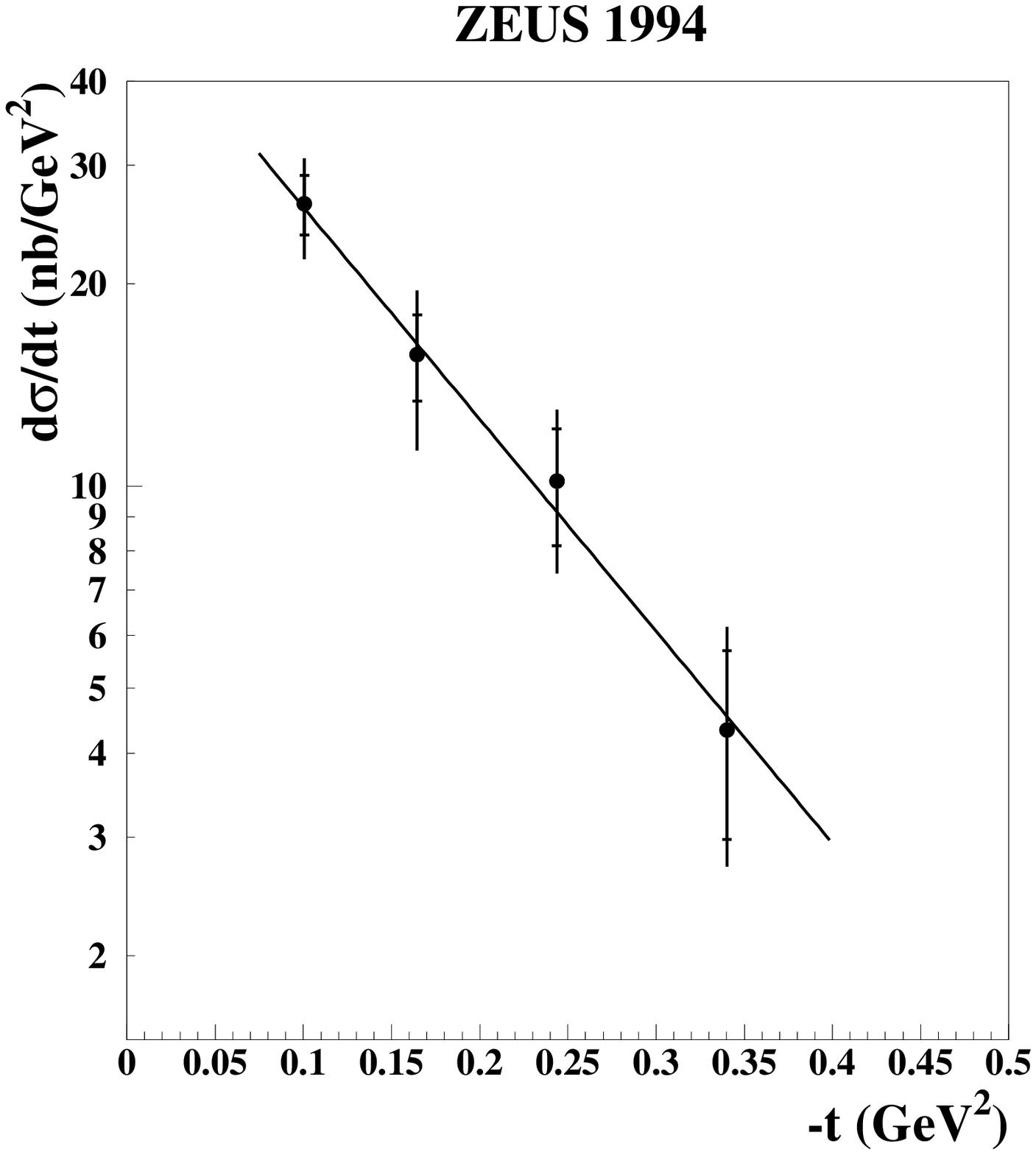}
{The differential cross section $d\!\sigma/d\!t$ for diffractive DIS
  events with a leading proton of $x_L>0.97$, in the range $5<Q^2<20
  \gevtwo$, $50<W<270 \gev$ and $0.015<\beta<0.5$. The line is the
  result of an exponential fit.}  {diff-tdislps}

For the 1994 DIS data, corresponding to an
integrated luminosity of $2.61 \pbi$, the ZEUS experiment has used the
$M_X$ method to determine the $W$ and $Q^2$ dependence of the $\gvp$
photon diffractive dissociation cross section~\cite{mxbrussels}.  The
measurements were performed for small masses, $M_X<15 \gev$, and cover
the range $7<Q^2<140 \gevtwo$ and $60 < W < 200 \gev$.  The results
are displayed in Fig.~\ref{fig:diff-sigmagp}. The differential cross
section $d\sigma/dM_X$ is shown as a function of $W$ in bins of $Q^2$
and $M_X$. The contribution of the photon dissociation accompanied by
small mass proton dissociation has not been subtracted. A strong
increase with $W$ is observed for masses $M_X<7.5 \gev$.

\epsfigure[width=0.8\hsize]{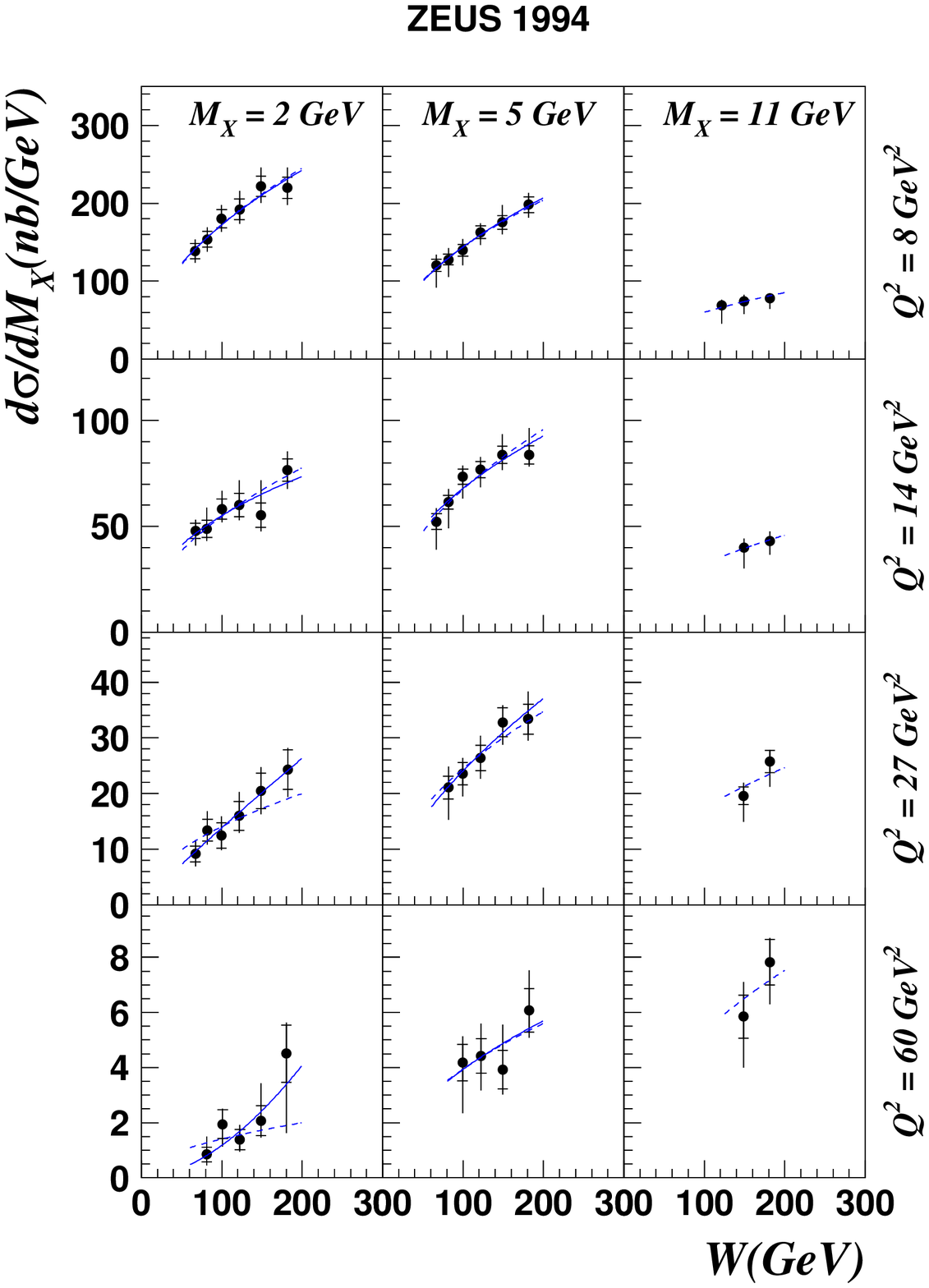}
{The differential cross section $d\!\sigma/d\!M_X$ as a function of
  the $\gvp$ center-of-mass of mass energy $W$ for fixed values of
  $Q^2$ and for ranges of $M_X$ as denoted in the figure. The solid
  line is the result of a fit with the function
  $d\!\sigma/d\!M_X\propto W^a$ in each of the bins separately. The
  dashed line is the result of the fit with $a$ the same for all bins.
  } {diff-sigmagp}

The corresponding $\ftwodthree$ shows a reasonable agreement with the
H1 measurements~\cite{h1_f2d2} and the LPS measurements in the overlap
region.

\paragraph{The $\xpom$ and $W$ dependence}

In the simple Regge model in which $\ftwodthree$ factorises into 
\begin{equation}
\ftwodthree=\pomfluxarg \ftwopomarg \, , 
\end{equation}the $\xpom$ 
dependence is expected to follow a $\xpom^{-n}$ dependence, where $n=2
\aveapom -1$, independently of $\beta$ and $Q^2$. Here $\aveapom$
stands for a value of $\apom(t)$ averaged over the $t$ distribution.
The value of $n$ can be also obtained from the $W$ dependence of the
$\gvp$ cross section for fixed $Q^2$ and $M_X$,
\begin{equation}
\sigma^D(\gvp) \prop Q^2 \ftwod \prop
Q^2 \xpom^{-n} \prop Q^2\left(\frac{W^2+Q^2}{M_X^2+Q^2}\right)^n \prop
W^{2n} \, , 
\end{equation} 
where in the last step the approximation $Q^2\ll W^2$ is made.

A fit to the H1 data~\cite{h1_f2d2} with a single power $n$ yields a poor
$\chi^2$. The exchange of a single factorisable Regge trajectory in
the $t$ channel does not provide an acceptable description. However if
one assumes in addition to the pomeron exchange the contribution of a
sub-leading reggeon trajectory, the quality of the fit improves
tremendously. In the fit the intercept of the pomeron $\apom(0)$ and
of the reggeon $\areg(0)$ are kept as free parameters, as well as
their relative contribution. The results are,
\begin{eqnarray}
\apom(0)&=&1.203 \pm 0.020  \pm 0.013
^{+ \textstyle 0.030}_{-\textstyle 0.035}  \, ,
\\ \label{apomh1}
\areg(0)&=&0.50 \pm 0.11 \pm 0.11 
^{+\textstyle 0.09}_{-\textstyle 0.10} \, , 
\label{aregh1}
\end{eqnarray} 
where the last error depends on the details of how the two
contributions are added. The results of one of the models, in which a
maximal interference between the pomeron and reggeon exchange is
assumed, is presented in Fig.~\ref{fig:diff-h1f2d3}.

The value of $\areg(0)$ agrees with the value
of $\simeq 0.55$ obtained in an analysis of total hadronic cross
sections~\cite{ref:DoLa_sigfit}. The value of $\apom(0)$ is
significantly larger than expected for the soft
pomeron~\cite{ref:DoLa_sigfit,ref:Cudell}.

\epsfigure[width=0.8\hsize]{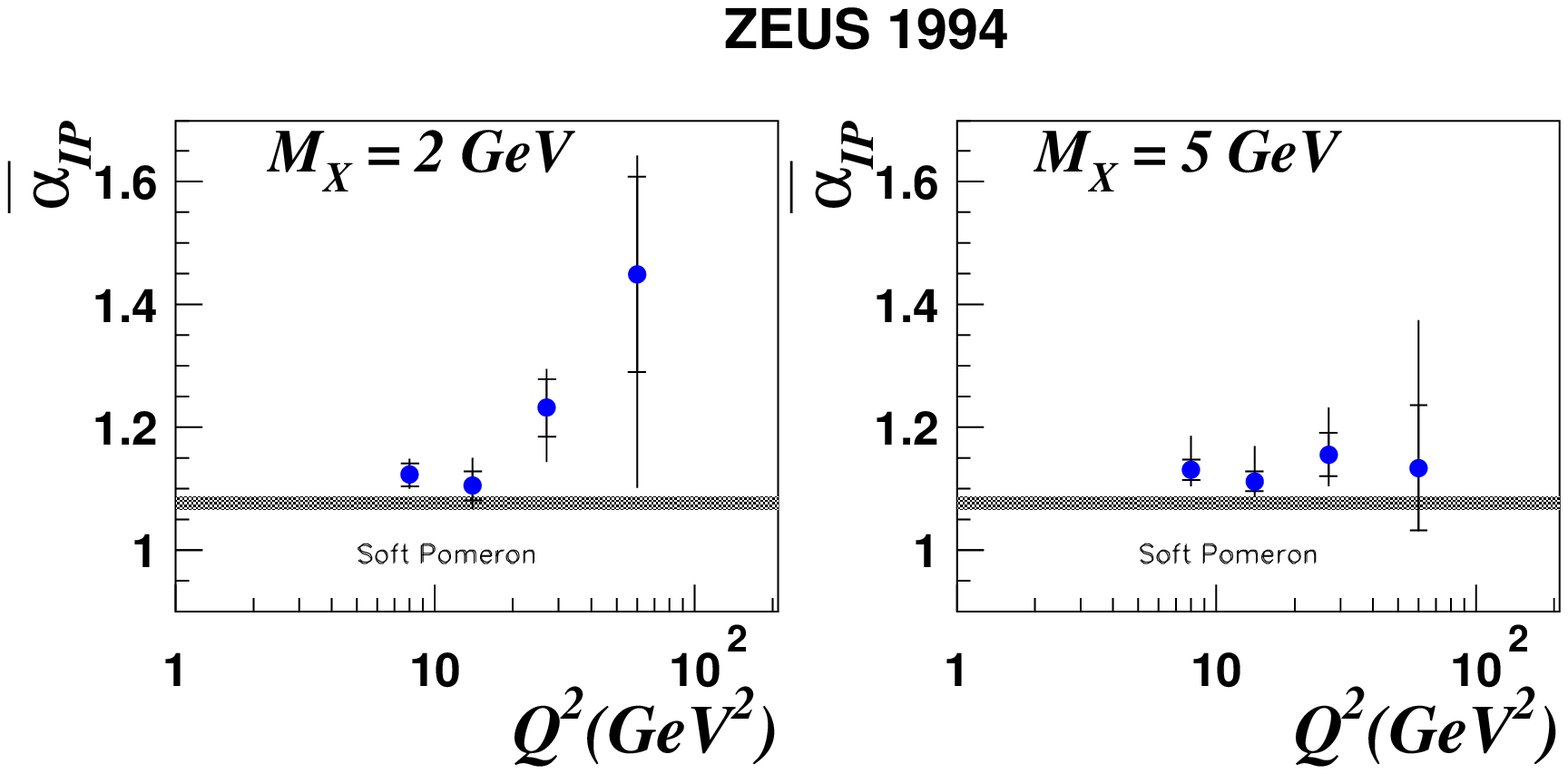} {The value of
  $\aveapom$ as a function of $Q^2$ and for two mass ranges $M_X$ as
  derived from the fit to the $W$ dependence of the cross section.
  Also shown are the result expected for the soft pomeron
  exchange~\protect\cite{ref:Cudell} assuming $\aprime=0.25 \gevmtwo$.
  } {diff-apomvsq2}

In the ZEUS data, $\apom(0)$ is derived from fits to the $W$
dependence of the $\gvp$ cross section, assumed to have the form
$W^a$, at each value of $Q^2$ and mass $M_X$~\cite{mxbrussels}. The
value of $a$ is related to $\apom(0)$ through the relation
\[a=4\left(\apom(0)-1-[0.03]\right) \, ,\] 
where the term in brackets is a correction due to averaging over the
$t$ distribution, assuming $\apom^\prime=0.25 \gevmtwo$ and $b=7
\gevmtwo$. The fitted curves are displayed in
Fig.~\ref{fig:diff-sigmagp} while the results of the fit for $\apom$
are presented in Fig.~\ref{fig:diff-apomvsq2} as a function of $Q^2$
for two ranges of masses. For all values of $Q^2$ and $M_X$ the value
of $\apom$ lies above the highest expectations from soft Regge
phenomenology~\cite{ref:Cudell}. Given the large systematic errors at
large $Q^2$ it is not possible to draw conclusions on any possible
$Q^2$ dependence. The value of $\apom$ averaged over the whole
measured kinematic range is,
\begin{equation}
\apom(0) = 1.16 \pm 0.01 ^{+\textstyle 0.04}_{-\textstyle 0.01}
 \, , 
\end{equation}
compatible within errors with the value obtained in the H1 analysis.
No correction for contribution of sub-leading trajectories was
attempted as the measurements correspond to $\xpom \lesssim 0.01$ in which
the pomeron exchange is dominant. The presence of sub-leading reggeon
contributions may however explain the relative flatness of the $\xpom
\ftwodfour$ distribution measured with the LPS tag at larger $\xpom$
values (see Fig.~\ref{fig:diff-f2d4lps}).

\epsfigure[width=0.8\hsize]{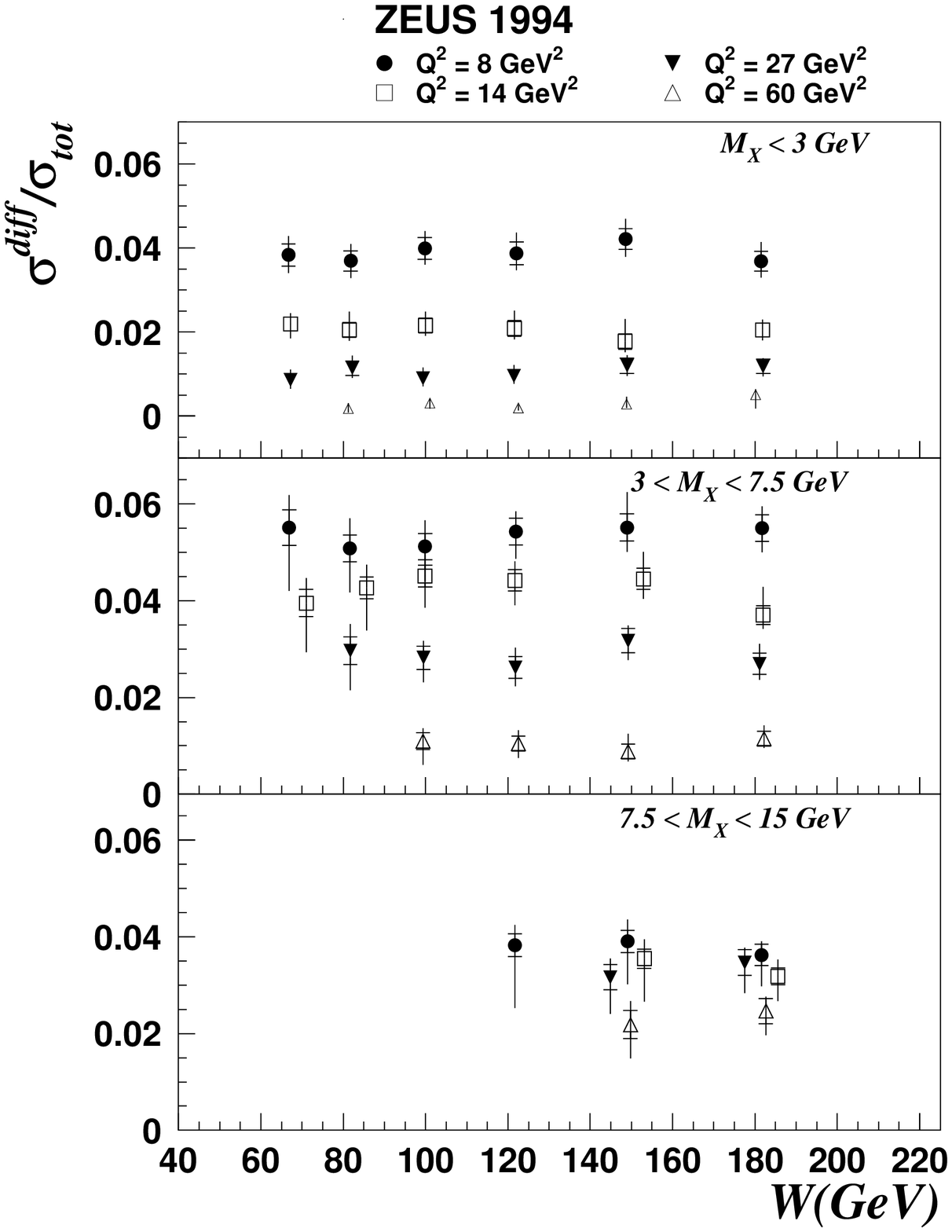}
{Ratio of the diffractive cross section $\sigma^{\rm diff}$,
  integrated over the mass range indicated in the figure, to the total
  DIS cross section $\sigma_{\rm tot}$ as a function of $W$ for
  different values of $Q^2$.  } {diff-ratio}

The $W$ dependence of the diffractive cross section as measured in
ZEUS had been compared to the $W$ dependence of the total $\gvp$ cross
section $\sigma_{\rm tot}$. The latter was derived from a
parameterization of the measured $F_2$ of the proton. The ratio $r_D$
defined as
\begin{equation}
r_D=\frac{1}{\sigma_{\rm tot}(\gvp)}\int_{M_X \mathrm{bin}}
\frac{d \sigma_D(\gvp)}{d\!M_X} d\!M_X \, ,
\label{eq-diff:ratio}
\end{equation}
is plotted in Fig.~\ref{fig:diff-ratio} as a function of $W$. It is
interesting to note that the ratio is within errors constant with $W$.
This is unlike hadron-hadron interactions for which this ratio
decreases strongly with $W$~\cite{goulianos3}.

\paragraph{The $Q^2$ dependence}

\epsfigure[width=0.9\hsize]{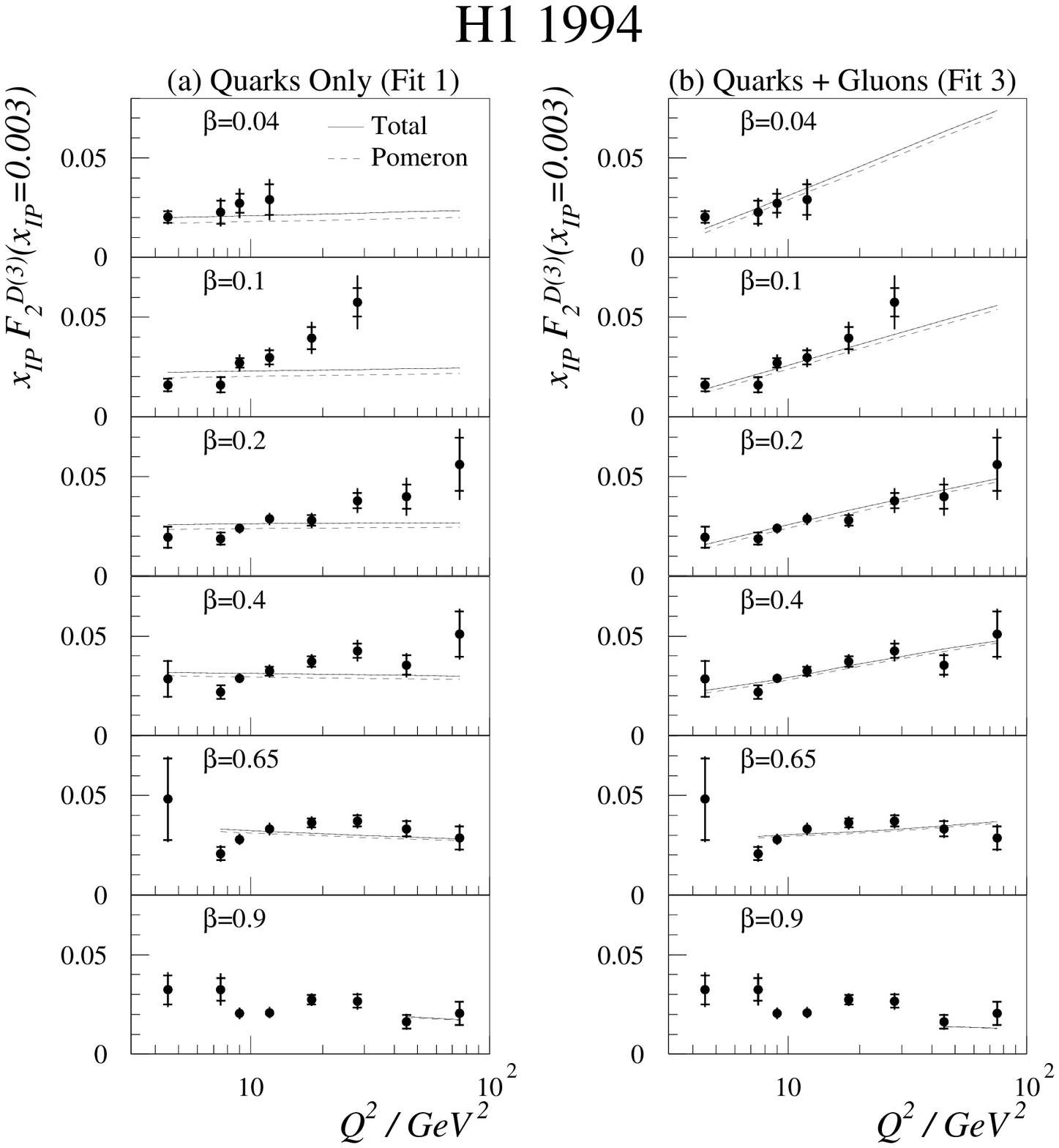}
{The structure function $\xpom \ftwodthree$ as a function of $Q^2$ for
  $\xpom = 0.003$ and for different values of $\beta$ as denoted in
  the figure. The solid line is the result of the QCD fit which
  includes the evolution of the pomeron and reggeon contributions. The
  dashed line represents the pomeron contribution only. Figures (a)
  and (b) differ in the assumptions about parton distributions in the
  pomeron at $Q^2_0=3 \gevtwo$ as described in the figure.  }
{diff-f2d3h1q2}

The $Q^2$ dependence of $\xpom \ftwodthree$ has been determined by H1
for $\xpom=0.003$ and is shown in Fig.~\ref{fig:diff-f2d3h1q2} for a
range of $\beta$ values. Scaling violation is observed for $\beta \leq
0.4$ where $\ftwodthree$ increases with $Q^2$. For larger $\beta$
values the $Q^2$ dependence is mild with a tendency of $\ftwodthree$
to decrease with increasing $Q^2$. The pattern of scaling violation is
different from the one observed for the inclusive $F_2$ of the proton,
especially at larger values of $\beta$. The slow decrease with $Q^2$
interpreted in terms of QCD evolution signals a substantial presence
of gluons at large $\beta$. The radiation of quarks by gluons
compensates the migration of quarks radiated by quarks towards lower
values of $\beta$.

The $Q^2$ dependence of $r_D$ at low mass is much stronger (see
Fig.~\ref{fig:diff-ratio}) than at larger masses where it almost
disappears, indicating that the diffractive production of a fixed low
mass is suppressed roughly by an extra power of $Q^2$ relative to the
total cross section.  This is not necessarily in variance with the
milder logarithmic dependence observed in the $H1$ data as it may
reflect the large $\beta$ dependence of $\ftwodthree$. For a fixed
$M_X$ the increase of $Q^2$ corresponds to an increase in $\beta$.
The $Q^2$ dependence observed here is not unlike the one observed in
exclusive vector meson production.

\paragraph{The $\beta$ dependence}

\epsfigure[width=0.9\hsize]{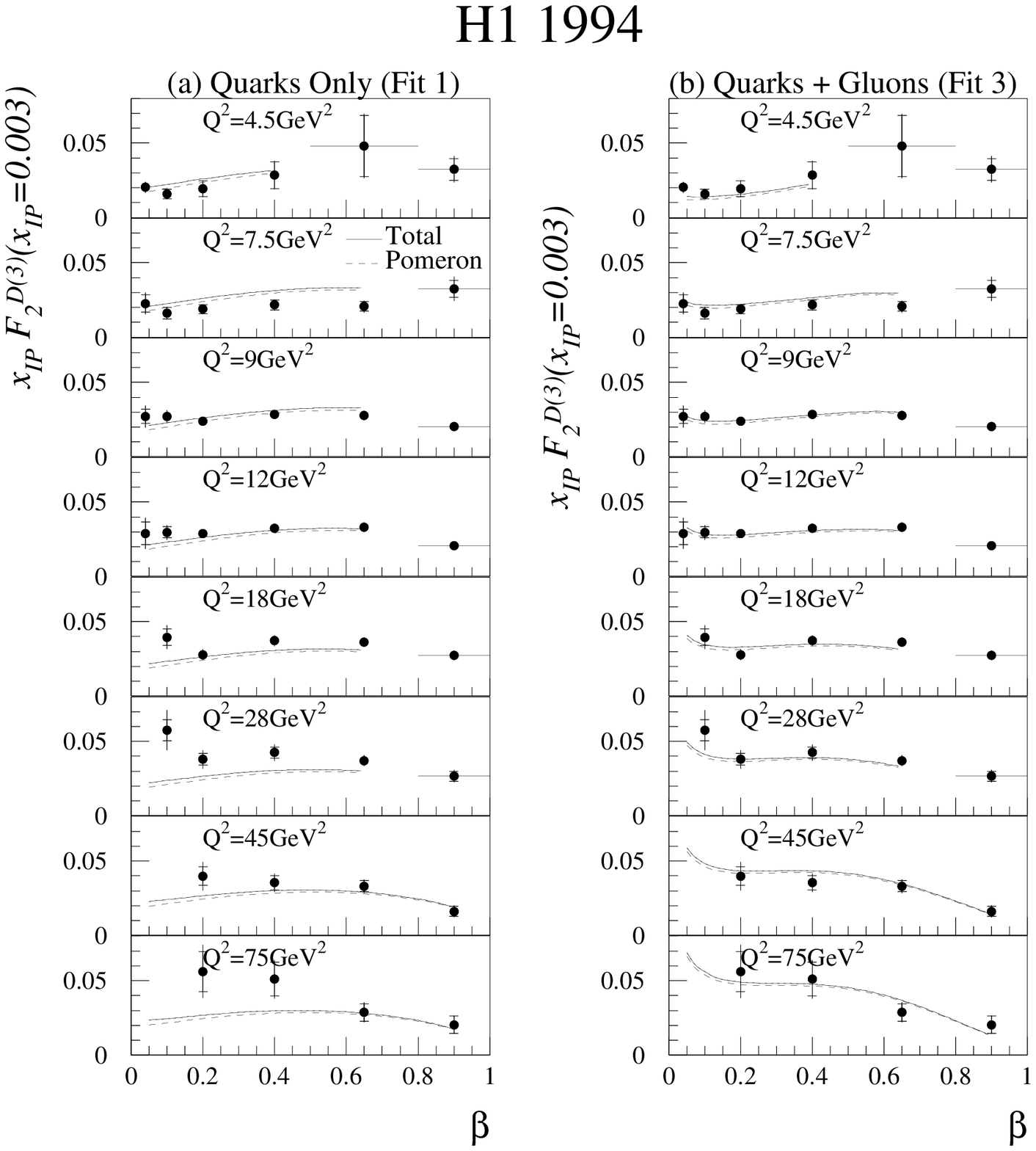}
{The structure function $\xpom \ftwodthree$ as a function of $\beta$
  for $\xpom = 0.003$ and for different values of $Q^2$ as denoted in
  the figure. The solid line is the result of the QCD fit which
  includes the evolution of the pomeron and reggeon contributions. The
  dashed line represents the pomeron contribution only. Figures (a)
  and (b) differ in the assumptions about parton distributions in the
  pomeron at $Q^2_0=3 \gevtwo$ as described in the figure.  }
{diff-f2d3h1beta}

The $\beta$ dependence of $\xpom \ftwodthree$ as determined by H1 for
$\xpom=0.003$ is shown in Fig.~\ref{fig:diff-f2d3h1beta} for different
$Q^2$ values. The $\beta$ dependence is relatively flat at small $Q^2$
and increases towards lower $\beta$ as $Q^2$ increases. The same
effect is observed by the ZEUS experiment.

\paragraph{The $t$ dependence}

The $t$ distribution measured by ZEUS (see
Fig.~\ref{fig:diff-tdislps}) has been fitted with a single
exponential function yielding a value of the slope $b$,
\begin{equation}
b=7.2 \pm 1.1  ^{+ \textstyle 0.7}_{- \textstyle 0.9} \gevmtwo \, .
\label{eq-diff:tslope}
\end{equation}
It is interesting to note that the slope of the $t$ distribution at an
average $Q^2 = 8 \gevtwo$ is larger than the slope of the $t$
distribution measured in this experiment for exclusive $J/\psi$
photoproduction~\cite{ref:ZEUS_BPC_rho} ($b=4.6 \pm 0.6~\gevmtwo$).
This may be an indication that in the mass range covered by the LPS
($0.015<\beta<0.5$)), soft processes contribute to virtual photon
dissociation.

\paragraph{Comparison with Models}

\epsfigure[width=0.8\hsize]{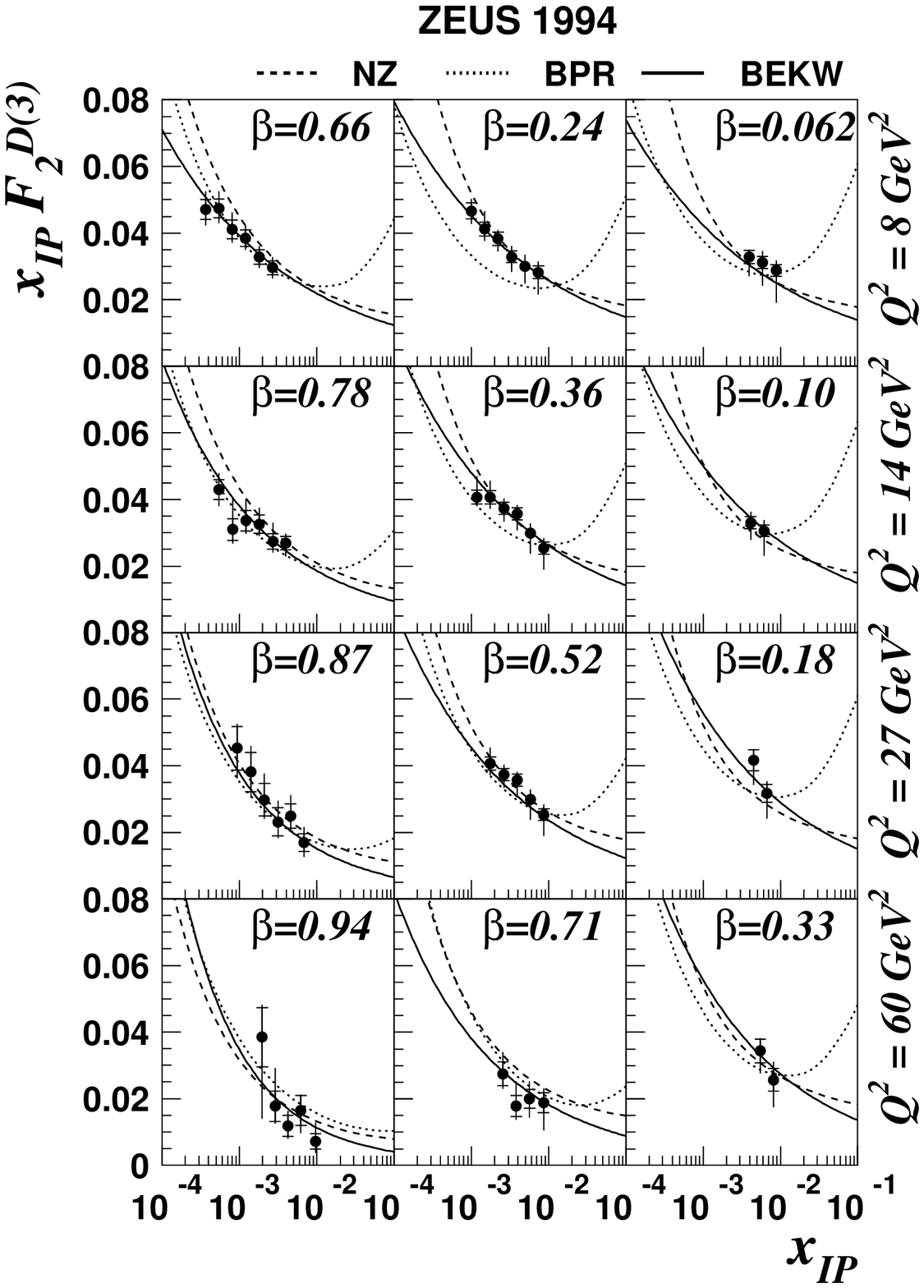}
{The structure function $\xpom \ftwodthree$ as a function of $\xpom$,
  obtained by ZEUS with the $M_X$ method (solid points) compared to
  expectations of QCD motivated models:
  BEKW~\protect\cite{bartels-kowalski}, NZ~\protect\cite{NikZak1},
  BP~\protect\cite{bialas2}. Note that the BP curve includes the
  contribution of the reggeon trajectory as determined by the H1
  experiment~\protect\cite{h1_f2d2}.}  {diff-f2d3theo}

The ZEUS measurements of $\ftwodthree$ have been compared to the
predictions of a selected number of models representative of the
present theoretical ideas behind inclusive DIS diffractive scattering:
\begin{enumerate} 
\item the model of~\citeasnoun{NikZak2};
\item the model of~\citeasnoun{bartels-kowalski}; 
\item the model of~\citeasnoun{bialas2}; 
\end{enumerate}
The predictions are shown in Fig.~\ref{fig:diff-f2d3theo}.

In the semi-classical approach to diffractive DIS~\cite{BH1},
for which there are no numerical predictions as yet, it is expected
that the $W$ dependence of diffractive dissociation will be similar to
that of the inclusive DIS, a feature which is borne in the data.

\subsubsection{Final states in diffractive DIS} 

The topological structure of the hadronic final states emerging from
the diffractive dissociation of virtual photons should reflect the
underlying production mechanism. It is convenient here to use the
language of photon partonic fluctuations. For diffractive scattering
dominated by the AJM configurations, one would expect the final state
in the $\gamma^* \pom$ centre of mass system to consists of two jets
of particles aligned along the $\gamma^*\pom$ collision axis. The
small size configurations would materialize as two jets aligned along
an axis rotated relative to the $\gamma^*\pom$ collision axis to
reflect the large $k_T$ of the quarks relative to the photon. Here an
analogy with the final states observed in $e^+e^-$ annihilation into
two quarks can be drawn. In models in which the pomeron is viewed as
consisting of quarks and gluons, large $k_T$ jets in the
$\gamma^*\pom$ system would be accompanied by the "remnant of the
pomeron".

\epsfigure[width=0.8\hsize]{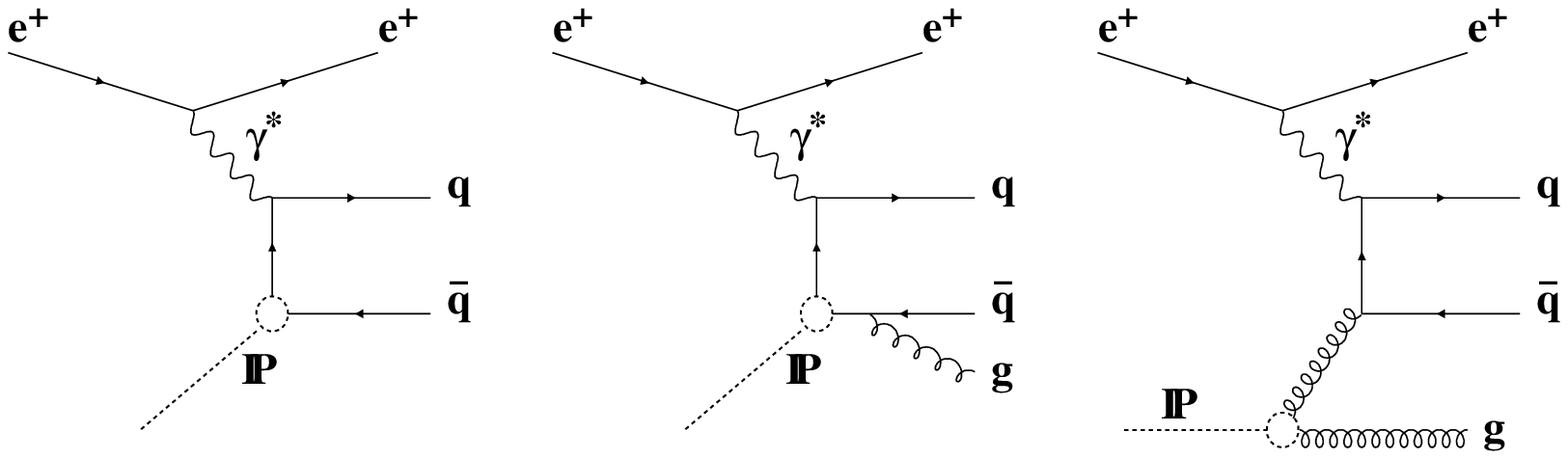}
{Diagrams representing possible partonic configurations of 
  the photon dissociation in DIS. From left to right, the Born term,
  QCD Compton process and boson-gluon fusion process.   }
{diff-hfsdiag}

Possible configurations of the final state are shown diagrammatically
in Fig.~\ref{fig:diff-hfsdiag}. In each configuration some $k_T$ can
be assigned to the partons, in particular for the QCD Compton and
boson-gluon fusion diagram. However, even the initial partons in the
pomeron could have large $k_T$.

The relatively small invariant mass $M_X$ of the dissociative system
makes a jet-search analysis of the final states very inefficient.
Topological studies can be carried out by invoking event shape
variables. These variables were very successful in describing the
partonic nature of the hadronic final states in $e^+e^-$ scattering at
centre of mass energies which were too low to resolve three-jet from
two-jet events~\cite{PLUTO} and which correspond to
the typical values of $M_X$ considered here.

Charm production is also believed to be a sensitive probe for the
mechanism underlying diffractive hard scattering~\cite{BDH2}. The mass
of the charm quark provides the hard scale for the perturbative
approach to be applicable. 

\paragraph{Event shape variables}

The momentum tensor for a state $X$, consisting of $N$ particles with
momentum vectors $\vec{p}_i$ in the rest frame of system $X$, is
defined as
\begin{equation}
Z_{m,n}=\sum_{i=1}^{N}p_{im}p_{in} \, ,
\end{equation}
where the subscripts $m$ and $n$ denote the three
coordinates of vector $\vec{p}$~\cite{momentumtensor}. The
diagonalization of $Z_{mn}$ yields three axes $\vec{n}_k$ and three
eigenvalues $\lambda_k = \sum_{i=1}^N(\vec{p}_i \cdot \vec{n}_k)^2$.
The normalized eigenvalues
$\Lambda_k=\lambda_k/\sum_{i=1}^N(\vec{p}_i)^2$ can be ordered in such
a way that $\Lambda_1<\Lambda_2<\Lambda_3$. The corresponding unit
vectors $\vec{n}_i$ define a reference frame in which vector
$\vec{n}_3$ is the unit vector along the principal axis, the so called
sphericity axis, which minimizes the sum of the squared transverse
momenta. The event plane is defined by $\vec{n}_2$ and $\vec{n}_3$,
while $\vec{n}_1$ defines the direction perpendicular to the event
plane. Sphericity $S$ is defined as
\begin{equation}
S=\frac{3}{2}(\Lambda_1+\Lambda_2)=\frac{3}{2}\min_{\vec{n}}
\frac{\sum_{i=1}^{n} p_{T_{i}}^{2}}{\sum_{i=1}^{n}
  p_i^{2}}\;\;\;\;\;\;\;\;(\vec{n} = \vec{n}_3).
\label{eq-diff:sphericity}
\end{equation} 
Sphericity indicates the total $p_T^2$ with respect to the event axis.
For isotropic distributions of particles in the phase space $S=1$. In
models with a constant limited transverse momentum of particles
relative to the interaction axis, the mean sphericity values are
inversely proportional to the centre of mass energy. For pencil like
two-jet events $S=0$.  For collimated jets, the polar angle $\theta_S$
of the sphericity axis with respect to the $\gamma^* \pom$ axis is a
measure of the alignment of the jets with respect to the interaction
axis.
    
Alternatively one can use the thrust variable
$T$~\cite{thrust2,thrust1}. The thrust axis, denoted by the unit
vector $\vec{n}_T$, is the direction in space which maximizes the
longitudinal momentum of particles.  Thrust is then defined as
\begin{equation}
T=\max_{\vec{n}} \frac{\sum_{i=1}^{n} | \vec{p}_i \cdot
  \vec{n} |} {\sum_{i=1}^{n} |\vec{p}_i |}
\;\;\;\;\;\;\;\;(\vec{n} = \vec{n}_T).
\end{equation}
Isotropic events are characterized by $T\simeq 0.5$, while for
collimated two-jet events $T\simeq 1$. For a symmetric three particle
configuration $T=2/3$ and the thrust axis is arbitrary, while for an
asymmetric topology the thrust axis is pointing in the direction of
the most energetic particle and $T>2/3$. 

\epsfigure[width=0.8\hsize]{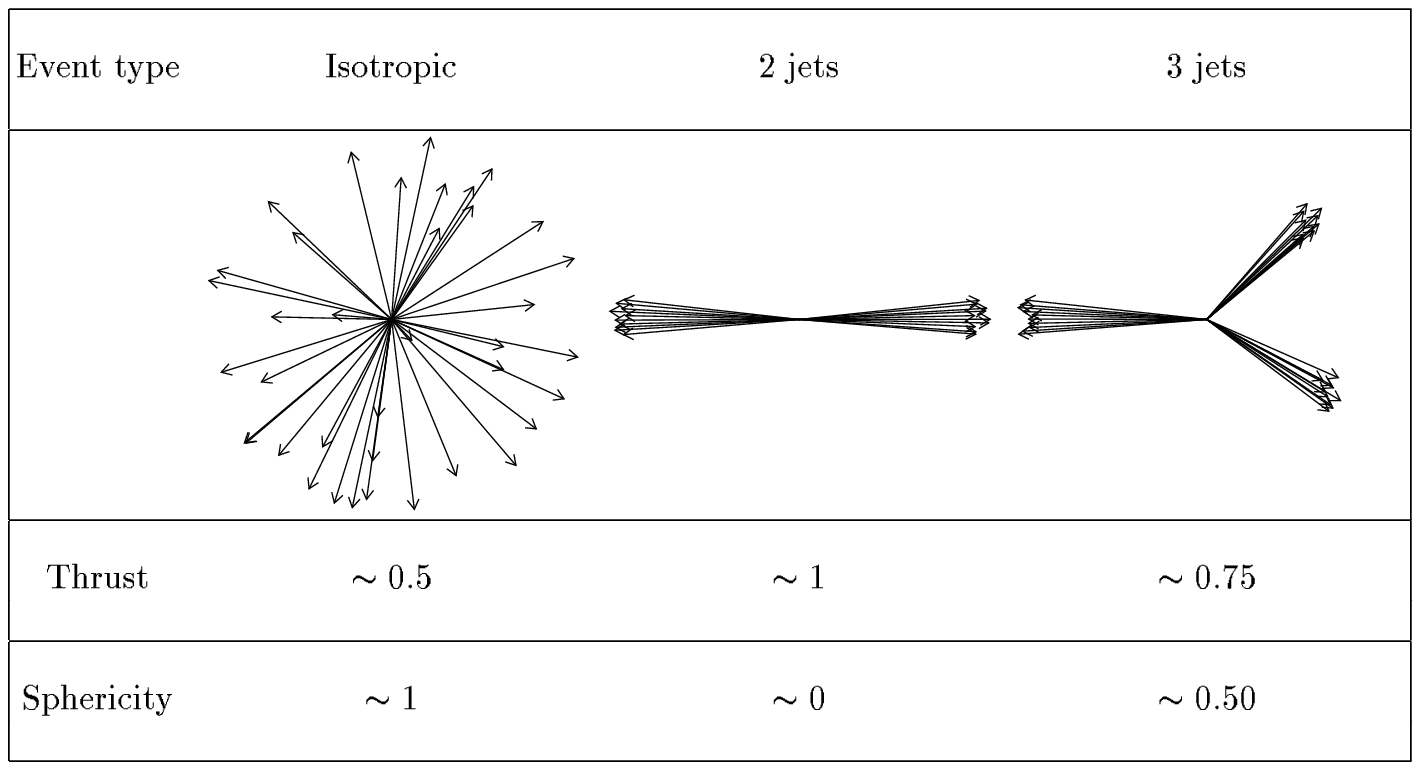} {Values of
  thrust and sphericity expected for different configurations of
  hadrons in the final state.}  {diff-st}

Once the thrust axis is found, the hadrons can be clustered into two
jets, based on the sign of their longitudinal momentum projected onto
$\vec{n}_T$. By definition the two jets will have equal in size and
opposite in direction momentum-vectors.  One can then defined the
thrust jet transverse momentum $P_T$, as the transverse momentum of
one of the jets relative to the $\gamma^*\pom$ axis. For large values
of $P_T$ the value of $T$ will differentiate between two partons with
large $k_T$ and a three parton configuration.  

The values of $T$ and $S$ expected for three selected configurations
of hadrons are summarized in Fig.~\ref{fig:diff-st}.

\paragraph{Event shapes in LRG events}

\epsfigure[width=0.8\hsize]{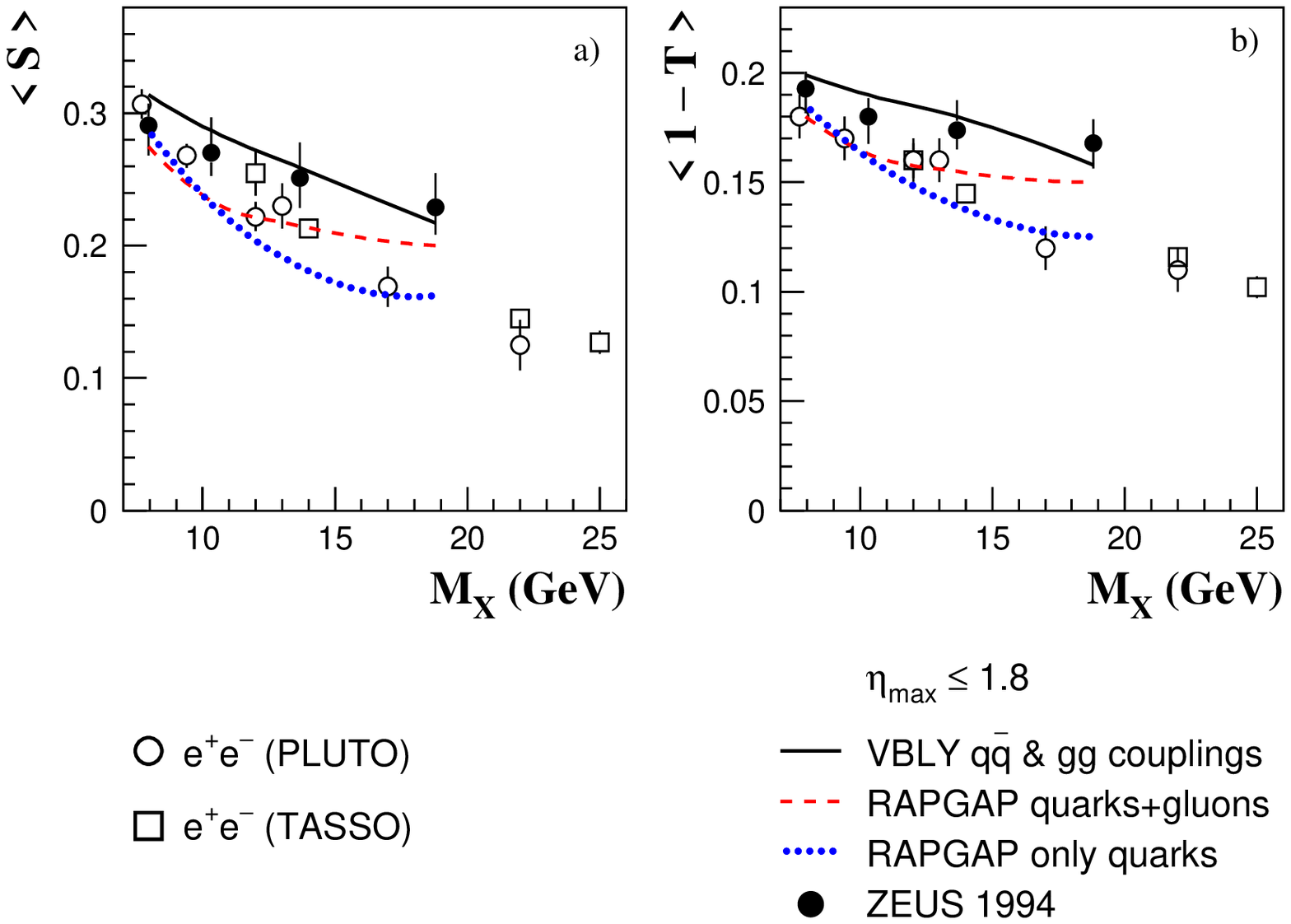} {The mean
  sphericity, $\langle S \rangle$, a) and one minus mean thrust,
  $\langle 1-T \rangle$, b) as a function of the mass $M_X$.  The ZEUS
  data (solid dots) selected with an $\etamax < 1.8$ cut cover the
  kinematic range $5\le Q^2 \le 185 \gevtwo$ and $160\le W \le 250
  \gev$. Also shown are data from $e^+e^-$
  annihilation~\protect\cite{PLUTO,TASSO1}. The lines represent the
  expectations of various MC models in the kinematic range of the ZEUS
  data.  } {diff-ZEUST}

\epsfigure[width=0.8\hsize]{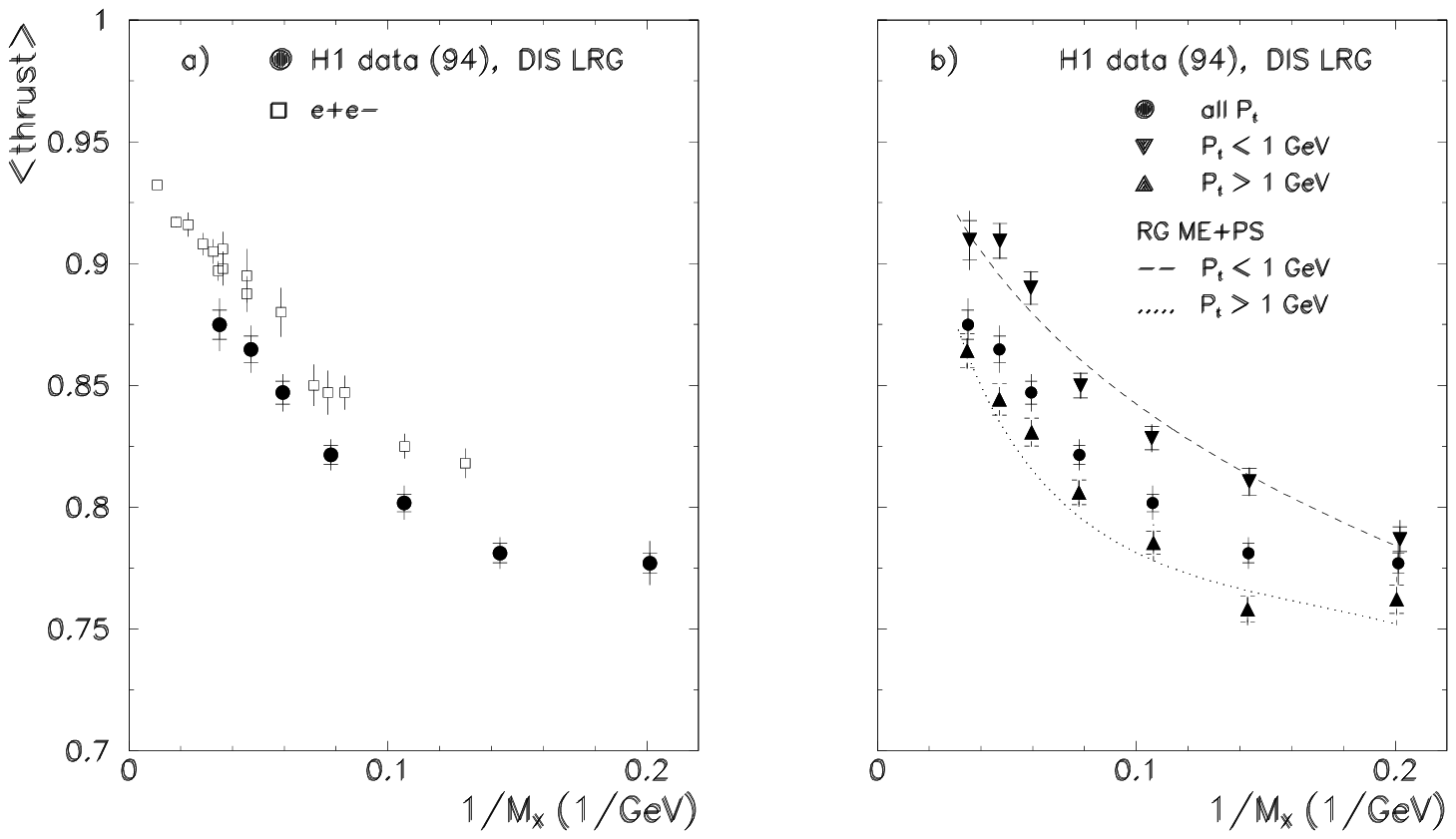} {The mean thrust
  as a function of the inverse mass $1/M_X$ for a) DIS events selected
  with a large rapidity gap in the region $10<Q^2<100 \gev2$, $\xpom
  <0.05$ (solid point) and $e^+e^-$ annihilation
  data~\protect\cite{PLUTO,MarkII,TASSO1,TASSO2,AMY,DELPHI} (empty
  squares).  In b), the data are compared with expectations of the
  RAPGAP MC model (RG ME+PS) for two samples od DIS events with
  different thrust transverse momentum $P_T$.}  {diff-H1T}

Both HERA experiments have studied the event shapes in the DIS photon
diffractive dissociation~\cite{zeus_evsh,h1_evsh}.  The studies of the
ZEUS experiment~\cite{zeus_evsh} in the region of $5\le Q^2 \le 185
\gevtwo$ are confined to a sample of events selected with requirements
of $\etamax\le 1.8$ and $7\le M_X \le 25 \gev$. To increase the
acceptance of diffractive events selected with these cuts the $W$
range is further restricted to $160\le W \le 250 \gev$. The selection
of H1 follows that for the $\ftwodthree$ analysis, where the system
$X$ is contained within the central calorimeter and a large rapidity
gap of about 4 units in the forward region is required. The analysis
is performed for $10<Q^2<100 \gevtwo$, $\xpom <0.05$ and $4<M_X<36
\gev$.

Both experiments observe that as the mass $M_X$ increases the hadronic
final state is more and more collimated along the thrust or sphericity
axis. This is shown in Fig.~\ref{fig:diff-ZEUST}a,b for the ZEUS data
and in Fig.~\ref{fig:diff-H1T}a for the H1 data. The sphericity is
seen to decrease, while the thrust is increasing with increasing
$M_X$. This is a sign of jet formation along the respective axes. The
level of collimation is compared to the one measured in $e^+e^-$
experiments, which is representative of a final state consisting of a
$q\bar{q}$ pair with possible extra gluon radiation.
Both experiments observe that the jets in diffractive events are less
collimated than in $e^+e^-$ interactions indicating that $q\bar{q}$
configurations alone cannot explain the observed features and higher
parton multiplicity are required.  This conclusion is supported by the
comparison of data with the RAPGAP MC model with final states
consisting of only $q\bar{q}$ pairs which does not give a good
representation of the data.

\epsfigure[width=0.8\hsize]{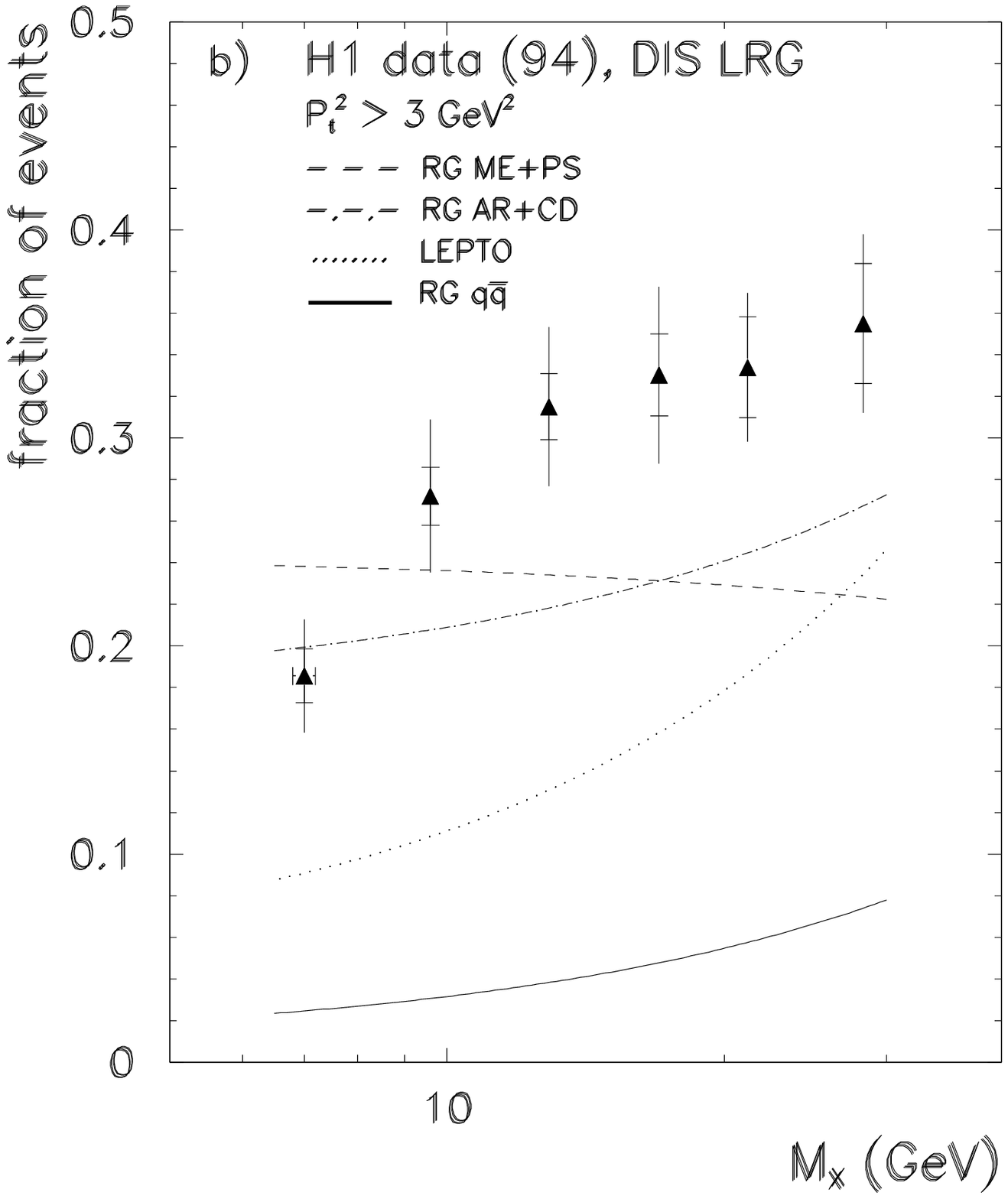} {Fraction of
  events with thrust transverse momentum $P_T^2>3 \gevtwo$ as a
  function of $M_X$ (data points), compared to the predictions of
  various MC models which differ in the treatment of the origin of the
  large rapidity gaps (RG - RAPGAP, factorizable partonic pomeron;
  LEPTO - soft color interactions) and in the final state
  fragmentation. ME+PS stands for matrix elements with parton showers
  for higher order corrections, while AR+CD stands for the ARIADNE
  color dipole model.}  {diff-H1frac}

The data were also compared to the expectations of the RAPGAP MC in
which both the quark and the gluon component of the pomeron, as well
as all the diagrams of Fig.~\ref{fig:diff-hfsdiag}, are included in
the generation of final states (see Fig.~\ref{fig:diff-ZEUST}a,b and
\ref{fig:diff-H1T}b). The MC samples used by the two experiments
differ in the description of the partonic content of the pomeron and
in the fragmentation of the final states.  The version used by the H1
experiment, which was tuned to reproduced best the cross section
measurements and the characteristics of the hadronic final states,
describes the $1/M_X$ dependence of $T$ very well.  However even in
this version of the RAPGAP model, the tail of the large $P_T^2$ is not
properly reproduced as can be seen in Fig.~\ref{fig:diff-H1frac}.

The ZEUS data have been compared to the VBLY model~\cite{VBLY} which
differs from the RAPGAP model in that the partons, quarks and gluons,
which couple to the photon, acquire a substantial $k_T$ as their
distribution is obtained assuming a point-like coupling to a scalar
particle. This leads effectively to larger $P_T$ values.  In the
particular region of phase space selected in the ZEUS experiment, the
VBLY model gives a better description of the ZEUS data than the RAPGAP
model.

Note that the $\etamax$ cut applied to select the diffractive sample
in the ZEUS experiment was not corrected for, as the correction was
found to be strongly model dependent~\cite{zeus_evsh}. For larger
masses, this cut may bias the hadronic configurations towards those
which have been produced with a large transverse momentum relative to
the $\gamma^*p$ axis. It may well be that the excess of events with
large $P_T$ over the RAPGAP expectations observed in the H1 sample and
the disagreement of the ZEUS data with RAPGAP have a common origin.

In conclusion the features observed in the study of event shape
variables cannot be explained by a pure $q\bar{q}$ configuration, even
assuming a large relative $k_T$ of the pair. A substantial
contribution of $q\bar{q}g$ or higher partonic multiplicities is
required in the final states.

\paragraph{Charm production}

The experimental signature for charm production is a particular decay
channel of the $D^{*\pm}$ mesons. The $D^*$ mesons are reconstructed
from their decay products through the chain $D^*\! \rightarrow
D^0\pi_s^+ \!  (K^-\pi^+)\pi_s^+$ (and the charge conjugate), where
the subscript $s$ stands for the slower pion of the pair. The small
mass difference $M(D^*) - M(D^0)=145.42 \pm 0.05 \mev$ yields a
prominent signal just above the threshold of the
$\Delta M=M(K\pi\pi_s)-M(K\pi)$ distribution, where the phase space
contribution is highly suppressed.

\epsfigure[width=0.8\hsize]{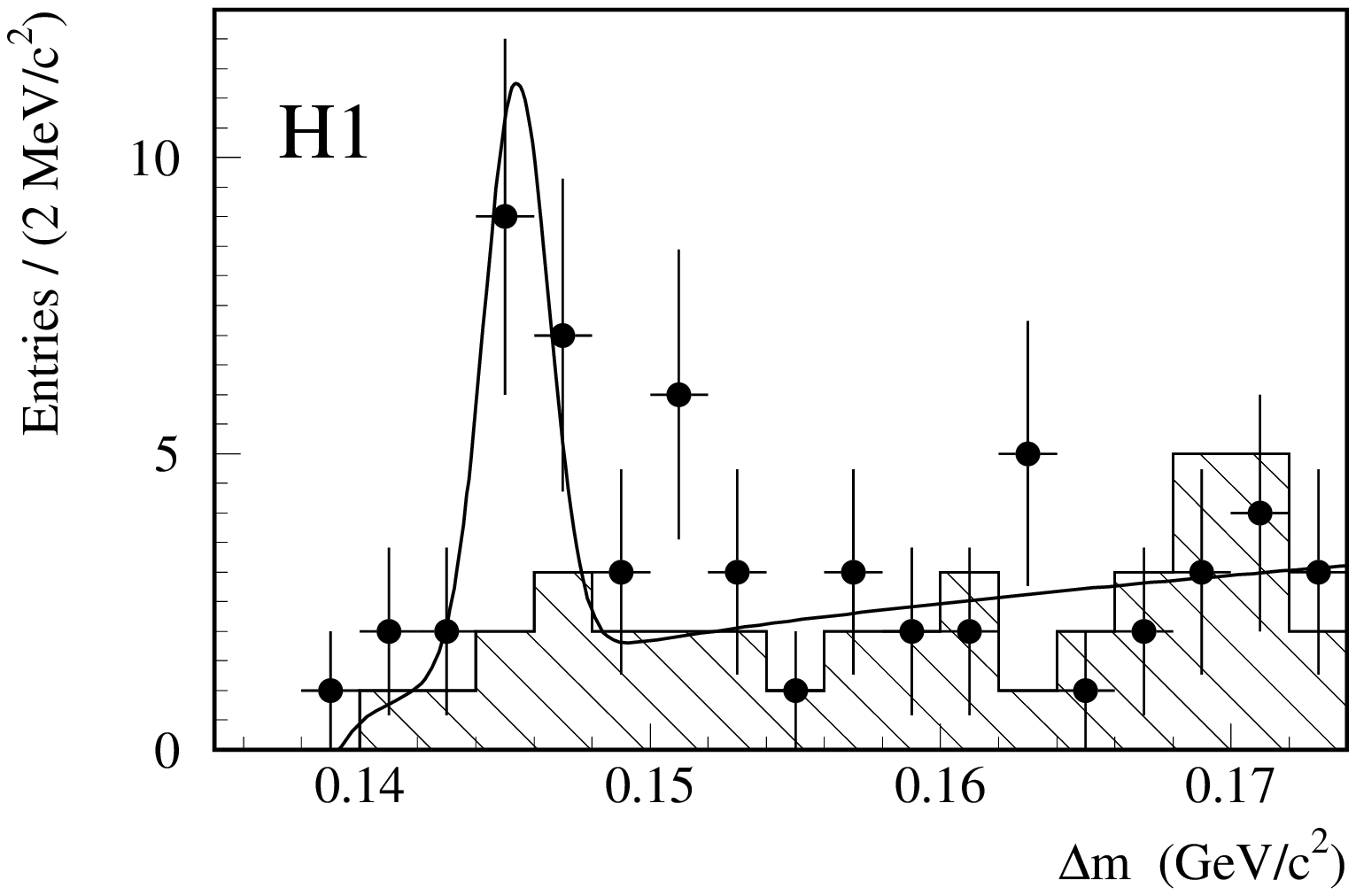} {Distribution of
  the mass difference between the $D^*$ and $D^0$ candidates, $\Delta
  M= M(D^*)-M(D^0)$, for the diffractive DIS sample selected with
  $\etamax<1.5$, $\xpom<0.012$ and $\beta<0.8$. The solid dots
  indicate the signal distribution, while the dashed histogram
  represents the background distribution from wrong charge
  combinations. The solid line is the result of fitting a Gaussian
  distribution and a threshold function of the form $a(\Delta M -
  m_{pi})^b$. } {diff-charmdis}

Charm production associated with large rapidity gaps has been searched
for both in deep inelastic
scattering~\cite{zeus_disc_diff,h1_disc_diff} and in hard
photoproduction~\cite{h1_phpc_diff}. The $\Delta M$ distribution for
events with a large rapidity gap is shown 
in Fig.~\ref{fig:diff-charmdis} for deep inelastic scattering with
$Q^2>3 \gevtwo$. The charm production cross section, integrated over
the phase space of the measurements are compatible with expectations
based on the RAPGAP MC model with gluons as obtained from the
$\ftwodthree$ measurements. In deep inelastic
scattering~\cite{zeus_disc_diff}, for $3<Q^2<150 \gevtwo$,
$0.02<y<0.7$, $p_T(D^*)>1.5 \gev$ and $|\eta^{D*}|<1.5$, the fraction
of diffractively produced events with $.002<\xpom<0.012$ was found to
be $7 \pm 2.2 \%$ of the total $D^*$ sample.

The available charm production studies suffer from large statistical
errors. However, the present results point to a substantial charm
production in hard diffractive processes and give a strong support to
models in which diffractive production is mediated through gluons.

\subsection{Hard diffraction in photoproduction} 

In hadron-hadron interactions two processes have been proposed to
study the nature of the pomeron. Both involve the production of large
$p_T$ jets as a trigger for the partonic nature of the interaction.
One process is the production of jets embedded in a hadronic final
state well separated in rapidity from the target hadron, originally
proposed by Ingelman and Schlein~\cite{ingelman-schlein} and the other
is the production of two large $p_T$ jets separated by a large
rapidity gap proposed by Bjorken~\cite{bjorkenLRG}. These two
types of diffractive processes may in principle be very different in
nature. In the first process, in which the hadron identity is to be
preserved, the momentum transfer square $t$ is small, while in the
second process, $t$ is large and the diffractive exchange takes place
between two partons.

Both processes can be studied at HERA in the interactions of
quasi-real photons with protons. The photon is known to display a
partonic structure in the presence of strong interactions (see
section~\ref{sec:photon_structure}). Two components can be identified in
the structure of the photon, the large size partonic configuration
where the photon behaves essentially as a vector meson state and the
small size configuration, called the anomalous contribution. The
presence of the latter makes the photon interactions different from
that of hadrons and is at the origin of the so called direct photon
contribution to hard processes.  In the spirit of QCD color
transparency phenomena, in which final states interactions are
suppressed for the interactions of small size configurations, the LRG
production could be enhanced in hard photoproduction processes.

\subsubsection{Diffractive jet production in photoproduction}

\epsfigure[width=0.8\hsize]{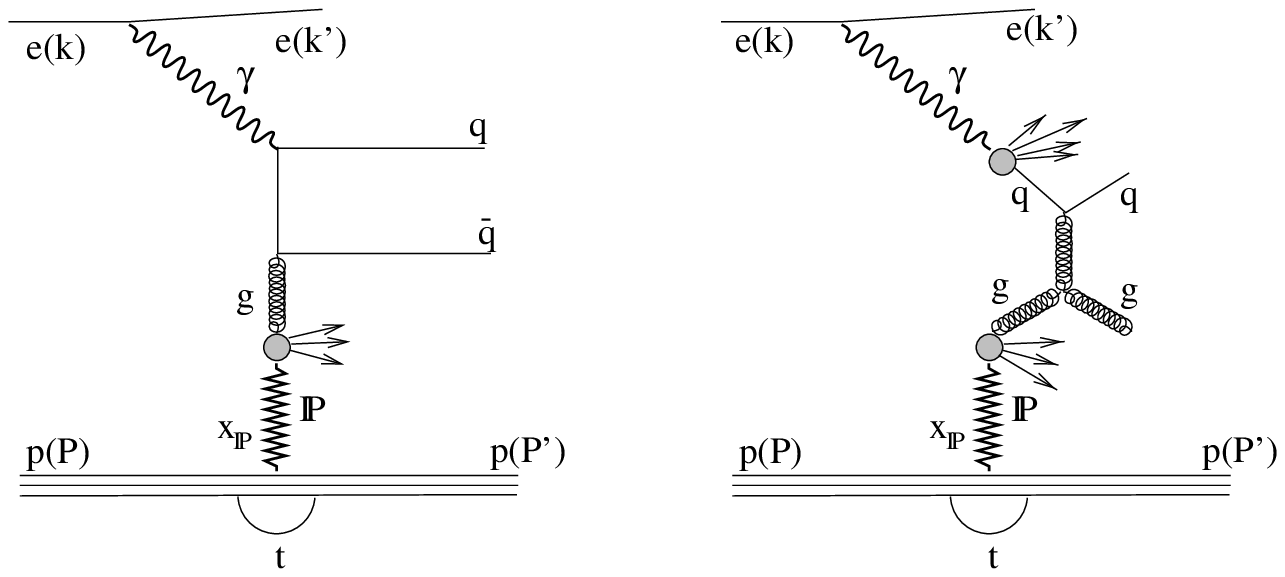} {Examples of
  diagrams contributing to dijet production in hard diffractive
  photoproduction. Left -- direct photon contribution, right --
  resolved photon contribution.  } {diff-dijetdiag}

In the Ingelman-Schlein model of hard diffractive scattering the
pomeron exchange also contributes to the production of large $p_T$
jets associated with a photon dissociation system. Examples of diagrams
for diffractive dijet production, one for a direct and one of a
resolved process, are shown in Fig.~\ref{fig:diff-dijetdiag}. The quark
content of the pomeron can be measured in DIS, while jet production in
photoproduction is sensitive to both the quark and gluon content of
the interacting particles.

\paragraph{Theoretical background}
The cross section for dijet production can be expressed through
universal parton distributions if use is made of the QCD factorization
theorem (see section~\ref{sec:photon_structure}). For the diffractive
dijet production,
\begin{equation}
ep \rightarrow ep + {\rm jet}_1 + {\rm jet}_2 + X^{\prime} \, ,
\end{equation}
where $X^\prime$ denotes the rest of the final state not associated
with jets,  the cross section can be symbolically written as
\begin{eqnarray}
\lefteqn{\sigma(y,\xpom,t:p_T^2,\hat{s}) =  } \ \ \ \ \ \\ 
& & f_{\gamma/e}(y) f_{\pomsub/p}(\xpom,t) 
\sum_{i,j} \sum_{k,l} f_{i/\gamma}(x_\gamma,p_T^2) 
f_{j/\pomsub}(\beta,p_T^2) 
\hat{\sigma}_{i+j \rightarrow k+l}(p_T^2,\hat{s}) \nonumber \, ,
\label{eq-diff:factorization}
\end{eqnarray}
where $f$ denotes the respective fluxes and Regge factorization is
assumed for the pomeron flux. The first sum runs over all possible
types of partons present in the photon and the pomeron and the second
sum runs over all possible types of final state partons. The cross
section $\hat{\sigma}(p_T^2,\hat{s})$ for two
body collisions, $i+j \rightarrow k+l$, depends on the centre of mass
energy, $\hat{s}= \xpom \beta x_{\gamma} W$, and the transverse
momentum of the outgoing partons, $p_T$. For large values of $p_T$ the
cross section can be reliably calculated in perturbative QCD.

The QCD factorization theorem has been proven to hold for diffractive
DIS but fails for hard diffractive processes in hadron-hadron
collisions~\cite{collins}. This would imply that
expression~(\ref{eq-diff:factorization}) is expected to hold for
direct photon processes (i.e.
$f_{i/\gamma}(x_\gamma,p_T^2)=\delta(1-x_\gamma)$) but not for resolved
photon processes.

The interest of studying jet production in diffractive photoproduction
in combination with the DIS measurements allows to address the issues
of the gluon content of the pomeron, the validity of QCD factorization
and the validity of Regge factorization.

\paragraph{HERA data}

\epsfigure[width=0.8\hsize]{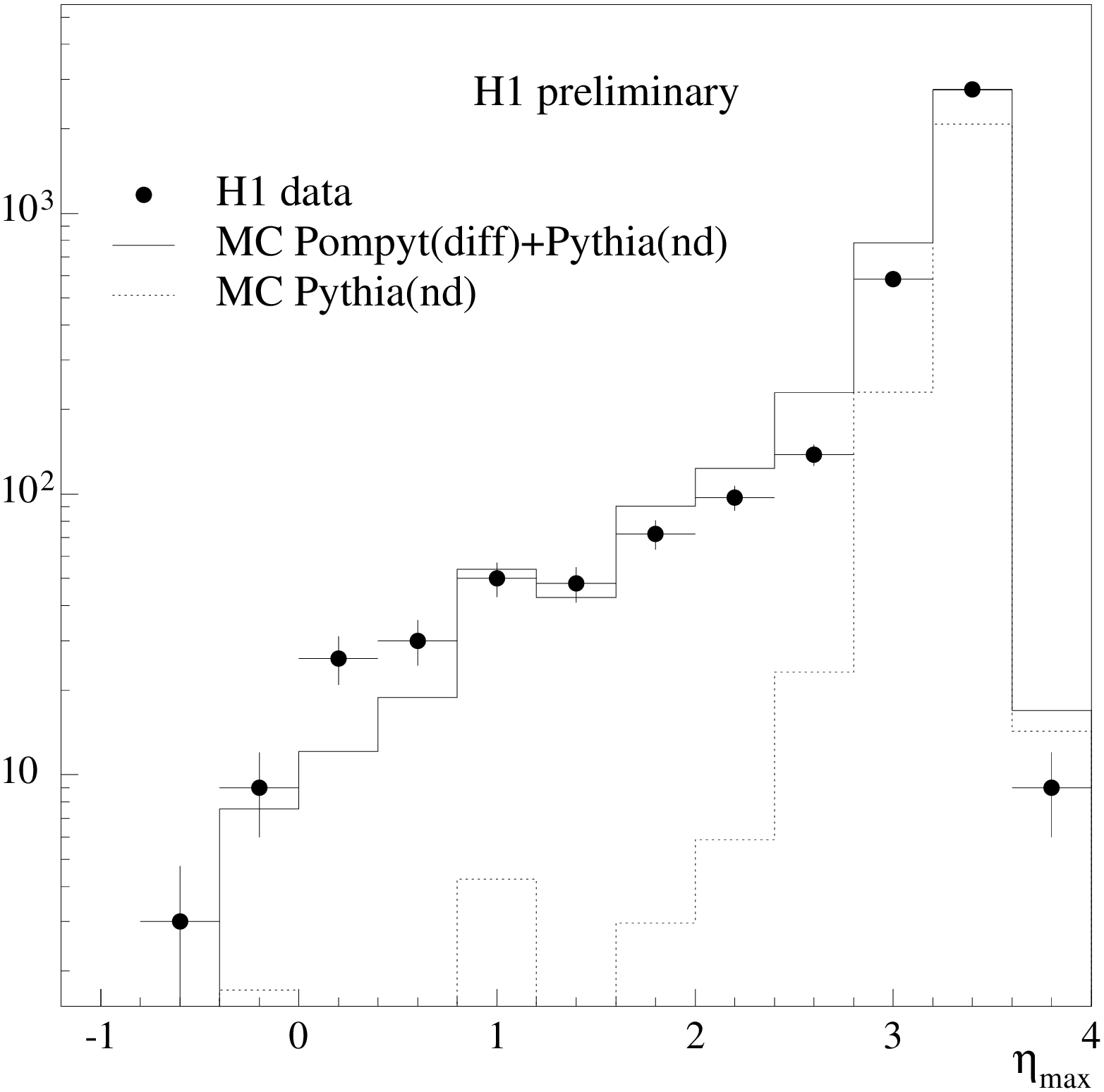} {Distribution of
  $\etamax$ in tagged $\gamma p$ interactions containing jets with
  transverse energy $E_T>5 \gev$ in the interval $-1.5<\eta_{jet}<2.5$
  (dots). The solid line shows MC expectations for a mixture of
  diffractive and non-diffractive processes. The dashed line denotes
  the non-diffractive contribution from the MC model
  (PYTHIA~\protect\cite{ref:PYTHIA}).}  {diff-etamaxjet}

The search for jet production associated with a large rapidity gap in
the proton fragmentation region ~\cite{zeus_dijet_diff1,h1_hph_diff}
has established the presence of hard diffractive scattering in
photoproduction. An example of the $\etamax$ distribution for the
inclusive sample of photoproduction events with a reconstructed jet
with $E_T> 4\gev$ is shown in Fig.~\ref{fig:diff-etamaxjet}. An excess
of events at small $\etamax$ over the expectations for non-diffractive
jet production is observed. The shape of the $\etamax$ distribution
can be well reproduced if the non-diffractive model is supplemented
with hard diffractive processes assuming a pomeron consisting of hard
gluons. The Ingelman-Schlein model is implemented in the POMPYT MC
generator~\cite{pompyt}.

To describe the measured inclusive diffractive jet production cross
section within the Ingelman-Schlein model a substantial hard gluon
content of the pomeron ($30\%$ to $80\%$) is
required~\cite{zeus_dijet_diff1}. The quark content is
constrained by the measurements of $\ftwodthree$~\cite{zeus_f2d1}.

The diffractive dijet cross sections have been measured both by the
ZEUS \cite{zeus_dijet_diff2} and the H1~\cite{h1_dijet_diff2}
experiments.  The samples were selected by requiring the presence of a
large rapidity gap. The ZEUS analysis, requiring $\etamax<1.8$, covers
the range $134<W<277 \gev$ for jets with transverse energy
$E_T^{jet}>6 \gev$ and $-1.5<\eta^{jet}<1$.  In case of H1 the
diffractive selection is similar to the one of the DIS inclusive
analysis~\cite{h1_f2d2} and covers the range $150<W<250 \gev$ and
$\xpom<0.05$. In addition the jets are required to have $E_T^{jet}>5
\gev$ and $-1<\eta^{jet}<2$. The measured $E_T^{jet}$ spectra exhibit a 
typical steep fall-off expected for hard parton-parton scattering.

The fraction of the photon, $x_\gamma$, and of the pomeron momentum,
$\beta$, carried by the partons taking part in the scattering cannot
be determined directly from the hadronic final state, due to higher
order QCD processes. Instead the variables $x_\gamma^{OBS}$ and
$\beta^{OBS}$ ($z^{jets}_{\xpom}$ for H1) are introduced.  At the
parton level $x_\gamma$ and $\beta$ are defined as
\begin{eqnarray}
x_\gamma &=& \frac{(p_1+p_2)\cdot (P-P^\prime)}{q \cdot (P-P^\prime)} \, ,
\\ \label{eq-diff:xgamma}
\beta &=& \frac{(p_1+p_2)\cdot q}{q \cdot (P-P^\prime)} \, ,
\label{eq-diff:betaph}
\end{eqnarray}
where $p_{1,2}$ are the momenta of the two final state partons, $P$
($P^\prime$) denotes the initial (final) proton momentum and $q$ is the
momentum of the photon. The approximations $(P-P^\prime)^2 \simeq 0$
and $q^2\simeq 0$ have been made.  The corresponding variables at the
hadron level are defined as
\begin{eqnarray}
x_\gamma^{OBS} &=& \frac {\sum_{jets}E_T^{jet} e^{-\eta^{jet}}}{2yE_e} \, ,
\\ \label{eq-diff:xgammaobs}
\beta^{OBS} &=& \frac {\sum_{jets}E_T^{jet} e^{\eta^{jet}}}{2 \xpom E_p} \, ,
\label{eq-diff:betaobs}
\end{eqnarray}
where $E_e$ and $E_p$ are the incident electron and proton energies
respectively and the sum runs over the two jets with highest
$E_T^{jet}$ in the event. In case of H1, the estimator of the pomeron
momentum invested in the interaction is defined as
\begin{equation}
z^{jets}_{\xpom} = \frac {\sum_{jets} (E_i+p_{z,i})}{ \sum_{X} (E_i+p_{z,i})},
\end{equation}
where the sums run either over the jets or over the full hadronic
system.

\epsfigure[width=0.8\hsize]{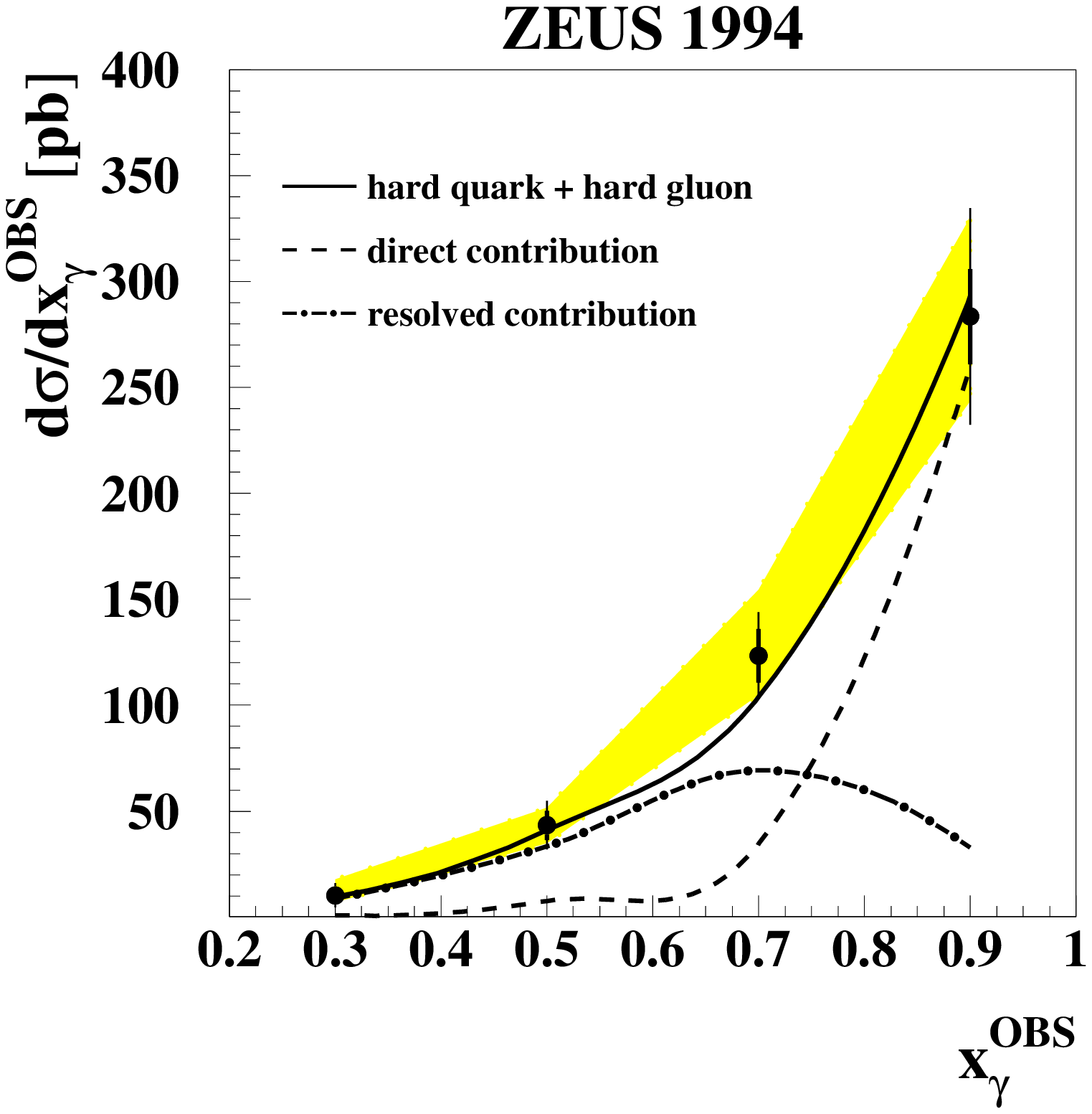} {Dijet
  production cross section in diffractive photoproduction in the range
  $134<W<277 \gev$, as a function of $x_{\gamma}^{OBS}$, for jets in
  the region $-1.5<\eta_{jet}<1$ and the most forward going hadron at
  $\etamax<1.8$ (solid dots). The solid line correspond to the
  Ingelman-Schlein model of the pomeron with a hard quark and gluon
  momentum distributions. Also shown is the resulting direct (dashed
  line) and resolved (dashed-dotted line) photon contributions.  The
  shaded area represents the systematic error due to energy scale
  uncertainty.  } {diff-xgamma}

The presence of both direct and resolved photon contributions has been
established by studying the $x_\gamma^{OBS}$ distribution shown in
Fig.~\ref{fig:diff-xgamma}. The direct processes populate the large
$x_\gamma^{OBS}$ region, while the resolved processes tend to populate
the lower $x_\gamma^{OBS}$ processes. A clear tail of the resolved
processes is observed.

\epsfigure[width=0.8\hsize]{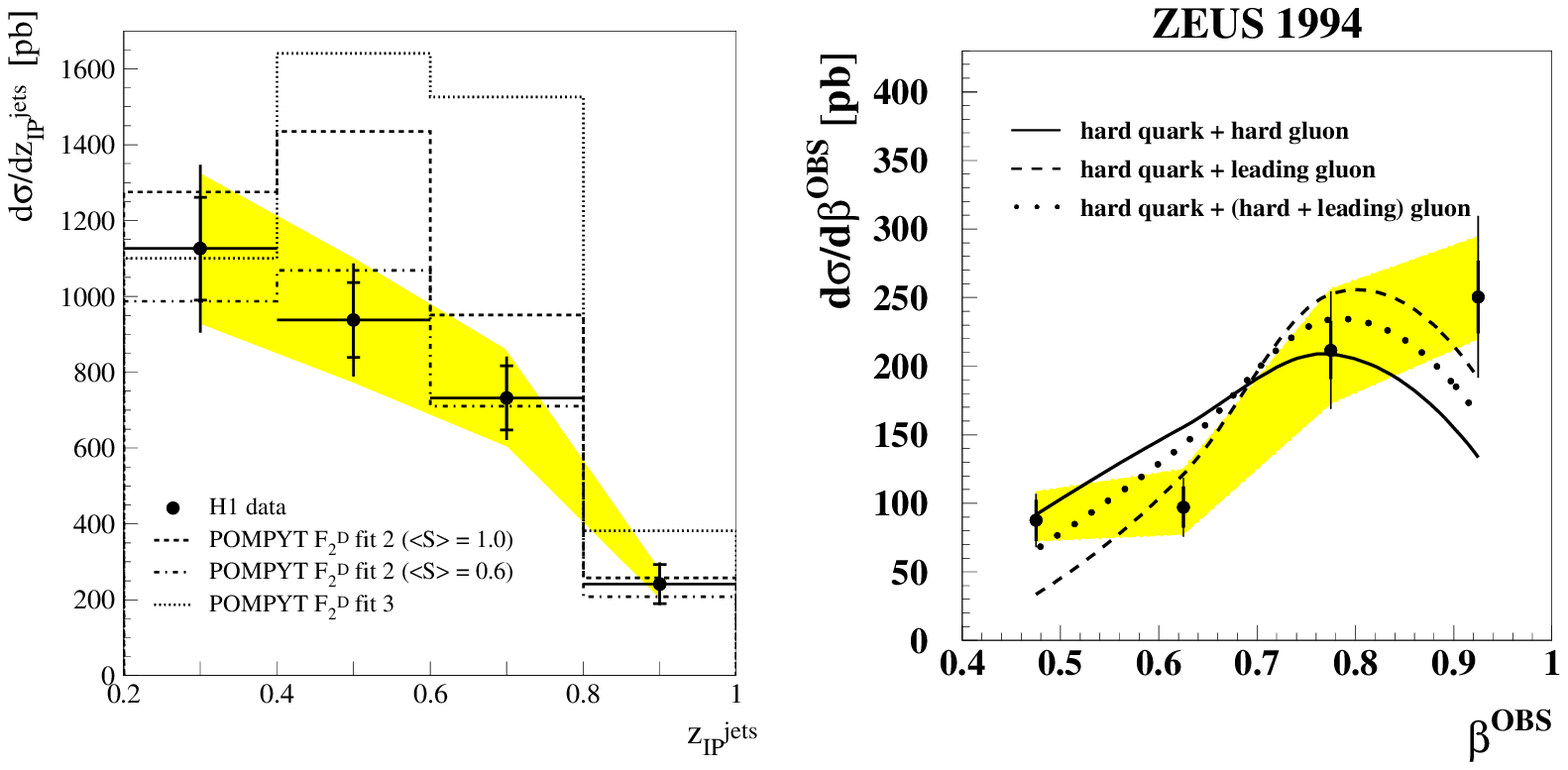} {Left - Jet
  production cross section in diffractive photoproduction in the range
  $150<W<250 \gev$ and $\xpom<0.05$, as a function of
  $z^{jets}_{\xpom}$. Right - Jet production cross section in
  diffractive photoproduction in the range $134<W<277 \gev$, as a
  function of $\beta^{OBS}$, for the most forward going hadron at
  $\etamax<1.8$ (solid dots).  The lines correspond to the
  Ingelman-Schlein model of the pomeron with various momentum
  distributions of gluons. Left - fit 2 denotes a flat gluon
  distribution at the starting scale of $3 \gevtwo$, while fit 3
  denotes a leading gluon type of distribution. $<S>$ stands for
  survival probability. Right - as denoted in the figure. The shaded
  area represents the systematic error due to energy scale
  uncertainty.  } {diff-betajet}

The $\beta^{OBS}$ and the $z^{jets}_{\xpom}$ distributions are shown
in Fig.~\ref{fig:diff-betajet}. The comparison between the two figures
demonstrates the restriction imposed by the $\etamax<1.8$ cut, which
limits the invariant mass and the $\xpom$ of the selected events. For
this restricted sample of events the distribution is seen to peak at
large $\beta^{OBS}$, indicating that for these events the whole of the
pomeron momentum is invested in the interaction.

The ZEUS results are well reproduced by a model with a soft
factorisable pomeron with parton distributions obtained as the result
of a combined fit to the DIS and photoproduction cross sections
assuming both QCD and Regge factorization. Within the relatively large
statistical and systematic errors, an acceptable description of the
data is obtained, provided a large fraction of the pomeron momentum is
carried by gluons. 

Among the parton distributions which fit the $\ftwodthree$
measurements of H1, the jet measurements prefer a gluon distribution
which is flat at a scale of about $3 \gevtwo$. However, even then the
measured cross section seems to be systematically lower than
expected. The agreement becomes better, if one assumes that the
survival probability of a large rapidity gap is about $60\%$ (for
discussion see section~\ref{sec:LRGbjets}). Note that for $Q^2>7.5
\gevtwo$~\cite{h1_dijet_diff2}, the dijet cross section as a function
of $z^{jets}_{\xpom}$ was found to be well reproduced, both in shape
and normalization, with the "flat" gluon distribution.

\subsubsection{Large rapidity gaps between jets} 
\label{sec:LRGbjets}

One of the challenges of experimental physics is the search for the
BFKL dynamics~\cite{ref:BFKL1,ref:BFKL2,ref:BFKL3}. It is by now well
understood that in the presence of the interplay of soft and hard QCD
phenomena it may be very difficult to uncover the BFKL dynamics if it
exists.  However, in hard reactions in which the $Q^2$ evolution is
suppressed, the BFKL dynamics may be enhanced. An example of such a
reaction is the exchange of a color-singlet between two jets with
large transverse momenta. The requirement that both $p_T$ be large and
approximately equal to $Q$ guarantees a large scale for
perturbative calculations to be applicable and at the same time, prevents
$Q^2$ evolution~\cite{mueller-tang,delduca-tang,lu}.

\paragraph{Theoretical background}

In high energy hadronic collisions, the dominant mechanism for jet
production is described by a hard scattering between partons from the
incoming hadrons via a quark or gluon propagator. This propagator
carries color charge. Since color confinement requires that the
final state contains only color singlet objects, the exchange of
color quantum numbers in the hard process means that a jet at some
later stage generally exchanges color with another jet or beam
remnant widely separated from it in rapidity.  However, if hard
scattering was mediated by the exchange of a color singlet
propagator in the $t$-channel, each jet would be color connected only
to the beam remnant closest in rapidity and the rapidity region
between the jets would contain few final-state
particles~\cite{dokshitzer}. The color singlet propagator could be an
electroweak gauge boson or a strongly interacting object, and the soft
gluon emission pattern produced in each case is
similar~\cite{zeppenfeld} but the rates could be very different.  In
order to determine the rate of color singlet exchange processes, it
has been proposed~\cite{bjorkenLRG} to study the multiplicity
distribution in pseudo-rapidity between the two jets.

\epsfigure[width=0.8\hsize]{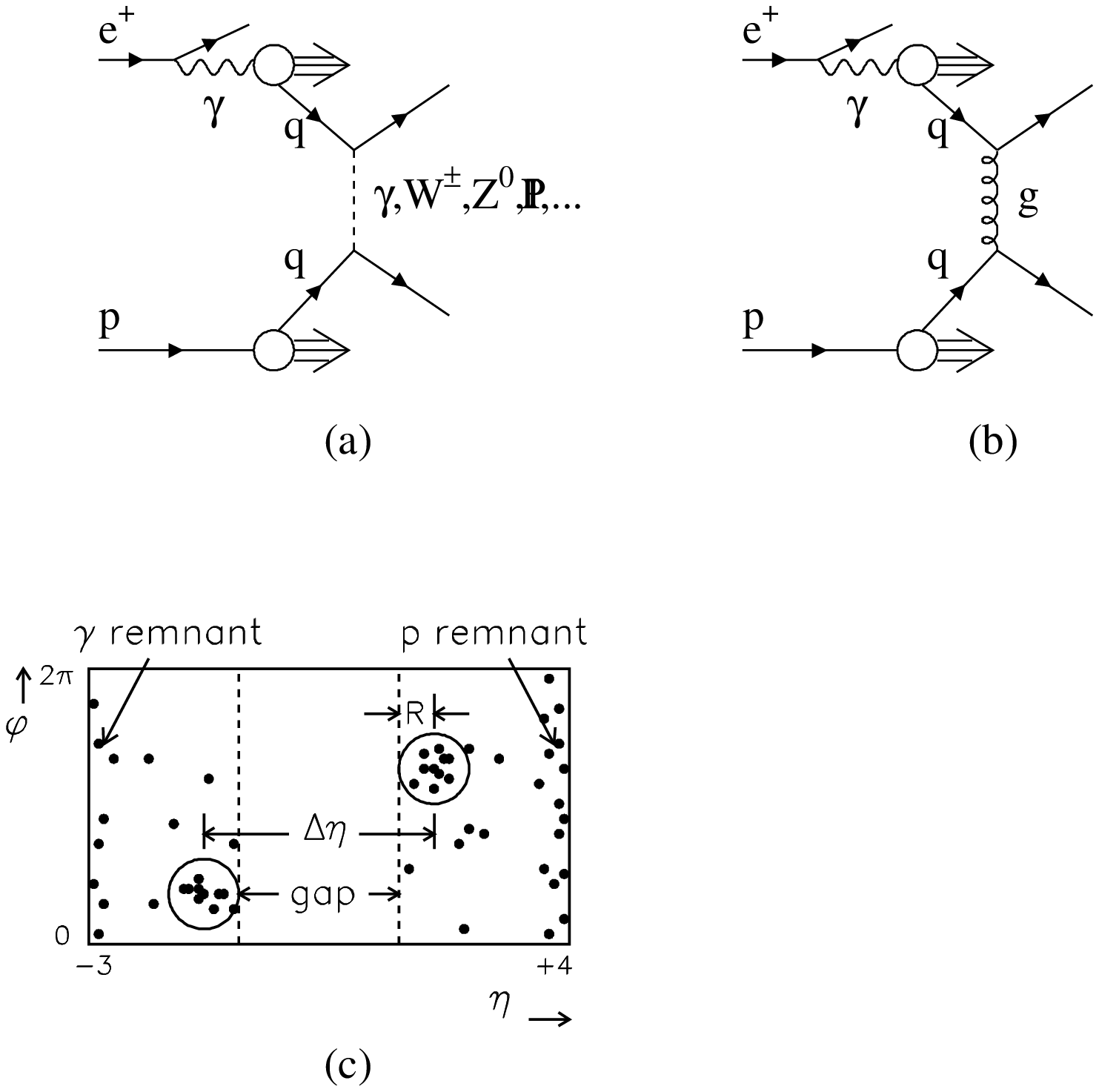} {Diagrams for
  resolved photoproduction of jets via (a) color singlet exchange and
  (b) color non-singlet exchange. The topology of an event with a
  large rapidity gap between jets is shown in (c) where black dots
  represent final state hadrons and the boundary illustrates the
  acceptance of the detector.  } {diff-laurel1}

An example of color singlet exchange in resolved photoproduction, in
which a parton in the photon scatters from a parton in the proton via
$t$-channel exchange of a color singlet object, is shown in
Fig.~\ref{fig:diff-laurel1}a. An example of the more common color
non-singlet exchange mechanism is shown in
Fig.~\ref{fig:diff-laurel1}b.  For high $E_T^{jet}$ dijet
production, the magnitude of the square of the four-momentum ($|t|$)
transferred by the color singlet object is large. Thus it is possible
to calculate in perturbative QCD the cross section for the exchange of
a strongly interacting color singlet
object~\cite{bjorkenLRG,mueller-tang,delduca-tang,lu}. For instance,
the ratio of the two-gluon color singlet exchange cross section to
the single gluon exchange cross section has been estimated to be about
0.1~\cite{bjorkenLRG}.

The event topology for the process of Fig.~\ref{fig:diff-laurel1}a is
illustrated in Fig.~\ref{fig:diff-laurel1}c. There are two jets in the
final state, shown as circles in ($\eta,\varphi$) space.  For the
color singlet exchange process, radiation into the region (labeled
``gap'') between the jet cones is suppressed, giving rise to the
rapidity gap signature.  For color non-singlet exchange, the
probability of finding no particles in the gap is expected to fall
exponentially with increasing $\deta$, the distance in $\eta$ between
the centres of the two jet cones.

\paragraph{HERA data}

The search for rapidity gaps between jets has been performed by the
ZEUS experiment for a sample of photoproduction events with at least
two jets of $E_T^{jet} > 6$~GeV in the $\gamma p$ centre of mass
energy range $135< W_{\gamma p}<280 \gev$. The two highest transverse
energy jets were required to have $\Delta\eta > 2$ (i.e. cones not
overlapping in $\eta$), $\eta^{jet} < 2.5$ and boost $|(\eta_1 +
\eta_2)|/2 = |\bar{\eta}| < 0.75$.  These conditions constrain the
jets to lie within the kinematic region where the detector and event
simulations are best understood.  The events with {\it no} particle of
transverse energy $E_T^{particle} > 300$~MeV within the $\eta$ space
between the edges of the two highest $E_T^{jet}$ jet cones, are called
gap events.

\epsfigure[width=0.8\hsize]{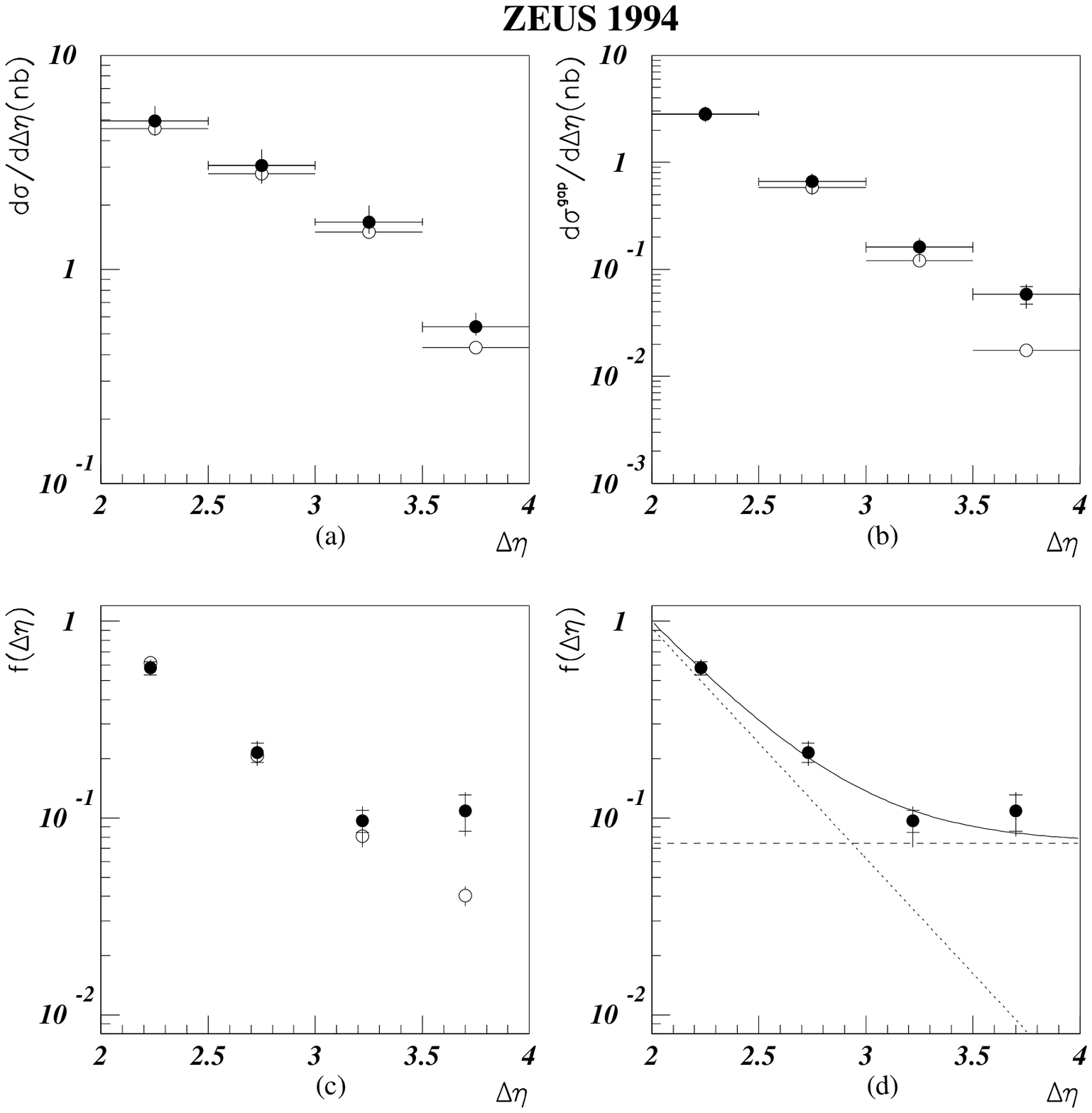} {The inclusive
  cross section as a function of the rapidity distance $\deta$ between
  jets (a) and for gap events (b). The data (black circles) are
  compared to MC expectations for a non-color singlet exchange in the
  jet production. The fraction is shown in (c) as a function of
  $\deta$ and redisplayed in (d) together with the results of a fit
  (full line) consisting of an exponential function (dotted line) and
  a constant (dashed line). } {diff-deta}

The inclusive cross section, $d\sigma / d\Delta\eta$, and the cross
section for events with a gap, $d\sigma_{gap} / d\Delta\eta$, are
presented in Fig.~\ref{fig:diff-deta}a and b as a function of
$\deta$. The cross sections are compared to MC expectations based on
the PYTHIA generator~\cite{ref:PYTHIA} which does not incorporate any
color singlet exchange mechanism. The model gives a good
description of the inclusive cross section at $\deta < 3.5$ but an excess
of events in the data is observed for larger $\deta$. The excess
becomes more pronounced when the requirement of a gap is applied to
the data and the MC model. 

The gap-fraction, $f(\Delta\eta)$, is defined as the ratio of the
number of dijet events at this $\deta$ which have a rapidity gap
between the jets to the total number of dijet events at this $\deta$.
The gap-fraction is shown in Fig.~\ref{fig:diff-deta}c and compared to
the MC model. The data were fitted with a sum of an exponential and
constant functions and the gap fraction for color singlet processes
was found to be $f_{\mathrm{gap}} = 0.07 \pm 0.02  ^{+0.01}
_{-0.02}$. The result of the fit is shown in
Fig.~\ref{fig:diff-deta}d.

The excess in the gap-fraction over the expectation from non-singlet
exchange may be interpreted as evidence for the exchange of a color
singlet object. The fraction of gap events in photoproduction is
larger than the values obtained for $p\bar{p}$
collisions~\cite{CDFgap,D0gap}, which are $\sim 0.01$.

The fraction of events due to color singlet exchange,
$\hat{f}(\Delta\eta)$, may be even higher than the measured excess.
Secondary interactions of the photon and proton remnant jets could
fill in the gap.  A survival probability, ${\cal P}$, has been
defined~\cite{bjorkenLRG} which represents the probability that a
secondary interaction does not occur.  Then $f(\Delta\eta) =
\hat{f}(\Delta\eta) \cdot {\cal P}$.  Estimates of the survival
probability for $p\bar{p}$ collisions at the Tevatron range from about
5\% to 30\%~\cite{bjorkenLRG,GLMgap,fletcher}.  The survival
probability at HERA could be higher due to the lower centre-of-mass
energy, or the fact that one remnant jet comes from a photon rather
than a proton, or the fact that the mean fraction of the photon energy
participating in the jet production in these events is high.
Therefore the effect observed in photoproduction and $p\bar{p}$ jet
production could arise from the same underlying process. The
percentage of gap events in photoproduction is compatible with $\sim
0.10$ expected for two-gluon exchange mechanism.

\subsection{Parton distributions in the pomeron}

The QCD factorization theorem has been proven to be valid for the
diffractive structure function $\ftwodthree$~\cite{collins}.  The
immediate consequence of this theorem is that the DGLAP evolution
equation should describe the scaling violations observed in
$\ftwodthree$ and therefore the diffractive parton distributions can
be derived, in a similar way as it is done for the inclusive $F_2$ of
the proton. There is however no constraints on the gluon momentum
distribution since the momentum sum rule, used for the proton
structure function, does not formally apply.

The H1 experiment performed a QCD fit to their $\ftwodthree$
measurements \cite{h1_f2d2}, assuming in addition the validity of
Regge factorization with flux factors for the pomeron and the reggeon
as determined from the data.  The reggeon was assumed to have the
parton content of the pion \cite{GRVpi}. The fit excluded the resonance
region, $M_X<2 \gev$ and the data points which could be affected by a
presence of a large longitudinal structure function component which
was neglected in the measurements. The results of the fit were
parameterizations of the parton distributions in the pomeron at a
starting scaling $Q_0^2=3 \gevtwo$.

\epsfigure[width=0.8\hsize]{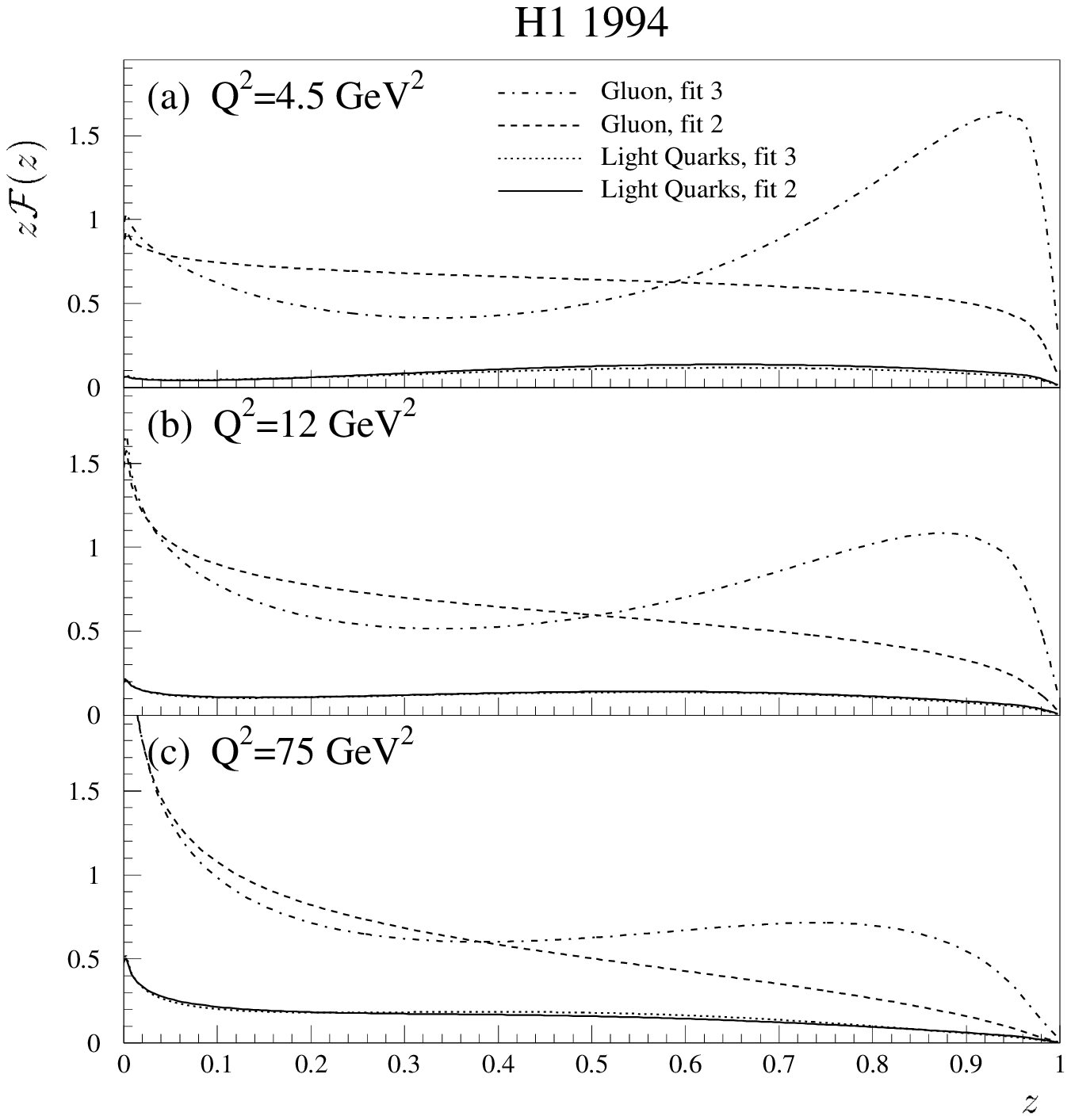} {Quark and
  gluon momentum distributions in the pomeron as a function of the
  fraction of the pomeron momentum, $z$, for different fits to the
  $\ftwodthree$ measurements.  The results are evolved to different
  $Q^2$ values as denoted in the figure.  } {diff-partonsH1}

The data cannot be described by a parameterization in which the
pomeron contains only quarks at $Q_0^2 = 3 \gevtwo$. Various forms of
the gluon distribution at the starting scale were explored. The best
fit is obtained with a gluon distribution peaking at large $\beta$ as
shown in Fig.~\ref{fig:diff-partonsH1}. A flatter gluon distribution,
albeit with worse $\chi^2$, can also describe the data. Note that the
latter is preferred by the measurements of the dijet cross
section~\cite{h1_dijet_diff2}. The quality of the fits can be seen in
Figs.~\ref{fig:diff-f2d3h1q2} and~\ref{fig:diff-f2d3h1beta} where the
measurements of $\xpom \ftwodthree(\xpom=.003)$ as a function of $Q^2$
for fixed $\beta$ values and as a function of $\beta$ for fixed $Q^2$
values are compared to the expectations derived from the fit. The
preferred solution is the one with a substantial gluon content in the
pomeron. The relative fraction of the pomeron momentum carried by
gluons is $\sim 90\%$ at $Q^2=4.5 \gevtwo$ and decreases to $\sim
80\%$ at $Q^2=75 \gevtwo$.

A similar conclusion has been reached by the ZEUS
experiment~\cite{zeus_dijet_diff2} from a QCD fit to the $\ftwodthree$
measurements and the diffractive dijet cross sections. The validity of
Regge factorization was assumed with the pomeron flux as determined
from soft interactions and in addition the QCD factorization theorem
was assumed to be valid for diffractive dijet photoproduction.  The
last assumption has no theoretical foundation, with that this approach
can be treated as a test for factorization breaking.

\epsfigure[width=0.8\hsize]{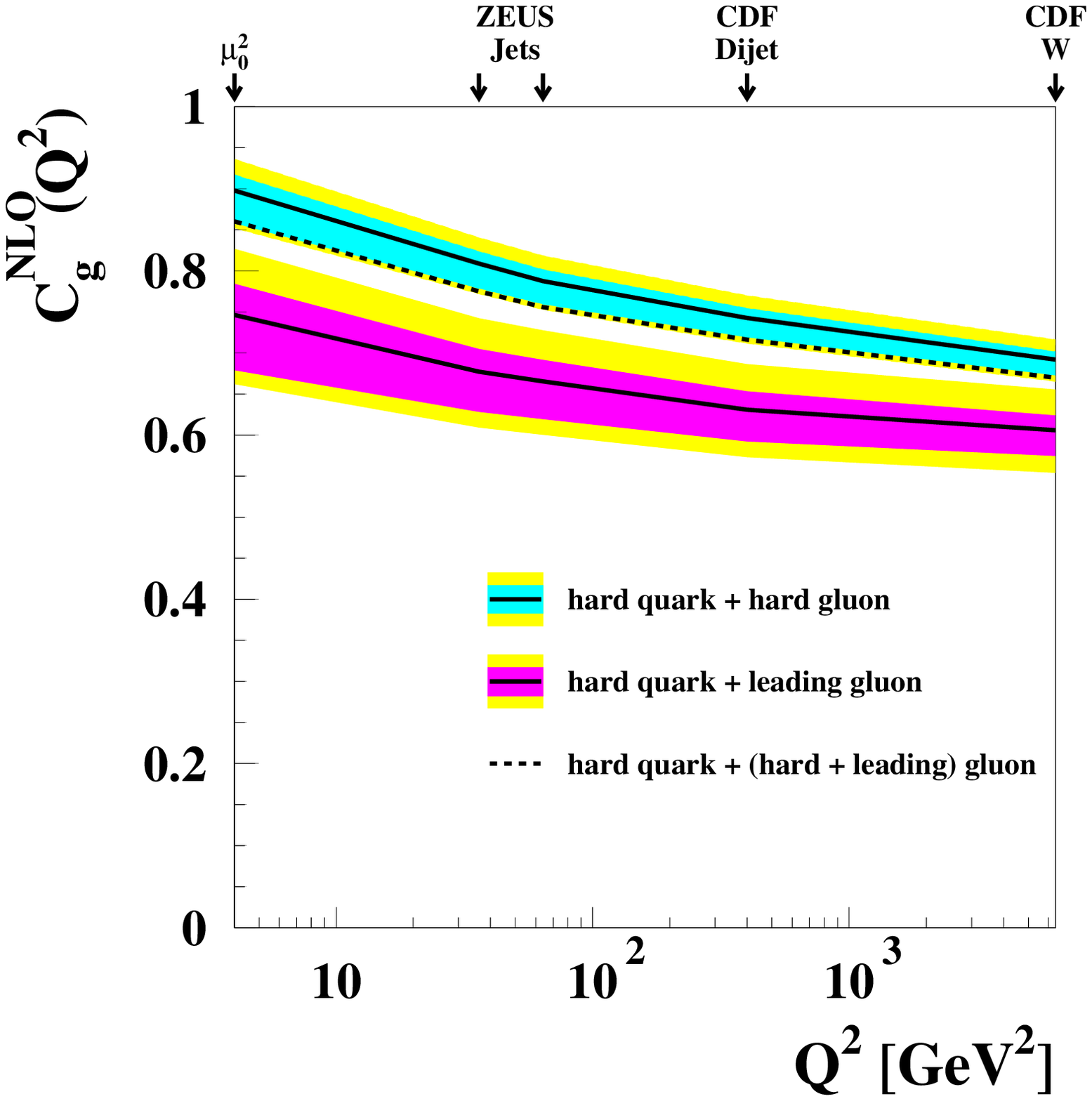} {Fraction of
  the pomeron momentum carried by gluons $c_g$ as a function of $Q^2$,
  determined from a NLO QCD fit to the $\ftwodthree$ measurements and
  the diffractive dijet cross section in photoproduction. }
{diff-ZEUSglu}

A good fit to the data is found with a gluon distribution peaking at
large $\beta$. Here the gluon distribution is constrained by the shape
of the measured $\beta^{OBS}$ distribution. The relative fraction of
the pomeron momentum carried by gluons, determined in $NLO$
approximation is shown in Fig.~\ref{fig:diff-ZEUSglu} as a function
of $Q^2$ for various assumptions on the shape of the gluon
distribution at a starting scale $Q_0^2=4 \gevtwo$. The data require the
fraction of the pomeron momentum carried by partons due to gluons to
lie in the range $0.64<c_g^{NLO}<0.94$. The sensitivity of the data is
not sufficient to prove or disprove the validity of the QCD
factorization theorem for diffractive hard photoproduction.

The extrapolation of the ZEUS results to the $Q^2$ values probed in
the measurements of hard diffractive processes in $p\bar{p}$
collisions is also consistent with the estimate of $70 \pm 20 \%$ by
the CDF experiment~\cite{CDFglu}. Note that this does not imply
agreement in the total fraction of the pomeron momentum carried by
partons, which is not constrained by any momentum sum rule and which
has been found to be lower in $p\bar{p}$ collisions~\cite{CDFglu}.

\subsection{Conclusions and outlook}

Hard diffractive interactions, in which the hadronic final state is
separated from the proton by a large rapidity gap, have been observed
in deep inelastic scattering and in high transverse momentum jet
production in photoproduction. The interpretation of these data is in
terms of a color singlet exchange, identified as the pomeron. The
presence of a large scale allows to study the nature of this exchange
in the language of partons and QCD. The measurements at HERA can be
accounted for by a factorizable pomeron with a predominantly gluonic
structure. Where available, also QCD inspired predictions describe the
data well. The final verdict will have to await more precise
measurements.

The quantitative and qualitative difference between diffraction in the
presence of a hard scale and soft diffraction, and between hard
diffraction in hadron-hadron and photon-hadron interactions may be the
first sign that partons are not distributed uniformly inside the
proton~\cite{Muellertalk,Predazzi}. In addition, the sensitivity of
hard diffractive processes to the transverse momenta of partons opens
the possibility of investigating a three-dimensional color
distribution in hadrons.  The processes probed till now in QCD were
only sensitive to the longitudinal dimension.

The diffractive phenomena have triggered a lot of new theoretical
developments in QCD. It has brought to light the importance of color
dynamics of QCD in understanding the nature of strong
interactions. New phenomena such as color transparency in interactions
of small size partonic configurations and color opacity in
interactions of large size configurations are presently being
discussed. They are essential in differentiating between hard
processes which can be calculated in QCD and those where soft
interactions still play an important role.

\section{Vector meson production} 
\label{sec:VM} 
There is a long experimental and theoretical history to the study of
vector meson production.  This study has found new vigor with the
advent of HERA.  On the experimental side, it is found that the cross
sections for exclusive vector meson production rise strongly with
energy when compared to fixed target experiments, if a hard scale is
present in the process.  This strong rise occurs although the
reactions are very far from threshold.  In the case of $J/\psi$
production, the strong rise of the cross section is measured directly
in the HERA data.  This steep rise in the cross sections was
anticipated by some authors based on QCD inspired calculations.  These
calculations indicate that the cross sections depend on the square of
the gluon density in the proton.  If higher order calculations become
available, the measurement of the energy dependence of the vector
meson cross section may be the ideal method for measuring the gluon
density in the proton.

In addition to being a probe of the gluon density, vector meson
production tests our understanding of QCD for exclusive reactions in a
domain where soft and hard physics merge.  This is a region of
fundamental importance, since basic physical realities such as the
confinement of color are not at all understood.  Vector meson
production also offers an opportunity to study the properties of
vacuum exchange in QCD, since no quantum numbers are exchanged in the
scattering process.  New experimental and theoretical results are
becoming available at a rapid pace, and the study of vector meson
production therefore promises to be a very fruitful one for the
development of our understanding of QCD dynamics.

A word on terminology: at HERA, the reaction, $ep \rightarrow epV$,
where $V$ represents a vector meson $(\rho,\; \omega,\; \phi,\;
J/\psi,\; \Upsilon)$, is often referred to as elastic scattering.
This is because the photon is viewed as fluctuating into a vector
meson before the interaction with the proton, followed by a $Vp
\rightarrow Vp$ scattering.  The vector meson has the same quantum
numbers as the photon, and is in some sense the ``same'' particle.
However, at finite $Q^2$ this picture is no longer valid and the
scattering is then no longer elastic.  In this case, the process is
often called exclusive vector meson production.

A comprehensive review of vector meson production at HERA has recently
been published by~\citeasnoun{ref:Crittenden}.  

\subsection{Expectations for soft elastic scattering}

\subsubsection{Vector meson dominance model}

\epsfigure[width=0.6\hsize]{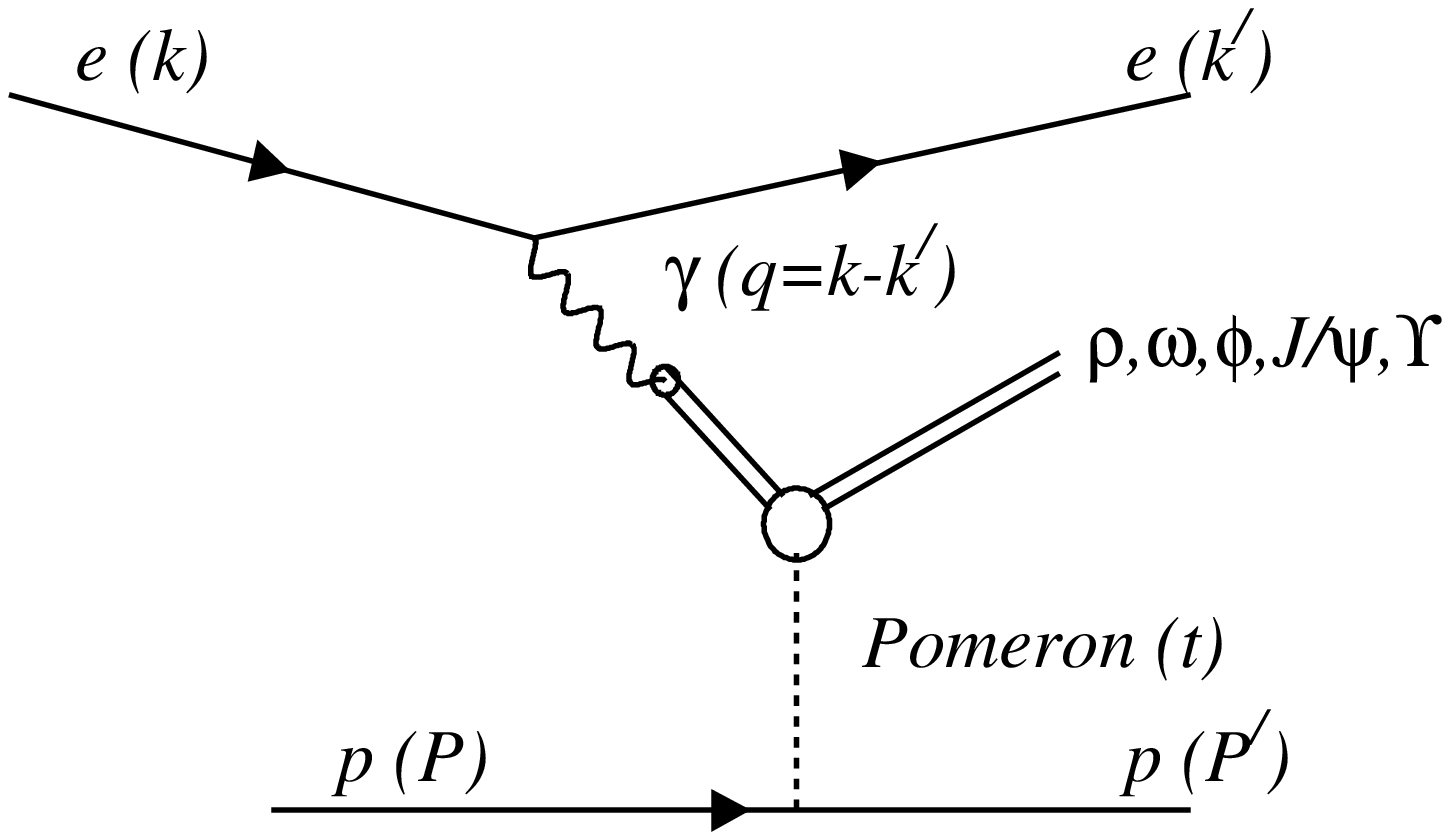} {Elastic vector
  meson production in the vector dominance model (VDM).  Here the
  photon is pictured as fluctuating into a vector meson which then
  scatters elastically from the proton via the exchange of a pomeron.}
{vdm}

Light vector meson leptoproduction  at small  photon virtuality 
and small $W$ has been successfully explained as a three step 
process (see Fig.~\ref{fig:vdm}):
\begin{description}
\item{I)} 
The incoming electron radiates a photon of small virtuality ($Q^2\approx 0$);
\item{II)} 
The photon fluctuates into a light vector meson, $V$, which carries
the same quantum numbers as the photon; i.e.,${\rho, \omega, \phi}$;
\item{III)} 
The vector meson scatters elastically off the incoming proton via 
pomeron exchange.
\end{description}

This model~\cite{ref:Sakurai}, the so-called ``Vector meson Dominance
Model'', or VDM, gives good results when used, for example, to explain
the total photon-proton cross section, $\sigma_{\gamma p}$, in terms
of the total vector meson-proton cross section, $\sigma_{Vp}$ (see,
e.g., \citeasnoun{ref:Bauer}),
\begin{equation}
\sigma_{\gamma p} \propto \sum_{\rho,\omega,\phi} A(f_V)\sigma_{Vp} \; ,
\end{equation}
where $A(f_V)$ is a known function of the coupling.

The $Q^2$ dependence of the $\gamma p \rightarrow V p$ cross section
can also be described by VDM,
\begin{equation}                              
\label{eq:VDMQ2}
  \frac{d\sigma_{\gamma p \rightarrow V p}}{d|t|} = 
\frac{d\sigma_0}{d|t|} \left (\frac{M_{V}^2}
  {M_{V}^2 + Q^2} \right)^2 \left(   {1+\epsilon \xi \frac{Q^2}
  {M_{V}^2}} \right) e^{-b|t|},
\end{equation} 
   where $\sigma_0$ is the cross section at $Q^2=0$ , $\xi$ is the
ratio of the longitudinal to transverse forward cross sections,
$\epsilon$ is the ratio of the longitudinal to transverse photon flux
of the virtual photon and $M_{V}$ is the mass of the vector meson; the
distribution of $t$, the square of the four-momentum transfer between
the photon and the vector meson, is experimentally described by a
single exponential dependence, in the range from $t$ = 0 to $t$ =
$-0.5$ GeV$^2$, with a slope parameter, $b\approx 7-12$ GeV$^{-2}$.

The differential cross section $d\sigma_{\gamma p \rightarrow V
p}/d|t|$ can be related to the total vector meson-proton cross section
$\sigma_{Vp}$ by applying the optical theorem,
\begin{eqnarray}
\left.\frac{ d\sigma_{V p \rightarrow V p}}{d|t|}\right|_{t=0}=
\frac{1+\eta ^2}{16\pi} \sigma_{Vp}^2,
\label{optical}
\end{eqnarray}
where $\eta$ is the ratio of the real to the imaginary part of the
forward $V p$ scattering amplitude. Within VDM, elastic vector meson
photoproduction is related to the elastic $Vp$ cross section; in
particular, for $t=0$
\begin{eqnarray}
\left. \frac{d\sigma_{\gamma p \rightarrow V p}}{d|t|}\right|_{t=0}=
\frac{4 \pi \alpha}{f_{V}^2}
\left. \frac{d\sigma_{Vp \rightarrow Vp}}{d|t|}\right |_{t=0},
\label{vmd1}
\end{eqnarray}
\noindent 
where $4 \pi \alpha/f_{V}^2$ is the probability for the $\gamma
\rightarrow V$ transition.

Assuming that the real part of the amplitude is zero, as expected for
purely diffractive scattering,
\begin{equation}
{\left.\frac{ d\sigma_{\gamma p \to Vp}}{d|t|}~\right|_{t=0}}
             =  \frac{4\pi \alpha}{f_V^2}~\frac{1}{16\pi}\,{\sigma}^2_{Vp}\, .
\label{eq:opttheo}
\end{equation}
A measurement of $d\sigma_{\gamma p \to Vp}/d|t|$ can therefore be used
to calculate $\sigma_{Vp}$. 

As discussed in section~\ref{sec:photoprod}, hadronic cross sections,
and the total photoproduction cross section, have been measured to
have an energy dependence
\begin{equation}
\sigma_{\gamma p} \sim \sigma_{\pi p} \sim W^{2\cdot0.08} \; .
\end{equation}
Assuming that the $Vp$ cross section has the typical behavior of
hadronic cross sections, VDM therefore predicts
\begin{equation}
{\left.\frac{ d\sigma_{\gamma p \to Vp}}{d|t|}~\right|_{t=0}}  \sim  \sigma_{Vp}^2 
\sim W^{0.32} \; .
\end{equation}

\subsubsection{Regge Theory expectations}

In Regge Theory, we expect the $\gamma p\rightarrow Vp$ cross section
to have the form
\begin{equation}
\frac{d\sigma_{\gamma p\rightarrow Vp}}{d|t|} \propto e^{-b_0|t|} 
\left(\frac{W^2}{W^2_0}\right)^{2(\alpha(t)-1)} \; ,
\end{equation}
where $\alpha(t)$ is the pomeron trajectory.  Phenomenological
fits~\cite{ref:DoLa_sigfit} to fixed target and hadron-hadron
scattering data have determined that $\alpha(t)$ can be parameterized
by a linear form,
\begin{equation}
\alpha(t) = \alpha_0 + \alpha't \; ,
\end{equation}
 with  parameters 
\begin{eqnarray}
\alpha_0 & = & 1.08 \; , \\
\alpha' & = & 0.25 \;{\rm GeV}^{-2} .
\end{eqnarray}
The fact that $\alpha'$ is non-zero means that the slope of the $t$
distribution will depend on the energy as
\begin{equation}
\label{eq:tshrinkage}
b = b_0 + 2\alpha' \ln\frac{W^2}{W^2_0} \; ,
\end{equation}
while the $W$ dependence will depend on $t$.  Denoting $\delta$ as the
power of $W$ ($d\sigma/dt \propto W^{\delta}$), we have
\begin{equation}
\label{eq:Wshrinkage}
\delta = 4(\alpha_0 + \alpha't - 1) \; .
\end{equation}
Integrating over $t$ gives
\begin{equation}
\sigma = \frac{W^{4(\alpha_0 -1)}}{b} \; ,
\end{equation}
which results in an effective power of the $W$ dependence of
\begin{equation}
\delta \approx 4(\alpha_0 - \frac{\alpha'}{b} -1) \; ,
\end{equation}
where $b$ depends on $W$ as shown above.  There is no prediction on
the value of $b$ from Regge theory - the parameters $b_0, \; W_0, {\rm
and} \; \alpha'$ could all be process dependent, and could also be
$Q^2$ dependent for a given process.  Typically, values of
$b=10$~GeV$^{-2}$ are found in soft processes, leading to an
expectation of $\delta=0.22$ for $\alpha'=0.25$~GeV$^{-2}$.  In
processes with a hard scale, measured $b$ values are considerably
smaller.  For $b=5$~GeV$^{-2}$, the expectation is $\delta=0.12$.

In a geometrical picture, the slope of the $t$ dependence can be
interpreted as giving the size of the scattering objects,
\begin{equation}
b  \propto {R_p^2} + {R_{\gamma}^2} \; ,
\end{equation}
where $R_p$ is the effective proton radius and $R_{\gamma}$ is the
effective size of the $\gamma$ induced state scattering on the proton.  
For $b=8$~GeV$^{-2}$, $\sqrt{R_p^2+R_{\gamma}^2} \approx 1$~fm.

The $t$ dependence has been studied experimentally as a function of
$Q^2$ and $W$.  It is seen that the slope $b$ decreases with $Q^2$,
indicating that the photon is becoming more point-like as the
virtuality increases.  The tendency for $b$ to increase with $W$ has
been called ``shrinkage'' although, in the geometrical picture, a
steeper $t$ dependence corresponds to the scattering of larger
objects.

\subsection{Expectations in the presence of a hard scale}
\subsubsection{Non-perturbative approach}

The non-perturbative methods are essentially attempts to extend Regge
Theory to reactions where a hard scale is present.  Many such models
have been constructed. \citeasnoun{ref:Crittenden} gives a review of
these.  We will focus in this review on pQCD inspired models as they
contain more dynamics, and generally make more quantitative
predictions. A focus of the HERA VM studies is to determine at which
scale, and for which reactions, the pQCD approach gives a better
description of the data than the pomeron model.

\subsubsection{Outline of pQCD approaches}

In the past few years, new calculations of diffractive vector meson
production, using perturbative QCD, have been performed by many
authors. We give here a description of the general features of the
pQCD models.

\epsfigure[width=0.6\hsize]{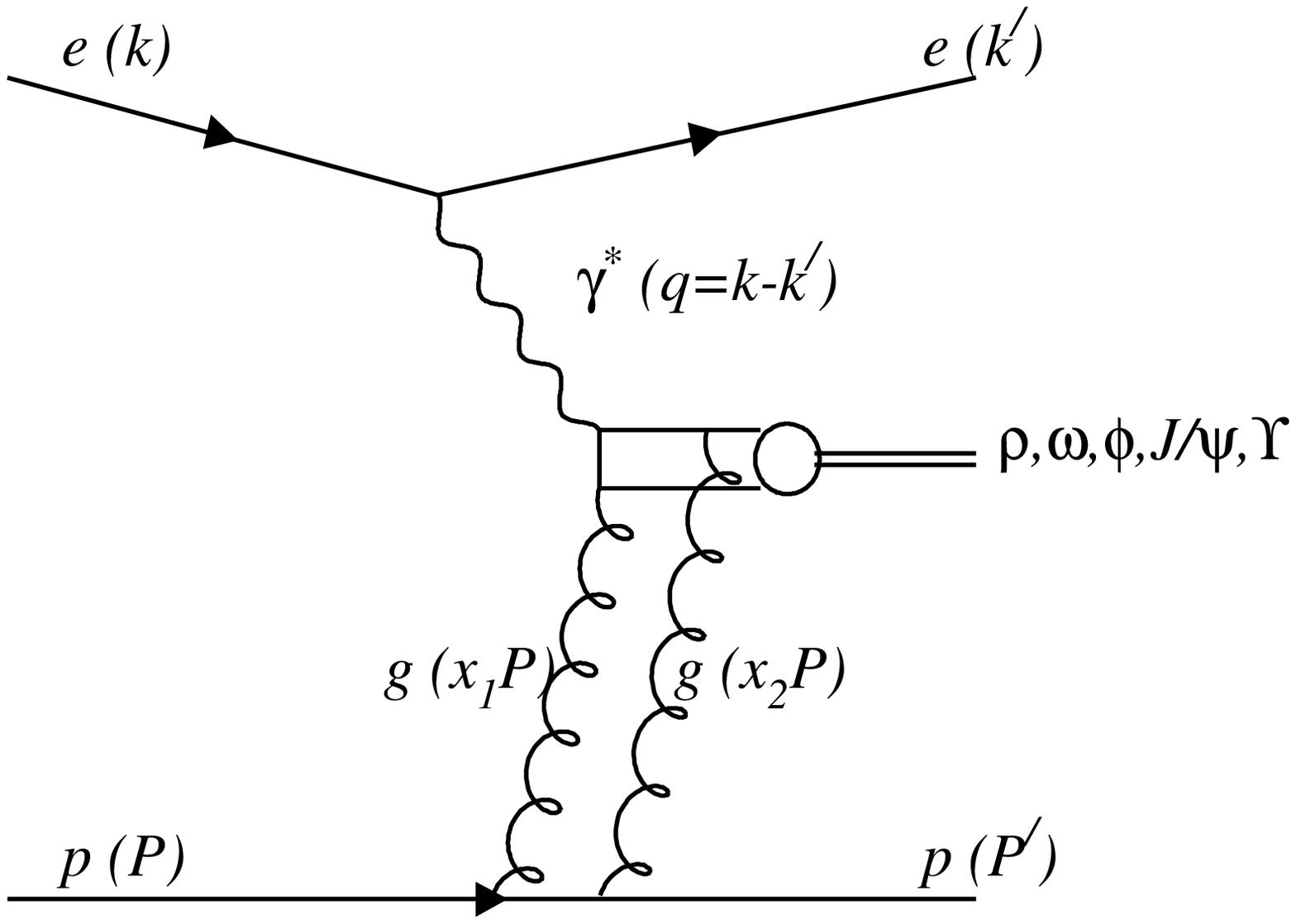} {Exclusive vector
  meson production in QCD based models.  Here the photon is viewed as
  fluctuating into a quark-antiquark pair.  These then couple to the
  proton via the exchange of two gluons (with momentum fractions
  $x_1,x_2$).  The vector meson is formed after the scattering has
  occurred.}  {qcdvm}

In pQCD models, the scattering (${\gamma p \rightarrow V p}$) is
viewed, in the proton rest frame, as a sequence of events very well
separated in time~\cite{ref:Brodsky}.  The process is depicted in
Fig.~\ref{fig:qcdvm}.  The steps are:
\begin{description}
\item{I)} The photon fluctuates into a $q\bar q$ state, 
\item{II)} the $q\bar q$  pair scatters on the proton target,
\item{III)} the scattered $q\bar q$  pair turns into a vector meson.
\end{description}

The proton target acts as a source of color fields and the interaction
with the quark-antiquark pair is mediated by the exchange of two
gluons in a color singlet state.  According to this picture,
diffractive production of vector mesons in the presence of a hard
scale probes the gluonic content of the proton.  The scale $\mu$ at which
$\as$ and the gluon density are evaluated can depend on the photon
virtuality $Q^2$, on the mass of the vector meson M$_V$ and on the
$4$-momentum transfer at the proton vertex, $t$;
\begin{equation}
 \mu^2 = f(M_V^2, Q^2, t) \; ,
\end{equation}
with different models using different ans\"atze.  This formalism can
be applied to photoproduction events if $M_V$ is large enough.

\paragraph{$J/\psi$ production in pQCD}

\citeasnoun{ref:Ryskin1} has calculated the cross section
for diffractive photo- and electroproduction of $J/\psi$ mesons within
the $\as \ln\frac{1}{x}$ approximation and in a constituent quark
model approximation for the wave function of the $J/\psi$.  In this
model, the scale of the interaction is given by
\begin{equation}
\mu^2 = \frac{Q^2+m_{J/\psi}^2+|t|}{4} \; .
\end{equation}
In photoproduction at small $t$, the scale would therefore take on the
value $\mu^2=2.4$~GeV$^2$.  Using a gluon density with an $x$
dependence given by $x^{-0.2}$ at small-$x$ would results in a cross
section dependence $\sigma_{\gamma p \rightarrow J/\psi p}
\propto W^{0.8}$ at large $W$, a much steeper dependence
than expected from Regge Theory.

The calculation was extended beyond leading log~\cite{ref:Ryskin2} and
compared to HERA data.  It was found that, although large
normalization uncertainties remain, this process is indeed very
sensitive to the form of the gluon density in the proton.

\paragraph{Electroproduction of vector mesons}

In DIS, the cross section is predicted to be dominated by
longitudinally polarized virtual photons scattering into
longitudinally polarized vector
mesons~\cite{ref:DoLa_VM,ref:BFGMS,ref:Ginzburg1}.  The cross section
has been calculated in leading $\as
\ln\frac{Q^2}{\Lambda^2}\ln\frac{1}{x}$ approximation~\cite{ref:BFGMS}
for vector mesons built of light flavors, and reads
\begin{equation}
\label{eq-cross}
\left.\frac{{\rm d}\sigma_L}{{\rm d}t}\right|_{t=0} = \frac{A}{Q^6}\as^2(Q^2)
 \left| \left( 1 + \frac{i\pi}{2} \frac{{\rm d}}{{\rm d}\ln x} \right) 
 xg(x,Q^2)\right|^2 \, ,
\end{equation}
where $A$ is a constant which depends on the VM wave function.  Within
this theoretical framework the measurement of $\left.{\rm
d}\sigma/{\rm d}t\right|_{t=0}$ for vector meson electroproduction
provides a probe for the gluon content in the proton, which is
sensitive to the square of the gluon density. Different types of
vector mesons can be used for independent measurements that can be
compared to the theoretical predictions.

We review some of the expectations of the cross sections:
\begin{itemize}
\item
The cross section contains a $1/Q^6$ factor.  However, the $Q^2$
dependencies of $\as$ and the gluon density also need to be taken into
account.  These compensate some of the fall off at small $x$.  The
effective $Q^2$ dependence using CTEQ3L~\cite{ref:CTEQ3} and the LO
form for $\as$ is found to be $d\sigma/dt \propto 1/Q^5$, with a weak
$x$ dependence.

The calculation presented in~\citeasnoun{ref:BFGMS} has been redone in
leading $\as \ln\frac{Q^2}{\Lambda^2}$ approximation~\cite{ref:Koepf}.
In this work, among other improvements, the Fermi motion of the quarks
in the vector meson has been considered.  The net effect is to reduce
the steepness of the $Q^2$ dependence and to delay the onset of the
asymptotic regime.  Precise measurements could therefore in principle
yield information on the wave function of the vector mesons.

\item
In the pQCD calculations, there is no coupling of the $t$ and $W$
dependences, such that no shrinkage is expected.  The lack of
shrinkage, along with the prediction of a steep $W$ dependence, are
telltale signs that the reaction is predominantly driven by
perturbative processes.
\item
The cross section presented above is for longitudinally polarized
photons. The authors~\cite{ref:BFGMS} expect that this is the dominant
contribution to the cross section in DIS.

The region of validity of the pQCD calculations has been investigated
be several authors.  In~\citeasnoun{ref:Ginzburg2}, it is argued that
the region of validity of the pQCD calculations is signaled by the
exclusively longitudinal polarization of the vector meson.  The reason
for this~\cite{ref:BFGMS} is that longitudinally polarized photons
will preferentially produce $q\bar{q}$ pairs which are symmetric in
longitudinal momentum, and therefore have large $k_T$, while
transverse photons will produce asymmetric pairs in longitudinal
momentum, and correspondingly small $k_T$.  In the latter case, the
scattering is dominantly soft (see the discussion in
section~\ref{sec:spacetime}).

A recent pQCD calculation for $\rho^0$
electroproduction~\cite{ref:Martin_Ryskin_Teubner} based on the open
production of light $q\overline{q}$ pairs and parton-hadron duality
gives an estimate of the transverse photon contribution to the
$\gamma^*p \rightarrow Vp$ cross section.  It is found that the
transverse cross section does not drop off as fast as would be
predicted from a convolution of the vector meson wave function.  In
fact, the authors claim that a convolution of the type
$<q\overline{q}|\rho^0>$ is wrong and cannot reproduce the data.
Rather, confinement forces the $q\overline{q}$ into a $\rho^0$ long
after the interaction with the proton.

\item
The interaction should be flavor independent at high enough scales.
From the quark charges of the vector mesons and a flavor independent
production mechanism, the exclusive production cross section is
expected to have relative size $9:1:2:8$ for
$\rho^0:\omega^0:\phi:J/\psi$. This expectation is badly broken at
small $Q^2$, where the heavier vector mesons are strongly suppressed.
The pQCD predictions change the ratio somewhat due to wave function
effects, such that the relative contribution from heavier vector
mesons is modified~\cite{ref:Koepf} at large $Q^2$.
\end{itemize}

\paragraph{Vector meson production at large $t$}

In~\cite{ref:Forshaw,ref:Bartels}, it is pointed out that the
hard scale necessary for the pQCD calculations to be valid can also be
provided by $t$, and that, at large $t$, the BFKL equation can be used
to predict the slope of both the $W$ and $t$ dependencies.  Recent QCD
calculations predict that light vector mesons produced at large $t$ in
proton dissociative production should have zero
helicity~\cite{ref:Ivanov}, independent of the initial photon
polarization.

\subsection{Pre-HERA experimental results}
A comprehensive review of results from photoproduction of vector
mesons prior to 1978 can be found in~\citeasnoun{ref:Bauer}.  These
measurements show, at energies up to $20$~GeV, a weak dependence of
the cross sections on energy, similar to those found in hadron-hadron
scattering.  The $t$ dependence of the data can be characterized by an
exponential at small $t$, $d\sigma/d|t| \propto exp(-b|t|)$, and the
$Q^2$ dependence is well described by VDM (see Eq.~(\ref{eq:VDMQ2})).
Also, the helicity of the vector meson is similar to that of the
incident photon, i.e. s-channel helicity is largely conserved (SCHC).

At larger $Q^2$, leptoproduction results have been presented by
EMC \cite{ref:EMC_VM1,ref:EMC_VM2}, NMC \cite{ref:NMC_VM1,ref:NMC_VM2}
and E665 \cite{ref:E665_VM}.  In these measurements, the $Q^2$ range
extends up to $25$~GeV$^2$ and $W$ extends to $28$~GeV.  The data are
generally consistent with a $1/Q^4$ behavior, the $|t|$ slope is found
to be much shallower than in photoproduction (the NMC
Collaboration~\cite{ref:NMC_VM2} has measured $b=4.6\pm0.8$~GeV$^{-2}$
for $Q^2>6$~GeV$^2$), the fraction of longitudinally polarized
$\rho^0$'s increases beyond $50$~\% at the highest $Q^2$ probed, and
no significant $W$ dependence of the cross sections is found.

\subsubsection{Comment on Vector Meson Data}

It should be mentioned that the experimental results on $\rho^0$
production are not always consistent.  For example, NMC
measurements~\cite{ref:NMC_VM2} disagree with EMC
measurements~\cite{ref:EMC_VM1,ref:EMC_VM2} as to the $Q^2$, $t$, and
angular dependencies of the data.  The recent cross section
measurements by E665 are considerably higher than those from the NMC
collaboration, and the H1 and ZEUS collaborations have widely
differing cross sections for $\rho^0$ electroproduction.  The problems
likely stem from two main sources:
\begin{enumerate}
\item
The difficulty in defining the $\rho^0$ signal.  The resonance is
broad, and must be integrated over a fixed range.  The contribution
from non-resonant $\pi^+\pi^-$ production or background under the
resonance is very difficult to estimate.
\item
The uncertainty in the contribution from proton dissociation
reactions.  This background is difficult to estimate since a large
fraction of the events with proton dissociation look in the detector
like elastic events, and there is no good model describing the
properties of the proton dissociation system.
\end{enumerate}

Given these difficulties, comparison of data sets from different
measurements should be performed with care, as they could lead to
erroneous conclusions.

\subsection{HERA results}
\subsubsection{Kinematics of vector meson production at HERA}

In addition to the standard DIS variables, several extra variables are
needed to describe vector meson production in the reaction,
\begin{equation}                                                      
      e~p \rightarrow e~V~+Y \, ,
\end{equation} 
\noindent
where $Y$ represents either a proton or a diffractively dissociated 
proton remnant of mass $M_Y$:
\begin{itemize}
\item
$ t^\prime=|t~-~t_{min}|$, where $t$ is the four-momentum transfer
squared, $t$ = $(q - v)^2 = (P-P^{\prime})^2$, from the photon to the
$\rho^0$ (with four-mqomentum $v$), $t_{min}$ is the minimum
kinematically allowed value of $t$ and $P^{\prime}$ is the
four-momentum of the outgoing proton.  The squared transverse momentum
$p_T^2$ of the $\rho^0$ with respect to the photon direction is a good
approximation to $t^\prime$,
\item
the three angles $\Phi$, $\phi_h$, and $\theta_h$, described below.
\end{itemize}
 
\epsfigure[height=0.8\hsize,angle=-90]{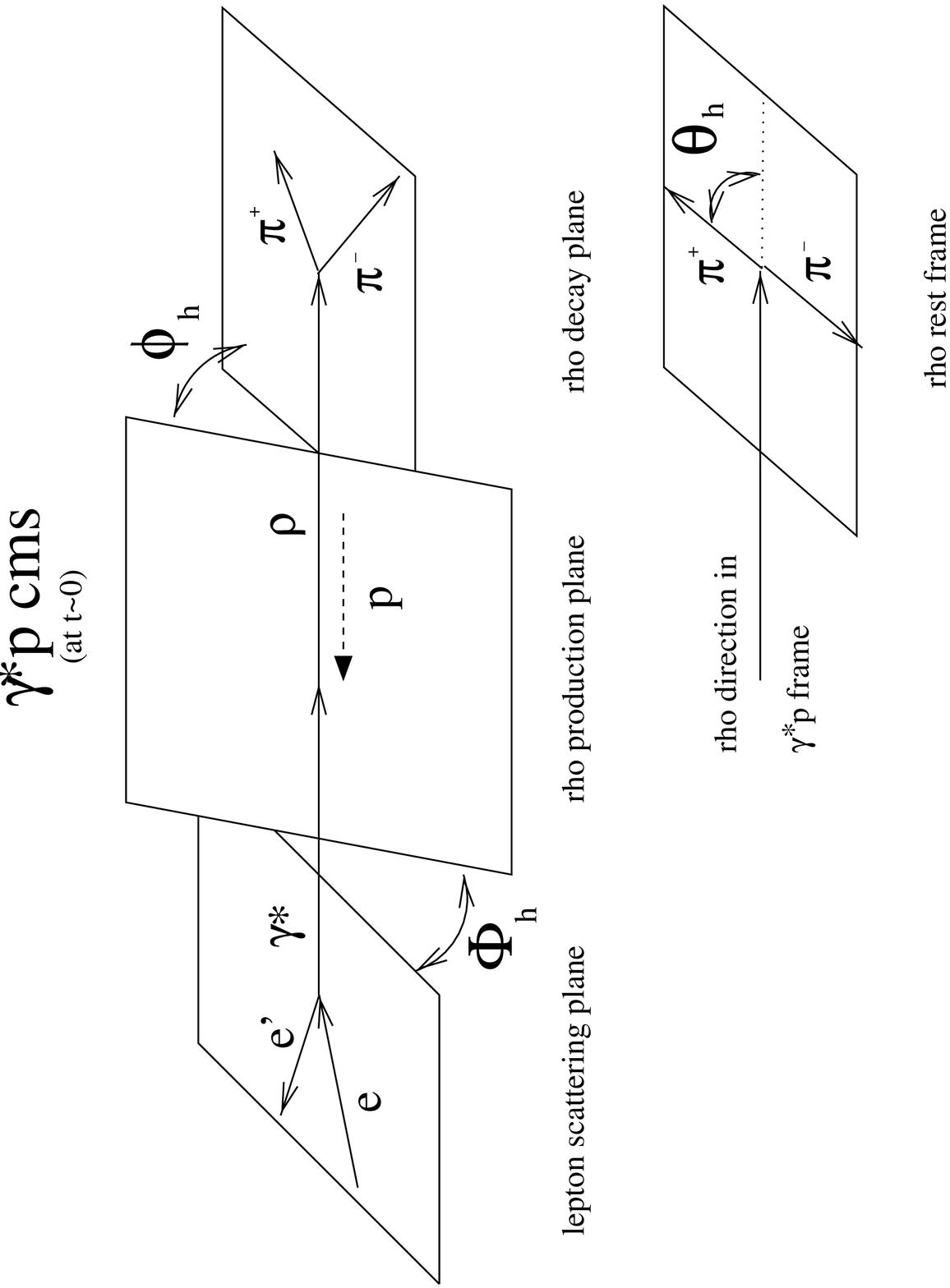} {A graphical
  description of the angles needed to analyze the helicity states of
  the vector meson (here denoted as $\rho$): $\theta_h$ is the polar
  angle in the helicity system, $\phi_h$ the angle between the $\rho$
  decay and production plane, and $\Phi_h$ the angle between the
  lepton scattering and $\rho$ production planes.}  {angle_defs}

The production and decay angles are usually defined in the s-channel
helicity frame~\cite{ref:Schilling_and_wolf}, as shown in
Fig.~\ref{fig:angle_defs}.  The vector meson direction in the
$\gamma^*p$ frame is taken as the quantization axis, and, in the case
of a two-body decay, the direction of one of the particles in the rest
frame of the vector meson is used to calculate $\theta_h, \phi_h$.  In
the case of a three-body decay, the normal to the decay plane is used
to define the decay angles.  The angle $\Phi$ between the vector meson
production plane and the electron scattering plane in the $\gamma^*p$
rest frame is also used.  The full decay angular distribution is
usually expressed in terms of $(\cos\theta_h,
\phi_h, \Phi)$.  In the case where the electron beam is not longitudinally
polarized, the electron beam energy is not varied, and s-channel
helicity conservation (SCHC) is assumed to apply, the decay distribution
reduces to (for a decay into two spin-0 particles),
\begin{eqnarray}
\lefteqn{W(\cos\theta_h,\psi_h) = \frac{3}{4\pi}\left[\frac{1}{2}
(1-r_{00}^{04})
+ \frac{1}{2}(3r_{00}^{04}-1)\cos^2\theta_h \right. } \ \ \ \ \ \ \ \ \ 
 \\
& & \! \! \! \! \! \left. \mbox{ } + 
\epsilon r_{1-1}^{1}\sin^2\theta_h\cos2\psi_h-2
\sqrt{\epsilon(1+\epsilon)}
Re[r_{10}^{5}]\sin2\theta_h\cos\psi_h \right], \nonumber 
\end{eqnarray}
where $\psi_h = \phi_h - \Phi$ and $r_{00}^{04},\; r_{1-1},\;
Re[r_{10}^{5}]$ are the non-zero combinations of the vector meson spin
density matrix elements.  The parameter $\epsilon$ is the ratio of the
longitudinal to transverse photon flux.  $r_{00}^{04}$ gives the
probability that the vector meson is produced in a helicity zero
state, and, if s-channel helicity holds, is related to the ratio of
the longitudinal to transverse photon cross sections as
\begin{eqnarray}
R & = & \frac{\sigma_L}{\sigma_T} \\
  & = & \frac{r_{00}^{04}}{\epsilon(1-r_{00}^{04})} \;\;\; .
\end{eqnarray}
The matrix element $r_{1-1}^1$ determines the anisotropy of the $\psi_h$
distribution, while $Re[r_{10}^5]$ is related to the interference between
the production amplitudes by longitudinal and transverse 
photons~\cite{ref:Joos}.  Taking the one dimensional projections in
$\cos\theta_h$ and $\phi_h$ gives
\begin{eqnarray}
\frac{1}{N} \frac{dN}{d\cos\theta_h}& =& \frac{3}{4}[1-r_{00}^{04}
+(3r_{00}^{04}-1)\cos^2\theta_h],\\
\frac{1}{N} \frac{dN}{d\psi_h} & = & \frac{1}{2\pi}(1+2\epsilon r^1_{1-1}
\cos2\psi_h) \;\;\; .
\end{eqnarray}
If, in addition to SCHC, the exchange has so-called natural parity, then
the further relationship holds
\begin{equation}
r^1_{1-1} = \frac{1}{2}(1-r_{00}^{04}) \, .
\end{equation}

\subsubsection{Experimental analysis}
The analysis of vector meson production is at first sight quite
straightforward.  Events are searched for with typically two or three
isolated tracks and no extra energy depositqs in the calorimeter (e.g.,
$ep \rightarrow e~\rho~p, \; \rho \rightarrow \pi^+ \pi^-$, where the
electron is in the detector in the case of DIS, and escapes down the
beampipe in the case of photoproduction.)  The momenta from the tracks
are then used to reconstruct an invariant mass, and events are kept
where the reconstructed mass is near the mass of the particle under
study.  The kinematics are generally reconstructed very precisely.  In
DIS, the variables ($Q^2,y,\; p_T^2,\; M_V,\; \Phi,\; \phi_h,
\; \theta_h$) can all be reconstructed, while
in photoproduction, the angle $\Phi$ is unmeasurable because the
scattered electron is typically not observed.  The complication in the
analysis comes from the fact that the acceptance depends, to
varying extent, on all kinematic variables.  It is therefore
important to have Monte Carlo simulations which are capable of
reproducing the data to have confidence in the results.

It is usually required to see the decay tracks of the vector meson in
the central tracking detectors to perform a full reconstruction of the
events.  This limits the $W$ range of the measurements to typically
$40 < W < 140$~GeV.  The $Q^2$ range up to $50$~GeV$^2$ has been
measured, limited by the steep drop of the cross section with $Q^2$.
The $t$ dependence is also steep, and at large $|t|$ backgrounds from
proton dissociation are important.  This generally limits analyses to
the range $|t|<0.6$~GeV$^2$.  Measurements have also been performed
for the proton dissociation events, in which case larger $|t|$ values
are used.

The cross sections of interest are the $\gamma^*p$ cross sections.  These
are derived from the $ep$ cross sections starting with the
differential form of the $ep$ cross section,
\begin{equation}
\frac{d\sigma_{ep\rightarrow epV}}{dydQ^2} = \Gamma (\sigma_T + 
\epsilon \sigma_L) \, ,
\end{equation}
where
\begin{equation}
\label{eq:fluxfactor}
\Gamma = \frac{\alpha}{2\pi y Q^2} \left(Y_+ - 2(1-y)\frac{Q^2_{min}}{Q^2}
\right)
\end{equation}
and
\begin{equation}
\epsilon = \frac{2(1-y)}{Y_+-2(1-y)\frac{Q^2_{min}}{Q^2}} \; .
\end{equation}
$Q^2_{min}=m_e^2\frac{y^2}{(1-y)}$ is the minimum possible $Q^2$ and
$Y_+=1+(1-y)^2$.

\begin{itemize}
\item
In photoproduction, the behavior of the cross section is assumed to
follow the VDM form, given in Eq.~(\ref{eq:VDMQ2}),
\begin{eqnarray}
\sigma_T(y,Q^2)&=& \left(\frac{M_V^2}{Q^2+M_V^2}\right)^2\sigma_T(y,Q^2=0)\, ,
\\
\sigma_L(y,Q^2)&=& \frac{Q^2}{M_V^2}\sigma_T(y,Q^2) \; .
\end{eqnarray}
The $y$ dependence is expected to be slow (recall that $y=W^2/s$).  In
this case, we can extract $\sigma_T$ from
\begin{eqnarray}
\lefteqn{\int_{y_{min}}^{y_{max}}dy \int_{Q^2_{min}}^{Q^2_{max}} dQ^2
\frac{d\sigma_{ep \rightarrow epV}}{dydQ^2} = } \ \ \ \ \ \\
& & \!\!\!\!\! \! \! \sigma_T(\bar{y},Q^2=0)\!\!\!
\int_{y_{min}}^{y_{max}}\! \! \!\!\!dy \int_{Q^2_{min}}^{Q^2_{max}}\!\!\! dQ^2
\Gamma \left(\!\!\frac{M_V^2}{Q^2+M_V^2}\!\!\right)^2\!\!\left(\!1+\epsilon
\frac{Q^2}{M_V^2}\!\right). \nonumber 
\end{eqnarray}
\item
In the case of heavy quark production, or vector meson production in DIS,
we can no longer assume that the $y$ (or $W$) dependence will be weak.
In this case, we need a model for the $\gamma p$ cross section, which has
to be tuned to the data, before the cross sections can be extracted.  This
procedure is very similar to the extraction of $F_2$.
\end{itemize}

\subsubsection{Backgrounds}

The final state for vector meson production is very clean, and the
background from non-vector meson final states is usually negligible.
However, the cross sections of interest are either the elastic or the
proton-dissociative cross sections.  These are very difficult to
disentangle, since the proton is often excited to a small mass state
which is not distinguishable from elastic reactions in the detectors.
Models are then needed to extrapolate from the high mass tail, which
can be identified in the detectors, to the full mass range for the
proton-dissociative system.  This correction can be quite large (of
order 25~\%), and can therefore lead to uncertainties in the overall
normalization of the elastic cross section.  The $t$ slopes of the two
reactions are also considerably different, such that the extracted $t$
dependence is also quite sensitive to the estimated contributions.

\subsubsection{Light vector meson photoproduction}

A large amount of data exists from pre-HERA times on light vector
meson photoproduction (see, e.g., \citeasnoun{ref:Bauer}).  It is
interesting to test how these results extrapolate to the much higher
center-of-mass energies available in photoproduction at HERA.  Results
exist from HERA on the photoproduction of light vector mesons for
$\rho^0$, $\omega^0$ and $\phi$.  The data sets are summarized in
Table~\ref{tab:VM_soft_summary}.  In the following sections, we
discuss the mass, $W$ and $t$ dependencies of the cross sections, and
review the tests of s-channel helicity conservation.

\begin{table} 
\tablecaption{Summary of results on cross sections and $t$ slopes 
from the elastic photoproduction of light vector mesons at small $t$ 
at HERA.  The different symbols are defined in the text.  The 
$\gamma p$ cross sections quoted are for the $t$ range specified 
in the table.}
\begin{center}  
\rotatebox{-90}{%
\label{tab:VM_soft_summary}
\begin{tabular}{c|c|c|c|c|p{4.1cm}}
Reaction &  $<\!W\!>$ & $\sigma_{\gamma p\rightarrow Vp}$ &
$|t|$-range & $b$  & comment\\ 
studied &(GeV) & ($\mu$b) & (GeV$^2$) & (GeV$^{-2}$) & \\
\hline
$\rho^0 \rightarrow \pi^+~\pi^-$  & 
$70$ & $14.7\pm0.4\pm2.4$ & $ <0.50$ & $9.9\pm1.2\pm1.4$ & 
\protect\cite{ref:ZEUS_photo_rho1} \\
&&&&&\\
$\rho^0 \rightarrow \pi^+~\pi^-$  & 
$55$ & $9.1\pm0.9\pm2.5$ & 
$ <0.50$ & $10.9\pm2.4\pm1.1$ & \protect\cite{ref:H1_photo_rho} \\ 
 & $187$ & $13.6\pm0.8\pm2.4$ & & &\\
&&&&&\\
$\rho^0 \rightarrow \pi^+~\pi^-$  & 
$73$ & $5.8\pm0.3\pm0.7$ & 
$0.073-0.40$ & $9.8\pm0.8\pm1.1$ &\protect\cite{ref:ZEUS_LPS_rho}
 \\
&&&&&\\
$\rho^0 \rightarrow \pi^+~\pi^-$  & 
$55$ & $10.9\pm0.2^{+1.5}_{-1.3}$ & $<0.50$ & $10.9\pm0.3^{+1.0}_{-0.5}$
& fit form  \\
 &$65$ & $10.8\pm0.2^{+1.3}_{-1.1}$ & & & $d\sigma/d|t|=e^{-b|t|+ct^2}$\\ 
 & $75$ & $11.4\pm0.3^{+1.0}_{-1.2}$ & & & \protect\cite{ref:ZEUS_photo_rho2}
\\ 
 & $90$ & $11.7\pm0.3^{+1.1}_{-1.3}$ & & & \\ 
&&&&&\\
\hline
&&&&&\\
$\omega^0 \rightarrow \pi^+~\pi^-~\pi^0$  & 
$80$ & $1.21\pm0.12\pm0.23$ & 
$ <0.60$ & $10.0\pm1.2\pm1.3$ &\protect\cite{ref:ZEUS_photo_omega}\\ 
&&&&&\\
\hline
&&&&&\\
$\phi \rightarrow K^+~K^-$  & 
$70$ & $0.96\pm0.19^{+0.21}_{-0.18}$ & 
$ < 0.50$ & &\protect\cite{ref:ZEUS_photo_phi}
\\ 
 & & & & &
\end{tabular}  
}
\end{center}
\end{table}

\paragraph{Mass spectra}

\epsfigure[width=0.6\hsize]{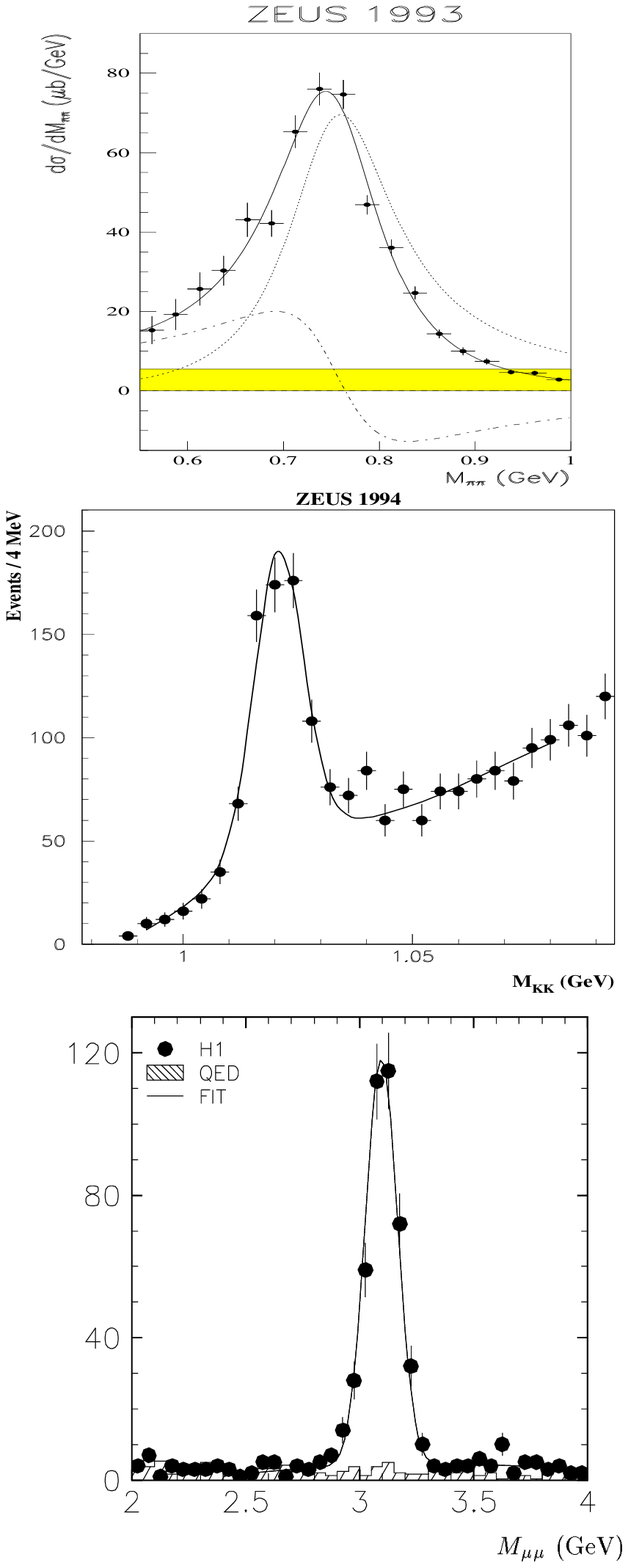} {Typical mass
  distribution observed in photoproduction for the $\rho^0\rightarrow
  \pi^+\pi^-$ signal (top), the $\phi \rightarrow K^+ K^-$ signal
  (center), and the $J/\psi \rightarrow \mu^+\mu^-$ signal (bottom).
  In the case of the $\rho^0$, the dotted curve represents the
  resonant Breit-Wigner contribution and the dot-dashed curve the
  interference term; the shaded band represents the size of the
  uncertainty on the background term.  The continuous curve is the sum
  of the contributions.}  {mass_spectra}

Typical mass distributions are shown in Fig.~\ref{fig:mass_spectra}.
Clear peaks are seen in all cases.  The $\phi$ and $J/\psi$ mass
distributions are narrow, and the backgrounds under the peak are
extracted in a straightforward way by fitting the distributions with
Gaussians or Gaussians+Breit-Wigner functions for the peaks and
polynomial functions for the background.  In the case of the $\rho^0$,
the mass spectrum is quite broad, and care has to be taken in
extracting the resonant contribution.

The mass distribution of the $\rho^0$ is skewed compared to a
Breit-Wigner distribution, as seen in Fig.~\ref{fig:mass_spectra}:
there is an enhancement of the small mass side and a suppression of
the high mass side. This distribution can be understood in terms of
the interference between the resonant $\pi^+ \pi^-$ production and a
non-resonant background as discussed by \citeasnoun{ref:Soeding}.  The
resonant production is described by a relativistic p-wave Breit-Wigner
function,
\begin{eqnarray}
BW_{\rho}(M_{\pi\pi}) = \frac{M_{\pi\pi} m_{\rho}
\Gamma_{\rho}(M_{\pi\pi})} {(M_{\pi\pi}^2-m_{\rho}^2)^2 + 
m_{\rho}^2 \Gamma_{\rho}^2(M_{\pi\pi})} \, ,
\label{breit}
\end{eqnarray}
 with a momentum dependent width~\cite{ref:Jackson}
\begin{eqnarray}
\Gamma_{\rho}(M_{\pi\pi}) = \Gamma_0 \left(\frac{p^*}{p^*_0}\right)^3 
\frac{m_{\rho}}{M_{\pi\pi}} \, ,
\label{gamma1}
\end{eqnarray}
where $\Gamma_0$ is the width of the $\rho^0$, $p^*$ is the $\pi$ 
momentum in the $\pi^+ \pi^-$ rest frame and $p^*_0$ is the value 
of $p^*$ at the $\rho^0$ nominal mass $m_{\rho}$.

Parameterizations are then used for the background and interference terms
which are added to the Breit-Wigner in order to describe the mass spectrum.
In this way, fits can be performed which reproduce the standard values for
the $\rho^0$ mass and width; viz., $m_{\rho}=0.770$~GeV and 
$\Gamma_0=150$~MeV~\cite{ref:EWpar}.  An example of such a fit is given in
Fig.~\ref{fig:mass_spectra}, where the different terms are described.

Different assumptions for the functional form of $d\sigma/d
M_{\pi\pi}$ are often used.  One common form is that proposed
\citeasnoun{ref:Stodolsky},
\begin{eqnarray}
\frac{d\sigma}{dM_{\pi\pi}} = f_{\rho} \cdot BW_{\rho}(M_{\pi\pi}) \cdot 
(m_{\rho}/M_{\pi\pi})^k + f_{PS},
\label{stodolsky}
\end{eqnarray}

\epsfigure[width=0.8\hsize]{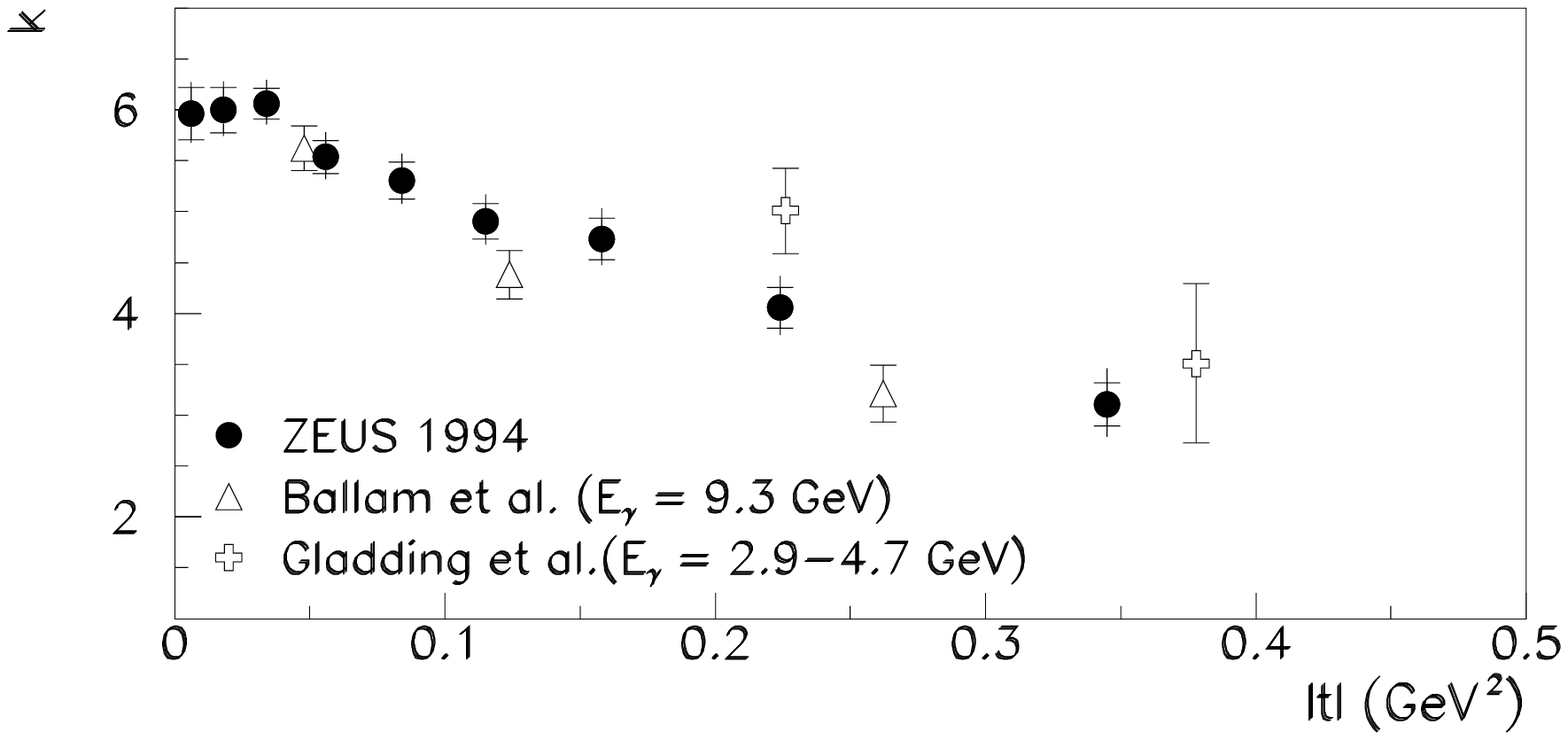} {The Ross-Stodolsky
  parameter $k$ as a function of $|t|$.  The ZEUS data are compared to
  results from fixed target experiments~\cite{ref:Ballam,ref:Gladding}
  and show the same trend.}  {RS_vs_t}

\noindent
where the factor $(m_{\rho}/M_{\pi\pi})^k$ accounts for the skewing of
the shape of the $\rho^0$ signal. The background term $f_{PS}$ is
usually taken to be constant. The value of $k$ is then a measure of
the skewing.  This value is given in Fig.~\ref{fig:RS_vs_t} as a
function of $|t|$.  It is seen that the skewing is reduced when $|t|$
is increased.

\paragraph{$t$  dependence}

\epsfigure[width=0.8\hsize]{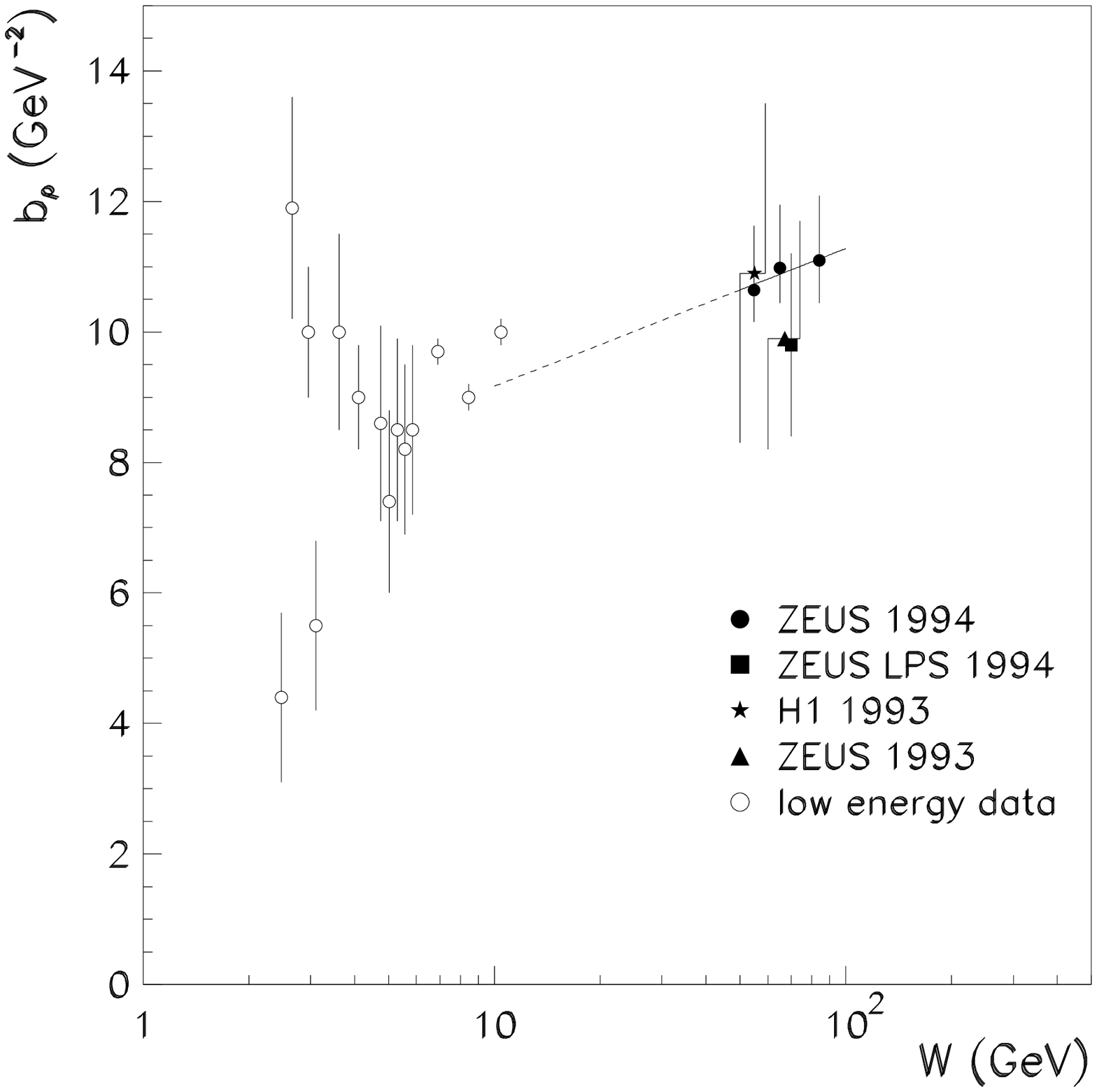} {The slope $b$ for
  elastic $\rho^0$ photoproduction.  The line is the result of a fit
  of the form $b=b_0+2\alpha'ln(W^2/W_0^2)$.  The fit is performed to
  the ZEUS data, and extrapolated down to the fixed target
  measurements.}  {photo_rho_bvsW}

The $t$ slopes are summarized in Table~\ref{tab:VM_soft_summary}.  As
can be seen, values around $b=10$~GeV$^{-2}$ are measured.  The
development of $b$ with $W$ is shown for $\rho^0$ photoproduction in
Fig.~\ref{fig:photo_rho_bvsW}.  The results from fixed target
experiments are shown along with the HERA data. A fit of the form
$b=b_0+2\alpha'\ln(W^2/W_0^2) $ was performed on the ZEUS
data~\cite{ref:ZEUS_photo_rho2}, resulting in
$\alpha'=0.23\pm0.15^{+0.10}_{-0.07} \gevmtwo$.  The result is
compatible with the presence of shrinkage, but the large errors on
$\alpha'$ precludes any definitive statement.

The $|t|$ slope has been measured as a function of the invariant mass
of the $\pi^+~\pi^-$, and found to decrease in accord with results
from previous measurements~\cite{ref:Bauer}.

\paragraph{$W$ dependence}

\epsfigure[width=0.8\hsize]{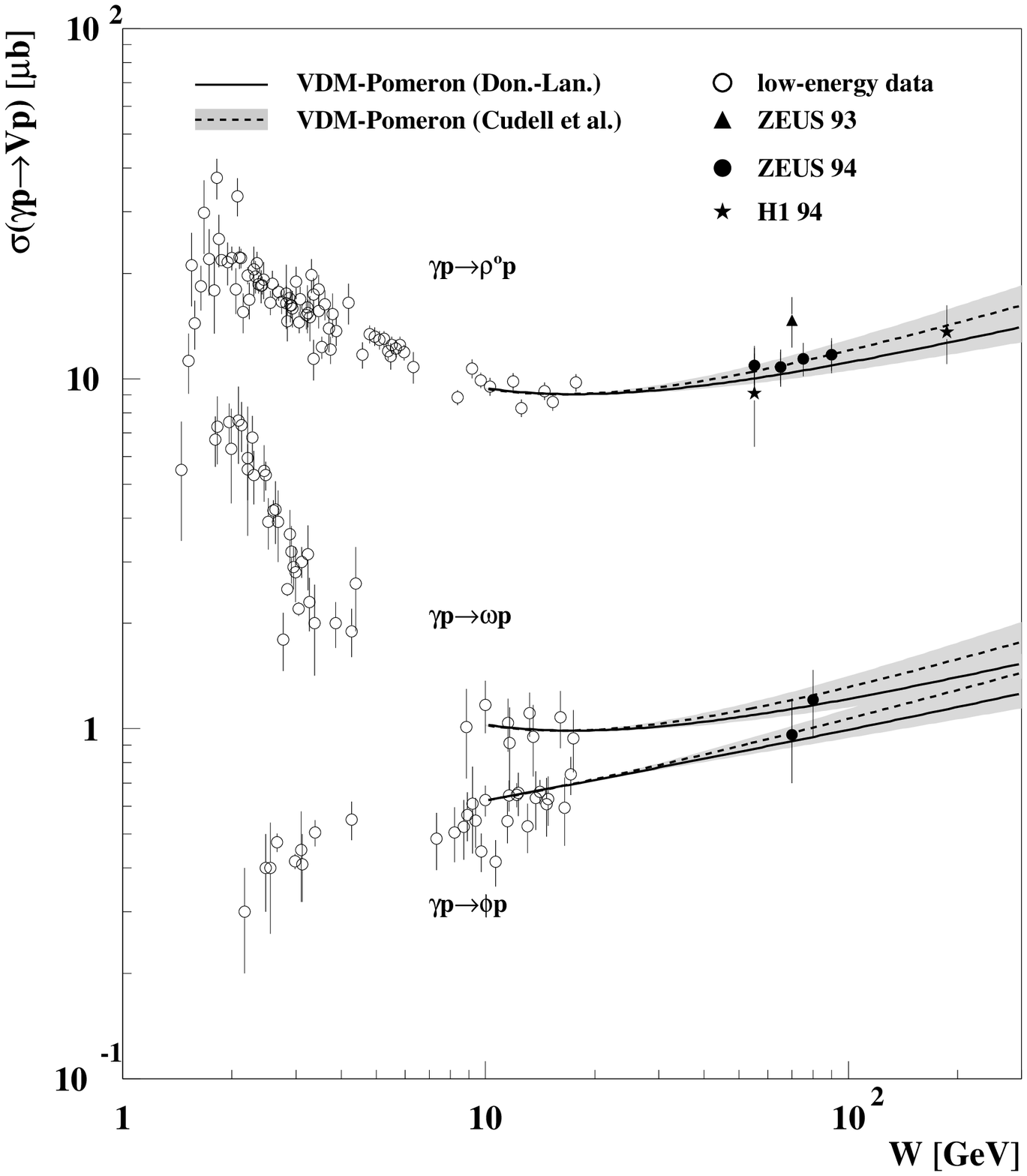} {The elastic vector
  meson photoproduction cross sections as functions of $W$ for
  $\rho^0,\; \omega$ and $\phi$.  The solid curve is for the pomeron
  model of~\citeasnoun{ref:DoLa_VM} while the dashed curve is based on
  an updated version by \citeasnoun{ref:Cudell}.  The error band
  specifies the range of expectations in the Cudell model.}
{photo_VM}

The $W$ dependence of the cross sections are shown in
Fig.~\ref{fig:photo_VM}. Given a typical $t$ slope of
$b=10$~GeV$^{-2}$, the Regge theory expectation for the exponent of
$W$ in the cross section is (see Eq.~(\ref{eq:Wshrinkage}))
\begin{equation}
\delta=4(1.08 - \frac{0.25}{10} -1) = 0.22
\end{equation}
If the more recent value of $\alpha_0=1.096^{+0.012}_{-0.009}$ from
\citeasnoun{ref:Cudell} is taken, then the expected $W$ slope
is $\delta=0.28^{+0.05}_{-0.04}$.  These two expectations are shown in
the figure. They are both in good agreement with the data.

\paragraph{Test of SCHC}

As described above, the angle $\Phi$ is not measured in photoproduction.
Integrating over this angle, the distribution in the remaining two angles
is
\begin{eqnarray}
\lefteqn{W(\cos\theta_h,\phi_h) = 
\frac{3}{4\pi}\left[ \frac{1}{2}(1-r^{04}_{00}) + 
                     \frac{1}{2}(3r^{04}_{00}-1)\cos^2\theta_h \right.} \ \ \ \ \ \ \nonumber\\ 
& & \mbox{} -\sqrt{2}Re[r^{04}_{10}]\sin2\theta_h\cos\phi_h
\left. \mbox{} -r^{04}_{1-1}\sin^2\theta_h \cos2\phi_h \right] \; .
\label{eq:photo_SCHC}
\end{eqnarray}
\epsfigure[width=0.8\hsize]{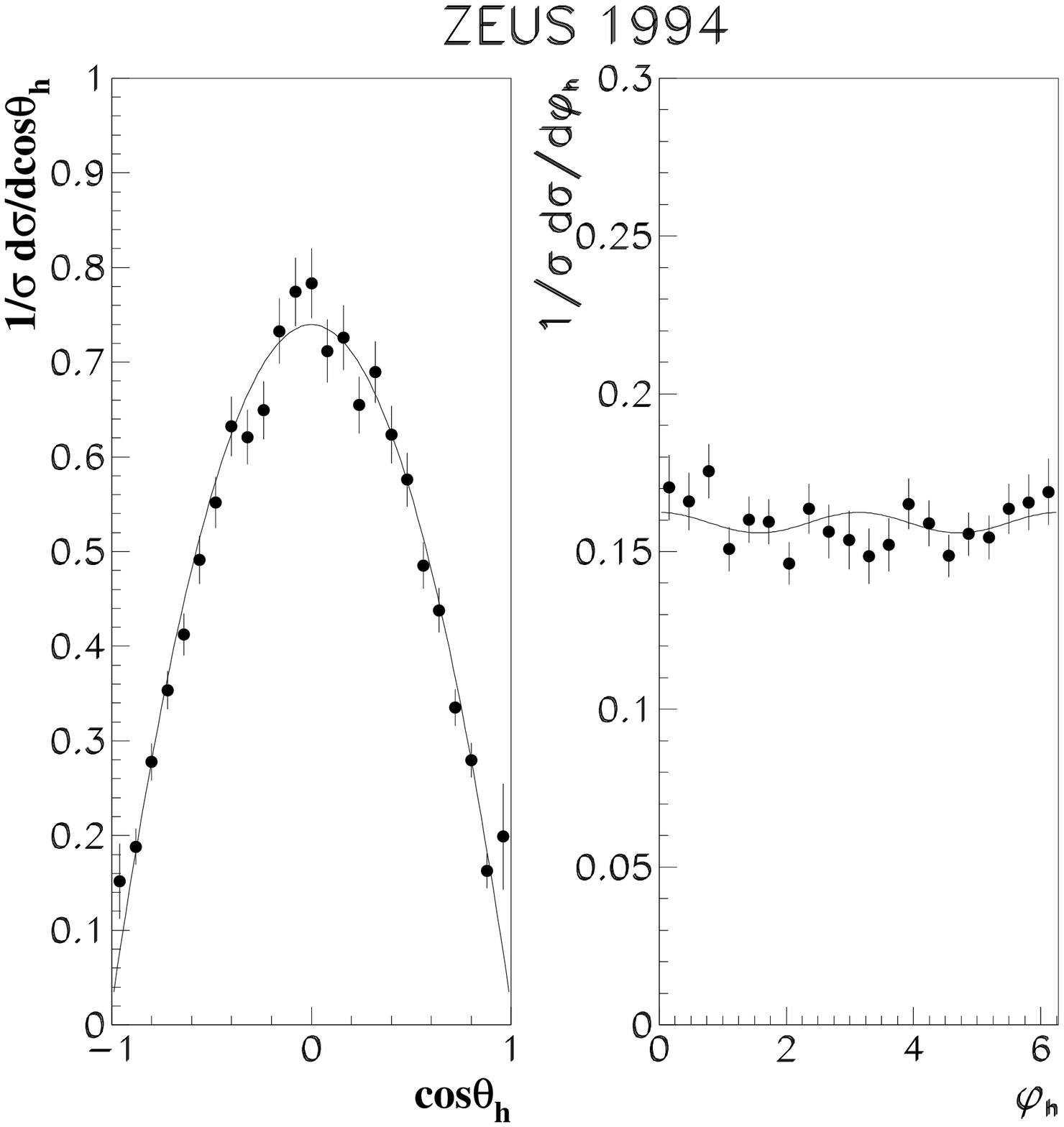} {The differential
  distributions in $cos\theta_h$ and $\phi_h$ for the reaction $\gamma
  p \rightarrow \rho^0 p$.}  {photo_angles}

The element $r^{04}_{1-1}$ is related to the size of the interference
between the helicity non-flip and double-flip amplitudes, while
$qRe[r^{04}_{10}]$ is related to the interference between the helicity
non-flip and single flip amplitudes.  As mentioned above,
$r^{04}_{00}$ is the probability to find the $\rho^0$ with helicity
$0$.  If SCHC holds, then both $r^{04}_{1-1}$ and $Re[r^{04}_{10}]$
should be zero.  In addition, $r^{04}_{00}$ is expected to be small
since the photon is predominantly transversely polarized.  Typical
distributions for $\cos \theta_h$ and $\phi_h$ are shown in
Fig.~\ref{fig:photo_angles}, along with the result of a fit using
Eq.~(\ref{eq:photo_SCHC}).  For this particular
result~\cite{ref:ZEUS_photo_rho2}, the fitted matrix elements are
\begin{eqnarray}
r^{04}_{00} & = & 0.01 \pm 0.01 \pm 0.02 \; , \\
r^{04}_{1-1} & = & -0.01 \pm 0.01 \pm 0.01 \; ,\\
Re[r^{04}_{10}] & = & 0.01 \pm 0.01 \pm 0.01 \; ,
\end{eqnarray}
in good agreement with expectations from SCHC.  This conclusion is
generally true for all photoproduction measurements.

\paragraph{Proton dissociation results}

The ZEUS experiment has also measured the $|t|$ dependence for $\gamma
p \rightarrow \rho^0 N$, where $N$ represents a small mass baryonic
state from proton dissociation.  In this case, the $t$ slope is
measured to be about $b=6$~GeV$^{-2}$~\cite{ref:ZEUS_photo_rho2};
i.e., about half the value found for elastic $\rho^0$ production.  A
smaller value of $b$ is expected since the reaction is now inelastic
on the proton side, with a smaller effective radius.

\subsubsection{$J/\psi$ and $\Upsilon$ photoproduction}

The reaction $\gamma p \rightarrow J/\psi Y$ has been measured at HERA
for several different cases:
\begin{enumerate}
\item The system Y consists of a single proton.  This is the
so-called elastic reaction.
\item The system Y consists of a small mass system stemming from
the dissociation of the proton.
\item Inelastic production, generated by photon-gluon fusion, or
resolved photon-parton scattering.
\end{enumerate}
Inelastic $J/\psi$ production has been described in
 section~\ref{sec:partons}.  We focus here on cases 1,2.  The recent
 results on $\Upsilon$ photoproduction are discussed at the end of the
 section.

A summary of the different data sets available on $J/\psi$
photoproduction from HERA is given in Table~\ref{tab:VM_psi_summary}.

\begin{table} 
\tablecaption{ Summary of results on cross sections and
$t$ slopes from the elastic (and proton dissociation)
 photoproduction of $J/\psi$ mesons at HERA.
 The comments apply to the measurement of $b$.}
\label{tab:VM_psi_summary}
\renewcommand\arraystretch{0.7}
\begin{center}  
\rotatebox{-90}{%
\begin{tabular}{c|c|c|c|c|p{4.2cm}}
Reaction &  $<\!W\!>$ & $\sigma_{\gamma p\rightarrow J/\psi N}$ 
& $|t|$-range & $b$  
& Comment \\ 
studied &(GeV) & (nb) & (GeV$^2$) & (GeV$^{-2}$) & \\
\hline
$\gamma p\rightarrow J/\psi p$ & & $56\pm13\pm14$ & $<0.75$ 
& $4.7\pm1.9$ &\cite{ref:H1_photo_psi_94}\\
$J/\psi \rightarrow l^+~l^-$  &&&&&  \\
$l=\mu,e$ &  &  & & & \\
&&&&& \\
$\gamma p\rightarrow J/\psi p$ &$67$  & $52^{+7}_{-12}\pm10$ & $<1.0$ & $5.0\pm1.4$ 
&\cite{ref:ZEUS_photo_psi_95}\\
$J/\psi \rightarrow l^+~l^-$ &$114$ & $71^{+13}_{-20}\pm12$ &&& \\ 
$l=\mu,e$  &&&&& \\
$\gamma p\rightarrow J/\psi p$ & $42$ & $36.8\pm3.9\pm6.6$ 
& $<1.0$ & $3.7\pm0.3\pm0.2$ & $30<W<90$~GeV \\
$J/\psi \rightarrow l^+~l^-$ &$72$ & $50.6\pm4.8\pm9.1$ &  &  & \\ 
$l=\mu,e$  & $102$ & $70.6\pm7.0\pm12.7$ & & $4.5\pm0.4\pm0.3$ & $90<W<150$~GeV \\
 & $132$ & $68.0\pm10.6\pm12.2$ & & & \cite{ref:H1_photo_psi_96} \\
&&&&& \\
$\gamma p\rightarrow J/\psi N$ & $42$ & $23.0\pm3.2\pm4.0$ & $<1.0$ & 
$1.6\pm0.3\pm0.1$ &\cite{ref:H1_photo_psi_96} \\
$J/\psi \rightarrow l^+~l^-$ &  $72$ & $63.5\pm5.8\pm11.4$ & & &\\
& $102$ & $62.7\pm7.4\pm11.2$ & &  &\\
& $132$ & $128.9\pm19.5\pm23.2$ & & &\\
&&&&&\\
$\gamma p\rightarrow J/\psi p$& $49.8$ & $30.4\pm3.4^{+2.9+3.2}_{-4.4-0.}$&$<1.0$ & 
$4.6\pm0.4^{+0.4}_{-0.6}$ &\cite{ref:ZEUS_photo_psi_97} \\
$J/\psi \rightarrow l^+~l^-$& $71.2$ & $42.9\pm4.5^{+4.1+4.1}_{-5.6-0.0}$ & & &\\ 
$l=\mu,e$ & $89.6$ & $57.7\pm5.8^{+5.3+5.8}_{-6.9-0.0}$ & & &\\
& $121$ & $66.5\pm6.8^{+6.4+6.8}_{-9.6-0.0}$ & & &\\
&&&&&\\
$\gamma p\rightarrow J/\psi p$ &$120-240$ & & & $\approx 4$ & $p_T^2<1$~GeV$^2$ \\
$J/\psi \rightarrow l^+~l^-$ &&&&&\cite{ref:H1_photo_psi_97}\\
$l=\mu,e$ &&&&& \\
&&&&&\\ \hline
$\gamma p\rightarrow \psi(2s) N$ &$80$ & $17.9\pm2.8\pm2.7\pm1.4$ & & & $z>0.95$ \\
$\psi(2s) \rightarrow l^+~l^-$, &&&&&\cite{ref:H1_photo_psi_97}\\
$ \rightarrow J/\psi\pi^+\pi^-$ & & & & & \\
\end{tabular}  
}
\end{center}
\end{table}

\paragraph{$t$  dependence}

The slope of the $t$ dependence for elastic production is in the range
$b=4.5-5$~GeV$^{-2}$, which is about a factor of two smaller than was
found in the photoproduction of $\rho^0$ mesons.  This indicates that
the photon is indeed point-like in these interactions, such that pQCD
calculations should be applicable.  The slope has also been measured
by the H1 Collaboration~\cite{ref:H1_photo_psi_97} for the proton
dissociation reaction.  In this case, the slope is reduced to
$b=1.6\pm0.3\pm0.1$~GeV$^{-2}$.

\paragraph{$W$  dependence}

\epsfigure[width=0.8\hsize]{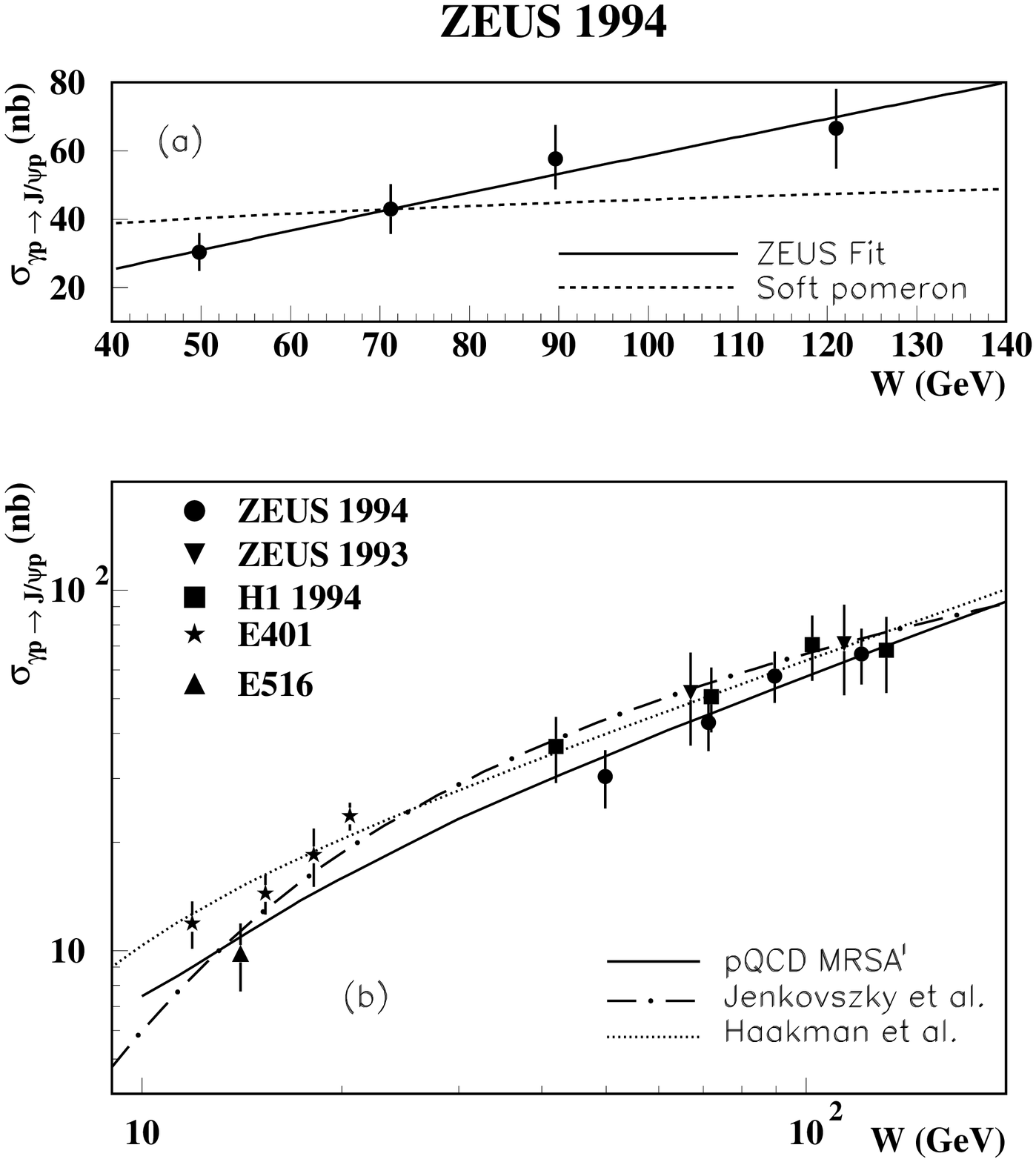} {The cross section
  for the reaction $\gamma p \rightarrow J/\psi p$ as a function of
  $W$.  The upper plot shows the latest ZEUS results, along with the
  results of a fit of the form $\sigma \propto W^{\delta}$, as
  described in the text.  The value of $\delta=0.92\pm0.14\pm0.10$ was
  obtained.  The dashed line shows the results expected for a
  dependence of the form $W^{0.22}$.  In the lower plot, the HERA data
  are compared to results from fixed target experiments. The results
  of a pQCD calculation~\cite{ref:Ryskin2} are also plotted, as well
  as the results from two other models
  ~\cite{ref:Jenkovszky,ref:Haakman}.}  {photo_Jpsi}

The cross section $\gamma p \rightarrow J/\psi p$ has a very steep $W$
dependence, as can be seen in Fig.~\ref{fig:photo_Jpsi}.  In the top
plot, the ZEUS data alone are fitted, giving a value of
$\delta=0.92\pm0.14\pm0.10$.  In the lower plot, the ZEUS and H1 data
are compared to results from fixed target experiments and curves from
different models, including the Ryskin model.  The H1 Collaboration
has recently~\cite{ref:H1_psi_eps} extended the measurements of
$J/\psi$ production to higher $W$ and find a value
$\delta=0.77\pm0.13$ from a fit to the H1 data in the range
$30<W<240$~GeV.  These results can be compared to pomeron model and
pQCD expectations:
\begin{itemize}
\item
Taking the value $b=4.7$~GeV$^{-2}$ and assuming that the values
$\alpha_0=1.08,\; \alpha'=0.25$ also apply to $J/\psi$
photoproduction, the prediction of $\delta=0.11$ is obtained from the
pomeron model~\cite{ref:DoLa_VM}.  This value is very far from the
data.  Elastic $J/\psi$ production clearly proceeds via a different
mechanism than that involved in soft interactions.
\item
The expectations from pQCD are for a steep $W$ dependence driven by
the gluon density.  For an effective scale around $2.5$~GeV$^2$, the
gluon density from the GRV(HO)~\cite{ref:GRV94} parameterization has
the form $xg(x) \propto x^{-0.18}$ at small $x$.  The $W$ dependence
is therefore expected to be approximately
\begin{equation}
\sigma \propto \left[xg(x,2.5 \rm{GeV}^2)\right]^2 
\propto x^{-0.36} \propto W^{0.72} \; .
\end{equation}
This is in much better agreement with the data than the pomeron model
expectation.  Elastic $J/\psi$ photoproduction therefore gives
evidence for significant contribution from perturbative QCD processes
in vector meson production.
\end{itemize}

The H1 collaboration~\cite{ref:H1_photo_psi_97} has also measured the
$W$ dependence for proton dissociation events.  The $W$ dependence has
been found to be even steeper, with $\delta=1.2\pm0.2$ being measured.
H1 argues that the steeper slope may be due to increasing phase space
for proton dissociation with increasing $W$.

\paragraph{Test of SCHC}

The analysis of the angular distributions of the leptons from the decay of
the $J/\psi$ have been performed by both the H1 and ZEUS collaborations.
All fit results are consistent with SCHC.

\paragraph{Photoproduction of $\psi(2S)$}

The H1 Collaboration has analyzed the reaction $\gamma p \rightarrow 
\psi(2S) X$ in the kinematic range $40<W<160$~GeV and $z>0.95$.  Recall
that $z$ gives the fraction of the photon energy transferred to the
$\psi(2S)$ in the proton rest frame.  This large value of $z$
indicates that the process is quasi-elastic.  The ratio of $\psi(2S)$
to $J/\psi$ production for this kinematic range is found to be $0.150
\pm 0.027 \pm 0.022$.  This measurement is in good agreement with
expectations from a pQCD calculation~\cite{ref:KZ,ref:KNNZ}.

Recently, both ZEUS~\cite{ref:ZEUS_upsilon} and
H1~\cite{ref:H1_upsilon} have reported the observation of $\Upsilon$
photoproduction (the $\Upsilon(1S),\Upsilon(2S),\Upsilon(3S)$ are not
resolved) decaying to $\mu^+\mu^-$.  The ZEUS signal consists of $17.1
\pm 7.5$ events. The $\gamma p$ cross section times branching ratio
into muons is $13.3\pm6.0^{+2.7}_{-2.3}$~pb at a mean photon-proton
center-of-mass energy of $120$~GeV.  The H1 result is quite similar,
and is based on $8.3\pm3.9$ events.  The $\gamma p$ cross section
times branching ratio into muons is $16.0\pm7.5\pm4.0$~pb.  Using CDF
data on cross section times branching ratio for the different
states~\cite{ref:CDF_upsilon} and the muonic branching ratio
$B(\Upsilon(1S)\rightarrow\mu^+\mu^-)=2.48\pm0.007$~\%~\cite{ref:EWpar}
yields the cross sections for $\sigma(\gamma p \rightarrow
\Upsilon(1S) p)$ shown in Fig.~\ref{fig:photo_upsilon}.  As is seen,
the measurements are considerably larger than the first pQCD
calculations for this process~\cite{ref:Strikfurt_upsilon}.

\epsfigure[width=0.8\hsize]{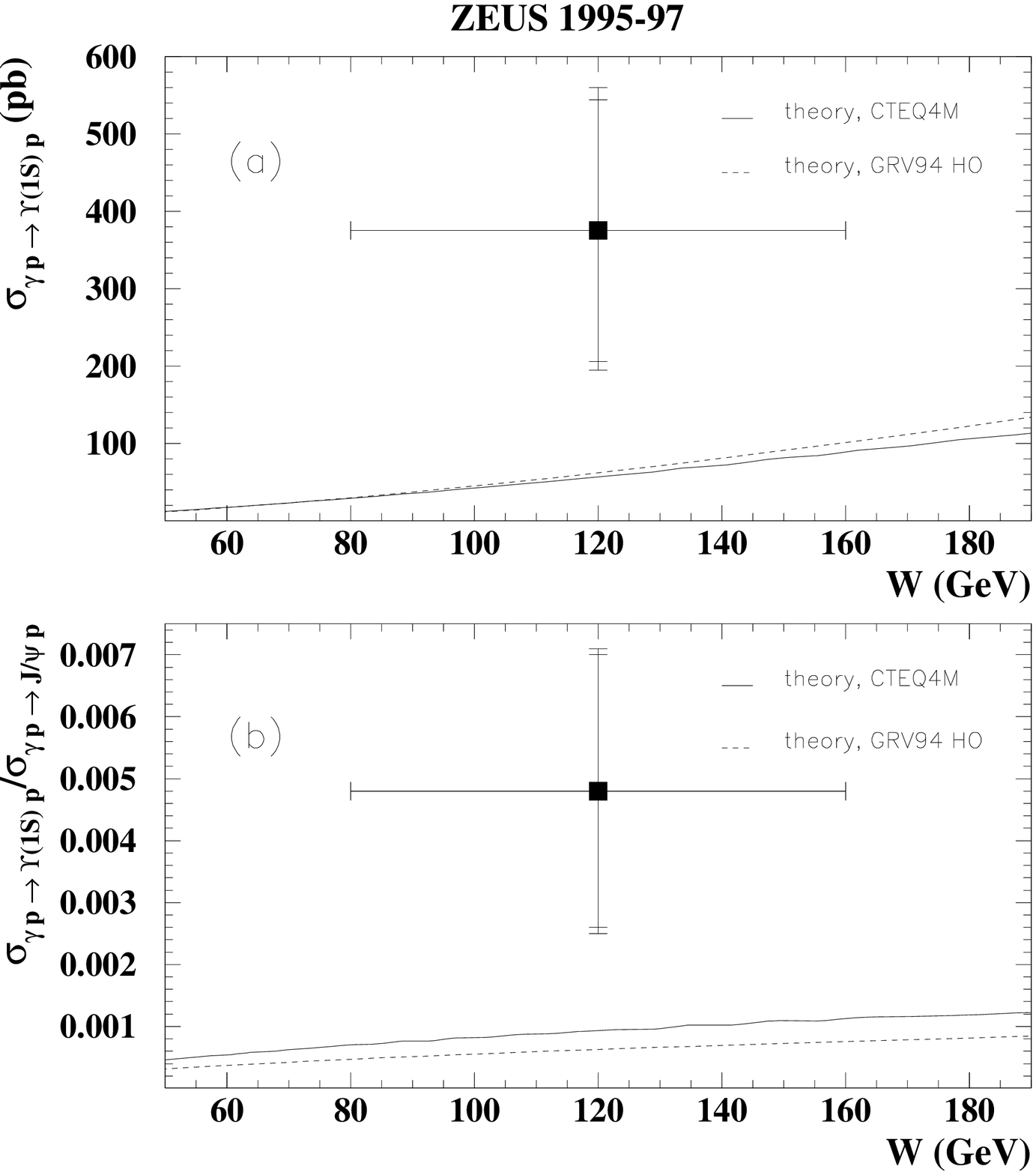} {Comparison of HERA
  data for $\sigma(\gamma p \rightarrow \Upsilon(1S) p)$ with the
  predictions of \citeasnoun{ref:Strikfurt_upsilon} using the
  GRV94(HO) and CTEQ4M parton density parameterizations.}
{photo_upsilon}

\subsubsection{Vector meson photoproduction at large $t$}

Results on vector meson photoproduction at large $t$ have been
presented for $\rho^0,\; \phi, \rm{and} \; J/\psi$ mesons.  The data
on $\rho^0$ and $\phi$ production~\cite{ref:ZEUS_EPS_larget} are at $W
\approx 100$~GeV and for $|t|$ values up to $13$~GeV$^2$.  For
$|t|>0.5$~GeV$^2$, it is found that the proton dissociative reaction
dominates the cross section.  The vector mesons are found to be
produced predominantly in the helicity $\pm1$, state, indicating that
the pQCD regime has not been reached.

\epsfigure[width=0.8\hsize]{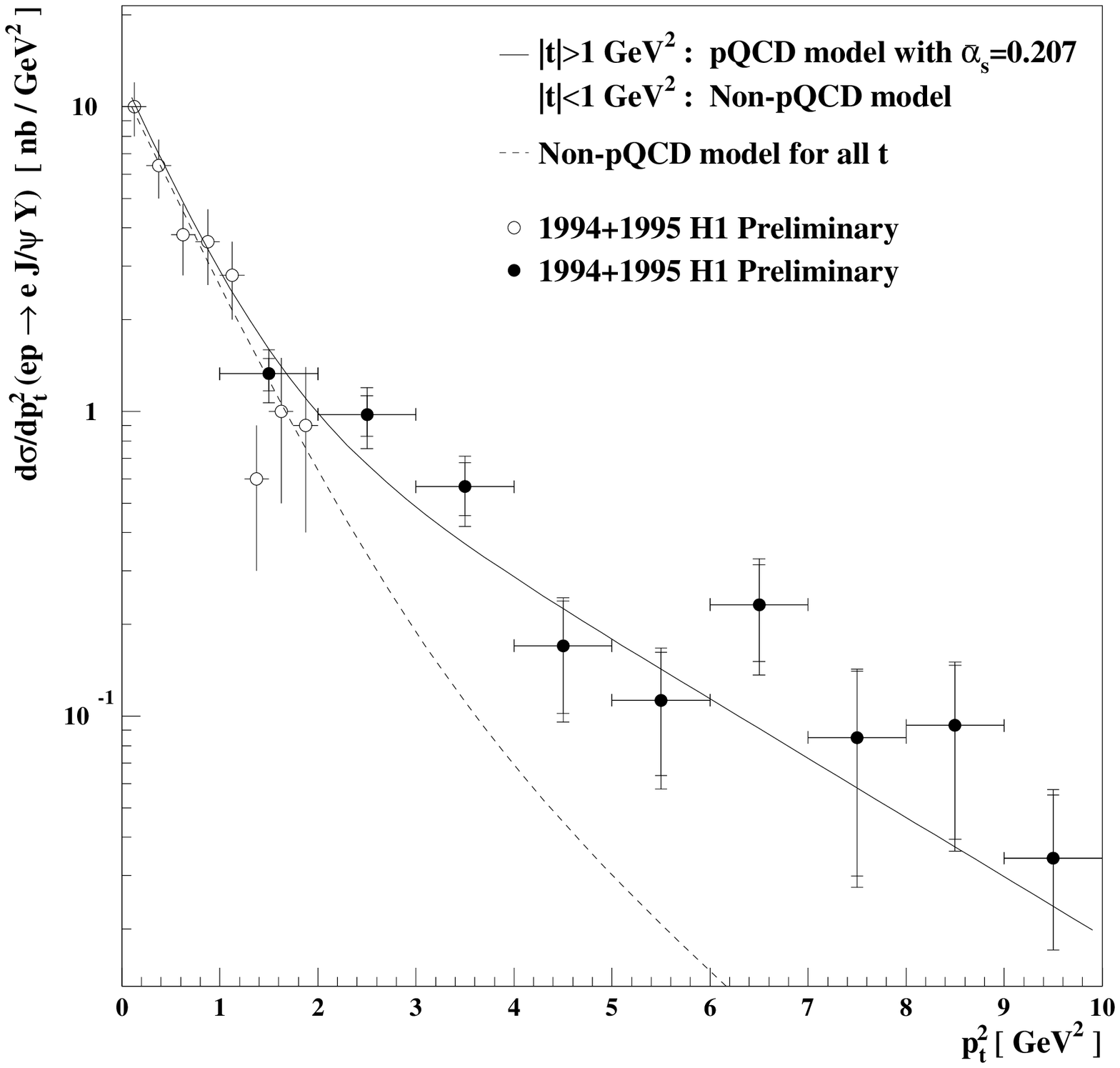} {Distribution of the
  differential cross section $d\sigma_{ep}/dp_T^2$ for $J/\psi$
  production with proton dissociation.  The solid line represents the
  combination of the pQCD calculations for $|t|>1$~GeV$^2$ and a
  non-perturbative model when $|t|<1$~GeV$^2$.  The dashed line
  represents the prediction from the non-perturbative model when the
  full $t$ range is used.}  {pt_psi}

The cross section for $J/\psi$ photoproduction with proton
dissociation has been measured~\cite{ref:H1_EPS_larget} in the
kinematic range $|t|>1$~GeV$^2$ and $30<W<150$~GeV.  qThe $p_T^2$
dependence is shown in Fig.~\ref{fig:pt_psi} and compared to the pQCD
expectations~\cite{ref:Bartels}.  The model is seen to give a good
representation of the $p_T$ spectrum.  It is also able to reproduce
the $W$ dependence of the data.

\subsubsection{Vector meson production in DIS}

The exclusive production of $\rho^0$, $\rho'$, $\phi$, and $J/\psi$
vector mesons has been measured in DIS at HERA.  The data sets are
summarized in Table~\ref{tab:VM_summary}.

\begin{table}
\tablecaption{ Summary of published results on cross sections and
$t$-slopes from the exclusive electroproduction of 
vector mesons at HERA.}
\label{tab:VM_summary}
\begin{center}  
\begin{tabular}{  c | c | c | c | c }
Reaction & $Q^2$ & $<W>$ & $t$-range & luminosity \\ 
studied &(GeV$^2$) & (GeV) & (GeV$^2$) & pb$^{-1}$ \\
\hline
$\rho^0 \rightarrow \pi^+~\pi^-$  & 
$7-25$ & $40-130$ & $< 0.60$ & $0.55$ \\
\cite{ref:ZEUS_DIS_rho} & & & & \\
&&&&\\

$\rho^0 \rightarrow \pi^+~\pi^-$  & 
$8-50$ & $40-140$ & $< 0.60$ & $3.1$ \\
\cite{ref:H1_DIS_rho} & & & & \\
&&&&\\

$\rho^0 \rightarrow \pi^+~\pi^-$ & 
$0.25-50$ & $20-167$ & $<0.6$ & $6.0$ \\ 
\cite{ref:ZEUS_VM95} & & &  &  \\
&&&&\\

$\rho^0 \rightarrow \pi^+~\pi^-$  & 
$7-35$ & $60-180$ & $<1.2$ & $2.8$ \\
\cite{ref:H1_DIS_phi} & & & &  \\
&&&&\\
\hline
&&&&\\
$\phi \rightarrow K^+~K^-$  & 
$7-25$ & $42-134$ & $< 0.60$ & $2.6$ \\
\cite{ref:ZEUS_DIS_phi} & &  &  &\\
&&& &\\

$\phi \rightarrow K^+~K^-$  & 
$6-20$ & $42-134$ & $<0.6$ & $2.8$ \\
\cite{ref:H1_DIS_phi} & &  & & \\
&&&&\\
\hline
&&&&\\
$J/\psi \rightarrow l^+~l^-$  & 
$8-40$ & $30-150$ & $<1.0$ & $3.1$  \\
\cite{ref:H1_DIS_rho} & & & &  \\
&&&&\\

$J/\psi \rightarrow l^+~l^-$  & 
$2-40$ & $50-150$ & $< 1$ & $6.0$ \\
\cite{ref:ZEUS_VM95} & & & &   
\end{tabular}  
\end{center}
\end{table}

\paragraph{Mass Spectra}

As with photoproduction, the mass distributions show very clear peaks,
and there is little background to the vector meson signal.  

\epsfigure[width=0.8\hsize]{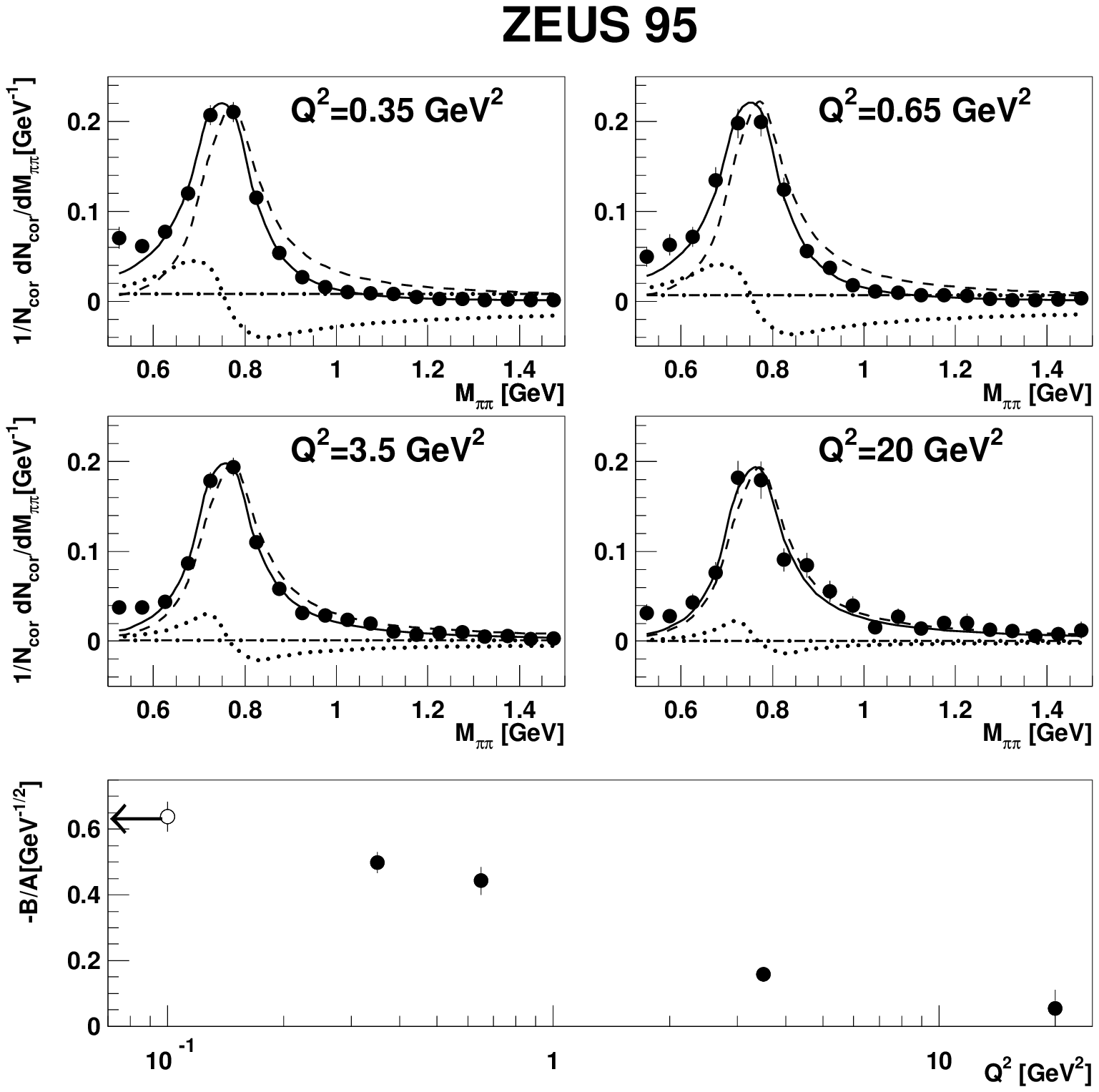} {The differential
  $dN/dM_{\pi\pi}$ distributions for the exclusive reaction $ep
  \rightarrow e \pi^+ \pi^- p$ normalized to unit area for different
  $Q^2$.  The line types show the different contributions from a
  S\"oding type fit.  The lower plot shows the ratio, $B/A$, of the
  background to the Breit-Wigner normalization as a function of
  $Q^2$.}  {DIS_rho_mass}

The shape of the $\rho^0$ mass distribution has been studied as a
function of $Q^2$.  The skewing of the mass spectrum is found to
decrease as the scale increases.  Figure~\ref{fig:DIS_rho_mass} shows
the mass spectrum as measured by ZEUS~\cite{ref:ZEUS_VM95} for
different values of $Q^2$, and the results of a fit with a S\"oding
form.  The ratio of the non-resonant background term to the
Breit-Wigner normalization, $B/A$, is seen to decrease to zero as $Q^2
\rightarrow 20$~GeV$^2$.  Recent H1 results give a similar
picture~\cite{ref:H1_VM96}.

\paragraph{$Q^2$ dependence of cross sections}

\epsfigure[width=0.8\hsize]{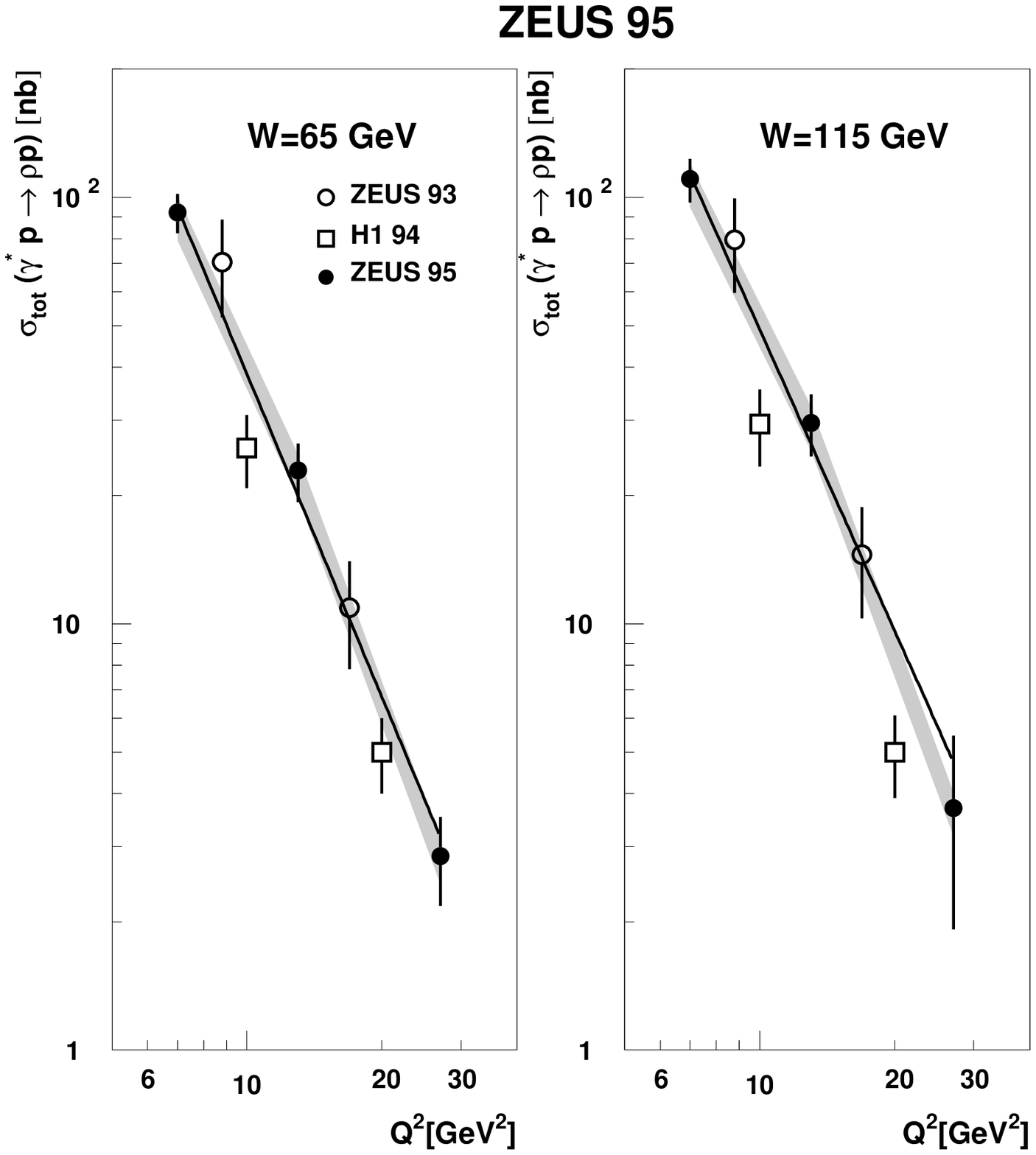} {The dependence of
  the $\gamma^*p \rightarrow \rho^0p$ cross section on $Q^2$.}
{rho_vs_Q2}

Published results exist starting from $Q^2= 0.25$~GeV$^2$ and
extending to $Q^2= 27$~GeV$^2$.  The cross sections are clearly much
smaller than those from photoproduction.  This is due to the steep
$Q^2$ dependence of the cross section, shown graphically in
Fig.~\ref{fig:rho_vs_Q2} for $\rho^0$ production.  Fits of the form $d
\sigma/dQ^2 \propto Q^{-2n}$ to the higher $Q^2$ data yield values of
$n = 2.1-2.5$, while the lower $Q^2$ data is well fit by a form of the
type $1/(Q^2+M_{\rho}^2)^2$.  This is more-or-less in line with
expectations from both soft exchange and perturbative two-gluon
exchange.

\paragraph{$W$ dependence of cross sections}

\epsfigure[width=0.8\hsize]{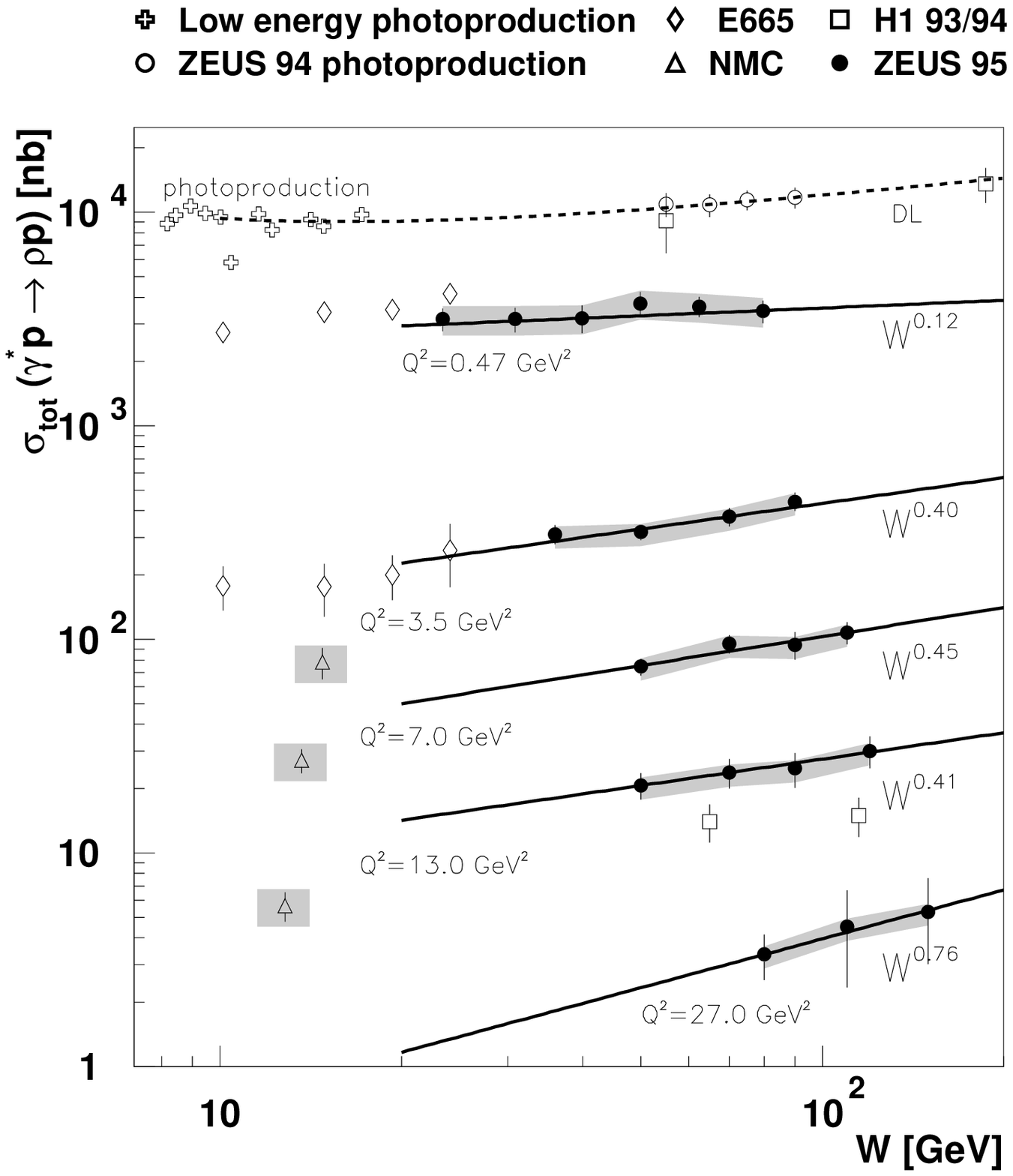} {The cross section
  for $\gamma^*p \rightarrow \rho^0 p$ as a function of $W$ for
  different $Q^2$ values.  The solid lines show the $W$ dependence
  from fits to ZEUS data. The dashed line is the prediction of
  \citeasnoun{ref:DoLa_sigfit} for photoproduction.  The
  NMC~\cite{ref:NMC_VM2}, E665~\cite{ref:E665_VM} and
  H1~\cite{ref:H1_DIS_rho} data were interpolated to the indicated
  $Q^2$ values.}  {DISrho_vs_W}

\epsfigure[width=0.8\hsize]{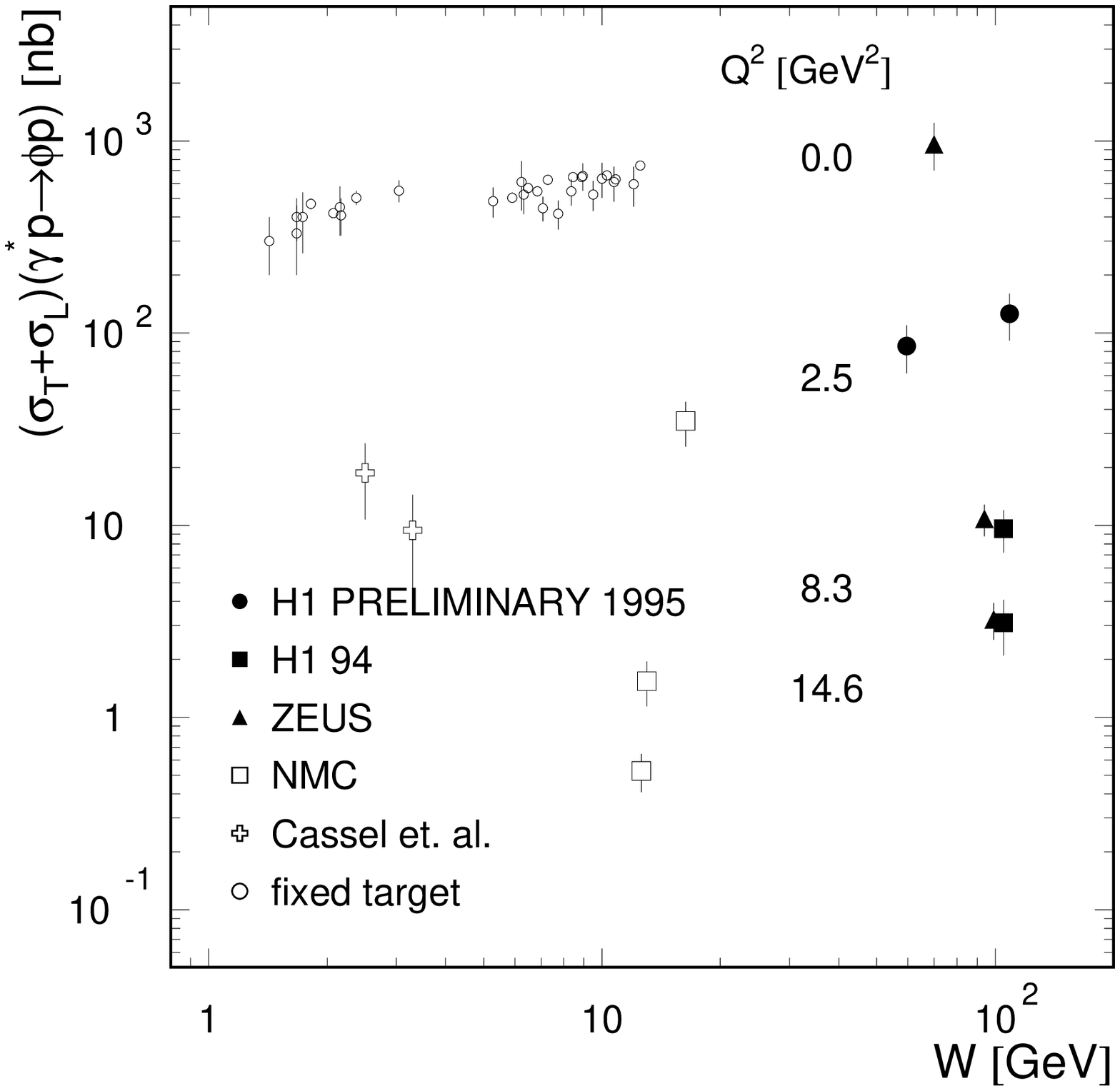} {The cross section for $\gamma^*p
  \rightarrow \phi p$ as a function of $W$ for photoproduction and
  DIS.}  {DISphi_vs_W}

\epsfigure[width=0.8\hsize]{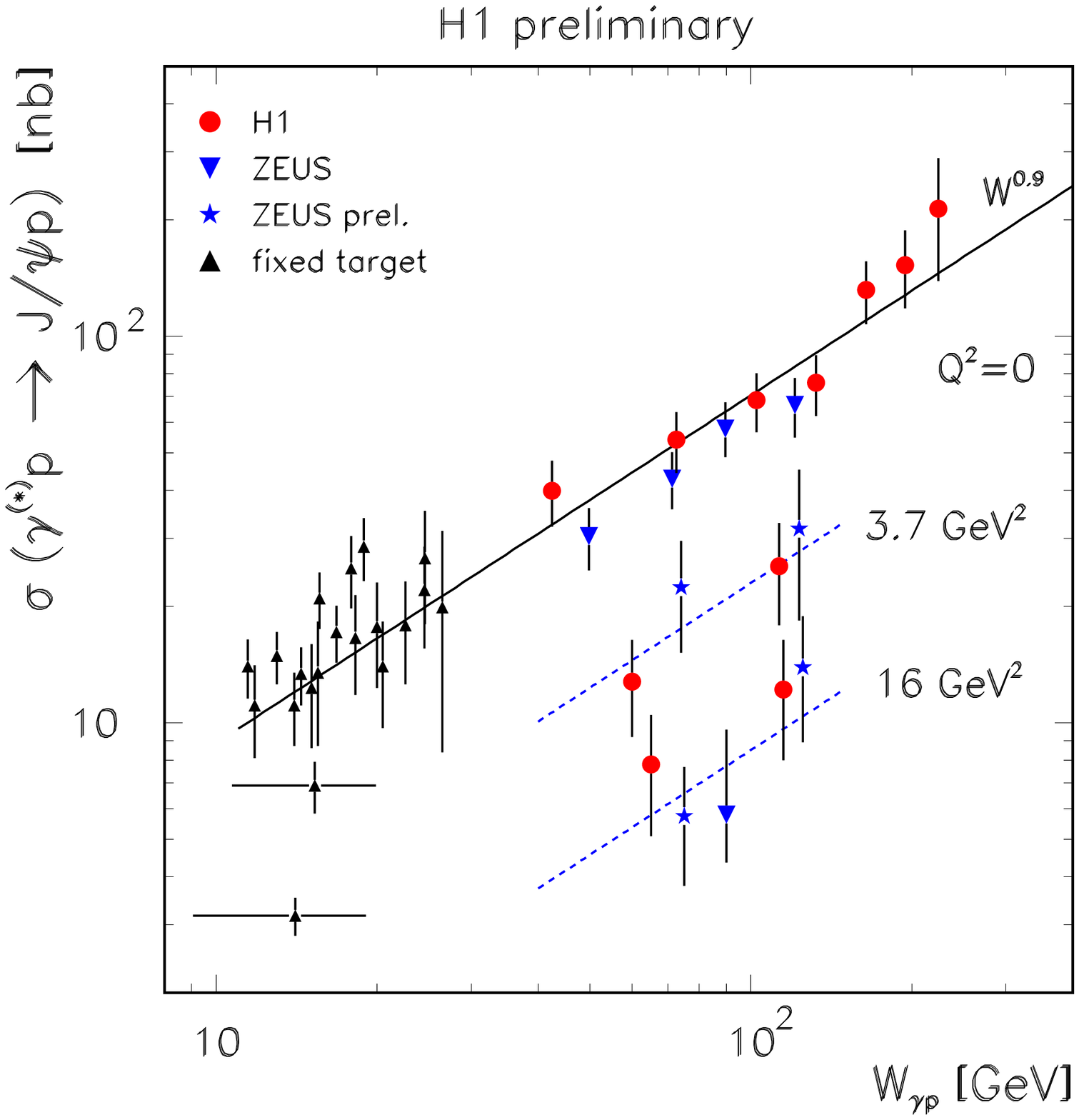} {The cross section for $\gamma^*p
  \rightarrow J/\psi p$ as a function of $W$ for photoproduction and
  DIS.  The lines indicate a $W^{0.9}$ dependence.}  {DISpsi_vs_W}

The cross section at fixed $Q^2$ shows an increase with $W$ when
compared to fixed target experiments.  This is shown in
Fig.~\ref{fig:DISrho_vs_W} for the $\rho^0$ meson, in
Fig.~\ref{fig:DISphi_vs_W} for the $\phi$ meson, and in
Fig.~\ref{fig:DISpsi_vs_W} for the $J/\psi$.  The ZEUS collaboration
has measured the $W$ dependence within its own data
~\cite{ref:ZEUS_BPC_rho} for $\rho^0$ electroproduction, and finds
\begin{equation}
\begin{array}{llrl}
Q^2 = &0.47~\rm{GeV}^2 & \delta = &0.12 \pm 0.03\pm 0.08 \; ,\\
      &3.5  &          &0.40 \pm 0.12\pm 0.12 \; , \\
      &7.0  &          &0.45 \pm 0.15\pm 0.07 \; , \\
      &13.0  &          &0.41 \pm 0.19\pm 0.10 \; ,\\
      &27.0  &          &0.76 \pm 0.55\pm 0.60 \; .
\end{array}
\end{equation}
These data are suggestive of an increasing steepness with $W$, but
clearly more precision is needed before any conclusion can be drawn.
Recent preliminary results for $\rho^0$ production from
H1~\cite{ref:H1_VM96} and $\phi$ production from
ZEUS~\cite{ref:ZEUS_VM97} show more conclusively the increasing
steepness of the $W$ dependence with $Q^2$.

The $W$ dependence cannot be measured in the HERA data alone for
$J/\psi$ production because of limited statistics.  Comparison with
the fixed target data indicates that the $\phi$ has a steeper $W$
dependence than the $\rho^0$.  The same statement applies for the DIS
production of $J/\psi$ mesons, where the slope is similar, within
errors, to that measured in photoproduction of $J/\psi$.

\paragraph{$t$ slopes versus $Q^2$ and $W$}

\epsfigure[width=0.8\hsize]{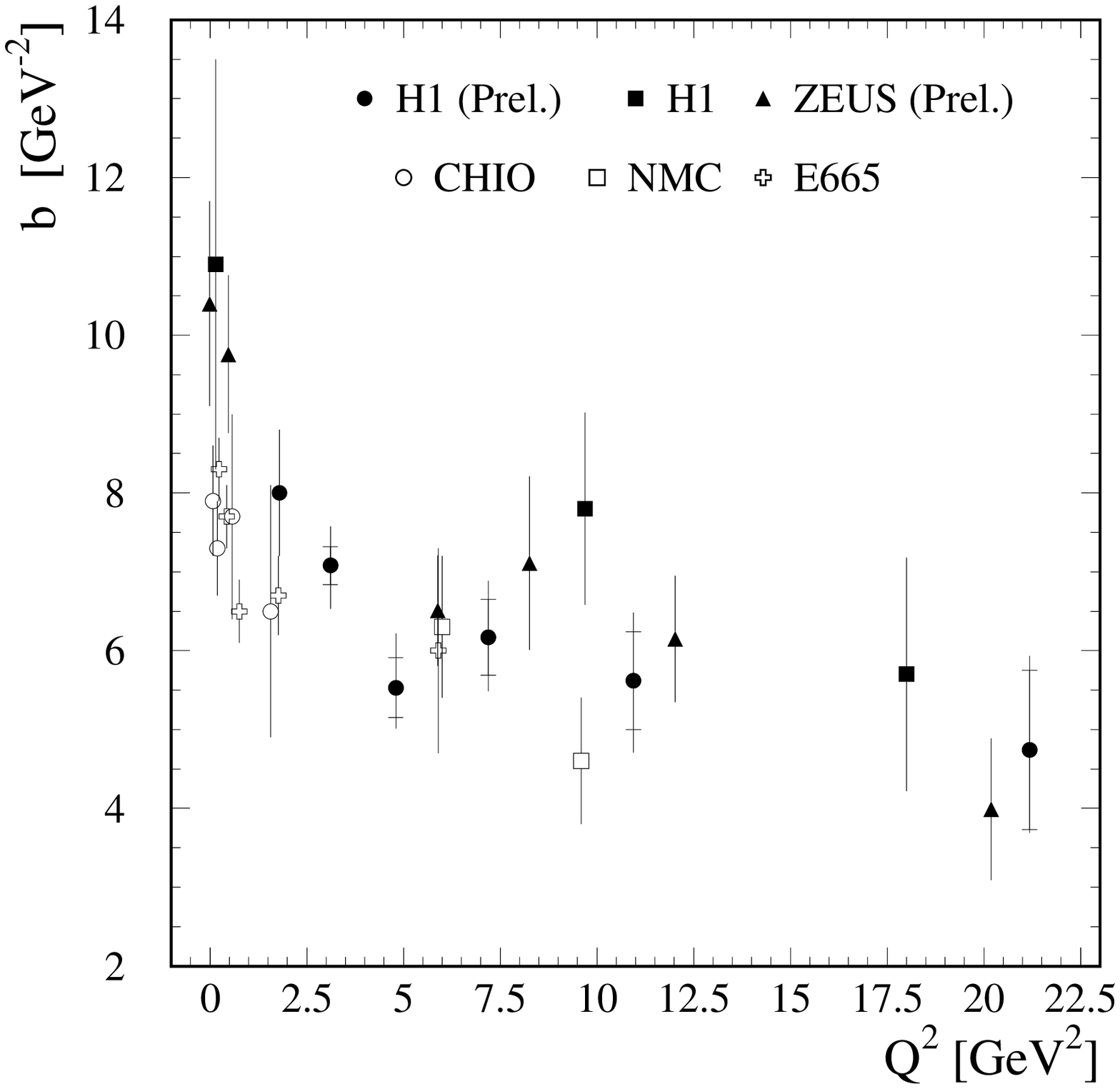} {The slope $b$ 
of the exponential
  $t$ dependence $\exp(-b|t|)$ for $\rho^0$ production as a function
  of the photon virtuality $Q^2$.}  {b_vs_Q2}

The measurements of the $t$ dependence are summarized in
Table~\ref{tab:VM_summary}.  The values of $b$ are smaller than those
measured in photoproduction, and show a trend to smaller values as
$Q^2$ increases.  This is expected since the contribution to $b$ from
the size of the photon should decrease with increasing $Q^2$.  The
$Q^2$ dependence of $b$ for $\rho^0$ production is shown in
Fig.~\ref{fig:b_vs_Q2}.  The data are currently not accurate enough to
determine whether there is any shrinkage of the forward peak with
energy, as expected in Regge Theory.

\paragraph{Tests of SCHC and $R$}

The spin density matrix elements have been determined for $\rho^0$
production under the hypothesis of SCHC.
They indicate that the fraction of $\rho^0$ mesons produced in
the helicity 0 state increases with $Q^2$.  If in addition to the SCHC
hypothesis, natural parity exchange in the $t$-channel is assumed, then
the $r^1_{1-1}$ matrix element can be expressed as
\begin{equation}
r^1_{1-1} = 0.5(1-r^{04}_{00}) \; .
\end{equation}
The values obtained are consistent with this relationship, indicating
that the assumptions of SCHC and natural parity exchange are
valid\footnote{ Recent preliminary results from H1~\cite{ref:H1_VM96}
and ZEUS~\cite{ref:ZEUS_VM97_helicity} show that SCHC does not hold
for electroproduction.  The extraction of $R$ is only slightly
affected by the small deviations observed from SCHC.}.  The transition
from helicity $\pm 1$ vector mesons to dominantly helicity $0$ vector
mesons as $Q^2$ increases is clearly seen in the data.  Assuming SCHC,
this translates into a strong dependence of $R$ on $Q^2$.  This
dependence is shown in Fig.~\ref{fig:R_rho} and compared to different
model predictions.

\epsfigure[width=0.8\hsize]{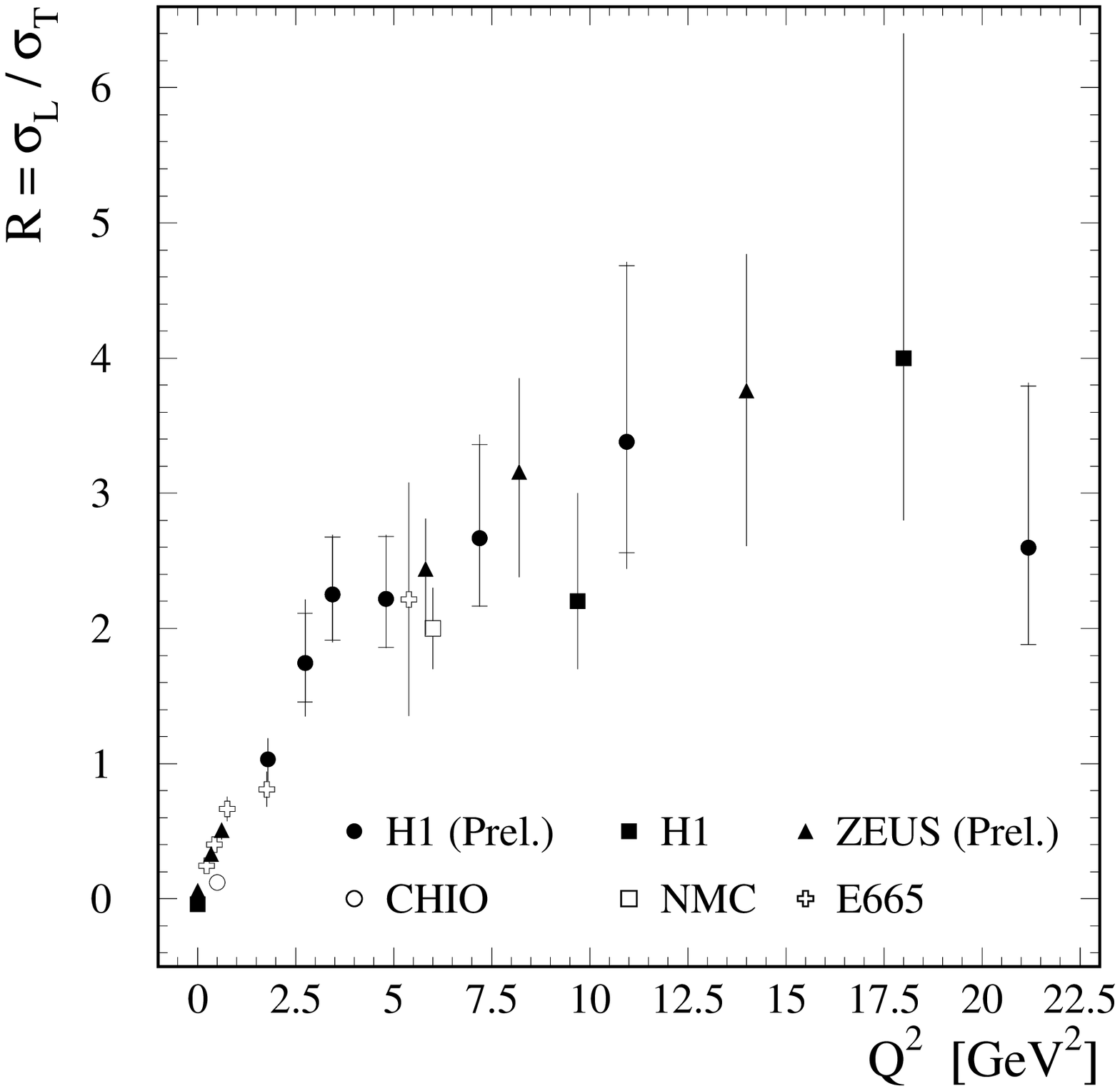} {The ratio
  $R=\sigma_L/\sigma_T$ for the reaction $\gamma^*p \rightarrow \rho^0
  p$ as a function of $Q^2$} {R_rho}

\paragraph{Results with proton dissociation}

The proton dissociation reaction $ep \rightarrow e\rho^0 N$, where $N$
is a small mass baryonic system, has also been measured at
HERA~\cite{ref:H1_DIS_phi,ref:ZEUS_rho_Warsaw}.  The $W$, $Q^2$ and
helicity angle distributions are found to be similar in shape to those
measured for the elastic process.  However, the $t$ slope is
considerably shallower, with measured $b\approx 2$~GeV$^{-2}$.  These
measurements are interesting in their own right, e.g., to test the
factorization of the cross section, but are also important for the
measurement of the elastic cross section since a fraction of the
proton dissociative cross section is indistinguishable from the
elastic events in the H1 and ZEUS detectors.

\subsubsection{Vector meson production ratios}

\epsfigure[width=0.8\hsize]{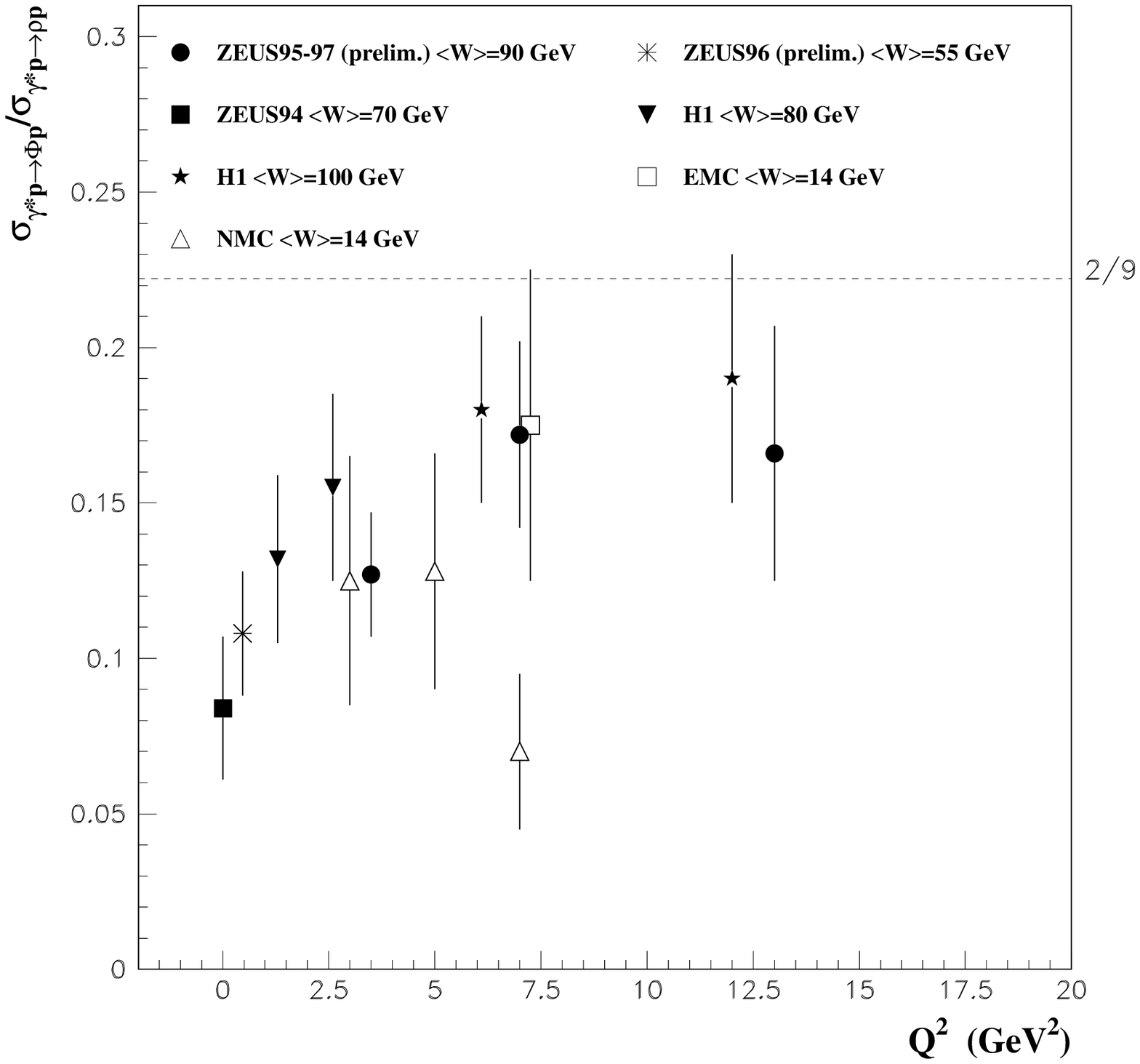} {Production rate of
  $\phi$ mesons relative to $\rho^0$ as a function of $Q^2$.  The
  dashed line corresponds to the ratio $2/9$ expected from the quark
  charges and a flavor symmetric coupling mechanism.}  {VM_ratio}

The measured cross section ratio for $\phi$ relative to $\rho^0$ is
shown as a function of $Q^2$ in Fig.~\ref{fig:VM_ratio}.  There is a
clear tendency for this ratio to increase with $Q^2$ to the values
expected from a flavor independent production mechanism.  This trend
has also been observed for $J/\psi$ production.

\subsubsection{Comparison with pQCD Models}

\epsfigure[width=0.8\hsize]{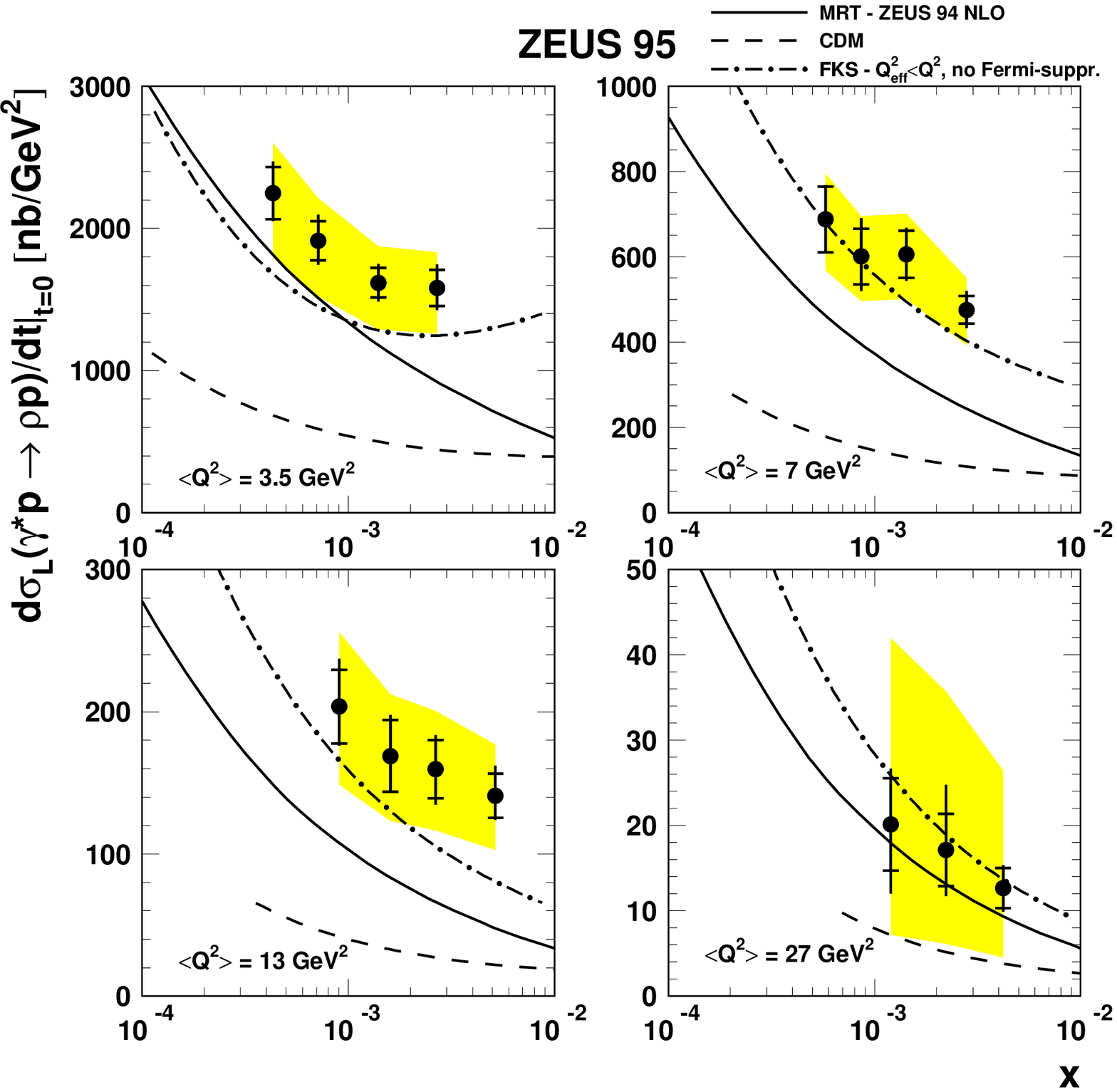} {Comparison of the
  forward longitudinal cross section, $d\sigma_L/d|t||_{t=0}$, to
  different pQCD models (see text).}  {pQCD_compare}

The comparison to pQCD calculations is performed for the forward
longitudinal cross section, $d\sigma_L/d|t||_{t=0}$.  The experimental
values for $R$ and $b$ must be used to convert the measured
$\gamma^*p$ cross sections to $d\sigma_L/d|t||_{t=0}$.  The resulting
cross sections tend to have large uncertainties, but nevertheless are
able to distinguish between different pQCD models.
Figure~\ref{fig:pQCD_compare} compares the calculations of the models
of 
\citeasnoun{ref:Koepf}, 
\citeasnoun{ref:Martin_Ryskin_Teubner} and 
Nemchik, Nikolaev, Predazzi and Zakharov
\cite{ref:Nemchik1,ref:Nemchik2}  to the
ZEUS data.  The first two models make use of the gluon density
extracted by ZEUS from NLO fits to $F_2$, and both reproduce the data
reasonably well given the remaining large normalization uncertainties.
The model of Nemchik et al. underestimates the cross section over the
entire kinematic range.

\subsection{Discussion}

Light vector meson photoproduction shows all the features of soft
diffractive scattering: the $W$ dependence is soft, the $t$ dependence
is steep, SCHC applies, and the $\rho^0$ mass spectrum is skewed.  The
situation is drastically different for $J/\psi$ production.  Here, the
$W$ dependence is steep, and the $t$ slope is a factor of two smaller.

As $Q^2$ is increased, the light vector mesons become progressively
more polarized in the helicity zero state, and the $t$ dependence of
the cross section becomes less steep.  The situation for the $W$
dependence is not so clear.  The data are not precise enough to draw
conclusions based on single data sets, and combining data sets is
difficult.  However, the data indicate a steeper dependence as $Q^2$
increases, and the steepness is more pronounced for the heavier vector
mesons.

First results exist on vector meson production at large $|t|$
which, in the case of the $J/\psi$, are compatible with pQCD expectations.

In which kinematic regime is the process dominated by perturbative two
gluon exchange ? The results show that $J/\psi$ production has a steep
$W$ dependence, in accord with pQCD expectations. Assuming that the
scale is set by $\mu^2=(Q^2+m_{J/\psi}^2+|t|)/4$, as prescribed in the
Ryskin model~\cite{ref:Ryskin1}, this would imply that light vector
meson production should also be perturbative at $Q^2 \sim m_{J/\psi}^2
= 10$~GeV$^2$.  The data for $\rho^0$ electroproduction prefer a
shallower $W$ dependence at this scale than is observed in $J/\psi$
photoproduction.  In addition, there is still a significant
contribution from $\sigma_T$ ($R=\sigma_L/\sigma_T\approx 3$) to the
$\rho^0$ cross section at $Q^2=10$~GeV$^2$.  This implies that this
value of the scale is not high enough for light vector mesons
production to be fully perturbative in nature, leading to the
conclusion that $M_V^2 \; {\rm and} \; Q^2$ are not symmetric in the
determination of the scale of the reaction.

Vector meson production has proven to be a very interesting process in
which to test perturbative QCD.  The different scales which enter into
the process, $M_V, \; Q^2 \; t$, and their interplay, provide a
challenge to our understanding of this basic physical process.  In
pQCD, the cross sections depend on the square of the gluon density,
making this process a very interesting one to study the parton
densities in the proton.  Clearly, further theoretical as well as
experimental work is needed.  On the theoretical side, higher order
calculations are needed to reduce the scale uncertainties and to allow
more quantitative predictions.  On the experimental side, further
extensions in the different scales $Q^2, \; |t|, \; M$ are needed to
understand the transition from soft to hard reactions.  This topic
will continue to be of interest as it provides a testing ground for
pQCD in exclusive reactions.


\section{Searches for new particles and new interactions}
\label{sec:newphys}
In the previous sections, we have reported on physics within the realm
of the Standard Model (SM) of electroweak+strong interactions.  The SM
has been exceptionally successful, and there are currently no known
violations of its predictions.  It is however expected that the SM
will not be the final theory, but rather a low energy manifestation of
a more general theory, perhaps encompassing gravity.  There are many
open issues in the Standard Model.
\begin{itemize}
\item Why are there three families of quarks and leptons ?  In the past, 
a large number of ``fundamental particles'' eventually led to a new
level substructure (atoms are composed of nuclei and electrons, nuclei
are composed of neutrons and protons, hadrons are composed of quarks).
It is therefore natural to speculate that new levels of substructure
exist within quarks and leptons.

\item The mass spectrum of known particles is not understood.  In the Standard 
Model, particles acquire their mass via the mechanism of spontaneous
 symmetry breaking, leading to the presence of a scalar particle, the
 Higgs particle.  This particle has not yet been observed.  Given the
 observation of the Higgs, the question of the immense gap in mass
 between the weak bosons, which set the scale of the weak
 interactions, and the Planck mass, which sets the scale of
 gravitational interactions, will still need clarification. A theory
 which explained the observed mass spectrum would be a great step
 forward in our understanding of nature.

\item The SM interactions violate chiral symmetry.  This is not currently 
understood, and theories postulating left-right symmetry at higher
 energies are certainly tempting to consider.

\item Gravity does not appear in the SM. In this context, supersymmetric 
theories are an exciting possibility since they naturally incorporate
  gravity.  Theories attempting to unify gravity with the strong and
  electroweak forces have to date not had any experimental
  verification.

\end{itemize}

The discovery of new interactions or particles beyond the SM would
certainly be welcome as they would lead to new theories and a new
conception of nature.  There are currently two results from HERA which
deviate from Standard Model expectations.  These are the observation
of excess events at large $x \; {\rm and} \; Q^2$, seen by both the
H1~\cite{ref:H1_highQ2} and ZEUS~\cite{ref:ZEUS_highQ2}
collaborations, and the anomalous rate of events with high $p_T$ muons
and large hadronic $p_T$ seen by the H1
collaboration~\cite{ref:H1_muon}.  These will be reviewed in some
detail below.  We also summarize searches which have been carried out
for new physics.

The HERA center-of-mass energy, $\sqrt{s}=300$~GeV, is intermediate
between those available at LEP (up to about $185$~GeV) and the
Tevatron ($1.8$~TeV).  The ``cleanliness'' of the final state is also
intermediate between the two since one of the colliding particles is
point-like.  An advantage of HERA over the Tevatron is that the SM DIS
cross sections at high $E_T$ can be calculated precisely.  Deviations
from expectations are therefore more easily spotted.

HERA is of course ideally suited to new states which couple directly
to leptons and quarks, such as leptoquarks or some supersymmetric
quarks.  Searches for excited fermions are also particularly clean,
and can be performed up to the kinematic limit.

In the following sections, we review:

\begin{itemize}

\item 

Direct searches for the production of new particles, either as
$s$-channel resonances or produced in the $t$-channel.  In the direct
$s$-channel formation of new particles, the incoming $e^{\pm}$ fuses
with a quark or gluon in the proton to form a resonance.  Examples of
these types of particles are leptoquarks, leptogluons and quarks in
R-parity violating supersymmetry.  These particles will appear as
resonances for $M < \sqrt{s}$.  Examples of $t$-channel processes are
the single production of excited fermions, heavy neutral leptons or
selectron+squark production in R-parity conserving supersymmetry.

\item 

Searches for new interactions at mass scales $\Lambda
\stackrel{>}{\sim} 1$~TeV.  These can be studied in the context of
4-fermion contact interactions.  In general, the presence of new
particles, or compositeness of any kind, are signals for new
interactions which in the low energy limit can be described as contact
interactions.  However, explicit searches for specific types of
particles or interactions are more sensitive than the general search
for contact interactions, such that the two types of searches are
complementary.

\item
Searches for interactions which violate lepton flavor.  Any reaction
resulting in flavor violation will be a clear violation of the SM.
These searches can be performed in a general way, and can be
interpreted as limits on different types of particles appearing in
specific models.

\item The observation, by the H1 collaboration, of events containing a
  high $p_T$ muon and a hadronic system with large $p_T$.  The first
  such event was observed by the H1 collaboration in
  1994~\cite{ref:H1_muon1,ref:H1_Rpvio}, and the analysis has since
  been updated.
\item The recent observation of events at large $x,Q^2$ observed by
  the H1 and ZEUS collaborations.  First results were presented in
  February, 1997, and have since been updated.

\end{itemize}

\subsection{Search for leptoquarks and leptogluons}
\label{sec:leptoquark}
Leptoquarks and leptogluons generally appear in theories which link
the leptons and quarks, such as grand unified theories, technicolor
models\,\cite{ref:bschrempp}, or models for compositeness.
Leptoquarks (LQ) are hypothetical color triplet bosons (spin $0$ or
$1$), with fractional electric charge, and non-zero lepton and baryon
numbers.  They couple to lepton-quark pairs, and therefore would
appear as s-channel resonances at HERA for $M_{LQ}<\sqrt{s}$. In this
case, $M_{LQ}$ is directly related to the fraction of the proton
momentum carried by the quark, $x$, as
\begin{equation}
M_{LQ}^2 = sx \; .
\end{equation} 
  The effects of higher mass leptoquarks can also be observed via a
deviation of the observed cross section from SM expectations.

Leptogluons are hypothetical color octet fermions (spin $1/2$) with
integer electric charge and non-zero lepton number.  They also often
appear in theoretical extensions of the SM.  Leptogluons would also
appear as a resonance in the $x$ distribution if their mass is less
than $\sqrt{s}$.

In general, a LQ can couple to multiple lepton and/or quark flavors,
thereby providing a mechanism for flavor violation.  The different
types of leptoquarks with $SU(3)\times SU(2) \times U(1)$ invariant
couplings have been worked out by \citeasnoun{ref:Buchmuller}.  They
are distinguished by the quark flavors to which they couple, the
lepton flavor to which they couple, their spin, and their chiral
properties.  Very strong limits exist on chirality violating
leptoquarks ($\lambda_L \cdot \lambda_R>0$, where $\lambda_L,\;
\lambda_R$ are the left and right-handed couplings of the LQ), so that
we will only consider chirality conserving LQ.  Given this constraint,
we can distinguish 14 species of LQ, grouped into two sets of 7
species by the fermion number $F = L + 3B$, where $L$ and $B$ denote
the lepton and baryon number, respectively.  These are listed in
Table~\ref{tab:lq}.  Note that LQ cross sections will be higher for
$F=0$ in $e^+p$ scattering since in this case a quark fuses with the
positron.  In $e^-p$ scattering, the cross sections are larger for
$F=2$.

\begin{table}
\tablecaption{Table showing the different species of leptoquarks and their 
couplings for those leptoquarks which can be produced at HERA.
The leptoquarks in the upper block are color anti-triplets and have fermion
number $F=L+3B=0$.  The leptoquarks in the lower block are color triplets and 
have $F=2$.  The LQ species are further divided according to their spin 
($S$ for scalar and $V$ for vector), their chirality ($L$ or $R$) and their
weak isospin ($0,1/2,1$).  The leptoquarks ${\tilde S}$ and ${\tilde V}$ differ
by two units of hypercharge from $S$ and $V$, respectively.
In addition, the electric charge $q$ of the leptoquarks, 
the production channel, as well as their allowed decay channels assuming,
assuming lepton flavor conservation are displayed.
The quantum numbers and decay channels are assuming an electron
type LQ.  For positrons, the sign of the electric charge is reversed, and
anti-quarks should be replaced by the corresponding quark. The nomenclature
follows the Aachen convention~\protect\cite{ref:bschrempp}.}
\label{tab:lq}
\begin{center}   \begin{tabular}{c|cccc}
LQ species & $q$ & Production & Decay & Branching ratio \\
\hline
$S_{1/2}^L$ & -5/3 & $e_L\bar{u}_R$ & $  e\bar{u}$ & 1 \\
$S_{1/2}^R$ & -5/3 & $e_R\bar{u}_L $ & $ e \bar{u}$ & 1 \\
          & -2/3 & $e_R\bar{d}_R $ & $ e\bar{d}$ & 1 \\
${\tilde S}_{1/2}^L$ & -2/3 & $e_L\bar{d}_L $ & $ e\bar{d}$ & 1 \\
$V_0^L$ & -2/3 & $e_L\bar{d}_R $ & $ e\bar{d}$ & 1/2 \\
      &      & $              $ & $ \nu_e\bar{u}$ & 1/2 \\
$V_0^R$ & -2/3 & $e_R\bar{d}_L $ & $ e\bar{d}$ & 1 \\
${\tilde V}_0^R$ & -5/3 & $e_R\bar{u}_L $ & $ e\bar{u}$ & 1 \\
$V_1^L$ & -5/3 & $e_L\bar{u}_R $ & $ e\bar{u}$ & 1 \\
      & -2/3 & $e_L\bar{d}_R $ & $ e\bar{d}$& 1/2 \\
      &      & $             $ & $ \nu_e\bar{u}$ & 1/2 \\
\hline
$S_{0}^L$ & -1/3 & $e_L{u}_L $ & $ e{u}$ & 1/2 \\
        &        &             & $ \nu_e{d}$ & 1/2 \\
$S_{0}^R$ & -1/3 & $e_R{u}_R $ & $ e{u}$ & 1 \\
${\tilde S}_{0}^R$ & -4/3 & $e_R{d}_R $ & $ e{d}$ & 1 \\
$S_1^L$  & -1/3 & $e_L{u}_L $ & $ e{u}$ & 1/2 \\
       &      & $ $ & $ \nu_e{d}$ & 1/2 \\
       & -4/3 & $e_L{d}_L $ & $ e{d}$ & 1 \\
$V_{1/2}^L$ &-4/3 & $e_L{d}_R $ & $ e{d}$ & 1 \\
$V_{1/2}^R$ & -4/3 & $e_R{d}_L $ & $ e{d}$ & 1 \\ 
          & -1/3 & $e_R{u}_L $ & $ e{u}$ & 1 \\ 
$\tilde{V}_{1/2}^L$ & -1/3 & $e_L{u}_R $ & $ e{u}$ & 1 
\end{tabular}   \end{center}
\end{table}

We can define many possible LQ types~\cite{ref:Davidson}, depending on
the flavors of the quarks and leptons involved in the production and
decay.  We will initially consider LQ which preserve lepton number,
and which couple to a single generation.  The more general case
involving lepton flavor violation is treated below.

\subsubsection{Searches for leptoquarks}

 \epsfigure[width=0.8\hsize]{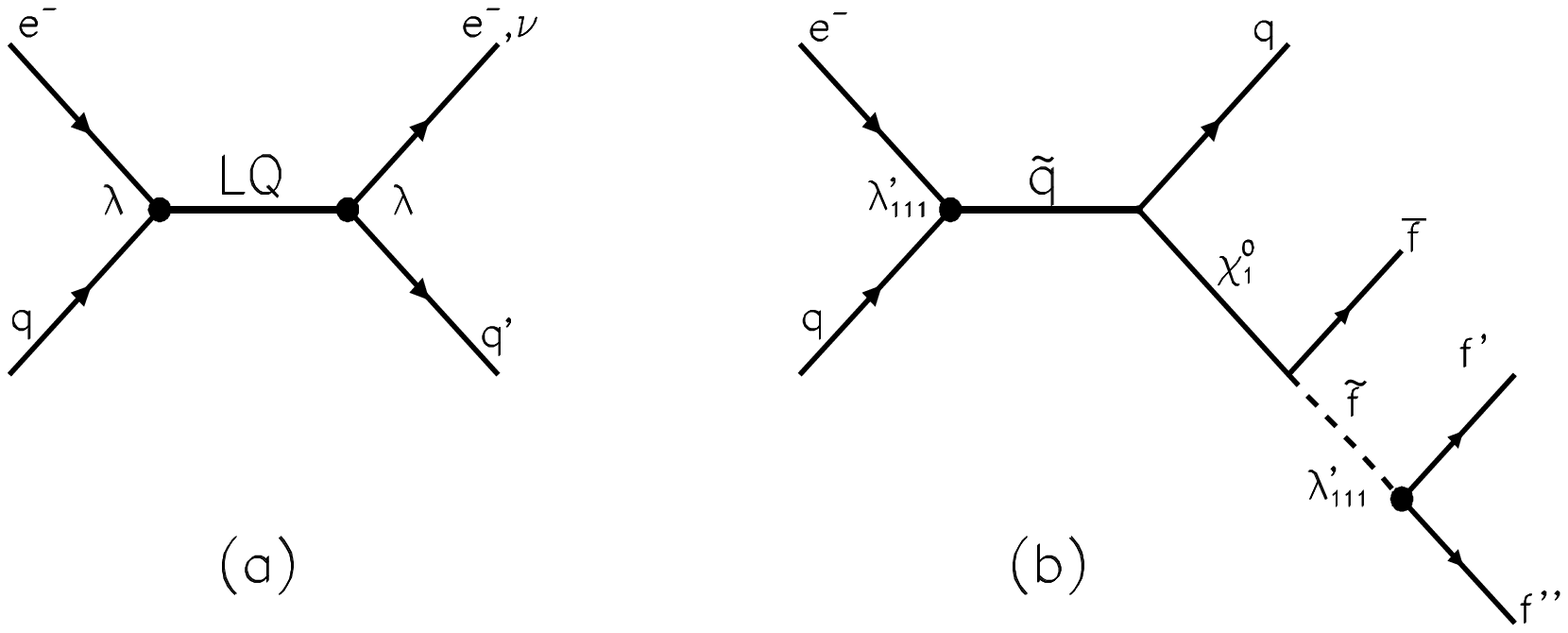} {Diagrams showing
   a) the s-channel production of a leptoquark in $ep$ scattering, and
   b) the s-channel production of a squark in R-parity violating SUSY,
   with subsequent decay.}  {schanlq}

Leptoquarks with masses below $\sqrt{s}$ are formed at HERA via
electron (or positron) fusion with a quark or antiquark from the
proton.  The leading order diagram for this process is shown in
Fig.~\ref{fig:schanlq}.

If the electron-quark coupling is not too large, then the mass of the
leptoquark is related to the center-of-mass energy via
\begin{equation}
x_0 \approx \frac{M_{LQ}^2}{s} \; \; .
\end{equation}
where $x_0$ is the fractional momentum of the proton carried by the
quark or anti-quark taking part in the interaction.  The width of the
leptoquark is given by
\begin{eqnarray}
\Gamma &  =  & \frac{\lambda^2}{16 \pi} M_{LQ} \; \; \; {\rm Spin}\; 0 \; ,\\
       &  =  & \frac{\lambda^2}{24 \pi} M_{LQ} \; \; \; {\rm Spin}\; 1 \; ,
\end{eqnarray}
where $\lambda$ is the coupling of the LQ to the electron-quark pair.
In the narrow width approximation~\cite{ref:Buchmuller}, the cross
section is given by
\begin{eqnarray}
\sigma(eq \rightarrow LQ) & \simeq & \frac{\pi}{4s} \lambda^2 q(x_0) 
\; \; \; {\rm Spin}\; 0 \; ,\\
                          & \simeq & \frac{2\pi}{4s} \lambda^2 q(x_0) 
\; \; \; {\rm Spin}\; 1 \; , 
\end{eqnarray}
where $q$ is the quark (or anti-quark) density in the proton.  Note that
the quark density is generally evaluated at the scale $\mu=M_{LQ}$.

As can be seen in Table~\ref{tab:lq}, the LQ can either decay to
electron+quark, or neutrino+quark.  The final states are
indistinguishable from normal DIS NC or CC events. The experimental
searches for leptoquarks therefore follow the standard neutral current
and charged current event selection procedures, and look for
enhancements in the $x$ or $M$ distributions, where $M$ is the mass of
the electron (neutrino) jet final state. The sensitivity to a
leptoquark signal can be enhanced by taking advantage of the $y$
distribution, which differs from that of DIS.  A scalar leptoquark,
which has an isotropic decay angular distribution in its rest frame,
will have a flat $y$ distribution, while vector leptoquarks will have
a $(1-y)^2$ distribution.  To date, no statistically significant
signal has been found in these direct searches (see however
section~\ref{sec:excess}), and 95~\% CL limits are set on the size of
the allowed lepton-quark coupling as a function of the mass of the
leptoquark.  The ZEUS limits~\cite{ref:ZEUS_LQ,ref:straub_lq,%
ref:ZEUS_LQ2} come from the 1992-3 $e^-p$ running (for $F=2$
leptoquarks) and 1994-97 $e^+p$ running (for $F=0$ leptoquarks).  The
limits are based on $0.55$~pb$^{-1}$ of $e^-p$ data and $47$~pb$^{-1}$
of $e^+p$ data.  H1 has published results based on the 1992-1993
$e^-p$ data~\cite{ref:H1_LQ_2} as well as the 1994 $e^+p$
data~\cite{ref:H1_LQ_3}, with luminosities of $0.43$~pb$^{-1}$ and
$2.83$~pb$^{-1}$ respectively.  They have also reported limits on
scalar leptoquarks from the full 1994-97 $e^+p$
data~\cite{ref:H1_LQ_5}.  The use of both $e^-p$ and $e^+p$ data
allows strong limits on both $F=0$ and $F=2$ leptoquarks.  H1 has
additionally set limits on the ratio $M_{LQ}/\lambda$ for
$M_{LQ}>\sqrt{s}$ from a contact term analysis~\cite{ref:H1_LQ_4}.
The limits achieved by the ZEUS experiment from their 1994-97 data
sets are shown in Fig.~\ref{fig:lqlimits}.  
 
 \epsfigure[width=0.8\hsize]{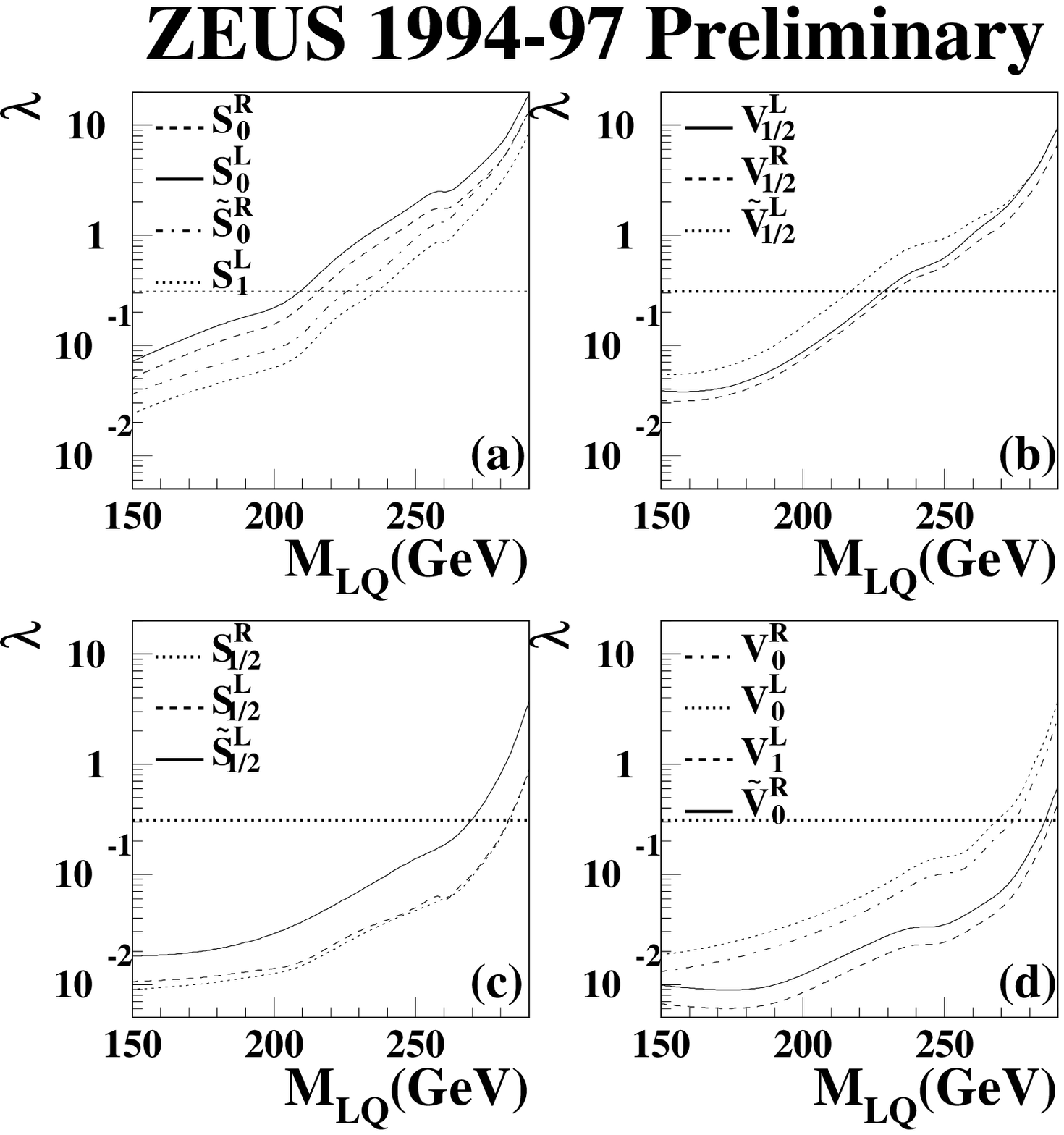} {Upper limits from
   the ZEUS experiment at 95~\% CL as a function of mass on the
   coupling $\lambda$ for scalar and vector leptoquarks which for an
   $e^+$ beam can decay into lepton+$\overline{q}$ (a,b) and
   lepton+$q$ (c,d).  The regions above the curves are excluded. The
   horizontal lines indicate the approximate coupling for an
   electroweak scale, $\sqrt{4\pi\alpha}=0.31$.}  {lqlimits}

The mass limits from both
experiments on the different leptoquark species are given in
Table~\ref{tab:lqlimits} for a coupling strength equivalent to that of
the electroweak interaction, $\lambda=\sqrt{4\pi\alpha}\approx 0.3$.
At such a coupling strength, LQ production is ruled out for
$M_{LQ}<200$~GeV.  The strongest limits are for those LQs which couple
to $u$ quarks as these have the largest density at large $x$.  For
these processes, the limits exceed $280$~GeV.  At a fixed mass of
$150$~GeV, the limits on $\lambda$ are in many cases below $0.01$.

\begin{table}
\tablecaption{Table showing the mass limits on the different types 
of leptoquarks 
for direct as well as indirect searches.  For the direct searches,
the lower limit on $M_{LQ}$ is
given for lepton-quark coupling of electromagnetic strength, 
$\lambda_{eq_1}=\sqrt{4\pi\alpha}$.  In the case  of indirect searches,
the limit is set on $M_{LQ}/\lambda$.
The results in column 3 are based on Fig.~2 of 
reference~\cite{ref:H1_LQ_3} and, for scalar leptoquarks with $F=2$,
on Fig.~11 of reference~\cite{ref:H1_LQ_5}.
The results in column 4 are from~\cite{ref:straub_lq} 
for $F=2$ leptoquarks, and from Fig.~6 of reference
\cite{ref:ZEUS_LQ2} for $F=0$ leptoquarks.}
\label{tab:lqlimits}
\begin{center}   \begin{tabular}{c|c|cc}
LQ    & indirect limit 
           & \multicolumn{2}{c}{limit on $M_{LQ}$ (GeV)} \\ 
species           & on $M_{LQ}/\lambda$ (GeV) 
           & \multicolumn{2}{c}{for $\lambda=0.31$} \\
  & H1 & H1 & ZEUS \\
\hline
$S_{1/2}^L$ &  & 275 & 282 \\
$S_{1/2}^R$ &  & 275 & 282 \\
${\tilde S}_{1/2}^L$ & 360  & 263 & 270 \\
$V_0^L$ &  & 250 & 268\\
$V_0^R$ &  & 235 & 273\\
${\tilde V}_0^R$ & 760  & 260 & 285\\
$V_1^L$ & 1020  & 270 & 287\\
\hline
$S_{0}^L$ &  & 240 & 242 \\
$S_{0}^R$ &  & 245 & 242 \\
${\tilde S}_{0}^R$ & 350  & 215 & 214 \\
$S_1^L$ & 340  & 240 & 245 \\
$V_{1/2}^L$ & 300  & 220 & 224\\
$V_{1/2}^R$ & 710  & 240 & 252\\
$\tilde{V}_{1/2}^L$ & 800  & 240 & 251 
\end{tabular}   \end{center}
\end{table}

Other bounds on leptoquarks exist from $e^+e^-$ scattering, from $p
\overline{p}$ scattering, from searches for forbidden decays, and from
the observation of lepton universality.  In $e^+e^-$ and $p
\overline{p}$ scattering the leptoquarks are produced via
electromagnetic or strong interactions which are independent of the
lepton-quark coupling strength. The strongest direct limits are from
the Tevatron and are currently $213$~GeV from CDF~\cite{ref:CDF_LQ}
and $225$~GeV from D0~\cite{ref:D0_LQ}.  These limits apply to scalar
leptoquarks assuming a branching ratio of 1.  They depend strongly on
the assumed branching ratios of the LQs to electron and neutrinos,
with smaller branching ratios giving weaker limits.  Weak
universality~\cite{ref:Davidson,ref:Leurer} imposes that $\lambda_L <
M/1.7$~(TeV).

\subsubsection{Searches for leptogluons}

The leptogluon decay width depends on the ratio $M_{LG}/\Lambda$,
where $\Lambda$ is a scale characterizing the underlying interaction,
as

\begin{equation}
\Gamma = \frac{\as}{4} \frac{M_{LG}^3}{\Lambda^2} \; ,
\end{equation}
where $\as$ is the strong coupling constant.  For $M_{LG} \ll
\Lambda$, the decay width is small and the total cross section is
approximately given by

\begin{equation}
\sigma(eg \rightarrow LG) \simeq  \frac{2\pi^2\as}{s}
(\frac{M_{LG}}{\Lambda})^2 g(x_0) \; .
\end{equation}

Leptogluons have spin $1/2$, and will therefore have a $y$
distribution of the form $(1-y)$.  The analysis proceeds along similar
lines to the search for leptoquarks.  Leptogluon limits as derived by
the H1 collaboration~\cite{ref:H1_estar1} from the 1992 $e^-p$ running
exclude at 95~\% C.L. scale parameters $\Lambda \leq 1.8$~TeV for a
leptogluon mass $M=100$~GeV and $\Lambda\leq200$~GeV at $M=200$~GeV.

\subsubsection{Flavor violating interactions}

\label{sec:flavio}

 \epsfigure[width=0.8\hsize]{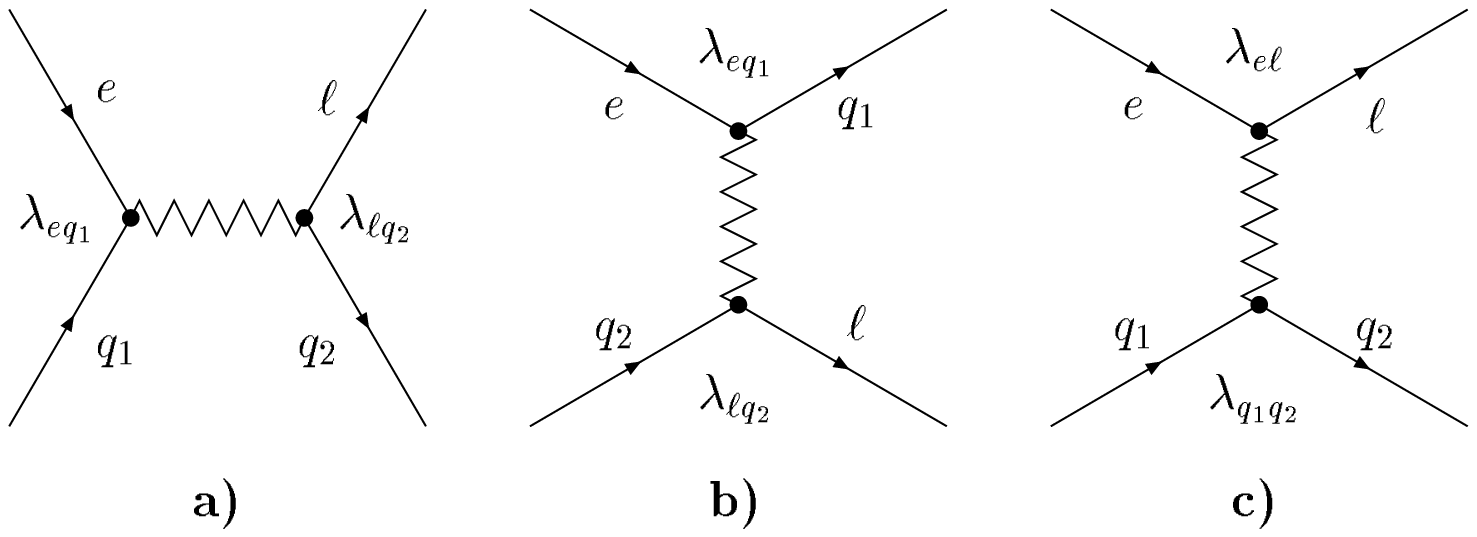} { The (a) $s$-, (b)
   $u$-, and (c) $t$-channel Feynman diagrams for LFV.  For the
   $s$-channel and $u$-channel diagrams, we denote the couplings as
   $\lambda_{\lepton q}$, where the indices refer to the lepton and
   quark flavors.}  {stugraph}

At HERA, lepton flavor violation (LFV) could occur via $s$-, $u$-, or
$t$-channel exchanges as shown in Fig.~\ref{fig:stugraph}. For the
$s$- and $u$-channel processes the exchanged particle has the quantum
numbers of a leptoquark or an $\rpvio$ squark.  For the case of
$t$-channel exchange, the process would be mediated by a
flavor-changing neutral boson.  The ZEUS~\cite{ref:ZEUS_LFV} and
H1~\cite{ref:H1_LQ_5,ref:H1_Rpvio} experiments have performed searches
for LFV by looking for events with high $p_T$ muons or taus. No
evidence for flavor violation was found.  This leads to model
dependent limits on possible types of leptoquarks or $\rpvio$ squarks.
We outline the results attained for flavor violating leptoquarks in
this section.  The results on flavor violation in the context of
$\rpvio$ squarks are given in section~\ref{sec:Rpvio}.

We can distinguish 252 different LQ scenarios leading to lepton flavor
violation.  These are defined by the 14 different LQ species described
in Table~\ref{tab:lq}, by the flavors of the quarks which couple to
the electron and the final state lepton, and by the final state lepton
flavor ($\mu$ or $\tau$).  The reactions are described by two
dimensionless couplings, $\lambda_{eq_1}$ and $\lambda_{lq_2}$ defined
in Fig.~\ref{fig:stugraph}.  The limit setting procedure depends on
whether the search is for LQ with $M_{LQ}<\sqrt{s}$ or
$M_{LQ}>\sqrt{s}$.

\epsfigure[width=0.95\hsize]{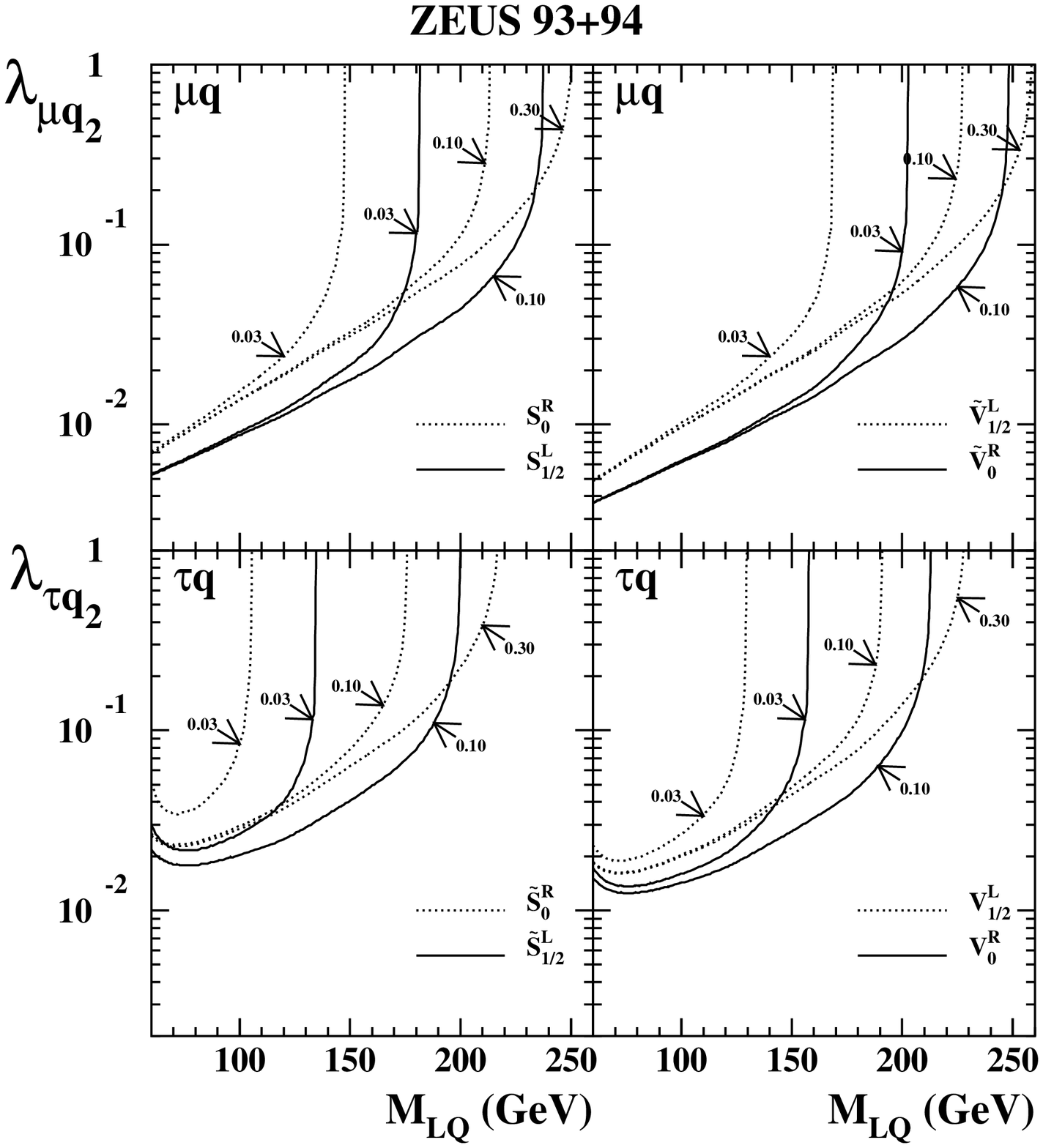} { The 95~\% CL
   upper limit on the coupling at the decay vertex {\it
     vs.}~leptoquark mass $M_{LQ}$, for several values of the
   first-generation coupling at the production vertex.  The dotted
   curves are for $F=2$ leptoquarks and the solid curves are for $F=0$
   leptoquarks.}  {ZEUS_LFV}

 \epsfigure[width=0.8\hsize]{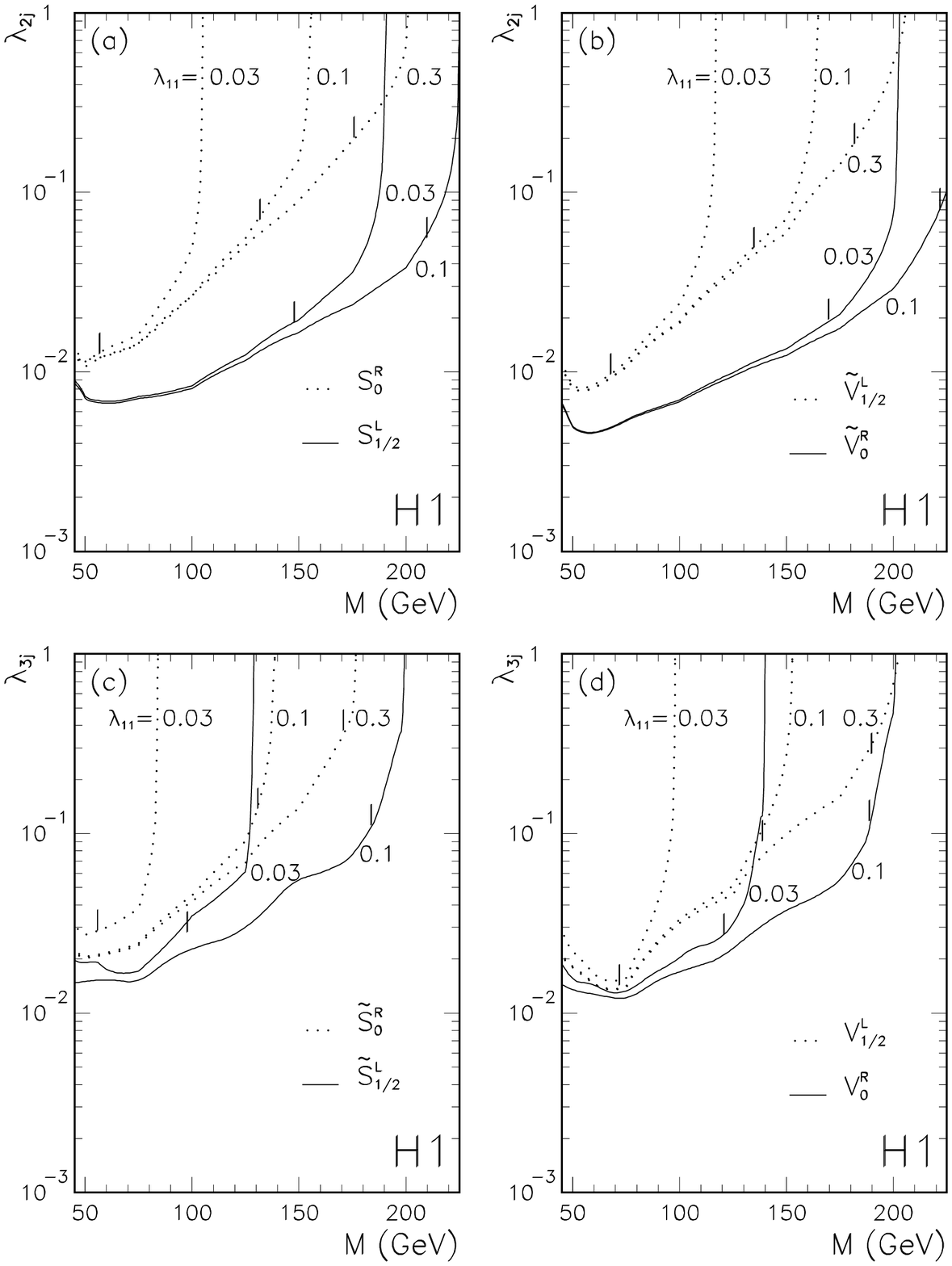} { The 95~\% CL
   upper limit on the coupling at the decay vertex {\it
     vs.}~leptoquark mass $M_{LQ}$, for several values of the
   first-generation coupling at the production vertex.  The dotted
   curves are for $F=2$ leptoquarks and the solid curves are for $F=0$
   leptoquarks.}  {H1_LFV}

\begin{itemize}
\item
For $M_{LQ}<\sqrt{s}$, limits can be set on $M_{LQ}$ for given
coupling strengths $\lambda_{eq_1}, \lambda_{lq_2}$ or
$\lambda_{eq_1}$ and branching ratio $B(lq_2) =
\lambda_{lq_2}^2/(\lambda_{eq_1}^2+\lambda_{lq_2}^2)$.  Limits on
$\lambda_{lq_2}$ versus $M_{LQ}$ are shown for fixed values of
$\lambda_{eq_1}$ in Fig.~\ref{fig:ZEUS_LFV} for the ZEUS analysis, and
Fig.~\ref{fig:H1_LFV} for the H1 analysis.  For this figure, only
leptoquarks with no neutrino decay modes are shown, and it is assumed
that the branching ratio for the decay $LQ \rightarrow l+q_2$ is given
by $B_{lq_2}=\lambda_{lq_2}^2/(\lambda_{eq_1}^2 + \lambda_{lq_2}^2)$.
For flavor violating couplings of electromagnetic strength, the ZEUS
collaboration sets 95~\% CL lower limits on leptoquark masses between
$207$~GeV and $272$~GeV, depending on the LQ species and final-state
lepton.  The recent H1 results~\cite{ref:H1_LQ_5} extend the limits
further.  For $\lambda_{eq_1}=\lambda_{\tauq_2}=0.03$, the excluded
mass range is extended by about $65$~GeV.
\item
For $M_{LQ}>\sqrt{s}$, the cross section is proportional to the square
of $\lambda_{eq_1}\lambda_{lq_2}/M_{LQ}^2$, and ZEUS has set upper
bounds on this quantity.  Many of the ZEUS limits supersede prior
limits~\cite{ref:Davidson}, particularly in the case where $e
\rightarrow \tau$.  The results for this case are summarized in
Tables~\ref{tab:LFV_tau_F0} and~\ref{tab:LFV_tau_F2}.
\end{itemize} 

\begin{table}
\tablecaption{The best upper bounds on 
$\lambda_{eq_1}\lambda_{\tau q_2}/M_{LQ}^2$
for $F=0$ leptoquarks, in units of
$10^{-4}$ GeV$^{-2}$. Each column corresponds to a given leptoquark species
and each row to the quark flavors $q_1$ and $q_2$ which couple to $e$ and 
$\tau$, the generation indices of which are specified in the first column.
The possible decay models of the LQ are listed, including decays not possible
at HERA but possible at other experiments.
The top line in each box gives the previous 
measurement~\protect\cite{ref:Davidson}
which had obtained the strictest limit. The limit from that experiment
is given on the second line in the box and the ZEUS limit, shown on
the third line, is printed in boldface if it supersedes the previous
limit. The asterisks denote those cases where lepton flavor violation
occurs only via processes involving top.}
\label{tab:LFV_tau_F0}
\begin{center}  
{\tiny 
\begin{tabular}{c||c|c|c|c|c|c|c}
\multicolumn{8}{c}{\rule{0mm}{8mm}\Large $e\leftrightarrow\tau$\hskip4cm$F=0$} \\ \hline
\rule{0mm}{8mm}& $S_{1/2}^L$   & $S_{1/2}^R$       & $\tilde{S}_{1/2}^L$ & $V_0^L$                & $V_0^R$           & $\tilde{V}_0^R$   & $V_1^L$           \\  
$(q_1q_2)$ &$e^-\ubar$         &$e^-(\ubar+\dbar)$ & $e^-\dbar$          & $e^-\dbar$             & $e^-\dbar$        & $e^-\ubar$        & $e^-(\sqrt2\ubar+\dbar)$ \\
           &$\nu\;\,\ubar$     &                   & $\nu\;\,\dbar$      & $\nu\;\,\ubar$         &                   &                   & $\nu \;\,(\ubar+\sqrt2\dbar)$ \\ \hline\hline
(11)       &  \taupie          &  \taupie          &  \taupie            & $G_F$                  &  \taupie          &  \taupie          & $G_F$             \\ 
           &  0.02             &  0.01             &  0.02               &  0.002                 &  0.01             &  0.01             &  0.002            \\
           &  0.11             &  0.09             &  0.18               &  0.11                  &  0.11             &  0.07             &  0.04             \\ \hline
(12)       &                   &  \tauKe           &  \Kpi               &  \tauKe                &  \tauKe           &                   & \Kpinunu          \\ 
           &                   &  0.05             & $2\times 10^{-5}$   &  0.03                  &  0.03             &                   & $5\times 10^{-6}$ \\
           &  {\bf 0.12 }      &  0.10             &  0.18               &  0.15                  &  0.15             &  {\bf 0.10 }      &  0.05             \\ \hline
(13)       &                   &  \BtaueX          &  \BtaueX            & \BlnuX                 &  \BtaueX          &                   & \BlnuX            \\ 
           &                   &  0.08             &  0.08               &  0.02                  &  0.04             &                   &  0.02             \\
           &  *                &  0.18             &  0.18               &  0.16                  &  0.16             &  *                &  0.16             \\ \hline
(21)       &                   &  \tauKe           &  \Kpi               &  \tauKe                &  \tauKe           &                   & \Kpinunu          \\ 
           &                   &  0.05             & $2\times 10^{-5}$   &  0.03                  &  0.03             &                   & $5\times 10^{-6}$ \\
           &  {\bf 0.34 }      &  0.26             &  0.39               &  0.14                  &  0.14             &  {\bf 0.10 }      &  0.05             \\ \hline
(22)       & \tauegam          & \tauegam          &                     &                        &                   &                   &                   \\ 
           &  0.2              &  0.2              &                     &                        &                   &                   &                   \\
           &  0.60             &  0.37             &  {\bf 0.48 }        &  {\bf 0.25 }           &  {\bf 0.25 }      &  {\bf 0.31 }      &  {\bf 0.13 }      \\ \hline
(23)       &                   &  \BtaueX          &  \BtaueX            & \BlnuX                 &  \BtaueX          &                   & \BlnuX            \\ 
           &                   &  0.08             &  0.08               &  0.02                  &  0.04             &                   &  0.02             \\
           &  *                &  0.50             &  0.50               &  0.33                  &  0.33             &  *                &  0.33             \\ \hline
(31)       &                   &  \BtaueX          &  \BtaueX            & $V_{bu}$               &  \BtaueX          &                   & $V_{bu}$          \\ 
           &                   &  0.08             &  0.08               &  0.002                 &  0.04             &                   &  0.002            \\
           &  *                &  0.47             &  0.47               &  0.15                  &  0.15             &  *                &  0.15             \\ \hline
(32)       &                   &  \BtaueX          &  \BtaueX            & \BlnuX                 &  \BtaueX          &                   & \BlnuX            \\ 
           &                   &  0.08             &  0.08               &  0.02                  &  0.04             &                   &  0.02             \\
           &  *                &  0.65             &  0.65               &  0.26                  &  0.26             &  *                &  0.26             \\ \hline
(33)       &                   &                   &                     &  \tauegam              &  \tauegam         &                   &                   \\ 
           &                   &                   &                     &   3.4                  &  3.4              &                   &                   \\
           &  *                &  {\bf 0.72 }      &  {\bf 0.72 }        &  {\bf 0.38 }           &  {\bf 0.38 }      &  *                & {\bf 0.38}        
\end{tabular}   
}
\end{center}
\end{table}                 

\begin{table}
\tablecaption{The best upper bounds on $\lambda_{eq_1}\lambda_{\tau q_2}/\MLQsq$
for $F=2$ leptoquarks, in units of
$10^{-4}$ GeV$^{-2}$. Each column corresponds to a given leptoquark species
and each row to the quark flavors $q_1$ and $q_2$ which couple to $e$ and $\tau$,
the generation indices of which are specified in the first column.
The top line in each box gives the previous measurement~\cite{ref:Davidson}
which had obtained the strictest limit. The limit from that experiment
is given on the second line in the box and the ZEUS limit, shown on
the third line, is printed in boldface if it supersedes the previous
limit. The asterisks denote those cases where lepton flavor violation
occurs only via processes involving top.}
\label{tab:LFV_tau_F2}
\begin{center}  
{\tiny
\begin{tabular}{c||c|c|c|c|c|c|c}
\multicolumn{8}{c}{\rule{0mm}{8mm}\Large $e\leftrightarrow\tau$\hskip4cm$F=2$} \\ \hline
\rule{0mm}{8mm}&$S_0^L$        & $S_0^R$           & $\tilde{S}_0^R$     & $S_1^L$                & $V_{1/2}^L$       & $V_{1/2}^R$       & $\tilde{V}_{1/2}^L$ \\ 
$(q_1q_2)$ & $e^-u$            & $e^-u$            & $e^-d$              & $e^-(u+\sqrt2d)$     & $e^-d$            & $e^-(u+d)$        & $e^-u$            \\
           &$\nu\;\, d$        &                   &                     &$\nu\;\, (\sqrt2u+d)$ & $\nu\;\,d$        &                   & $\nu\;\,u$        \\ \hline\hline
(11)       & $G_F$             &  \taupie          &  \taupie            & $G_F$                  &  \taupie          &  \taupie          &  \taupie          \\ 
           &  0.003            &  0.02             &  0.02               &  0.003                 &  0.01             &  0.005            &  0.01             \\
           &  0.15             &  0.15             &  0.23               &  0.09                  &  0.09             &  0.05             &  0.06             \\ \hline
(12)       &  \Kpi             &                   &  \tauKe             &  \Kpi                  &  \Kpi             &  \tauKe           &                   \\ 
           & $2\times 10^{-5}$ &                   &  0.05               & $2\times 10^{-5}$      & $10^{-5}$         &  0.03             &                   \\
           &  0.20             &  {\bf 0.20 }      &  0.27               &  0.11                  &  0.19             &  0.13             &  {\bf 0.16 }      \\ \hline
(13)       & $V_{bu}$          &                   &  \BtaueX            & $V_{bu}$               & \BtaueX           &  \BtaueX          &                   \\ 
           &  0.004            &                   &  0.08               &  0.004                 &  0.04             &  0.04             &                   \\
           &                   &  *                &  0.28               &  0.14                  &  0.23             &  0.23             &   *               \\ \hline
(21)       &  \Kpi             &                   &  \tauKe             &  \Kpi                  &  \Kpi             &  \tauKe           &                   \\ 
           & $2\times 10^{-5}$ &                   &  0.05               & $2\times 10^{-5}$      & $10^{-5}$         &  0.03             &                   \\
           &  0.22             &  {\bf 0.22 }      &  0.31               &  0.12                  &  0.09             &  0.05             &  {\bf 0.06 }      \\ \hline
(22)       & \tauegam          & \tauegam          & \tauegam            & \tauegam               &                   &                   &                   \\ 
           &  0.5              &  0.5              &  0.3                &  0.1                   &                   &                   &                   \\
           &  0.60             &  0.60             &  0.48               &  0.22                  &  {\bf 0.25 }      &  {\bf 0.19 }      &  {\bf 0.31 }      \\ \hline
(23)       & \BlnuX            &                   &  \BtaueX            & \BlnuX                 &  \BtaueX          &  \BtaueX          &                   \\ 
           &  0.04             &                   &  0.08               &  0.04                  &  0.04             &  0.04             &                   \\
           &                   &  *                &  0.50               &  0.25                  &  0.33             &  0.33             &   *               \\ \hline
(31)       & \BlnuX            &                   &  \BtaueX            & \BlnuX                 &  \BtaueX          &  \BtaueX          &                   \\ 
           &  0.04             &                   &  0.08               &  0.04                  &  0.04             &  0.04             &                   \\
           &                   &  *                &  0.34               &  0.17                  &  0.10             &  0.10             &   *               \\ \hline
(32)       & \BlnuX            &                   &  \BtaueX            & \BlnuX                 &  \BtaueX          &  \BtaueX          &                   \\ 
           &  0.04             &                   &  0.08               &  0.04                  &  0.04             &  0.04             &                   \\
           &                   &  *                &  0.65               &  0.32                  &  0.26             &  0.26             &   *               \\ \hline
(33)       &                   &                   & \tauegam            & \tauegam               &                   &                   &                   \\ 
           &                   &                   &  0.3                &  0.1                   &                   &                   &                   \\
           &                   &  *                &  0.72               &  0.36                  &  {\bf 0.38 }      &  {\bf 0.38 }      &   *              
\end{tabular}
}    
\end{center}
\end{table}

\subsection{Supersymmetry at HERA}
\label{sec:susy}
Supersymmetry (SUSY)~\cite{ref:susyrefs1,ref:susyrefs2} is widely
acknowledged as a leading candidate theory to describe possible
physics not included in the Standard Model.  It relates the properties
of fermions to those of bosons (e.g., in SUSY, selectrons, $\widetilde
e$ are scalar partners of electrons, $e$) and locally supersymmetric
theories naturally incorporate gravity.  SUSY also resolves in a
natural way, in the unbroken version, the hierarchy (or naturalness)
problem; i.e., why the ratio $M_W/M_{Planck}$ is so small.  It is
therefore of great interest to search for signs of this new theory.
To date, no evidence for supersymmetry has been found.

The fact that the superpartners to the existing particles have not
been observed implies that supersymmetry is broken.  The breaking of
supersymmetry introduces many new parameters into the theory which
makes the classification of possible SUSY scenarios rather forbidding.
In general, assumptions must be made to reduce the number of
parameters and thereby allow for reasonably compact classification
schemes~\cite{ref:baeretal}.  In the sections which follow, we will
only consider the minimal supersymmetric model (MSSM) and $R$-parity
violating supersymmetry.

\subsubsection{The minimal supersymmetric model - MSSM}
\label{sec:mssm}

The simplest supersymmetric extension of the Standard Model is denoted
MSSM (for minimal supersymmetric model).  It is a direct
generalization of the Standard Model, with electroweak symmetry
breaking occurring via vacuum expectation values of two different
Higgs superfields, denoted $H^1$ and $H^2$.  Baryon number and lepton
number are separately conserved.  This is most easily expressed in
terms of the multiplicative quantum number $R$-parity, defined as
$R_P=(-1)^{3B+L+2S}$ where $B$ is the baryon number, $L$ is the lepton
number and $S$ the spin of the particle, so that $R_P=+1$ for SM
particles and $R_P=-1$ for SUSY particles.  The MSSM contains the
smallest number of new particles and new interactions compatible with
the Standard Model.  If $R$-parity were conserved, SUSY particles
would be produced in pairs and would ultimately decay into the
lightest supersymmetric particle (LSP), which would be stable and
neutral.  It is therefore expected in MSSM that signatures with
missing energy would arise since the LSP would escape the detector
unseen.

The spectrum of particles expected in the MSSM, in addition to those
already known in the SM, includes scalar squarks and sleptons,
spin-$1/2$ charginos, neutralinos, the gluino and five different Higgs
bosons.  Charginos and neutralinos are charged and neutral mass
eigenstates of the (mixed) supersymmetric partners of the
$W^{\pm},Z^0,\gamma$ (gauginos) and the two Higgs doublets
(higgsinos).  The LSP is constrained to be neutral for cosmological
reasons, and it is therefore generally assumed that the lightest
neutralino, $\chi_1^0$, is the LSP.  There are in addition many
possible couplings.  The MSSM therefore contains many new parameters
and further assumptions are necessary to reduce these to a more
tractable number.  This necessarily introduces model dependence in the
interpretation of searches for new particles.  A common approach is to
use constraints from supergravity models~\cite{ref:baeretal} to
express the various SUSY parameters in terms of a common scalar mass
$m_0$, a common gaugino mass $m_{1/2}$, the ratio of the vacuum
expectation value of the two Higgs superfields $\tan \beta$, a common
trilinear interaction $A_0$, and a sign parameter $sign(\mu)$.

 \epsfigure[width=0.6\hsize]{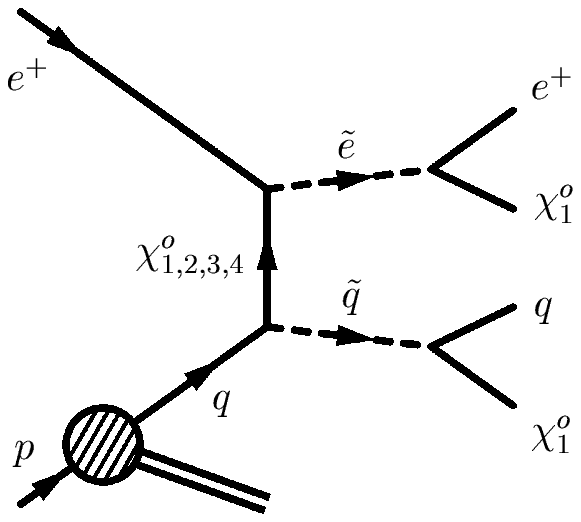} {Feynman diagram
   for selectron-squark production via neutralino exchange and the
   subsequent decays into the lightest supersymmetric particle
   $\chi_1^0$.}  {mssmselsq}

The dominant MSSM process expected at HERA is the production of a 
selectron and a squark via neutralino exchange, $ep \rightarrow 
\widetilde e \widetilde q X$, depicted in Fig.~\ref{fig:mssmselsq}.   
The H1 and ZEUS collaborations have performed searches for this
channel~\cite{ref:H1selsq,ref:ZEUS_MSSM}.  The selectron and squark
are taken to decay into the lightest neutralino plus SM particles, and
the appropriate branching ratios are used.  The limits are found under
the following assumptions:
\begin{itemize}
\item $M_{{\widetilde e}_L} = M_{{\widetilde e}_R}$;
\item $M_{{\widetilde q}_L} = M_{{\widetilde q}_R}$;
\item the four lightest squarks are assumed to be mass degenerate;
\item the GUT relation $M_1 = 5/3 M_2 \tan^2 \theta_W$ holds.
\end{itemize}

\epsfigure[width=0.95\hsize]{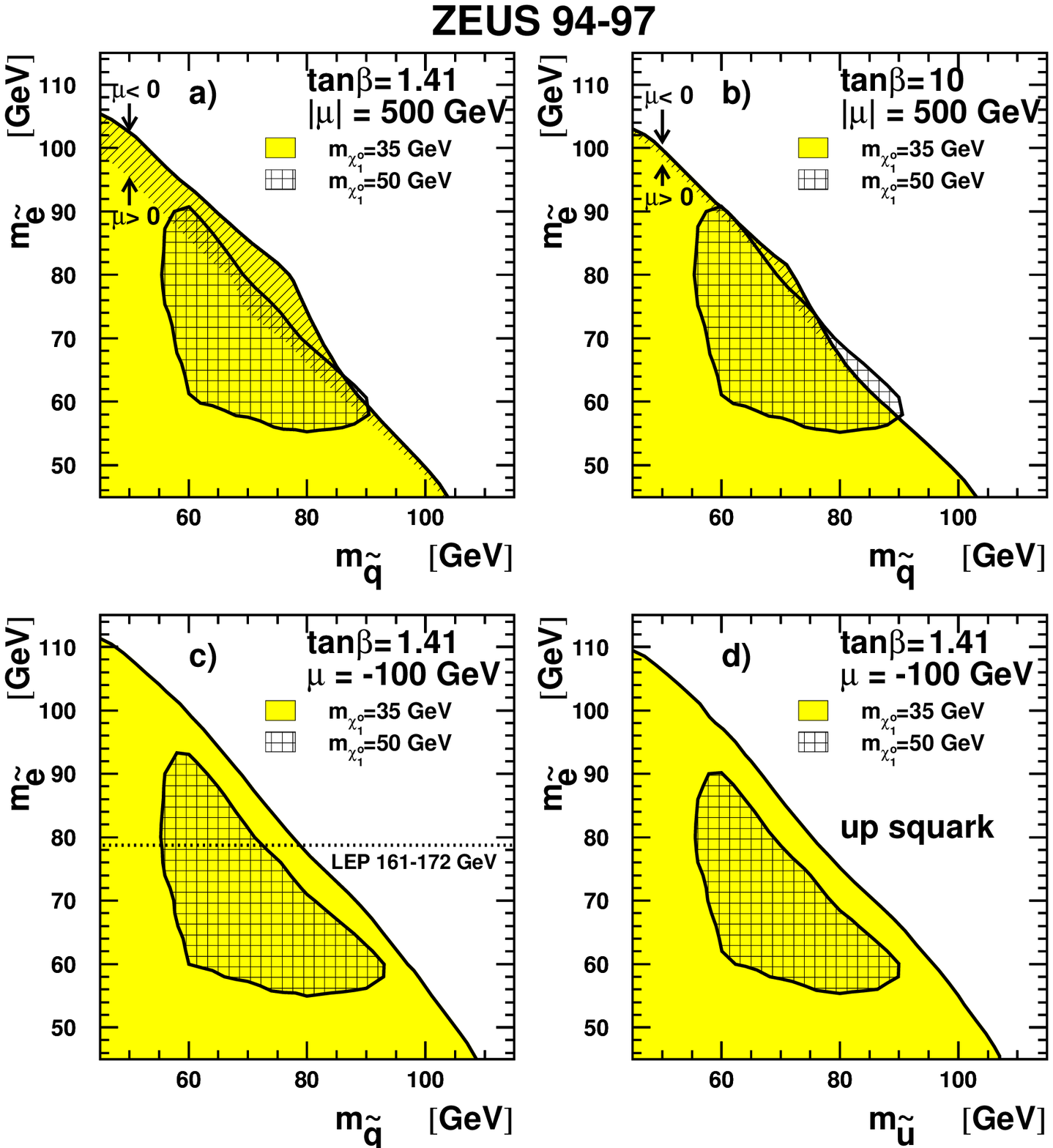} {Excluded region at
   95~\% CL in the plane defined by the selectron and the squark mass,
   for two different values of the neutralino mass,
   $m_{\chi_1^0}=35,50$~GeV.  In a) and b) the limits are for
   $|\mu|=500$~GeV and $\tan\beta=1.41,10$.  For $\mu<0$, the excluded
   region also includes the single-hatched area.  The limits obtained
   for $\mu=-100$~GeV and $\tan\beta = 1.41$ are shown in c) along
   with the LEP limits on $m_{\tilde e}$.  The limits for the up
   squark alone are shown in d).}  {ZEUS_MSSM}

An integrated luminosity of $6.38$~pb$^{-1}$ of $e^+p$ collisions was
used in the H1 search and $47$~pb$^{-1}$ for the ZEUS search, and
signatures based principally on missing energy and missing transverse
momentum were searched for.  For $\tan\beta=\sqrt{2}$, masses below
$(M_{\widetilde e}+M_{\widetilde q})/2 = 77$~GeV for
$M_{\chi^0_1}=40$~GeV are excluded at $95$~\% confidence level.  The
limits are compared to contemporaneous LEP
results~\cite{ref:ALEPH_SUSY_1,ref:L3_SUSY_1,ref:OPAL_SUSY_1} in
Fig.~\ref{fig:ZEUS_MSSM}.  The HERA results exceed beyond the LEP
results for squark masses below about $80$~GeV.  LEP
experiments~\cite{ref:ALEPH_SUSY_2,ref:OPAL_SUSY_2,ref:L3_SUSY_2} have
also reported limits on stop and on sbottom of about $75$~GeV.  The
limits from HERA, for $M_{\widetilde e}=M_{\widetilde q}$ are at the
same level. Since the production at HERA is principally off up quarks,
the HERA limits on $(M_{\widetilde e}+M_{\widetilde u})/2>75$~GeV are
almost as high as considering all squarks.

Strong limits on the squark mass have been obtained at the 
Tevatron \cite{ref:CDF_SUSY,ref:D0_SUSY} which are complementary to those
obtained at HERA.  The Tevatron squark limits are in general at several
hundred GeV.  However, they are obtained using GUT inspired assumptions
not required by the HERA experiments 
and they are less sensitive for small mass
differences $M_{\widetilde q}-M_{\chi_1^0}$. The HERA analyses probe
different regions of MSSM parameter space where the gluinos are heavier
than the squarks and the squark masses are independent of the selectron
mass.

\subsubsection{R parity violating SUSY}
\label{sec:Rpvio}

 $R$-parity need not be conserved in supersymmetric theories.  The
general superpotential allows for Yukawa couplings between the
Standard Model fermions and $\squark$ or $\slepton$.  However,
superpotential terms which violate $R$-parity can in some cases lead
to proton decay, which has very stringent experimental constraints, so
that the appropriate couplings must be small.  Of particular interest
for $ep$ collisions are $R$-parity violating superpotential terms of
the form $\lambda'_{ijk}L^i_LQ^j_L{\overline
D}^k_R$~\cite{ref:Butterworth_Dreiner}.  Here $L_L$, $Q_L$, and
${\overline D}_R$ denote left-handed lepton and quark doublets and the
right handed $d$-quark singlet chiral superfields respectively, and
the indices $i$, $j$, and $k$ label their respective
generations. Expanded into four-component Dirac notation, the
corresponding terms of the Lagrangian are
\begin{eqnarray}
{\cal L}&=&\lambda'_{ijk}\left[
   {\tilde\nu}^i_L {\overline d}^k_R d^j_L
 + {\tilde d}^j_L  {\overline d}^k_R \nu^i_L
 + ({\tilde d}^k_R)^* ({\overline\nu}^i_L)^c d^j_L \right. \nonumber \\
& & \mbox{ } - \left. {\tilde e}^i_L {\overline d}^k_R u^j_L
 - {\tilde u}^j_L {\overline d}^k_R e^i_L
 - ({\tilde d}^k_R)^* ({\overline e}^i_L)^c u^j_L \right] + {\rm h.~c.} 
\label{eq:Lsquark}
\end{eqnarray}

\begin{table}
\tablecaption{Table showing the different single squark production processes 
via an $\rpvio$ coupling at HERA with an $e^+$ beam.}
\label{tab:Rpvio_squarks}
\begin{center}   \begin{tabular}{c|cc}
$\lambda_{1jk}'$ & \multicolumn{2}{c}{Production process} \\
& {down type squarks} & {up type squarks} \\
\hline
$111$ & $e^+ + \bar{u} \rightarrow \bar{{\tilde d}}_R $ &  
$e^+ + d \rightarrow {\tilde u}_L $ \\
$112$ & $e^+ + \bar{u} \rightarrow \bar{{\tilde s}}_R $ &  
$e^+ + s \rightarrow {\tilde u}_L $ \\
$113$ & $e^+ + \bar{u} \rightarrow \bar{{\tilde b}}_R $ &  
$e^+ + b \rightarrow {\tilde u}_L $ \\
$121$ & $e^+ + \bar{c} \rightarrow \bar{{\tilde d}}_R $ &  
$e^+ + d \rightarrow {\tilde c}_L $ \\
$122$ & $e^+ + \bar{c} \rightarrow \bar{{\tilde s}}_R $ &  
$e^+ + s \rightarrow {\tilde c}_L $ \\
$123$ & $e^+ + \bar{c} \rightarrow \bar{{\tilde b}}_R $ &  
$e^+ + b \rightarrow {\tilde c}_L $ \\
$131$ & $e^+ + \bar{t} \rightarrow \bar{{\tilde d}}_R $ &  
$e^+ + d \rightarrow {\tilde t}_L $ \\
$132$ & $e^+ + \bar{t} \rightarrow \bar{{\tilde s}}_R $ &  
$e^+ + s \rightarrow {\tilde t}_L $ \\
$133$ & $e^+ + \bar{t} \rightarrow \bar{{\tilde b}}_R $ &  
$e^+ + b \rightarrow {\tilde t}_L $ 
\end{tabular}   \end{center}
\end{table}

\begin{table}
\tablecaption{95\% CL mass limits (GeV) from ZEUS 
for squarks with $\rpvio$-couplings, 
of electromagnetic strength 
($\lambda_{11k}^2=\lambda_{ijk}^2=4\pi\alpha=4\pi/128$),
for different masses of the lightest SUSY particle (LSP), denoted
$\chi_1^0$.
The mixing angle of the stop is assumed to be $\cos^2\theta_t=0.5$.
The limits for the $m_{\chi_1^0}=0$ case are somewhat weaker than those
for a heavy LSP because ZEUS did not search for the gauge decay
$\squark\to q\chi_1^0$.}
\label{tab:LFV_squark}
\begin{center}   \begin{tabular}{c|c|c|c|c|c|c}
\rule{0mm}{8mm}
&
$\sdown\rightarrow\mu q$ &
$\suup\rightarrow\mu q$ &
$\sTop\rightarrow\mu q$ &
$\sdown\rightarrow\tau q$ &
$\suup\rightarrow\tau q$ &
$\sTop\rightarrow\tau q$ \\ \hline \hline
$m_{\chi_1^0}=0$                   & 229 & 229 &  -  & 221 & 222 &  -  \\ \hline
$m_{\chi_1^0}+m_q\ge m_{\tilde q}$ & 231 & 234 & 223 & 223 & 228 & 216 
\end{tabular}   \end{center}
\end{table}

For $i=1$, which is the case at HERA, the last two terms will result
in ${\tilde u}$ and ${\tilde d}$ type squark production in $ep$
collisions.  All possible right-handed down type squarks and
left-handed up type squarks can be produced, as listed in
Table~\ref{tab:Rpvio_squarks}.  In $e^+p$ collisions these are
produced off up type antiquarks and down type quarks, respectively.
There are thus 18 possible production couplings probed in $e^+p$
collisions.  For production and decay via $\lambda'_{ijk}$, the
squarks behave as scalar leptoquarks and the final state will be
indistinguishable, event-by-event, from Standard Model neutral and
charged current events.  However, as with the scalar leptoquarks, the
angular distributions of the final state lepton and quark will be
different and this fact can be exploited in performing searches.

 \epsfigure[width=0.8\hsize]{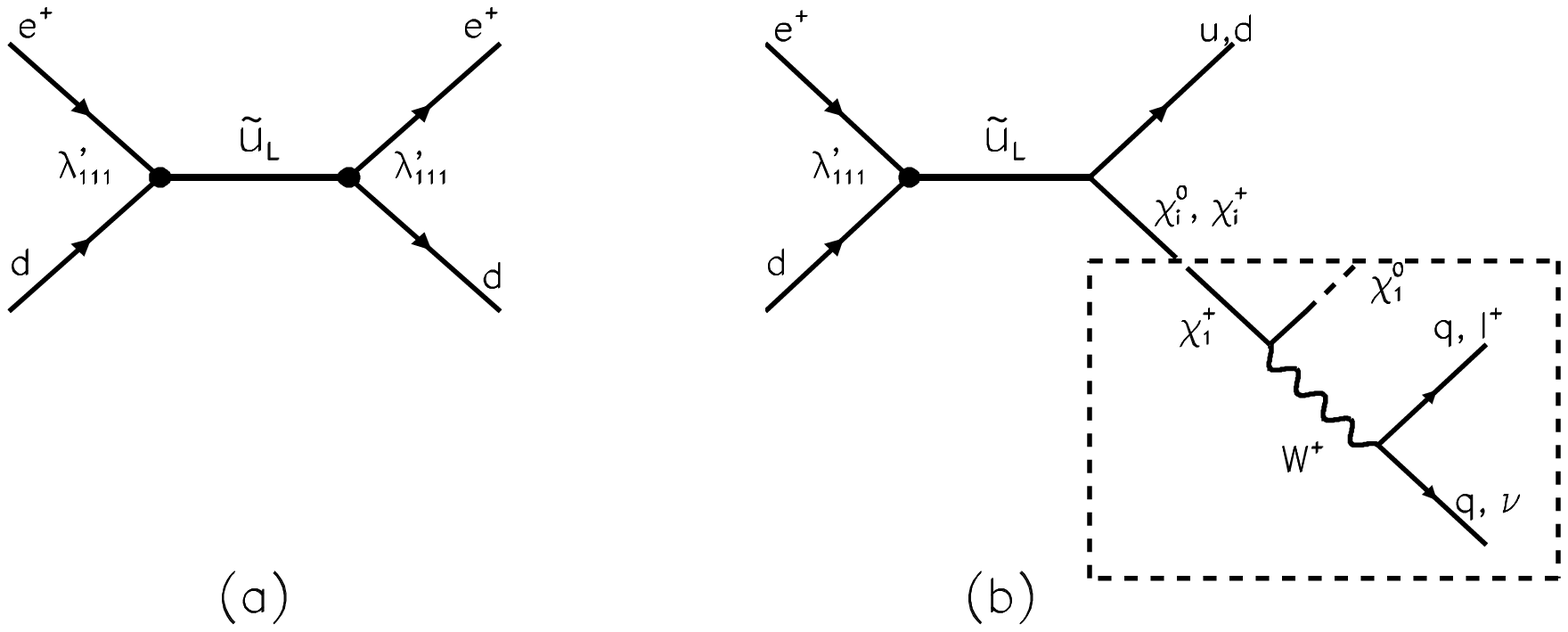} {Lowest order
   $s$-channel diagrams for first generation squark production via
   $e^+d$ scattering at HERA followed by (a) $\rpvio$ decay and (b)
   gauge decay.  In (b), the emerging neutralino or chargino might
   subsequently undergo $\rpvio$ decays of which an examples is shown
   in the dotted box for the $\chi_1^+$.}  {Rpvio}

In addition to the Yukawa couplings, there are also gauge couplings
such that the $\squark$ can decay by radiating a neutralino or chargino.
The neutralino or chargino can then subsequently decay.  The final
state signature will depend on the properties of the neutralino or
chargino involved.  Figure~\ref{fig:Rpvio} shows examples of both
Yukawa type decays and gauge type decays.  For the gauge decays,
examples are given of subsequent decays of the neutralino or chargino.
The H1 collaboration has 
searched for $\rpvio$ squark production~\cite{ref:H1_Rpvio,ref:H1_Rpvio_2} 
under the following assumptions:
\begin{itemize}
\item  One $\lambda'_{ijk}$ dominates;
\item  $M_{\widetilde{q_R}} \approx M_{\widetilde{q_L}}$ for the first
two generations;
\item The lightest neutralino, $\chi_1^0$, is the lightest SUSY particle;
\item $M_{\widetilde{g}} > M_{\widetilde {q}}$.
\end{itemize}
The stop is a special case since it is not ruled out that it can be 
lighter than the corresponding Standard Model particle, the top quark.

 \epsfigure[width=0.95\hsize]{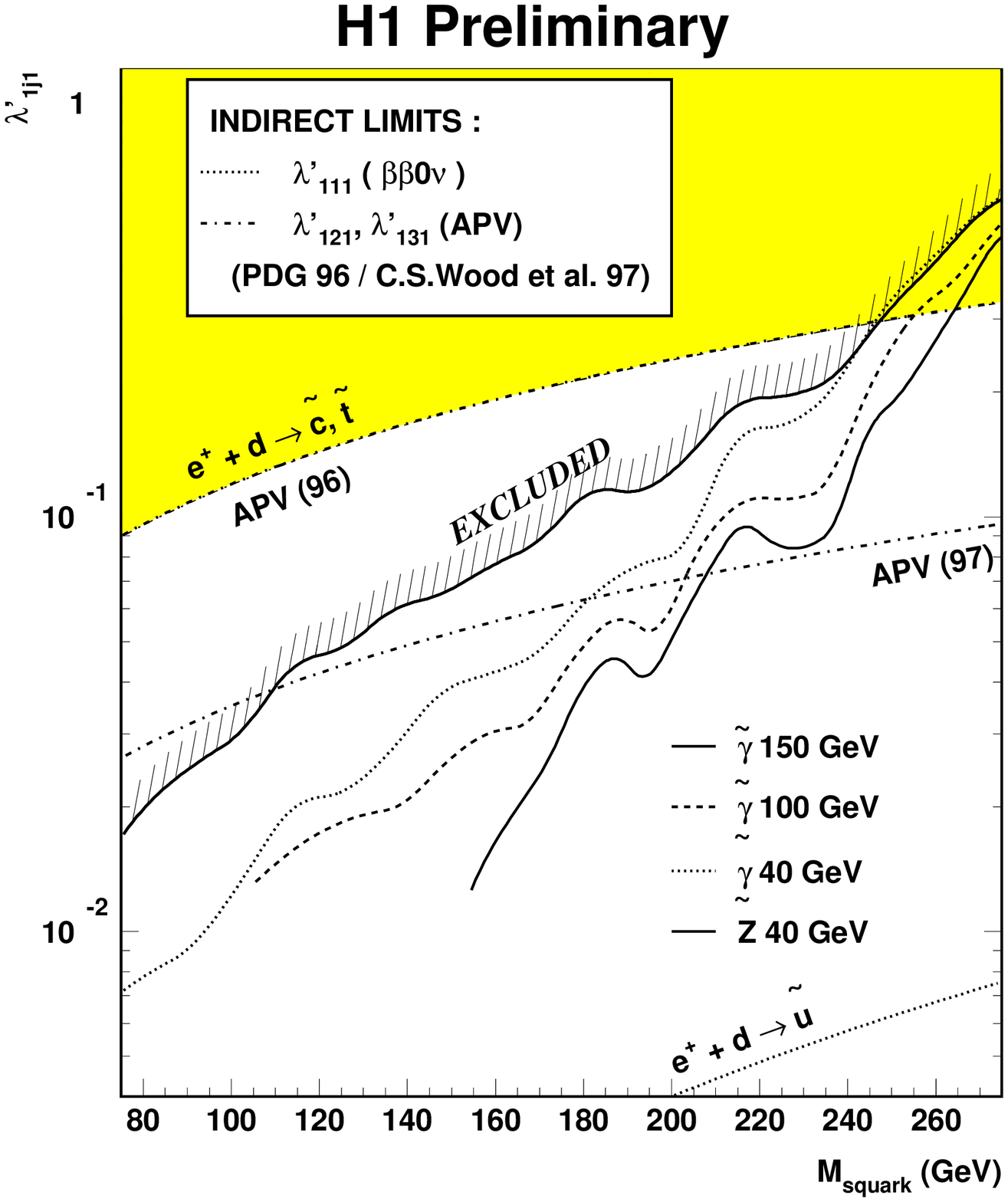}
{Exclusion upper limits at 95~\% CL for the coupling
$\lambda_{1j1}'$ as a function of the squark mass for various
masses and mixtures of the $\chi_1^0$;  also represented are the
most stringent indirect limits on $\lambda_{111}'$ and 
$\lambda_{1j1}',j=2,3$.}{H1_Rpvio_photino}  

A systematic analysis of the possible final states resulting from the
different production and decay modes was performed, and the search
procedure was optimized for each.  No significant evidence for SUSY
particle production was found, and the H1 collaboration excludes
$\suup$ and $\sdown$ under various model assumptions.  In
Fig.~\ref{fig:H1_Rpvio_photino}, exclusion limits on $\lambda_{1j1}'$
are shown as a function of the squark mass for different assumptions
on the $\chi_1^0$ mass and composition. Under these conditions,
squarks up to masses of $262$~GeV are ruled out at $95$~\% CL for
$\lambda_{1j1}^{'2} = 4\pi\alpha$.

The Tevatron leptoquark limits rule out $\suup_L^j$ squark masses
below $200$~GeV for branching fractions into $e^+q$ greater than
50~\%.  However, the limits degrade quickly as the branching lower
becomes smaller, which is a natural feature in R-parity violating
SUSY.

The indirect limits from atomic parity violation~\cite{ref:APV_1} and
from the non-observation of neutrinoless double beta
decay~\cite{ref:DBD1,ref:DBD2} are compared to the HERA limits in
Fig.~\ref{fig:H1_Rpvio_photino}.  The most stringent limits come from
the neutrinoless double beta decay searches but only concern the
$\lambda'_{111}$ coupling.  For the couplings
$\lambda'_{121},\lambda'_{131}$ the limits are comparable to the HERA
results.

\subsubsection{R-parity violation and lepton flavor violation}
\label{sec:Rpvio_LFV}

\epsfigure[width=0.8\hsize]{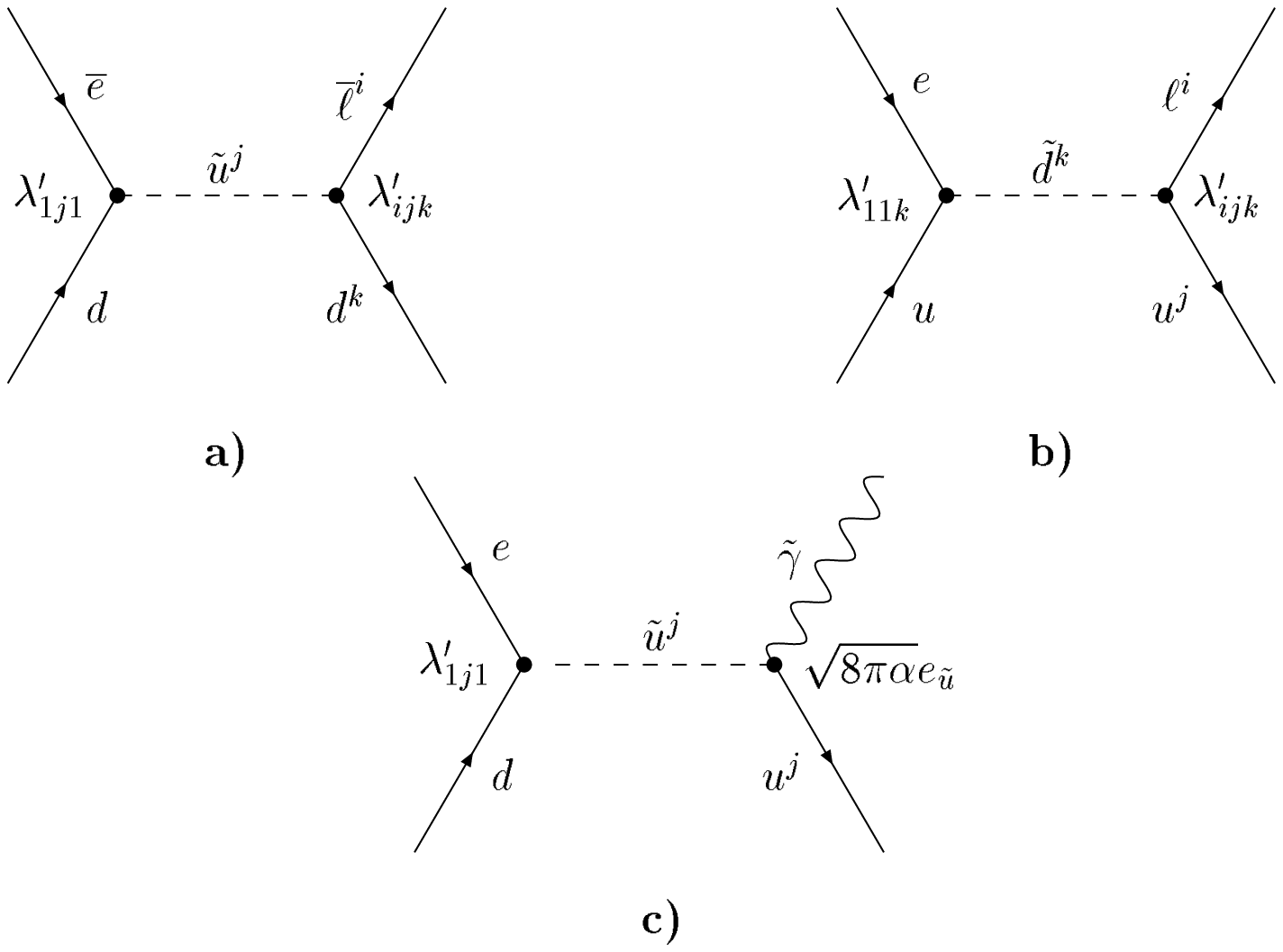} { $R_P$ violating
  single squark production in $ep$ collisions.  Diagrams a) and b)
  show production of ${\tilde u}$ and ${\tilde d}$ squarks with
  leptoquark-like $\rpvio$ decays, where $\lepton^i$ denotes the
  final-state charged lepton of generation $i$. The indices $j$ and
  $k$ denote the generations of up-type and down-type (s)quarks
  respectively.  Diagram c) shows ${\tilde u}$ production with an
  $R_P$-conserving decay.}  {susygraph}
  
Lepton-flavor violating $ep$ interactions would occur in a model with
two non-zero couplings $\lambda'_{ijk}$ which involve different lepton
generations.  For example, the process $e^+d\to\suup^j\to \mu^+ d^k$
involves the couplings $\lambda'_{1j1}$ and $\lambda'_{2jk}$.
Similarly, non-zero values for $\lambda'_{11k}$ and $\lambda'_{3jk}$
would lead to the reaction $e^-u\to\sdown^k\to\tau^- u^j$ (see
Fig.~\ref{fig:susygraph}).  Down-type squarks have the additional
decay ${\tilde d}^k\to\nu^id^j$, a mode unavailable to up-type
squarks.

As pointed out above, the difference between mechanisms involving
$R$-parity violating squarks and leptoquarks is that the squarks have
the additional decay mode ${\tilde q}\to q\chi_1^0$ shown in
Fig.~\ref{fig:Rpvio}. The branching ratios $B_{q\chi_1^0}$ for the
$R_P$-conserving decay $\squark\to q\chi_1^0$ and $B_{ijk}'$ for any
$\rpvio$ decay mode are related~\cite{ref:Butterworth_Dreiner} by
\begin{equation}
   {{B_{ijk}'}\over{(\lambda_{ijk}')^2}} =
  {{B_{q\chi_1^0}}\over{ 8\pi\alpha}e_{\squark}^2(1-
{m_{\chi_1^0}^2}/{m_{\squark}^2)^2}},  \label{eq:translqsq}
\end{equation}
where $\lambda_{ijk}'$ is the $\rpvio$ coupling at the decay vertex,
$\alpha$ is the electromagnetic coupling, $e_{\tilde q}$ is the squark
charge in units of the electron charge and the LSP and squark masses
are $m_{\chi_1^0}$ and $m_{\tilde q}$, respectively.

Coupling limits for LFV decays of an $S_0^L$ leptoquark can be
interpreted as ${\tilde d}^k$ coupling limits through the
correspondence $\lambda_{eq_1}\sqrt{B_{\lepton
q_2}}=\lambda_{11k}'\sqrt{B_{ijk}'}$ where $i$ and $j$ are the
generations of the LQ decay products $\lepton$ and $q_2$.  Similarly,
coupling limits on the $\tilde{S}_{1/2}^L$ LQ can be converted to
limits on couplings to ${\tilde u}^j$ via
$\lambda_{eq_1}\sqrt{B_{\lepton q_2}}=\lambda_{1j1}'\sqrt{B_{ijk}'}$,
where $i$ and $k$ are the generations of $\lepton$ and $q_2$.

If the stop ($\sTop$) \cite{ref:stop} is lighter than the top quark,
then the $R_P$-conserving decay $\sTop\rightarrow t\chi_1^0$ will not
exist. In the case of $\sTop$, the correspondence with the coupling
limit on $\tilde{S}_{1/2}^L$ is given by $\lambda_{ed}\sqrt{B_{\lepton
q_2}}=\cos\theta_t\lambda_{131}'\sqrt{B_{i3k}'}$ where $\theta_t$ is
the mixing angle between the SUSY partners of the left- and
right-handed top quarks. Over a broad range of possible stop masses,
it is expected that $\cos^2\theta_t\sim0.5$ \cite{ref:stop}.

\epsfigure[width=0.95\hsize]{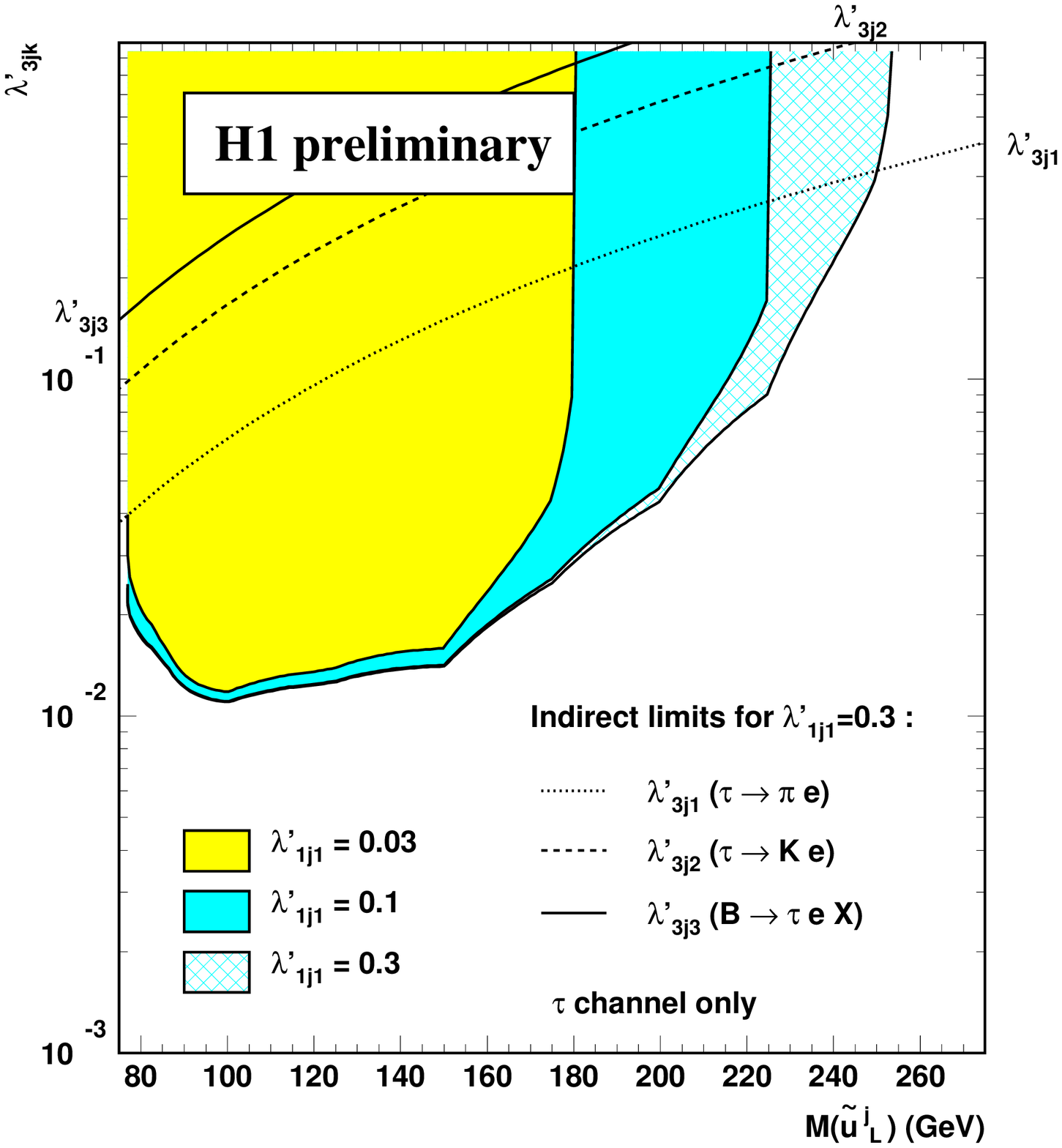} {Upper limits at the
  95~\% CL for the coupling $\lambda_{3jk}'$ as a function of squark
  mass for different values of $\lambda_{1j1}'$.  The best indirect
  limits are also shown.}  {H1_LFV2}
  
The ZEUS~\cite{ref:ZEUS_LFV} and H1~\cite{ref:H1_Rpvio_2}
collaborations have set limits on $\rpvio$-squarks leading to LFV.
Table~\ref{tab:LFV_squark} gives the lower limits from the ZEUS
experiment for $\sdown$, $\suup$ and $\sTop$ quarks assuming that the
couplings at the production and decay vertex are equal to the
electromagnetic coupling.  The H1 collaboration has set limits on
$\lambda_{3jk}$ assuming different values for $\lambda_{1j1}$, and
assuming that gauge decays of squarks are forbidden.  These limits are
shown in Fig.~\ref{fig:H1_LFV2}.  The limits from indirect searches
\cite{ref:Davidson} are also given in the figure, and are seen to
be substantially weaker.  The Tevatron limits, resulting from searches
for $\tau\tau bb$ final states~\cite{ref:CDF_LFV} and
for $\nu\nu bb$ final states~\cite{ref:D0_LFV} extend to about
$99$~GeV.  The mass limits from HERA are substantially better
than those achieved by previous experiments, particularly in the case
where the $\tau$ is involved

\subsection{Search for excited fermions}
\label{sec:estar}

\epsfigure[width=0.8\hsize]{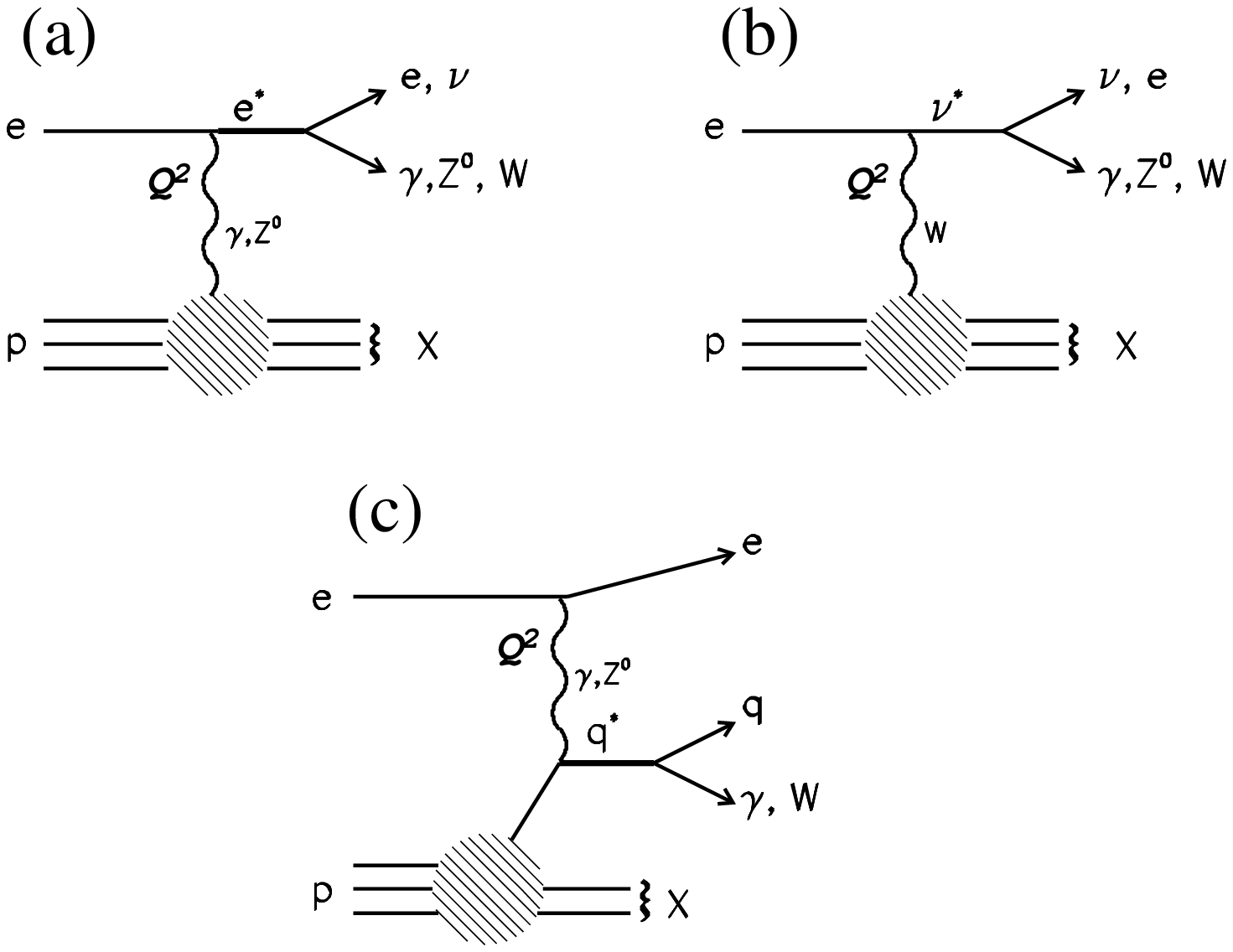} {Diagrams for
  the production of a) excited electrons, b) excited neutrinos, and c)
  excited quarks in ep collisions.  Only some of the possible decay
  modes are shown.}  {fstar}

The existence of excited fermions would be direct evidence for a new
level of substructure.  Excited electrons, quarks, or neutrinos could
be potentially produced at HERA by $\gamma, Z$ or $W$ exchange.  The
decay products of these excited particles would include these gauge
bosons.  The search procedure therefore involves identifying either
photons or weak bosons in the final state.  Example diagrams for
production and decay of excited fermions are shown in
Fig.~\ref{fig:fstar}.

The production cross section for excited fermion production via
$\gamma$ exchange has the form~\cite{ref:hagiwara}
\begin{equation}
\sigma(ep \rightarrow f^*X) = \frac{\mid c \mid^2+\mid d \mid^2}{
\Lambda^2}\sigma_0(M_f^*) \; ,
\end{equation}
where $c$ is the vector coupling constant, $d$ the axial vector
coupling constant, $\Lambda$ the compositeness scale and $\sigma_0$ is
a reference cross section.  The production cross section via $Z$ and
$W$ exchange also include terms of the form $(cd^*+c^*d)/\Lambda^2$.
The agreement between the precise measurements of electron/muon $g-2$
and theoretical predictions implies that $|c|=|d|$ for compositeness
scales less than 10-100 TeV~\cite{ref:cdlim1,ref:cdlim2,ref:cdlim3}.
In other cases, specific models allow to relate these coupling
constants, so that the cross sections can be described by a single
parameter $f/\Lambda$, with dimensions GeV$^{-1}$.  Both HERA
experiments have set limits on excited fermion
production~\cite{ref:ZEUS_estar2,ref:ZEUS_estar3,ref:H1_estar2,ref:H1_estar3}.

ZEUS and H1 have searched for the following decay modes:
\begin{equation}
\begin{array}{ccccccccc}
e^* & \rightarrow & e\gamma & q^* \rightarrow & q\gamma & \nu^* & \rightarrow & \nu\gamma \\
    &             & eZ      &                 &         &      &             & \nu Z  \\
    &             & \nu W      &              & qW      &      &             & e W  \\
\end{array}
\end{equation}

The expected signature is often striking, and the event selection is
therefore very clean.  For example, excited electrons are expected to
be produced elastically 50~\% of the time.  The branching ratios are
model dependent, but it is expected that $e^*$ decays to $e\gamma$ a
significant fraction of the time.  This would produce a signature in
the detector consisting only of a high energy electron and a high
energy photon.

The search strategy depends on the individual decay mode.  E.g., the
search for $e^* \rightarrow e\gamma$ required two electromagnetic
clusters in the detector each with transverse energy $E_T>15$~GeV.
The searches for final states with $Z$ or $W$ bosons required large
transverse energy and invariant mass in the hadronic system.  For
instance, in the ZEUS search, the mode $e \rightarrow eZ$ required, in
addition to the electron candidate ($E_T^{had}>60$~GeV and
$M^{had}>60$~GeV) or ($E_T^{had}>70$~GeV and $M^{had}>40$~GeV).  The
H1 collaboration has specifically looked for leptonic decays of the
$Z$ in $e \rightarrow eZ$.  The searches with neutrinos in the final
state required an imbalance of at least $15$~GeV ($20$~GeV) in the
transverse momentum of the reconstructed particles in the searches
performed by ZEUS(H1).

\epsfigure[width=0.8\hsize]{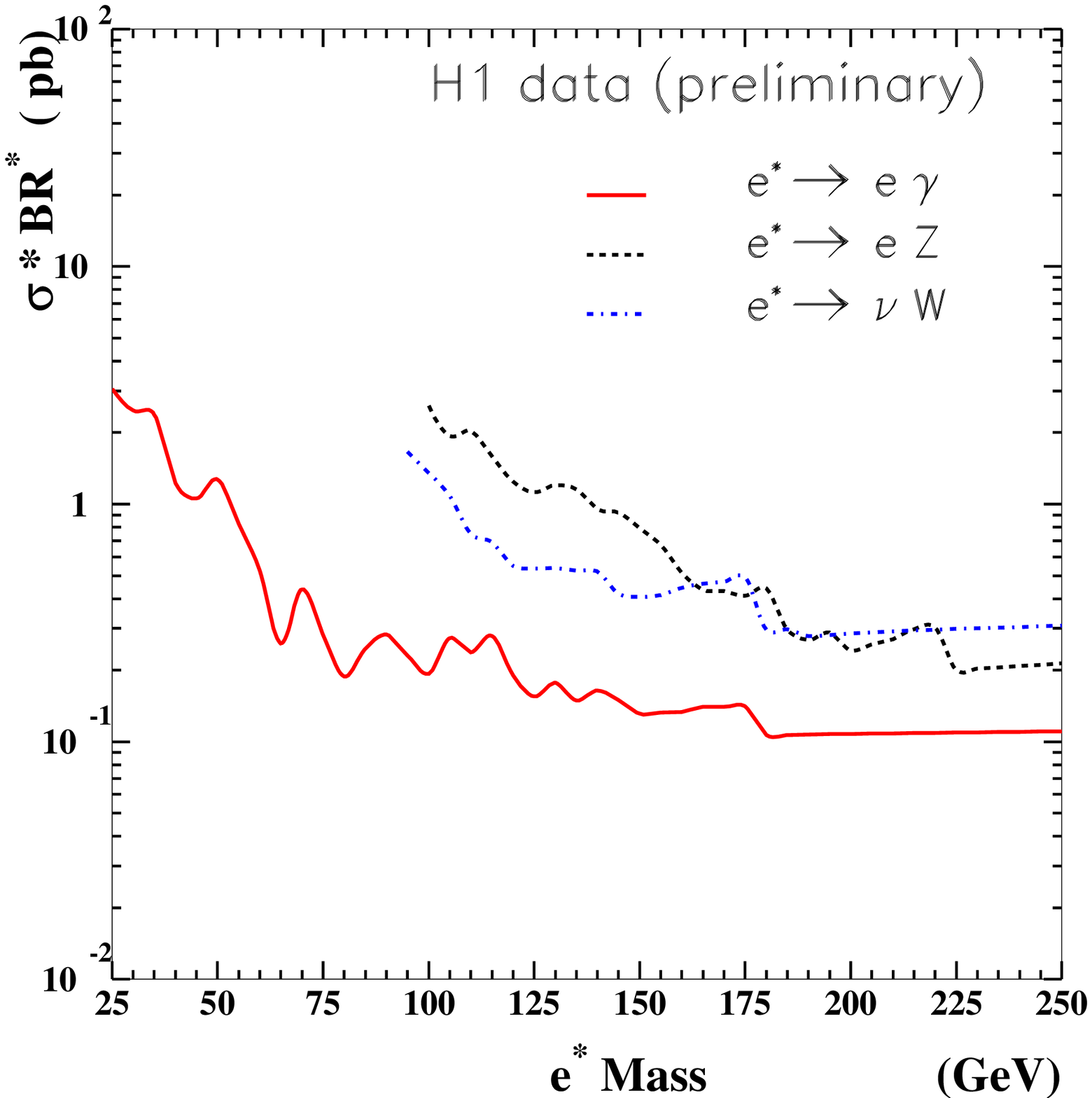} {Upper limits at the
  95~\% CL from the H1 collaboration on the product of the production
  cross section (in pb) and the branching ratio for excited electron
  production.}  {fstarlim}

The latest limits attained by the H1 collaboration on the product of
the production cross section times the branching ratio for excited
electrons are summarized in Fig.~\ref{fig:fstarlim}. These limits are
based on $37$~pb$^{-1}$ of $e^+p$ data accumulated from 1994-97.
Similar limits for the decay $e^*\rightarrow e\gamma$ exist from
ZEUS. The limits attained for excited neutrinos or squarks are roughly
a factor of two worse than those for excited electrons due to the more
difficult experimental search procedure.  For final states with
photons, the sensitivity is approximately $0.1$~pb.  These cross
section limits are independent of the particular model chosen.

Results obtained at
LEP~\cite{ref:ALEPH_estar,ref:DELPHI_estar,ref:L3_estar,ref:OPAL_estar}
and the Tevatron~\cite{ref:CDF_qstar,ref:D0_qstar} can be compared to
the HERA results for the quantity $f/\Lambda$.  The LEP limits for
$e^*$ and $\nu^*$ are more stringent by about a factor of two than
those from HERA, but are limited to masses of $180$~GeV.  The results
on $q^*$ from the Tevatron are very strong, but require a nonzero
strong coupling for the excited quarks. The HERA limits are therefore
complementary.

\subsection{Search for right handed currents}
\label{sec:heavyl}
  One of the particuliarities of the SM is that it is not left-right
symmetric.  It does not contain right-handed gauge bosons or
neutrinos.  It is important to determine whether this is only a low
energy phenomena, with left-right symmetry restoration occurring at
higher energies. A possible extension of the SM which is left-right
symmetric is obtained via the replacement
\begin{equation}
SU(2)_L \times U(1)_Y \rightarrow SU(2)_L \times SU(2)_R \times U(1)' \; \; .
\end{equation} 
The relationship between the electromagnetic and weak coupling
constants are given in this theory by~\cite{ref:leftright}
\begin{equation}
\label{eq:leftright}
\frac{1}{e^2} = \frac{1}{g_L^2} + \frac{1}{g_R^2} + \frac{1}{g'^2} \; \; ,
\end{equation}
where $g_L, g_R$ are the couplings associated with the $W_L$ and
$W_R$, respectively, and $e=g_L \sin\theta_W$.  The coupling constant
$g'$ is associated with the $U(1)'$ generator.  In the SM, $e=g_Y
\cos\theta_W$, where $g_Y$ is the coupling constant of $U(1)_Y$.

Equation~\ref{eq:leftright} gives the constraint
\begin{equation}
g_R \geq g_L \tan \theta_W
\end{equation}
or, $g_R \geq 0.55 g_L$ .  The model can therefore be invalidated if
an experiment is sensitive to right-handed couplings below this value.

\epsfigure[width=0.8\hsize]{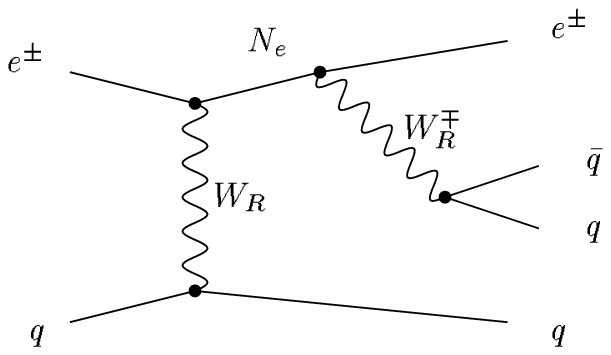} {Feynman diagram for
  the production of a heavy right-handed neutrino via the $t$-channel
  exchange of a right-handed $W$ in $e^{\pm}p$ collisions at HERA.
  The heavy neutrino subsequently decays into a $W^{\pm}$ and an
  $e^{\pm}$.}  {rightneutrino}

  The ZEUS collaboration has searched for heavy right-handed
neutrinos, $N_e$, produced in $e^-p$ and $e^+p$ collisions at HERA.
The Feynman diagram for the process is shown in
Fig.~\ref{fig:rightneutrino}.  Significant limits exist on
right-handed neutrinos and $W$'s from other
experiments~\cite{ref:ppright1,ref:ppright2,ref:rareright1,ref:rareright2},
but these are either very model dependent, or apply to heavy $W_R$'s.
In general, the parameter space given by
\begin{eqnarray}
M_{W_R} & < & 100 \; \; {\rm GeV} \; , \\
M_{W_R} & < & M_{N_e} \; ,
\end{eqnarray}
is largely unexplored by these limits.  The ZEUS analysis has focused on this 
region~\cite{ref:ZEUSright,ref:Larry}.

\epsfigure[width=0.8\hsize]{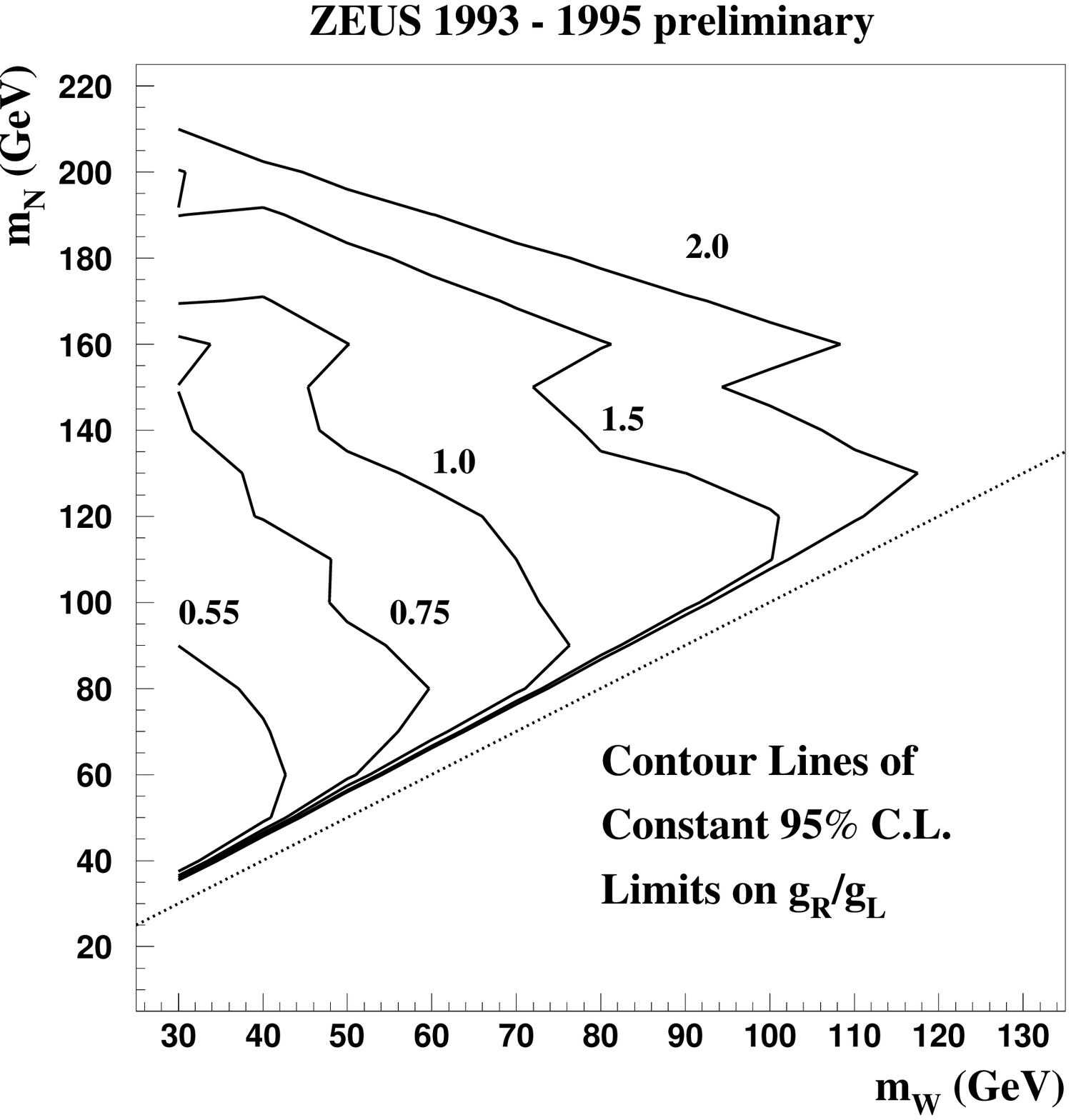} {The contours of
  constant 95~\% CL upper limits on $g_R/g_L$ in the $M_{W_R},M_{N_e}$
  plane.  The areas left of the curves are excluded.  For masses where
  ZEUS is sensitive to $g_R/g_L < 0.55$, the gauge structure
  $SU(2)_L\times SU(2)_R \times U(1)$ is ruled out.  The dotted line
  indicates the kinematic limit of the ZEUS search.}  {grlimits}

        The experimental search generally followed the lines of the NC
DIS analysis.  The heavy neutrino decays into a $W_R$ and an electron.
Evidence of the presence of a right-handed $W$ was searched for in the
jet-jet invariant mass spectrum, while the presence of the
right-handed neutrino was searched for in jet-jet-$e^{\pm}$ invariant
mass. A well reconstructed electron was required in the final state,
accompanied by at least $2$ jets. No significant signal was found, and
upper limits were placed on $M_{W_R}, M_{N_e}$ for different values of
$g_R/g_L$.  These limits are shown in Fig.~\ref{fig:grlimits}.  For
$M_{W_R} \leq 40$~GeV and $M_{N_e} \approx 60$~GeV, $g_R < 0.55 g_L$
is ruled out.  For $g_R = g_L$, $W_R$ masses up to $80$~GeV and $N_e$
masses up to $170$~GeV are ruled out.

\subsection{Search for new interactions}
\label{sec:newint}
In this section, we review the search for new interactions occurring
at high mass scales $\Lambda \gg \sqrt{s}$. These interactions
effectively appear as four-fermion contact interactions at HERA energy
scales, in analogy with Fermi's four-fermion contact interaction model
for the weak force. A contact interaction would produce a deviation of
the observed cross section, particularly at large $Q^2$, from the SM
expectations.

If the scale of the new physics is well beyond the center-of-mass
energy available, its effect can be parameterized as a four-fermion
interaction.  The simplest $lq\rightarrow lq$ contact interactions
which conserve $SU(3)\times U(1)$ can be represented as additional
terms to the Standard Model Lagrangian, $L_{SM}$, as


\begin{equation}
\begin{array}{lrlr}
L = L_{SM} + &{ {\displaystyle g^2} \over {\displaystyle \Lambda^2}} [ &
 \eta _s\left( {\bar e_Le_R} \right)\left( {\bar q_Lq_R} \right)+
\eta'_s\left( {\bar e_Le_R} \right)\left( {\bar q_Rq_L} 
\right)+h.c. & {\rm scalar}\mbox{} \\
& + & \eta _{LL}\left( {\bar e_L\gamma ^\mu e_L} \right)\left({\bar q_L\gamma_\mu q_L} 
\right)+\eta _{LR}\left({\bar e_L\gamma ^\mu e_L}\right)\left({\bar q_R\gamma _\mu q_R} 
\right) &    \\
  \mbox{} & + & \eta _{RL}\left( {\bar e_R\gamma ^\mu e_R} \right)\left(
{\bar q_L\gamma_\mu q_L} \right)+\eta _{RR}\left( {\bar e_R\gamma ^\mu e_R}
\right)\left({\bar q_R\gamma _\mu q_R} \right) &  {\rm vector} \\
  \mbox{} & + & \eta _T\left( {\bar e_L\sigma ^{\mu \nu }e_R} \right)\left(
{\bar q_L\sigma_{\mu \nu }q_R} \right)+h.c. ] , &   {\rm tensor}
\end{array}
\end{equation}

\noindent where $g$ is the coupling, $\Lambda$ is the effective mass 
scale, and $\eta$ determines the relative size and sign of the
individual terms.

Strong limits exist on the scalar and tensor
terms~\cite{ref:contlagrang}, so we focus here on the vector terms.
The effect of the vector interactions on the cross section is the
addition of an extra term to the vector and axial vector couplings for
each quark flavor (see Eq.~(\ref{eq:va})),

\begin{equation}
\begin{array}{lllll}
V^L&=&V_{SM}^L&+&{{Q^2} \over {8\pi \alpha }}{{g^2} \over {2\Lambda^2}}
       \left( {\eta _{LL}+\eta _{LR}}
\right) \; ,\\ 
  V^R&=&V_{SM}^R&+&{{Q^2} \over {8\pi \alpha }}{{g^2} \over {2\Lambda^2}}
       \left( {\eta _{RL}+\eta _{RR}}
\right) \; ,\\
  A^L&=&A_{SM}^L&+&{{Q^2} \over {8\pi \alpha }}{{g^2} \over {2\Lambda^2}}
       \left( {\eta _{LL}-\eta _{LR}}
\right) \; ,\\
  A^R&=&A_{SM}^R&+&{{Q^2} \over {8\pi \alpha }}{{g^2} \over {2\Lambda^2}}
       \left( {\eta _{RL}-\eta _{RR}}
\right) \; .\\
\end{array}
\label{eq:contact}
\end{equation}
Note that the contact terms enter only as a function of $Q^2$, and as
a consequence their relative contributions to the cross section at
fixed $Q^2$ are approximately independent of $x$.

A commonly used convention is to set $g=\sqrt{4\pi}$ and to
use $\eta$ as a sign parameter.  The relative strengths of the 
different terms is then placed in $\Lambda$.  In this convention,
we write the contact interaction Lagrangian (keeping only the vector
terms) as
\begin{equation}
\label{eq:contact2}
L_{CI} = \sum_{i,j=L,R} \frac{4\pi}{(\Lambda_{ij}^q)^2} \eta_{ij}^q
(\bar{e}_i\gamma_{\mu}e_i)(\bar{q}_j\gamma_{\mu}q_j) \; ,
\end{equation}
where $q$ represents the quark generation.  Assuming only valence
quarks participate, there are 16 different scales (2 each for
helicity of lepton, helicity of quark, quark flavor, and sign of $\eta$.)
The limits are then placed on $\Lambda_{ij}^q$.

\epsfigure[width=0.95\hsize]{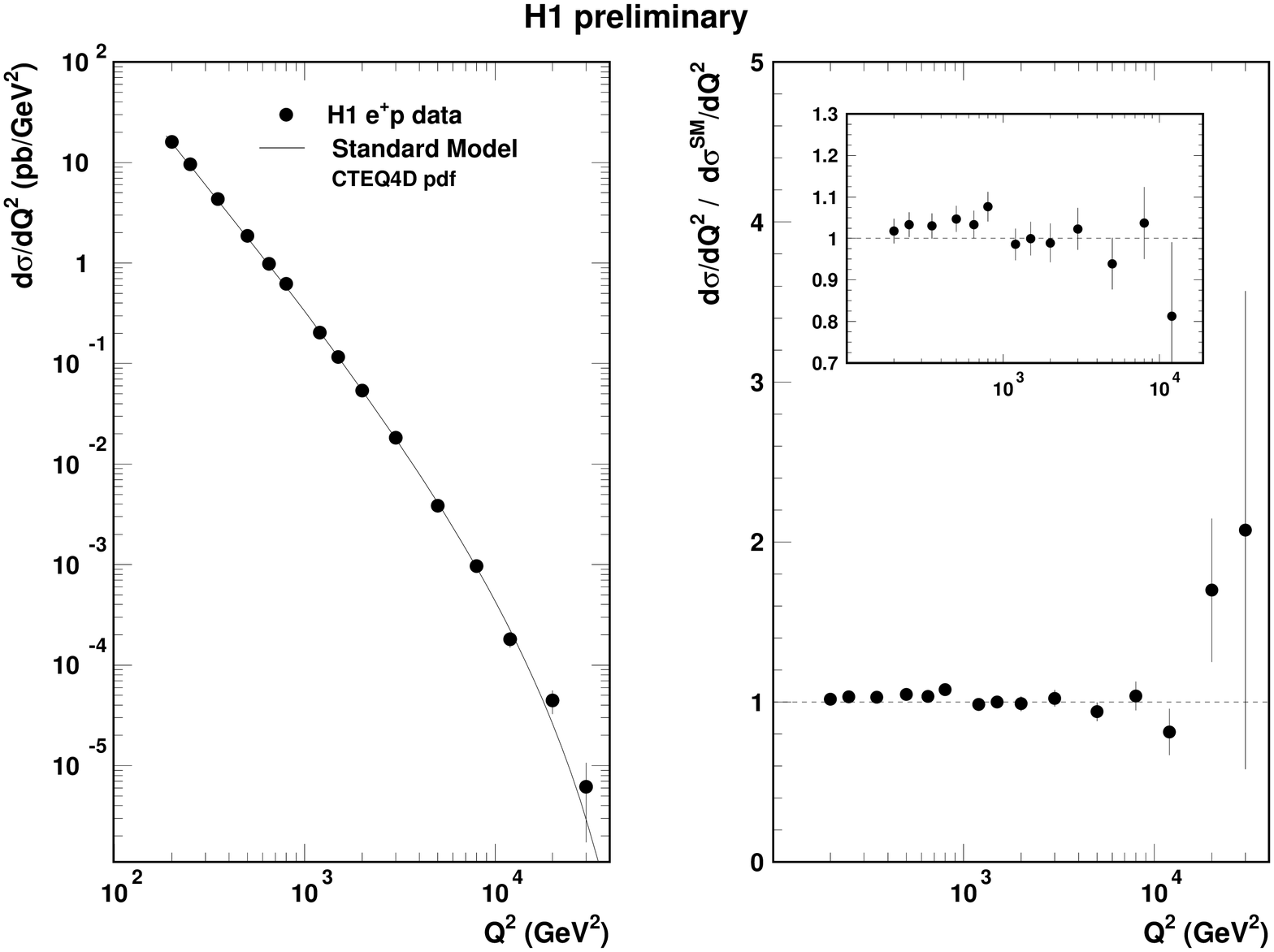} {The differential NC
  $e^+p$ cross section versus $Q^2$ compared with Standard Model
  expectations.  The right-hand plot gives the ratio to the
  expectations.}  {H1contact}

The H1~\cite{ref:H1_contact} and ZEUS~\cite{ref:ZEUS_contact}
collaborations have searched for contact interactions based on their
1994-97 data sets.  The measurement consists effectively of comparing
the measured differential cross section $d\sigma/dQ^2$ with SM
expectations.  Figure~\ref{fig:H1contact} shows the H1 measurement,
and the ratio to the SM expectation. No significant deviation is
found, and limits can be placed on the scale of possible new physics.
The limits attained by ZEUS and H1 are summarized in
Table~\ref{tab:contact} and compared to results from
LEP~\cite{ref:ALEPH_contact,ref:L3_contact,ref:OPAL_contact} and the
Tevatron~\cite{ref:CDF_contact}.  The limits are found to be
comparable in strength.

\begin{table} [p]
\tablecaption{
Selected limits on possible new contact interaction terms.
The class denotes the type of
interaction: $VV$, all $\eta^q_{ij}$ have the same sign, given by $+a$;
$AA$, $\eta^q_{LL}$ and $\eta^q_{RR}$  have sign $+a$, while 
$\eta^q_{LR}$ and $\eta^q_{RL}$ have sign $-a$; $VA$, $\eta^q_{iL}$ have
sign $+a$ while  $\eta^q_{iR}$ have sign $-a$; $X3$, $\eta^q_{LL}$ 
and $\eta^q_{RR}$ have sign $+a$ while other terms are zero; $X4$, the 
complement
of $X3$; and $U4$, where $\eta^u_{LR}$ and  $\eta^u_{RL}$ have sign $+a$ and
other terms are zero.}
\label{tab:contact}
\vskip 0.5 cm
\renewcommand{\arraystretch}{1.0}
\begin{center}   \begin{tabular}{  l | r | c | c | c | c | c | c }
     &     & {\bf Limit} & {\bf Limit} & Limit & Limit & Limit & Limit \\
     &     &  from     &   from      &  from &  from &  from   &  from \\
     &     &  H1       &   ZEUS      &  CDF  &  OPAL &  ALEPH  &  L3 \\
Class & $a$ & $\Lambda$(TeV) & $\Lambda$(TeV)
            & $\Lambda$(TeV) & $\Lambda$(TeV)
            & $\Lambda$(TeV) & $\Lambda$(TeV) \\
\hline
$VV$ & $+1$ & 4.5 & 4.9 & 3.5 & 3.3 & 4.0 & 3.2 \\
$VV$ & $-1$ & 2.5    & 4.6 & 5.2 & 4.3 & 5.2 & 3.9 \\
\hline
$AA$ & $+1$ &  2.0     & 2.0 & 3.8 & 4.9 & 5.6 & 4.3 \\
$AA$ & $-1$ &  3.8    & 4.0 & 4.8 & 3.1 & 3.7 & 2.9 \\
\hline
$VA$ & $+1$ &  2.6     & 2.8 & $-$ & $-$ & $-$ & $-$ \\
$VA$ & $-1$ &  2.8     & 2.8 & $-$ & $-$ & $-$ & $-$ \\
\hline
$X3$ & $+1$ &          & 2.8 & $-$ & 3.5 & 4.1 & 3.2 \\
$X3$ & $-1$ &          & 1.5 & $-$ & 2.9 & 3.6 & 2.8 \\
\hline
$X4$ & $+1$ &          & 4.5 & $-$ & 2.5 & 3.0 & 2.4 \\
$X4$ & $-1$ &          & 4.1 & $-$ & 4.1 & 4.9 & 3.7 \\
\hline
$U4$ & $+1$ &          & 4.6 & $-$ & 2.0 & 2.1 & 1.8 \\
$U4$ & $-1$ &          & 4.4 & $-$ & 2.3 & 2.6 & 2.2 
\end{tabular}   \end{center}
\end{table}

Other limits come from several sources.  Atomic parity violation
experiments place stringent limits on possible contact
interactions~\cite{ref:contact_APV}.  These can however be avoided if
$\eta_{LL} + \eta_{LR} - \eta_{RL} -\eta_{RR}=0$.  A comprehensive
review can be found in \citeasnoun{ref:Barger}.

\subsection{Observation of high $p_T$ lepton events}
\label{sec:lepton}

\epsfigure[width=0.9\hsize]{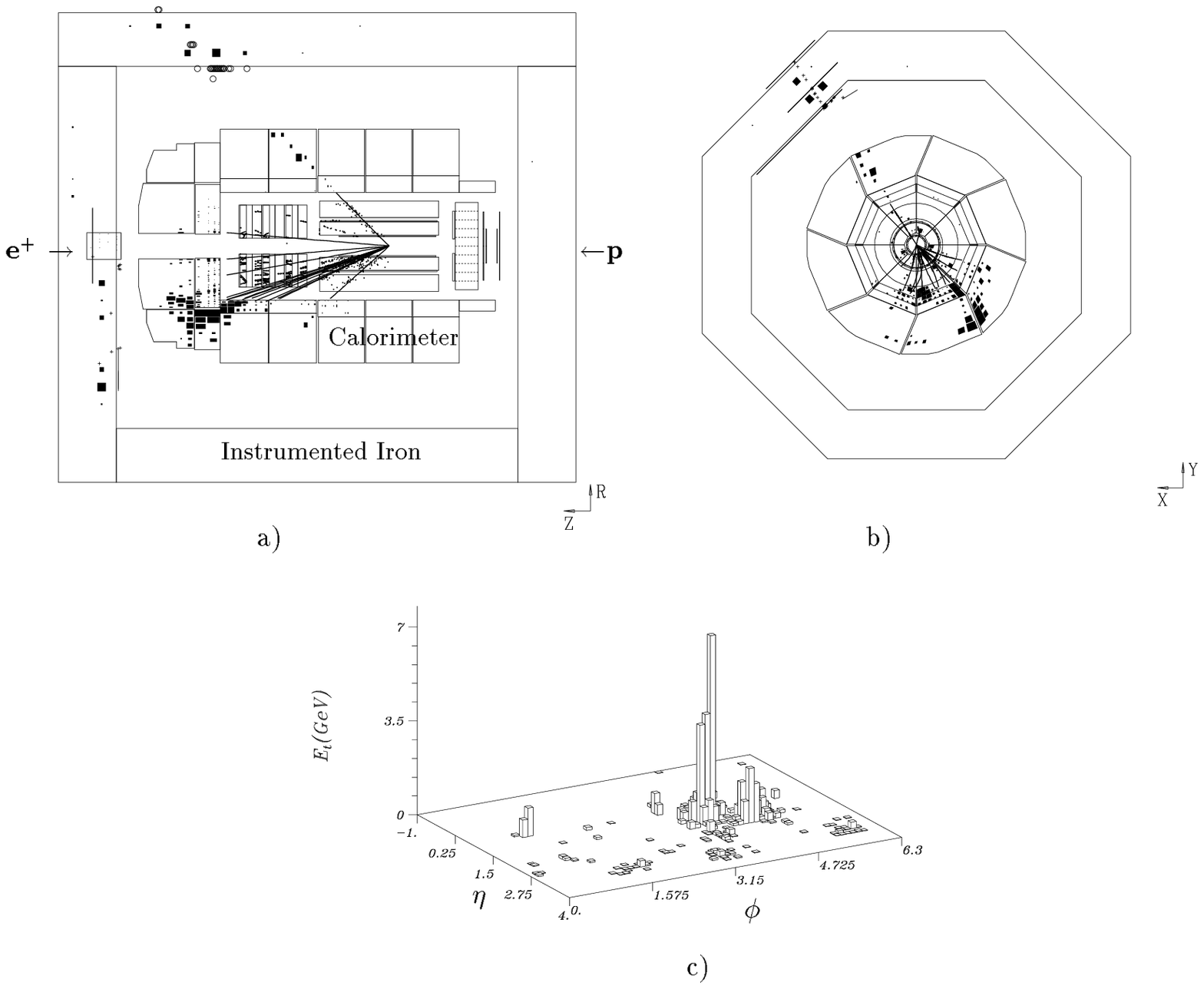} {Display of the
  $\mu^+$ event (MUON-1).  Shown are a side view (left) and a
  transverse view of the event.  The ``Legoplot'' shows the observed
  calorimetric energy depositions in the $\eta-\phi$ plane. } {H1muon}

The H1 collaboration has observed an excess of events consisting of
an isolated electron or muon of high transverse momentum and a
high $p_T$ jet.  The events also have a large imbalance in
transverse momentum.  One such event had previously been reported
based on the analysis of the 1994 data~\cite{ref:H1_muon1,ref:H1_Rpvio}.
This event is shown in Fig.~\ref{fig:H1muon}.  Analysis of
all the data from 1994-1997, giving an integrated luminosity of 
$36.5$~pb$^{-1}$, has resulted in the observation of an additional 5
events~\cite{ref:H1_muon3}.  The ZEUS collaboration has so far no
evidence for this type of event~\cite{ref:ZEUS_muon}.

\subsubsection{Analysis procedure}
The event selection follows the line of the standard H1
charged current event selection~\cite{ref:H1_CC}.  The
basic requirement is a large imbalance in the transverse
energy as measured in the calorimeter, $p_T^{calo}>25$~GeV.
Additionally, a tracking vertex is required, and topological
and timing filters are applied to remove backgrounds
from the sample.  This selection produces a clean sample of
charged current events, which contain typically one high
$p_T$ jet. Events are then searched for with tracks
with transverse momentum, $p_T>10$~GeV.  Six events
are observed where the track is far from the jet and any
other track in the event.  These tracks all correspond to
either an electron or a muon.  The event shown in
Fig.~\ref{fig:H1muon} is typical of these events.  The
events are very clean and well reconstructed.

\subsubsection{Results}
The event properties are listed in Table~\ref{tab:muon_events}.  Four
of the events contain a single high $p_T$ muon (MUON1, MUON2, MUON4,
MUON5), one contains a single high $p_T$ electron, and one contains a
high $p_T$ muon (MUON3) and a low $p_T$ electron. The five muon events
have hadronic transverse momentum ranging from $27 - 67$~GeV.  The
electron event has a hadronic $p_T$ of only $8$~GeV.

\begin{table}
\tablecaption{Reconstructed event kinematics for H1 events containing a
high $p_T$ lepton and  large missing 
transverse momentum . Energies, momenta and
masses are given in GeV and angles in degrees. For the  charge of the
high-$P_T$ lepton the significance of the determination is  given.
In case of event MUON-5 $2 \sigma$ limits are quoted for the muon momentum 
and derived quantities.}
\label{tab:muon_events}
\begin{center}                
{\tiny
\begin{tabular} {l|c|c|c|c|c|c}
 & {\bf ELECTRON} & {\bf MUON-1} & {\bf MUON-2} & {\bf MUON-3} *) &
{\bf MUON-4} & {\bf MUON-5} \\ 
 \hline
\multicolumn{7}{l}{}\\
\multicolumn{7}{c}{\bf The isolated high-$P_T$ lepton }\\
\multicolumn{7}{l}{}\\
\hline &&&&&&\\
Charge &Neg.( $ 5\sigma$) &  Pos.( $ 4\sigma$)&Pos.( $ 4\sigma$)&Neg.( $ 4\sigma$)&Neg.( $ 2\sigma$)&unmeasured\\  &&& &&&\\           
  $P_T^l$&$37.6^{+1.3}_{-1.3}$  &  $23.4^{+7.5}_{-5.5} $ &$28.0^{+8.7}_{-5.4}$&
$38.6^{+12.0}_{-7.4}$&$81.5^{+75.2}_{-26.4}$&$ > 44$
\\ &&& &&&\\
 $\theta^l$ & $27.3\pm 0.2$&  $ 46.2 \pm 0.1 $ &$28.9\pm 0.1$& $35.5 \pm 0.1$&$28.5 \pm 0.1 $ &$31.0\pm 0.1$\\
&&& &&&\\
 \hline
\multicolumn{7}{l}{}\\
\multicolumn{7}{c}{\bf The hadronic system }\\
\multicolumn{7}{l}{}\\
\hline &&&&&&\\
$P_T^X$&$8.0\pm0.8$&$42.2\pm3.8$&$67.4\pm5.4$&$27.4\pm2.7$&$59.3\pm5.9$&$30.0\pm3.0$\\ &&& &&&\\
$P_{\|}^X$ &$-7.2\pm0.8$&$-42.1\pm3.8$&$-61.9\pm4.9$&$-12.5\pm2.1$&$-57.0\pm5.5$&$-28.6\pm3.1$\\ &&& &&&\\
$P_{\bot}^X$ &$-3.4\pm0.9$&$-2.7\pm1.8$&$26.8\pm2.7$&$-24.3\pm2.5$&$-16.3\pm3.2$&$-9.1\pm2.3$\\ &&& &&&\\
$P_z^X$ &$79.9\pm4.4$&$153.1\pm9.1$&$247.0\pm18.9$&$183.7\pm13.6$&$118.9\pm12.1$&$145.4\pm8.2$\\ &&& &&&\\
$E^X$ &$81.1\pm4.5$&$162.0\pm10.0$&$256.9\pm19.5$&$186.8\pm14.0$&$141.7\pm13.7$&$154.8\pm9.1$\\ &&& &&&\\
\hline
\multicolumn{7}{l}{}\\
\multicolumn{7}{c}{\bf Global properties }\\
\multicolumn{7}{l}{}\\
\hline &&&&&&\\
 $ P_T^{miss}$ &$30.6\pm 1.5$& $18.9^{+6.6}_{-8.3}$ &$43.2^{+6.1}_{-7.7}$&$42.1^{+10.1}_{-5.9}$&$29.4^{+71.8}_{-13.9}$&$>18$\\ &&& &&&\\
  $\delta $&$10.4\pm 0.7$
 & $18.9^{+3.9}_{-3.2} $&$17.1^{+2.5}_{-1.7}$ &$26.9^{+4.2}_{-2.9}$&$43.5^{+19.3}_{-7.2}$&$>22$\\ &&& &&&\\
 $M_T^{l\nu}$ &$67.7\pm 2.7$&$3.0^{+1.5}_{-0.9}$ & $22.8^{+6.7}_{-4.2}$&$75.8^{+23.0}_{-14.0}$
&$94^{+157}_{-54}$&$>54$\\
 &&&&&&\\
\hline
\multicolumn{7}{l}{}\\
\multicolumn{7}{c}{$^{*)}$ Positron in MUON-3 :}\\
\multicolumn{7}{c} { $P_T^e = 6.7\pm 0.4$ ,
 $P_{\|}^e = 6.1 \pm 0.4$,
 $P_{\bot}^e = -2.8\pm 0.2$ , $P_z^e = -3.7 \pm 0.2 $}  \\ 
\multicolumn{7}{c}{}
\end{tabular}
}
 \end{center}
\end{table}

The main process expected to yield high $p_T$ leptons in events with
missing transverse energy is photoproduction of $W$ bosons, with
subsequent leptonic decays.  The $W$ is predominantly produced
radiatively from the scattered quark.  This process is calculable in
LO and has been quantitatively evaluated~\cite{ref:Wprod}, yielding a
total cross section of about $70$~fb per lepton species and charge.
The hadronic $p_T$ expected from $W$ production is generally small,
while the transverse lepton-$\nu$ mass has a Jacobian peak near the
$W$ mass.

\epsfigure[width=0.95\hsize]{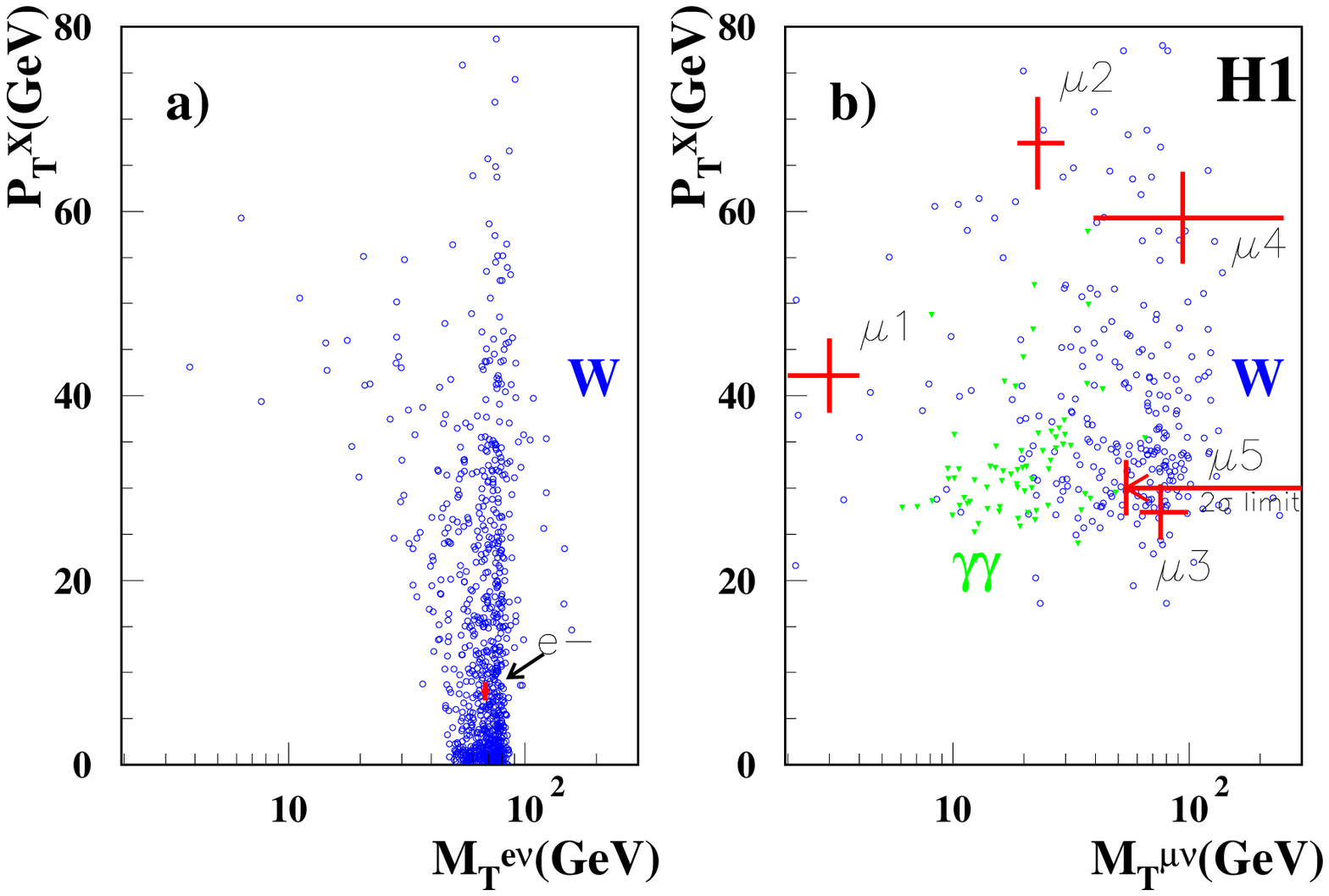} {Distribution of the
  six events in $p_T^X$ and $M_T^{l\nu}$:a) electron channel; b) muon
  channel.  The crosses correspond to the 1-sigma uncertainty on the
  measured parameters of each event.  The points show the standard
  model expectation for $W$ production for an accumulated luminosity a
  factor $500$ higher than in the data.}  {PtMt}

The events are compared to expectations from a $W$ Monte Carlo sample
corresponding to a luminosity 500 times that of the data in
Fig.~\ref{fig:PtMt}.  The muon events MUON1, MUON2 and MUON4 are in
regions with small expectations.  After applying the analysis cuts, H1
expects $2.4\pm0.5$($0.8\pm0.2$) events in the $e^{\pm}$($\mu^{\pm}$)
channels respectively, of which $1.7\pm0.5$($0.5\pm0.1$) are expected
from the LO calculations for $W$ production.  Two of the observed
events are consistent with the $W$ interpretation (ELECTRON, MUON3). A
second muon event, MUON5, is compatible, within the large measurement
errors, with $W$ production.

The ZEUS experiment has reported on a search for these
events using a very similar analysis 
procedure~\cite{ref:ZEUS_muon}, but has found no electron or
muon events with large missing $p_T$ and large hadronic $p_T$.

\subsubsection{Discussion}

The events observed by the H1 collaboration are clean and well
reconstructed. The ZEUS detector has a very similar acceptance
and reconstruction capability for these types of events, and should
therefore also have seen a signal if there was a large cross section for
new physics.  The non-observation of events by ZEUS can be accomodated
with reasonable probability  by assuming that the
true expectation is roughly $1/2$ what H1 measures.  It remains to be
seen whether such a cross section can be reproduced by higher order
calculations for $W$ production, or whether new physics mechanisms beyond
the Standard Model are necessary.

\subsection{Observation of events at large $x$ and large $Q^2$}
\label{sec:excess}
The H1 and ZEUS Collaborations reported an excess of events at large
$x$ (or $M_{eq}$) and large $Q^2$ in early
1997~\cite{ref:H1_highQ2,ref:ZEUS_highQ2}.  In the case of H1, 12
neutral current events were observed for $Q^2>15000$~GeV$^2$ where
$4.71\pm0.76$ were expected.  In the same $Q^2$ range, H1 observed $4$
charged current events where $1.77\pm0.87$ were expected.  For the NC
case, the excess is most prominent around $M_{eq}=200$~GeV, where 7
events were observed with $y>0.4$ in a mass window of width $25$~GeV,
where only $0.95\pm0.18$ events were expected.  ZEUS observed two
events for $Q^2>35000$~GeV$^2$ where $0.145\pm0.013$ were expected,
while for $x>0.55$ and $y>0.25$, four events were observed where
$0.91\pm0.08$ events were expected. The analyses, while each not
statistically compelling, generated interest because they occured in
previously unexplored kinematic regions.  At large $Q^2$, non-DIS
backgrounds are estimated to be below 1~\% and the uncertainty of the
SM predictions is below 10~\%.  The samples of events are background
free, and can be compared to precise predictions of the Standard
Model.  This is an ideal experimental situation.

\subsubsection{Interpretations of the HERA events}
The HERA observations have generated intense theoretical analysis of
possible underlying causes.  We briefly review a selection of these
analyses.
\begin{itemize}
\item
Several authors have attempted to modify the parton distributions at 
large-$x$ within the experimentally allowed range to see how big a
variation could be produced at large $x\; {\rm and}\; Q^2$.  
As mentioned above, ZEUS
and H1 concluded that the SM predictions were quite precise.  This was
based on existing fixed target measurements of structure functions,
conventional forms for the parton density parameterizations, and NLO DGLAP
evolution of the parton densities.  Some of these assumptions can be
questioned, such as the functional form of the quark densities at
large $x$.  For example, \citeasnoun{ref:Kuhlmann}
have added an extra component to the valence $u$-quark density of the
form $0.02(1-x)^{0.1}$, which would be compatible with existing
data.  While they find a $30$~\% effect for $Q^2=40000$~GeV$^2$ at
$x=0.7$, the effect in the $x$ range relevant for the HERA effect
($x=0.4-0.6$) is considerably smaller.  \citeasnoun{ref:Gunion} 
point out that intrinsic charm in the proton is
expected to give a contribution at large $x$, with a peak in the
vicinity of $M=200$~GeV.  However, existing data constrains the
cross section enhancement due to intrinsic charm to 15~\%.

The use of the NLO DGLAP evolution to predict the cross sections at
large $Q^2$ have also been called into question. \citeasnoun{ref:Kochelev}
argues that instanton induced quark-gluon interactions, a non-perturbative 
effect, will be proportionally more important at large $Q^2$.  The
cross sections estimated with the NLO DGLAP equations alone would
therefore be an underestimate, possibly by as much as 50~\%.

\item
The possibility of a leptoquark as the source of the events has been
considered by many authors.  For example,
see~\cite{ref:Bluemlein,ref:BuchmuellerWyler,ref:Altarelli}.  The
clustering of H1 events near $200$~GeV make the LQ interpretation an
intriguing possibility.  The combined ZEUS+H1 cross section for the
events near $200$~GeV is roughly
\begin{equation}
\sigma \approx \frac{10\;\rm{event}}{35\;\rm{pb}^{-1}} 
\approx 0.3 \rm{pb} \;\; .
\end{equation}
This can be used to estimate the size of the LQ coupling strength to the
lepton-quark pair, as discussed in section~\ref{sec:leptoquark}.  The cross
section is very strongly dependent on the quark or antiquark density 
involved.  We assume the production is off valence quarks given the large
suppression of sea quarks at the large values of $x$ involved.  Taking an
up quark density of about $0.15$ and a down quark density $4$ times smaller
(see Fig.~\ref{fig:highQ2_quarks}) yields a value of 
$\lambda\approx 0.025/\sqrt{B}$ for an $e^+u$ scalar LQ, where $B$ is the 
branching fraction into $e^+q$.  The corresponding coupling for an $e^+d$
LQ would be twice as large.

Note that LQ production can proceed via $q\bar{q}$ or $gg$ fusion at
the Tevatron independently of the coupling $\lambda$.  This allows
D0~\cite{ref:D0_LQ} and CDF~\cite{ref:CDF_LQ} to set limits on a
scalar LQ mass of $225$~GeV and $213$~GeV, respectively.  The bounds
assume a 100~\% branching ratio to $eq$, and are considerably reduced
for smaller branching ratios.  The cross section for vector LQ
production in $p\bar{p}$ collisions is model dependent, but is
expected to be much larger than that for scalar
production~\cite{ref:Bluemlein,ref:Altarelli}.  It is therefore
expected that vector LQ are unlikely.

\item
$\rpvio$ squarks have been proposed as an alternative mechanism for
producing a resonance near $200$~GeV (see, for example, 
\cite{ref:Altarelli,ref:Dreiner}).  
As described in section~\ref{sec:Rpvio}, $\rpvio$ squarks can be
produced by the fusion of a positron with a $d$-type quark to a
$u$-type antiquark.  Given the large $x$ values involved, we consider
only production from $d$-quarks.  $\rpvio$-squark production can be
differentiated from LQ production since different decay modes are
possible.  In particular, in addition to the Yukawa decay $\suup
\rightarrow e^+d$, the squark can also decay via gauge couplings
$\suup \rightarrow c\chi^0_{i(i=1,4)}, s\chi^+_{j(j=1,2)}$.

Strong limits exist from neutrinoless double beta
decay~\cite{ref:DBD1,ref:DBD2} which make the $\suup_L$ scenario very
unlikely.  The production of $\widetilde{t}_L$ is limited by atomic
parity violation~\cite{ref:APV}.  A detailed
analysis~\cite{ref:Altarelli} leads to the $\widetilde{c}_L$ squark as
the most likely candidate.  It is found that a branching ratio to
$e^+d$ near one is favored, so that any sizeable signal in the CC mode
would make this scenario unlikely.  No signal would be expected in
$e^-p$ collisions.  Also, other reactions such as $K^+\rightarrow
\pi^+\nu\bar{\nu}$ should become visible at experiments with increased
sensitivity.  This scenario should therefore be within experimental
reach very soon.

\item
Finally, contact interactions (see section~\ref{sec:newint})
representing the effect of some new interaction at a mass scale
$\Lambda\gg\sqrt{s}$ have been analyzed as a possible source of excess
events at large $Q^2$ by many authors.  In particular,
\citeasnoun{ref:Barger} have made a global study of all data relevant
to $eeqq$ contact interactions, including deep inelastic scattering
data, atomic parity violation experiments, polarized $e^-$ scattering
on nuclear targets, Drell-Yan production at the Tevatron, the total
hadronic cross section at LEP and neutrino-nucleon scattering
data. They conclude that contact interaction terms of the type
$\eta^{eu}_{LR}$ and $\eta^{eu}_{RL}$ give the best fit to the HERA
data while the $eeuu$ contact interactions are severely limited by the
Drell-Yan data.  Using a global fit to all the data, the authors find
that all contact interaction interpretations are strongly constrained.
\end{itemize}

Many more scenarios have been proposed beyond those described above.  Clearly,
more experimental data is needed to resolve which, if any, of these
approaches is correct, or if the effect is due to a statistical
fluctuation.

\subsubsection{Data sets}

Both collaborations have updated their analyses using the 1997
data. We review the status of the analyses as presented at the LP97
Symposium~\cite{ref:Bruce}.  These results from the H1 and ZEUS
collaborations include data up to the end of June, 1997.  The
corresponding integrated luminosities are $23.7$~pb$^{-1}$ for H1 and
$33.5$~pb$^{-1}$ for ZEUS.

\subsubsection{Charged current analysis}

\epsfigure[width=0.8\hsize]{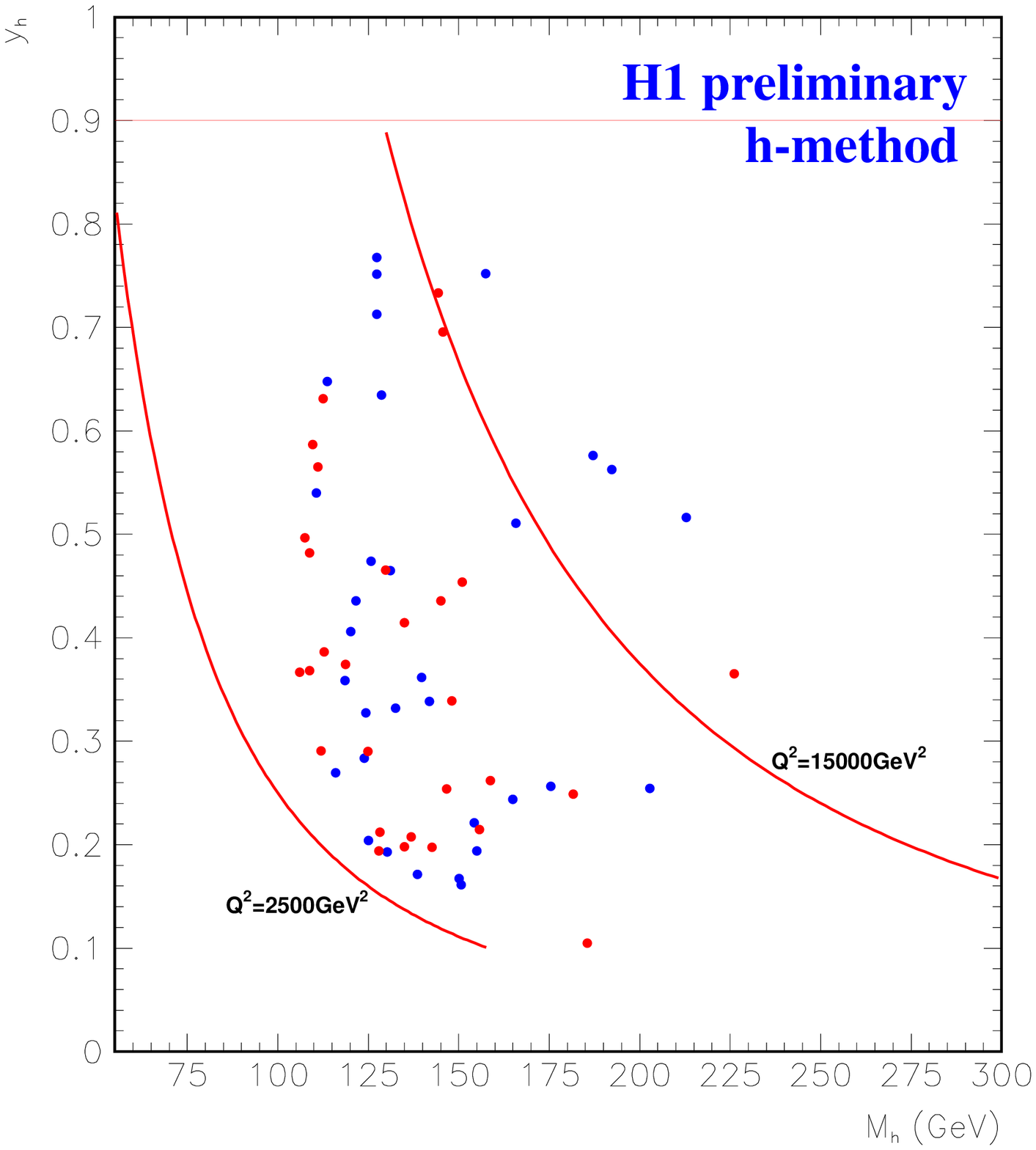} { The distribution
  in $M$ and $y$ of the H1 charged current event sample.  The curves
  show contours of constant $Q^2$.}  {H1_CC}

\epsfigure[width=0.8\hsize]{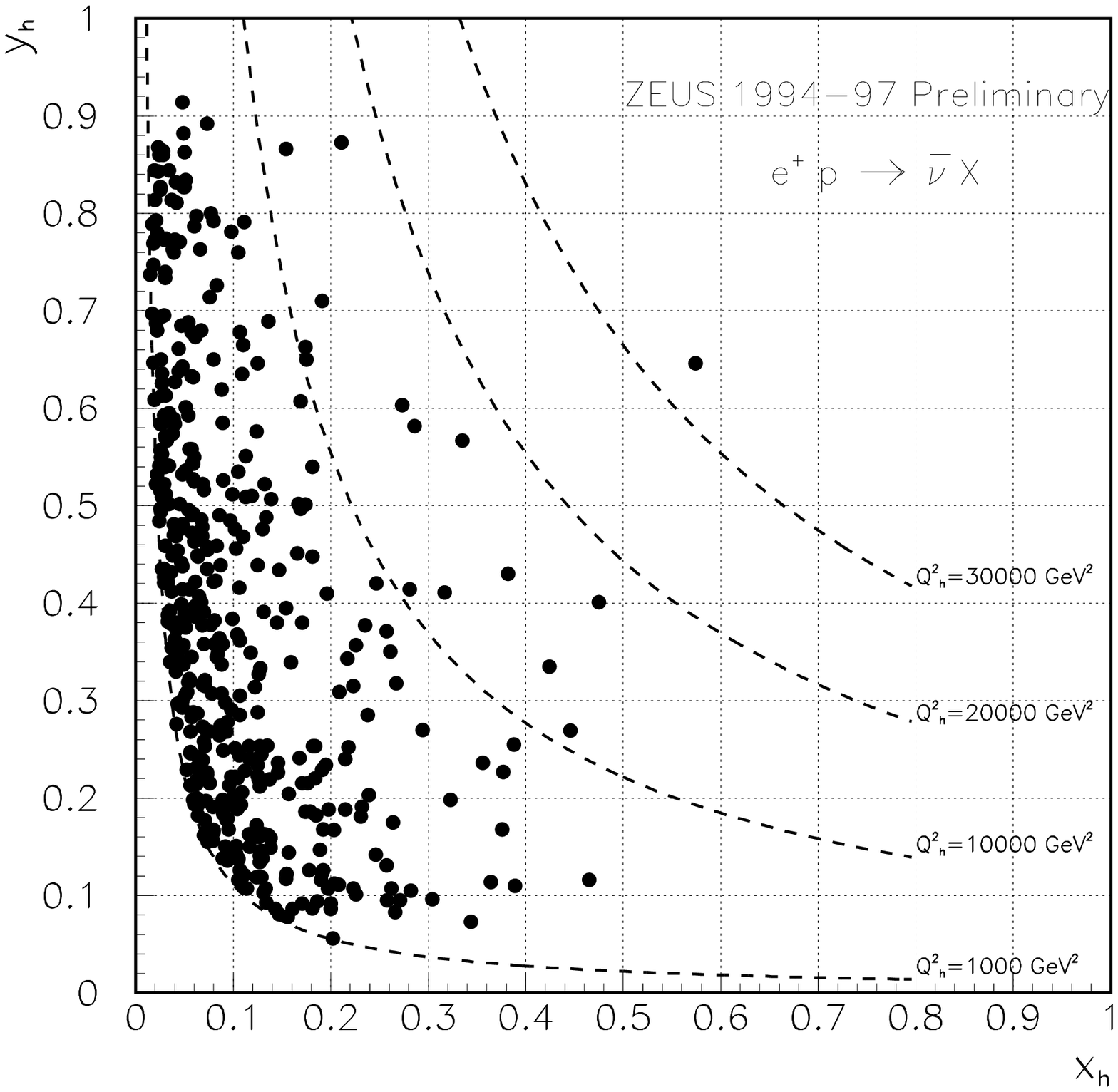} { The distribution
  in $x$ and $y$ of the ZEUS charged current event sample.  The curves
  show contours of constant $Q^2$.}  {ZEUS_CC}

The charged current (CC) analysis relies primarily on the selection of
events with large missing transverse momentum.  H1 requires events to
have at least $50$~GeV of missing $p_T$, while ZEUS requires
$p_T>15$~GeV and $Q^2>1000$~GeV$^2$.  With these cuts, the backgrounds
are found to be negligible.  The distribution of events observed by H1
is plotted in the plane of $y$ versus $M$ in Fig.~\ref{fig:H1_CC},
while the ZEUS event distribution in shown in the plane of $y$ versus
$x$ in Fig.~\ref{fig:ZEUS_CC}.  A typical large $Q^2$ CC event as
observed in the H1 detector is shown in Fig.~\ref{fig:CC_event}.  The
event consists of a single high energy jet and energy deposits from
the proton remnant.

\epsfigure[height=0.95\hsize,angle=90]{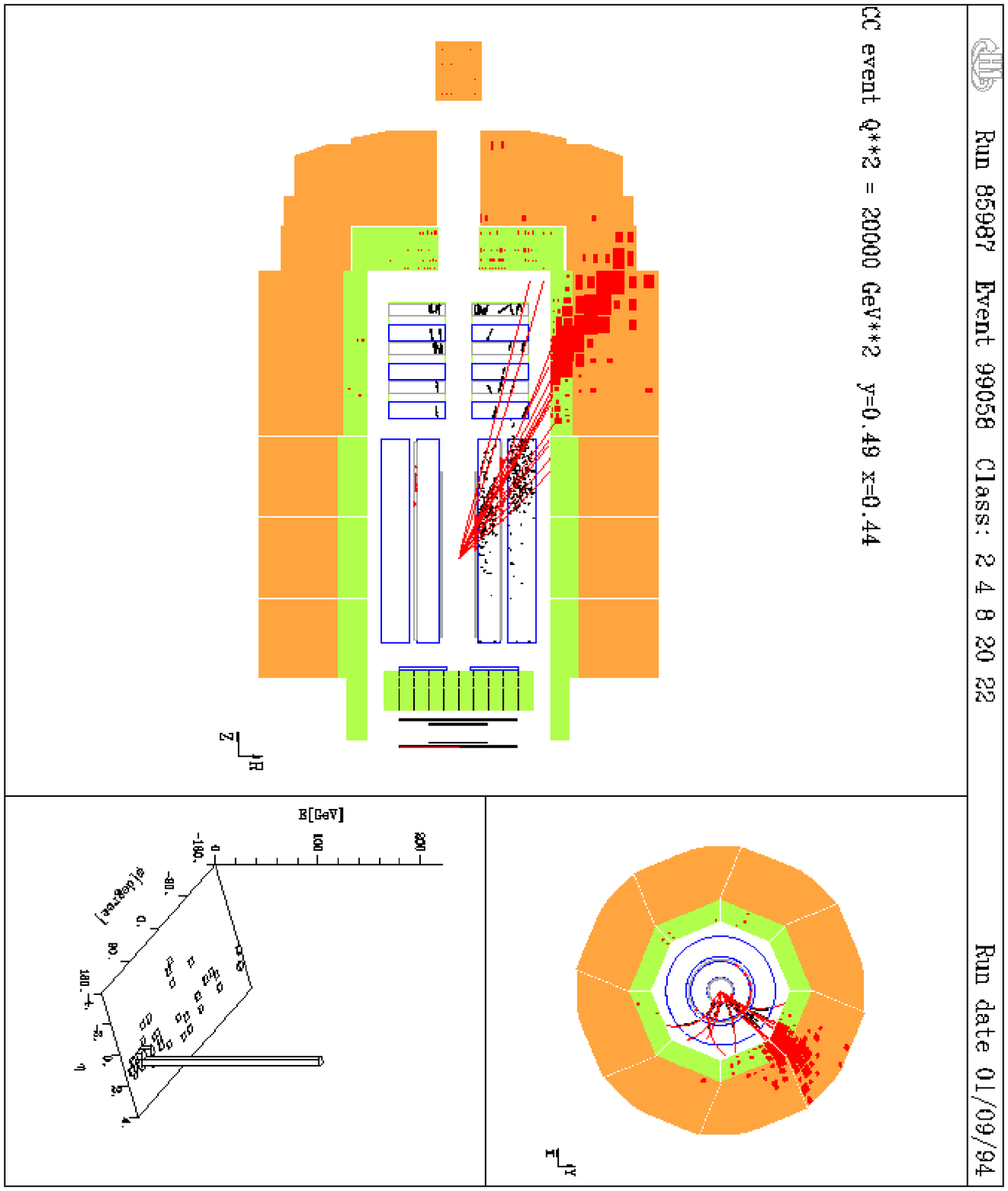} {A high $Q^2$
  charged current event, as observed in the H1 detector.  The left
  part shows the H1 inner tracking system and the calorimeter.  The
  filled rectangles in the calorimeter denote energy deposits which
  are above the noise thresholds.  The upper right display shows a
  projection onto a plane perpendicular to the beam axis, where only
  barrel calorimeter energy deposits are shown.  The lower right part
  of the figure shows the calorimeter transverse energy deposits in
  the $\eta-\phi$ plane.  The scattered quark jet dominates the event.
  The event parameters are given in the display.}  {CC_event}

The numbers of observed and expected events are compared in
Table~\ref{tab:CC_compare}. The data lie above expectations for
$Q^2>10000$~GeV$^2$.  For $Q^2>15000$~GeV$^2$, a total of $11$ events
are observed where $4.9\pm1.7$ are expected.  For $Q^2>30000$~GeV$^2$,
ZEUS observes one event where $0.03\pm0.04$ are expected.  Note that
the number of expected and observed events in ZEUS and H1 do not scale
with the luminosity at large $Q^2$.  This is due to differences in the
analysis.  The ZEUS data are partially unfolded for detector smearing,
using a CC MC, and this tends to reduce both the expectations and the
number of observed events at large $Q^2$.

\begin{table}
\tablecaption{Preliminary 1994-1997 results as of the Lepton-Photon 
97 Symposium
on the numbers of charged current events
observed ($N_{obs}$) by H1 and ZEUS above various $Q^2$ cuts
compared to SM expectations ($N_{exp}$).
The columns labeled $\delta N_{exp}$ give the uncertainty on the 
SM expectations.
For ZEUS separate uncertainties on the SM expectation are quoted 
for the uncertainty on the energy scale and the uncertainty
in the parton density functions.}\label{tab:CC_compare}
\renewcommand{\arraystretch}{1.2}
\begin{center}   \begin{tabular}{l||r|rl||r|rll} 
$Q^2$ cut  & \multicolumn{3}{c||}{H1}
           & \multicolumn{4}{c}{ZEUS} \\ \cline{2-8}
(GeV$^2$)  & $N_{obs}$ & $N_{exp}$ & $\pm\delta N_{exp}$
          & $N_{obs}$ & $N_{exp}$ & $\pm\delta N_{exp}$ &$\pm\delta N_{exp}$ \\
          &           &           &
          &           &           &       (E)           &      (PDF)    \\
 \hline
$Q^2 >1000 $ &    &      &          &455 & 419 & $\pm 13$            &$\pm33$    \\ \hline
$Q^2 >2500 $ & 61 & 56.3 &$\pm$ 9.40&192 & 178 & $\pm 13$            &$\pm17$    \\ \hline
$Q^2 >5000 $ & 42 & 34.7 &$\pm$ 6.90& 63 & 58.5& $\pm9.0$            &$\pm7.3$   \\ \hline
$Q^2 >10000$ & 14 & 8.33 &$\pm$ 3.10& 15 & 9.4 & $\pm2.5$            &$\pm1.6$   \\ \hline
$Q^2 >15000$ &  6 & 2.92 &$\pm$ 1.44&  5 & 2.0 & $^{+0.81}_{-0.54}$  &$\pm0.4$   \\ \hline
$Q^2 >20000$ &  3 & 1.21 &$\pm$ 0.64&  1 & 0.46& $^{+0.28}_{-0.16}$  &$\pm0.10$  \\ \hline
$Q^2> 30000$ &    &      &          &  1 & 0.034&$^{+0.037}_{-0.016}$&$\pm0.008$ 
\end{tabular}   \end{center}
\end{table}

\subsubsection{Neutral current analysis}
The neutral current (NC) event selection relies primarily on the
observation of an isolated high energy electron and conservation of
longitudinal ($E-P_Z$) and transverse momentum.  H1 considers events
for which $Q^2>2500$~GeV$^2$, while ZEUS starts at $Q^2>5000$~GeV$^2$.
Both collaborations conclude that the data sets are background free.

ZEUS and H1 have used different reconstruction methods.  H1 has chosen
the electron method as the primary method, with the double angle
method as check, while ZEUS has chosen the double angle method as
primary method, with the electron method as check.  The two methods
each give good resolutions at large $y$, with rapidly degrading
resolution as $y\rightarrow 0$.  The methods behave differently in the
presence of initial state QED radiation from the electron.  The double
angle method is more sensitive to ISR, and the $x$ is always
overestimated in the presence of radiation from the electron. These
effects are however expected to be small.  ZEUS achieves a resolution
of $5-10$~\% in $x$ and 5~\% in $Q^2$, while H1 finds a resolution of
about $2/y$~\% for $x$.

\epsfigure[width=0.8\hsize]{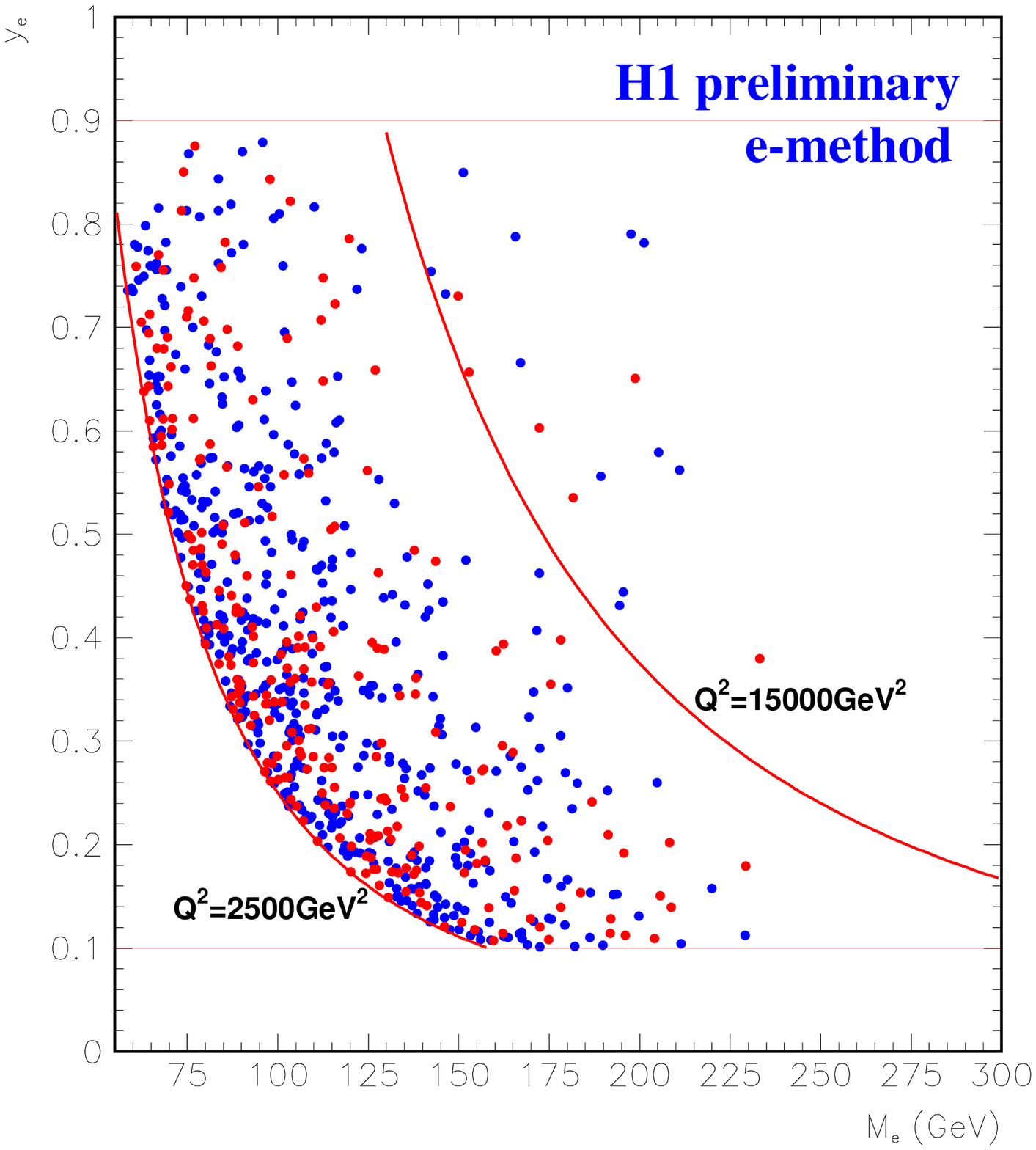} {The distribution in
  $M$ and $y$ of the H1 neutral current sample.  The curves show
  contours of constant $Q^2$.}  {H1_NC}

\epsfigure[width=0.8\hsize]{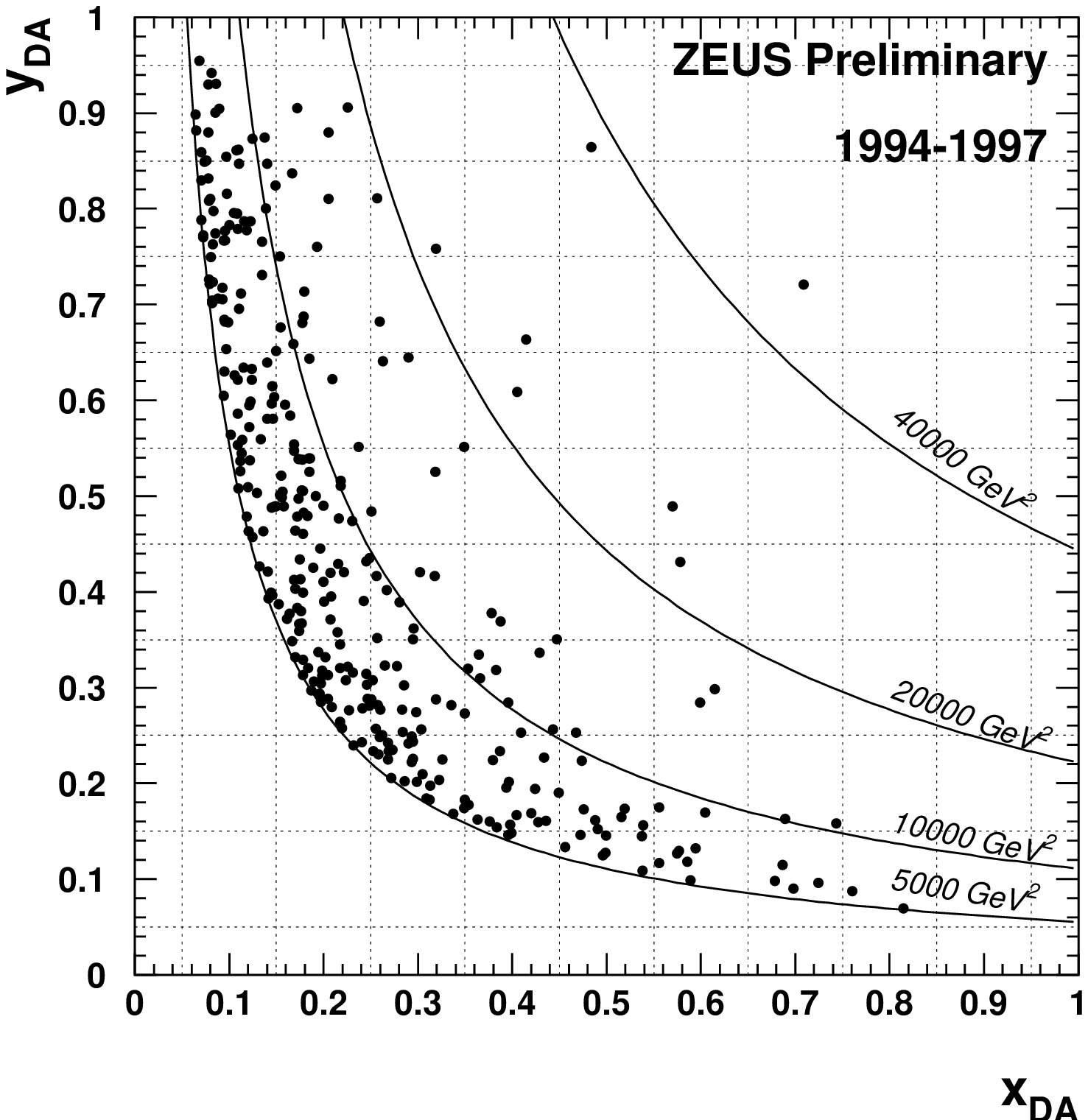} {The distribution in
  $x$ and $y$ of the ZEUS neutral current sample.  The curves show
  contours of constant $Q^2$.}  {ZEUS_NC}

The distribution of events measured by the H1 collaboration are shown
in the $M,y$ plane in Fig.~\ref{fig:H1_NC}, while the ZEUS results are
shown in the $x,y$ plane in Fig.~\ref{fig:ZEUS_NC}.  A typical large
$Q^2$ event as measured in the ZEUS detector is shown in
Fig.~\ref{fig:NC_event}.

\epsfigure[height=0.95\hsize,angle=-90]{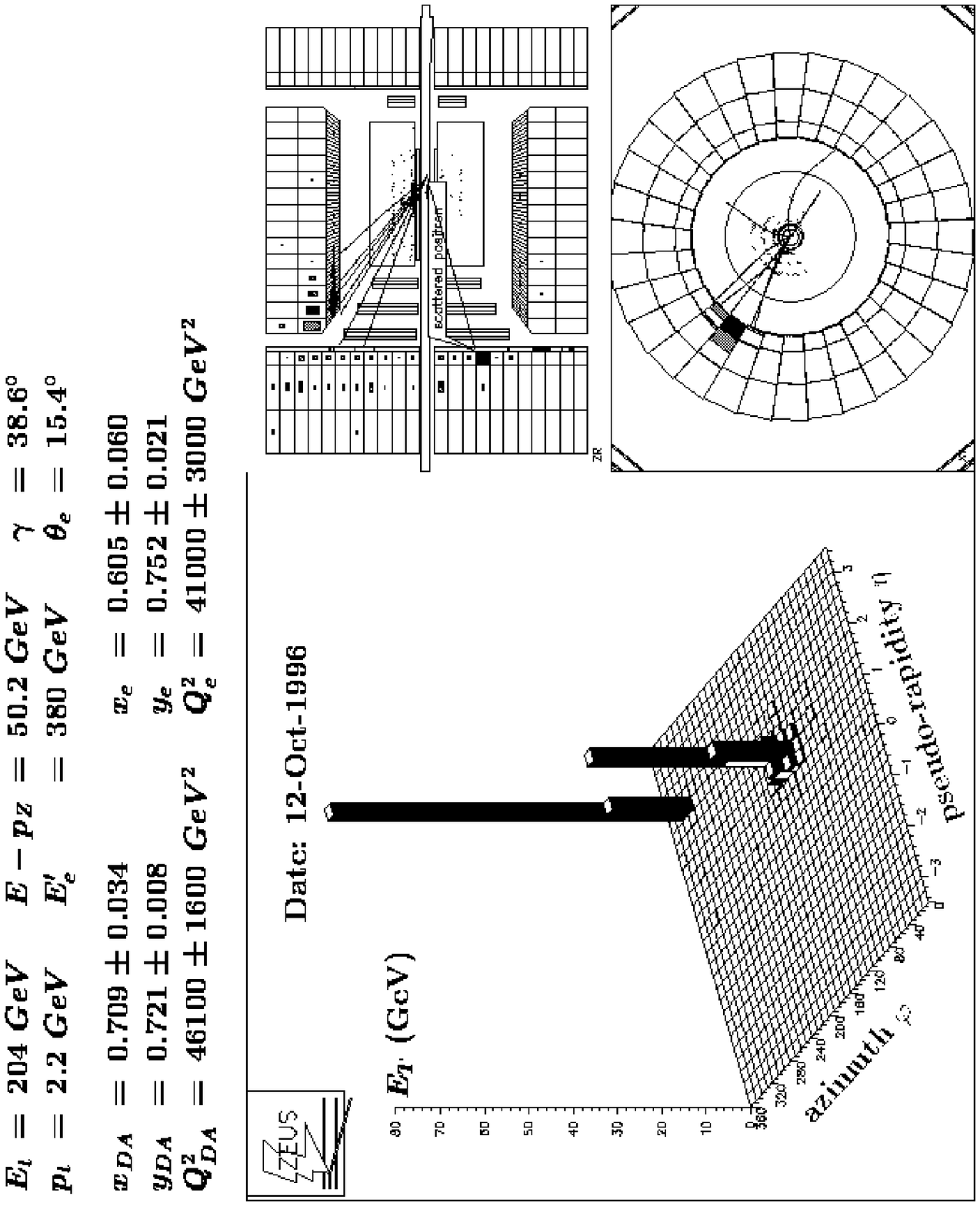} {The
  largest $Q^2$ neutral current event recorded at HERA to date, as
  observed in the ZEUS detector.  The top right part shows the ZEUS
  inner tracking system and the calorimeter.  The filled rectangles in
  the calorimeter denote energy deposits which are above the noise
  thresholds.  The bottom right display shows a projection onto a
  plane perpendicular to the beam axis, where only barrel calorimeter
  energy deposits are shown.  The left part of the figure shows the
  calorimeter transverse energy deposits in the $\eta-\phi$ plane, and
  demonstrates that the electron and the scattered quark jet dominate
  the event.  The event parameters are given above the event display.}
{NC_event}

\begin{table}
\tablecaption{Number of events observed ($N_{obs}$) by H1 
compared to SM NC DIS expectations ($N_{exp}$)
in several kinematic regions.
The first term in each sum gives the 1994-1996 value and the second term
gives the 1997 value as of the Lepton-Photon 97 Symposium.}
\label{tab:H1_NC}
\begin{center}   \begin{tabular}{c|c|c} 
region    & $N_{obs}$ & $N_{exp}$  \\ \hline
$Q^2> 2500$\, GeV$^2$    & 443+281=724 &  427+287=714$\pm$69     \\ \hline
$Q^2>15000$\, GeV$^2$    & 12+6=18     &  4.7+3.3=8.0$\pm$1.2    \\ \hline
$187.5<M<212.5$\, GeV    & 7+1=8       & 0.95+0.58=1.53$\pm$0.29 \\ 
and  $y>0.4$             &             &                         
\end{tabular}   \end{center}
\end{table}

The numbers of events in different kinematic regions as measured by H1
are summarized in Table~\ref{tab:H1_NC}.  The numbers of observed and
expected events are given separately for the 1994-96 running period,
and for the partial 1997 data set.  As is seen, the excess in the mass
window centered on $200$~GeV did not increase in significance with the
addition of the 1997 data.  However, the number of large $Q^2$ events
continued to accumulate at a rate higher than expected from the
SM. The $Q^2$ and $M$ distributions as measured by H1 are given in
Fig.~\ref{fig:H1_NC_Q2} and Fig.~\ref{fig:H1_NC_M}.  

\epsfigure[width=0.8\hsize]{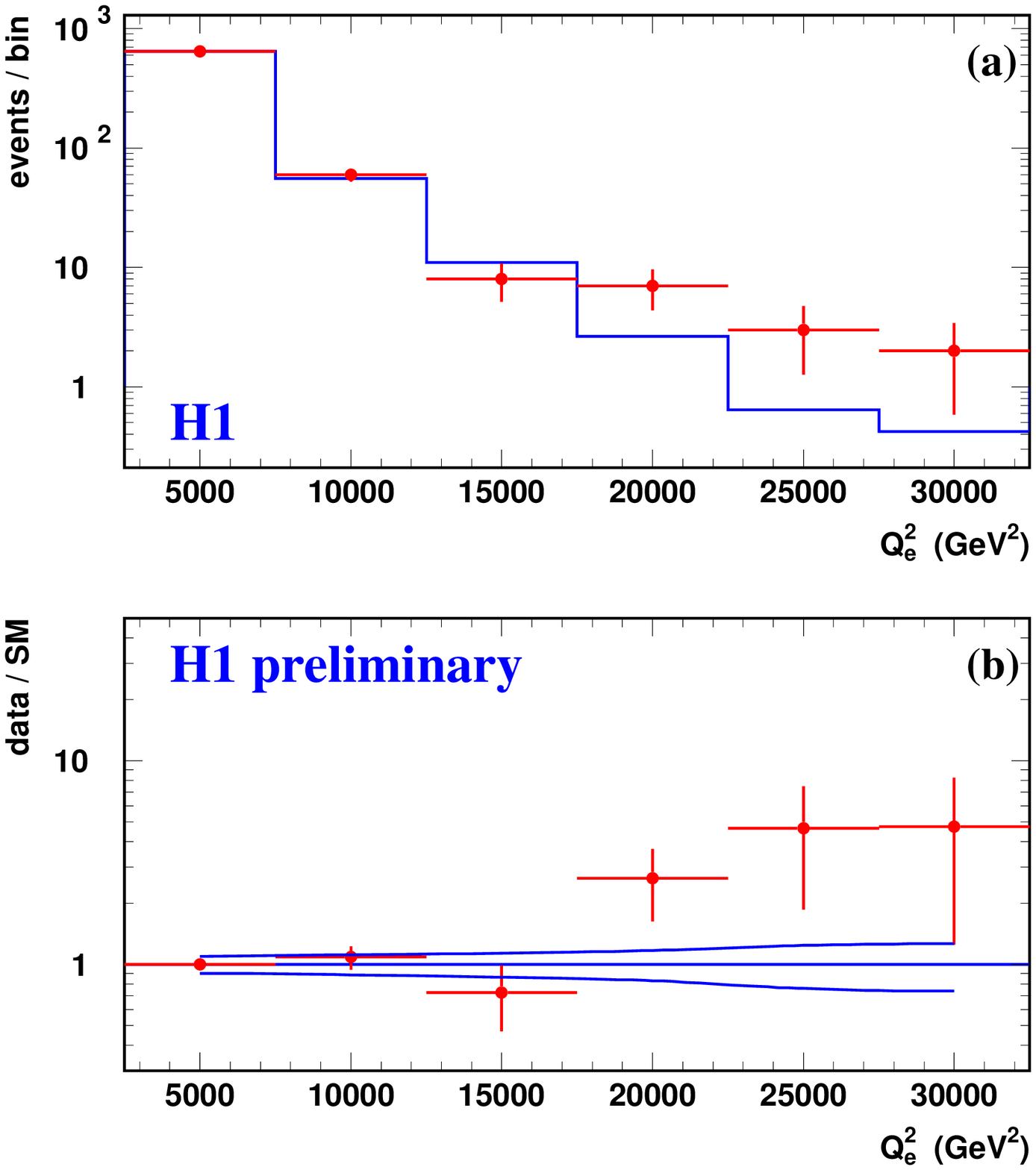} {The upper plot
  shows the $Q^2$ spectrum for the H1 neutral current sample.  The
  points with error bars indicate the data and the histogram shows the
  NC DIS expectations.  The lower plot shows the $Q^2$ spectrum
  divided by the SM expectations.  The smooth curves indicate the
  uncertainty in the expectations.}  {H1_NC_Q2}

\epsfigure[width=0.95\hsize]{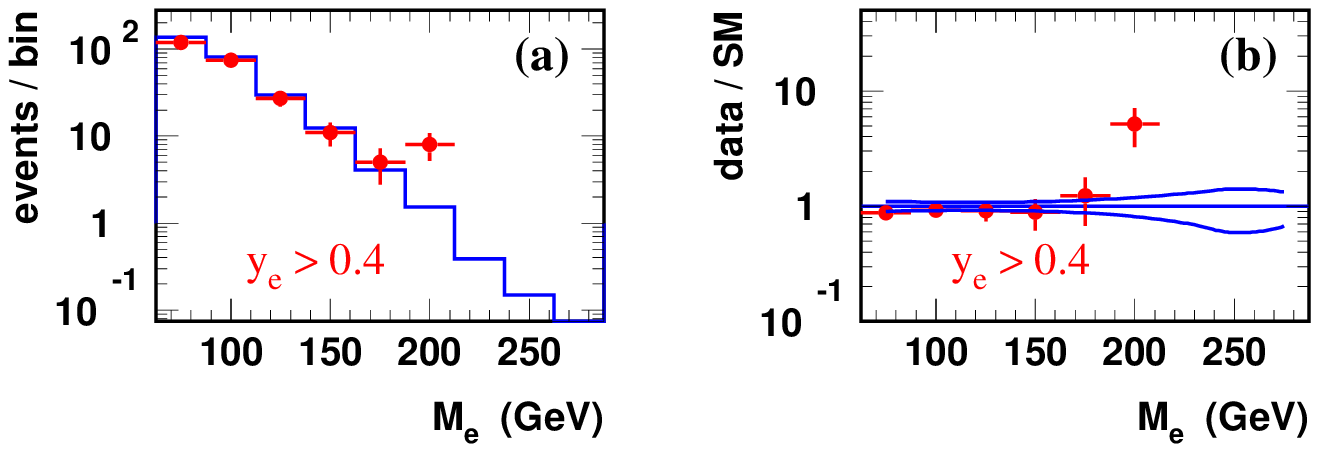} {The upper plot
  shows the $M$ spectrum for $y>0.4$ for the H1 neutral current
  sample. The points with error bars indicate the data and the
  histogram shows the NC DIS expectations. In the lower plot the
  points give the ratio of the number of events observed to the NC DIS
  expectations and the smooth curves give the uncertainty in the
  expectations.}  {H1_NC_M}

\noindent
The combined H1
data sample show a clear excess of observed events over expectations
starting at $Q^2>15000$~GeV$^2$.  For $Q^2>15000$ GeV$^2$, 18 events
are observed where $8.0\pm1.2$ are expected, corresponding to a
probability of $0.34~\%$.  The excess of events near $M=200$~GeV
remains prominent.  For 187.5 GeV $<M<$ 212.5 GeV, 8 events are
observed while $1.53\pm0.29$ are expected, corresponding to a
probability of $3.3\times10^{-4}$. The probability to see an excess
this significant in {\it some} $M$ window is $\sim1\%$.

\epsfigure[width=0.8\hsize]{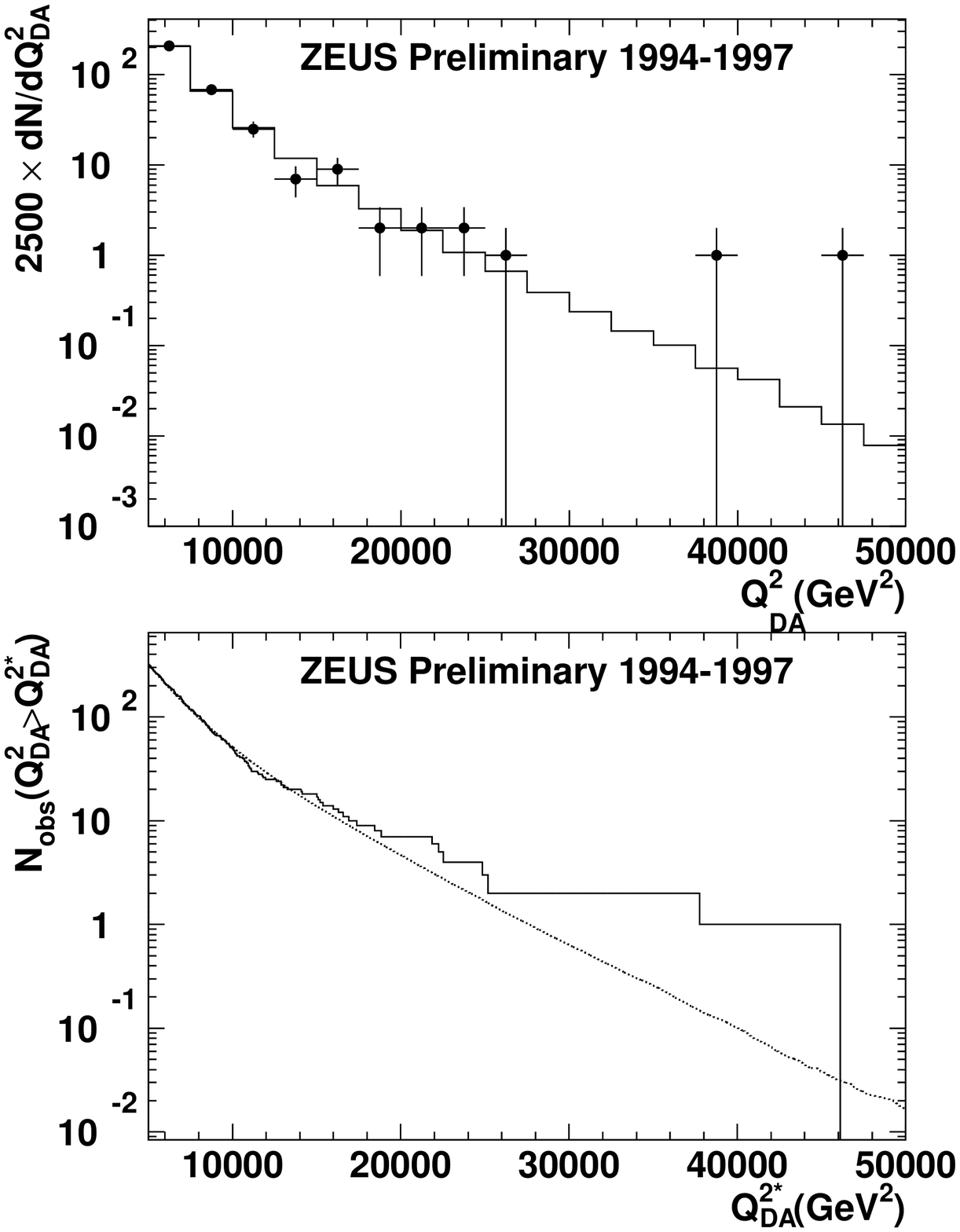} {The upper plot
  shows the $Q^2$ spectrum for the ZEUS neutral current sample.  The
  points with error bars indicate the data and the histogram shows the
  NC DIS expectations.  In the lower plot the dotted curve and the
  solid curve show respectively the number of events expected and the
  number observed with $Q^2>Q^{2*}$ {\it vs.} $Q^{2*}$.}  {ZEUS_NC_Q2}

\epsfigure[width=0.8\hsize]{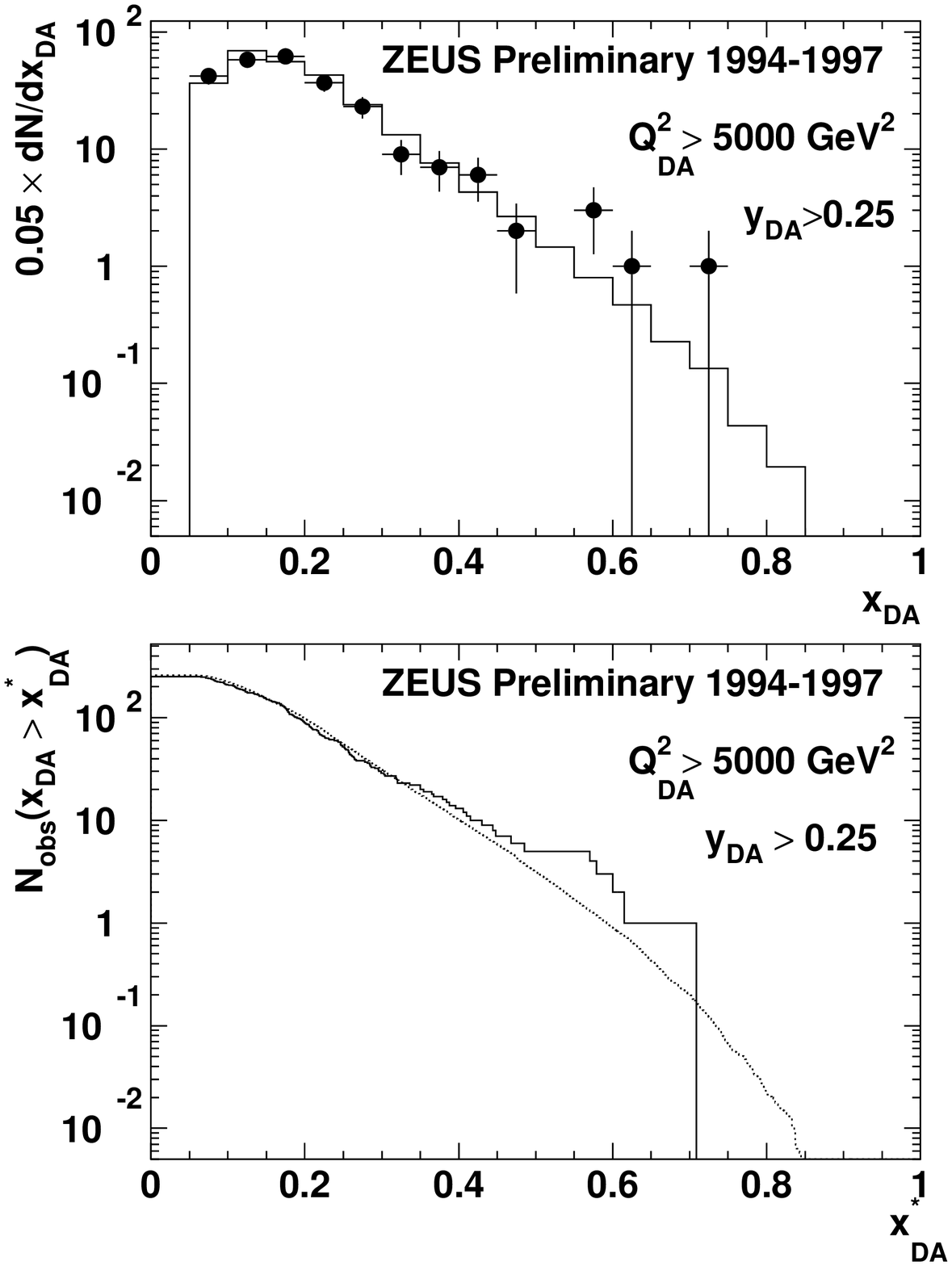} {The upper plot
  shows the $x$ spectrum for $y>0.25$ of the ZEUS neutral current
  sample.  The points with error bars indicate the data and the
  histogram shows the NC DIS expectations.  In the lower plot the
  dotted curve and the solid curve show respectively the number of
  events expected and the number observed with $x>x^*$ {\it vs.}
  $x^*$.}  {ZEUS_NC_X}

\begin{table}
\tablecaption{Number of events observed ($N_{obs}$) by ZEUS
compared to SM NC DIS expectations ($N_{exp}$)
in several kinematic regions.
The first term in each sum gives the 1994-1996 value and the second term
gives the 1997 value as of the Lepton-Photon 97 Symposium.}
\label{tab:ZEUS_NC}
\begin{center}   \begin{tabular}{c|c|c} 
region    & $N_{obs}$ & $N_{exp}$ \\ \hline
$Q^2> 5000$\, GeV$^2$ & 191+135=326 & 197+130=328$\pm$15          \\ \hline
$Q^2>15000$\, GeV$^2$ & 13+5=18     & 8.66+5.77=14.4$\pm$1.1      \\ \hline
$Q^2>35000$\, GeV$^2$ & 2+0=2       & 0.145+0.097=0.242$\pm$0.017 \\ \hline
$x>0.55,y>0.25$    & 4+1=5       & 0.91+0.61=1.51$\pm$0.13     
\end{tabular}   \end{center}
\end{table}

The numbers of events measured by ZEUS in the signal region defined in
the publication based on the 1994-96 publication are summarized in
Table~\ref{tab:ZEUS_NC}. As with H1, the numbers of observed and
expected events are given separately for the 1994-96 running period,
and for the partial 1997 data set. The numbers of events observed by
ZEUS in these kinematic regions are close to the SM expectations.  The
$Q^2$ and $x$ distributions as measured by ZEUS are given in
Fig.~\ref{fig:ZEUS_NC_Q2} and Fig.~\ref{fig:ZEUS_NC_X}.  The ZEUS
measured $Q^2$ spectrum is in reasonable agreement with the SM up to
about $Q^2=30000$~GeV$^2$.  For $Q^2>35000$ GeV$^2$, 2 events are
observed and $0.24\pm0.02$ are expected, corresponding to a
probability of $2.5\%$.  The $x$ spectrum shows an enhancement near
$x=0.55$.  For $x>0.55$ and $y>0.25$, 5 events are observed while
$1.51\pm0.13$ are expected, corresponding to a probability of $1.9\%$.

\paragraph{Comparison of H1 and ZEUS $x(M)$ Spectra}

For $y>0.4$, H1 has observed $8$ events where $1.53\pm0.29$ were
expected in the mass window $187.5<M<212.5$~GeV.  This corresponds to
the $x$ range $0.389<x<0.500$.  In this range, ZEUS observes $3$
events where $2.92\pm0.24$ events are expected.  In the region
$x>0.55,y>0.25$, ZEUS observes $5$ events where $1.51\pm0.13$ are
expected.  In this region, corresponding to $M>223$~GeV, H1 observes
one event, where $0.75\pm0.30$ are expected.  I.e., combining the H1
and ZEUS results does not increase the significance of either effect.
A study of the possible systematic shifts in the reconstruction has
led to the conclusion that the ZEUS and H1 events are unlikely to come
from a single narrow resonance in $x(M)$~\cite{ref:Bruce}.

\subsubsection{Conclusion}

The full 1994-97 data has recently become available.  No further
evidence was seen for either an excess near
$200$~GeV~\cite{ref:H1_LQ_5,ref:ZEUS_LQ2} nor for an enhanced cross
section at high
$Q^2$~\cite{ref:H1_NCCChighQ2,ref:ZEUS_CChighQ2,ref:ZEUS_NChighQ2}.
Both H1 and ZEUS have therefore proceeded to set limits on leptoquark
and R-parity violating squark production, as reported in previous
sections.  The NC and CC cross sections have been discussed in
section~\ref{sec:high_Q2}.

The 1997 data sets did not increase the significance of the results
based on the 1994-1996 data.  However, there is a general tendency for
the HERA data to lie above Standard Model expectations at large $Q^2$.
This is true for the charged current as well as the neutral current
events.  The probability to observe the combined counting rates are a
few~\% in each case.  It is clear that the confirmation or ruling out
of possible new physics will need substantially increased
luminosities.  The future HERA running program, outlined below, should
provide the data necessary for this.

\section{Outlook for HERA}
\label{sec:future}
\subsection{The HERA luminosity upgrade}
The luminosity of HERA has reached instantaneous values of
$L \simeq 1.4\cdot 10^{31}$~cm$^{-2}$s$^{-1}$, which is close to
the design value.
The progress over the years has been steady, as can be seen in
Fig.~\ref{fig:HERA_lumi}.  This luminosity has allowed the HERA 
experiments to perform many important measurements, including the 
observation of the rise of the structure function $F_2$ with decreasing
$x$, the observation and study of events with a large rapidity gap
in the hadronic final state, the observation of resolved photon interactions,
and many others.  It must however be noted that most of the physics done 
so far is `small $Q^2$ physics'
compared with the electroweak scale of $Q^2 = m_W^2$. The
physics on the electroweak scale has so far been barely touched.
This is the region HERA was built to explore. The unique features
of HERA, such as electron polarization and the capability to compare
electron and positron 
interactions
are relevant only in the electro-weak regime.
A luminosity upgrade bringing integrated
luminosities to the level of $1$~fb$^{-1}$ by 2005 would open the possibility 
for many more exciting measurements.  The excess of events observed
by both the ZEUS and H1 experiments at large $x$ and $Q^2$ has put even
more priority on this luminosity upgrade.

Given this strong motivation, the HERA machine group, in conjunction with
members of the ZEUS and H1 experiment, have  studied different 
alternatives to achieve the desired luminosity upgrade. 
The solution chosen is to install 
superconducting combined function (focusing and bending) magnets
in the interaction regions and
to rebuild focusing magnets near the interaction regions, thereby
significantly reducing the beam cross sections.  Instantaneous luminosities
of $7.4\cdot 10^{31}$~cm$^{-2}$~s$^{-1}$ are expected in this configuration,
leading to delivered integrated luminosities of about $150$~pb$^{-1}$ per 
year per experiment.

  In addition to the luminosity upgrade machine elements, spin rotators will 
also be installed as planned. This will allow data taking with different 
electron (or positron) longitudinal polarization. 

\subsection{Future physics at HERA}

\subsubsection{Excess at large {$\boldmath x, Q^2$}}

The ZEUS and H1 experiments both reported an excess of large-$x, Q^2$
events as described in section~\ref{sec:excess}. The combined ZEUS and H1
cross section for $Q^2>15000$~GeV$^2$ reported at the LP97 conference
\cite{ref:Bruce} is $0.71^{+0.14}_{-0.12}$~pb, compared
to the Standard Model expectation of $0.49$~pb.  For $Q^2>30000$~GeV$^2$,
the combined cross section is $0.098^{+0.059}_{-0.042}$~pb, compared
to the Standard Model expectation of $0.023$~pb. 

\epsfigure[width=0.8\hsize]{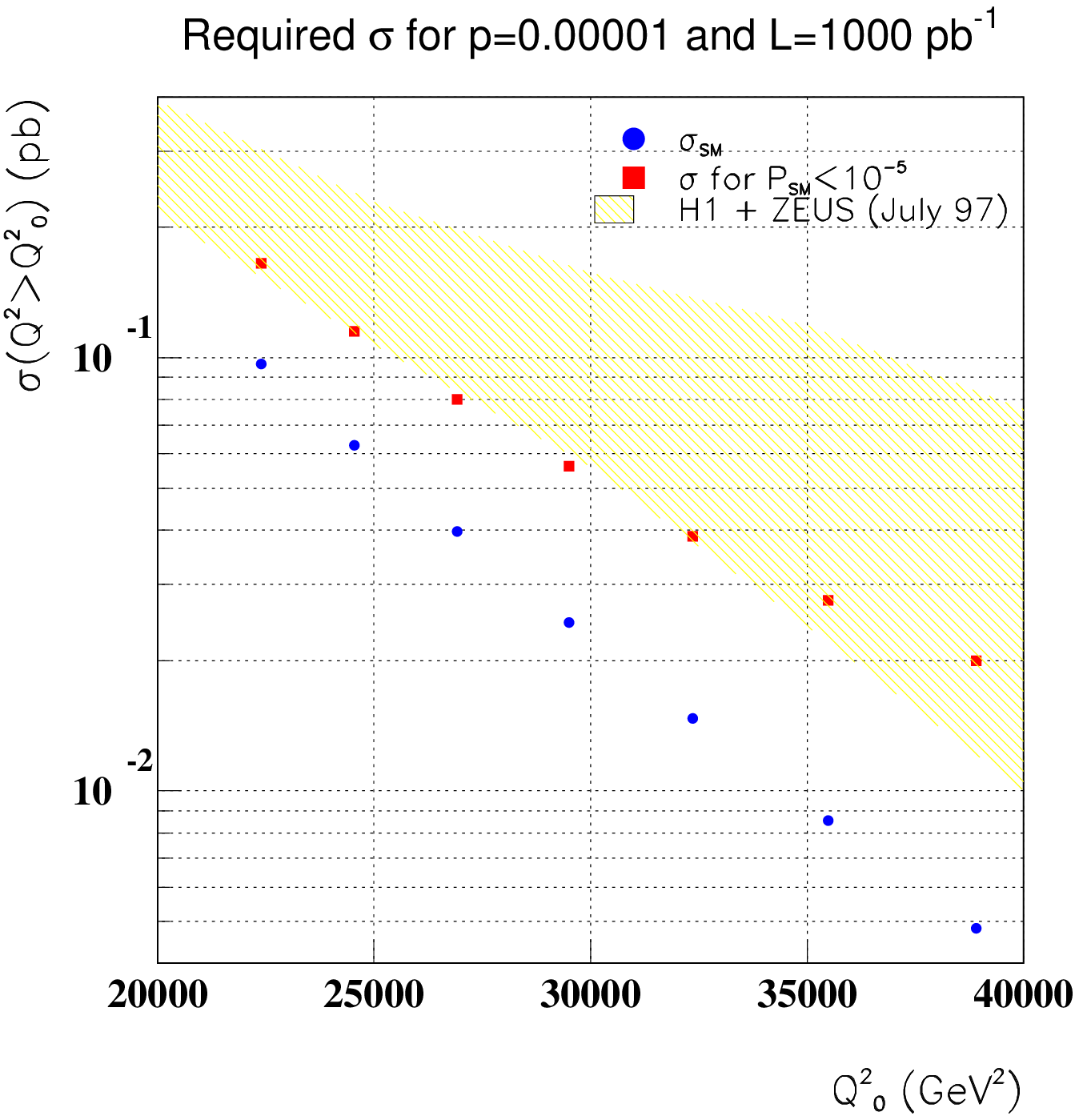} {The cross section
  necessary to produce a deviation from the Standard Model at the
  $10^{-5}$ level is plotted, and compared to the Standard Model cross
  section.  Note that the integrated cross section above $Q^2_0$ is
  shown.  These cross sections are compared to the current (LP97
  status) cross sections measured by ZEUS and H1.}  {discover}

If the excess at large $Q^2$ is confirmed, it signals new physics
beyond the Standard Model. It will be very difficult to confirm the
effect with the present HERA luminosity.  However, with an accumulated
$1$~fb$^{-1}$ of luminosity, a significant deviation from Standard
Model expectations will be measured if the current cross section is
close to the true value.  This is shown in Fig.~\ref{fig:discover},
where the cross section required for a significant deviation from the
Standard Model (probability for a fluctuation less than $10^{-5}$,
including an estimate of the systematic errors on the cross section
measurements) is compared to both the Standard Model expectations and
the cross section measurements from ZEUS and H1.  As can be seen from
this figure, the present cross section levels would result in a very
significant effect.

If the central value of the cross section currently measured by ZEUS and H1
were accurate, then a sample of order $500$ events will be 
available per experiment for $Q^2>15000$~GeV$^2$, allowing the study of a new
field of physics.

The luminosity increase can be compared to an increase of the beam energies.
For example, a luminosity increase
by a factor 4 is roughly equivalent to an increase of the proton
beam energy from 820 GeV to 1160 GeV.  This type of energy increase
of the machine is prohibitively expensive.  The planned increase in
luminosity is therefore the best way to access new physics at HERA.
The proton beam energy is expected to increase 
to a value above 900\,GeV, which can be
done with minimum impact on the current machine.  In fact, it is
hoped that this higher energy running will already take place in 
1998~\footnote{HERA is running with $E_p=920$~GeV since August 1998.}.
This improvement
will add to the physics capability of the machine. As an
example the production cross section of 240\,GeV leptoquarks is increased
by a factor of 2 as the proton beam energy is increased from 820\,GeV
to 920\,GeV.  Increasing the electron (positron) beam energy has also been
discussed but is determined to be too expensive.

\subsubsection{Physics reach with the luminosity upgrade}

In the following, we review some of the conclusions of
the workshop `Future Physics at HERA' (ed. Ingelman, De Roeck and Klanner).
The workshop consisted of several working groups: Structure Functions,
Electroweak Physics, Beyond the Standard Model, Heavy Quark Production and
Decay, Jets and High-$E_T$ Phenomena, Diffractive Hard Scattering, Polarized
Protons and Electrons, Light and Heavy Nuclei, and a group studying the HERA
upgrade.  The results are summarized in a 1200 page report.  
It is obviously
not possible to do justice to the full range of physics topics covered in
this document.  A few examples are chosen to give an indication of the
physics reach achievable with the upgraded luminosity.

\epsfigure[width=0.8\hsize]{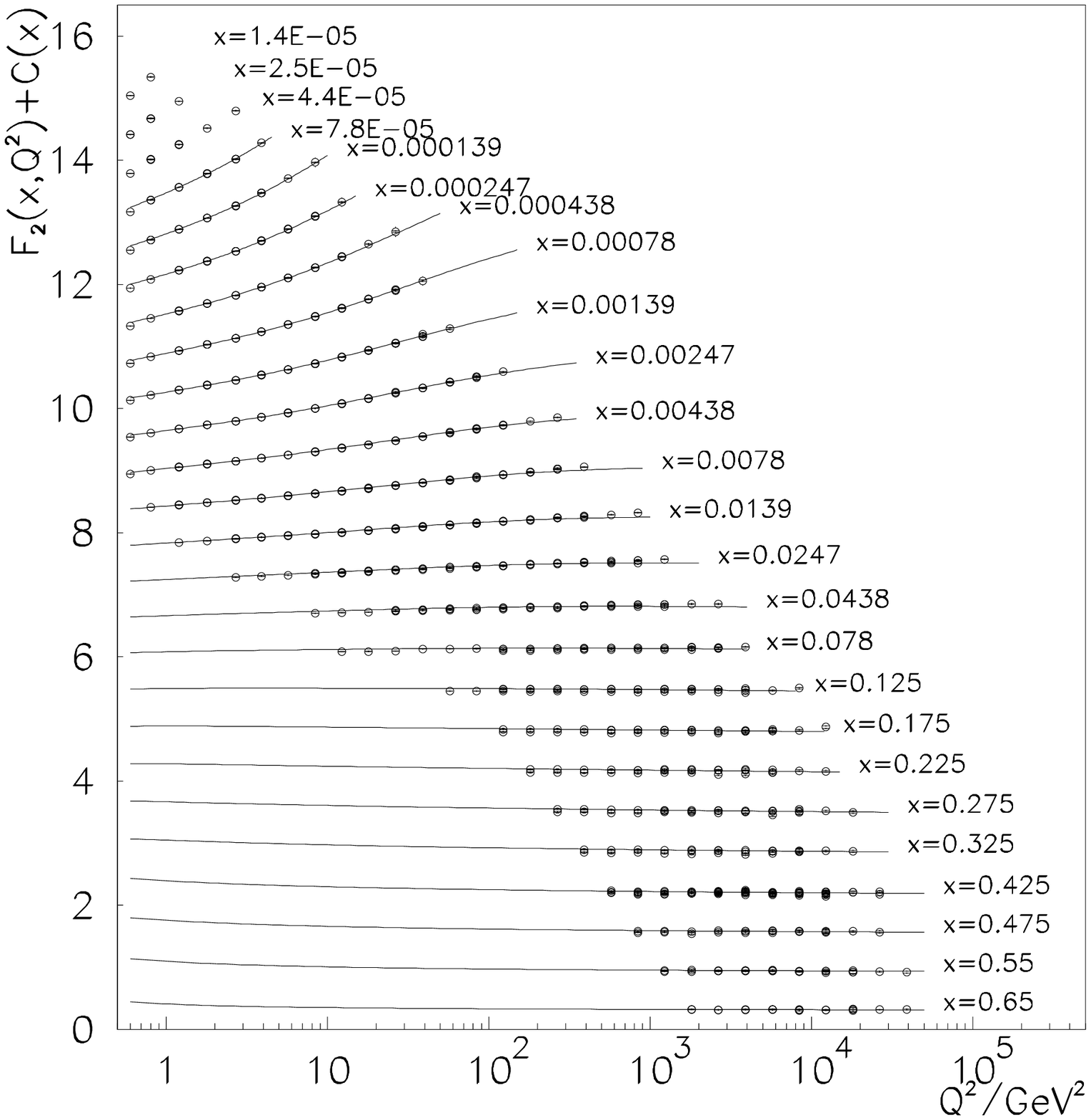} {Simulated structure
  function data sets. The luminosity of 1~fb$^{-1}$ will lead to
  precise data even at very large $Q^{2}$. For $Q^{2} \geq
  10000$~GeV$^{2}$ about 2000 events will be available.  The curve
  represents a NLO QCD fit. The large $x$, small $Q^{2}$ region can
  not be accessed with HERA but is almost completely covered by the
  fixed target experiment data, not shown here.}  {F2_future}

\epsfigure[width=0.8\hsize]{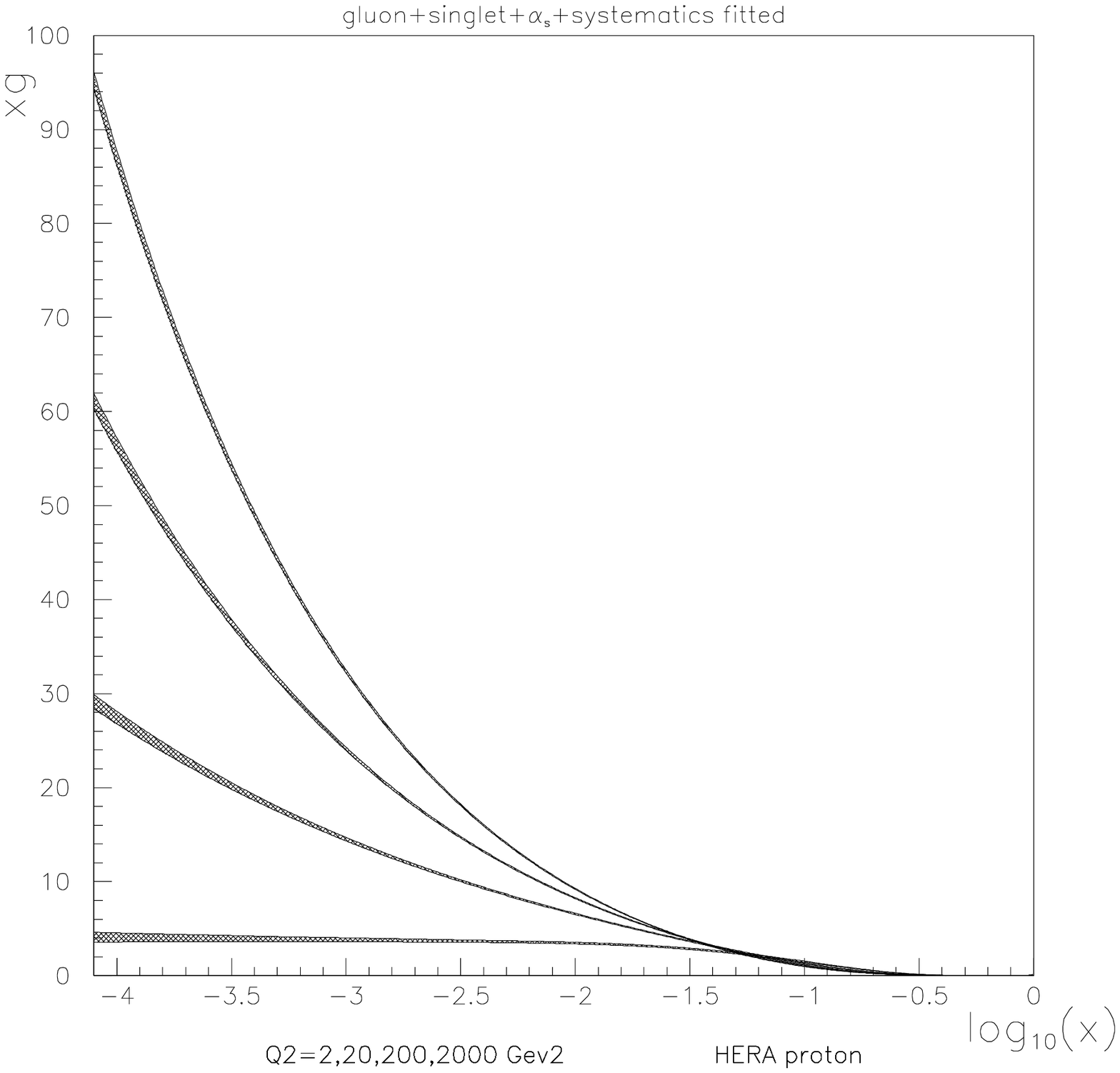} {Determination of
  gluon distribution using future $F_{2}$ data from electron-proton
  scattering.  Note that for simplicity the gluon is shown also
  outside the allowed region of $x \leq Q^{2}(\gevtwo)/10^{5}$.}
{gluon_future}

The exploration of the proton structure and QCD can be continued with
much larger data sets and better tools. The structure function $F_2$
will be measured to higher precision, over an extended kinematic
region, allowing an extraction of the gluon density with an
experimental precision of $1$~\%.  An example of what can be achieved
in the measurement of $F_2$ is shown in Fig.~\ref{fig:F2_future}.  The
corresponding gluon density is shown in Fig.~\ref{fig:gluon_future}.
These very precise measurements will provide stringent tests of the
QCD evolution equations.  They will also provide important
experimental input for cross section calculations for future
accelerators utilizing hadrons. Experimental precision levels of
$0.001$ on $\as(m_Z^2)$ will be possible by combining HERA structure
function data with fixed target data.

The ability to run HERA with both electrons and positrons will allow
an extraction of $F_3$, and therefore will give a measurement of the
valence quark distributions. If HERA is also run with deuterons, then the 
experiments can make direct measurements of the (up - down) 
quark distribution, which has so far not been measured.  Other important
QCD tests will consist of precision measurements of charm and beauty
production.
The study of diffraction will be significantly enhanced with increased
statistics.  The luminosity upgrade will allow the study of diffractive
processes, such as exclusive vector meson production, at large $Q^2$
where perturbative QCD calculations are applicable.  The study of jets
will be performed at high $E_T$, resulting in measurements of $\as$
at different 
scales.  Sensitive searches for novel phenomena predicted in QCD such as 
instanton production will also be performed.  It is clear that a high
luminosity HERA will provide precise measurements of many cross sections
which should be calculable in QCD, thereby providing stringent tests
of our theoretical understanding of the strongest force in nature.

\epsfigure[width=0.95\hsize]{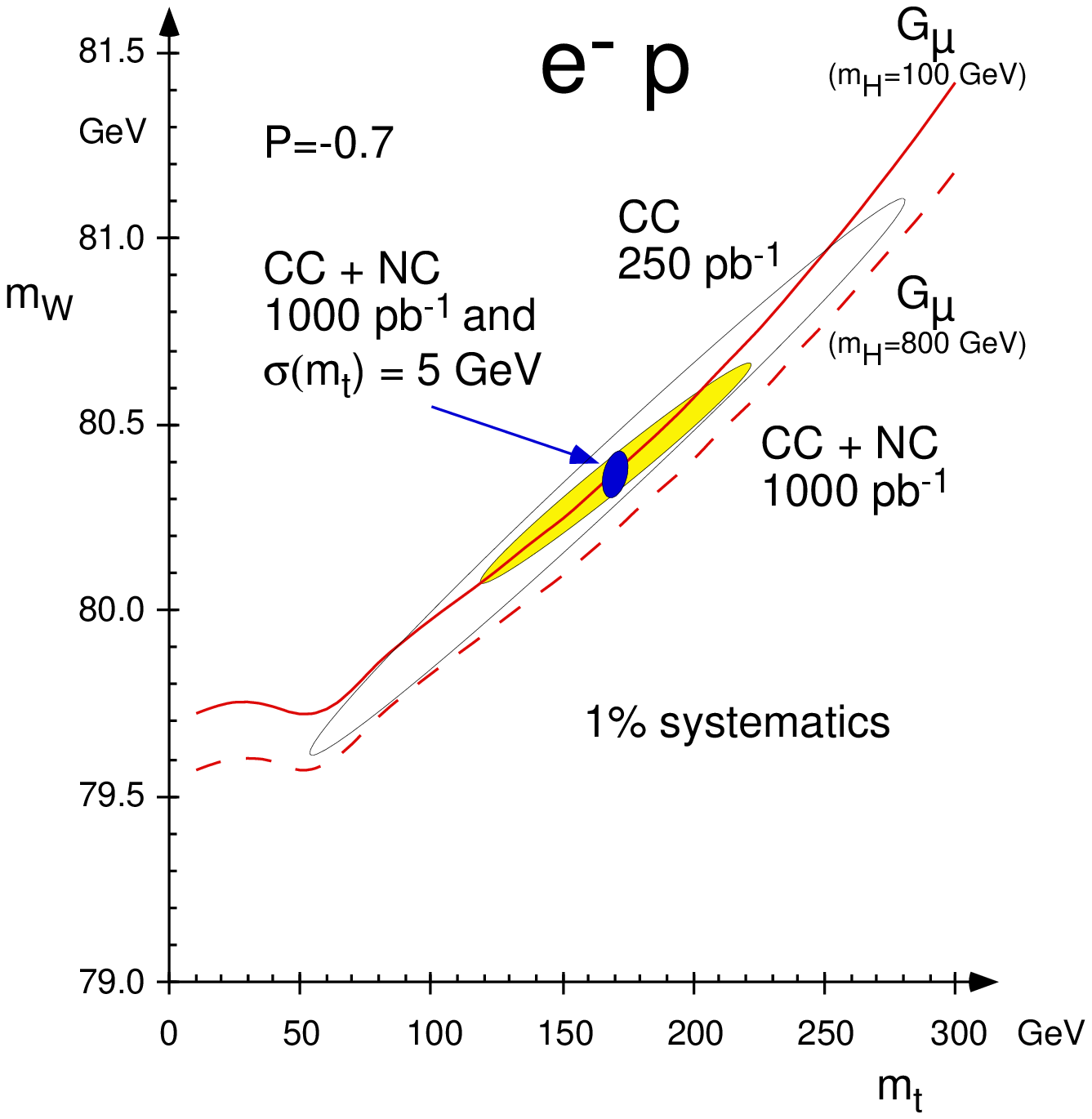} {The
  1$\sigma$-confidence contours in the $(m_W,m_t)$ plane from
  polarized electron scattering (P=-0.7), utilizing charged current
  scattering at HERA only with an integrated luminosity of
  $250$~pb$^{-1}$ (large ellipse), neutral and charged current
  scattering with $1000$~pb$^{-1}$ (shaded ellipse), and the
  combination of the latter with a direct top mass measurement with a
  precision $\sigma_{m_t}=5$~GeV (full ellipse).  The $m_W$-$m_t$
  relation from the $G_{\mu}$ constraint is also shown for two values
  of $m_H$.}  {EWtests}

Sensitive tests of the electroweak sector of the Standard Model will
be possible, which are complementary to those at LEP and the Tevatron.
As an example, the constraints imposed on the Standard Model by the
HERA data can be translated into an effective measurement of the W
mass.  As is shown in Fig.~\ref{fig:EWtests}, the $W$ mass will be
probed at the level of $50$~MeV given a precision on the top quark
mass of $5$~GeV.

In addition, HERA will be competitive in tests of anomalous $WW\gamma$ and
$Z\gamma\gamma$ couplings.  The neutral current couplings of the
light quarks can also be untangled from comparisons of 
neutral current and charged current cross sections with different beam
polarizations.

Many different possibilities for new physics, including  contact interactions,
compositeness, lepton flavor violation,
heavy neutral leptons, supersymmetry and other new particles have been
studied.  It was determined that HERA would produce
the most sensitive searches for many of these processes.
Examples of new physics for which HERA is well-suited are R-parity
violating supersymmetry, lepton flavor violation, and the production of 
excited fermions or heavy leptons.  HERA already sets the best limits in
many instances for these searches, and the large luminosity increase would
further enhance the sensitivity.

The luminosity upgrade clearly gives HERA a bright future, which should
extend at least until the middle of the next decade.

\clearpage

\section*{Acknowledgements}

We would like to thank all our colleagues from the ZEUS and H1
collaborations who have assisted us in preparing this report.  In
particular, we are grateful to Aharon Levy, Bruce Straub, and Rik
Yoshida for locating errors in different sections of this report.
Sabine Lammers kindly assisted us in the difficult task of getting the
references right.  All remaining errors and omissions are of course
the responsibility of the authors.

Both of us would like to thank the Alexander von Humboldt Foundation
for supporting our work at DESY over the years.  The participation of
Halina Abramowicz in the scientific program of HERA was partly
supported by the Israel Science Foundation, the German-Israeli
Foundation, the US-Israel Binational Science Foundation, the Minerva
Foundation and the Israel Ministry of Science.  The participation of
Allen Caldwell was partly supported by the US National Science
Foundation.

One of us, Allen Caldwell, would like to especially thank Juliane and Emma, for
their tolerance of the extra hours at the office needed to write this
report.

Halina Abramowicz dedicates this report to the memory of Professor
Judah Eisenberg of Tel Aviv University, a mentor and a friend, whose
invaluable support in research and beyond will be dearly missed.

\bibliographystyle{rmp}
\bibliography{FINAL}

\end{document}